\documentclass[
headinclude=true, 
twoside=true,
BCOR=5mm, 
    fontsize=11pt,
paper=21cm:24cm, 
pagesize, 
DIV=13, 
headings=normal, 
appendixprefix=true,
numbers=noendperiod, 
toc=bibliography, 
parskip=false, 
version=last 
]{scrbook}

\makeatletter
\renewcommand{\@chapapp}{}

\makeatother

\usepackage{graphicx,ragged2e}
\usepackage{mathtools}
\usepackage{longtable}
\usepackage{makeidx}
\usepackage{amsmath}
\usepackage{amssymb}
\usepackage{setspace}
\usepackage{epsfig}
\usepackage[percent]{overpic}
\usepackage{subcaption}
\usepackage{hyperref}

\usepackage{color}

\usepackage{mathrsfs}

\usepackage[toc]{appendix}

\def\[{\left [}
\def\]{\right ]}
\def\({\left (}
\def\){\right )}

\def\r{\rho}

\def\r2{\sqrt{2}}

\newcommand{\bea}{\begin{eqnarray}}
\newcommand{\eea}{\end{eqnarray}}

\newcommand{\beq} {\begin{equation}}
\newcommand{\eeq} {\end{equation}}
\newcommand{\beqa} {\begin{eqnarray}}
\newcommand{\eeqa} {\end{eqnarray}}

\newcommand{\beqn}{\begin{eqnarray}}
\newcommand{\eeqn}{\end{eqnarray}}

\begin{document}


\thispagestyle{empty}
\vspace*{5cm}
{\centering
{\huge  \textbf{Stochastic Mechanics Without \\ Ad Hoc Quantization:\\ Theory And Applications To \\ \vspace{0.3cm}Semiclassical Gravity}}\vspace{1.15cm}\\
}

\newpage
\thispagestyle{empty}
{\centering
{\huge \textbf{Stochastic Mechanics Without \\ Ad Hoc Quantization:\\ Theory And Applications To \\ \vspace{0.3cm}Semiclassical Gravity}}\vspace{1.15cm}\\

{\LARGE \textbf{Stochastische mechanica zonder \\ ad hoc quantisatie:\\ theorie en toepassingen op \\ \vspace{0.3cm} semi-klassieke zwaartekracht}}\vspace{0.45cm}\\

 (met een samenvatting in het Nederlands)\vspace{1cm}\\
 
{\Large \textbf{Proefschrift}}\vspace{0.4cm}\\
\par} 
{\large{\centering{ \noindent ter verkrijging van de graad van doctor\\ aan de Universiteit Utrecht\\ op gezag van de rector magnificus, prof. dr. G. J. van der Zwaan,\\ ingevolge het besluit van het college voor promoties\\ in het openbaar te verdedigen\\ op donderdag 19 oktober 2017 des middags te 12.45 uur\\}}}\vspace{0.5cm}
{\centering
{\large door}\vspace{0.5cm}\\
{\Large \textbf{Maaneli Derakhshani}}\vspace{1cm}\\
{\large geboren op 3 oktober 1985\\ te Rancho Palos Verdes, Verenigde Staten van Amerika}\par}
 \newpage \thispagestyle{empty}
 Promotoren: Prof. dr. R. B. Mann 
 
 \indent \indent \indent \indent \indent \indent \indent Prof. dr. L. T. G. Theunissen 
 
 \indent Copromotor: Dr. G. Bacciagaluppi
 \vfill





%
\frontmatter


%
%
%
%





\chapter*{Papers and Publications}

\begin{itemize}
\item \textbf{Chapter 2}\\
 A Suggested Answer To Wallstrom's Criticism: Zitterbewegung Stochastic
Mechanics I\\
 Maaneli Derakhshani\\
 \textit{\emph{https://arxiv.org/abs/1510.06391}} 
\item \textbf{Chapter 3}\\
 A Suggested Answer To Wallstrom's Criticism: Zitterbewegung Stochastic
Mechanics II\\
 Maaneli Derakhshani\\
\emph{ }\textit{\emph{https://arxiv.org/abs/1607.08838}}
\item \textbf{Chapter 4 }\\
 Semiclassical Newtonian Field Theories Based On Stochastic Mechanics
I\\
 Maaneli Derakhshani\\
\emph{ }\textit{\emph{https://arxiv.org/abs/1701.06893}}
\item \textbf{Chapter 5 }\\
 Semiclassical Newtonian Field Theories Based On Stochastic Mechanics
II\\
 Maaneli Derakhshani\\
\emph{ }\textit{\emph{https://arxiv.org/abs/1702.02472}}
\end{itemize}
Work which is not directly part of this dissertation (but is referenced
throughout):
\begin{itemize}
\item Probing Gravitational Cat States In Canonical Quantum Theory vs. Objective
Collapse Theories\\
 Maaneli Derakhshani\\
\emph{ }https://arxiv.org/abs/1609.01711
\item Probing a Gravitational Cat State: Experimental Possibilities\\
 Maaneli Derakhshani\\
\emph{ Journal of Physics: Conference Series, Volume 1, conference
1 (2016)}\\
\emph{ }https://arxiv.org/abs/1603.04430
\item Newtonian Semiclassical Gravity in the Ghirardi-Rimini-Weber Theory
with Matter Density Ontology\\
 Maaneli Derakhshani\\
\emph{ Physics Letters A, Volume 378, Issue 14-15 (2014)}\\
\emph{ }https://arxiv.org/abs/1304.0471
\item The Newtonian Limit Of Stochastic Gravity Doesn't Entail The Many-Body
Stochastic Schr\"{o}dinger -Newton Equations Or Gravitationally-Induced
Wavefunction Collapse \\
 Maaneli Derakhshani\\
\emph{ (In preparation)}
\item On Multi-Time Correlations And Effective Collapse In Stochastic Mechanics\\
 Maaneli Derakhshani and Guido Bacciagaluppi\\
\emph{ (In preparation)}
\end{itemize}
\tableofcontents{}

\chapter*{Preface}

This thesis is the culmination of a nearly decade-long journey. The journey
started when I was a sophomore undergraduate at Stony Brook University
(SBU), excitedly learning about Nelson's stochastic mechanics and
the ``Wallstrom criticism'' thereof from the papers of Dr. Guido
Bacciagaluppi, and from discussions with Prof. Sheldon (Shelly) Goldstein
at Rutgers University. Having found stochastic mechanics a compelling
way to make sense of quantum mechanics, I was determined to find an
answer to the criticism. I explored various possibilities to no avail
until, one day, out of sheer luck, I stumbled upon a dusty, yellow-paged
book entitled ``Observation and Interpretation: A Symposium of Philosophers
and Physicists'', in the bookcase of Dr. John No\'{e}, the Director of
the Laser Teaching Center at SBU. (Thank you for the book, John.) 

This book, published in 1957, had among its contributions a little-known
paper by David Bohm entitled ``A proposed explanation of quantum
theory in terms of hidden variables at a sub-quantum-mechanical level''.
In this paper, Bohm sketches a model - which he credits to Louis de
Broglie (in fact, Bohm's model was just a slight reformulation of
an idea suggested by de Broglie in the latter's Ph.D thesis, an idea
which also happens to be the precursor to the de Broglie-Bohm pilot-wave
theory) - of an elementary particle as a localized periodic process
of fixed frequency in a hypothetical sub-quantum medium. Invoking
the Lorentz transformation of special relativity, Bohm shows how the
model recovers a quantization condition of Bohr-Sommerfeld type for
the phase of the periodic process, and how this quantization condition
is related to the single-valuedness condition on wavefunctions in
quantum mechanics. In fact, the quantization condition obtained in
the model of de Broglie and Bohm is precisely what's needed for stochastic
mechanical theories to recover the Schr\"{o}dinger equation
of non-relativistic quantum mechanics (as I will explain in the thesis).
And it was precisely the lack of justification for this quantization
condition in stochastic mechanics that Wallstrom emphasized and criticized
in the late 80's and early 90's. So it became clear to me how to answer
the criticism - reformulate Nelson's stochastic mechanics so as to
consistently incorporate the model of de Broglie and Bohm.

Throughout my circuitous path through graduate school, I persistently
worked on this problem, hoping to base my Ph.D thesis on
it. Eventually I would find the opportunity to do so under the supervision
of Guido (mentioned above), Prof. Robb Mann (at the University of
Waterloo), Prof. Bert Theunissen (at Utrecht University), and the
flagship of Utrecht University. Along the way, I extended the
reformulated stochastic mechanics to the domain of semiclassical Newtonian
gravity, finding that it has advantages over other approaches to semiclassical Newtonian gravity (e.g., the Schr\"{o}dinger-Newton equation). The result is the thesis before you. \\
 \\
\\
 Maaneli (Max) Derakhshani\\
 Utrecht, \today.\\

\mainmatter 

\chapter*{Acknowledgements}

So many people to thank. \textbf{Guido Bacciagaluppi}, for taking a chance on me as his first Ph.D candidate at Utrecht University, for always being a sounding board to my thoughts and ideas, for sharpening my thinking and writing, for helping me learn more about the foundations of physics than I could have on my own, for his open-mindedness, and just generally being one of the most kind and fair individuals I have ever known. \textbf{Bert Theunissen}, for his immense help with all things administrative, and his general enthusiasm and hospitality at the Director of the Descartes Centre. \textbf{Dennis Dieks}, for many enjoyable discussions about the foundations of physics, for helpful critical feedback on my work, for encouragement, for advice on funding possibilities in The Netherlands, and for his help in general with the Ph.D process (including being part of my thesis committee). \textbf{F. A. (Fred) Muller}, for his enthusiasm for foundations of physics, for his hospitality, and for enjoyable discussions. \textbf{Robb Mann}, for taking a chance on me when I approached him about being my co-supervisor many years ago, for hosting me one month in his group at the University of Waterloo in the summer of 2014, for inviting me up to his cottage for a weekend with his other students, for allowing me to write an ambitious grant proposal with him, for being a sounding board to my thoughts and ideas, and for his inspirational breadth of physics knowledge. \textbf{Bei-Lok Hu}, for taking a chance on me many years ago when I was an isolated physics graduate student, for his consistent and enthusiastic support of my career, for writing a paper with me despite not being a student at his academic institution, for being a sounding board for my thoughts and ideas, for integrating me into his network of colleagues (from whom I have learned an immense amount of physics), for inspiring me with his physics work from the time when I was an undergraduate, and for taking time out of his very busy schedule to be on my thesis committee. The other members of my thesis committee - \textbf{Ward Struyve}, \textbf{Nino Zangh\`{i}}, and \textbf{Harvey Brown}, thank you for taking the time to do this. I also want to thank \textbf{Ward} for many helpful discussions over the years about physics and career moves, for inspiring me with his work and persistence since the time I was an undergraduate, for his admirable patience and generosity, and for being a most reliable friend and colleague. 

My office mates, \textbf{Fedde Benedictus}, \textbf{Ivan Flis}, \textbf{Noortje Jacobs}, \textbf{Jesper Oldenberger}, and \textbf{Steven Van Der Laan}, for many enjoyable discussions about science, philosophy, history, Dutch culture and politics, and American culture and politics. My Utrecht/Amsterdam/Rotterdam colleagues working in the foundations of physics, \textbf{Fedde Benedictus}, \textbf{Sebastian De Haro}, \textbf{Gijs Leegwater}, \textbf{Leon Loveridge}, and \textbf{Ruward Mulder}, for many enjoyable discussions about the foundations of physics. Special thanks to \textbf{Ruward Mulder} for doing the Dutch translation of the thesis summary (samenvatting) in Chapter 9. And my other colleagues in the Descartes Centre HPS Masters program, for providing a supportive, lively, and friendly atmosphere to work in.   

Physics mentors and Professors who I'd like to recognize from \underline{Nebraska}: \textbf{Herman Batelaan}, for taking a chance on me, for his stead-fast moral and career support, and for his infectious enthusiasm for physics and life. \textbf{Brad Shadwick}, for many enjoyable and enlightening discussions about physics, for support of my career, and for teaching me plasma physics. \textbf{Kees Uiterwaal}, for his administrative and moral support. \underline{Clemson}. \textbf{Dieter Hartmann}, for his infectious enthusiasm for physics and astronomy, for his invaluable personal and career advice, for his consistent moral and career support, for allowing me to conduct an unorthodox 'reading course' on quantum field theory when I was a graduate student at Clemson University, and for the kinship we had during my time at Clemson (which was invaluable to me during those years). \textbf{Catalina Marinescu}, for taking a chance on me, for her moral support, and for understanding that I chose a different path. \underline{Stony Brook}: \textbf{Hal Metcalf}, for being one of my inspirations when I was an undergraduate physics student at Stony Brook, for taking a chance on me when I approached him as a bright-eyed freshman eager to do physics research, for his encouragement to pursue my ideas and interests, for facilitating invaluable educational and research opportunities that most undergrads never get, for his moral and career support, and for his understanding that I took a different career path in physics. \textbf{John No\'{e}}, for craftily recruiting me into his 2005 summer group at the Laser Teaching Center, for giving me my first hands-on experience with physics research, for believing I had potential in physics and helping me hone that potential, for being a sounding board to my half-baked ideas during my undergrad years, for allowing me to live out one of my dreams in working on a sonoluminescence apparatus, for training me how to think on my feet with physics, for giving me invaluable experience tutoring and interviewing high school students in the Laser Teaching Center, for allowing me to practically live at the Laser Teaching Center all throughout my undergraduate years, and for the book that eventually led me to write this thesis. \textbf{Fred Goldhaber}, for allowing me to drop in on his office hours (even when I wasn't taking any courses with him) and ask random physics questions for hours on end, for patiently indulging my ill-formed thoughts and ideas at the time, for helping me make one of my dreams a reality by co-creating and co-teaching (along with Bob Crease) a course on the philosophy of quantum mechanics and allowing me to play a role in it, for his moral support and guidance during my undergrad years, and for his support of my career over the years. \textbf{Martin Ro\u{c}ek}, for offering to do reading courses with me on advanced physics topics when I was a Freshman undergrad, and for giving me valuable first-hand exposure to the working mind of a theoretical physicist. \textbf{William Divine Linch III}, for countless impromptu sessions of teaching me advanced physics topics for several hours on end in his office, for endless discussions over beer about life in general, and for being the first example I had seen of a highly skilled theoretical physicist with impressive fashion style. \underline{Miscellaneous}: \textbf{Shelly Goldstein}, for generously tutoring me in the foundations of quantum mechanics, for being a role model and the first example I had seen of someone who could synthesize world-class expertise in physics, philosophy, and mathematics, for generously allowing me to visit him at Rutgers countless times, for integrating me into his network of colleagues, for morally encouraging me when I felt intellectually isolated at my home institution, and for his care and concern on a personal level. \textbf{Rodi Tumulka}, for also teaching me about the foundations of quantum mechanics, for his immense patience and generosity, for his inspiring knowledge of physics, mathematics, and philosophy, and for many enjoyable days visiting Rutgers. \textbf{Michael Kiessling}, for many enjoyable and informative discussions about the foundations of physics. \textbf{Detlef D\"{u}rr}, for the offer long ago to visit LMU and work with him and his students on Wheeler-Feynman electrodynamics and Bohmian mechanics. \textbf{Seth Putterman}, for allowing me an unforgettable 2 month experience working on sonoluminescence-related experiments in his labs at UCLA, and for understanding that I chose a different career path. \textbf{Dan Cole}, for taking a chance on me when I was a high school student, for teaching me about stochastic electrodynamics, for many enjoyable discussions about physics and philosophy, and for allowing me to work with him and his Ph.D student (\textbf{Wilson Zou}) at Boston University for a summer. And many thanks to \textbf{Wilson} for patiently working with me during that summer. \textbf{Michael Ibison}, for many enjoyable physics discussions and consistent support of my career goals. And last but not least, \textbf{Hal Puthoff}, for being the very first physicist to mentor me, for encouraging my interests in fundamental and foundational physics questions, and for inspiring me with his wide-ranging and creative work.  

Other philosophy mentors: \textbf{Tim Maudlin}, for being an inspiration, for being my first exposure to a world-class philosopher of physics, and for coming to Stony Brook to give a talk when I invited him as President of the Stony Brook Math Club (the best talk in the history of the Math Club). \textbf{Patrick Grim}, for countless discussions in his office during my undergrad years at Stony Brook, for moral support, for encouraging my interest in philosophy of physics, and for being my first professor of philosophy. \textbf{Bob Crease}, for encouraging my interest in philosophy of physics, and for co-creating and co-teaching (along with \textbf{Fred Goldhaber}) the philosophy of quantum mechanics course that I wanted so badly. 

Other teachers, colleagues, and friends who've supported me in one way or another over the years, and from whom I've benefitted from physics and philosophy-related discussions. \underline{Nebraska}: Salem Elzway, Eric Jones, Sam Keramati, and Omid Zandi. \underline{Clemson}: Eugen Dumitrescu, Chris Grau, Dhruva Kulkarni, Todd May, and Courtney McGahee. \underline{Stony Brook}: Azure Hansen, Brendan Keller, the various members of the Math Club crew, and Michael Schwartz. \underline{Miscellaneous}: Niayesh Afshordi, Charis Anastopoulos, Johann Baptista, Stanley Brodsky, Jeffrey Bub, Philip Chew, Afshin Goodarzi, Sara Goodarzi, Ed Gruber, Bassam Helou, Eric Dennis, Lajos Di\'{o}si, Jonathan Inbal, William Michael Kallfelz, Owen Maroney, James Mattingly, George Musser Jr, Shapour Neshatfar, Travis Norsen, Huw Price, Alexander Smith, Antoine Tilloy, Steve Weinstein, Hans Westman, Ken Wharton, Howard Wiseman, and many others which time prevents me from mentioning.
 
My neurosurgeons, \textbf{Dr. Stefanie Rifkinson-Mann}, \textbf{Dr. Neil Feldstein}, and \textbf{Dr. Saadi Ghatan}, for life-saving surgeries. 

My parents, \textbf{Saba} and \textbf{Reza}, my brother, \textbf{Kaaran}, and my aunt, \textbf{Fari}. Who always believed in my abilities, never pushed me into a life that I didn't want, and patiently supported me all these years as I pursued my Ph.D. And finally my late uncle, \textbf{Zabi}, who enthusiastically supported my scientific and philosophical interests, and taught me at an early age how to think critically and skeptically about the world.

\chapter*{Quotes}

These quotes have inspired me scientifically and philosophically over
the years.

\section*{Albert Einstein}
\begin{quotation}
I have no special talents. I am only passionately curious. - \emph{Einstein
to Carl Seelig (1952)}

\bigskip{}

It is difficult to believe that this {[}the quantum mechanical{]}
description is complete. It seems to make the world quite nebulous
unless somebody, like a mouse, is looking at it. The problem is to
understand that one can observe the particle with a lantern. - \emph{Einstein,
lecture at the J. A. Wheeler relativity seminar, 1954}

\bigskip{}

I fully agree with you about the significance and educational value
of methodology as well as history and philosophy of science. So many
people today - and even professional scientists - seem to me like
somebody who has seen thousands of trees but has never seen a forest.
A knowledge of the historic and philosophical background gives that
kind of independence from prejudices of his generation from which
most scientists are suffering. This independence created by philosophical
insight is - in my opinion - the mark of distinction between a mere
artisan or specialist and a real seeker after truth. \emph{Einstein
to Thornton, 7 December 1944, EA 61-574}

\bigskip{}

One is struck [by the fact] that the theory (except for the four dimensional
space) introduces two kinds of physical things, i.e.,
(1) measuring rods and clocks, (2) all other things, e.g., the electromagnetic
field, the material point, etc. This, in a certain sense, is
incoherent; strictly speaking measuring rods and clocks would
have to be represented as solutions of the basic equationsÉ, not, as
it were, as theoretically self-sufficient entities.17 [There was the]
obligation, however, of eliminating [this incoherence] at a later
stage of the theory. But one must not legalize the mentioned sin so
far as to imagine that intervals are physical entities of a special
type, essentially different from other physical variables (``reducing
physics to geometry", etc.). \emph{Einstein,
Autobiographical Notes, 1949}

\bigskip{}

How does it happen that a properly endowed natural scientist comes
to concern himself with epistemology? Is there no more valuable work
in his specialty? I hear many of my colleagues saying, and I sense
it from many more, that they feel this way. I cannot share this sentiment.
When I think about the ablest students whom I have encountered in
my teaching, that is, those who distinguish themselves by their independence
of judgment and not merely their quick-wittedness, I can affirm that
they had a vigorous interest in epistemology. They happily began discussions
about the goals and methods of science, and they showed unequivocally,
through their tenacity in defending their views, that the subject
seemed important to them. Indeed, one should not be surprised at this.
\emph{Einstein, ``Ernst Mach'', Physikalische Zeitschrift (1916)}

\bigskip{}

It has often been said, and certainly not without justification, that
the man of science is a poor philosopher. Why then should it not be
the right thing for the physicist to let the philosopher do the philosophizing?
Such might indeed be the right thing to do a time when the physicist
believes he has at his disposal a rigid system of fundamental laws
which are so well established that waves of doubt can't reach them;
but it cannot be right at a time when the very foundations of physics
itself have become problematic as they are now. At a time like the
present, when experience forces us to seek a newer and more solid
foundation, the physicist cannot simply surrender to the philosopher
the critical contemplation of theoretical foundations; for he himself
knows best and feels more surely where the shoe pinches. In looking
for an new foundation, he must try to make clear in his own mind just
how far the concepts which he uses are justified, and are necessities. \emph{Einstein, ``Physics and Reality'' in the Journal of the
Franklin Institute Vol. 221, Issue 3 (March 1936)}

\bigskip{}

Roughly stated the conclusion is this: Within the framework of statistical quantum theory there is no such thing as a complete description of the individual system. More cautiously it might be put as follows: The attempt to conceive the quantum-theoretical description as the complete description of the individual systems leads to unnatural theoretical interpretations, which become immediately unnecessary if one accepts the [p. 672] interpretation that the description refers to ensembles of systems and not to individual systems. In that case the whole ``egg-walking" performed in order to avoid the ``physically real" becomes superfluous. There exists, however, a simple psychological reason for the fact that this most nearly obvious interpretation is being shunned. For if the statistical quantum theory does not pretend to describe the individual system (and its development in time) completely, it appears unavoidable to look elsewhere for a complete description of the individual system in doing so it would be clear from the very beginning that the elements of such a description are not contained within the conceptual scheme of the statistical quantum theory. With this one would admit that, in principle, this scheme could not serve as the basis of theoretical physics. Assuming the success of efforts to accomplish a complete physical description, the statistical quantum theory would, within the framework of future physics, take an approximately analogous position to the statistical mechanics within the framework of classical mechanics. I am rather firmly convinced that the development of theoretical physics will be of this type; but the path will be lengthy and difficult. \emph{Einstein, 
in P. A. Sclipp, Albert-Einstein: Philosopher-Scientist}

\bigskip{}

The reciprocal relationship of epistemology and science is of noteworthy
kind. They are dependent upon each other. Epistemology without contact
with science becomes an empty scheme. Science without epistemology
is - insofar as it is thinkable at all - primitive and muddled. However,
no sooner has the epistemologist, who is seeking a clear system, fought
his way through to such a system, than he is inclined to interpret
the thought-content of science in the sense of his system and to reject
whatever does not fit into his system. The scientist, however, cannot
afford to carry his striving for epistemological systematic that far.
He accepts gratefully the epistemological conceptual analysis; but
the external conditions, which are set for him by the facts of experience,
do not permit him to let himself be too much restricted in the construction
of his conceptual world by the adherence to an epistemological system.
He therefore must appear to the systematic epistemologist as a type
of unscrupulous opportunist: he appears as realist insofar as he seeks
to describe a world independent of the acts of perception; as idealist
insofar as he looks upon the concepts and theories as the free inventions
of the human spirit (not logically derivable from what is empirically
given); as positivist insofar as he considers his concepts and theories
justified only to the extent to which they furnish a logical representation
of relations among sensory experiences. He may even appear as Platonist
or Pythagorean insofar as he considers the viewpoint of logical simplicity
as an indispensable and effective tool of his research. \emph{Einstein,
in P. A. Sclipp, Albert-Einstein: Philosopher-Scientist}
\end{quotation}

\section*{John Stewart Bell}
\begin{quotation}
It would seem that the theory {[}quantum mechanics{]} is exclusively
concerned about ``results of measurement'', and has nothing to say
about anything else. What exactly qualifies some physical systems
to play the role of ``measurer''? Was the wavefunction of the world
waiting to jump for thousands of millions of years until a single-celled
living creature appeared? Or did it have to wait a little longer,
for some better qualified system ... with a Ph.D.? If the theory is
to apply to anything but highly idealized laboratory operations, are
we not obliged to admit that more or less ``measurement-like'' processes
are going on more or less all the time, more or less everywhere. Do
we not have jumping then all the time? The first charge against ``measurement'',
in the fundamental axioms of quantum mechanics, is that it anchors
the shifty split of the world into ``system'' and ``apparatus''.
A second charge is that the word comes loaded with meaning from everyday
life, meaning which is entirely inappropriate in the quantum context.
When it is said that something is ``measured'' it is difficult not
to think of the result as referring to some preexisting property of
the object in question. This is to disregard Bohr's insistence that
in quantum phenomena the apparatus as well as the system is essentially
involved. If it were not so, how could we understand, for example,
that ``measurement'' of a component of ``angular momentum'' ...
in an arbitrarily chosen direction ... yields one of a discrete set
of values? When one forgets the role of the apparatus, as the word
``measurement'' makes all too likely, one despairs of ordinary logic
... hence ``quantum logic''. When one remembers the role of the
apparatus, ordinary logic is just fine. In other contexts, physicists
have been able to take words from ordinary language and use them as
technical terms with no great harm done. Take for example the ``strangeness'',
``charm'', and ``beauty'' of elementary particle physics. No one
is taken in by this ``baby talk''. ... Would that it were so with
``measurement''. But in fact the word has had such a damaging effect
on the discussion, that I think it should now be banned altogether
in quantum mechanics. - \emph{Bell, Against Measurement (1990)} 

\bigskip{}

Is it not clear from the smallness of the scintillation on the screen
that we have to do with a particle? And is it not clear, from the
diffraction and interference patterns, that the motion of the particle
is directed by a wave? De Broglie showed in detail how the motion
of a particle, passing through just one of two holes in screen, could
be influenced by waves propagating through both holes. And so influenced
that the particle does not go where the waves cancel out, but is attracted
to where they cooperate. This idea seems to me so natural and simple,
to resolve the wave-particle dilemma in such a clear and ordinary
way, that it is a great mystery to me that it was so generally ignored.
- \emph{Bell, Speakable and Unspeakable in Quantum Mechanics}

\bigskip{}

... in physics the only observations we must consider are position
observations, if only the positions of instrument pointers. It is
a great merit of the de Broglie-Bohm picture to force us to consider
this fact. If you make axioms, rather than definitions and theorems,
about the ``measurement'' of anything else, then you commit redundancy
and risk inconsistency. - \emph{Bell, Speakable and Unspeakable in
Quantum Mechanics}
\end{quotation}

\section*{Richard Feynman}
\begin{quotation}
{[}After a discussion of the measurement problem in quantum mechanics{]}
This is all very confusing, especially when we consider that even
though we may consistently consider ourselves always to be outside
observers when we look at the rest of the world, the rest of the world
is at the same time observing us ... . Does this mean that my observations
become real only when I observe an observer observing something as
it happens? This is an horrible viewpoint. Do you seriously entertain
the thought that without observer there is no reality? Which observer?
Any observer? Is a fly an observer? Is a star an observer? Was there
no reality before 109 B.C. before life began? Or are you the observer?
Then there is no reality to the world after you are dead? I know a
number of otherwise respectable physicists who have bought life insurance.
By what philosophy will the universe without man be understood? In
order to make some sense here, we must keep an open mind about the
possibility that for sufficiently complex systems, amplitudes become
probabilities.... - \emph{Feynman, Lecture Notes on Gravitation }
\end{quotation}

\section*{Imre Lakatos}
\begin{quotation}
In the new, post-1925 quantum theory the `anarchist\textquoteright{}
position became dominant and modern quantum physics, in its `Copenhagen
interpretation\textquoteright , became one of the main standard bearers
of philosophical obscurantism. In the new theory Bohr\textquoteright s
notorious `complementarity principle\textquoteright{} enthroned {[}weak{]}
inconsistency as a basic ultimate feature of nature, and merged subjectivist
positivism and antilogical dialectic and even ordinary language philosophy
into one unholy alliance. After 1925 Bohr and his associates introduced
a new and unprecedented lowering of critical standards for scientific
theories. This led to a defeat of reason within modern physics and
to an anarchist cult of incomprehensible chaos. - \emph{Lakatos, Falsification
and the Methodology of Scientific Research Programs (1970)}
\end{quotation}

\section*{Leo Tolstoy}
\begin{quotation}
The most difficult subjects can be explained to the most slow-witted
man if he has not formed any idea of them already; but the simplest
thing cannot be made clear to the most intelligent man if he is firmly
persuaded that he knows already, without a shadow of doubt, what is
laid before him. - \emph{Tolstoy, The Kingdom of God is Within You
(1894)}

\bigskip{}

I know that most men, including those at ease with prolems of the
highest complexity, can seldom accept even the simplest and most obvious
truth if it be such as would oblige them to admit the falsity of conclusions
which they have delighted in explaining to colleagues, which they
have proudly taught to others, and which they have woven, thread by
thread, into the fabric of their lives. - \emph{Tolstoy, What is Art?
(1897)}
\end{quotation}

\section*{Steve Jobs}
\begin{quotation}
Your time is limited, so don't waste it living someone else's life.
Don't be trapped by dogma - which is living with the results of other
people's thinking. Don't let the noise of other's opinions drown out
your own inner voice. And most important, have the courage to follow
your heart and intuition. They somehow already know what you truly
want to become. Everything else is secondary. - \emph{Jobs, Stanford
Commencement Address (2005)}

\bigskip{}

... almost everything - all external expectations, all pride, all
fear of embarrassment or failure - these things just fall away in
the face of death, leaving only what is truly important. Remembering
that you are going to die is the best way I know to avoid the trap
of thinking you have something to lose. You are already naked. There
is no reason not to follow your heart. - \emph{Jobs, Stanford Commencement
Address (2005)}
\end{quotation}

\chapter{Introduction }

In the foundations of quantum mechanics, it is recognized
that there are three logically distinct possibilities for solving
(or dissolving) the quantum measurement problem \cite{Maudlin1995,Schlosshauer2004a,Allori2012,Wallace2012,Albert2015}
- (1) the wavefunction evolves linearly and deterministically but
is not the complete description of a quantum system; (2) the wavefunction
may or may not be the complete description of a quantum system, but
the linear and deterministic evolution of the wavefunction is not
exact; and (3) the wavefunction may or may not be the complete description, but its
linear and deterministic evolution is exact, and `measurements' of
quantum systems don't have determinate outcomes (despite appearances).
Historically, the dominant theoretical instantiations of these three
respective possibilities have been (1) the de Broglie-Bohm pilot wave
theory (a.k.a. Bohmian mechanics), (2) dynamical collapse theories
(such as the GRW, CSL, and Di\'{o}si-Penrose theories), and (3) Everett's
many-worlds theory (and variants thereof). 

What these three dominant approaches have in common is that they all
posit the time-dependent Schr\"{o}dinger equation (or some slight stochastic nonlinear
modification thereof) as part of the fundamental dynamical laws of
physics (or derivative from a time-independent Schr\"{o}dinger-like equation such as the Wheeler-DeWitt equation), and the universal wavefunction (corresponding to either the time-dependent \emph{N}-particle wavefunction or the Wheeler-DeWitt wavefunctional) as part of either the fundamental
ontology or the fundamental physical laws \cite{Holland1993,BellQMCosmo2004,Bohm1995,DGZ1995,Duerr2009,Allori2009,GoldsteinZanghi2011,Ghirardi2011,Wallace2012,Goldstein2013,Bassi2013,Vaidm2014,Albert2015,VassalloDeckertEsfeld2016}. On the one
hand, this is a methodologically straightforward approach to modifying or supplanting
standard quantum mechanics with an empirically viable non-relativistic
quantum theory that's free of the measurement problem. Indeed, as
long as the aforementioned approaches involve an appropriate set of
`local beables' (i.e., objectively existing physical variables on
space-time) \cite{BellTLB}, the dynamics of which supervene on the
evolution of the universal wavefunction (or on time-dependent wavefunctions for subsystems, defined in terms of the universal wavefunction), all of these approaches (with the possible
exception of many-worlds theories, in my view) give mathematically and conceptually
clear accounts of how determinate measurement outcomes arise (or
appear to arise) from the space-time histories of the local beables, when `microscopic' quantum systems interact
with `macroscopic' quantum systems. On the other hand, it is not clear
that it is \emph{necessary} to posit the time-dependent Schr\"{o}dinger equation (or
some slight stochastic nonlinear modification thereof) as part of
the fundamental dynamical laws (or as derivative from a time-independent Schr\"{o}dinger-like equation), and the universal
wavefunction as part of the fundamental ontology (i.e., as a fundamental, \emph{nonlocal} beable) or the fundamental physical laws.
One might even question (as I would) whether it is viable to regard
the universal wavefunction (a complex-valued or real field on an extremely
high-dimensional space corresponding to configuration space) as part of the
fundamental ontology or the fundamental physical laws. \footnote{This is beyond the scope of the Introduction, but I shall nevertheless
elaborate a bit: In my view, it is not clear what it means to say
that configuration space, and the universal wavefunction on configuration space, have observer-independent
existences `out there' in the physical world, in parallel with or more fundamental than 3-space and the material
objects in the 3-space of everyday experience, as in certain readings
of the de Broglie-Bohm theory and dynamical collapse theories \cite{Holland1993,Albert1996,Albert2013,Albert2015,Ghirardi2011,Goldstein2013,Bassi2013}.
I also regard the functionalist-emergence arguments of Albert \cite{Albert1996,Albert2013,Albert2015}
and Wallace \cite{Wallace2012} as problematic on conceptual and technical
grounds, making it difficult for me to accept intelligibility of the
claim that the universal wavefunction on configuration space, perhaps in conjunction
with a world particle at a point in configuration space (as in Albert's
version of the de Broglie-Bohm theory), constitutes the fundamental ontology
of the physical world. With regard to nomic interpretations of the
universal wavefunction in theories such as de Broglie-Bohm \cite{DGZ1995,GoldsteinZanghi2011,VassalloDeckertEsfeld2016},
I am of the view that a nomic interpretation only really makes sense
if the universal wavefunction ends up being time-independent, unique,
and uncontrollable (by us or anything else), as suggested by D\"{u}rr-Goldstein-Zangh\`{i}
\cite{DGZ1995,GoldsteinZanghi2011}; however, for certain reasons,
I tend to be skeptical that this will pan out, as I tend to be skeptical
(for a variety of reasons) that canonical approaches
to quantum gravity, where the universal wavefunction is indeed time-independent
and uncontrollable (though not necessarily unique!), are the correct
ways to `quantize' Einstein gravity (if that is even necessary at
all, and in my view that is not yet clear). } At the very least, it seems fair to say that it is still an open
question whether the aforementioned interpretations of the universal wavefunction
(within the various solutions to the measurement problem where they're
applied) are in fact viable, with no consensus on this issue among
specialists in the foundations of quantum mechanics. (And let me emphasize
that, although the above discussion is primarily couched in the language of
non-relativistic quantum mechanics, all of it can be carried over
to the context of quantum field theory, more or less unchanged.)

Perhaps, instead, there exists a theoretical framework that makes
it possible to understand the Schr\"{o}dinger equation
as a phenomenological equation, and the wavefunction as a derived
quantity that (in some sense) reflects an array of local beables over and
above the local beables that directly determine the outcomes of measurements.
In this way, the fundamental ontology of the physical world would
involve only local beables, and the wavefunction and Schr\"{o}dinger 
equation would simply be effective descriptions of these local beables
and their more fundamental dynamical laws. If such a framework exists,
it is surely a scientifically and philosophically worthwhile project to develop it, work out its consequences,
and see how it compares to the more `standard' options for addressing
the measurement problem, particularly in cases where the more standard
options are known to still have difficulties or ambiguities (e.g.,
the domains of semiclassical gravity and quantum gravity). 

The stochastic mechanics framework, initiated by F\'{e}nyes in 1952 \cite{Fenyes1952},
rediscovered by Nelson in 1966 \cite{Nelson1966}, and developed by
legions of physicists and mathematicians up until the '80's and early
'90's \cite{Nelson1985nopagelist,Kyprianidis1992,Wallstrom1994,Derakhshani2016a,Derakhshani2016b,Derakhshani2017,Derakhshani2017b},
has, since its inception, been one of the leading candidates for a
theoretical framework of the type explained above. In terms of solving
the measurement problem, it is a version of option (1) insofar as
it aims to recover the wavefunction and deterministic Schr\"{o}dinger 
evolution as a universally valid, effective statistical description
of a classical-like ether medium on space-time that interacts with
point (or point-like) masses immersed in the ether, causing the latter
to undergo a classical Markovian diffusion process that conserves
the total energy of the particles on the average. In stochastic mechanics,
it is the conservative diffusions of the positions of the particles,
in conjunction with a decoherence-driven dynamical process known as
`effective collapse', that determines the outcomes of measurements
in accord with probabilities given by the Born rule \cite{Goldstein1987,Jibu1990,Blanchard1992,Peruzzi1996}.
Conversely, once the wavefunction of a quantum system is known in
stochastic mechanics, one can construct the corresponding diffusion
process for the particles, along with the assumption that the initial
particle positions are randomly distributed according to the Born
rule (though this assumption can be justified on other physical grounds). 

In fact, the stochastic mechanics framework was regarded by Edward
Nelson \cite{Nelson1985nopagelist,Nelson1986,Nelson2005}, perhaps
its most influential contributor, as a phenomenological stepping-stone
to an eventual physical theory of the ether and its interaction with
point masses. In his monograph ``Quantum Fluctuations" \cite{Nelson1985}, Nelson anticipated that this physical
theory would describe the ether as a classical electromagnetic background
field that interacts locally and deterministically with point charges,
and that the stochasticity of the evolution of the point charges would
arise as a result of imposing infrared and UV cutoffs on the charge-field
coupling, and taking the cutoffs to infinity. Nelson also argued that this charge-field coupling with
infrared and UV cutoffs (and the cutoffs taken to infinity) could violate Bell's local causality, even though
it would not violate what Nelson called the ``locality principle'',
i.e., that ``if we couple the {[}electromagnetic{]} field to a {[}charge{]}
current in a {[}spacetime{]} region, only the behavior of the field
in the future light cone of that region will be affected'' \cite{Nelson1985}.
However, as Nelson acknowledged decades later {[}personal communication{]},
he was never able to make his suggestion work. (And, I must admit,
it was never clear to me how the theory he sketched
could violate local causality without violating what he called the
locality principle.) Moreover Nelson abandoned stochastic mechanics
in the '80's because he realized that the Markovian nature of the
conservative diffusions (under the assumption that the diffusions
indeed correspond to single-valued wavefunctions) entails dynamical
nonlocality for the evolutions of the stochastic mechanical particles
in a multi-particle system. To quote him, 
\begin{quote}
If something is physically real, then it cannot be affected instantaneously
by a widely separated perturbation. This is the locality principle,
and it poses a severe challenge to stochastic mechanics. This is because
the diffusion occurs on configuration space, and if we have several
particles, possibly widely separated, the component of the drift for
any particular particle will in general be a function of the positions
of all the particles. \cite{Nelson1985}
\end{quote}
Furthermore, Nelson believed that ``a theory that violates locality
is untenable'' \cite{Nelson1985}. In addition to these objections,
Nelson also claimed that stochastic mechanics predicts different multi-time
correlation functions than standard quantum mechanics, and he questioned
why anyone should believe the stochastic mechanical prediction over
the standard quantum mechanical prediction \cite{Nelson1985nopagelist,Nelson2005}. 

Of course, as is well-known, Bell's theorem implies that any theory
that's in agreement with the empirical predictions of standard quantum mechanics
for Bell-type experiments, must violate locality in Bell's sense (unless,
perhaps, if one denies that measurements have determinate outcomes,
as in many-worlds theories). (Let me also emphasize here that Bell's
notion of local causality is neither the same as nor in conflict with
the `local commutativity' condition in standard quantum field theory,
as Bell emphasized \cite{BellLNC2004}. That is why a theory
can be nonlocal in the sense of Bell, and still satisfy local commutativity,
as is the case with standard quantum field theory.) As is also well-known,
experiments have repeatedly confirmed the violation of Bell's inequality.
So if stochastic mechanics does entail dynamical nonlocality as described
in the above quote, and if this dynamical nonlocality entails violation
of Bell's inequality in exact agreement with standard quantum mechanics
(hence experiment), this would seem like just what the doctor ordered,
rather than a reason to reject stochastic mechanics. From this point
of view, it seems fair to say that Nelson's reasons for abandoning
stochastic mechanics were misguided. This being said, I do not think
it is misguided to hope for a physical model of the stochastic mechanical
ether as a field/medium on space-time (in fact, this would arguably be the most natural way to understand the stochastic mechanical picture of the world). After all, the fact that a
theory may not be locally causal doesn't logically entail that some or
all of the beables of the theory must live on a high-dimensional space
like configuration space. On the contrary, it is entirely possible
to have a non-local causal, empirically viable, non-relativistic theory
of exclusively local beables, as demonstrated by Norsen \cite{Norsen2010,Norsen2014}
in the context of the de Broglie-Bohm theory. And as I will argue
in Chapter 3, there is even reason to think that a non-Markovian extension
of stochastic mechanics may allow for a reformulation of the theory
exclusively in terms of a finite number of local beables (in contrast
to the de Broglie-Bohm model of Norsen, which requires a countable
infinity of local beables or an ad hoc truncation thereof), while
recovering Markovian stochastic mechanics in a certain limit. 

As for Nelson's claim that stochastic mechanics predicts different
multi-time correlations than standard quantum mechanics, it has been
shown by Blanchard et al. \cite{Blanchard1986} that Nelson's analysis
was mistaken - in repeated ideal position measurements of a single-particle
stochastic mechanical system, the stochastic process changes because
the stochastic mechanical drifts are functions of the wavefunction,
and the wavefunction undergoes collapse in each measurement. Blanchard
et al. interpret the collapse of the wavefunction as taking the post-measurement
evolution of the wavefunction to be governed by the usual Schr\"{o}dinger 
equation, but with the initial condition that the wavefunction is
a delta function at the point where the system particle is found.
In this way, they show that the stochastic mechanical multi-time correlations
are in exact agreement with the standard quantum mechanical multi-time
correlations. Of course, a more proper treatment of this problem would
make use of the effective collapse process mentioned earlier; in other
words, the post-measurement wavefunction would correspond to the component
of the system-apparatus-environment entangled state that the system
particle has occupied during the decoherence process corresponding to the position measurement. It will be shown in future work that effective
collapse indeed resolves the apparent disagreement between multi-time
correlations in stochastic mechanics vs. standard quantum mechanics. 

Arguably the only substantive objection that's been raised against
the viability of stochastic mechanics is due to Wallstrom, who pointed
out in the late '80's \cite{Wallstrom1989} and early '90's \cite{Wallstrom1994}
that extant stochastic mechanical theories face one of two problems
- either they allowed for fewer solutions than the set of single-valued
solutions of the Schr\"{o}dinger  equation, or they allowed for more solutions.
The reason, in essence, is that stochastic mechanical theories derive
the `Madelung equations' for a pair of fields, $S$ and $\rho$, where
$S$ is a velocity potential that generates the current velocity field
of the diffusion process, and $\rho$ is the single-time probability
density for the diffusion process. These fields are then combined
via the `Madelung transformation' into a wavefunction $\psi=\sqrt{\rho}exp(iS/\hbar)$
that's assumed to satisfy the time-dependent Schr\"{o}dinger  equation. However, if the
$S$ field is assumed to be single-valued (as in certain versions
of stochastic mechanics), then while $\psi$ will also be single-valued,
this will exclude wavefunctions with angular momentum, i.e., wavefunctions
with phase factors of the form $exp(i\mathrm{m}\varphi)$, where $\mathrm{m}$
is integral and $S=\mathrm{m}\varphi$ is a multi-valued function.
(If we permit $S$ to have jump discontinuities, then it can be shown
that $\nabla\psi$ will develop a singularity, which is not permissible
on physical grounds.) Alternatively, if $S$ field is allowed to be
multi-valued (as in most versions of stochastic mechanics), then there
is no why it should satisfy the Bohr-Sommerfeld quantization condition
$\oint_{L}\nabla S\cdot d\mathbf{x}=nh$, where $L$ is any closed
loop, $n$ is an integer, and $h$ is Planck's constant. Yet, in standard
quantum mechanics, this quantization condition is exactly what follows
from requiring that wavefunctions be single-valued while allowing
multi-valued phases; and whereas there are natural physical justifications
for requiring $\psi$ to be single-valued (e.g., that $\left|\psi\right|^{2}$
has the interpretation of a probability density, and that $\psi$
satisfies the linear superposition principle), those justifications
do not carry over to the Madelung equations. In addition, if one allows
$S$ to be arbitrarily multi-valued, then it can be shown that there
exists a continuum of solutions to the Madelung equations that don't
correspond to any single-valued solution of the Schr\"{o}dinger  equation,
as in case of the central potential problem. These issues will be
discussed in greater detail in Chapter 2. 

(As an interesting historical aside, Shelly Goldstein {[}personal
communication{]} told me that he recognized the issues raised by
Wallstrom years before Wallstrom did; but I am unaware of Shelly writing about it prior to Wallstrom's papers, apart from pointing out in \cite{Goldstein1987} that there are two cases in which conservative diffusions don't correspond to any single-valued solution of Schr\"{o}dinger's equation: (i) conservative diffusions corresponding to the excited energy levels of the hydrogen atom, where the diffusions are decomposed into conservative diffusions separated by the nodal surfaces of the excited levels; and (ii) conservative diffusions in a multiply-connected configuration space, such as the configuration space in the Aharonov-Bohm effect situation. Wallstrom, in turn, told me {[}personal communication{]}
that David Hestenes claims to have observed the issues raised by Wallstrom back in the 1960's, but never bothered to publish about it. Wallstrom {[}personal communication{]} also gives priority of credit to Takehiko Takabayasi, who in 1952
made the point about the mathematical inequivalence between Schr\"{o}dinger's equation and the Madelung
equations without the quantization condition
on $S$, and was apparently the first to do so in the historical record \cite{Takabayasi1952}. Interestingly, though, Takabayasi pointed out this inequivalence not in the context of stochastic mechanical theories but rather versions of quantum mechanics that take the Madelung equations as primitive, such as Madelung's 1926 interpretation and Bohm's 1952 reformulation of pilot-wave theory. And yet, the first version of stochastic mechanics was proposed by F\'{e}nyes in 1952 \cite{Fenyes1952}, which Takabayasi was aware of \cite{Takabayasi1952} and critiqued on completely different grounds!
It's also interesting to mention that, despite his other criticisms
of stochastic mechanics, Nelson never commented on Wallstrom's criticism
in print, and I do not know what Nelson thought of it. And it is unfortunately
too late to ask Nelson, who passed away in 2014.) 

As I will discuss near the end of Chapter 3, many have attempted to
answer the Wallstrom criticism over the years; but, for different
reasons, none of the answers presented thus far have been satisfactory.
The lack of a satisfactory answer to Wallstrom's criticism is what
motivated me to search for an answer many years ago. As I explained
in the Preface, it turns out that de Broglie suggested a model in
his 1923 Ph.D thesis \cite{Broglie1925,Darrigol1994,BacciagaluppiV2010} that seems just right for addressing the criticism,
if imported into certain versions of stochastic mechanics (the versions
that allow $S$ to be multi-valued, such as Nelson's). 

In a nutshell, de Broglie suggested
that each elementary particle of rest mass $m_{0}$ could be thought
of as a spatially localized periodic process (or `clock particle')
of constant angular frequency $\omega_{0}$ in the translational rest
frame of the particle, with the relation between $m_{0}$ and $\omega_{0}$
given by $\hbar\omega_{0}=m_{0}c^{2}$. The precisely physical nature
of this localized periodic process was left unspecified, but de Broglie
hypothesized that there exists a ``phase wave'' in 3-space that
oscillates in step with the localized periodic process at the same
location (although not necessarily driving the periodic process).
Then, applying a Lorentz transformation to the lab frame, de Broglie
found that the phase of this periodic process has space and time dependence,
and by the ``theorem of the harmony of phases'', remains in step
with the phase of the accompanying wave, the latter of which travels
in the same direction as the particle with superluminal phase velocity
$V=c^{2}/v$, where $v$ is the speed of the particle in the lab frame.
(The superluminality of the wave is the reason it is called a ``phase
wave'' - the wave is viewed as a ``distribution in space of the
phases of a phenomenon'' \cite{Broglie1925}, rather than a wave
that carries energy.) Correspondingly, de Broglie showed that when
many phase waves of nearby frequencies have phase velocities in the
direction of motion of the particle, and the velocities of the phase
waves vary with the frequencies of the waves, then the group velocity
of the superposition of these phase waves equals the subluminal velocity
of the particle in the lab frame; hence the energy of the particle
plus its accompanying phase waves always propagates at subluminal
speed. De Broglie further went on to show that Maupertuis' Principle
applied to the particle coincides with Fermat's Principle applied
to the accompanying phase wave, and hence that the possible trajectories
of the particle correspond to the rays of the phase wave. He also
showed by explicit example that these results hold for the particle
and accompanying phase wave propagating through external fields. Finally,
from these results, de Broglie showed that the phase of the periodic
process comprising the particle changes around a closed space-time
orbit by integer multiples of $2\pi$, and that this corresponds exactly
to the Bohr-Sommerfeld quantization condition. The argument was as
follows: given that a trajectory of a moving particle is identical
to a ray of a phase wave, where the frequency is constant (because
total energy is constant) but the velocity is variable, the propagation
of the particle plus phase wave is analogous to ``a liquid wave in
a channel closed on itself but of variable depth'' \cite{Broglie1925}.
Then, in order to have a stable regime, the length $l$ of the channel
must be resonant with the wave. This leads to the resonance condition
$l=n\lambda$ if the wavelength is constant, and $\oint(v/V)dl=n$
in the general case, where \emph{n} is an integer. 

As I also explained in the Preface, I first learned about de Broglie's
model through an obscure 1957 paper \cite{Bohm1957} by Bohm, who
suggested a slight variation of de Broglie's model. In Bohm's model,
an elementary particle is hypothesized to be a spatially localized
mean periodic process, of fixed mean frequency $\omega_{0}$, at the
``sub-quantum level'', where again the precise physical nature of
the periodic process (as well as the precise physical nature of the
sub-quantum level) is left unspecified. Bohm then shows that, in a
fixed coordinate frame where the particle has constant speed $v$,
the Lorentz-transformed mean phase of this particle takes the form
of the phase of a free particle wavefunction at a particular space-time
point, i.e., $\delta\phi=\omega_{0}\delta t\rightarrow\delta\phi(\mathbf{x},t)=(\omega_{0}/m_{0}c^{2})[E\delta t-\mathbf{p}\cdot\delta\mathbf{x}]$,
where $E=\gamma m_{0}c^{2}$ and $\mathbf{p}=\gamma m_{0}\mathbf{v}$,
with $\gamma$ being the gamma factor. Then, Bohm writes, ``Let us
consider how $\phi$ changes as one goes around a virtual circuit
(in which the time as well as the position may change). If we add
up all the phase changes in such a circuit the consistency of the
theory requires that $\oint\delta\phi$ shall be an integral multiple
of $2\pi$, and likewise for a circuit in which time is held fixed
(otherwise, we will contradict the hypothesis that there is a well-defined
mean phase at each point in space)'' \cite{Bohm1957}. Moreover,
because the phase of this particle corresponds to the relativistic
action of the particle, which can be noticed by defining $S=-\hbar\phi$
with $\hbar\coloneqq m_{0}c^{2}/\omega_{0}$, Bohm finds that $\oint\delta S=nh$,
which is again just the Bohr-Sommerfeld quantization condition. Bohm
then observes that this condition is equivalent for the condition
of single-valuedness of the wavefunction in quantum mechanics, using
the definition $\psi=Rexp(iS/\hbar)$. A variation of this model was
also given by Bohm in 1985 \cite{Bohm2002}, but I won't review it
here. 

Thus the primary objective of this thesis is to reformulate Nelson's
stochastic mechanics so as to consistently incorporate (with appropriate
modifications) the above model(s) of de Broglie and Bohm, and thereby
explain how the condition $\oint_{L}\nabla S\cdot d\mathbf{x}=nh$
could arise naturally, rather than imposed ad hoc or by making logically-circular
appeals to the single-valuedeness requirement for wavefunctions. This
is taken up in Chapters 2 and 3, with each Chapter accompanied by
an abstract.

A secondary objective of this thesis is to: (i) use the reformulated
stochastic mechanics to formulate fundamentally-semiclassical theories
of Newtonian gravity and electrodynamics; (ii) compare these theories
to existing formulations of semiclassical Newtonian gravity and electrodynamics;
(iii) show that the stochastic mechanical theories are consistent,
empirically viable theories of (fundamentally-)semiclassical Newtonian
gravity and electrodynamics, with certain conceptual and technical
advantages over extant semiclassical theories based on either standard
quantum theory or measurement-problem-free alternative quantum theories;
and (iv) show that the stochastic mechanical theories can recover
classical Newtonian gravity under certain (physically reasonable)
conditions. The reason for doing all this (apart from it being intrinsically
interesting to me) is that semiclassical Newtonian gravity is becoming
a hot topic theses days in the physics literature, with the vast majority
of discussions centered around the Schr\"{o}dinger-Newton equations \cite{AnastopoulosHu2014,DerakProbingGravCat2016,Derakhshani2017}
and models of fundamentally-semiclassical gravity based on dynamical
collapse theories \cite{Derakhshani2014,DerakProbingGravCat2016,Tilloy2016,TilloyDiosi2017}.
There's also the interesting and long-standing question as to whether
or not a consistent and empirically viable version of fundamentally-semiclassical
gravity can be constructed (and, relatedly, whether gravity needs
to be quantized at all). To show that stochastic mechanics can contribute
something novel and useful to these discussions should (hopefully)
boost general interest in stochastic mechanics among physicists and
philosophers of physics, apart from showing that the Wallstrom criticism
is no longer a (seemingly) decisive objection to it. These tasks are
taken up in Chapters 4 and 5, and each Chapter is again accompanied
by an abstract.

The thesis closes with a Summary and Outlook section, a samenvatting
in het Nederlands, and a brief curriculum vitae.

\chapter{A Suggested Answer To Wallstrom's Criticism: ZSM I}

Wallstrom's criticism of existing formulations of stochastic mechanics
is that they fail to derive quantum theory because they require an
ad hoc quantization condition on the postulated velocity potential,
\emph{S}, in order to derive single-valued Schr\"{o}dinger
wave functions. We propose an answer to this criticism by modifying
the Nelson-Yasue formulation of non-relativistic stochastic mechanics
for a spinless particle with the following hypothesis: a spinless
Nelson-Yasue particle of rest mass $m$ continuously undergoes a driven
steady-state oscillation of `zitterbewegung' (\emph{zbw}) frequency,
$\omega_{c}=\left(1/\hbar\right)mc^{2}$, in its instantaneous mean
forward (and backward) translational rest frame. With this hypothesis
we show that, in the lab frame, \emph{S} arises from imposing the
constraint of conservative diffusions on the time-symmetrized steady-state
phase of the \emph{zbw} particle, satisfies the required quantization
condition, and evolves in time by the Hamilton-Jacobi-Madelung equations
(when generalized to describe a statistical ensemble of \emph{zbw}
particles). The paper begins by reviewing Nelson-Yasue stochastic
mechanics and Wallstrom's criticism, after which we develop a classical
model of a particle of rest mass \emph{m} constrained to undergo the
hypothesized \emph{zbw} oscillation, with the purpose of making clear
the physical assumptions of the \emph{zbw} model without the added
complications of stochastic mechanics. We develop the classical model
for the spinless one-particle case, without and with field interactions,
and then carry out the analogous developments for the Nelson-Yasue
version of this model. Using this `zitterbewegung stochastic mechanics'
(ZSM), we readily derive the single-valued wave functions of non-relativistic
quantum mechanics for a spinless particle in the analyzed cases. We
also apply ZSM to the case of a central potential and show that it
predicts angular momentum quantization. This paper sets the foundation
for Part II, which will (primarily) work out the many-particle version
of ZSM.

\section{Introduction}

Since its introduction by F\'{e}nyes in 1952 \cite{Fenyes1952},
the goal of the stochastic mechanics research program has been to
derive quantum theory from a classical-like statistical mechanics
of particles undergoing Brownian motion. Towards this end, non-relativistic
and relativistic models of stochastic mechanics have been constructed
for both spin-0 particles \cite{Fenyes1952,Nelson1966,Nelson1967,Nelson1985,Yasue1978a,Yasue1979,Yasue1981,Yasue1981a,Yasue1977,Guerra1981,Guerra1983,Davidson1979,Davidson(2006),Nagasawa1996,Cufaro-Petroni1995,Cufaro-Petroni1997,Bacciagaluppi2003,Bacciagaluppi2012,Kobayashi2011,Zastawniak1990,Dohrn1978,Dohrn1979,Dohrn1985,Smolin1986,Smolin(2002),Markopoulou2004,Serva1988,Marra1990,Aldrovandi1990,Aldrovandi1992,Garbaczewski1992,Morato1995}
and spin-$1/2$ particles \cite{Dankel1970,Dohrn1979,Faris1982,Angelis1986,Garbaczewski1992}.
A non-relativistic theory of single-time and multi-time measurements
has also been developed \cite{Blanchard1986,Goldstein1987,Jibu1990,Blanchard1992,Peruzzi1996},
as have extensions of non-relativistic stochastic mechanics to finite
temperature and non-equilibrium open systems \cite{Yasue1978a,Roy2009,S.Bhattacharya2011,Kobayashi2011,Koide2013}.
Field theoretic generalizations also exist, for the cases of scalar
fields \cite{Guerra1973,Yasue1978,Davidson1980,Guerra1981,Koide2015a},
Maxwell fields \cite{Guerra1979,Koide2014}, vector-meson fields \cite{Siena1983},
the linearized gravitational field \cite{Davidson1982}, coupling
to dissipative environments \cite{Yasue1978,Lim1987}, non-Abelian
gauge theory \cite{Yasue1979}, bosonic string theory \cite{Santos99},
M-theory \cite{Smolin(2002)}, and background-independent quantum
gravity \cite{Markopoulou2004}. However, Wallstrom \cite{Wallstrom1989,Wallstrom1994}
pointed out that extant formulations of stochastic mechanics ultimately
fail to derive quantum mechanics because they require an ``ad hoc''
quantization condition on the postulated velocity potential, \textit{S},
in order to recover single-valued Schr\"{o}dinger
wave functions. Moreover, this criticism appears to generalize to
the field-theoretic and quantum gravitational versions of stochastic
mechanics developed before, during, and after Wallstrom's publications,
insofar as they require analogous quantization conditions and don't
seem to give non-circular justifications for them.

Since Wallstrom, sporadic attempts have been made to answer his criticism
\cite{Carlen1989,Wallstrom1994,Bacciagaluppi2005,Smolin06,Fritsche2009,Catich2011,Schmelzer(2011),Groessing2011}.
However, in our view, all these attempts are either problematic or
limited in their applicability to stochastic mechanics (the follow
up paper, Part II, will give a discussion). Nevertheless, if a convincing
answer can be found, stochastic mechanics may once again be viewed
as a viable research program, and one that (in our view) offers elegant
solutions to many of the foundational problems with quantum mechanics.
As examples, stochastic mechanics would provide: (1) an unambiguous
solution to the quantum measurement problem (the local beables of
the theory on which measurement outcomes depend are point masses with
definite trajectories at all times) \cite{Goldstein1987,Jibu1990,Blanchard1992,Peruzzi1996};
(2) a novel and unambiguous physical interpretation of the wave function
(it is epistemic in the sense of being defined from field variables
describing a fictitious ensemble of point masses undergoing conservative
diffusions; and it has ontic properties in the specific sense that
the evolutions of said variables are constrained by beables over and
above the point masses) \cite{Nelson1985,Wallstrom1994,Derakhshani2016b};
(3) an explanation for why the position basis is preferred in decoherence
theory (the form of the Schr\"{o}dinger
Hamiltonian is a consequence of the particle diffusion process happening
in position space) \cite{Bacciagaluppi2005,Kobayashi2011}; and (4)
a justification for the symmetry postulates for wave functions of
identical particles (they arise from natural symmetry conditions on
the particle trajectories, with the possibility of parastatistics
being excluded) \cite{Nelson1985,Goldstein1987,Bacciagaluppi2003}. 

In this connection, it is worth mentioning that some of the aforementioned
virtues of stochastic mechanics, such as (1) and (4), are shared by
de Broglie-Bohm theories \cite{Holland1993,Bohm1995,Duerr2009,OriolsMompart2012,Goldstein2013,Oriols2016,Derakhshani2017b};
conversely, virtually all of the technical results obtained from de
Broglie-Bohm theories can be directly imported into stochastic mechanics
(basically because stochastic mechanics contains the dynamical equations
of de Broglie-Bohm theories as a subset). 

This being said, stochastic mechanics (if viable) has a notably significant
difference from the `standard' approaches to interpreting or reformulating
or replacing the quantum formalism in a realist way that solves the
measurement problem, those being many-worlds theories \cite{Allori2009,Wallace2012,Vaidm2014},
de Broglie-Bohm theories \cite{Holland1993,Bohm1995,Duerr2009,OriolsMompart2012,Goldstein2013,Oriols2016},
and dynamical collapse theories \cite{Ghirardi2011,Adler2012,Bassi2013}.
In all these approaches, the wave function is interpreted as fundamental
and ontic (or as some kind of physical law \cite{GoldsteinZanghi2011,Goldstein2013,VassalloDeckertEsfeld2016}),
and the Schr\"{o}dinger equation (or
some nonlinear modification of it) is taken as a dynamical law. So
if stochastic mechanics succeeds in deriving the Schr\"{o}dinger
equation and wave function, it constitutes (arguably) the first example
of a measurement-problem-free ontological reconstruction of quantum
mechanics in which the wave function could be considered (in a well-defined
sense) as genuinely derived and epistemic, and the Schr\"{o}dinger
evolution as phenomenological rather than law-like\footnote{The recent ``Many-Interacting-Worlds'' (MIW) theory of Hall, Deckert,
and Wiseman \cite{Hall2014}, shares some of these features in that
it recovers the Schr\"{o}dinger wave
function as an effective, mean-field description of a large number
of real classical worlds interacting through a non-classical (quantum)
force. On the other hand, it seems that their approach is also subject
to Wallstrom's criticism in that they also have to assume the quantization
condition (or something like it) on the dynamics of their classical
worlds. Similar comments apply to the ``Prodigal QM'' theory of
Sebens \cite{Sebens2015}.

Similarly, the ``Trace Dynamics'' theory of Steven Adler \cite{Adler2002,Adler2012,Adler2013,Bassi2013}
aims to derive the quantum formalism as an approximation to the thermodynamic
limit of a statistical mechanical description of Grassmannian matrices
living on space-time. However, Trace Dynamics requires certain ad
hoc assumptions, namely that the state-vector in the thermodynamic
description has a norm-preserving nonlinear stochastic evolution.
Such an assumption is ad hoc because it seems to have no justification
from within the assumptions of Trace Dynamics, whereas it presumably
should have such a justification in order to sustain the claim that
Trace Dynamics derives the quantum formalism in a certain approximation.
(This view is also espoused by Bassi et al. in \cite{Bassi2013}.)
In this sense, it seems fair to say that the norm-perserving assumption
is to Trace Dynamics what the quantization condition is to (extant
formulations of) stochastic mechanics. }. Thus stochastic mechanics would (if viable) constitute a couterexample
to an implicit assumption that motivates the aforementioned standard
approaches - that the wave function and Schr\"{o}dinger
equation must be part of the fundamental ontology (or laws) and dynamical
laws, respectively, in order to have a realist alternative to standard
quantum theory that solves the measurement problem, is empirically
adequate, and has a coherent physical/ontological interpretation.

It is also noteworthy that, as a dynamical theory of particle motion
in which probabilities play no fundamental role, stochastic mechanics
shares with de Broglie-Bohm theories the ability to justify the ``quantum
equilibrium'' density $|\psi|^{2}$ from typicality arguments \cite{Duerr1992}
and from dynamical relaxation of non-equilibrium densities to future
equilibrium \cite{Goldstein1987,Cufaro-Petroni1995,Cufaro-Petroni1997,Bacciagaluppi2012}.
As a result, stochastic mechanics can, on its own terms, be regarded
as a more general physical theory that contains quantum mechanics
as a fixed point - and outside this fixed point, it admits the possibility
of non-equilibrium physics, e.g., measurements more precise than the
uncertainty principle allows and superluminal signaling \cite{Nelson1985,Bohm1952I,Bohm1952II,Pearle2006}.
We will also argue in Part II \cite{Derakhshani2016b} that quantum
non-equilibrium states are more plausibly motivated in stochastic
mechanics than in deterministic de Broglie-Bohm theories. 

For all these reasons and more, it seems worthwhile to consider whether
the central obstacle for the stochastic mechanics research program
- Wallstrom's criticism - can be surmounted. The objective of this
series of papers is to suggest how non-relativistic stochastic mechanics
for spinless particles can be modified to provide a non-ad-hoc physical
justification for the required quantization condition on $S$, and
thereby recover all and only the single-valued wave functions of non-relativistic
quantum mechanics. In this paper, we propose to modify the Nelson-Yasue
formulation \cite{Nelson1985,Yasue1981} of non-relativistic stochastic
mechanics for a spinless particle with the following hypothesis: a
spinless particle of rest mass, $m$, bounded to a harmonic potential
of natural frequency, $\omega_{c}=\left(1/\hbar\right)mc^{2}$, and
immersed in Nelson's hypothetical ether medium (appropriately modified
in its properties), undergoes a driven steady-state oscillation of
`zitterbewegung' (\emph{zbw}) frequency, $\omega_{c}$, in its instantaneous
mean forward (and backward) translational rest frame. With this hypothesis
we show that, in the lab frame, the stochastic mechanical velocity
potential, \emph{S}, arises from imposing the constraint of conservative
diffusions on the time-symmetrized steady-state phase of the \emph{zbw}
particle, implies the needed quantization condition, and evolves by
the stochastically derived Hamilton-Jacobi-Madelung equations (when
generalized to describe a statistical ensemble of \emph{zbw} particles).
This modification of Nelson-Yasue stochastic mechanics (NYSM), which
we term `zitterbewegung stochastic mechanics' (ZSM), then allows us
to derive the single-valued wave functions of non-relativistic quantum
mechanics for a spinless particle. The problem of justifying the quantization
condition is thereby reduced to justifying the zitterbewegung hypothesis.
Accordingly, it is among the tasks of Part II to argue that the hypothesis
can be justified in terms of physical/dynamical models and can be
plausibly generalized to particles with spin as well as relativistic
particles and fields.

The outline of this paper is as follows. In section 2, we give a concise
review of the formal derivation of the Schr\"{o}dinger
equation from NYSM for a single, spinless particle in an external
scalar potential. (Such a review will be useful for the reader who
is unfamiliar with NYSM, and essential for following the logic and
presentation of the arguments later in the paper.) In section 3, we
review the Wallstrom criticism. In section 4, we introduce a classical
model of a spinless zitterbewegung particle which implies the quantization
condition for the phase of its oscillation, excluding and including
interactions with external fields. In each case, we extend the model
to a classical Hamilton-Jacobi statistical mechanics involving a Gibbsian
ensemble of such particles, with the purpose of making as clear as
possible the physical assumptions of the model in a well-established
classical physics framework that has conceptual and mathematical similarities
to stochastic mechanics. In section 5, we construct a Nelson-Yasue
stochastic mechanics for the zitterbewegung particle (ZSM), excluding
and including field interactions. In this way we derive one-particle
Schr\"{o}dinger equations with single-valued
wave functions that have (generally) multi-valued phases, and use
the hydrogen-like atom as a worked example.

This paper lays the foundation for Part II, where we will: (1) develop
the (non-trivial) many-particle cases of ZSM, (2) explicate the beables
of ZSM, (3) assess the plausibility and generalizability of the zitterbewegung
hypothesis, and (4) compare ZSM to other proposed answers to Wallstrom's
criticism.

\section{Nelson-Yasue Stochastic Mechanics}

In Edward Nelson's non-relativistic stochastic mechanics \cite{Nelson1966,Nelson1967,Nelson1985},
it is first hypothesized that the vacuum is pervaded by a homogeneous
and isotropic ``ether'' fluid with classical stochastic fluctuations
of uniform character. \footnote{The microscopic constituents of this ether are left unspecified by
Nelson; however, he suggests by tentative dimensional arguments relating
to the choice of diffusion constant in Eq. (2.3) (namely, that we
can write $\hbar=e^{2}/\alpha c$, where $\alpha$ is the fine-structure
constant and $e$ the elementary charge) that it may have an electromagnetic
origin \cite{Nelson1985}. } To ensure that observers in the ether can't distinguish absolute
rest from uniform motion, it is further hypothesized that the interaction
of a point mass with the ether is a frictionless diffusion process.
\footnote{Nelson points out \cite{Nelson1985} that this frictionless diffusion
process is an example of ``conservative diffusions'', or diffusions
in which the ensemble-averaged energy of the particle is conserved
in time (for a time-independent external potential). In other words,
on the (ensemble) average, there is no net transfer of energy between
the particle and the fluctuating ether, in contrast to classical Brownian
diffusions which are fundamentally dissipative in character.} Accordingly, a point particle of mass \emph{m} within this frictionless
ether will in general have its position 3-vector $\mathbf{q}(t)$
constantly undergoing diffusive motion with drift, as modeled by the
first-order stochastic differential equation,
\begin{equation}
d\mathbf{q}(t)=\mathbf{b}(\mathbf{q}(t),t)dt+d\mathbf{W}(t).
\end{equation}
The vector $\mathbf{b}(\mathbf{q}(t),t)$ is the deterministic ``mean
forward'' drift velocity of the particle, and $\mathbf{W}(t)$ is
the Wiener process modeling the effect of the particle's interaction
with the fluctuating ether.

The Wiener increment, $d\mathbf{W}(t)$, is assumed to be Gaussian
with zero mean, independent of $d\mathbf{q}(s)$ for $s\leq t$, and
with covariance,
\begin{equation}
\mathrm{E}_{t}\left[d\mathbf{W}_{i}(t)d\mathbf{W}_{j}(t)\right]=2\nu\delta_{ij}dt,
\end{equation}
where $\mathrm{E}_{t}$ denotes the conditional expectation at time
\emph{t}. 

Note that although Equations (2.1-2) are formally the same as those
used for the kinematical description of classical Brownian motion
in the Einstein-Smoluchowski (ES) theory, the physical context is
different; the ES theory uses (2.1-2) to model the Brownian motion
of macroscopic particles in a classical fluid in the large friction
limit \cite{Nelson1967}, whereas Nelson uses (2.1-2) to model frictionless
stochastic motion (i.e., ``conservative diffusions'' \cite{Nelson1985})
for elementary particles interacting with a fluctuating ether fluid
that permeates the vacuum.

In this connection, it is further hypothesized that the magnitude
of the diffusion coefficient $\nu$ is proportional to the reduced
Planck's constant, and inversely proportional to the particle mass
\emph{m} so that
\begin{equation}
\nu=\frac{\hbar}{2m}.
\end{equation}

In addition to (2.1), the particle's trajectory $\mathbf{q}(t)$ can
also satisfy the time-reversed equation 
\begin{equation}
d\mathbf{q}(t)=\mathbf{b}_{*}(\mathbf{q}(t),t)dt+d\mathbf{W}_{*}(t),
\end{equation}
where $\mathbf{b}_{*}(\mathbf{q}(t),t)$ is the mean backward drift
velocity, and $d\mathbf{W}_{*}(t)=d\mathbf{W}(-t)$ is the backward
Wiener process. The $d\mathbf{W}_{*}(t)$ has all the properties of
$d\mathbf{W}(t)$, except that it is independent of $d\mathbf{q}(s)$
for $s\geq t$. With these conditions on $d\mathbf{W}(t)$ and $d\mathbf{W}_{*}(t)$,
(2.1) and (2.4) respectively define forward and backward Markov processes
on $\mathbb{R}^{3}$. 

The forwards and backwards transition probabilities defined by (2.1)
and (2.4), respectively, should be understood, in some sense, as ontic
probabilities \cite{Uffink2006,BacciagaluppiProbab2011}. (Generally
speaking, `ontic probabilities' can be understood as probabilities
about objective physical properties of the \emph{N}-particle system, as opposed
to `epistemic probabilities' \cite{Arntzenius1995} which are about
our ignorance of objective physical properties of the \emph{N}-particle system.)
Just how `ontic' these transition probabilities should be is an open
question. One possibility is that these transition probabilities should
be viewed as phenomenologically modeling complicated deterministic
interactions of a massive particle (or particles) with the fluctuating
ether, in analogy with how equations such as (2.1) and (2.4) are used
in the ES to phenomenologically model the complicated deterministic
interactions of a macroscopic particle immersed in a fluctuating classical
fluid of finite temperature \cite{Nelson1967}. Another possibility
is that the fluctuations of the ether are irreducibly stochastic,
and this irreducible stochasticity is 'transferred' to a particle
immersed in and interacting with the ether. We prefer the former possibility,
but acknowledge that the latter possibility is also viable. \footnote{Concerning whether or not the forward and backwards transition probabilities
should be understood as `objective' (i.e., as chances governed by
natural law) versus `subjective' (i.e., encoding our expectations
or degrees of belief) \cite{FriggHoefer2010,Glynn2010,Emery2015},
this seems to depend on whether the transition probabilities are merely
phenomenological (in which case they would seem to be subjective)
or reflect irreducible stochasticity in the ether (in which case they
would seem to be objective). Our preference for viewing the transition
probabilities as phenomenological seems to commit us to the subjective
view, but the objective view also seems viable (the objective view
is taken by Bacciagaluppi in \cite{Bacciagaluppi2005,Bacciagaluppi2012}).
It is worth noting that, under the objective view, the backwards transition
probabilities can be regarded as being just as objective/law-like
as the forwards transition probabilities (but see \cite{Arntzenius1995}
for a different view).}

Associated to the trajectory $\mathbf{q}(t)$ is the probability density
$\rho(\mathbf{q},t)=n(\mathbf{q},t)/N$, where $n(\mathbf{q},t)$
is the number of particles per unit volume and $N$ is the total number
of particles in a definite region of space. Corresponding to (2.1)
and (2.4), then, are the forward and backward Fokker-Planck equations,
\begin{equation}
\frac{\partial\rho(\mathbf{q},t)}{\partial t}=-\nabla\cdot\left[\mathbf{b(}\mathbf{q},t)\rho(\mathbf{q},t)\right]+\frac{\hbar}{2m}\nabla^{2}\rho(\mathbf{q},t),
\end{equation}
and
\begin{equation}
\frac{\partial\rho(\mathbf{q},t)}{\partial t}=-\nabla\cdot\left[\mathbf{b}_{*}(\mathbf{q},t)\rho(\mathbf{q},t)\right]-\frac{\hbar}{2m}\nabla^{2}\rho(\mathbf{q},t),
\end{equation}
where we require that $\rho(\mathbf{q},t)$ satisfies the normalization
condition, 
\begin{equation}
\int\rho_{0}(\mathbf{q})d^{3}q=1.
\end{equation}

We emphasize that, in contrast to the transition probabilities defined
by (2.1) and (2.4), the probability distributions satisfying (2.5)
and (2.6) are epistemic distributions in the sense that they are distributions
over a Gibbsian ensemble of identical systems (i.e., the distributions
reflect our ignorance of the actual positions of the particles). Nevertheless,
for an epistemic distribution satisfying (2.5) or (2.6) at time $t$,
its subsequent evolution will be determined by the ontic transition
probabilities so that the distribution at later times will partly
come to reflect ontic features of the \emph{N}-particle system, and may asymptotically
become independent of the initial distribution. \footnote{I thank Guido Bacciagaluppi for emphasizing this point.}
Of course, the asymptotic distribution would still be epistemic in
the sense of encoding our ignorance of the actual particle positions,
even though it would be determined by the ontic features of the system. 

A frictionless (hence energy-conserving or conservative) diffusion
process such as Nelson's should have a time-symmetric probability
density evolution. The Fokker-Planck equations (2.5-6), on the other
hand, describe time-asymmetric evolutions in opposite time directions.
The reason is that, given all possible solutions to (2.1), one can
define as many forward processes as there are possible initial distributions
satisfying (2.5); likewise, given all possible solutions to (2.4),
one can define as many backward processes as there are possible `initial'
distributions satisfying (2.6). Consequently, the forward and backward
processes are both underdetermined, and neither (2.1) nor (2.4) has
a well-defined time-reversal. We must therefore restrict the diffusion
process to simultaneous solutions of (2.5) and (2.6). 

Note that the sum of (2.5) and (2.6) gives the continuity equation
\begin{equation}
\frac{\partial\rho({\normalcolor \mathbf{q}},t)}{\partial t}=-\nabla\cdot\left[\mathbf{v}(\mathbf{q},t)\rho(\mathbf{q},t)\right],
\end{equation}
where 
\begin{equation}
\mathbf{v}(\mathbf{q},t)\coloneqq\frac{1}{2}\left[\mathbf{b}(\mathbf{q},t)+\mathbf{b}_{*}(\mathbf{q},t)\right]
\end{equation}
is called the ``current velocity'' field. As it stands, this current
velocity field could have vorticity. But if vorticity is allowed,
then the time-reversal operation on (5.8) will change the orientation
of the curl, thus distinguishing time directions \cite{PenaCetto1982,CaliariInversoMorato2004,Bacciagaluppi2012}.
So we impose
\begin{equation}
\mathbf{v}(\mathbf{q}.t)=\frac{\nabla S(\mathbf{q},t)}{m},
\end{equation}
or that the current velocity field is irrotational. Accordingly, (2.8)
becomes 
\begin{equation}
\frac{\partial\rho({\normalcolor \mathbf{q}},t)}{\partial t}=-\nabla\cdot\left[\frac{\nabla S(\mathbf{q},t)}{m}\rho(\mathbf{q},t)\right],
\end{equation}
a time-reversal invariant evolution equation for the single-time density
$\rho({\normalcolor \mathbf{q}},t)$.

Physically speaking, the \textit{S} function in (2.10-11) has the
interpretation of a velocity potential connected with a Gibbsian ensemble
of fictitious, non-interacting, identical particles with density $\rho(\mathbf{q},t)$,
where each particle in the ensemble differs from the other in its
initial position (hence the dependence of $S$ on the generalized
coordinate $\mathbf{q}$) and initial irrotational velocity given
by (2.10). \footnote{Of course, one can still add to $\nabla S$ a solenoidal vector field
of any magnitude and, upon insertion into (2.8), recover the same
continuity equation \cite{Bacciagaluppi1999,HollandPhilipp2003}.
But the assumption of only irrotational flow velocity is the simplest
one, and as we already mentioned, it follows from the requirement
of time symmetry for the $\rho({\normalcolor \mathbf{q}},t)$ of the
diffusion process. } It is thereby analogous to the \textit{S} function in the Hamilton-Jacobi
formulation of classical statistical mechanics for a single point
particle \cite{Schiller1962,Rosen1964,Holland1993,Ghose2002,Nikolic2006,Nikolic2007}. 

Note also that subtracting (2.5) from (2.6) yields equality on the
right hand side of
\begin{equation}
\mathbf{u}(\mathbf{q},t)\coloneqq\frac{1}{2}\left[\mathbf{b}(\mathbf{q},t)-\mathbf{b}_{*}(\mathbf{q},t)\right]=\frac{\hbar}{2m}\frac{\nabla\rho(\mathbf{q},t)}{\rho(\mathbf{q},t)},
\end{equation}
where $\mathbf{u}(\mathbf{q},t)$ is called the ``osmotic velocity''
field (because it has the same dependence on the density as the velocity
acquired by a classical Brownian particle in equilibrium with respect
to an external force, in the ES theory \cite{Nelson1966,Nelson1967,Nelson1985}). 

As a consequence of (2.9), (2.10), and (2.12), we have that $\mathbf{b}=\mathbf{v}+\mathbf{u}$
and $\mathbf{b}_{*}=\mathbf{v}-\mathbf{u}$, which when inserted back
into (2.5) and (2.6), respectively, reduce both Fokker-Planck equations
to the time-reversal invariant continuity equation (2.11). So the
combination of (2.9), (2.10), and (2.12) fixes $\rho$ as the common,
single-time, `equilibrium' probability density (in analogy with a
thermal equilibrium density) for solutions of (2.1) and (2.4), even
though it is a time-dependent density.

In our view, the physical meaning of (2.12) has been misconstrued
by some researchers \cite{Bohm1989,Kyprianidis1992,Smolin06,Smolin2012}
to imply that $\rho$ must be interpreted as the physical cause of
the osmotic velocity of Nelson's particle. We want to stress that
this is not the case, and that such an interpretation would be logically
and physically inconsistent with the definition of $\rho$ as a probability
density. Instead, Nelson physically motivates his osmotic velocity
by analogy with the osmotic velocity in the ES theory \cite{Nelson1966,Nelson1967}
- essentially, he postulates the presence of an external (i.e., not
sourced by the particle) potential, $U(\mathbf{q},t)$, which couples
to the particle via some coupling constant, $\mu$, such that $R(\mathbf{q}(t),t)=\mu U(\mathbf{q}(t),t)$
defines a `potential momentum' for the particle. \footnote{It should be emphasized that $U(\mathbf{q},t)$ is not defined over
an ensemble of systems, but is a real physical field on 3-space analogous
to the classical external potential, $V(\mathbf{q},t)$, that causes
the osmotic velocity of a Brownian particle in the E-S theory. Nelson
does not specify whether $U(\mathbf{q},t)$ is sourced by the ether
or is an independently existing field on space-time, nor does he specify
whether the coupling $\mu$ corresponds to any of the fundamental
force interactions of the Standard Model. These elements of his theory
are phenomenological hypotheses that presumably should be made more
precise in a `deeper' extension of stochastic mechanics. Nonetheless,
as we will see in Part II, the many-particle extension of stochastic
mechanics puts additional constraints on how the osmotic potential
should be understood.} (Hereafter we shall permit ourselves to refer to $U(\mathbf{q},t)$
and $R(\mathbf{q},t)$ interchangeably as the `osmotic potential'.)
When $U(\mathbf{q},t)$ is spatially varying, it imparts to the particle
a momentum, $\nabla R(\mathbf{q},t)|_{\mathbf{q}=\mathbf{q}(t)}$,
which is then counter-balanced by the ether fluid's osmotic impulse
pressure, $\left(\hbar/2m\right)\nabla\ln[n(\mathbf{q},t)]|_{\mathbf{q}=\mathbf{q}(t)}$.
This leads to the equilibrium condition $\nabla R/m=\left(\hbar/2m\right)\nabla\rho/\rho$
(using $\rho=n/N$), which implies that $\rho$ depends on $R$ as
$\rho=e^{2R/\hbar}$ for all times. Hence, the \textit{physical cause}
of $\mathbf{u}$ is $R$ (or technically \emph{U}), and (2.12) is
just a mathematically equivalent and convenient rewriting of this
relation.

So far our discussion has been restricted to the first-order stochastic
differential equations for Nelson's particle, and the associated Fokker-Planck
evolutions. In order to discuss the second-order dynamics for Nelson's
particle, we must first motivate Nelson's analogues of the Ornstein-Uhlenbeck
mean derivatives. In the It\^{o} calculus, the mean forward and backward
derivatives of a solution $\mathbf{q}(t)$ satisfying (2.1) and (2.4)
are respectively defined as
\begin{equation}
D\mathbf{q}(t)=\underset{_{\Delta t\rightarrow0^{+}}}{lim}\mathrm{E_{t}}\left[\frac{q(t+\Delta t)-q(t)}{\Delta t}\right],
\end{equation}
and
\begin{equation}
D_{*}\mathbf{q}(t)=\underset{_{\Delta t\rightarrow0^{+}}}{lim}\mathrm{E_{t}}\left[\frac{q(t)-q(t-\Delta t)}{\Delta t}\right].
\end{equation}
Because $d\mathbf{W}(t)$ and $d\mathbf{W}_{*}(t)$ are Gaussian with
zero mean, it follows that $D\mathbf{q}(t)=\mathbf{b}(\mathbf{q}(t),t)$
and $D_{*}\mathbf{q}(t)=\mathbf{b}_{*}(\mathbf{q}(t),t)$. To compute
the second mean derivative, $D\mathbf{b}(\mathbf{q}(t),t)$ (or $D_{*}\mathbf{b}(\mathbf{q}(t),t)$),
we must expand $\mathbf{b}$ in a Taylor series up to terms of order
two in $d\mathbf{q}(t)$:
\begin{equation}
d\mathbf{b}(\mathbf{q}(t),t)=\frac{\partial\mathbf{b}(\mathbf{q}(t),t)}{\partial t}dt+d\mathbf{q}(t)\cdot\nabla\mathbf{b}(\mathbf{q}(t),t)+\frac{1}{2}\underset{i,j}{\sum}d\mathbf{\mathit{q}}_{i}(t)d\mathbf{\mathit{q}}_{j}(t)\frac{\partial^{2}\mathbf{b}(\mathbf{q}(t),t)}{\partial\mathbf{\mathit{q}}_{i}\partial\mathit{q}_{j}}+\ldots.
\end{equation}
From (2.1), we can replace $dx_{i}(t)$ by $dW_{i}(t)$ in the last
term, and when taking the conditional expectation in (2.13), we can
replace $d\mathbf{q}(t)\cdot\nabla\mathbf{b}(\mathbf{q}(t),t)$ by
$\mathbf{b}(\mathbf{q}(t),t)\cdot\nabla\mathbf{b}(\mathbf{q}(t),t)$
since $d\mathbf{W}(t)$ is independent of $\mathbf{q}(t)$ and has
mean 0. Using (2.2-3), we then obtain
\begin{equation}
D\mathbf{b}(\mathbf{q}(t),t)=\left[\frac{\partial}{\partial t}+\mathbf{b}(\mathbf{q}(t),t)\cdot\nabla+\frac{\hbar}{2m}\nabla^{2}\right]\mathbf{b}(\mathbf{q}(t),t),
\end{equation}
and likewise
\begin{equation}
D_{*}\mathbf{b}_{*}(\mathbf{q}(t),t)=\left[\frac{\partial}{\partial t}+\mathbf{b}_{*}(\mathbf{q}(t),t)\cdot\nabla-\frac{\hbar}{2m}\nabla^{2}\right]\mathbf{b}_{*}(\mathbf{q}(t),t).
\end{equation}
Using (2.16-17), along with Newton's 2nd law, Nelson wanted to construct
an expression for the `mean acceleration' of the particle consistent
with the principle of time-symmetry. He proposed
\begin{equation}
m\mathbf{a}(\mathbf{q}(t),t)=\frac{m}{2}\left[D_{*}D+DD_{*}\right]\mathbf{q}(t)=-\nabla V(\mathbf{q},t)|_{\mathbf{q}=\mathbf{q}(t)}.
\end{equation}
Physically, this equation says that the mean acceleration Nelson's
particle feels in the presence of an external (conservative) force
is the equal-weighted average of its mean forward drift $\mathbf{b}$
transported backwards in time, with its mean backward drift $\mathbf{b}_{*}$
transported forwards in time. It is this peculiar mean dynamics that
preserves the time-symmetry of Nelson's diffusion process.

Of course, other time-symmetric mean accelerations are possible. For
example, $(1/2)[D^{2}+D_{*}^{2}]\mathbf{q}(t)$, or any weighted average
of this with (2.18). So it may be asked: what other physical principles
(if any) privilege Nelson's choice? As first shown by Yasue \cite{Yasue1981,Yasue1981a}
and later adopted by Nelson \cite{Nelson1985}, a physically well-motivated
stochastic variational principle can give (2.18). \footnote{Another widely used stochastic variational principle is the one due
to Guerra and Morato \cite{Guerra1983}. We don't use their approach because it entails an $S$ function that's globally single-valued, which excludes the possibility of systems with angular momentum \cite{Wallstrom1989} and therefore will not be applicable to our proposed answer to Wallstrom's criticism.} Consider the ensemble-averaged, time-symmetric mean action
\begin{equation}
\begin{aligned}J & =\mathrm{E}\left[\int_{t_{i}}^{t_{f}}\left\{ \frac{1}{2}\left[\frac{1}{2}m\mathbf{b}(\mathbf{q}(t),t)^{2}+\frac{1}{2}m\mathbf{b}_{*}(\mathbf{q}(t),t)^{2}\right]-V(\mathbf{q}(t),t)\right\} dt\right]\\
 & =\mathrm{E}\left[\int_{t_{i}}^{t_{f}}\left\{ \frac{1}{2}m\mathbf{v}^{2}+\frac{1}{2}m\mathbf{u}^{2}-V\right\} dt\right].
\end{aligned}
\end{equation}
In other words, for a particle in a (possibly) time-dependent potential
$V$, undergoing the Markov processes given by (2.1) and (2.4) with
the restriction to simultaneous solutions of the Fokker-Planck equations
via (2.9), (2.10), and (2.12), a time-symmetric mean Lagrangian can
be defined by averaging together the mean Lagrangians associated with
the forward and backward processes. The \textcolor{black}{ensemble
averaged action obtained from} this time-symmetric mean Lagrangian
then corresponds to (2.19), where $\mathrm{E}\left[...\right]$ denotes
the absolute expectation. Upon imposing the conservative diffusion
condition through the variational principle,
\begin{equation}
J=\mathrm{E}\left[\int_{t_{i}}^{t_{f}}\left\{ \frac{1}{2}m\mathbf{v}^{2}+\frac{1}{2}m\mathbf{u}^{2}-V\right\} dt\right]=extremal,
\end{equation}
a straightforward computation (see Appendix 6.1) shows that this implies
(2.18) as the equation of motion. If, instead, we had allowed the
mean kinetic energy terms in (2.19) to not be positive-definite and
used the alternative time-symmetric mean kinetic energy, $(1/2)m\mathbf{b}\mathbf{b}_{*}=(1/2)m(\mathbf{v}^{2}-\mathbf{u}^{2})$,
then it can be shown \cite{Hasegaw1986,Kyprianidis1992,Davidson(2006)}
that imposing (2.20) would give the alternative time-symmetric mean
acceleration involving the derivatives $[D^{2}+D_{*}^{2}]$. \footnote{Additionally, Davidson \cite{Davidson(2006)} showed that by defining
a Lagrangian of the form $(1/2)m\left[(1/2)\left(\mathbf{b}^{2}+\mathbf{b}_{*}^{2}\right)-(\beta/8)(\mathbf{b}-\mathbf{b}_{*})^{2}\right]$,
where $\beta$ is a constant, the resulting equation of motion is
also equivalent to the usual Schr\"{o}dinger  equation, provided that the
diffusion coefficient $\nu=(1/\sqrt{1-\beta/2})\frac{\hbar}{2m}$.
We can see, however, that our criterion of restricting the kinetic
energy terms in the Lagrangian to only terms that are positive-definite,
excludes Davidson's choice of Lagrangian too.} So Nelson's mean acceleration choice is justified by the principle
of time-symmetry \textit{and} the natural physical requirement that
all the contributions to the mean kinetic energy of the Nelsonian
particle should be positive-definite.

By applying the mean derivatives in (2.18) to $\mathbf{q}(t)$, using
that $\mathbf{b}=\mathbf{v}+\mathbf{u}$ and $\mathbf{b}_{*}=\mathbf{v}-\mathbf{u}$,
and removing the dependence of the mean acceleration on the actual
particle trajectory $\mathbf{q}(t)$ so that $\mathbf{a}(\mathbf{q}(t),t)$
gets replaced by the mean acceleration field $\mathbf{a}(\mathbf{q},t)$,
a straightforward computation gives
\begin{equation}
\begin{aligned}m\mathbf{a}(\mathbf{q},t) & =m\left[\frac{\partial\mathbf{v}(\mathbf{q},t)}{\partial t}+\mathbf{v}(\mathbf{q},t)\cdot\nabla\mathbf{v}(\mathbf{q},t)-\mathbf{u}(\mathbf{q},t)\cdot\nabla\mathbf{u}(\mathbf{q},t)-\frac{\hbar}{2m}\nabla^{2}\mathbf{u}(\mathbf{q},t)\right]\\
 & =\nabla\left[\frac{\partial S(\mathbf{q},t)}{\partial t}+\frac{\left(\nabla S(\mathbf{q},t)\right)^{2}}{2m}-\frac{\hbar^{2}}{2m}\frac{\nabla^{2}\sqrt{\rho(\mathbf{q},t)}}{\sqrt{\rho(\mathbf{q},t)}}\right]=-\nabla V(\mathbf{q},t).
\end{aligned}
\end{equation}
The mean acceleration field $\mathbf{a}(\mathbf{q},t)$ describes
the possible mean accelerations of the actual particle given all of
the possible spatial locations that the actual particle can occupy
at time \emph{t}. In other words, $\mathbf{a}(\mathbf{q},t)$ is the
mean acceleration field connected with the set of fictitious particles
forming the Gibbsian ensemble that reflects our ignorance of the actual
trajectory $\mathbf{q}(t)$ \cite{Holland1993}. Integrating both
sides of (2.21), and setting the arbitrary integration constants equal
to zero, we then obtain the Quantum Hamilton-Jacobi equation,
\begin{equation}
-\frac{\partial S(\mathbf{q},t)}{\partial t}=\frac{\left(\nabla S(\mathbf{q},t)\right)^{2}}{2m}+V(\mathbf{q},t)-\frac{\hbar^{2}}{2m}\frac{\nabla^{2}\sqrt{\rho(\mathbf{q},t)}}{\sqrt{\rho(\mathbf{q},t)}},
\end{equation}
which describes the total energy field over the possible positions
of the actual\emph{ }point mass, and upon evaluation at $\mathbf{q}=\mathbf{q}(t),$
the total energy of the point mass along its actual trajectory.

Although the last term on the right hand side of (2.22) is often called
the ``quantum potential'', we note that it arises here from the
osmotic kinetic energy term in (2.19). So the quantum potential must
be physically understood in stochastic mechanics as a kinetic energy
field (which hereafter we prefer to call the `quantum kinetic' for
accuracy of meaning) arising from the osmotic velocity field.

The pair of nonlinear equations coupling the evolution of $\rho$
and $S$, as given by (2.11) and (2.22), are generally known as the
Hamilton-Jacobi-Madelung (HJM) equations, and can be formally identified
with the imaginary and real parts of the Schr\"{o}dinger  equation under
polar decomposition \cite{Takabayasi1952,Holland1993}. Therefore,
(2.11) and (2.22) can be formally rewritten as the Schr\"{o}dinger  equation,
\begin{equation}
i\hbar\frac{\partial\psi(\mathbf{q},t)}{\partial t}=-\frac{\hbar^{2}}{2m}\nabla^{2}\psi(\mathbf{q},t)+V(\mathbf{q},t)\psi(\mathbf{q},t),
\end{equation}
where $\psi(\mathbf{q},t)=\sqrt{\rho(\mathbf{q},t)}e^{iS(\mathbf{q},t)/\hbar}$.
In contrast to other ontological formulations of quantum mechanics,
this wave function must be interpreted as an epistemic field in the
sense that it encodes information about the possible position and
momenta states that the actual particle can occupy at any instant,
since it is defined in terms of the ensemble variables $\rho$ and
$S$. \footnote{Though it may not be obvious here, this interpretation of the Nelson-Yasue
wave function is not undermined by the Pusey-Barrett-Rudolph theorem
\cite{Pusey2012}. Whereas this theorem assumes factorizability/separability
of the ``ontic state space'', the ontic osmotic potential, $U$,
which is encoded in the amplitude of the wave function via $R$ and
plays a role in the particle dynamics via (21), is in general not
separable when extended to the \emph{N}-particle case (as will be
shown in Part II \cite{Derakhshani2016b}). } Although the treatment here did not include coupling to electromagnetic
potentials, it is straightforward to do so \cite{Nelson1985} (see
also Appendix 6.1 ).

\section{Wallstrom's Criticism}

In the previous section, we referred to the correspondence between
the HJM equations and (2.23) as only formal because we had not considered
the boundary conditions that must be imposed on solutions of the Schr\"{o}dinger 
equation and the HJM equations, respectively, in order for mathematical
equivalence to be established. In standard quantum mechanics, it is
well-known that physical wave functions satisfying the Schr\"{o}dinger 
equation are required to be single-valued. For the HJM equations,
it was not specified in the Nelson-Yasue derivation whether \textit{S}
is assumed to be single-valued, arbitrarily multi-valued, or multi-valued
in accordance with a quantization condition. Wallstrom \cite{Wallstrom1989,Wallstrom1994}
showed that for all existing formulations of stochastic mechanics,
all these possible conditions on $S$ are problematic in one way or
another.

If $S$ is constrained to be single-valued, then stochastic mechanical
theories exclude single-valued Schr\"{o}dinger  wave functions with angular
momentum. This is so because single-valued wave functions with angular
momentum have phase factors of the form $\exp\left(i\mathrm{m}\varphi\right),$
where $\mathrm{m}$ is an integer and $\varphi$ is the azimuthal
angle, which implies that $S(\varphi)=\mathrm{m}\hbar\varphi$. By
contrast, if $S$ is assumed to be arbitrarily multi-valued, they
produce all the single-valued wave functions of the Schr\"{o}dinger  equation,
along with infinitely many multi-valued `wave functions', which smoothly
interpolate between the single-valued wave functions. This can be
seen by comparing solutions of the Schr\"{o}dinger  and HJM equations for
a two-dimensional central potential, $V(\mathbf{r})$ \cite{Wallstrom1989}.
The Schr\"{o}dinger  equation with $V(\mathbf{r})$ has single-valued wave
functions of the form $\psi_{\mathrm{m}}(\mathbf{r},\varphi)=R_{\mathrm{m}}(\mathrm{\mathbf{r}})exp(i\mathrm{m}\varphi),$
where $\psi_{\mathrm{m}}(\mathrm{\mathbf{r}},\varphi)=\psi_{\mathrm{m}}(\mathrm{\mathbf{r}},\varphi+2\pi n)$,
implying that $\mathrm{m}$ is an integer. For the HJM equations,
however, the solutions $\rho_{\mathrm{m}}=|R_{\mathrm{m}}(\mathrm{\mathbf{r}})|^{2}$
and $\boldsymbol{\mathbf{\mathrm{v}}}_{\mathrm{m}}=\left(\mathrm{m}\hbar/mr\right)\hat{\varphi}$
don't require $\mathrm{m}$ to be integral. To see this, consider
the effective central potential, $V_{a}(\boldsymbol{\mathrm{r}})=V(\boldsymbol{\mathrm{r}})+a/r^{2}$,
where $a$ is a positive real constant. For this potential, consider
the Schr\"{o}dinger  equation with stationary solution $\psi_{a}(\mathbf{r},\varphi)=R_{\mathrm{a}}(\mathrm{\mathbf{r}})exp(i\varphi)$,
where $\mathrm{m}=1$ and radial component corresponding to the ground
state solution of the radial equation. This wave function yields osmotic
and current velocities, $\mathbf{u}{}_{a}$ and $\mathrm{\mathbf{v}}{}_{a}$,
which satisfy (2.11) and (2.21) with the potential $V_{a}$:
\begin{equation}
0=\frac{\partial\rho{}_{a}}{\partial t}=-\nabla\cdot\left(\mathbf{v}{}_{a}\rho{}_{a}\right),
\end{equation}
\begin{equation}
0=\frac{\partial\mathbf{v}{}_{a}}{\partial t}=-\nabla\left(V+\frac{a}{r^{2}}\right)-\mathbf{v}{}_{a}\cdot\nabla\mathbf{v}{}_{a}+\mathbf{u}{}_{a}\cdot\nabla\mathbf{u}{}_{a}+\frac{\hbar^{2}}{2m}\nabla^{2}\mathbf{u}{}_{a}.
\end{equation}
Using $\mathbf{v}{}_{a}=\left(\hbar/mr\right)\hat{\varphi}$ and $\mathbf{v}{}_{a}\cdot\nabla\mathbf{v}{}_{a}=\nabla\left[m\mathbf{v}{}_{a}^{2}/2\right]$,
we can then rewrite (2.25) as
\begin{equation}
\begin{aligned}0 & =-\nabla V-\nabla\left(\frac{a}{r^{2}}+\frac{1}{2}m\mathbf{v}{}_{a}^{2}\right)+\mathbf{u}{}_{a}\cdot\nabla\mathbf{u}{}_{a}+\frac{\hbar^{2}}{2m}\nabla^{2}\mathbf{u}{}_{a}\\
 & =-\nabla V-\frac{m}{2}\nabla\left(\frac{2ma}{\hbar^{2}}+1\right)\mathbf{v}{}_{a}^{2}+\mathbf{u}{}_{a}\cdot\nabla\mathbf{u}{}_{a}+\frac{\hbar^{2}}{2m}\nabla^{2}\mathbf{u}{}_{a}.
\end{aligned}
\end{equation}
This gives us $\mathbf{v}'_{a}=\mathbf{v}{}_{a}\sqrt{2ma/\hbar^{2}+1}$
and $\mathbf{u}'_{a}=\mathbf{u}{}_{a}$. Note that since $a$ is a
constant that can take any positive real value, $\mathbf{v}'_{a}$
is not quantized, and yet it is a solution of the HJM equations. By
contrast, in the quantum mechanical version of this problem, we would
have $V_{a}(\boldsymbol{\mathrm{r}})=V(\boldsymbol{\mathrm{r}})+\mathrm{m}^{2}/2r^{2}$,
where $\mathrm{m}=\sqrt{2ma/\hbar^{2}+1}$ would be integral due to
the single-valuedness condition on $\psi_{\mathrm{m}}$. In other
words, the $\mathbf{v}{}_{a}$ and $\mathbf{u}{}_{a}$ in stochastic
mechanics only correspond to a single-valued wave function when $a$
is an integer, and this is true of all systems of two dimensions or
higher. Equivalently, we may say that the HJM equations predict a
continuum of energy and momentum states for the particle, which smoothly
interpolate between the quantized energy and momentum eigenvalues
predicted by the quantum mechanical case. \footnote{Before Wallstrom's critiques, it was pointed out by Albeverio and
Hoegh-Krohn \cite{Albeverio1974} as well as Goldstein \cite{Goldstein1987}
that, for the cases of stationary bound states with nodal surfaces
that separate the manifold of diffusion into disjoint components,
Nelson's equations (the HJM equations and his stochastic differential
equations) contain more solutions than Schr\"{o}dinger 's equation. In
addition, Goldstein \cite{Goldstein1987} was the first to point out
that solutions exist to the HJM equations which don't correspond to
any single-valued solution of the Schr\"{o}dinger  equation, for the case
of a multiply-connected configuration space. Nevertheless, Wallstrom's
example of extraneous solutions is of a more general nature, as it
applies to a simply-connected space where the diffusion process is
not separated into disjoint components.}

The only condition on $S$ (and hence the current velocity $\mathbf{v}{}_{a}$)
that allows stochastic mechanics to recover all and only the single-valued
wave functions of the Schr\"{o}dinger  equation is the condition that the
change in $S$ around any closed loop $L$ in space (with time held
constant) is equal to an integer multiple of Planck's constant, \footnote{Wallstrom notes that Takabayasi \cite{Takabayasi1952} was first to
recognize the necessity of this quantization condition and suggests
{[}private communication{]} that priority of credit for this discovery
should go to him \cite{Wallstrom1994}. However, it seems that Takabayasi
only recognized this issue in the context of Bohm's 1952 hidden-variables
theory, even though F\'{e}nyes proposed the first formulation of stochastic
mechanics that same year \cite{Fenyes1952}. Wallstrom appears to
have been the first in the literature to recognize and discuss the
full extent of this inequivalence in the context of stochastic mechanical
theories.} or
\begin{equation}
\oint_{L}dS=\oint_{L}\nabla S\cdot d\mathbf{q}=nh.
\end{equation}

But this condition is arbitrary, Wallstrom argued, as there's no reason
in stochastic mechanics why the change in $S$ along $L$ should be
constrained to an integer multiple of $h$. Indeed, assuming this
condition amounts to assuming that wave functions are single-valued,
which amounts to assuming that the solution space of the Nelson-Yasue
stochastic mechanical equations is equivalent to the solution space
of the quantum mechanical Schr\"{o}dinger  equation. Such an assumption
cannot be made, however, in a theory purporting to \textit{\textcolor{black}{derive}}
the Schr\"{o}dinger  equation of quantum mechanics.

These arguments notwithstanding, one might question whether the requirement
of single-valued wave functions in quantum mechanics is any less arbitrary
than imposing (2.27) in stochastic mechanics. This is not the case.
The single-valuedness condition, as usually motivated, is a consequence
of imposing two completely natural boundary conditions on solutions
of (2.23): (a) that the solutions satisfy the linear superposition
principle \cite{Schroedinger1938,Wallstrom1989}, and (b) that $|\psi|^{2}$
can be physically interpreted as a probability density \cite{Bohm1951,Merzbacher1962,Matthews1979}.
\footnote{Henneberger et al. \cite{Henneberger1994} argue that the single-valuedness
condition on wave functions is strictly a consequence of the linear
superposition principle. However, this nuance is inessential to our
arguments. } Condition (a) is natural to the single-valuedness requirement because
of the linearity of the Schr\"{o}dinger  equation, and condition (b) is
natural to it because a probability density is, by definition, a single-valued
function on its sample space. Moreover, it can be shown that if (a)
doesn't hold then (b) doesn't hold for any linear superposition of
two or more solutions. To illustrate this, consider the free particle
Schr\"{o}dinger  equation on the unit circle, $\textrm{S}^{1}$: \footnote{This argument was relayed to the author by T. Wallstrom {[}private
communication{]}.}
\begin{equation}
-\frac{\hbar^{2}}{2m}\frac{1}{r^{2}}\frac{\partial^{2}\psi}{\partial\theta^{2}}=E\psi.
\end{equation}
The un-normalized wave function satisfying this equation is of the
form $\psi(\theta)=Ne^{ik\theta}$, where $k=\frac{r}{\hbar}\sqrt{2mE}$.
For this wave function to satisfy (b), $k$ (and hence the energy
$E$) can take any positive value among the real numbers since obviously
$|\psi|^{2}=N^{2}$. Consider now a superposition of the form $\psi_{s}(\theta)=N\left(e^{ik_{1}\theta}+e^{ik_{2}\theta}\right)$,
which leads to the density
\begin{equation}
|\psi_{s}|^{2}=2N^{2}\left(1+cos\left[(k_{1}-k_{2})\theta\right]\right).
\end{equation}

If $k_{1}$ and $k_{2}$ are allowed to take non-integer values, then
$(k_{1}-k_{2})$ can also take non-integer values, and the density
formed from the superposition can be multi-valued, thereby violating
(b). Condition (a) will also be violated since, although a single
wave function in the superposition satisfies (b), the superposition
does not; so the set of wave functions of the form $\psi(\theta)=Ne^{ik\theta}$,
where $k$ can take non-integer values, does not form a linear space.
If, however, $k_{1}$ and $k_{2}$ are integers, then so is $(k_{1}-k_{2})$,
and conditions (a) and (b) will be satisfied since $|\psi_{s}|^{2}$
will always be single-valued. Correspondingly, it follows that the
energy and momentum of the particle on the unit circle will be quantized
with $e^{i2\pi\frac{r}{\hbar}\sqrt{2mE}}=1=e^{i2\pi n}$ yielding
$E_{n}=\frac{p_{\theta}^{2}}{2mr^{2}}=\frac{n^{2}\hbar^{2}}{2mr^{2}}$,
where $n$ is an integer.

The wave functions constructed from stochastic mechanics will therefore
satisfy only (b) if $S$ is arbitrarily multi-valued, while they will
satisfy (a) and (b) together only when (2.27) is imposed. But as previously
mentioned, (2.27) is ad hoc in stochastic mechanics, and assuming
it to obtain only single-valued wave functions is logically circular
if the objective of stochastic mechanics is to derive quantum mechanics.
The challenge then is to find a physically plausible justification
for (2.27) strictly within the assumptions of existing formulations
of stochastic mechanics, or otherwise some new formulation. Accordingly,
we shall now begin the development of our proposed justification through
a reformulation of Nelson-Yasue stochastic mechanics (NYSM).

\section{Classical Model of Constrained Zitterbewegung Motion}

Here we develop a classical model of a particle of mass \emph{m} constrained
in its rest frame to undergo a simple harmonic oscillation of (electron)
Compton frequency, and show that it gives rise to a quantization condition
equivalent to (2.27). Our model motivates the quantization condition
from essentially the same physical arguments used by de Broglie in
his ``phase-wave'' model \cite{Broglie1925,Darrigol1994} and by
Bohm in his subquantum field-theoretic models \cite{Bohm1957,Bohm2002}.
However, it differs from both de Broglie's model and Bohm's models
in that we do not need to refer to fictitious ``phase waves'', nor
assume that our particle is some localized distribution of a (hypothetical)
fluctuating subquantum field \cite{Bohm1957}, nor assume a non-denumerable
infinity of ``local clocks'' at each point in space-time \cite{Bohm2002}.
We start by developing the free particle case, extend it to a classical
Hamilton-Jacobi (HJ) statistical mechanical description, and repeat
these steps with the inclusion of interactions with external fields.

The purpose of this section is three-fold: (i) to explicitly show,
without the added conceptual complications of stochastic mechanics,
the basic physical assumptions underlying our particle model; (ii)
to show how our model can be consistently generalized to include interactions
with external fields; (iii) to show, using a well-established formulation
of classical statistical mechanics that has conceptual and mathematical
similarities to stochastic mechanics, how our model can be consistently
generalized to a statistical ensemble description (which will also
be necessary in the stochastic mechanical case), and how doing so
gives a quantization condition equivalent to (2.27) for a `classical'
wave function satisfying a nonlinear Schr\"{o}dinger  equation. No attempt
will be made here to suggest a physical/dynamical model for the zitterbewegung
motion. A framework for a physical model is given in section 5, while
a discussion of possible physical models is reserved for Part II.

\subsection{One free particle}

Suppose that a classical particle of rest mass $m$ is rheonomically
constrained to undergo a periodic process with constant angular frequency,
$\omega_{0}$, about some fixed point in 3-space, $\mathbf{q}_{0}$,
in a Lorentz frame where the particle has translational velocity $\mathbf{v}=d\mathbf{q}_{0}/dt=0$.
The exact nature of this process is not important for the argument
that follows, as long as it is periodic. For example, this process
could be an oscillation or (if the particle is spinning) a rotation.
But since we are considering the spinless case, we will take the periodic
process to be some kind of oscillation. The constancy of $\omega_{0}$
implies that the oscillation is simply harmonic with phase $\theta=\omega_{0}t_{0}+\phi$.
Although the assumption of simple harmonic motion implies that $\theta$
is a continuous function of the particle's position, in the translational
rest frame, it must be the case that the phase change $\delta\theta$
at any fixed instant $t_{0}$ will be zero for some translational
displacement $\delta\mathbf{q}_{0}$. Otherwise, such a displacement
would define a preferred direction in space given by $\nabla\theta(\mathbf{q}_{0})$.
Hence, in the translational rest frame, we can write
\begin{equation}
\delta\theta=\omega_{0}\delta t_{0},
\end{equation}
where $\delta t_{0}$ is the change in proper time.

If we Lorentz transform to the lab frame where the particle has constant
translational velocity, $\mathbf{v}$, and undergoes a displacement
$\delta\mathbf{q}(t)$ in $\delta t$, then $\delta t_{0}=\gamma\left(\delta t-\mathbf{v}\cdot\delta\mathbf{q}(t)/c^{2}\right)$
and (2.30) becomes
\begin{equation}
\delta\theta(\mathbf{q}(t),t)=\omega_{0}\gamma\left(\delta t-\frac{\mathbf{v}\cdot\delta\mathbf{q}(t)}{c^{2}}\right),
\end{equation}
where $\gamma=1/\sqrt{\left(1-\mathbf{v}^{2}/c^{2}\right)}$. Recalling
that for a relativistic free particle we have $E=\gamma mc^{2}$ and
$\mathbf{p}=\gamma m\mathbf{v}$, (2.31) can be equivalently expressed
as
\begin{equation}
\delta\theta(\mathbf{q}(t),t)=\frac{\omega_{0}}{mc^{2}}\left(E\delta t-\mathbf{p}\cdot\delta\mathbf{q}(t)\right).
\end{equation}
Suppose now that the oscillating particle is physically or virtually
\footnote{Because we permit a virtual displacement where time changes, we cannot
use the definition of a virtual displacement often found in textbooks
\cite{H.Goldstein2011,JoseSaletan} (which assumes time is fixed under
the displacement). Instead, we use the more refined definition of
virtual displacements proposed by Ray \& Shamanna \cite{Ray2006},
namely that a virtual displacement is the difference between any two
(unequal) ``allowed displacements'', or $\delta\mathbf{q}_{k}=d\mathbf{q}_{k}-d\mathbf{q}_{k}^{'}$,
where $k=1,2,...,N,$ and an allowed displacement is defined as $d\mathbf{q}_{k}=\mathbf{v}_{k}dt,$
where $\mathbf{v}_{k}$ are the ``virtual velocities'', or the velocities
allowed by the mechanical constraints of a given system.} displaced around a closed loop $L$ (i.e., a continuous, non-self-intersecting
loop that is otherwise arbitrary) in which both position and time
can vary. The consistency of the model requires that the accumulated
phase change be given by
\begin{equation}
\oint_{L}\delta\theta(\mathbf{q}(t),t)=\frac{\omega_{0}}{mc^{2}}\oint_{L}\left(E\delta t-\mathbf{p}\cdot\delta\mathbf{q}(t)\right)=2\pi n,
\end{equation}
where $n$ is an integer. This follows from the assumption that the
oscillation is simply harmonic in the particle's rest frame, which
makes $\theta$ in the lab frame a single-valued function of $\mathbf{q}(t)$
(up to an additive integer multiple of $2\pi$). Indeed, if (2.33)
were not true, we would contradict our hypothesis that the oscillating
particle has a well-defined phase at each point along its space-time
trajectory.

If we further make the `zitterbewegung' (\emph{zbw}) hypothesis that
$m=m_{e}=9.11\times10^{-28}g$ and $\omega_{0}/m_{e}c^{2}=1/\hbar$
so that $\omega_{0}=\omega_{c}=7.77\times10^{20}rad/s,$ which is
the electron Compton angular frequency, then we can define $\bar{\theta}\eqqcolon-\frac{1}{\hbar}S$
and (2.33) can be rewritten as
\begin{equation}
\oint_{L}\delta S(\mathbf{q}(t),t)=\oint_{L}\left(\mathbf{p}\cdot\delta\mathbf{q}(t)-E\delta t\right)=nh.
\end{equation}
Finally, for the special case of loop integrals in which time is held
fixed ($\delta t=0$), (2.34) reduces to
\begin{equation}
\oint_{L}\mathbf{p}\cdot\delta\mathbf{q}(t)=nh,
\end{equation}
which we may observe is formally identical to the Bohr-Sommerfeld-Wilson
quantization condition.

By integrating (2.32) and using
the Legendre transformation, it can be shown that the phase of the
free \emph{zbw} particle is, equivalently, its relativistic action
up to an additive constant, or $S(\mathbf{q}(t),t)=\mathbf{p}\cdot\mathbf{q}(t)-Et-\hbar\phi=-mc^{2}\int_{t_{i}}^{t}dt'/\gamma+C$.
\footnote{The proof is as follows. From $L=-mc^{2}/\gamma$, the Legendre transform
gives $E=\mathbf{p}\cdot\mathbf{v}-L=\gamma mv^{2}+mc^{2}/\gamma=\gamma mc^{2}$
and $L=\mathbf{p}\cdot\mathbf{v}-E.$ So for the free \emph{zbw} particle,
$S=\int Ldt+C=\int\left(\mathbf{p}\cdot\mathbf{v}-E\right)dt+C=\int\left(\mathbf{p}\cdot d\mathbf{q}-Edt\right)+C=\mathbf{p}\mathbf{\cdot q}-Et+C$
(absorbing the integration constants arising from $d\mathbf{q}$ and
$dt$ into \emph{C}).}, where $\phi$ is the initial phase constant. Recognizing also that $\mathbf{p}=\hbar\gamma\omega_{c}\mathbf{v}/c^{2}=\hbar\gamma\mathbf{k}$
and $E=\hbar\gamma\omega_{c}$, the translational 3-velocity of the
particle can be obtained from $S(\mathbf{q}(t),t)$ as $\mathbf{v}=(1/\gamma m)\nabla S(\mathbf{q},t)|_{\mathbf{q}=\mathbf{q}(t)}$,
and the total relativistic energy as $E=-\partial_{t}S(\mathbf{q},t)|_{\mathbf{q}=\mathbf{q}(t)}$.
It follows then that $S(\mathbf{q}(t),t)$ is a solution of the classical
relativistic Hamilton-Jacobi equation,
\begin{equation}
-\partial_{t}S(\mathbf{q},t)|_{\mathbf{q}=\mathbf{q}(t)}=\sqrt{m^{2}c^{4}+\left(\nabla S(\mathbf{q},t)\right)^{2}c^{2}}|_{\mathbf{q}=\mathbf{q}(t)}.
\end{equation}
In the non-relativistic limit, $v\ll c$, $S(\mathbf{q}(t),t)\approx m\mathbf{v}\cdot\mathbf{q}(t)-\left(mc^{2}+\frac{mv^{2}}{2}\right)t-\hbar\phi$,
and (2.36) becomes
\begin{equation}
-\partial_{t}S(\mathbf{q},t)|_{\mathbf{q}=\mathbf{q}(t)}=\frac{\left(\nabla S(\mathbf{q},t)\right)^{2}}{2m}|_{\mathbf{q}=\mathbf{q}(t)}+mc^{2},
\end{equation}
where $\mathbf{v}=\left(1/m\right)\nabla S|_{\mathbf{q}=\mathbf{q}(t)}=(1/m)\hbar\mathbf{k}$
and satisfies the trivial classical Newtonian equation
\begin{equation}
m\mathbf{a}=\left(\frac{\partial}{\partial t}+\mathbf{v}\cdot\nabla\right)\nabla S=0.
\end{equation}
We find then that, in the non-relativistic limit, the oscillation
frequency of the \emph{zbw} particle has two parts - a low frequency
oscillation, $\omega_{dB}=\hbar k^{2}/2m$, which modulates the high
frequency oscillation $\omega_{c}$.

Evidently (2.37) has the form of the non-relativistic dispersion relation
$E=\hbar^{2}k^{2}/2m+mc^{2}$, which naively suggests that one can
obtain the free-particle Schr\"{o}dinger  equation for a plane wave by
introducing operators $\hat{p}=-i\hbar\nabla$ and $\hat{E}=i\hbar\partial_{t}$
such that $\hat{p}\psi=\hbar k\psi$, $\hat{E}\psi=\hbar\omega\psi$,
and $i\hbar\partial_{t}\psi=-\left(\hbar^{2}/2m\right)\nabla^{2}\psi$
for $\psi(\mathbf{q},t)=Ae^{i\left(\mathbf{p}\cdot\mathbf{q}-Et\right)/\hbar}.$
However, there is no physical wave for such a plane wave to be identified
with in our model. Such a plane wave and Schr\"{o}dinger  equation are
nothing more than abstract, mathematically equivalent re-writings
of the \emph{zbw} particle energy equation (2.37). On the other hand,
as we will see next, a \textit{\textcolor{black}{nonlinear}} Schr\"{o}dinger 
equation that describes the dynamical evolution of a statistical ensemble
of identical \emph{zbw} particles is derivable from the classical
HJ description of the ensemble.

\subsection{Classical Hamilton-Jacobi statistical mechanics for one free particle}

Suppose that the actual position and momentum of a \emph{zbw} particle,
$\left(\mathbf{q}(t),\mathbf{p}(t)\right)$, are unknown. Then we
must resort to the description of a classical (i.e., Gibbsian) statistical
ensemble of fictitious, identical, non-interacting \emph{zbw} particles
\cite{Holland1993}, which differ from each other only by virtue of
their initial positions, velocities, and (possibly) phases. (Consideration
of this in the classical context will be helpful for seeing how our
model can be incorporated into stochastic mechanics.) In terms of
the \emph{zbw} phase, this change in description corresponds to replacing
$\delta S(\mathbf{q}(t),t)$ by $dS(\mathbf{q},t)=\mathbf{p}(\mathbf{q},t)\cdot\mathbf{\mathit{d}q}-E(\mathbf{q},t)dt$,
which we obtained from replacing $\mathbf{q}(t)$ by $\mathbf{q}$,
where $\mathbf{q}$ labels a \textit{\textcolor{black}{possible}}
position in 3-D space that the actual \emph{zbw} particle could occupy
at time $t$. Integrating $dS(\mathbf{q},t)$ then gives $S(\mathbf{q},t)=\int\mathbf{p}(\mathbf{q},t)d\mathbf{q}-\int E(\mathbf{q},t)dt+C$,
where $C=\hbar\phi$ is just the initial phase constant. So $S(\mathbf{q},t)$
is a phase \textit{\textcolor{black}{field}} connected with the ensemble,
$\mathbf{p}(\mathbf{q},t)=\nabla S(\mathbf{q},t)$ is the corresponding
translational momentum field, and $E(\mathbf{q},t)=-\partial_{t}S(\mathbf{q},t)$
is the total energy field. Note that, for any initial $\mathbf{q}$
and $t$, the constant $\phi$ can be given any value on the interval
$\left[0,2\pi\right]$; i.e., the initial phase constant associated
with any member of the ensemble can be freely specified on that interval.
(Of course, this phase constant does not affect the momentum field
or the total energy field, as these fields are obtained from space-time
derivatives of the phase field. Thus there are many phase fields corresponding
to a unique momentum field and total energy field.)

Now, in the specific case of the free \emph{zbw} particle, $p=const$
and $E=const$ for each member of the ensemble. So the infinitesimal
phase change connected with the ensemble is just $dS(\mathbf{q},t)=\mathbf{p}\cdot\mathbf{\mathit{d}q}-Edt$,
yielding $S(\mathbf{q},t)=\mathbf{p}\cdot\mathbf{q}-Et+C$ upon integration.

With this phase field in hand, we can now construct a classical HJ
statistical mechanics for our \emph{zbw} particle. Essentially, $S(\mathbf{q},t)$
and $\nabla S(\mathbf{q},t)$ will respectively satisfy the classical
Hamilton-Jacobi equation,
\begin{equation}
-\partial_{t}S(\mathbf{q},t)=\frac{\left(\nabla S(\mathbf{q},t)\right)^{2}}{2m}+mc^{2},
\end{equation}
and the trivial classical Newtonian equation,
\begin{equation}
m\mathbf{a}(\mathbf{q},t)=\left(\frac{\partial}{\partial t}+\mathbf{v}(\mathbf{q},t)\cdot\nabla\right)\nabla S(\mathbf{q},t)=0.
\end{equation}
If we now suppose that the density of ensemble particles per unit
volume in an element $d^{3}q$ surrounding the point $\mathbf{q}$
at time $t$ is given by the function $\rho(\mathbf{q},t)\geq0$,
which satisfies the normalization condition $\int\rho_{0}(\mathbf{q})d^{3}q=1$,
then it is straightforward to show \cite{Holland1993} that $\rho(\mathbf{q},t)$
evolves in time by the continuity equation
\begin{equation}
\frac{\partial\rho({\normalcolor \mathbf{q},t})}{\partial t}=-\nabla\cdot\left[\mathbf{\frac{\nabla\mathrm{\mathit{S\mathrm{(\mathbf{q},\mathit{t})}}}}{\mathit{m}}}\rho(\mathbf{q},t)\right].
\end{equation}
Accordingly, $\rho(\mathbf{q},t)$ carries the interpretation of the
probability density for the actual \emph{zbw} particle position $\mathbf{q}(t)$.
And since $S(\mathbf{q},t)$ is a field over the possible positions
that the actual \emph{zbw} particle can occupy at time \emph{t}, where
for each possible position the actual \emph{zbw} particle's phase
will satisfy the relation (2.35), $S(\mathbf{q},t)$ will be a single-valued
function of $\mathbf{q}$ and $t$ (up to an additive integer multiple
of $2\pi$) and satisfy
\begin{equation}
\oint_{L}dS(\mathbf{q},t)=\oint_{L}\nabla S(\mathbf{q},t)\cdot d\mathbf{q}=nh.
\end{equation}
The use of exact differentials in (2.42) indicates that the loop integral
is now an integral of the momentum field along any closed \textit{mathematical}
loop in 3-space with time held constant; that is, a closed loop around
which the actual particle with momentum $\mathbf{p}$ \emph{could
potentially be displaced}, starting from any possible position $\mathbf{q}$
it can occupy at fixed time $t$. This tells us that the circulation
of the momentum field is quantized, in contrast to an ordinary classical
statistical mechanical ensemble for which the momentum field circulation
need not satisfy (2.42).

Finally, we can combine (2.39) and (2.41) into the nonlinear Schr\"{o}dinger 
equation \cite{Schiller1962,Rosen1964,Holland1993,Ghose2002,Nikolic2006,Nikolic2007},
\begin{equation}
i\hbar\frac{\partial\psi(\mathbf{q},t)}{\partial t}=-\frac{\hbar^{2}}{2m}\nabla^{2}\psi(\mathbf{q},t)+\frac{\hbar^{2}}{2m}\frac{\nabla^{2}|\psi(\mathbf{q},t)|}{|\psi(\mathbf{q},t)|}\psi(\mathbf{q},t)+mc^{2}\psi(\mathbf{q},t),
\end{equation}
with general solution $\psi(\mathbf{q},t)=\sqrt{\rho_{0}(\mathbf{q}-\mathbf{v}_{0}t)}e^{iS(\mathbf{q},t)/\hbar}$,
which is single-valued because of (2.42). (Note that $C$ will contribute
a global phase factor, $e^{iC/\hbar}$, which cancels out from both
sides.) As an example of a specific solution, the complex phase $e^{iS/\hbar}$
takes the form of a plane-wave, $S=\mathbf{p}\cdot\mathbf{q}-Et+\hbar\phi$,
while the initial probability density, $\rho_{0}$, can take the form
of a Gaussian that propagates with fixed profile and speed $v$ (in
contrast to a Gaussian density in free particle quantum mechanics,
which disperses over time).

We have thereby shown that extending our free \emph{zbw} particle
model to a classical HJ statistical mechanics allows us to derive
a nonlinear Schr\"{o}dinger  equation with single-valued wave functions.
Next we will incorporate interactions of the \emph{zbw} particle with
external fields.

\subsection{One particle interacting with external fields}

To describe the interaction of our \emph{zbw} particle with fields,
let us reconsider the change in the \emph{zbw} phase in the rest frame.
In terms of the rest energy of the \emph{zbw} particle, we can rewrite
(2.30) as
\begin{equation}
\delta\theta=\omega_{c}\delta t_{0}=\frac{1}{\hbar}\left(mc^{2}\right)\delta t_{0}.
\end{equation}
Any additional contribution to the energy of the particle, such as
from a weak external gravitational field (e.g. the Earth's gravitational
field) coupling to the particle's mass $m$ via $\Phi_{g}=\mathbf{g\cdot q}$,
will then modify (2.44) as
\begin{equation}
\delta\theta=\left(\omega_{c}+\kappa(\mathbf{q})\right)|_{\mathbf{q}=\mathbf{q}_{0}}\delta t_{0}=\frac{1}{\hbar}\left(mc^{2}+m\Phi_{g}(\mathbf{q})\right)|_{\mathbf{q}=\mathbf{q}_{0}}\delta t_{0},
\end{equation}
where $\kappa=\omega_{c}\Phi_{g}/c^{2}$. In other words, the gravitational
field shifts the \emph{zbw} frequency in the rest frame by a very
small amount. For example, if $|\mathbf{g}|=\mathrm{10^{3}\mathit{cm/s^{2}}}$
and is in the $\mathbf{\hat{z}}$ direction, and we take $|\mathbf{q}|=100cm$,
then $\kappa\approx\omega_{c}\times10^{-16}.$ Here we have approximated
the point at which the \emph{zbw} particle interacts with the external
gravitational field to be just its equilibrium position, $\mathbf{q}_{0}$,
because the displacement $|\mathbf{q}|\gg\lambda_{c},$ allowing us
to approximate the interaction with the mass as point-like. \footnote{This appears to be the same assumption made by de Broglie for his
equivalent model, although he never explicitly says so. Bohm, to the
best of our knowledge, never extended his models to include field
interactions.}

In addition, we could allow the \emph{zbw} particle to carry charge
$e$ (so that it now becomes a classical charged oscillator, subject
to the hypothetical constraint that it does not radiate electromagnetic
energy in its rest frame, or the constraint that the oscillation of
the charge is radially symmetric so that there is no net energy radiated
\cite{Schott1,Schott2,Schott3}, or constrained to correspond to one
of the non-spherically-symmetric charge distributions considered by
Bohm and Weinstein \cite{BohmWeinstein1948} for which the retarded
self-fields cause the charge distribution to oscillate at a fixed
frequency without radiating) which couples to an external (and possibly
space-time varying) electric field such that $\Phi_{e}=\mathbf{E(q},t)\cdot\mathbf{q}$,
where \textbf{$\mathbf{q}$} is the displacement vector in some arbitrary
direction from the field source. Here again we can make the point-like
approximation, as in laboratory experiments the displacement of a
particle from a field source is typically on the centimeter scale,
making $|\mathbf{q}|\gg\lambda_{c}$). Then
\begin{equation}
\delta\theta=\left(\omega_{c}+\kappa(\mathbf{q}_{0})+\varepsilon(\mathbf{q}_{0},t_{0})\right)\delta t_{0}=\frac{1}{\hbar}\left(mc^{2}+m\Phi_{g}(\mathbf{q}_{0})+e\Phi_{e}(\mathbf{q}_{0},t_{0})\right)\delta t_{0},
\end{equation}
where $\varepsilon=\omega_{c}\left(e/mc^{2}\right)\Phi_{e}$. Assuming
$\mathbf{E}$ has an experimental value of $\sim10^{5}V/cm\approx.03stV/cm$,
which is the upper limit laboratory field strength that can be produced
in Stark effect experiments \cite{Bichsel2007}, and $|\mathbf{q}|=1cm$,
then $\varepsilon\approx\omega_{c}\times10^{-5}$, which is also a
very small shift.

If we now transform to the laboratory frame where the \emph{zbw} particle
has nonzero but variable translational velocity, (2.46) becomes
\begin{equation}
\begin{aligned}\delta\theta(\mathbf{q}(t),t) & =\left[\left(\omega_{dB}+\kappa(\mathbf{q})+\varepsilon(\mathbf{q})\right)\gamma\left(\delta t-\frac{\mathbf{v}_{0}(\mathbf{q},t)\cdot\delta\mathbf{q}}{c^{2}}\right)\right]_{\mathbf{q}=\mathbf{q}(t)}\\
 & =\frac{1}{\hbar}\left[\left(\gamma mc^{2}+\gamma m\Phi_{g}(\mathbf{q})+e\Phi_{e}(\mathbf{q},t)\right)\delta t\right.\\
 & \left.-\left(\gamma mc^{2}+\gamma m\Phi_{g}(\mathbf{q})+e\Phi_{e}(\mathbf{q},t)\right)\frac{\mathbf{v}_{0}(\mathbf{q},t)\cdot\delta\mathbf{q}}{c^{2}}\right]|_{\mathbf{q}=\mathbf{q}(t)}\\
 & =\frac{1}{\hbar}\left(E(\mathbf{q}(t),t)\delta t-\mathbf{p}(\mathbf{q}(t),t)\cdot\delta\mathbf{q}(t)\right),
\end{aligned}
\end{equation}
where $E=\gamma mc^{2}+\gamma m\Phi_{g}+e\Phi_{e}$ and $\mathbf{p}=m\mathbf{v}=\left(\gamma mc^{2}+\gamma m\Phi_{g}+e\Phi_{e}\right)\left(\mathbf{v}_{0}/c^{2}\right)$.
(Note that the term $e\Phi_{e}$ is unaffected by the Lorentz transformation
because it doesn't involve the particle's rest mass.) Here the velocity
$\mathbf{v}_{0}$ is that of a free particle, while $\mathbf{v}$
is the adjusted velocity due to the presence of external potentials.
In this moving frame, we can also have the \emph{zbw} particle couple
to an external magnetic vector potential \footnote{We could of course also include a gravitational vector potential,
but for simplicity we'll just stick with the magnetic version. } such that $\mathbf{v}\rightarrow\mathbf{v}'=\mathbf{v}+e\mathbf{A}_{ext}/\gamma mc$
(and $\gamma$ depends on $v$). Although the physical influence of
the fields now allows the $\omega$ and $\mathbf{k}$ of the particle
to vary as a function of position and time, the phase of the oscillation
is still a well-defined function of the particle's space-time location;
so if we displace the oscillating particle around a closed loop, the
phase change is still given by
\begin{equation}
\oint_{L}\delta\theta(\mathbf{q}(t),t)=\frac{1}{\hbar}\oint_{L}\left(E(\mathbf{q}(t),t)\delta t-\mathbf{p}'(\mathbf{q}(t),t)\cdot\delta\mathbf{q}(t)\right)=2\pi n,
\end{equation}
or
\begin{equation}
\oint_{L}\delta S(\mathbf{q}(t),t)=\oint_{L}\left(\mathbf{p}'(\mathbf{q}(t),t)\cdot\delta\mathbf{q}(t)-E(\mathbf{q}(t),t)\delta t\right)=nh.
\end{equation}
For the special case of a loop in which time is held fixed, we then
have
\begin{equation}
\oint_{L}\nabla S(\mathbf{\mathbf{q}},t)|_{\mathbf{q}=\mathbf{q}(t)}\cdot\delta\mathbf{q}(t)=\oint_{L}\mathbf{p}'(\mathbf{q}(t),t)\cdot\delta\mathbf{q}(t)=nh,
\end{equation}
or
\begin{equation}
\oint_{L}m\mathbf{v}(\mathbf{\mathbf{q}}(t),t)\cdot\delta\mathbf{q}(t)=nh-\frac{e}{c}\oint_{L}\mathbf{A}_{ext}(\mathbf{q}(t),t)\cdot\delta\mathbf{q}(t),
\end{equation}
where the last term on the right hand side of (2.51) is, by Stokes'
theorem, the magnetic flux enclosed by the loop.

We can also integrate (2.47) and rewrite in terms of $S(\mathbf{q}(t),t)$
to obtain
\begin{equation}
S(\mathbf{q}(t),t)=\int_{\mathbf{q}_{i}(t_{i})}^{\mathbf{q}(t)}\mathbf{p}'(\mathbf{q}(s),s)\cdot d\mathbf{q}(s)-\int_{t_{i}}^{t}E(\mathbf{q}(s),s)ds-\hbar\phi,
\end{equation}
where $\phi$ is the initial phase constant and (2.52) is equivalent
(up to an additive constant) to the relativistic action of a particle
in the presence of external fields. \footnote{The proof is as follows. From $L=-mc^{2}/\gamma-\gamma m\Phi_{g}-e\Phi_{e}+e\mathbf{\frac{v}{c}}\cdot\mathbf{A}_{ext}$,
the Legendre transform gives $E=\mathbf{p}'\cdot\mathbf{v}-L=\gamma mv{}^{2}+mc^{2}/\gamma+\gamma m\Phi_{g}+e\Phi_{e}=\gamma mc^{2}+\gamma m\Phi_{g}+e\Phi_{e}$
and $L=\mathbf{p}'\cdot\mathbf{v}-E$. So, $S=\int Ldt+C=\int\left(\mathbf{p}'\cdot\mathbf{v}-E\right)dt+C=\int\left(\mathbf{p}'\cdot d\mathbf{q}-Edt\right)+C$.} As before, the translational kinetic 3-velocity of the particle can
be obtained from $S(\mathbf{q}(t),t)$ as $\mathbf{v}(\mathbf{q}(t),t)=\mathbf{p}(\mathbf{q}(t),t)/\gamma m=(1/\gamma m)\nabla S(\mathbf{q},t)|_{\mathbf{q}=\mathbf{q}(t)}-e\mathbf{A}_{ext}(\mathbf{q}(t),t)/\gamma mc$,
and the total relativistic energy as $E(\mathbf{q}(t),t)=-\partial_{t}S(\mathbf{q},t)|_{\mathbf{q}=\mathbf{q}(t)}$.
It then follows that $S(\mathbf{q}(t),t)$ is a solution of the classical
relativistic Hamilton-Jacobi equation
\begin{equation}
-\partial_{t}S(\mathbf{q},t)|_{\mathbf{q}=\mathbf{q}(t)}=\sqrt{m^{2}c^{4}+\left(\nabla S(\mathbf{q},t)-\frac{e}{c}\mathbf{A}_{ext}(\mathbf{q},t)\right)^{2}c^{2}}|_{\mathbf{q}=\mathbf{q}(t)}+\gamma m\Phi_{g}(\mathbf{q}(t))+e\Phi_{e}(\mathbf{q}(t),t).
\end{equation}
When $v\ll c,$
\begin{equation}
\begin{aligned}S(\mathbf{q}(t),t) & \approx\int_{\mathbf{q}_{i}(t_{i})}^{\mathbf{q}(t)}m\mathbf{v'(q}(s),s)\cdot d\mathbf{q}(s)-\\
 & -\int_{t_{i}}^{t}\left(mc^{2}+\frac{1}{2m}\left[\mathbf{p}(\mathbf{q}(s),s)-\frac{e}{c}\mathbf{A}_{ext}(\mathbf{q}(s),s)\right]^{2}+m\Phi_{g}(\mathbf{q}(s))+e\Phi_{e}(\mathbf{q}(s),s)\right)ds-\hbar\phi,
\end{aligned}
\end{equation}
and (2.53) becomes
\begin{equation}
-\partial_{t}S(\mathbf{q},t)|_{\mathbf{q}=\mathbf{q}(t)}=\frac{\left(\nabla S(\mathbf{q},t)-\frac{e}{c}\mathbf{A}_{ext}(\mathbf{q},t)\right)^{2}}{2m}|_{\mathbf{q}=\mathbf{q}(t)}+mc^{2}+m\Phi_{g}(\mathbf{q}(t))+e\Phi_{e}(\mathbf{q}(t),t),
\end{equation}
with $\mathbf{v}(\mathbf{q}(t),t)=(1/m)\nabla S(\mathbf{q},t)|_{\mathbf{q}=\mathbf{q}(t)}-e\mathbf{A}_{ext}(\mathbf{q}(t),t)/mc$
and satisfies the classical Newtonian equation of motion,
\begin{equation}
\begin{aligned}m\mathbf{a}(\mathbf{q}(t),t) & =\left(\frac{\partial}{\partial t}+\mathbf{v}(\mathbf{q}(t),t)\cdot\nabla\right)\left[\nabla S(\mathbf{q},t)-\frac{e}{c}\mathbf{A}_{ext}(\mathbf{q},t)\right]|_{\mathbf{q}=\mathbf{q}(t)}\\
 & =-\nabla\left[m\Phi_{g}(\mathbf{q})+e\Phi_{e}(\mathbf{q},t)\right]|_{\mathbf{q}=\mathbf{q}(t)}-\frac{e}{c}\frac{\partial\mathbf{A}_{ext}(\mathbf{q},t)}{\partial t}|_{\mathbf{q}=\mathbf{q}(t)}+\frac{e}{c}\mathbf{v(q}(t),t)\times\mathbf{B}_{ext}(\mathbf{q}(t),t).
\end{aligned}
\end{equation}

Incidentally, if we choose $\Phi_{e}$ as the Coulomb potential for
the hydrogen atom and set $\mathbf{B}_{ext}=0$, then our model is
empirically equivalent to the Bohr model of the hydrogen atom (the
demonstration of this can be found in Appendix 6.2). As in the previous
section, we now want to extend our model to a classical HJ statistical
mechanics.

\subsection{Classical Hamilton-Jacobi statistical mechanics for one particle
interacting with external fields}

Suppose now that, in the lab frame with
$v\ll c$, we do not know the actual position $\mathbf{q}(t)$ of
the \emph{zbw} particle. Then the phase (2.54) becomes the phase field
\begin{equation}
\begin{aligned}S(\mathbf{q},t) & =\int_{\mathbf{q}(t_{i})}^{\mathbf{q}(t)}m\mathbf{v'(q}(s),s)\cdot\mathbf{\mathit{d}q}(s)|_{\mathbf{q}(t)=\mathbf{q}}\\
 & -\int_{t_{i}}^{t}\left(mc^{2}+\frac{1}{2m}\left[\mathbf{p}(\mathbf{q}(s),s)-\frac{e}{c}\mathbf{A}_{ext}(\mathbf{q}(s),s)\right]^{2}+m\Phi_{g}(\mathbf{q}(s))+e\Phi_{c}(\mathbf{q}(s),s)\right)ds|_{\mathbf{q}(t)=\mathbf{q}}-\hbar\phi.
\end{aligned}
\end{equation}
To obtain the equations of motion for $\ensuremath{S(\mathbf{q},t)}$
and $\ensuremath{\mathbf{v}(\mathbf{q},t)}$ we will apply the classical
analogue of Yasue's variational principle, in anticipation of the
method we will use for constructing ZSM.

First we introduce the ensemble-averaged action/phase functional (inputting
limits between initial and final states),
\begin{equation}
\begin{aligned}J & =\mathrm{E}\left[\int_{\mathbf{q}(t_{i})}^{\mathbf{q}(t_{f})}m\mathbf{v'}\cdot\mathbf{\mathit{d}q}(t)-\int_{t_{i}}^{t_{f}}\left(mc^{2}+\frac{1}{2m}\left[\mathbf{p}-\frac{e}{c}\mathbf{A}_{ext}\right]^{2}+m\Phi_{g}+e\Phi_{e}\right)dt-\hbar\phi\right]\\
 & =\mathrm{E}\left[\int_{t_{i}}^{t_{f}}\left\{ \frac{1}{2}m\mathbf{v}^{2}+\frac{e}{c}\mathbf{A}_{ext}\cdot\mathbf{v}-mc^{2}-m\Phi_{g}-e\Phi_{e}\right\} dt-\hbar\phi\right],
\end{aligned}
\end{equation}
where the equated expressions are related by the usual Legendre transformation.
Imposing the variational constraint,
\begin{equation}
J=extremal,
\end{equation}
a straightfoward computation exactly along the lines of that in Appendix
6.1  yields (2.56), which, upon replacing $\mathbf{q}(t)$ by $\mathbf{q}$,
corresponds to the classical Newtonian equation,
\begin{equation}
\begin{aligned}m\mathbf{a}(\mathbf{q},t) & =\left(\frac{\partial}{\partial t}+\mathbf{v}(\mathbf{q},t)\cdot\nabla\right)\left[\nabla S(\mathbf{q},t)-\frac{e}{c}\mathbf{A}_{ext}(\mathbf{q},t)\right]\\
 & =-\nabla\left[m\Phi_{g}(\mathbf{q})+e\Phi_{e}(\mathbf{q},t)\right]-\frac{e}{c}\frac{\partial\mathbf{A}_{ext}(\mathbf{q},t)}{\partial t}+\frac{e}{c}\mathbf{v(q},t)\times\mathbf{B}_{ext}(\mathbf{q},t),
\end{aligned}
\end{equation}
where $\mathbf{v}(\mathbf{q},t)=(1/m)\nabla S(\mathbf{q},t)-e\mathbf{A}_{ext}(\mathbf{q},t)/mc$
corresponds to the kinetic velocity field. By integrating both sides
and setting the integration constant equal to the rest mass, we then
obtain the classical Hamilton-Jacobi equation for (2.57),
\begin{equation}
-\partial_{t}S(\mathbf{q},t)=\frac{\left(\nabla S(\mathbf{q},t)-\frac{e}{c}\mathbf{A}_{ext}(\mathbf{q},t)\right)^{2}}{2m}+mc^{2}+m\Phi_{g}(\mathbf{q})+e\Phi_{e}(\mathbf{q},t).
\end{equation} 
Because the momentum field couples to the vector potential, it can
be readily shown that $\rho(\mathbf{q},t)$ now evolves by the modified
continuity equation
\begin{equation}
\frac{\partial\rho({\normalcolor \mathbf{q},t})}{\partial t}=-\nabla\cdot\left[\left(\mathbf{\frac{\nabla\mathrm{\mathit{S\mathrm{(\mathbf{q},\mathit{t})}}}}{\mathit{m}}}-\frac{e}{mc}\mathbf{A}_{ext}(\mathbf{q},t)\right)\rho(\mathbf{q},t)\right],
\end{equation}
which preserves the normalization, $\int\rho_{0}(\mathbf{q})d^{3}q=1$.
As before, $S(\mathbf{q},t)$ is a field over the possible positions
that the actual \emph{zbw} particle can occupy at time \emph{t}. Since
for each possible position the actual \emph{zbw} particle's phase
will satisfy the relation (2.50), $S(\mathbf{q},t)$ will be a single-valued
function of $\mathbf{q}$ and $t$ (up to an additive integer multiple
of $2\pi$) and
\begin{equation}
\oint_{L}\nabla S(\mathbf{q},t)\cdot d\mathbf{q}=nh.
\end{equation}
Finally, we can combined (2.61) and (2.62) into the nonlinear Schr\"{o}dinger 
equation,
\begin{equation}
i\hbar\frac{\partial\psi}{\partial t}=\frac{\left[-i\hbar\nabla-\frac{e}{c}\mathbf{A}_{ext}\right]^{2}}{2m}\psi+\frac{\hbar^{2}}{2m}\frac{\nabla^{2}|\psi|}{|\psi|}\psi+m\Phi_{g}\psi+e\Phi_{e}\psi+mc^{2}\psi,
\end{equation}
with wave function $\psi(\mathbf{q},t)=\sqrt{\rho(\mathbf{q},t)}e^{iS(\mathbf{q},t)/\hbar}$,
which is single-valued because of (2.63). (Again, $C$ will contribute
a global phase $e^{iC/\hbar}$ which drops out.)

\section{Zitterbewegung Stochastic Mechanics}

We are now ready to extend the classical \emph{zbw} model developed
in section 4 to Nelson-Yasue stochastic mechanics for all the same
cases. In doing so, we will show how this `zitterbewegung stochastic
mechanics' (ZSM) avoids the Wallstrom criticism and explain the `quantum-classical
correspondence' between the ZSM equations and the classical HJ statistical
mechanical equations. We will also apply ZSM to the central potential
problem considered by Wallstrom, to demonstrate how angular momentum
quantization emerges and therefore that the solution space of ZSM's
HJM equations is equivalent to the solution space of the quantum mechanical
Schr\"{o}dinger  equation.

\subsection{One free particle}

As in NYSM, we take as our starting point that a particle of rest
mass $m$ is immersed in Nelson's hypothesized ether and has a 3-space
coordinate $\mathbf{q}(t)$ undergoes a frictionless diffusion process
according to the stochastic differential equations,
\begin{equation}
d\mathbf{q}(t)=\mathbf{b}(\mathbf{q}(t),t)dt+d\mathbf{W}(t),
\end{equation}
for the forward-time direction, and
\begin{equation}
d\mathbf{q}(t)=\mathbf{b}_{*}(\mathbf{q}(t),t)dt+d\mathbf{W}_{*}(t),
\end{equation}
for the backward-time direction. As in NYSM, $d\mathbf{W}$ is the
Wiener process satisfying $\mathrm{E}_{t}\left[d\mathbf{W}\right]=0$
and $\mathrm{E}_{t}\left[d\mathbf{W}^{2}\right]=\left(\hbar/m\right)dt$.
Now, in order to incorporate the \emph{zbw} oscillation as a property
of the particle, we must amend Nelson's original phenomenological
hypotheses about his ether and particle with the following additional
hypotheses of phenomenological character: \footnote{Meaning, we will follow Nelson's approach of provisionally not offering
an explicit physical model of the ether, and de Broglie-Bohm's approach
of provisionally not offering an explicit physical model for the \emph{zbw}
particle, beyond the hypothetical characteristics listed here. However,
these characteristics should be regarded as general constraints on
any future physical model of Nelson's ether, the \emph{zbw} particle,
and the dynamical coupling between the two.} 
\begin{enumerate}
\item Nelson's ether is not only a stochastically fluctuating medium in
space-time, but an oscillating medium with a spectrum of angular frequencies
superposed at each point in 3-space. More precisely, we imagine the
ether as a continuous (or effectively continuous) medium composed
of a countably infinite number of fluctuating, stationary, spherical
waves \footnote{These ether waves could be fundamentally continuous field variables
or perhaps collective modes arising from nonlinear coupling between
(hypothetical) discrete constituents of the ether. Both possibilities
are logically compatible with what follows.} superposed at each point in space, with each wave having a different
(constant) angular frequency, $\omega_{0}^{k}$, where $k$ denotes
the \emph{k}-th ether mode. (If we assume an upper frequency cut-off
for our modes as the inverse Planck time, this will imply an upper
bound on the Compton frequency of an elementary particle immersed
in the ether, as we will see from hypothesis 3 below.) The relative
phases between the modes are taken to be random so that each mode
is effectively uncorrelated with every other mode. 
\item The particle of rest mass \emph{m}, located in its instantaneous mean
forward translational rest frame (IMFTRF), i.e., the frame in which
$D\mathbf{q}(t)=\mathbf{b}(\mathbf{q}(t),t)=0$, at some point $\mathbf{q}_{0}$,
is bounded to a harmonic oscillator potential with fixed natural frequency
$\omega_{0}=\omega_{c}=\left(1/\hbar\right)mc^{2}$. In keeping with
the phenomenological approach of ZSM and the approach taken by de
Broglie and Bohm with their \emph{zbw} models, we need not specify
the precise physical nature of this harmonic oscillator potential.
This is task is left for a future physical model of the ZSM particle. 
\item The particle's center of mass, as a result of being immersed in the
ether, undergoes an approximately frictionless translational Brownian
motion (due to the homogeneous and isotropic ether fluctuations that
couple to the particle by possibly electromagnetic, gravitational,
or some other means), as already described by (2.65-66); and, in its
IMFTRF, undergoes a driven oscillation about $\mathbf{q}_{0}$ by
coupling to a narrow band of ether modes that resonantly peaks around
the particle's natural frequency. However, in order that the oscillation
of the particle doesn't become unbounded in its kinetic energy, there
must be some mechanism by which the particle dissipates energy back
into the ether modes so that, on the average, a steady-state equilibrium
regime is reached for the oscillation. That is to say, on some hypothetical
characteristic short time-scale, $\tau$, the average energy absorbed
from the driven oscillation by the resonant ether modes equals the
average energy dissipated back to the ether by the particle. We note
that the average, in the present sense, would be over the random phases
of the ether modes. (Here we are taking inspiration from stochastic
electrodynamics \cite{Boyer1975,Boyer1980}, where it has been shown
that a classical charged harmonic oscillator immersed in a classical
electromagnetic zero-point field has a steady-state regime where the
phase-averaged power absorbed by the oscillator balances the phase-averaged
power radiated by the oscillator back to the zero-point field, yielding
a steady-state oscillation at the natural frequency of the oscillator
\cite{Boyer1975,Boyer1980,Puthoff1987,HuangBatelaan2013,HuangBatelaan2015,Puthoff2016}.
However, in keeping with our phenomenological approach, we will not
propose a specific mechanism for this energy exchange in ZSM, only
provisionally assume that it occurs somehow.) Accordingly, we suppose
that, in this steady-state regime, the particle undergoes undergoes
a steady-state \emph{zbw} oscillation of angular frequency $\omega_{c}$
about $\mathbf{q}_{0}$ in its IMFTRF, as characterized by the `fluctuation-dissipation'
relation, $<H>_{steady-state}=\hbar\omega_{c}=mc^{2}$, where $<H>_{steady-state}$
is the conserved random-phase-average energy associated with the steady-state
oscillation. 
\end{enumerate}
It follows then that, in the IMFTRF, the mean forward steady-state
\emph{zbw} phase change is given by
\begin{equation}
\delta\bar{\theta}_{0+}\coloneqq\omega_{c}\delta t_{0}=\frac{mc^{2}}{\hbar}\delta t_{0},
\end{equation}
and the cumulative forward steady-state \emph{zbw} phase, obtained
from the indefinite integral of (2.67), is
\begin{equation}
\bar{\theta}_{0+}=\omega_{c}t_{0}+\phi=\frac{mc^{2}}{\hbar}t_{0}+\phi_{+},
\end{equation}
where $\phi_{+}$ is the initial (forward) phase constant.

The reason for starting our analysis with the IMFTRF goes back to
the fact that, before constraining the diffusion process to simultaneous
solutions of the forward and backward Fokker-Planck equations associated
to (2.65-66), neither the forward nor the backward stochastic differential
equations (2.65-66) have well-defined time reversals. So the forward
and backward stochastic differential equations describe independent,
time-asymmetric diffusion processes in opposite time directions, and
we must start by considering the steady-state \emph{zbw} phase in
each time direction separately. We chose to start with the more intuitive
forward time direction. 

For the \emph{zbw} particle in the instantaneous mean backward translational
rest frame (IMBTRF), i.e., the frame defined by $D_{*}\mathbf{q}(t)=\mathbf{b}_{*}(\mathbf{q}(t),t)=0$,
its mean backward steady-state \emph{zbw} phase change is given by
\begin{equation}
\delta\bar{\theta}_{0-}\coloneqq-\omega_{c}\delta t_{0}=-\frac{mc^{2}}{\hbar}\delta t_{0},
\end{equation}
and
\begin{equation}
\bar{\theta}_{0-}=\left(-\omega_{c}t_{0}\right)+\phi=\left(-\frac{mc^{2}}{\hbar}t_{0}\right)+\phi_{-}.
\end{equation}

Note that, in the above construction, both the diffusion coefficient
$\nu=\hbar/2m$ and the (reduced) \emph{zbw} period $T_{c}=1/\omega_{c}=\hbar/mc^{2}$
are scaled by $\hbar$. This is consistent with our hypothesis that
the ether is the common physical cause of both the frictionless diffusion
process and the steady-state \emph{zbw} oscillation. Had we not proposed
Nelson's ether as the physical cause of the steady-state \emph{zbw}
oscillation as well as the frictionless diffusion process, the occurrence
of $\hbar$ in both of these particle properties would be inexplicable
and compromising for the plausibility of our proposed modification
of NYSM.

It should be stressed here that it is not possible to talk of the
\emph{zbw} phase in a rest frame other than the IMFTRF or IMBTRF of
the \emph{zbw} particle, as we cannot transform to a frame in which
$d\mathbf{q}(t)/dt=0$, since such an expression is undefined for
the (non-differentiable) Wiener process.

Now suppose we Lorentz transform back to the lab frame. For the forward
time direction, this corresponds to a boost of (2.67) by $-\mathbf{b}(\mathbf{q}(t),t)$
where $\mathbf{b}(\mathbf{q}(t),t)\neq0$. Approximating the transformation
for non-relativistic velocities so that $\gamma=1/\sqrt{\left(1-\mathbf{b}^{2}/c^{2}\right)}\approx1+\mathbf{b}^{2}/2c^{2},$
the forward steady-state \emph{zbw} phase change (2.67) becomes
\begin{equation}
\begin{aligned}\delta\bar{\theta}_{+}(\mathbf{q}(t),t) & \coloneqq\frac{\omega_{c}}{mc^{2}}\mathrm{E}_{t}\left[E_{+}(D\mathbf{q}(t))\delta t-mD\mathbf{q}(t)\cdot\left(D\mathbf{q}(t)\right)\delta t\right]\\
 & =\frac{\omega_{c}}{mc^{2}}\mathrm{E}_{t}\left[E_{+}\delta t-m\mathbf{b}(\mathbf{q}(t),t)\cdot\delta\mathbf{q}_{+}(t)\right],
\end{aligned}
\end{equation}
where
\begin{equation}
E_{+}(D\mathbf{q}(t))=mc^{2}+\frac{1}{2}m\left|D\mathbf{q}(t)\right|^{2}=mc^{2}+\frac{1}{2}m\mathbf{b}^{2},
\end{equation}
neglecting the momentum term proportional to $\mathbf{b}^{3}/c^{2}$,
and where $\delta\mathbf{q}_{+}(t)$ in (2.71) corresponds to the
physical, translational, mean forward displacement of the \emph{zbw}
particle, defined by 
\begin{equation}
\delta\mathbf{q}_{+}(t)=\left[D\mathbf{q}(t)\right]\delta t=\mathbf{b}(\mathbf{q}(t),t)\delta t.
\end{equation}
The first line on the right hand side of (2.71) is the straightforward
stochastic generalization of the Lorentz-transformed classical \emph{zbw}
phase (just as Yasue's mean action functional (2.19) is the straightforward
stochastic generalization of the ordinary action functional in classical
mechanics \cite{Yasue1981a}) for non-relativistic velocities. Note,
however, that the conditional expectation $\mathrm{E}_{t}[...]$ in
(2.71) is redundant since the right hand side of (2.71) involves terms
depending only on the mean forward velocity $D\mathbf{q}(t)=\mathbf{b}(\mathbf{q}(t),t)$,
where $D$ already involves taking a conditional expectation (see
the definitions (2.13) and (2.14) in section 2). However, in the more
general case of a \emph{zbw} particle in an external potential $V_{ext}$,
a case we will consider in the next section, the conditional expectation
cannot be dropped since there will be an external-potential-dependent
term in $E_{+}$ that will depend directly on $\mathbf{q}(t)$ via
$V_{ext}(\mathbf{q}(t))$. The expectation will also be useful for
giving a natural connection between the integral of the time-symmetrized
analogue of (2.71) (which we will introduce shortly) and Yasue's mean
action functional, as we will show later in this section.

For the backward time direction, the Lorentz transformation to the
lab frame corresponds to a boost of (2.69) by $-\mathbf{b}_{*}(\mathbf{q}(t),t)$
where $\mathbf{b}_{*}(\mathbf{q}(t),t)\neq0$. Then the backward steady-state
\emph{zbw} phase change (2.69) becomes
\begin{equation}
\begin{aligned}\delta\bar{\theta}_{-}(\mathbf{q}(t),t) & \coloneqq\frac{\omega_{c}}{mc^{2}}\mathrm{E}_{t}\left[-E_{-}(D_{*}\mathbf{q}(t))\delta t+mD_{*}\mathbf{q}(t)\cdot\left(D_{*}\mathbf{q}(t)\right)\delta t\right]\\
 & =\frac{\omega_{c}}{mc^{2}}\mathrm{E}_{t}\left[-E_{-}\delta t+m\mathbf{b}_{*}(\mathbf{q}(t),t)\cdot\delta\mathbf{q}_{+}(t)\right],
\end{aligned}
\end{equation}
where
\begin{equation}
E_{-}(D_{*}\mathbf{q}(t))=mc^{2}+\frac{1}{2}m\left|D_{*}\mathbf{q}(t)\right|^{2}=mc^{2}+\frac{1}{2}m\mathbf{b}_{*}^{2},
\end{equation}
and where $\delta\mathbf{q}_{-}(t)$ in (2.74) corresponds to the
physical, translational, mean backward displacement of the \emph{zbw}
particle, defined by 
\begin{equation}
\delta\mathbf{q}_{-}(t)=\left[D_{*}\mathbf{q}(t)\right]\delta t=\mathbf{b}_{*}(\mathbf{q}(t),t)\delta t.
\end{equation}
(Notice that $\delta\mathbf{q}_{+}(t)$ and $\delta\mathbf{q}_{-}(t)$
are not equal in general since $\delta\mathbf{q}_{+}(t)-\delta\mathbf{q}_{-}(t)=(\mathbf{b}-\mathbf{b}_{*})\delta t\neq0$
in general.) Since, at this stage, the forward and backward steady-state
\emph{zbw} phase changes, (2.71) and (2.74), are independent of one
another, each must equal $2\pi n$ when integrated along a closed
loop $L$ in which both time and position change. Otherwise we will
contradict our hypothesis that, up to this point, the \emph{zbw} particle
has a well-defined steady-state phase at each point along its mean
space-time trajectory in the forward or backward time direction.

In the lab frame, the forward and backward stochastic differential
equations for the \emph{zbw} particle's translational motion are as
before
\begin{equation}
d\mathbf{q}(t)=\mathbf{b}(\mathbf{q}(t),t)dt+d\mathbf{W}(t),
\end{equation}
and
\begin{equation}
d\mathbf{q}(t)=\mathbf{b}_{*}(\mathbf{q}(t),t)dt+d\mathbf{W}_{*}(t),
\end{equation}
with corresponding Fokker-Planck equations
\begin{equation}
\frac{\partial\rho(\mathbf{q},t)}{\partial t}=-\nabla\cdot\left[\mathbf{b(}\mathbf{q},t)\rho(\mathbf{q},t)\right]+\frac{\hbar}{2m}\nabla^{2}\rho(\mathbf{q},t),
\end{equation}
and
\begin{equation}
\frac{\partial\rho(\mathbf{q},t)}{\partial t}=-\nabla\cdot\left[\mathbf{b}_{*}(\mathbf{q},t)\rho(\mathbf{q},t)\right]-\frac{\hbar}{2m}\nabla^{2}\rho(\mathbf{q},t).
\end{equation}
Restricting the diffusion process to simultaneous solutions of (2.79)
and (2.80) via
\begin{equation}
\mathbf{v}\coloneqq\frac{1}{2}\left[\mathbf{b}+\mathbf{b}_{*}\right]=\frac{\nabla S(\mathbf{q},t)}{m}
\end{equation}
and
\begin{equation}
\mathbf{u}\coloneqq\frac{1}{2}\left[\mathbf{b}-\mathbf{b}_{*}\right]=\frac{\hbar}{2m}\frac{\nabla\rho(\mathbf{q},t)}{\rho(\mathbf{q},t)}
\end{equation}
reduces the forward and backward Fokker-Planck equations to
\begin{equation}
\frac{\partial\rho({\normalcolor \mathbf{q}},t)}{\partial t}=-\nabla\cdot\left[\mathbf{\frac{\nabla\mathrm{\mathit{S\mathrm{(\mathbf{q},\mathit{t})}}}}{\mathit{m}}}\rho(\mathbf{q},t)\right],
\end{equation}
with $\mathbf{b}=\mathbf{v}+\mathbf{u}$ and $\mathbf{b}_{*}=\mathbf{v}-\mathbf{u}$.
We also follow Nelson in postulating the presence of an external osmotic
potential $U(\mathbf{q},t)$ which couples to the \emph{zbw} particle
as $R(\mathbf{q},t)=\mu U(\mathbf{q},t)$, and by the same reasoning
discussed in section 2, imparts an osmotic velocity $\nabla R/m=\left(\hbar/2m\right)\nabla\rho/\rho$.
We then have $\rho=e^{2R/\hbar}$ for all times. 

To obtain the 2nd-order time-symmetric mean dynamics for the translational
motion of the \emph{zbw} particle, we will use the variational principle
of Yasue. To do this, we must first define the time-symmetrized steady-state
phase change of the \emph{zbw} particle in the lab frame, via a symmetric
combination of the forward and backward steady-state \emph{zbw} phase
changes (2.71) and (2.74). This is natural to do since (2.71) and
(2.74) correspond to the same frame (the lab frame), and since (2.71)
and (2.74) are no longer independent of one another as a result of
the constraints (2.81-82). Taking the difference between (2.74) and
(2.71), we obtain (replacing $\delta t\rightarrow dt$, hence $\delta\mathbf{q}_{+,-}(t)\rightarrow d\mathbf{q}_{+,-}(t)$)
\begin{equation}
\begin{aligned}d\bar{\theta}(\mathbf{q}(t),t) & \coloneqq\frac{1}{2}\left[d\bar{\theta}_{+}(\mathbf{q}(t),t)-d\bar{\theta}_{-}(\mathbf{q}(t),t)\right]\\
 & =\frac{\omega_{c}}{mc^{2}}\mathrm{E}_{t}\left[E(D\mathbf{q}(t),D_{*}\mathbf{q}(t))dt-\frac{m}{2}\left(\mathbf{b}(\mathbf{q}(t),t)\cdot d\mathbf{q}_{+}(t)+\mathbf{b}_{*}(\mathbf{q}(t),t)\cdot d\mathbf{q}_{-}(t)\right)\right]\\
 & =\frac{\omega_{c}}{mc^{2}}\mathrm{E}_{t}\left[Edt-\frac{m}{2}\left(\mathbf{b}\cdot\frac{d\mathbf{q}_{+}(t)}{dt}+\mathbf{b}_{*}\cdot\frac{d\mathbf{q}_{-}(t)}{dt}\right)dt\right]\\
 & =\frac{\omega_{c}}{mc^{2}}\mathrm{E}_{t}\left[\left(E-\frac{m}{2}\left(\mathbf{b}\cdot\frac{d\mathbf{q}_{+}(t)}{dt}+\mathbf{b}_{*}\cdot\frac{d\mathbf{q}_{-}(t)}{dt}\right)\right)dt\right]\\
 & =\frac{\omega_{c}}{mc^{2}}\mathrm{E}_{t}\left[\left(E-\frac{m}{2}\left(\mathbf{b}^{2}+\mathbf{b}_{*}^{2}\right)\right)dt\right]\\
 & =\frac{\omega_{c}}{mc^{2}}\mathrm{E}_{t}\left[\left(E-\left(m\mathbf{v}\cdot\mathbf{v}+m\mathbf{u}\cdot\mathbf{u}\right)\right)dt\right]\\
 & =\frac{\omega_{c}}{mc^{2}}\mathrm{E}_{t}\left[\left(mc^{2}-\frac{1}{2}m\mathbf{v}^{2}-\frac{1}{2}m\mathbf{u}^{2}\right)dt\right],
\end{aligned}
\end{equation}
where, from (2.72) and (2.75), we have defined
\begin{equation}
E=\frac{1}{2}\left(E_{+}+E_{-}\right)=mc^{2}+\frac{1}{2}\left[\frac{1}{2}m\mathbf{b}^{2}+\frac{1}{2}m\mathbf{b}_{*}^{2}\right]=mc^{2}+\frac{1}{2}m\mathbf{v}^{2}+\frac{1}{2}m\mathbf{u}^{2},
\end{equation}
and where we have used (2.73) and (2.76) in (2.84). 

It is important to note that because $\bar{\theta}_{+}$ and $\bar{\theta}_{-}$
are no longer independent of one another, it is no longer the case
that $\delta\bar{\theta}_{+}$ and $\delta\bar{\theta}_{-}$ will
each equal $2\pi n$ when integrated along a closed loop $L$ in which
both time and position change. However, the consistency of our theory
does require that $\oint_{L}\delta\bar{\theta}=2\pi n$, otherwise
we would contradict our hypothesis that the \emph{zbw} particle, after
restricting to simultaneous solutions of (2.79) an (2.80), has a well-defined
and unique steady-state phase at each 3-space location it can occupy
at each time, regardless of time-direction. Note also that, without
the constraints (2.81-82), we would always have $\oint_{L}\delta\bar{\theta}_{+}=2\pi n$
and $\oint_{L}\delta\bar{\theta}_{-}=2\pi n$, hence $\oint_{L}\delta\bar{\theta}=0$.
In other words, a time-symmetrized ``phase'' defined from the subtractive
combination of $\bar{\theta}_{+}$ and $\bar{\theta}_{-}$, without
the constraints (2.81-82), would be globally single-valued instead
of single-valued up to an additive integer multiple of $2\pi$. 

Now, from the last line of (2.84), we can integrate and define the
time-symmetric steady-state phase-principal function as 
\begin{equation}
I(\mathbf{q}(t),t)=-\hbar\bar{\theta}(\mathbf{q}(t),t)\coloneqq\mathrm{E}\left[\int_{t_{i}}^{t}\left(\frac{1}{2}m\mathbf{v}^{2}+\frac{1}{2}m\mathbf{u}^{2}-mc^{2}\right)dt'-\hbar\phi\left|\mathbf{q}(t)\right.\right],
\end{equation}
where the expectation on the right hand side is now conditional on
the Nelsonian path $\mathbf{q}(t)$. (Note that the interchangeability
of the expectation and the time integral follows from Fubini's theorem
in stochastic calculus, since the integral of the conditional expectation
and the conditional expectation of the integral are both required
to be finite quantities here \cite{Klebaner2005}.) We note that (2.86)
is formally identical to the \emph{W} function introduced by Yasue
in \cite{Yasue1981a}, and from which Yasue shows that the variation
$\delta W/\delta\mathbf{q}(t)$ implies the current velocity relation
(2.81) with \emph{W} in place of \emph{S}. The latter result also
applies to (2.86), given the formal identicality between \emph{I}
and \emph{W}, however we will use a different approach to connect
$\nabla I$ with the current velocity (2.81). Also, whereas Yasue's
\emph{W} function isn't constrained to satisfy $\oint_{L}\delta W=nh$,
(2.86) does satisfy $\oint_{L}\delta I=nh$ since it is explicitly
defined in terms of the phase function $\bar{\theta}$. 

By a slight modification of (2.86), we can also define the steady-state
phase-action functional
\begin{equation}
\begin{aligned}J & \coloneqq I_{if}=\mathrm{E}\left[\int_{t_{i}}^{t_{f}}\left[\frac{1}{2}m\mathbf{v}^{2}+\frac{1}{2}m\mathbf{u}^{2}-mc^{2}\right]dt-\hbar\phi\right],\end{aligned}
\end{equation}
where $\phi$ is the initial phase constant, and where (2.87) differs
from (2.86) by the end-point at $t_{f}$ being fixed and $\mathrm{E}[...]$
being the absolute expectation. It is easily seen that (2.87) is just
Yasue's time-symmetric ensemble-averaged action functional, Eq. (2.19)
in section 2, with $V=0$, inclusion of the rest-energy term $-mc^{2}$,
and inclusion of the initial phase constant $\phi$. 

Note, also, that from the second to last line of (2.84), we can obtain
the cumulative, time-symmetric, steady-state phase at a time \emph{t}
as
\begin{equation}
\begin{aligned}\bar{\theta}(\mathbf{q}(t),t) & =\frac{\omega_{c}}{mc^{2}}\mathrm{E}\left[\int_{t_{i}}^{t}\left(E-\left(m\mathbf{v}\cdot\mathbf{v}+m\mathbf{u}\cdot\mathbf{u}\right)\right)dt'\left|\mathbf{q}(t)\right.\right]+\phi\\
 & =\frac{\omega_{c}}{mc^{2}}\mathrm{E}\left[\int_{t_{i}}^{t}\left(\left(E-m\mathbf{u}\cdot\mathbf{u}\right)-m\mathbf{v}\cdot\mathbf{v}\right)dt'\left|\mathbf{q}(t)\right.\right]+\phi\\
 & =\frac{\omega_{c}}{mc^{2}}\mathrm{E}\left[\int_{t_{i}}^{t}\left(H-m\mathbf{v}\cdot\mathbf{v}\right)dt'\left|\mathbf{q}(t)\right.\right]+\phi\\
 & =\frac{\omega_{c}}{mc^{2}}\mathrm{E}\left[\int_{t_{i}}^{t}\left(H-\frac{m}{4}\left(D\mathbf{q}(t')+D_{*}\mathbf{q}(t')\right)\cdot\left(D+D_{*}\right)\mathbf{q}(t')\right)dt'\left|\mathbf{q}(t)\right.\right]+\phi\\
 & =\frac{\omega_{c}}{mc^{2}}\mathrm{E}\left[\int_{t_{i}}^{t}Hdt'-\int_{\mathbf{q}(t_{i})}^{\mathbf{q}(t)}\frac{m}{2}\left(D\mathbf{q}(t')+D_{*}\mathbf{q}(t')\right)\cdot\mathrm{D}\mathbf{q}(t')\left|\mathbf{q}(t)\right.\right]+\phi,
\end{aligned}
\end{equation}
where 
\begin{equation}
H\coloneqq E-m\mathbf{u}\cdot\mathbf{u}=mc^{2}+\frac{1}{2}m\mathbf{v}^{2}-\frac{1}{2}m\mathbf{u}^{2},
\end{equation}
and where we have used the fact that $0.5\left(D+D_{*}\right)\mathbf{q}(t)=\left(\partial_{t}+\mathbf{v}\cdot\nabla\right)\mathbf{q}(t)$,
and $\mathbf{v}(\mathbf{q}(t),t)=\left(\partial_{t}+\mathbf{v}\cdot\nabla\right)\mathbf{q}(t)\eqqcolon\mathrm{D}\mathbf{q}(t)/\mathrm{D}t$,
and and $\mathrm{D}\mathbf{q}(t)=\left(\mathrm{D}\mathbf{q}(t)/\mathrm{D}t\right)dt$.
Now, given an integral curve $\mathbf{Q}(t)$ of the current velocity/momentum
field, i.e., a solution of
\begin{equation}
m\frac{d\mathbf{Q}(t)}{dt}=m\mathbf{v}(\mathbf{Q}(t),t)=\mathbf{p}(\mathbf{Q}(t),t)=\nabla S(\mathbf{q},t)|_{\mathbf{q}=\mathbf{Q}(t)},
\end{equation}
and given that $\bar{\theta}(\mathbf{q},t)=\bar{\theta}|_{\mathbf{q}(t)=\mathbf{q}}$
is a field on 3-space representing the possible phases that the actual
\emph{zbw} particle could have at a point $\mathbf{q}$ at time \emph{t}
(up to addition of a constant), we can also evaluate $\bar{\theta}(\mathbf{q},t)$
with respect to $\mathbf{Q}(t)$, which allows us to drop the conditional
expectation (since $\mathbf{Q}(t)$ is deterministic) to obtain 
\begin{equation}
\begin{aligned}\bar{\theta}(\mathbf{Q}(t),t) & =\frac{\omega_{c}}{mc^{2}}\int_{t_{i}}^{t}\left[H-m\mathbf{v}(\mathbf{Q}(t'),t')\cdot\frac{d\mathbf{Q}(t')}{dt'}\right]dt'+\phi\\
 & =\frac{\omega_{c}}{mc^{2}}\left[\int_{t_{i}}^{t}Hdt'-\int_{\mathbf{Q}(t_{i})}^{\mathbf{Q}(t)}\mathbf{p}\cdot d\mathbf{Q}(t')\right]+\phi.
\end{aligned}
\end{equation}
Here (2.91) corresponds to the time-symmetrized steady-state phase
of the \emph{zbw} particle in the lab frame, evaluated along the \emph{zbw}
particle's `time-symmetric mean trajectory', where the time-symmetric
mean trajectory corresponds to an integral curve of the current velocity
field, i.e., (2.90). That the time-symmetric mean trajectories should
correspond to integral curves of the current velocity field can be
seen from the fact that the single-time probability density $\rho(\mathbf{q},t)$,
after imposing (2.81-82), is a solution of the continuity equation
(2.83), from which it follows that the possible mean trajectories
of the \emph{zbw} particle are the flow lines of the probability current
$\rho\mathbf{v}$, i.e., the solutions of (2.90) for all possible
initial conditions $\mathbf{Q}(0)$.

Now, taking the total differential of the left hand side of (2.91)
gives
\begin{equation}
d\bar{\theta}=\nabla\bar{\theta}|_{\mathbf{q}=\mathbf{Q}(t)}d\mathbf{Q}(t)+\partial_{t}\bar{\theta}|_{\mathbf{q}=\mathbf{Q}(t)}dt.
\end{equation}
This allows us to identify 
\begin{equation}
\mathbf{p}(\mathbf{Q}(t),t)=-\left(\frac{mc^{2}}{\omega_{c}}\right)\nabla\bar{\theta}|_{\mathbf{q}=\mathbf{Q}(t)}=\nabla S|_{\mathbf{q}=\mathbf{Q}(t)},
\end{equation}
where we have used (2.92) along with (2.91) and (2.90). Thus the current
velocity of the \emph{zbw} particle can be identified with the gradient
of the \emph{zbw} particle's time-symmetrized steady-state phase with
respect to the location of the \emph{zbw} particle at time \emph{t}
in the lab frame, given the assumption that the current velocity is
integrable, i.e., given (2.81) and (2.90). Accordingly, the $S$ function
can be identified with (2.91). In addition, (2.92) along with (2.91)
relates the \emph{H} function to $\bar{\theta}$ (hence \emph{S})
by 
\begin{equation}
H(\mathbf{Q}(t))=\left(\frac{mc^{2}}{\omega_{c}}\right)\partial_{t}\bar{\theta}|_{\mathbf{q}=\mathbf{Q}(t)}=-\partial_{t}S|_{\mathbf{q}=\mathbf{Q}(t)}.
\end{equation}
From (2.94), (2.93), and (2.91), it follows that 
\begin{equation}
\begin{aligned}S(\mathbf{Q}(t),t) & =\int_{\mathbf{Q}(t_{i})}^{\mathbf{Q}(t)}\mathbf{p}\cdot d\mathbf{Q}(t')-\int_{t_{i}}^{t}Hdt'-\hbar\phi\\
 & =\int_{t_{i}}^{t}\left[\frac{1}{2}m\mathbf{v}(\mathbf{Q}(t'),t')^{2}+\frac{1}{2}m\mathbf{u}(\mathbf{Q}(t'),t')^{2}-mc^{2}\right]dt'-\hbar\phi=I(\mathbf{Q}(t),t),
\end{aligned}
\end{equation}
and
\begin{equation}
\oint_{L}\delta S(\mathbf{Q}(t),t)=\left(-\frac{mc^{2}}{\omega_{c}}\right)\oint_{L}\delta\bar{\theta}(\mathbf{q}(t),t)=\oint_{L}\left[\mathbf{p}\cdot\delta\mathbf{Q}(t)-H\delta t\right]=nh.
\end{equation}
We will use these last two expressions for later comparisons.

As an aside, let us recall that after restricting the forward and
backward diffusions to simultaneous solutions of (2.79-80), we had
$\mathbf{b}=\mathbf{v}+\mathbf{u}$ and $\mathbf{b}_{*}=\mathbf{v}-\mathbf{u}$.
So the IMFTRF and the IMBTRF will not coincide since for $\mathbf{b}=\mathbf{v}+\mathbf{u}=0$
it will not generally be the case that $\mathbf{b}_{*}=\mathbf{v}-\mathbf{u}=0$.
Nevertheless, we can define an instantaneous mean (time-)symmetric
rest frame (IMSTRF) as the frame in which $\mathbf{b}+\mathbf{b}_{*}=2\mathbf{v}=0$.
In the IMSTRF, (2.88) or (2.91) or (2.95) reduces to $\bar{\theta}=(\omega_{c}/mc^{2})\left[\left(mc^{2}-\frac{1}{2}m\mathbf{u}^{2}\right)t+\phi\right]$,
since $\mathbf{v}=0$ and $\partial_{t}\rho=0$. This shows that the
kinetic energy term due to the osmotic velocity contributes a tiny
shift to the steady-state \emph{zbw} phase (2.88) or (2.91) or (2.95)
in the IMSTRF (since, in the non-relativistic regime, $\mathbf{u}^{2}/c^{2}\ll1$). 

Returning now to (2.87), the imposition of the conservative-diffusions
constraint implies extremality of (2.87), which further implies (see
Appendix 6.1) Nelson's mean acceleration equation,
\begin{equation}
m\mathbf{a}(\mathbf{q}(t),t)=\frac{m}{2}\left[D_{*}D+DD_{*}\right]\mathbf{q}(t)=0.
\end{equation}
Computing the derivatives in (2.97), and using that $\mathbf{b}=\mathbf{v}+\mathbf{u}$
and $\mathbf{b}_{*}=\mathbf{v}-\mathbf{u}$, we obtain
\begin{equation}
\begin{aligned}m\mathbf{a}(\mathbf{q}(t),t) & =m\left[\frac{\partial\mathbf{v}(\mathbf{q},t)}{\partial t}+\mathbf{v}(\mathbf{q},t)\cdot\nabla\mathbf{v}(\mathbf{q},t)-\mathbf{u}(\mathbf{q},t)\cdot\nabla\mathbf{u}(\mathbf{q},t)-\frac{\hbar}{2m}\nabla^{2}\mathbf{u}(\mathbf{q},t)\right]|_{\mathbf{q}=\mathbf{q}(t)}\\
 & =\nabla\left[\frac{\partial S(\mathbf{q},t)}{\partial t}+\frac{\left(\nabla S(\mathbf{q},t)\right)^{2}}{2m}-\frac{\hbar^{2}}{2m}\frac{\nabla^{2}\sqrt{\rho(\mathbf{q},t)}}{\sqrt{\rho(\mathbf{q},t)}}\right]|_{\mathbf{q}=\mathbf{q}(t)}=0.
\end{aligned}
\end{equation}
Integrating both sides of (2.98) gives the total translational energy
of the \emph{zbw} particle along the stochastic trajectory $\mathbf{q}(t)$:
\begin{equation}
\tilde{E}(\mathbf{q}(t),t)=-\frac{\partial S(\mathbf{q},t)}{\partial t}|_{\mathbf{q}=\mathbf{q}(t)}=mc^{2}+\frac{\left(\nabla S(\mathbf{q},t)\right)^{2}}{2m}|_{\mathbf{q}=\mathbf{q}(t)}-\frac{\hbar^{2}}{2m}\frac{\nabla^{2}\sqrt{\rho(\mathbf{q},t)}}{\sqrt{\rho(\mathbf{q},t)}}|_{\mathbf{q}=\mathbf{q}(t)},
\end{equation}
where we have set the integration constant equal to the \emph{zbw}
particle's rest energy. Alternatively, we can again consider integral
curves of the current velocity/momentum field, but where now the integral
curves are obtained from solutions of 
\begin{equation}
m\frac{d^{2}\mathbf{Q}(t)}{dt^{2}}=m\left(\partial_{t}\mathbf{v}+\mathbf{v}\cdot\nabla\mathbf{v}\right)|_{\mathbf{q}=\mathbf{Q}(t)}=-\nabla\left(-\frac{\hbar^{2}}{2m}\frac{\nabla^{2}\sqrt{\rho(\mathbf{q},t)}}{\sqrt{\rho(\mathbf{q},t)}}\right)|_{\mathbf{q}=\mathbf{Q}(t)},
\end{equation}
i.e., the mean acceleration equation (2.98), rewritten so that only
the \textbf{v}-dependent terms are kept on the left hand side. Then
we can replace $\mathbf{q}(t)$ in (2.99) with $\mathbf{Q}(t)$ to
obtain the total translational energy associated with the \emph{zbw}
particle's time-symmetric mean trajectory, i.e., $\tilde{E}(\mathbf{Q}(t),t)$.
Moreover, we can express the solution of (2.99) in terms of $\mathbf{Q}(t)$,
thereby obtaining
\begin{equation}
\begin{aligned}S(\mathbf{Q}(t),t) & =\int_{\mathbf{Q}(t_{i})}^{\mathbf{Q}(t)}\mathbf{p}\cdot d\mathbf{Q}(t')-\int_{t_{i}}^{t}\tilde{E}dt'-\hbar\phi\\
 & =\int_{t_{i}}^{t}\left[\frac{1}{2}m\mathbf{v}(\mathbf{Q}(t'),t')^{2}-\left(-\frac{\hbar^{2}}{2m}\frac{\nabla^{2}\sqrt{\rho(\mathbf{Q}(t'),t')}}{\sqrt{\rho(\mathbf{Q}(t'),t')}}\right)-mc^{2}\right]dt'-\hbar\phi\\
 & =\int_{t_{i}}^{t}\left[\frac{1}{2}m\mathbf{v}^{2}+\frac{1}{2}m\mathbf{u}^{2}+\frac{\hbar}{2}\nabla\cdot\mathbf{u}-mc^{2}\right]dt'-\hbar\phi.
\end{aligned}
\end{equation}
We identify (2.101) as the conservative-diffusion-constrained, time-symmetrized,
steady-state phase (action) of the \emph{zbw} particle in the lab
frame, evaluated along an integral curve $\mathbf{Q}(t)$ obtained
from (2.100). 

Notice that the last line of (2.101) differs from the last line of
(2.95) only by addition of the term involving $\nabla\cdot\mathbf{u}$.
(The equality between the last two lines of (2.101) follows from the
well-known fact that the quantum kinetic can be decomposed as $\left(-\hbar^{2}/2m\right)\rho^{-1/2}\nabla^{2}\rho^{1/2}=0.5m\mathbf{u}^{2}-\left(\hbar^{2}/4m\right)\rho^{-1}\nabla^{2}\rho$
\cite{Bohm1952I}, and by the product rule, $0.5m\mathbf{u}^{2}-\left(\hbar^{2}/4m\right)\rho^{-1}\nabla^{2}\rho=-0.5m\mathbf{u}^{2}-m\left(\hbar/2m\right)\nabla\cdot\mathbf{u}$.)

Notice also that the equation of motion for (2.101) differs from the
equation of motion for the classical \emph{zbw} particle phase by
the presence of the quantum kinetic entering into (2.98-99). The two
phases might appear to be connected by the `classical limit' $(\hbar/2m)\rightarrow0$,
but this is only a formal connection since such a limit corresponds
to deleting the presence of the ether, thereby also deleting the physical
mechanism that we hypothesize to cause the \emph{zbw} particle to
oscillate at its Compton frequency. The physically realistic `classical
limit' for (2.98-99) corresponds to situations where the quantum kinetic
and quantum force are negligible. Such situations will arise (as in
the dBB theory) whenever the center of mass of a system of particles
is sufficiently large and environmental decoherence is appreciable
\cite{Allori2001,Bowm2005,Oriols2016,Derakhshani2017b}. 

Inasmuch as (2.101) is a well-defined phase function of the \emph{zbw}
particle's time-symmetric mean trajectory $\mathbf{Q}(t)$ in the
lab frame (because it was derived from applying the variational principle
to (2.87), the latter of which was defined in terms of (2.84), which
we argued must satisfy $\oint_{L}\delta\bar{\theta}=2\pi n$), if
we integrate $\delta S(\mathbf{Q}(t),t)$ around a closed loop $L$
in which time and position may change, we will have
\begin{equation}
\oint_{L}\delta S(\mathbf{Q}(t),t)=\oint_{L}\left[\mathbf{p}\cdot\delta\mathbf{Q}(t)-\tilde{E}\delta t\right]=nh,
\end{equation}
and for a special loop in which time is held fixed,
\begin{equation}
\oint_{L}\nabla S|_{\mathbf{q}=\mathbf{Q}(t)}\cdot\delta\mathbf{Q}(t)=\oint_{L}\mathbf{p}\cdot\delta\mathbf{Q}(t)=nh.
\end{equation}
Otherwise, we would contradict our hypothesis that the \emph{zbw}
particle still has a well-defined, time-symmetrized, steady-state
phase at each 3-space location it can occupy along a mean trajectory
$\mathbf{Q}(t)$ in either time direction, after the constraint of
conservative diffusions has been imposed. (Notice that (2.102) differs
from (2.96) by $\tilde{E}$ replacing $H$, and that $\tilde{E}-H=-(\hbar/2)\nabla\cdot\mathbf{u}$.)
If we also consider the time-symmetrized steady-state phase field,
$S(\mathbf{q},t)$, which is a field over the possible locations of
the actual \emph{zbw} particle (as described in section 4.2), then
by applying the same physical reasoning above to each possible initial
position that the \emph{zbw} particle can occupy, it follows that
the net change of the phase field along any mathematical loop in space
(with time held fixed) will be
\begin{equation}
\oint_{L}dS(\mathbf{q},t)=\oint_{L}\mathbf{p}\cdot d\mathbf{q}=nh.
\end{equation}
(The justification for (2.104) where $\rho=0$ is discussed in section
5.2, since such ``nodal points'' commonly arise in the presence
of bound states.) 

The total energy field $\tilde{E}(\mathbf{q},t)$ will correspondingly
be given by (2.99) when $\mathbf{Q}(t)$ is replaced by $\mathbf{q}$.
So with (2.104), (2.99), and (2.83), we can construct the 1-particle
Schr\"{o}dinger equation,
\begin{equation}
i\hbar\frac{\partial\psi(\mathbf{q},t)}{\partial t}=-\frac{\hbar^{2}}{2m}\nabla^{2}\psi(\mathbf{q},t)+mc^{2}\psi(\mathbf{q},t),
\end{equation}
where $\psi(\mathbf{q},t)=\sqrt{\rho(\mathbf{q},t)}e^{iS(\mathbf{q},t)/\hbar}$
is a single-valued wave function as a result of (2.104). As in the
classical case, the constant $C=\hbar\phi$ will contribute a global
phase factor $e^{iC/\hbar}$ which cancels out from both sides of
(2.105). We thereby have a formulation of free-particle ZSM that recovers
the usual free-particle Schr\"{o}dinger
equation.

\subsection{One particle interacting with external fields}

Suppose again that the particle undergoes a steady-state \emph{zbw}
oscillation in the IMFTRF, but now carries charge $e$ so that it
is a classical charged harmonic oscillator of some type (subject again
to the hypothetical constraint of no electromagnetic radiation emitted
when there is no translational motion; or the constraint that the
oscillation of the charge is radially symmetric so that there is no
net energy radiated; or, if the ether turns out to be electromagnetic
in nature as Nelson suggested \cite{Nelson1985}, then that the steady-state
\emph{zbw} oscillation is due to a balancing between the random-phase-averaged
electromagnetic energy absorbed from the charged harmonic oscillator's
driven oscillation, and the random-phase-averaged electromagnetic
energy radiated back to the ether, much like in stochastic electrodynamics
\cite{Boyer1975,Boyer1980,Puthoff1987,HuangBatelaan2013,HuangBatelaan2015,Puthoff2016}).
Then, in the presence of an external electric potential $\Phi_{e}(\mathbf{q}_{0}(t_{0}),t_{0})=\mathbf{E}_{ext}(\mathbf{q}_{0}(t_{0}),t_{0})\cdot\mathbf{q}_{0}(t_{0})$,
where \textbf{$\mathbf{q}_{0}(t_{0})$} is the positional displacement
of the \emph{zbw} particle in some arbitrary direction from the field
source (again making the point-like approximation for $|\mathbf{q}_{0}|\gg\lambda_{c}$)
and satisfies the forward stochastic differential equation (2.77)
with $\mathbf{b}=0$, the\emph{ zbw} phase change in this IMFTRF is
shifted by
\begin{equation}
\delta\bar{\theta}_{0+}=\mathrm{E}_{t}\left[\left(\omega_{c}+\varepsilon(\mathbf{q}_{0}(t_{0}),t_{0})\right)\delta t_{0}\right]=\frac{1}{\hbar}\left(mc^{2}\delta t_{0}+\mathrm{E}_{t}\left[e\Phi_{e}(\mathbf{q}_{0}(t_{0}),t_{0})\delta t_{0}\right]\right),
\end{equation}
where $\varepsilon(\mathbf{q}_{0}(t_{0}),t_{0})=\omega_{c}\left(e/mc^{2}\right)\Phi_{e}(\mathbf{q}_{0}(t_{0}),t_{0})$.
Direct integration gives 
\begin{equation}
\begin{aligned}\bar{\theta}_{0+} & =\mathrm{E}\left[\int_{t_{a}}^{t_{0}}\left(\omega_{c}+\varepsilon(\mathbf{q}_{0}(t'_{0}),t'_{0})\right)dt'_{0}\left|\mathbf{q}_{0}(t_{0})\right.\right]\\
 & =\frac{1}{\hbar}\left(mc^{2}t_{0}+\mathrm{E}\left[e\int_{t_{a}}^{t_{0}}\Phi_{e}(\mathbf{q}_{0}(t'_{0}),t'_{0})dt'_{0}\left|\mathbf{q}_{0}(t_{0})\right.\right]\right)+\phi.
\end{aligned}
\end{equation}
In the IMBTRF, 
\begin{equation}
\delta\bar{\theta}_{0-}=-\mathrm{E}_{t}\left[\left(\omega_{c}+\varepsilon(\mathbf{q}_{0}(t_{0}),t_{0})\right)\delta t_{0}\right]=-\frac{1}{\hbar}\left(mc^{2}\delta t_{0}+\mathrm{E}_{t}\left[e\Phi_{e}(\mathbf{q}_{0}(t_{0}),t_{0})\delta t_{0}\right]\right).
\end{equation}
Direct integration gives
\begin{equation}
\begin{aligned}\bar{\theta}_{0-} & =-\mathrm{E}\left[\int_{t_{a}}^{t_{0}}\left(\omega_{c}+\varepsilon(\mathbf{q}_{0}(t'_{0}),t'_{0})\right)dt'_{0}\left|\mathbf{q}_{0}(t_{0})\right.\right]\\
 & =-\frac{1}{\hbar}\left(mc^{2}t_{0}+\mathrm{E}\left[e\int_{t_{a}}^{t_{0}}\Phi_{e}(\mathbf{q}_{0}(t'_{0}),t'_{0})dt'_{0}\left|\mathbf{q}_{0}(t_{0})\right.\right]\right)+\phi.
\end{aligned}
\end{equation}
Now suppose we Lorentz transform back to the lab frame. For the forward
time direction, this corresponds to a boost of (2.106) by $-\mathbf{b}(\mathbf{q}(t),t)$
where $\mathbf{b}(\mathbf{q}(t),t)\neq0$. Approximating the transformation
for non-relativistic velocities so that $\gamma=1/\sqrt{\left(1-\mathbf{b}^{2}/c^{2}\right)}\approx1+\mathbf{b}^{2}/2c^{2},$
(2.106) becomes
\begin{equation}
\delta\bar{\theta}_{+}(\mathbf{q}(t),t)=\frac{\omega_{c}}{mc^{2}}\mathrm{E}_{t}\left[E_{+}(\mathbf{q}(t),D\mathbf{q}(t),t)\delta t-m\mathbf{b}(\mathbf{q}(t),t)\cdot\delta\mathbf{q}_{+}(t)\right],
\end{equation}
where
\begin{equation}
E_{+}(\mathbf{q}(t),D\mathbf{q}(t),t)=mc^{2}+\frac{1}{2}m\mathbf{b}^{2}+e\Phi_{e},
\end{equation}
neglecting the momentum term proportional to $\mathbf{b}^{3}/c^{2}$.
Again we take $\delta\mathbf{q}_{+}(t)$ to correspond to (2.73).
For the backward time direction, we have a boost of (2.108) by $-\mathbf{b}_{*}(\mathbf{q}(t),t)$
where $\mathbf{b}_{*}(\mathbf{q}(t),t)\neq0$, hence
\begin{equation}
\delta\bar{\theta}_{-}(\mathbf{q}(t),t)=\frac{\omega_{c}}{mc^{2}}\mathrm{E}_{t}\left[-E_{-}(\mathbf{q}(t),D_{*}\mathbf{q}(t),t)\delta t+m\mathbf{b}_{*}(\mathbf{q}(t),t)\cdot\delta\mathbf{q}_{-}(t)\right],
\end{equation}
where
\begin{equation}
E_{-}(\mathbf{q}(t),D_{*}\mathbf{q}(t),t)=mc^{2}+\frac{1}{2}m\mathbf{b}_{*}^{2}+e\Phi_{e}.
\end{equation}
Again we take $\delta\mathbf{q}_{-}(t)$ to correspond to (2.76).

As in the free particle case, at this stage, the forward and backward
steady-state \emph{zbw} phase changes, (2.110) and (2.112), are independent
of one another. So both (2.110) and (2.112) must equal $2\pi n$ when
integrated along a closed loop $L$ in which both time and position
change. Otherwise we will contradict our hypothesis that, up to this
point, the \emph{zbw} particle has a well-defined mean forward or
backward steady-state phase at each point along its mean forward or
backward space-time trajectory.

In the lab frame, the forward and backward stochastic differential
equations for the translational motion are once again
\begin{equation}
d\mathbf{q}(t)=\mathbf{b}(\mathbf{q}(t),t)+d\mathbf{W}(t),
\end{equation}
and
\begin{equation}
d\mathbf{q}(t)=\mathbf{b}_{*}(\mathbf{q}(t),t)+d\mathbf{W}_{*}(t),
\end{equation}
with corresponding Fokker-Planck equations
\begin{equation}
\frac{\partial\rho(\mathbf{q},t)}{\partial t}=-\nabla\cdot\left[\mathbf{b(}\mathbf{q},t)\rho(\mathbf{q},t)\right]+\frac{\hbar}{2m}\nabla^{2}\rho(\mathbf{q},t),
\end{equation}
and
\begin{equation}
\frac{\partial\rho(\mathbf{q},t)}{\partial t}=-\nabla\cdot\left[\mathbf{b}_{*}(\mathbf{q},t)\rho(\mathbf{q},t)\right]-\frac{\hbar}{2m}\nabla^{2}\rho(\mathbf{q},t).
\end{equation}
Let us now suppose that an external magnetic field $\mathbf{B}_{ext}(\mathbf{q},t)=\nabla\times\mathbf{A}_{ext}(\mathbf{q},t)$
is also present. Then, restricting ourselves to simultaneous solutions
of (2.116-117) via
\begin{equation}
\mathbf{v}\coloneqq\frac{1}{2}\left[\mathbf{b}+\mathbf{b}_{*}\right]=\frac{\nabla S}{m}-\frac{e}{mc}\mathbf{A}_{ext}
\end{equation}
and
\begin{equation}
\mathbf{u}\coloneqq\frac{1}{2}\left[\mathbf{b}-\mathbf{b}_{*}\right]=\frac{\hbar}{2m}\frac{\nabla\rho}{\rho}
\end{equation}
entails that (2.116-117) reduce to
\begin{equation}
\frac{\partial\rho}{\partial t}=-\nabla\cdot\left[\left(\mathbf{\frac{\nabla\mathrm{\mathit{S}}}{\mathit{m}}}-\frac{e}{mc}\mathbf{A}_{ext}\right)\rho\right].
\end{equation}
We can then write $\mathbf{b}'=\mathbf{v}'+\mathbf{u}$ and $\mathbf{b}'_{*}=\mathbf{v}'-\mathbf{u}$,
where we recall that $\mathbf{v}'=\mathbf{v}+(e/mc)\mathbf{A}_{ext}$,
implying $\mathbf{b}=\mathbf{b}'-(e/mc)\mathbf{A}_{ext}$ and $\mathbf{b}_{*}=\mathbf{b}'_{*}-(e/mc)\mathbf{A}_{ext}$.
Once again the osmotic potential $R(\mathbf{q},t)=\mu U(\mathbf{q},t)$
imparts to the particle an osmotic velocity $\nabla R/m=\left(\hbar/2m\right)\nabla\rho/\rho$
(see section 2), implying $\rho=e^{2R/\hbar}$ for all times.

As in the free particle case, we can obtain the 2nd-order time-symmetric
mean dynamics from Yasue's variational principle. 

Since (2.110) and (2.112) correspond to the same (lab) frame and are
no longer independent because of (2.118-119), it is natural to define
the time-symmetric steady-state \emph{zbw} particle phase in the lab
frame by taking the difference between (2.110) and (2.112) (under
the replacements $\mathbf{b}\rightarrow\mathbf{b}'$ and $\mathbf{b}_{*}\rightarrow\mathbf{b}'_{*}$
in the mean forward and mean backward momentum contributions to the
phases):
\begin{equation}
\begin{aligned}d\bar{\theta}(\mathbf{q}(t),t) & \coloneqq\frac{1}{2}\left[d\bar{\theta}_{+}(\mathbf{q}(t),t)-d\bar{\theta}_{-}(\mathbf{q}(t),t)\right]\\
 & =\frac{\omega_{c}}{mc^{2}}\mathrm{E}_{t}\left[E(\mathbf{q}(t),D\mathbf{q}(t),D_{*}\mathbf{q}(t),t)dt-\frac{m}{2}\left(\mathbf{b}'(\mathbf{q}(t),t)\cdot d\mathbf{q}_{+}(t)+\mathbf{b}'_{*}(\mathbf{q}(t),t)\cdot d\mathbf{q}_{-}(t)\right)\right]+\phi\\
 & =\frac{\omega_{c}}{mc^{2}}\mathrm{E}_{t}\left[Edt-\frac{m}{2}\left(\mathbf{b}'\cdot\frac{d\mathbf{q}_{+}(t)}{dt}+\mathbf{b}'_{*}\cdot\frac{d\mathbf{q}_{-}(t)}{dt}\right)dt\right]+\phi\\
 & =\frac{\omega_{c}}{mc^{2}}\mathrm{E}_{t}\left[\left(E-\frac{m}{2}\left(\mathbf{b}'\cdot\frac{d\mathbf{q}_{+}(t)}{dt}+\mathbf{b}'_{*}\cdot\frac{d\mathbf{q}_{-}(t)}{dt}\right)\right)dt\right]+\phi\\
 & =\frac{\omega_{c}}{mc^{2}}\mathrm{E}_{t}\left[\left(E-\frac{m}{2}\left(\mathbf{b}'\cdot\mathbf{b}+\mathbf{b}'_{*}\cdot\mathbf{b}_{*}\right)\right)dt\right]+\phi\\
 & =\frac{\omega_{c}}{mc^{2}}\mathrm{E}_{t}\left[\left(E-\frac{m}{2}\left(\mathbf{b}{}^{2}+\frac{e}{mc}\mathbf{b}\cdot\mathbf{A}_{ext}+\mathbf{b}_{*}^{2}+\frac{e}{mc}\mathbf{b}_{*}\cdot\mathbf{A}_{ext}\right)\right)dt\right]+\phi\\
 & =\frac{\omega_{c}}{mc^{2}}\mathrm{E}_{t}\left[\left(E-\frac{m}{2}\left(\mathbf{b}{}^{2}+\mathbf{b}_{*}^{2}\right)-\frac{e}{c}\left(\frac{\mathbf{b}+\mathbf{b}_{*}}{2}\right)\cdot\mathbf{A}_{ext}\right)dt\right]+\phi\\
 & =\frac{\omega_{c}}{mc^{2}}\mathrm{E}_{t}\left[\left(E-\left(m\mathbf{v}\cdot\mathbf{v}+m\mathbf{u}\cdot\mathbf{u}\right)-\frac{e}{c}\mathbf{v}\cdot\mathbf{A}_{ext}\right)dt\right]+\phi\\
 & =\frac{\omega_{c}}{mc^{2}}\mathrm{E}_{t}\left[\left(mc^{2}+e\Phi_{e}-\frac{1}{2}m\mathbf{v}^{2}-\frac{1}{2}m\mathbf{u}^{2}-\frac{e}{c}\mathbf{v}\cdot\mathbf{A}_{ext}\right)dt\right]+\phi.
\end{aligned}
\end{equation}
where, using (2.111) and (2.113), along with the constraints (2.118)
and (2.119), we have defined
\begin{equation}
\begin{aligned}E(\mathbf{q}(t),D\mathbf{q}(t),D_{*}\mathbf{q}(t),t) & =mc^{2}+\frac{1}{2}\left[\frac{1}{2}m\mathbf{b}^{2}+\frac{1}{2}m\mathbf{b}_{*}^{2}\right]+e\Phi_{e}\\
 & =mc^{2}+\frac{1}{2}m\mathbf{v}^{2}+\frac{1}{2}m\mathbf{u}^{2}+e\Phi_{e}.
\end{aligned}
\end{equation}

As in the free particle case, the consistency of our theory requires
that the time-symmetrized steady-state \emph{zbw} phase change of
the \emph{zbw} particle in the lab frame, (2.121), satisfies $\oint_{L}\delta\bar{\theta}=2\pi n$.
Otherwise we would contradict our hypothesis that the \emph{zbw} particle,
under the time-symmetric constraints (2.118-119), has a well-defined
and unique steady-state phase at each 3-space location it can occupy
at each time, regardless of time direction.

Using the integral of (2.121) in the definition of the steady-state
phase-principal function
\begin{equation}
I=-\frac{mc^{2}}{\omega_{c}}\bar{\theta}=\mathrm{E}\left[\int_{t_{i}}^{t}\left(\frac{1}{2}m\mathbf{v}^{2}+\frac{1}{2}m\mathbf{u}^{2}+\frac{e}{c}\mathbf{v}\cdot\mathbf{A}_{ext}-mc^{2}-e\Phi_{e}\right)dt'\left|\mathbf{q}(t)\right.\right]-\hbar\phi,
\end{equation}
we can define the steady-state phase-action functional as
\begin{equation}
\begin{aligned}J & =I_{if}=\mathrm{E}\left[\int_{t_{i}}^{t_{f}}\left(\frac{1}{2}m\mathbf{v}^{2}+\frac{1}{2}m\mathbf{u}^{2}+\frac{e}{c}\mathbf{v}\cdot\mathbf{A}_{ext}-mc^{2}-e\Phi_{e}\right)dt'\right]-\hbar\phi.\end{aligned}
\end{equation}
Equation (2.124) is just Yasue's mean action functional, Eq. (6.1)
in Appendix 6.1, but with the inclusion of the rest-energy term $-mc^{2}$
and the time-symmetrized initial phase constant $\phi$. 

Note, also, that from the second to last line of (2.121), we can write
the cumulative, time-symmetric, steady-state phase at a time \emph{t}
as
\begin{equation}
\begin{aligned}\bar{\theta}(\mathbf{q}(t),t) & =\frac{\omega_{c}}{mc^{2}}\mathrm{E}\left[\int_{t_{i}}^{t}\left(E-\left(m\mathbf{v}\cdot\mathbf{v}+m\mathbf{u}\cdot\mathbf{u}\right)-\frac{e}{c}\mathbf{v}\cdot\mathbf{A}_{ext}\right)dt'\left|\mathbf{q}(t)\right.\right]+\phi\\
 & =\frac{\omega_{c}}{mc^{2}}\mathrm{E}\left[\int_{t_{i}}^{t}\left(\left(E-m\mathbf{u}\cdot\mathbf{u}\right)-m\mathbf{v}\cdot\mathbf{v}-\frac{e}{c}\mathbf{v}\cdot\mathbf{A}_{ext}\right)dt'\left|\mathbf{q}(t)\right.\right]+\phi\\
 & =\frac{\omega_{c}}{mc^{2}}\mathrm{E}\left[\int_{t_{i}}^{t}\left(H-m\mathbf{v}\cdot\mathbf{v}-\frac{e}{c}\mathbf{v}\cdot\mathbf{A}_{ext}\right)dt'\left|\mathbf{q}(t)\right.\right]+\phi\\
 & =\frac{\omega_{c}}{mc^{2}}\mathrm{E}\left[\int_{t_{i}}^{t}Hdt'-\int_{\mathbf{q}(t_{i})}^{\mathbf{q}(t)}\left(m\mathbf{v}(\mathbf{q}(t'),t')+\frac{e}{c}\mathbf{A}_{ext}(\mathbf{q}(t'),t')\right)\cdot\mathrm{D}\mathbf{q}(t')\left|\mathbf{q}(t)\right.\right]+\phi,
\end{aligned}
\end{equation}
where 
\begin{equation}
H\coloneqq E-m\mathbf{u}\cdot\mathbf{u}=mc^{2}+\frac{1}{2}m\mathbf{v}^{2}-\frac{1}{2}m\mathbf{u}^{2}+e\Phi_{e}.
\end{equation}
Now, given an integral curve $\mathbf{Q}(t)$ obtained from 
\begin{equation}
m\frac{d\mathbf{Q}(t)}{dt}=\mathbf{p}(\mathbf{Q}(t),t)=\nabla S(\mathbf{q},t)|_{\mathbf{q}=\mathbf{Q}(t)}-\frac{e}{c}\mathbf{A}_{ext}(\mathbf{Q}(t),t),
\end{equation}
we can replace (2.125) with 
\begin{equation}
\begin{aligned}\bar{\theta}(\mathbf{Q}(t),t) & =\frac{\omega_{c}}{mc^{2}}\int_{t_{i}}^{t}\left(H-m\mathbf{v}\cdot\frac{d\mathbf{Q}(t')}{dt'}-\frac{e}{c}\frac{d\mathbf{Q}(t')}{dt'}\cdot\mathbf{A}_{ext}(\mathbf{Q}(t'),t')\right)dt'+\phi\\
 & =\frac{\omega_{c}}{mc^{2}}\left[\int_{t_{i}}^{t}Hdt'-\int_{\mathbf{Q}(t_{i})}^{\mathbf{Q}(t)}\left(\mathbf{p}+\frac{e}{c}\mathbf{A}_{ext}\right)\cdot d\mathbf{Q}(t')\right]+\phi.
\end{aligned}
\end{equation}
The total differential of the left hand side of (2.128) gives
\begin{equation}
d\bar{\theta}=\nabla\bar{\theta}|_{\mathbf{q}=\mathbf{Q}(t)}d\mathbf{Q}(t)+\partial_{t}\bar{\theta}|_{\mathbf{q}=\mathbf{Q}(t)}dt.
\end{equation}
Hence, 
\begin{equation}
\mathbf{p}(\mathbf{Q}(t),t)+\frac{e}{c}\mathbf{A}_{ext}(\mathbf{Q}(t),t)=-\left(\frac{mc^{2}}{\omega_{c}}\right)\nabla\bar{\theta}|_{\mathbf{q}=\mathbf{Q}(t)}=\nabla S|_{\mathbf{q}=\mathbf{Q}(t)}.
\end{equation}
Thus the current velocity, plus the correction due to the external
vector potential, corresponds the gradient of the \emph{zbw} particle's
time-symmetrized steady-state phase at the location of the \emph{zbw}
particle, and $S$ can again be identified with the time-symmetrized
steady-state action/phase function of the \emph{zbw} particle in the
lab frame. Along with 
\begin{equation}
H(\mathbf{Q}(t),t)=\left(\frac{mc^{2}}{\omega_{c}}\right)\partial_{t}\bar{\theta}|_{\mathbf{q}=\mathbf{Q}(t)}=-\partial_{t}S|_{\mathbf{q}=\mathbf{Q}(t)},
\end{equation}
it follows that 
\begin{equation}
\begin{aligned}S(\mathbf{Q}(t),t) & =\int_{t_{i}}^{t}\left(\mathbf{p}+\frac{e}{c}\mathbf{A}_{ext}\right)\cdot d\mathbf{Q}(t')-\int_{t_{i}}^{t}Hdt'-\hbar\phi\\
 & =\int_{t_{i}}^{t}\left[\frac{1}{2}m\mathbf{v}^{2}+\frac{1}{2}m\mathbf{u}^{2}+\frac{e}{c}\mathbf{v}\cdot\mathbf{A}_{ext}-mc^{2}-e\Phi_{e}\right]dt'-\hbar\phi=I(\mathbf{Q}(t),t),
\end{aligned}
\end{equation}
and
\begin{equation}
\oint_{L}\delta S(\mathbf{Q}(t),t)=\left(-\frac{mc^{2}}{\omega_{c}}\right)\oint_{L}\delta\bar{\theta}(\mathbf{q}(t),t)=\oint_{L}\left[\mathbf{p}'\cdot\delta\mathbf{Q}(t)-H\delta t\right]=nh.
\end{equation}

Recall that after restricting the forward and backward diffusions
to simultaneous solutions of (2.116-117), we have $\mathbf{b}=\mathbf{v}+\mathbf{u}$
and $\mathbf{b}_{*}=\mathbf{v}-\mathbf{u}$. So the IMFTRF and the
IMBTRF will not coincide since, for $\mathbf{b}=\mathbf{v}+\mathbf{u}=0$,
it will generally not be the case that $\mathbf{b}_{*}=\mathbf{v}-\mathbf{u}=0$.
This motivates defining an instantaneous mean (time-)symmetric rest
frame (IMSTRF) as the frame in which $\mathbf{b}+\mathbf{b}_{*}=2\mathbf{v}=0$.
In the IMSTRF, (2.128) reduces to $\bar{\theta}=(\omega_{c}/mc^{2})\left[\left(mc^{2}-\frac{1}{2}m\mathbf{u}^{2}\right)t+\int_{t_{i}}^{t}e\Phi_{e}(\mathbf{Q}_{0},t')dt'\right]+\phi$,
since $\mathbf{v}=0$ and $\partial_{t}\rho=0$. So the external potential
contributes a tiny shift to the time-symmetrized steady-state \emph{zbw}
phase in the IMSTRF, along with the kinetic energy term involving
the osmotic velocity.

Applying the conservative diffusion constraint to the steady-state
phase/action functional (2.124), we recover the mean acceleration
equation
\begin{equation}
m\mathbf{a}(\mathbf{q}(t),t)=\frac{m}{2}\left[D_{*}D+DD_{*}\right]\mathbf{q}(t)=e\left[-\frac{1}{c}\frac{\partial\mathbf{A}_{ext}}{\partial t}-\nabla\Phi_{e}+\frac{\mathbf{v}}{c}\times\mathbf{B}_{ext}\right]|_{\mathbf{q}=\mathbf{q}(t)}.
\end{equation}
Applying the mean derivatives in (2.133), we find
\begin{equation}
\begin{aligned}m\mathbf{a}(\mathbf{q}(t),t) & =m\left[\frac{\partial\mathbf{v}}{\partial t}+\mathbf{v}\cdot\nabla\mathbf{v}-\mathbf{u}\cdot\nabla\mathbf{u}-\frac{\hbar}{2m}\nabla^{2}\mathbf{u}\right]|_{\mathbf{q}=\mathbf{q}(t)}\\
 & =e\left[-\frac{1}{c}\frac{\partial\mathbf{A}_{ext}}{\partial t}-\nabla\Phi_{e}+\frac{\mathbf{v}}{c}\times\mathbf{B}_{ext}\right]|_{\mathbf{q}=\mathbf{q}(t)}.
\end{aligned}
\end{equation}
Integrating both sides gives 
\begin{equation}
\tilde{E}(\mathbf{q}(t),t)=-\frac{\partial S(\mathbf{q},t)}{\partial t}|_{\mathbf{q}=\mathbf{q}(t)}=mc^{2}+\left[\frac{\left(\nabla S-\frac{e}{c}\mathbf{A}_{ext}\right)^{2}}{2m}+e\Phi_{e}-\frac{\hbar^{2}}{2m}\frac{\nabla^{2}\sqrt{\rho}}{\sqrt{\rho}}\right]|_{\mathbf{q}=\mathbf{q}(t)},
\end{equation}
where we have fixed the integration constant equal to the particle
rest energy. Alternatively, we can again consider integral curves
of the current velocity/momentum field, but where now the integral
curves are obtained from solutions of
\begin{equation}
\begin{aligned}m\frac{d^{2}\mathbf{Q}(t)}{dt^{2}} & =m\left(\partial_{t}\mathbf{v}+\mathbf{v}\cdot\nabla\mathbf{v}\right)|_{\mathbf{q}=\mathbf{Q}(t)}\\
 & =-\nabla\left(-\frac{\hbar^{2}}{2m}\frac{\nabla^{2}\sqrt{\rho(\mathbf{q},t)}}{\sqrt{\rho(\mathbf{q},t)}}\right)|_{\mathbf{q}=\mathbf{Q}(t)}+e\left[-\frac{1}{c}\partial_{t}\mathbf{A}_{ext}-\nabla\Phi_{e}+\frac{\mathbf{v}}{c}\times\mathbf{B}_{ext}\right]|_{\mathbf{q}=\mathbf{Q}(t)},
\end{aligned}
\end{equation}
i.e., the mean acceleration equation (2.98), rewritten so that only
the \textbf{v}-dependent terms are kept on the left hand side. Then
we can replace $\mathbf{q}(t)$ in (2.136) with $\mathbf{Q}(t)$ to
obtain $\tilde{E}(\mathbf{Q}(t),t)$. The corresponding general solution,
i.e., the time-symmetrized steady-state phase/action of the \emph{zbw}
particle in the lab frame, after having imposed the conservative diffusion
constraint on (2.124), is of the form 
\begin{equation}
\begin{aligned}S(\mathbf{Q}(t),t) & =\int_{\mathbf{Q}(t_{i})}^{\mathbf{Q}(t)}\left(\mathbf{p}+\frac{e}{c}\mathbf{A}_{ext}\right)\cdot d\mathbf{Q}(t')-\int_{t_{i}}^{t}\tilde{E}dt'-\hbar\phi\\
 & =\int_{t_{i}}^{t}\left[\frac{1}{2}m\mathbf{v}^{2}-\left(-\frac{\hbar^{2}}{2m}\frac{\nabla^{2}\sqrt{\rho}}{\sqrt{\rho}}\right)+\frac{e}{c}\mathbf{v}\cdot\mathbf{A}_{ext}-mc^{2}-e\Phi_{e}\right]dt'-\hbar\phi\\
 & =\int_{t_{i}}^{t}\left[\frac{1}{2}m\mathbf{v}^{2}+\frac{1}{2}m\mathbf{u}^{2}+\frac{\hbar}{2}\nabla\cdot\mathbf{u}+\frac{e}{c}\mathbf{v}\cdot\mathbf{A}_{ext}-mc^{2}-e\Phi_{e}\right]dt'-\hbar\phi.
\end{aligned}
\end{equation}
Notice that the last line of (2.138) differs from the last line of
(2.132) only by addition of the term involving $\nabla\cdot\mathbf{u}$.

As also in the free particle case, the equation of motion for (2.138)
differs from the equation of motion for the classical \emph{zbw} particle
phase by the presence of the quantum kinetic in (2.135-136). Our earlier
discussion of the quantum-classical correspondence applies here as
well.

Insofar as (2.138) is a well-defined phase function, if we integrate
$\delta S(\mathbf{Q}(t),t)$ around a closed loop $L$ in which time
and position may change, we will have
\begin{equation}
\oint_{L}\delta S(\mathbf{Q}(t),t)=\oint_{L}\left[\mathbf{p}'\cdot\delta\mathbf{Q}(t)-\tilde{E}\delta t\right]=nh,
\end{equation}
and for a special loop in which time is held fixed,
\begin{equation}
\oint_{L}\delta S(\mathbf{Q}(t))=\oint_{L}\nabla S|_{\mathbf{q}=\mathbf{Q}(t)}\cdot\delta\mathbf{Q}(t)=\oint_{L}\mathbf{p}'\cdot\delta\mathbf{Q}(t)=nh.
\end{equation}
Considering also the \emph{zbw} phase field $S(\mathbf{q},t)$, which
we recall is a field over the possible locations of the actual \emph{zbw}
particle, and applying the same physical reasoning above to each possible
initial position that the \emph{zbw} particle can occupy, it follows
that the net phase change along any mathematical loop in space (with
time held fixed) will be given by
\begin{equation}
\oint_{L}\nabla S\cdot d\mathbf{q}=\oint_{L}\mathbf{p}'\cdot d\mathbf{q}=nh.
\end{equation}
The corresponding total energy field $E(\mathbf{q},t)$ is given by
(2.136) when $\mathbf{Q}(t)$ is replaced by $\mathbf{q}$. From (2.141),
(2.136), and (2.120), we can then construct the 1-particle Schr\"{o}dinger
equation in external fields as
\begin{equation}
i\hbar\frac{\partial\psi}{\partial t}=\frac{\left[-i\hbar\nabla-\frac{e}{c}\mathbf{A}_{ext}\right]^{2}}{2m}\psi+e\Phi_{e}\psi+mc^{2}\psi,
\end{equation}
where $\psi(\mathbf{q},t)=\sqrt{\rho(\mathbf{q},t)}e^{iS(\mathbf{q},t)/\hbar}$
is a single-valued wave function as a consequence of (2.141).

At this point, it is worth observing an important difference between
the (time-symmetrized steady-state \emph{zbw}) phase field evolving
by (2.136) and the classical \emph{zbw} phase field evolving by Eq.
(2.61) in section 4.4. In the former case, the nonlinear coupling
to the density $\rho$ via the quantum kinetic implies that, at nodal
points (i.e., where $\rho=\psi=0$), such as found in excited states
of the hydrogen atom or quantum harmonic oscillator, the phase field
develops a singularity where both $\mathbf{v}=\nabla S$ and $\mathbf{u}=\left(\hbar/2m\right)\nabla\ln\rho$
diverge. Moreover, (2.141) implies that the phase field along a closed
loop \emph{L} undergoes a discontinuous jump of magnitude $nh$ if
the loop happens to cross a nodal point. Neither of these observations
are inconsistent with our hypothesis that the steady-state phase of
the actual \emph{zbw} particle is a well-defined function of the actual
particle's mean space-time trajectory (or any mean space-time trajectory
it can potentially realize), since it can be readily shown that the
particle's actual (mean or stochastic) trajectory never hits a nodal
point \cite{Nelson1966,Nelson1985,Blanchard1987,Wallstrom1990,Holland1993}.
\footnote{A simple proof \cite{Holland1993} of this for the actual mean trajectory
can be given as follows. First, note that the actual particle's initial
mean position, $\mathbf{q}(0)$, can never be at nodal points (since
this would contradict the physical meaning of $\rho$ as the probability
density for the particle to be at position $\mathbf{q}$ at time $t$).
Now, rewrite $\partial_{t}\rho=-\nabla\cdot\left(\mathbf{v}\rho\right)$
as $\left(\partial_{t}+\mathbf{v}\cdot\nabla\right)\rho=-\rho\nabla\cdot\mathbf{v}$.
Along the actual mean trajectory, $\mathbf{q}(t$), we then have $(d/dt)ln[\rho(\mathbf{q}(t),t)]=-\nabla\cdot\mathbf{v}|_{\mathbf{q}=\mathbf{q}(t)}$.
Solving this gives $\rho(\mathbf{q}(t),t)=\rho_{0}(\mathbf{q}_{0})exp[-\int_{0}^{t}\left(\nabla\cdot\mathbf{v}\right)|_{\mathbf{q}=\mathbf{q}(t')}dt']$,
which implies that if $\rho_{0}(\mathbf{q}_{0})>0$, then $\rho(\mathbf{q}(t),t)>0$
for all times. Correspondingly, from $\rho(\mathbf{q}(t),t)$ we obtain
$R(\mathbf{q}(t),t)=R_{0}(\mathbf{q}_{0})-(\hbar/2)\int_{0}^{t}\left(\nabla\cdot\mathbf{v}\right)|_{\mathbf{q}=\mathbf{q}(t')}dt'$,
which never becomes undefined if $R_{0}(\mathbf{q}_{0})$ is not undefined.} Indeed, if the phase field would not undergo the discontinuous jump
at a nodal point, then this would imply that there are mean trajectories
near nodes for which the actual particle does not have a well-defined
mean phase, thereby contradicting our hypothesis. By contrast, for
the classical \emph{zbw} phase field, there is no reason for it to
be undefined at nodal regions since there is no nonlinear coupling
to the (inverse of the) probability density. Rather, the fact that
the classical phase field also satisfies a condition of the form (2.141)
implies that it changes discontinuously across a discontinuity in
the external potential, \emph{V}, and takes discrete values for changes
along a closed loop \emph{L} encircling the discontinuity in \emph{V}
(as demonstrated for the hydrogen-like atom in Appendix 6.2 ). 

We thus have a formulation of ZSM in external fields that avoids the
Wallstrom criticism and is ready to be applied to the central potential
example considered in section 3.

\subsection{The central potential revisited}

With ZSM in hand, we can now return to the central potential example
considered by Wallstrom, and show how ZSM gives the same result as
quantum mechanics.

For the effective central potential, $V_{a}(\boldsymbol{\mathrm{r}})=V(\boldsymbol{\mathrm{r}})+a/r^{2}$,
we found that the HJM equations implied $\mathbf{v}'_{a}=\mathbf{v}{}_{a}\sqrt{2ma/\hbar^{2}+1}$
and $\mathbf{u}'_{a}=\mathbf{u}{}_{a}$, where $\mathbf{v}{}_{a}=\left(\hbar/mr\right)\hat{\varphi}$.
The problem in standard NYSM was that the constant $a$ could take
any positive real value, making $\mathbf{v}'_{a}$ not quantized.
By contrast, in the quantum mechanical version, $\mathrm{m}=\sqrt{2ma/\hbar^{2}+1}$
would be integral due to the single-valuedness condition on $\psi_{\mathrm{m}}$.

In the ZSM version of this problem, the \emph{zbw} phase field in
the lab frame, $S_{a}=\hbar\varphi$, satisfies
\begin{equation}
\oint\frac{dS_{a}}{d\varphi}d\varphi=\oint\hbar d\varphi=\mathrm{m}h,
\end{equation}
as a consequence of the reasoning used in section 5.2. Accordingly,
for the effective \emph{zbw} phase field, $S'_{a}=\hbar\sqrt{2ma/\hbar^{2}+1}\varphi=\hbar\varphi'$,
we will also have
\begin{equation}
\oint\hbar\sqrt{2ma+1}d\varphi=\oint\hbar d\varphi'=\mathrm{m}h,
\end{equation}
where $\mathrm{m}=\sqrt{2ma/\hbar^{2}+1}$ is integral. So ZSM predicts
quantized energy-momentum in the central potential case, in accordance
with quantum mechanics.

\section{Conclusion}

To answer Wallstrom's criticism, we first developed a classical \emph{zbw}
model (based on the earlier models of de Broglie and Bohm) which implies
a quantization condition reminiscent of the Bohr-Sommerfeld-Wilson
condition. We did this excluding and including interactions with external
fields, and formulated the classical Hamilton-Jacobi statistical mechanics
of each case. We then extended this model to Nelson-Yasue stochastic
mechanics - which we termed zitterbewegung stochastic mechanics (ZSM)
- and showed, using the same two cases, that it allows us to recover
the Schr\"{o}dinger  equation for single-valued wave functions with (in
general) multi-valued phases. Finally, we showed that ZSM works for
the concrete case of a two-dimensional central potential.

In Part II, our approach will be generalized to the case of many \emph{zbw}
particles, excluding and including (external and inter-particle) field
interactions, the latter of which turns out to be a non-trivial task.
We will also: (i) elaborate on the beables of ZSM, (ii) assess the
plausibility and generalizability of the \emph{zbw} hypothesis, and
(iii) compare ZSM to other (previously) proposed answers to Wallstrom's
criticism.

\section{Acknowledgments}

It is a pleasure to acknowledge helpful discussions with Guido Bacciagaluppi,
Dieter Hartmann, and Herman Batelaan. I am also gateful to Guido for
carefully reading an earlier draft of this paper and making useful
suggestions for improvements. Lastly, I thank Mike Towler for inviting
me to talk on an earlier incarnation of this work at his 2010 de Broglie-Bohm
research conference in Vallico Sotto, Tuscany, Italy.

\chapter{A Suggested Answer To Wallstrom's Criticism: ZSM II}

The ``zitterbewegung stochastic mechanics'' (ZSM) answer to Wallstrom's
criticism, introduced in Part I \cite{Derakhshani2016}, is extended
to many particles. We first formulate the many-particle generalization
of Nelson-Yasue stochastic mechanics (NYSM), incorporating external
and classical interaction potentials. Then we formulate the many-particle
generalization of the classical zitterbewegung \emph{zbw} model introduced
in Part I, for the cases of free particles, particles interacting
with external fields, and classically interacting particles. On the
basis of these developments, ZSM is constructed for classically free
particles, as well as for particles interacting both with external
fields and through inter-particle scalar potentials. Throughout, the
beables of ZSM (based on the many-particle formulation) are made explicit.
Subsequently, we assess the plausibility and generalizability of the\emph{
zbw} hypothesis. We close with an appraisal of other proposed answers,
and compare them to ZSM.

\section{Introduction}

This paper is a direct continuation of the preceding paper, Part I
\cite{Derakhshani2016}. There we proposed an answer to the Wallstrom
criticism of stochastic mechanical theories by modifying Nelson-Yasue
stochastic mechanics (NYSM) for a single non-relativistic particle
with the following hypothesis: Nelson's hypothetical stochastic ether
medium that drives the conservative diffusions of the particle, also
induces steady-state harmonic oscillations of zitterbewegung ($\emph{zbw}$)
frequency in the particle's instantaneous mean forward/backward translational
rest frame. We then showed that, in the lab frame, the function \emph{S}
arises from imposing the constraint of conservative diffusions on
the time-symmetrized steady-state phase of the \emph{zbw} particle,
satisfies the required quantization condition, and evolves in time
by the Hamilton-Jacobi-Madelung equations (when generalized to describe
a statistical ensemble of \emph{zbw} particles). This allowed us to
recover the Schr\"{o}dinger equation
for single-valued wavefunctions with (potentially) multi-valued phases,
for the cases of a free particle and a particle interacting with external
fields (the latter of which we illustrated with the two-dimensional
central potential problem). We termed this modification of NYSM ``zitterbewegung
stochastic mechanics'' or ZSM.

The approach of this paper is similar to that of Part I. In section
2, we formulate the many-particle generalization of NYSM and point
out where in the derivation of the many-particle Schr\"{o}dinger
equation the Wallstrom criticism applies. Section 3 discusses how
to properly physically interpret the wavefunction in NYSM. Section
4 formulates the classical model of constrained zitterbewegung motion
for the cases of many free particles, many particles interacting with
external fields, and classically interacting particles. Section 5
generalizes ZSM to the cases of many free particles, many particles
interacting with external fields, and classically interacting particles;
throughout, the beables \footnote{This term was coined by J.S. Bell \cite{BellTLB} as a play on
``observables\textquotedblright{} in standard quantum mechanics.
It refers to ``those elements which might correspond to elements
of reality, to things which exist. Their existence does not depend
on `observation.' Indeed observation and observers must be made out
of beables\textquotedblright{} \cite{BellBQFT}.} of ZSM are made explicit. Section 6 assesses the plausibility and
generalizability of the $\emph{zbw}$ hypothesis through multiple
considerations. Finally, Section 7 appraises other proposed answers
to Wallstrom's criticism, and compares them to ZSM.

\section{Nelson-Yasue Stochastic Mechanics for Many Particles}

The first non-relativistic, $N$-particle, stochastic mechanical reconstruction
of the \emph{N}-particle Schr\"{o}dinger  equation was given by Loffredo and Morato
\cite{M.I.Loffredo2007}, who used the Guerra-Morato variational formulation.
\footnote{Prior to Loffredo-Morato, Nelson \cite{Nelson1985} and Bacciagaluppi
\cite{Bacciagaluppi2003} employed \emph{N}-particlee xtensions of stochastic
mechanics for scalar particles. However, they did so by assuming (rather
than reconstructing) the \emph{N}-particle Schr\"{o}dinger  equation, and constructing
\emph{N}-particle extensions of the stochastic mechanical equations of motion
from solutions of the \emph{N}-particle Schr\"{o}dinger  equation. The \emph{N}-particle
stochastic mechanical equations obtained by Nelson and Bacciagaluppi
are formally the same as those obtained by Loffredo-Morato.} However, as noted in footnote 9 of Part I \cite{Derakhshani2016},
the the Guerra-Morato formulation is not applicable to ZSM because the Guerra-Morato variational principle entails a globally single-valued $S$ function, and this excludes the possibility of single-valued wavefunctions with multi-valued phases, which excludes the possibility of single-valued wavefunctions with multi-valued phases (as in systems with angular momentum \cite{Wallstrom1989,Wallstrom1994}. Koide \cite{Koide2015} has given a brief two-particle extension
of the non-relativistic Nelson-Yasue formulation, for the case of
a classical interaction potential, but otherwise no comprehensive
\emph{N}-particle Nelson-Yasue reconstruction of the \emph{N}-particle Schr\"{o}dinger 
equation has been given (to the best of our knowledge). Accordingly,
we shall develop the \emph{N}-particle extension of NYSM before extending
ZSM to the many-particle case. This will also be useful for identifying
the various points of demarcation between NYSM and ZSM in the many-particle
formulation. For completeness, we will incorporate coupling of the
particles to external (scalar and vector) potentials and to each other
through scalar interaction potentials.

As in the single-particle formulation of NYSM \cite{Nelson1966,Nelson1967,Nelson1985},
we hypothesize that the vacuum of 3-D space is pervaded by a homogeneous
and isotropic ether fluid with classical stochastic fluctuations that
impart a frictionless, conservative diffusion process to a point particle
of mass $m$ and charge $e$ immersed within the ether. Accordingly,
for $N$ point particles of masses $m_{i}$ and charges $e_{i}$ immersed
in the ether, each particle will in general have its position 3-vector
$\mathbf{q}_{i}(t)$ constantly undergoing diffusive motion with drift,
as modeled by the first-order forward stochastic differential equations
\begin{equation}
d\mathbf{q}_{i}(t)=\mathbf{b}_{i}(q(t),t)dt+d\mathbf{W}_{i}(t).
\end{equation}
Here $q(t)=\{\mathbf{q}_{1}(t),\mathbf{q}_{2}(t),...,\mathbf{q}_{N}(t)\}$
$\in$ $\mathbb{R}^{3N}$, $\mathbf{b}_{i}(q(t),t)$ is the deterministic
mean forward drift velocity of the $i$-th particle (which in general
may be a function of the positions of all the other particles, such
as in the case of particles interacting with each other gravitationally
and/or electrostatically), and $\mathbf{W}_{i}(t)$ is the Wiener
process modeling the $i$-th particle's interaction with the ether
fluctuations. (Recall that ``mean'', here, refers to averaging over
the Wiener process in the sense of the conditional expectation at
time \emph{t}.)

The Wiener increments $d\mathbf{W}_{i}(t)$ are assumed to be Gaussian
with zero mean, independent of $d\mathbf{q}_{i}(s)$ for $s\leq t$,
and with variance
\begin{equation}
\mathrm{E}_{t}\left[d\mathbf{W}_{in}(t)d\mathbf{W}_{im}(t)\right]=2\nu_{i}\delta_{nm}dt,
\end{equation}
where $\mathrm{E}_{t}$ denotes the conditional expectation at time
\emph{t}. We then hypothesize that the magnitudes of the diffusion
coefficients $\nu_{i}$ are given by
\begin{equation}
\nu_{i}=\frac{\hbar}{2m_{i}}.
\end{equation}

In addition to (3.1), we also have the backward stochastic differential
equations
\begin{equation}
d\mathbf{q}_{i}(t)=\mathbf{b}_{i*}(q(t),t)dt+d\mathbf{W}_{i*}(t),
\end{equation}
where $\mathbf{b}_{i*}(q(t),t)$ are the mean backward drift velocities,
and $d\mathbf{W}_{i*}(t))$ are the backward Wiener processes. As
in the single-particle case, the $d\mathbf{W}_{i*}(t)$ have all the
properties of $d\mathbf{W}_{i}(t)$ except that they are independent
of the $d\mathbf{q}_{i}(s)$ for $s\geq t$. With these conditions
on $d\mathbf{W}_{i}(t)$ and $d\mathbf{W}_{i*}(t)$, Eqs. (3.1) and
(3.4) respectively define forward and backward Markov processes for
$N$ particles on $\mathbb{R}^{3}$ (or, equivalently, for a single
particle on $\mathbb{R}^{3N}$).

The forwards and backwards transition probabilities defined by (3.1)
and (3.4), respectively, should be understood, in some sense, as ontic
probabilities \cite{Uffink2006,BacciagaluppiProbab2011}. (Broadly
speaking, `ontic probabilities' can be understood as probabilities
about objective physical properties of the \emph{N}-particle system, as opposed
to `epistemic probabilities' \cite{Arntzenius1995} which are about
our ignorance of objective physical properties of the \emph{N}-particle system.)
Just how `ontic' these transition probabilities should be is an open
question. One possibility is that these transition probabilities should
be viewed as phenomenologically modeling the complicated deterministic
interactions of a massive particle (or particles) with the fluctuating
ether, in analogy with how equations such as (3.1) and (3.4) are used
in the Einstein-Smoluchowski theory \cite{Nelson1967} to phenomenologically
model the complicated deterministic interactions of a macroscopic
particle immersed in a fluctuating classical fluid of finite temperature.
Another possibility is that the fluctuations of the ether are irreducibly
stochastic, and this irreducible stochasticity is `transferred' to
a particle immersed in and interacting with the ether. We prefer the
former possibility, but acknowledge that the latter possibility is
also viable. \footnote{Concerning whether or not the forward and backwards transition probabilities
should be understood as `objective' (i.e., as chances governed by
natural law) versus `subjective' (i.e., encoding our expectations
or degrees of belief) \cite{FriggHoefer2010,Glynn2010,Emery2015},
this seems to depend on whether the transition probabilities are merely
phenomenological (in which case they would seem to be subjective)
or reflect irreducible stochasticity in the ether (in which case they
would seem to be objective). Our preference for viewing the transition
probabilities as phenomenological seems to commit us to the subjective
view, but the objective view also seems viable (the objective view
is taken by Bacciagaluppi in \cite{Bacciagaluppi2005,Bacciagaluppi2012}).
It is worth noting that, under the objective view, the backwards transition
probabilities can be regarded as being just as objective/law-like
as the forwards transition probabilities (but see \cite{Arntzenius1995}
for a different view).}

Associated to the trajectories $\mathbf{q}_{i}(t)$ is the $N$-particle
probability density $\rho(q,t)=n(q,t)/N$ where $n(q,t)$ is the number
of particles per unit volume. Corresponding to (3.1) and (3.4), then,
are the $N$-particle forward and backward Fokker-Planck equations
\begin{equation}
\frac{\partial\rho(q,t)}{\partial t}=-\sum_{i=1}^{N}\nabla_{i}\cdot\left[\mathbf{b}_{i}(q,t)\rho(q,t)\right]+\sum_{i=1}^{N}\frac{\hbar}{2m_{i}}\nabla_{i}^{2}\rho(q,t),
\end{equation}
and
\begin{equation}
\frac{\partial\rho(q,t)}{\partial t}=-\sum_{i=1}^{N}\nabla_{i}\cdot\left[\mathbf{b}_{i*}(q,t)\rho(q,t)\right]-\sum_{i=1}^{N}\frac{\hbar}{2m_{i}}\nabla_{i}^{2}\rho(q,t),
\end{equation}
where we assume that the solutions $\rho(q,t)$ in each time direction
satisfy the normalization condition 
\begin{equation}
\int_{\mathbb{R}^{3N}}\rho_{0}(q)d^{3N}q=1.
\end{equation}
In contrast to the transition probabilities defined by (3.1) and (3.4),
the probability distributions satisfying (3.5) and (3.6) are epistemic
distributions in the sense that they are distributions over a Gibbsian
ensemble of identical systems (i.e., the distributions reflect our
ignorance of the actual positions of the particles). Nevertheless,
for an epistemic distribution satisfying (3.5) or (3.6) at time $t$,
its subsequent evolution will be determined by the ontic transition
probabilities so that the distribution at later times will partly
come to reflect ontic features of the \emph{N}-particle system, and may asymptotically
become independent of the initial distribution. \footnote{I thank Guido Bacciagaluppi for emphasizing this point.}
Of course, the asymptotic distribution would still be epistemic in
the sense of encoding our ignorance of the actual particle positions,
even though it would be determined by the ontic features of the system. 

Up to this point, (3.5) and (3.6) correspond to independent diffusion
processes in opposite time directions. \footnote{In fact, given all possible solutions to (3.1), one can define as
many forward processes as there are possible initial distributions
satisfying (3.5); likewise, given all possible solutions to (3.4),
one can define as many backward processes as there are possible `initial'
distributions satisfying (3.6). Consequently, the forward and backward
processes are both underdetermined, and neither (3.1) nor (3.4) has
a well-defined time-reversal.} To fix the diffusion process uniquely for both time directions, we
must constrain the diffusion process to simultaneous solutions of
(3.5) and (3.6).

Note that the sum of (3.5) and (3.6) yields the $N$-particle continuity
equation
\begin{equation}
\frac{\partial\rho({\normalcolor q},t)}{\partial t}=-\sum_{i=1}^{N}\nabla_{i}\cdot\left[\mathbf{v}_{i}(q.t)\rho(q,t)\right],
\end{equation}
where
\begin{equation}
\mathbf{v}_{i}(q.t)\coloneqq\frac{1}{2}\left(\mathbf{b}_{i}(q,t)+\mathbf{b}_{i*}(q,t)\right)
\end{equation}
is the current velocity field of the $i$-th particle. We shall also
require that $\mathbf{v}_{i}(q.t)$ is equal to the gradient of a
scalar potential $S(q,t)$ (since, if we allowed $\mathbf{v}_{i}(q.t)$
a non-zero curl, then the time-reversal operation would change the
orientation of the curl, thus distinguishing time directions \cite{PenaCetto1982,Bacciagaluppi2012}).
And for particles classically interacting with an external vector
potential $\mathbf{A}_{i}^{ext}(\mathbf{q}_{i},t)$, the current velocities
get modified by the usual expression
\begin{equation}
\mathbf{v}_{i}(q.t)=\frac{\nabla_{i}S(q,t)}{m_{i}}-\frac{e_{i}}{m_{i}c}\mathbf{A}_{i}^{ext}(\mathbf{q}_{i},t).
\end{equation}
So (3.8) becomes
\begin{equation}
\frac{\partial\rho({\normalcolor q},t)}{\partial t}=-\sum_{i=1}^{N}\nabla_{i}\cdot\left[\left(\mathbf{\frac{\nabla_{\mathit{i}}\mathrm{\mathit{S\mathrm{(\mathit{q},\mathit{t})}}}}{\mathit{m}_{\mathit{i}}}}-\frac{e_{i}}{m_{i}c}\mathbf{A}_{i}^{ext}(\mathbf{q}_{i},t)\right)\rho(q,t)\right],
\end{equation}
which is now a time-reversal invariant evolution equation for $\rho$. 

The function \textit{S} is an $N$-particle velocity potential, defined
here as a field over the possible positions of the particles (hence
the dependence of $S$ on the generalized coordinates $\mathbf{q}_{i}$),
and generates different possible initial irrotational velocities for
the particles via (3.10). We make no assumptions at this level as
to whether or not $S$ can be written as a sum of single-particle
velocity potentials. Rather, this will depend on the initial conditions
and constraints specified for a system of $N$ Nelsonian particles,
as well as the dynamics we obtain for $S$. For example, for $N$
particles constrained to interact with each other through a classical
Newtonian gravitational and/or electrostatic potential, and $S$ evolving
by the $N$-particle generalization of the quantum Hamilton-Jacobi
equation (which will turn out to be the case), we will find that $S$
won't be decomposable into a sum as long as the interactions are appreciable.
On the other hand, for $N$ non-interacting particles, we will find
that $S$ evolving by the quantum Hamilton-Jacobi equation can (in
certain cases) be written as $\sum_{i=1}^{N}S_{i}(\mathbf{q}_{i},t)$.

Note also that subtracting (3.5) from (3.6) yields the equality on
the right hand side of 
\begin{equation}
\mathbf{u}_{i}(q,t)\coloneqq\frac{1}{2}\left[\mathbf{b}_{i}(q,t)-\mathbf{b}_{i*}(q,t)\right]=\frac{\hbar}{2m_{i}}\frac{\nabla_{i}\rho(q,t)}{\rho(q,t)},
\end{equation}
where $\mathbf{u}_{i}(q,t)$ is the osmotic velocity field of the
$i$-th particle. From (3.10) and (3.12), we then have $\mathbf{b}_{i}=\mathbf{v}_{i}+\mathbf{u}_{i}$
and $\mathbf{b}_{i*}=\mathbf{v}_{i}-\mathbf{u}_{i}$, which when inserted
back into (3.5) and (3.6), respectively, returns (3.11). Thus $\rho$
is fixed as the unique, single-time, `quantum equilibrium' distribution
for the solutions of (3.1) and (3.4), and evolves by (3.11). Moreover,
the epistemic probabilities associated with $\rho$ are now fully
determined by the ontic transition probabilities corresponding to
solutions of (3.1) and (3.4).

As in the single-particle case, we can give physical meaning to the
osmotic velocities by analogy with the Einstein-Smoluchowski theory:
We postulate the presence of an external ``osmotic'' potential (which
we will formally write as a field on the $N$-particle configuration
space, in analogy with a classical $N$-particle external potential),
$U(q,t)$, which couples to the $i$-th particle as $R(q(t),t)=\mu U(q(t),t)$
(we assume that the coupling constant $\mu$ is identical for particles
of the same species), and imparts to the $i$-th particle a momentum,
$\nabla_{i}R(q,t)|_{\mathbf{q}_{j}=\mathbf{q}_{j}(t)}$. This momentum
then gets counter-balanced by the ether fluid's osmotic impulse pressure,
$\left(\hbar/2m_{i}\right)\nabla_{i}\ln[n(q,t)]|_{\mathbf{q}_{j}=\mathbf{q}_{j}(t)}$.
This leads to the equilibrium condition $\nabla_{i}R/m_{i}=\left(\hbar/2m_{i}\right)\nabla_{i}\rho/\rho$
(using $\rho=n/N$), which implies that $\rho$ depends on $R$ as
$\rho=e^{2R/\hbar}$ for all times. So the osmotic velocity of the
\emph{i}-th particle is the `equilibrium velocity' that the $i$-th
particle would acquire in the absence of any current velocity $\nabla_{i}S/m_{i}$.
(Note that the sense here in which the osmotic velocity is an equilibrium
velocity is different from the sense in which $\nabla_{i}S$ is an
equilibrium velocity; the latter is an equilibrium velocity in the
sense that it's the velocity that transports the quantum equilibrium
distribution $\rho$ on configuration space via the continuity equation
(3.11).)

It might be thought that, as an external potential (in the sense of
a potential not sourced by the particle), it should be reasonable
to assume that $R$ is a separable function of the $N$ coordinates
so that we can write $R(q,t)=\sum_{i=1}^{N}R_{i}(\mathbf{q}_{i},t)$.
However, we know from the single-particle case that the evolution
of $R$ depends on the evolution of $S$ (through the continuity equation
for $\rho$), and that the evolution of $S$ depends on the classical
potential $V$. Since, for many particles, $V$ can be an interaction
potential (such as an \emph{N}-particle Coulomb potential), and since
we expect to find that the \emph{N}-particle evolution equations for
$R$ and $S$ are the \emph{N}-particle generalizations of the HJM
equations, we should expect $R$ to possibly depend on the positions
of all the other particle coordinates as a consequence of its nonlinear
coupling to $S$.

From a more physical point of view, it would be reasonable to expect
that $R$ functionally depends on the coordinates of all the other
particles if either (i) the source of the potential $U$ dynamically
couples to all the particles in such a way that the functional dependence
of $U$ is determined by the magnitude of inter-particle physical
interactions, or (ii) $U$ is an independently existing field in space-time
that directly exchanges energy-momentum with the particles. Since,
by Nelson's hypothesis, each particle undergoes a conservative diffusion
process through the ether, on the (ensemble) average, the energy-momentum
of each particle is a constant (assuming no time-dependent classical
external potentials are present). This suggests that the source of
$U$ should be Nelson's ether \footnote{So the idea would be that the ether fluid produces a potential field
$U$ that imparts a momentum of $\nabla_{i}(\mu U)$ to each particle,
causing the particles to scatter through the ether constituents and
thereby experience a counter-balancing osmotic impulse pressure of
magnitude $\left(\hbar/2m_{i}\right)\nabla_{i}ln[n]$.} (otherwise the diffusions would not be conservative). So the functional
dependence of $U$ must be determined by the (hypothetical) dynamical
coupling of the ether to the particles, and whether or not the particles
classically interact with one another. In this way, it is conceivable
how $U$ could have a non-separable functional dependence on the coordinates
associated with all the particles. Moreover, we should expect the
`strength' of the non-separability (i.e., the inter-particle correlations)
of $U$ to be proportional to the strength of the classical interactions
between the particles. (As it turns out, a dust grain undergoing Brownian
motion in a nonequilibrium plasma induces an electrostatic osmotic
potential from the plasma through an analogous mechanism to what we've
sketched here \textcolor{black}{\cite{Lev2009}}; moreover, the corresponding
Fokker-Planck equation for the stationary probability distribution
in velocity space is formally equivalent to Eq. (3.5) here.)

Since we do not at present have a physical model for Nelson's ether
and its dynamical interactions with the particles, in practice, hypothesis
(i) in the previous paragraph gets implemented via Eq. (3.11) (which,
as we've noted, equivalently describes the time-evolution of $R$
and thereby the time-evolution of the coupling of the particles to
$U$) and Yasue's stochastic variational principle for the particles.
Thus, for $N$ particles constrained to interact with each other through
a classical Newtonian gravitational and/or electrostatic potential,
and $R$ coupled to $S$ by the $N$-particle HJM equations, we will
indeed see that $R$ (and hence $\rho$) is not separable, from which
we can deduce that $U$ will also be non-separable. On the other hand,
in the case of non-interacting particles, we will find that it is
possible to have $R(q,t)=\sum_{i=1}^{N}R_{i}(\mathbf{q}_{i},t)$ (hence
$\rho(q,t)=\prod_{i=1}^{N}\rho_{i}(\mathbf{q}_{i},t)$). So, for now,
we will keep writing the general form $R=R(q,t)$.

In order to formulate the second-order dynamics of the particles,
we need to construct the $N$-particle generalizations of Nelson's
mean forward and backward derivatives. This generalization is straightforwardly
given by
\begin{equation}
D\mathbf{q}_{i}(t)=\underset{_{\Delta t\rightarrow0^{+}}}{lim}\mathrm{E}_{t}\left[\frac{q_{i}(t+\Delta t)-q_{i}(t)}{\Delta t}\right],
\end{equation}
and
\begin{equation}
D_{*}\mathbf{q}_{i}(t)=\underset{_{\Delta t\rightarrow0^{+}}}{lim}\mathrm{E}_{t}\left[\frac{q_{i}(t)-q_{i}(t-\Delta t)}{\Delta t}\right].
\end{equation}
By the Gaussianity of $d\mathbf{W}_{i}(t)$ and $d\mathbf{W}_{i*}(t)$,
we obtain $D\mathbf{q}_{i}(t)=\mathbf{b}_{i}(q(t),t)$ and $D_{*}\mathbf{q}_{i}(t)=\mathbf{b}_{i*}(q(t),t)$.
To compute $D\mathbf{b}_{i}(q(t),t)$ (or $D_{*}\mathbf{b}_{i}(q(t),t)$),
we expand $\mathbf{b}_{i}$ in a Taylor series up to terms of order
two in $d\mathbf{q}_{i}(t)$:
\begin{equation}
\begin{aligned}d\mathbf{b}_{i}(q(t),t) & =\frac{\partial\mathbf{b}_{i}(q(t),t)}{\partial t}dt+\sum_{i=1}^{N}d\mathbf{q}_{i}(t)\cdot\nabla_{i}\mathbf{b}_{i}(q,t)|_{\mathbf{q}_{j}=\mathbf{q}_{j}(t)}\\
 & +\sum_{i=1}^{N}\frac{1}{2}\underset{n,m}{\sum}d\mathbf{\mathit{q}}_{in}(t)d\mathbf{\mathit{q}}_{im}(t)\frac{\partial^{2}\mathbf{b}_{i}(q,t)}{\partial\mathbf{\mathit{q}}_{in}\partial\mathit{q}_{im}}|_{\mathbf{q}_{j}=\mathbf{q}_{j}(t)}+\ldots.
\end{aligned}
\end{equation}
From (3.1), we can replace $dq_{i}(t)$ by $dW_{i}(t)$ in the last
term, and when taking the conditional expectation at time \emph{t}
in (3.13), we can replace $d\mathbf{q}_{i}(t)\cdot\nabla_{i}\mathbf{b}_{i}|_{\mathbf{q}_{j}=\mathbf{q}_{j}(t)}$
by $\mathbf{b}_{i}(\mathbf{q}(t),t)\cdot\nabla_{i}\mathbf{b}_{i}|_{\mathbf{q}_{j}=\mathbf{q}_{j}(t)}$
since $d\mathbf{W}_{i}(t)$ is independent of $\mathbf{q}_{i}(t)$
and has mean 0. From (3.2), we then obtain
\begin{equation}
D\mathbf{b}_{i}(q(t),t)=\left[\frac{\partial}{\partial t}+\sum_{i=1}^{N}\mathbf{b}_{i}(q(t),t)\cdot\nabla_{i}+\sum_{i=1}^{N}\frac{\hbar}{2m_{i}}\nabla_{i}^{2}\right]\mathbf{b}_{i}(q(t),t),
\end{equation}
and likewise
\begin{equation}
D_{*}\mathbf{b}_{i*}(q(t),t)=\left[\frac{\partial}{\partial t}+\sum_{i=1}^{N}\mathbf{b}_{i*}(q(t),t)\cdot\nabla_{i}-\sum_{i=1}^{N}\frac{\hbar}{2m_{i}}\nabla_{i}^{2}\right]\mathbf{b}_{i*}(q(t),t).
\end{equation}
Using (3.16-17), and assuming the particles also couple to an external
electric potential, $\Phi_{i}^{ext}(\mathbf{q}_{i}(t),t)$, as well
as to each other by the Coulomb interaction potential $\Phi_{c}^{int}(\mathbf{q}_{i}(t),\mathbf{q}_{j}(t))=\frac{1}{2}\sum_{j=1}^{N(j\neq i)}\frac{e_{j}}{|\mathbf{q}_{i}(t)-\mathbf{q}_{j}(t)|}$
we can then construct the \emph{N}-particle generalization of Yasue's
ensemble-averaged, time-symmetric mean action:
\begin{equation}
\begin{aligned}J & =\mathrm{E}\left[\int_{t_{I}}^{t_{F}}\sum_{i=1}^{N}\left\{ \frac{1}{2}\left[\frac{1}{2}m_{i}\mathbf{b}_{i}^{2}+\frac{1}{2}m_{i}\mathbf{b}_{i*}^{2}\right]+\frac{e_{i}}{c}\mathbf{A}_{i}^{ext}\cdot\frac{1}{2}\left(D+D_{*}\right)\mathbf{q}_{i}(t)-e_{i}\left[\Phi_{i}^{ext}+\Phi_{c}^{int}\right]\right\} dt\right]\\
 & =\mathrm{E}\left[\int_{t_{I}}^{t_{F}}\sum_{i=1}^{N}\left\{ \frac{1}{2}m_{i}\mathbf{v}_{i}^{2}+\frac{1}{2}m_{i}\mathbf{u}_{i}^{2}+\frac{e_{i}}{c}\mathbf{A}_{i}^{ext}\cdot\mathbf{v}_{i}-e_{i}\left[\Phi_{i}^{ext}+\Phi_{c}^{int}\right]\right\} dt\right],
\end{aligned}
\end{equation}
where $\mathrm{E}\left[...\right]$ denotes the absolute expectation,
and we note that $\mathbf{v}_{i}(q(t),t)=\frac{1}{2}\left(D+D_{*}\right)\mathbf{q}_{i}(t)$.

Upon imposing the conservative diffusion constraint through the \emph{N}-particle
generalization of Yasue's variational principle
\begin{equation}
J=extremal,
\end{equation}
a straightforward computation (see Appendix 7.1) shows that (3.19)
implies
\begin{equation}
\sum_{i=1}^{N}\frac{m_{i}}{2}\left[D_{*}D+DD_{*}\right]\mathbf{q}_{i}(t)=\sum_{i=1}^{N}e_{i}\left[-\frac{1}{c}\partial_{t}\mathbf{A}_{i}^{ext}-\nabla_{i}\left(\Phi_{i}^{ext}+\Phi_{c}^{int}\right)+\frac{\mathbf{v}_{i}}{c}\times\left(\nabla_{i}\times\mathbf{A}_{i}^{ext}\right)\right]|_{\mathbf{q}_{j}=\mathbf{q}_{j}(t)}.
\end{equation}
Moreover, since the $\delta\mathbf{q}_{i}(t)$ are independent (as
we show in Appendix 7.1), it follows from (3.20) that we have the
equations of motion
\begin{equation}
\begin{aligned}m_{i}\mathbf{a}_{i}(q(t),t) & =\frac{m_{i}}{2}\left[D_{*}D+DD_{*}\right]\mathbf{q}_{i}(t)\\
 & =\left[-\frac{e_{i}}{c}\partial_{t}\mathbf{A}_{i}^{ext}-e_{i}\nabla_{i}\left(\Phi_{i}^{ext}+\Phi_{c}^{int}\right)+\frac{e_{i}}{c}\mathbf{v}_{i}\times\left(\nabla_{i}\times\mathbf{A}_{i}^{ext}\right)\right]|_{\mathbf{q}_{j}=\mathbf{q}_{j}(t)},
\end{aligned}
\end{equation}
for $i=1,...,N$. Applying the mean derivatives in (3.20), using that
$\mathbf{b}_{i}=\mathbf{v}_{i}+\mathbf{u}_{i}$ and $\mathbf{b}_{i*}=\mathbf{v}_{i}-\mathbf{u}_{i}$,
and replacing $q(t)$ with $q$ in the functions on both sides, straightforward
manipulations show that (3.20) turns into
\begin{equation}
\begin{aligned} & \sum_{i=1}^{N}m_{i}\left[\partial_{t}\mathbf{v}_{i}+\mathbf{v}_{i}\cdot\nabla_{i}\mathbf{v}_{i}-\mathbf{u}_{i}\cdot\nabla_{i}\mathbf{u}_{i}-\frac{\hbar}{2m_{i}}\nabla_{i}^{2}\mathbf{u}_{i}\right]\\
 & =\sum_{i=1}^{N}\left[-\frac{e_{i}}{c}\partial_{t}\mathbf{A}_{i}^{ext}-e_{i}\nabla_{i}\left(\Phi_{i}^{ext}+\Phi_{c}^{int}\right)+\frac{e_{i}}{c}\mathbf{v}_{i}\times\left(\nabla_{i}\times\mathbf{A}_{i}^{ext}\right)\right].
\end{aligned}
\end{equation}
Using (3.10) and (3.12), integrating both sides of (3.22), and setting
the arbitrary integration constants equal to zero, we then obtain
the \emph{N}-particle quantum Hamilton-Jacobi equation
\begin{equation}
\begin{aligned}-\partial_{t}S(q,t) & =\sum_{i=1}^{N}\frac{\left[\nabla_{i}S(q,t)-\frac{e_{i}}{c}\mathbf{A}_{i}^{ext}(\mathbf{q}_{i},t)\right]^{2}}{2m_{i}}\\
 & +\sum_{i=1}^{N}e_{i}\left[\Phi_{i}^{ext}(\mathbf{q}_{i},t)+\Phi_{c}^{int}(\mathbf{q}_{i},\mathbf{q}_{j})\right]-\sum_{i=1}^{N}\frac{\hbar^{2}}{2m_{i}}\frac{\nabla_{i}^{2}\sqrt{\rho(q,t)}}{\sqrt{\rho(q,t)}},
\end{aligned}
\end{equation}
which describes the total energy of the possible mean trajectories
of the \emph{zbw} particles, and, upon evaluation at $q=q(t),$ the
total energy of the actual particles along their mean trajectories.
So (3.11) and (3.23) together define the \emph{N}-particle HJM equations.

Note that, as a consequence of the non-separability of $\Phi_{c}^{int}(\mathbf{q}_{i},\mathbf{q}_{j})$,
we will not be able to write (3.23) as a sum of total energies for
each particle (unless the particles are sufficiently spatially separated
from each other that we can effectively neglect this interaction term),
which means $S(q,t)\neq\sum_{i=1}^{N}S_{i}(\mathbf{q}_{i},t)$. Indeed,
as a consequence of this non-separability, we can now see from the
coupling of (3.11) and (3.23) that $R$ (and hence $U$) will also
be non-separable since its evolution depends on $\nabla_{i}S$ through
(3.11). We can make this more explicit by writing the general solutions,
$S$ and $R$, to (3.23) and the differentiated form of (3.11), respectively.
For (3.23), the general solution takes the form 
\begin{equation}
\begin{aligned}S(q,t) & =\sum_{i=1}^{N}\int\mathbf{p}_{i}(q,t)\cdot d\mathbf{q}_{i}\\
 & -\sum_{i=1}^{N}\int\left[\frac{\left[\mathbf{p}_{i}(q,t)-\frac{e_{i}}{c}\mathbf{A}_{i}^{ext}(\mathbf{q}_{i},t)\right]^{2}}{2m_{i}}+e_{i}\left[\Phi_{i}^{ext}(\mathbf{q}_{i},t)+\Phi_{c}^{int}(\mathbf{q}_{i},\mathbf{q}_{j})\right]-\frac{\hbar^{2}}{2m_{i}}\frac{\nabla_{i}^{2}\sqrt{\rho(q,t)}}{\sqrt{\rho(q,t)}}\right]dt.
\end{aligned}
\end{equation}

For the differentiated form of (3.11), the general solution $R$ can
be found most easily by first solving (3.11) directly in terms of
$\rho$ and then using the relation $\rho=e^{2R/\hbar}$. Rewriting
(3.11) as $\left(\partial_{t}+\sum_{i}^{N}\mathbf{v}_{i}\cdot\nabla_{i}\right)\rho=-\rho\sum_{i}^{N}\nabla_{i}\cdot\mathbf{v}_{i}$,
we have $(d/dt)ln[\rho]=-\sum_{i}^{N}\nabla_{i}\cdot\mathbf{v}_{i}$.
Solving this last expression yields 
\begin{equation}
\rho(q,t)=\rho_{0}(q_{0})exp[-\int_{0}^{t}\left(\sum_{i=1}^{N}\nabla_{i}\cdot\mathbf{v}_{i}\right)dt'.
\end{equation}
The osmotic potential obtained from $\rho$ then takes the form
\begin{equation}
R(q,t)=R_{0}(q_{0})-(\hbar/2)\int_{0}^{t}\left(\sum_{i=1}^{N}\nabla_{i}\cdot\mathbf{v}_{i}\right)dt'.
\end{equation}
Accordingly, we see clearly that $R$ depends on $S$ through $\mathbf{v}_{i}$,
and that $S$ depends on $R$ through the quantum kinetic. So the
non-separability of $\Phi_{c}^{int}$ alone entails non-factorizability
of $S(q,t)$, which entails non-factorizability of $R(q,t)$, which
entails non-factorizability of the quantum kinetic. \footnote{In Part I, we explained that we prefer to call the ``quantum potential''
the ``quantum kinetic'' in order to emphasize its physical origin
in the kinetic energy term associated with the osmotic velocity of
a Nelsonian particle. } That is, the nonlinear coupling between (3.24) and (3.26) entails
that $S$ is actually non-separable by virtue of the non-separability
of $\Phi_{c}^{int}$ \emph{and} (as a consequence thereof) that the
quantum kinetic is non-separable. Thus we've explicitly shown, from
the \emph{N}-particle HJM equations, that the presence of classical
interactions between Nelsonian particles means that the \emph{N}-particle
osmotic potential cannot be written as a separable sum of \emph{N}
osmotic potentials associated to each particle.

Let us now combine (3.11) and (3.23) into an \emph{N}-particle Schr\"{o}dinger 
equation and write down the most general form of the \emph{N}-particle
wavefunction. To do this, we first need to impose the \emph{N}-particle
generalization of the quantization condition 
\begin{equation}
\sum_{i=1}^{N}\oint_{L}\nabla_{i}S(q,t)\cdot d\mathbf{q}_{i}=nh,
\end{equation}
which, by (3.26), also constrains the osmotic potential sourced by
the ether. Then we can combine (3.11) and (3.23) into
\begin{equation}
i\hbar\frac{\partial\psi(q,t)}{\partial t}=\sum_{i=1}^{N}\left[\frac{\left[-i\hbar\nabla_{i}-\frac{e_{i}}{c}\mathbf{A}_{i}^{ext}(\mathbf{q}_{i},t)\right]^{2}}{2m_{i}}+e_{i}\left(\Phi_{i}^{ext}(\mathbf{q}_{i},t)+\Phi_{c}^{int}(\mathbf{q}_{i},\mathbf{q}_{j})\right)\right]\psi(q,t),
\end{equation}
where the single-valued \emph{N}-particle wavefunction in polar form
is $\psi(q,t)=\sqrt{\rho(q,t)}e^{iS(q,t)/\hbar}$.

\section{Interpretation of the Nelson-Yasue wavefunction}

How should we understand the NYSM-derived wavefunction satisfying
(3.28)? Is it part of NYSM's physical ontology, i.e., is it a beable?
Or should it be viewed as strictly epistemic, i.e., strictly reflecting
our ignorance about ontic aspects of an \emph{N}-particle NYSM system?

Straight off the bat, we can see that $\psi(q,t)$ is defined in terms
of $\rho(q,t)$ and $S(q,t)$. As noted in section 2, $\rho(q,t)$
is an epistemic distribution in that it reflects our ignorance of
the actual positions of the particles; hence $\rho(q,t)$ is not a
beable. As also noted in section 2, $S(q,t)$ is a field over the
possible positions of the actual particles and describes the possible
current velocities that the actual particles can have at each possible
point in 3-space they can occupy at time $t$; hence $S(q,t)$ is
also not a beable. Since $\psi(q,t)$ is defined in terms of $\rho(q,t)$
and $S(q,t)$, we must conclude that $\psi(q,t)$ is also not a beable
in NYSM. Rather, $\psi(q,t)$ can be said to be epistemic in the precise
sense that it's defined in terms of $\rho(q,t)$ and $S(q,t)$, and
these latter two variables reflect our ignorance about ontic properties
of the actual particles (their actual positions and velocities). In
other words, $\psi(q,t)$ ``represents our knowledge of the underlying
reality'' \cite{Leifer2014}, rather than being an element of the
underlying reality. 

However, even though $\psi(q,t)$ is not a beable, it does indirectly
reflect certain ontic aspects of the \emph{N}-particle system in NYSM. In
particular, the evolution of $\rho(q,t)$ depends on the evolution
of $R(q,t)$ via $\rho=e^{2R/\hbar}$, where $R(q,t)=\mu U(q,t)$
and where $U(q,t)$ is a beable. So $\rho(q,t)$ reflects an ontic
aspect of the system, namely the system's osmotic potential field
$U(q,t)$, and by extension so does $\psi(q,t)$ through its modulus.
Additionally, recall from section 2 that the introduction of $S(q,t)$
through the constraint $\mathbf{v}_{i}=m_{i}^{-1}\nabla_{i}S$ implies
that the ether, which is a beable of NYSM, is irrotational, and this
irrotationality is an ontic property of the ether. $S(q,t)$ also
encodes the presence of classical fields in the system (which can
be reasonably regarded as beables, in the sense that the electromagnetic
field is typically regarded as a beable) via the quantum Hamilton-Jacobi
equation (3.23-24), while also satisfying an ontic (law-like) constraint
via the quantization condition (3.27). So insofar as $S(q,t)$ reflects
ontic aspects of the system, namely the irrotationality of the ether,
the presence of classical fields in the system, and the quantization
constraint on the current velocities of the particles, so does $\psi(q,t)$
through its complex phase. 

It is worth emphasizing the significant conceptual differences between
$S(q,t)$ and $U(q,t)$, despite their formal mathematical similarities:
Even though both are fields on configuration space $\mathbb{R}^{3N}$,
and even though both enter into the stochastic differential equations
of motion (3.1) and (3.4) - $S(q,t)$ generating the current velocities,
and $U(q,t)$ generating the osmotic velocities - one field (the $U(q,t)$
field) is a beable and the other (the $S(q,t)$ field) isn't (though
it reflects ontic aspects/properties of a beable, the ether). Additionally,
$S(q,t)$ is subject to the quantization condition (3.27), which only
indirectly constrains the evolution of $U(q,t)$ via (3.26). 

It is also worth emphasizing that the epistemic features of the \emph{N}-particle
NYSM wavefunction are not in logical contradiction with the Pusey-Barrett-Rudolph
(PBR) theorem \cite{Pusey2012}: One of the assumptions of the PBR
theorem is that it is possible to prepare N systems independently,
with quantum states $\psi_{q_{1},...,}\psi_{q_{N}}$, which results
in ontic states $\lambda_{1},...,\lambda_{N}$ distributed according
to the product distribution $\mu_{q_{1}}(\lambda_{1})\mu_{q_{2}}(\lambda_{2})...\mu_{q_{N}}(\lambda_{N})$.
However, the ontic states of \emph{N}-particle NYSM, which include
the \emph{N}-particle osmotic potential, will in general not conform
to this `independence assumption', because the \emph{N}-particle osmotic
potential is in general non-separable, as we will show later on in
this section. In the special cases where the PBR independence assumption
is effectively satisfied in \emph{N}-particle NYSM, the NYSM wavefunction
qualifies as (effectively) ``psi-ontic'' (to use PBR's terminology)
in the precise sense that (effectively) distinct pure states satisfying
the \emph{N}-particle Schr\"{o}dinger
equation would (effectively) have non-overlapping distributions for
$\lambda$. Yet, it seems clear that the NYSM wavefunction being psi-ontic
in PBR's sense would not be logically inconsistent with the NYSM wavefunction
not being a beable (in Bell's sense, see footnote 2) and having epistemic
features in the precise sense we've already explained.

To see why the \emph{N}-particle osmotic potential is in general non-separable,
and to get a better feel for the conceptual and technical interplay
between the $\psi$, $R$, and $S$ fields, it is worth considering
a concrete example involving an entangled state. 

Consider the case of 2 distinguishable particles, where particle 1
is associated with a wavepacket $\psi_{A}$ and particle 2 is associated
with a packet $\psi_{B}$. If, initially, the particles are classically
non-interacting and there are no correlations between them, then the
joint wavefunction is the product state (suppressing the $t$ variable
for simplicity)
\begin{equation}
\psi_{f}(\mathbf{q}_{1},\mathbf{q}_{2})\coloneqq\psi_{A}(\mathbf{q}_{1})\psi_{B}(\mathbf{q}_{2}).
\end{equation}
We can also construct a non-factorizable solution of (3.28) by writing
\begin{equation}
\psi_{nf}(\mathbf{q}_{1},\mathbf{q}_{2})\coloneqq Norm\left[\psi_{A}(\mathbf{q}_{1})\psi_{B}(\mathbf{q}_{2})+\psi_{C}(\mathbf{q}_{1})\psi_{D}(\mathbf{q}_{2})\right].
\end{equation}
If the summands in (3.30) negligibly overlap by virtue of either $\psi_{A}\cap\psi_{C}\approx\varnothing$
or $\psi_{B}\cap\psi_{D}\approx\varnothing$ (\emph{Norm} = normalization
factor), then the system wavefunction is `effectively factorizable';
that is, the 2-particle wavefunction associated with the actual particles
at time $t$ is effectively either $\psi_{f}=\psi_{A}(\mathbf{q}_{1})\psi_{B}(\mathbf{q}_{2})$
or $\psi_{f}=\psi_{C}(\mathbf{q}_{1})\psi_{D}(\mathbf{q}_{2})$. On
the other hand, if we `turn on' the classical interaction $\Phi_{c}^{int}$,
evolution by (3.28) will make the overlap of the summands non-negligible,
and the system wavefunction will not be effectively factorizable \cite{Holland1993}.
Consequently, from (3.30), we will have a non-separable 2-particle
velocity potential given by
\begin{equation}
S_{nf}(\mathbf{q}_{1},\mathbf{q}_{2},)\coloneqq-\frac{i\hbar}{2}\ln\left(\frac{\psi_{nf}(\mathbf{q}_{1},\mathbf{q}_{2})}{\psi_{nf}^{\ast}(\mathbf{q}_{1},\mathbf{q}_{2})}\right).
\end{equation}
The probability density will also be non-factorizable since it becomes 
\begin{equation}
\begin{aligned}\rho_{nf}\left(\mathbf{q}_{1},\mathbf{q}_{2}\right) & \coloneqq|\psi_{nf}(\mathbf{q}_{1},\mathbf{q}_{2})|^{2}=Norm^{2}\left\{ e^{2\left(R_{A1}+R_{B2}\right)/\hbar}+e^{2\left(R_{C1}+R_{D2}\right)/\hbar}\right.\\
 & \left.+2e^{\left(R_{A1}+R_{C1}+R_{B2}+R_{D2}\right)/\hbar}cos\left[\left(S_{A1}+S_{B2}-S_{C1}-S_{D2}\right)/\hbar\right]\right\} .
\end{aligned}
\end{equation}
And the corresponding non-separable 2-particle osmotic potential takes
the form
\begin{equation}
R_{nf}(\mathbf{q}_{1},\mathbf{q}_{2})\coloneqq\hbar\ln\left(|\psi_{nf}(\mathbf{q}_{1},\mathbf{q}_{2})|\right),
\end{equation}
where $|\psi_{nf}(\mathbf{q}_{1},\mathbf{q}_{2})|=\sqrt{\rho_{nf}\left(\mathbf{q}_{1},\mathbf{q}_{2}\right)}$.

By the mathematical equivalence of (3.28) with the equation set (3.11)-(3.23)-(3.27),
we can see that (3.33) and (3.31) will be coupled solutions of (3.11)
and (3.23), respectively. On the other hand, when the summands of
$\psi_{nf}$ have effectively disjoint support in configuration space
(e.g., in the case of particles sufficiently separated that their
classical interaction can be neglected), the system wavefunction becomes
effectively factorizable again. In this case, the system velocity
potential is either $S_{f}=S_{A1}+S_{B2}$ or $S_{f}=S_{C1}+S_{D2}$,
the probability density reduces to $\rho_{f}\approx N^{2}\left(e^{2\left(R_{A1}+R_{B2}\right)/\hbar}+e^{2\left(R_{C1}+R_{D2}\right)/\hbar}\right)$,
and the system osmotic potential is either $R_{f}=R_{A1}+R_{B2}$
or $R_{f}=R_{C1}+R_{D2}$.

Incidentally, this latter case most clearly illustrates how, from
the stochastic mechanics viewpoint, the wavefunction plays the role
of an epistemic variable while also reflecting some of the ontic properties
of the physical system: The modulus-square of the factorizable two-particle
wavefunction describes the position density for a statistical ensemble
of two-particle systems, while the $R$ and $S$ functions encoded
in the factorizable two-particle wavefunction represent the \emph{possible}
$R$ and $S$ functions that the actual particles actually `have'
at time $t$; concurrently, the possible $R$ and $S$ functions for
the two-particle system reflect objectively real properties of Nelson's
ontic ether, insofar as $R_{A1}$ ($R_{B2}$) and $R_{C1}$ ($R_{D2}$)
correspond to (effectively) disjoint regions of the ontic osmotic
potential sourced by the ether $U_{A1}$ ($U_{B2}$) and $U_{C1}$
($U_{D2}$), and insofar as $S_{A1}$ ($S_{B2}$) and $S_{C1}$ ($S_{D2}$)
reflect the irrotationality of the ether in regions \emph{A} and \emph{B}
and regions $C$ and $D$. This confirms the properties of the osmotic
potential and its relation to the velocity potential that we observed
from the solutions of the \emph{N}-particle HJM equations, for the
cases of classically interacting and non-interacting distinguishable
particles.

However, we should note that the linearity of (3.28) entails non-factorizable
solutions for the case of classically non-interacting identical bosons
or fermions. (To justify the symmetrization postulates, we can import
Bacciagaluppi's finding \cite{Bacciagaluppi2003} that the symmetrization
postulates are derivable from the assumption of symmetry of the Nelsonian
particle trajectories in configuration space.) For identical bosons
or fermions, we simply replace $\psi_{C}(\mathbf{q}_{1})\psi_{D}(\mathbf{q}_{2})$
in (3.30) with $\pm\psi_{A}(\mathbf{q}_{2})\psi_{B}(\mathbf{q}_{1})$,
and similarly for $S_{nf}$, $\rho_{nf}$, and $R_{nf}$. Then, if
particle 1 and particle 2 start out without any classical interaction,
we will initially have $\psi_{A}\cap\psi_{B}\approx\varnothing$ (approximately,
because the wavepackets never have completely disjoint support in
configuration space, even in the non-interacting case); if the packets
of these particles then move towards each other and overlap such that
$\left(<\mathbf{q}_{1}>-<\mathbf{q}_{2}>\right)^{2}\leq\sigma_{A}^{2}+\sigma_{B}^{2}$,
where $\sigma_{A}$ and $\sigma_{B}$ are the widths of the packets,
the resulting wavefunction of the 2-particle system will be given
by (3.30) with $\psi_{A}\cap\psi_{B}\neq\varnothing$ \cite{Holland1993}.
Physically, the appreciable overlap of the wavepackets implies that
the initially independent osmotic potentials possibly associated with
particle 1 ($R_{A1}$ or $R_{B1}$) and particle 2 ($R_{A2}$ or $R_{B2}$),
respectively, become non-separable by virtue of their joint support
in configuration space becoming non-negligible. So the resulting motion
of particle 1 will have a non-separable physical dependence on part
of the osmotic potentials possibly associated with particle 2 (and
vice versa), a dependence which is instantaneous between the particles
in 3-space (since the $N$-particle quantum kinetic in (3.23) acts
instantaneously on the two particles at time $t$). Of course, for
classically non-interacting identical particles, the 2-particle wavefunction
will satisfy $\psi_{A}\cap\psi_{B}=\varnothing$ again once the wavepackets
pass each other and their overlap becomes negligible; but if the particles
are classically interacting via $\Phi_{c}^{int}$ the non-separability
will persist until the particles are sufficiently spatially separated
that $\Phi_{c}^{int}\approx0$.

Thus the linearization of the HJM equations into Schr\"{o}dinger's
equation, through the use of condition (3.27), makes possible non-separable/non-local
correlations between (distinguishable or identical) particles not
admitted by the HJM equations alone (since the solutions of the HJM
equations don't generally satisfy the superposition principle without
(3.27), as we know from Wallstrom \cite{Wallstrom1989}). \footnote{To be clear, we are not claiming that the HJM equations, without the
quantization condition, do not admit solutions that make possible
EPR-type correlations between particles. It seems plausible that they
do, considering that classical Liouville statistical mechanics (with
an epistemic restriction akin to the Heisenberg uncertainty principle)
does so \cite{Bartlett2012}, and that even without the quantization
condition stochastic mechanics reproduces the uncertainty relations.
But whether solutions exist that are non-local enough to entail violations
of the continuous-variable Bell inequality \cite{Cavalcanti2007}
seems unclear. Answering this question requires a detailed mathematical
study of the analytic solutions of the HJM equations, without the
quantization condition imposed. To the best of our knowledge, this
has yet to be done.} In fact, such solutions tell us that the two-particle wavefunction
for identical bosons (interacting or non-interacting) must always
be given by (3.30), where the joint support of the summands never
completely vanishes and can increase appreciably due to (classical
or non-classical) interactions between the particles \cite{Holland1993}.

This last realization complicates the interpretation of the space
in which Nelson's ether lives versus the space in which the particles
live: we started out by postulating that the ether lives in 3-D space,
but have found that once the constraints (3.19) and (3.27) are imposed,
the $R$ and $S$ functions (which, as we've seen, reflect objectively
real properties of the ether) are in general not separable, and thus
(mathematically) always live in $3N$-dimensional configuration space.
If we take this mathematical non-factorizability of $R$ and $S$
as a literal indication about the ontic nature of the ether, then
this would seem to force us to infer that the ether must actually
live in $3N$-dimensional configuration space, and therefore regard
configuration space as an ontic space in its own right. We could then
say (to whatever extent one finds this plausible) that the ether and
osmotic potential live in configuration space, but that there are
still \emph{N} ontic particles living in an (also) ontic 3-D space,
and postulate that the two sets of beables can somehow causally interact
with each other via the set (3.1)-(3.4)-(3.21), despite living in
independent ontic spaces. (This situation is analogous to a common
interpretation of the de Broglie-Bohm theory, where the fundamental
ontology consists of an ontic wavefunction living in an ontic $3N$-dimensional
configuration space, and $N$ ontic particles living in an ontic 3-D
space; one then postulates a one-way causal relationship between the
wavefunction and the \emph{N} particles via the ``guiding equation''
\cite{Holland1993,Bohm1995,BellQMCosmo2004}.)

Alternatively, if one finds it unintelligible to say that beables
living in two independent ontic spaces can causally interact (or even
that one set of beables merely naturally supervenes on the other set),
we could suppose (in analogy with Albert's ``flat-footed'' interpretation
of the de Broglie-Bohm theory \cite{Albert1996,Albert2013}) that
the representation of \emph{N} particles in 3-D space is a mathematical
fiction and that the ontic description is actually a single particle
in $3N$-dimensional configuration space. This has the virtue that
it is straightforward to assert that this single particle causally
interacts with Nelson's ether (since they both live in the same ontic
space). The cost is that one now has to employ a complicated (philosophical)
functional analysis \cite{Albert1996,Albert2013,Albert2015} of how
the form of the interaction potential $\Phi_{c}^{int}(\mathbf{q}_{i},\mathbf{q}_{j})$
in the quantum Hamilton-Jacobi equation (3.23) makes it possible to
recover \emph{N} particles in 3-D space as an emergent ontology; additionally,
this view seems logically inconsistent with the fact that the non-separable
\emph{R} and \emph{S} functions are \emph{consequences} \emph{of}
extremizing the action (3.18), defined in terms of $N$ contributions,
if there aren't really $N$ particles diffusing in 3-D space to which
those \emph{N} contributions correspond.

A third possibility is that the configuration-space representation
of $R$ and $S$ is somehow just an abstract encoding of a complicated
array of ontic fields in space-time that nonlocally connect the motions
of the particles. In practice, we might implement this by analogy
with Norsen's ``TELB'' approach to the de Broglie-Bohm theory \cite{Norsen2010,Norsen2014}:
Taylor-expand the $R$ and $S$ functions in configuration space into
$N$ one-particle $R$ and $S$ functions, each coupled to a countably
infinite hierarchy of ``entanglement fields'' in space-time that
implement the nonlocal connections between the motions of the particles.
The upshot of this approach is that one can maintain that Nelson's
ether lives in plain-old 3-D space along with \emph{N} particles.
A drawback is the immense complexity of positing a countable infinity
of ontic fields in space-time, in order to reproduce all the information
encoded in the \emph{R} and \emph{S} functions in configuration space.
To be sure, this last possibility is more speculative than the former
two (since it would be non-trivial to actually construct such a variant
of NYSM); but we think it is ultimately the most intelligible and
fruitful one for stochastic mechanics (for reasons discussed in sections
4 and 5).

Of course, the validity of constructing the non-separable solutions
(3.30-33) in NYSM depends on the plausibility of imposing (3.27).
But such a condition is arbitrary from the point of view of (3.11)
and (3.23), insofar as we have reconstructed those equations from
the Nelson-Yasue assumptions. This, in essence, is Wallstrom's criticism
applied to the $N$-particle case. Our task then is to reformulate
\emph{N}-particle NYSM into \emph{N}-particle ZSM.

\section{Classical Model of Constrained Zitterbewegung Motion for Many Particles}

In developing $N$-particle ZSM, it will be helpful to first develop
the $N$-particle version of our classical $\emph{zbw}$ model, for
free particles, particles interacting with external fields, and particles
interacting with each other through Coulomb forces. As we will see,
even at the classical level, the $N$-particle extension turns out
to be non-trivial.

\subsection{Free \emph{zbw} particles}

Let us now suppose we have $N$ identical, non-interacting \emph{zbw}
particles in space-time, and no external fields present. In other
words, the $i$-th particle has rest mass $m_{i}$ (taking $i=1,...,N$)
and is rheonomically constrained to undergo an unspecified oscillatory
process with constant angular frequency $\omega_{ci}$ about some
fixed point in 3-space $\mathbf{q}_{0i}$ in a Lorentz frame where
$\mathbf{v}_{i}=d\mathbf{q}_{0i}/dt=0$. Then, in a fixed Lorentz
frame where $\mathbf{v}_{i}\neq0$, the $zbw$ phase for the $i$-th
free particle takes the form (using $\theta_{i}\eqqcolon-\frac{\omega_{ci}}{m_{i}c^{2}}S_{i}=-\frac{1}{\hbar}S_{i}$)
\begin{equation}
\delta S_{i}(\mathbf{q}_{i}(t),t)=\left(\mathbf{p}_{i}\cdot\delta\mathbf{q}_{i}(t)-E_{i}\delta t\right),
\end{equation}
where $E_{i}=\gamma_{i}m_{i}c^{2}$. So for each particle, we will
have
\begin{equation}
\oint_{L}\delta S_{i}(\mathbf{q}_{i}(t),t)=\oint_{L}\left(\mathbf{p}_{i}\cdot\delta\mathbf{q}_{i}(t)-E_{i}\delta t\right)=nh,
\end{equation}
which implies
\begin{equation}
\sum_{i=1}^{N}\oint_{L}\delta S_{i}(\mathbf{q}_{i}(t),t)=\sum_{i=1}^{N}\oint_{L}\left(\mathbf{p}_{i}\cdot\delta\mathbf{q}_{i}(t)-E_{i}\delta t\right)=nh.
\end{equation}
In the non-relativistic limit, the $i$-th $zbw$ phase is
\begin{equation}
S_{i}(\mathbf{q}_{i}(t),t)\approx m_{i}\mathbf{v}_{i}\cdot\mathbf{q}_{i}(t)-\left(m_{i}c^{2}+\frac{m_{i}v_{i}(\mathbf{q}_{i}(t),t)^{2}}{2}\right)t+\hbar\phi_{i},
\end{equation}
and satisfies the classical HJ equation
\begin{equation}
E_{i}(\mathbf{q}_{i}(t),t)=-\partial_{t}S_{i}(\mathbf{q}_{i},t)|_{\mathbf{q}_{j}=\mathbf{q}_{j}(t)}=\frac{\left(\nabla_{i}S_{i}(\mathbf{q}_{i},t)\right)^{2}}{2m_{i}}|_{\mathbf{q}_{j}=\mathbf{q}_{j}(t)}+m_{i}c^{2}.
\end{equation}
We can also define the total system energy as the sum of the individual
energies of each \emph{zbw} particle:
\begin{equation}
E(q(t),t)=-\partial_{t}S(q,t)|_{\mathbf{q}_{j}=\mathbf{q}_{j}(t)}=\sum_{i=1}^{N}\frac{\left(\nabla_{i}S(q,t)\right)^{2}}{2m_{i}}|_{\mathbf{q}_{j}=\mathbf{q}_{j}(t)}+\sum_{i=1}^{N}m_{i}c^{2},
\end{equation}
where we have used $E=-\partial_{t}S=\sum_{i=1}^{N}E_{i}=-\sum_{i=1}^{N}\partial_{t}S_{i}=-\partial_{t}\sum_{i=1}^{N}S_{i}$.
Accordingly, we can define the `joint phase' of the \emph{N}-particle
system as the sum
\begin{equation}
S(q(t),t)=\sum_{i=1}^{N}S_{i}(\mathbf{q}_{i}(t),t)\approx\sum_{i=1}^{N}m_{i}\mathbf{v}_{i}(q(t),t)\cdot\mathbf{q}_{i}(t)-\left(\sum_{i=1}^{N}m_{i}c^{2}+\sum_{i=1}^{N}\frac{m_{i}v_{i}(q(t),t)^{2}}{2}\right)t+\hbar\sum_{i=1}^{N}\phi_{i},
\end{equation}
which satisfies (3.39). Correspondingly, we can rewrite (3.36) as
\begin{equation}
\sum_{i=1}^{N}\oint_{L}\nabla_{i}S|_{\mathbf{q}_{j}=\mathbf{q}_{j}(t)}\cdot\delta\mathbf{q}_{i}(t)=nh,
\end{equation}
for displacements along closed loops with time held fixed. We are
now ready to formulate the HJ statistical mechanics for \emph{N} free
particles.

\subsection{Classical Hamilton-Jacobi statistical mechanics for free \emph{zbw}
particles}

If the actual positions of the $zbw$ particles are unknown, then
$\mathbf{q}_{i}(t)$ gets replaced by $\mathbf{q}_{i}$, and the non-relativistic
joint \emph{zbw} phase becomes a field over the possible positions
of the actual \emph{zbw} particles, namely
\begin{equation}
S(q,t)\approx\sum_{i=1}^{N}m_{i}\mathbf{v}_{i}(q,t)\cdot\mathbf{q}_{i}-\sum_{i=1}^{N}\left(m_{i}c^{2}+\frac{m_{i}v_{i}(q,t)^{2}}{2}\right)t+\sum_{i=1}^{N}\hbar\phi_{i},
\end{equation}
where $\mathbf{v}_{i}(q,t)=\nabla_{i}S(q,t)/m_{i}$ and satisfies
\begin{equation}
\sum_{i=1}^{N}\oint_{L}\nabla_{i}S\cdot d\mathbf{q}_{i}=nh,
\end{equation}
and
\begin{equation}
E(q,t)=-\partial_{t}S=\sum_{i=1}^{N}\left[\frac{\left(\nabla_{i}S\right)^{2}}{2m_{i}}+m_{i}c^{2}\right].
\end{equation}
The physical independence of the particles further implies 
\begin{equation}
E_{i}=-\partial_{t}S_{i}=\frac{\left(\nabla_{i}S_{i}\right)^{2}}{2m_{i}}+m_{i}c^{2},
\end{equation}
where
\begin{equation}
S(q,t)=\sum_{i=1}^{N}S_{i}(\mathbf{q}_{i},t),
\end{equation}
and
\begin{equation}
\oint_{L}\nabla_{i}S_{i}\cdot d\mathbf{q}_{i}=nh.
\end{equation}

As (3.42) is defined from the sum of \emph{N} independent phase fields,
Eq. (3.46), the corresponding velocity fields, $\mathbf{v}_{i}(q,t)$,
are also physically independent of one another. Consequently, for
the trajectory fields obtained from integrating $\mathbf{v}_{i}(q,t)$,
the associated \emph{N}-particle probability density $\rho(q,t)=n(q,t)/N$
can be taken in most cases to be factorizable into a product of \emph{N}
independent probability densities (for simplicity, we ignore the special
case of classical correlations corresponding to when $\rho$ is a
mixture of factorizable densities; but see \cite{Bacciagaluppi2012}
for a discussion of classical correlations in a related context):
\begin{equation}
\rho(q,t)=\prod_{i}^{N}\rho_{i}(\mathbf{q}_{i},t),
\end{equation}
where (3.48) satisfies $\rho(q,t)\geq0$, the normalization condition
$\int_{\mathbb{R}^{3N}}\rho_{0}(q)d^{3N}q=1$, and evolves by the
\emph{N}-particle continuity equation
\begin{equation}
\frac{\partial\rho}{\partial t}=-\sum_{i=1}^{N}\nabla_{i}\cdot\left[\left(\mathbf{\frac{\nabla_{\mathit{i}}\mathrm{\mathit{S}}}{\mathit{m_{i}}}}\right)\rho\right],
\end{equation}
which by (3.48) implies 
\begin{equation}
\frac{\partial\rho_{i}}{\partial t}=-\nabla_{i}\cdot\left[\left(\mathbf{\frac{\nabla_{\mathit{i}}\mathrm{\mathit{S_{i}}}}{\mathit{m_{i}}}}\right)\rho_{i}\right].
\end{equation}

We can then combine (3.44) and (3.49) to obtain a single-valued \emph{N}-particle
classical wavefunction $\psi(q,t)=\sqrt{\rho_{0}(\mathbf{q}_{1}-\mathbf{v}_{1}t,...,\mathbf{q}_{N}-\mathbf{v}_{N}t)}e^{iS(q,t)/\hbar}$
satisfying the \emph{N}-particle nonlinear Schr\"{o}dinger  equation
\begin{equation}
i\hbar\frac{\partial\psi}{\partial t}=\sum_{i=1}^{N}\left[-\frac{\hbar^{2}}{2m_{i}}\nabla_{i}^{2}+\frac{\hbar^{2}}{2m_{i}}\frac{\nabla_{i}^{2}|\psi|}{|\psi|}+m_{i}c^{2}\right]\psi,
\end{equation}
which implies
\begin{equation}
i\hbar\frac{\partial\psi_{i}}{\partial t}=\left[-\frac{\hbar^{2}}{2m_{i}}\nabla_{i}^{2}+\frac{\hbar^{2}}{2m_{i}}\frac{\nabla_{i}^{2}|\psi_{i}|}{|\psi_{i}|}+m_{i}c^{2}\right]\psi_{i},
\end{equation}
since 
\begin{equation}
\psi(q,t)=\prod_{i}^{N}\psi_{i}(\mathbf{q}_{i},t).
\end{equation}
Having completed the description of \emph{N} free particles, we now
develop the slightly less trivial case of \emph{zbw} particles interacting
with external fields.

\subsection{External fields interacting with \emph{zbw} particles}

To describe the interaction of our \emph{zbw} particles with external
fields, consider first the change in the \emph{zbw} phase of the $i$-th
particle in its rest frame:
\begin{equation}
\delta\theta_{i}(t_{0})=\omega_{ci}\delta t_{0}=\frac{1}{\hbar}\left(m_{i}c^{2}\right)\delta t_{0}.
\end{equation}
The coupling of the particle to (say) the Earth's external gravitational
field leads to a small correction (in the now instantaneous rest frames
of the particles) as follows:
\begin{equation}
\delta\theta_{i}(\mathbf{q}_{0i},t_{0})=\left[\omega_{ci}+\kappa_{i}(\mathbf{q}_{0i})\right]\delta t_{0}=\frac{1}{\hbar}\left[m_{i}c^{2}+m_{i}\Phi_{gi}^{ext}(\mathbf{q}_{0i})\right]\delta t_{0},
\end{equation}
where $\kappa_{i}=\omega_{ci}\Phi_{gi}^{ext}/c^{2}$. As in the single
particle case, we have approximated the coupling as point-like since
we assume $|\mathbf{q}_{i}|\gg\lambda_{ci}$. Supposing also that
the \emph{zbw} particles carry charge $e_{i}$ (so that they now become
classical charged oscillators of some identical type), their point-like
couplings to a space-time varying external electric field lead to
additional (small) phase shifts of the form
\begin{equation}
\delta\theta_{i}(\mathbf{q}_{0i},t_{0})=\left[\omega_{ci}+\kappa_{i}(\mathbf{q}_{0i})+\varepsilon_{i}(\mathbf{q}_{0i},t_{0})\right]\delta t_{0}=\frac{1}{\hbar}\left[m_{i}c^{2}+m_{i}\Phi_{gi}^{ext}(\mathbf{q}_{0i})+e_{i}\Phi_{ei}^{ext}(\mathbf{q}_{0i},t_{0})\right]\delta t_{0},
\end{equation}
where $\varepsilon_{i}=\omega_{ci}\left(e_{i}/m_{i}c^{2}\right)\Phi_{ei}^{ext}$.

Transforming to the lab frame where the $i$-th \emph{zbw} particle
has nonzero but variable translational velocity, (3.56) becomes
\begin{equation}
\begin{aligned}\delta\theta_{i}(\mathbf{q}_{i}(t),t) & =\left[\left(\omega_{dBi}+\kappa_{i}(\mathbf{q}_{i}(t))+\varepsilon_{i}(\mathbf{q}_{i}(t),t)\right)\gamma_{i}\left(\delta t-\frac{\mathbf{v}_{0i}(\mathbf{q}_{i}(t),t)\cdot\delta\mathbf{q}_{i}(t)}{c^{2}}\right)\right]\\
 & =\frac{1}{\hbar}\left[\left(\gamma_{i}m_{i}c^{2}+\gamma_{i}m_{i}\Phi_{gi}^{ext}+e_{i}\Phi_{ei}^{ext}\right)\delta t-\left(\gamma_{i}m_{i}c^{2}+\gamma_{i}m_{i}\Phi_{gi}^{ext}+e_{i}\Phi_{ei}^{ext}\right)\frac{\mathbf{v}_{0i}\cdot\delta\mathbf{q}_{i}(t)}{c^{2}}\right]\\
 & =\frac{1}{\hbar}\left(E_{i}\delta t-\mathbf{p}_{i}\cdot\delta\mathbf{q}_{i}(t)\right),
\end{aligned}
\end{equation}
where $E_{i}=\gamma_{i}m_{i}c^{2}+\gamma_{i}m_{i}\Phi_{gi}^{ext}+e_{i}\Phi_{ei}^{ext}$
and $\mathbf{p}_{i}=m_{i}\mathbf{v}_{i}=\left(\gamma_{i}m_{i}c^{2}+\gamma_{i}m_{i}\Phi_{gi}^{ext}+e_{i}\Phi_{ei}^{ext}\right)\left(\mathbf{v}_{0i}/c^{2}\right)$.
Incorporating coupling to an external vector potential, we have $\mathbf{v}_{i}\rightarrow\mathbf{v}_{i}'=\mathbf{v}_{i}+e_{i}\mathbf{A}_{i}^{ext}/\gamma_{i}m_{i}c$
(where $\gamma_{i}$ depends on the time-dependent $v_{i}$).

Now, even under the physical influence of the external fields, the
phase of the $i$-th particle's oscillation is a well-defined function
of its space-time location. Thus, if we displace the $i$-th particle
around a closed loop, the phase change is still given by
\begin{equation}
\oint_{L}\delta\theta_{i}=\frac{1}{\hbar}\oint_{L}\left[E_{i}\delta t-\mathbf{p}_{i}'\cdot\delta\mathbf{q}_{i}(t)\right]=2\pi n,
\end{equation}
or
\begin{equation}
\oint_{L}\delta S_{i}=\oint_{L}\left[\mathbf{p}_{i}'\cdot\delta\mathbf{q}_{i}(t)-E_{i}\delta t\right]=nh.
\end{equation}
Accordingly, we will also have
\begin{equation}
\sum_{i=1}^{N}\oint_{L}\delta S_{i}=\sum_{i=1}^{N}\oint_{L}\left[\mathbf{p}_{i}'\cdot\delta\mathbf{q}_{i}(t)-E_{i}\delta t\right]=nh.
\end{equation}
Moreover, for the special case of a loop in which time is held fixed,
we have
\begin{equation}
\oint_{L}\nabla_{i}S_{i}|_{\mathbf{q}_{i}=\mathbf{q}_{i}(t)}\cdot\delta\mathbf{q}_{i}(t)=\oint_{L}\mathbf{p}_{i}'\cdot\delta\mathbf{q}_{i}(t)=nh,
\end{equation}
or
\begin{equation}
\oint_{L}m_{i}\mathbf{v}_{i}\cdot\delta\mathbf{q}_{i}(t)=nh-\frac{e_{i}}{c}\oint_{L}\mathbf{A}_{i}^{ext}\cdot\delta\mathbf{q}_{i}(t).
\end{equation}
Likewise
\begin{equation}
\sum_{i=1}^{N}\oint_{L}\nabla_{i}S_{i}|_{\mathbf{q}_{i}=\mathbf{q}_{i}(t)}\cdot\delta\mathbf{q}_{i}(t)=\sum_{i=1}^{N}\oint_{L}\mathbf{p}_{i}'\cdot\delta\mathbf{q}_{i}(t)=nh,
\end{equation}
which is equivalent to
\begin{equation}
\sum_{i=1}^{N}\oint_{L}m_{i}\mathbf{v}_{i}\cdot\delta\mathbf{q}_{i}(t)=nh-\sum_{i=1}^{N}\frac{e_{i}}{c}\oint_{L}\mathbf{A}_{i}^{ext}\cdot\delta\mathbf{q}_{i}(t).
\end{equation}

Integrating (3.57) and rewriting in terms of $S_{i}$, we obtain
\begin{equation}
S_{i}=\int\left[\mathbf{p}_{i}'\cdot d\mathbf{q}_{i}(t)-E_{i}dt\right]-\hbar\phi_{i},
\end{equation}
and thus
\begin{equation}
S=\sum_{i=1}^{N}S_{i}=\sum_{i=1}^{N}\int\left[\mathbf{p}_{i}'\cdot d\mathbf{q}_{i}(t)-E_{i}dt\right]-\sum_{i=1}^{N}\hbar\phi_{i}.
\end{equation}
When $v_{i}\ll c$
\begin{equation}
\begin{aligned}S & \approx\sum_{i=1}^{N}\int m_{i}\mathbf{v}_{i}'\cdot d\mathbf{q}_{i}(t)-\\
 & -\sum_{i=1}^{N}\int\left(m_{i}c^{2}+\frac{1}{2m_{i}}\left[\mathbf{p}_{i}-\frac{e_{i}}{c}\mathbf{A}_{i}^{ext}\right]^{2}+m_{i}\Phi_{gi}^{ext}+e_{i}\Phi_{ei}^{ext}\right)dt-\sum_{i=1}^{N}\hbar\phi_{i},
\end{aligned}
\end{equation}
and satisfies
\begin{equation}
-\partial_{t}S|_{\mathbf{q}_{j}=\mathbf{q}_{j}(t)}=\sum_{i=1}^{N}\frac{\left(\nabla_{i}S-\frac{e_{i}}{c}\mathbf{A}_{i}^{ext}\right)^{2}}{2m_{i}}|_{\mathbf{q}_{j}=\mathbf{q}_{j}(t)}+\sum_{i=1}^{N}\left[m_{i}c^{2}+m_{i}\Phi_{gi}^{ext}+e_{i}\Phi_{ei}^{ext}\right],
\end{equation}
where the kinetic velocity, $\mathbf{v}_{i}=(1/m_{i})\nabla_{i}S|_{\mathbf{q}_{j}=\mathbf{q}_{j}(t)}-e_{i}\mathbf{A}_{i}^{ext}/m_{i}c$,
satisfies the classical Newtonian equation of motion
\begin{equation}
\begin{aligned}m_{i}\ddot{\mathbf{q}}_{i}(t) & =\left(\frac{\partial}{\partial t}+\mathbf{v}_{i}\cdot\nabla_{i}\right)\left[\nabla_{i}S-\frac{e_{i}}{c}\mathbf{A}_{i}^{ext}\right]|_{\mathbf{q}_{j}=\mathbf{q}_{j}(t)}\\
 & =-\nabla_{i}\left[m_{i}\Phi_{gi}^{ext}+e_{i}\Phi_{ei}^{ext}\right]|_{\mathbf{q}_{j}=\mathbf{q}_{j}(t)}-\frac{e_{i}}{c}\frac{\partial\mathbf{A}_{i}^{ext}}{\partial t}|_{\mathbf{q}_{j}=\mathbf{q}_{j}(t)}+\frac{e_{i}}{c}\mathbf{v}_{i}\times\mathbf{B}_{i}^{ext}.
\end{aligned}
\end{equation}

As in the previous section, we now want to extend our model to a classical
HJ statistical mechanics for $N$-particles.

\subsection{Classical Hamilton-Jacobi statistical mechanics for \emph{zbw} particles
interacting with external fields}

If in the lab frame we do not know the actual positions of the \emph{zbw}
particles, then $\mathbf{q}_{i}(t)$ gets replaced by $\mathbf{q}_{i}$,
and the phase (3.67) becomes a field over the possible positions of
the \emph{zbw} particles. In the $v_{i}\ll c$ approximation
\begin{equation}
\begin{aligned}S(q,t) & =\sum_{i=1}^{N}\int_{\mathbf{q}_{i}(t_{i})}^{\mathbf{q}_{i}(t)}m_{i}\mathbf{v}_{i}'(q(s),s)\cdot\mathbf{\mathit{d}q}_{i}(s)|_{\mathbf{q}_{j}(t)=\mathbf{q}_{j}}\\
 & -\sum_{i=1}^{N}\int_{t_{i}}^{t}\left(m_{i}c^{2}+\frac{1}{2m_{i}}\left[\mathbf{p}_{i}(q(s),s)-\frac{e_{i}}{c}\mathbf{A}_{i}^{ext}(q(s),s)\right]^{2}\right.\\
 & \left.+m_{i}\Phi_{gi}^{ext}(\mathbf{q}_{i}(s))+e_{i}\Phi_{ei}^{ext}(\mathbf{q}_{i}(s),s)\right)ds|_{\mathbf{q}_{j}(t)=\mathbf{q}_{j}}-\sum_{i=1}^{N}\hbar\phi_{i}.
\end{aligned}
\end{equation}
To obtain the equations of motion for $\ensuremath{S}$ and $\ensuremath{\mathbf{v}_{i}}$
we will now apply the classical analogue of Yasue's $N$-particle
variational principle, in anticipation of the method we will use for
constructing $N$-particle ZSM (we did not do this in the free-particles
case because there the dynamics of the particles is trivial).

First we define the ensemble-averaged $N$-particle phase/action (inputting
limits between initial and final states),
\begin{equation}
\begin{aligned}J & =\mathrm{E}\left[\sum_{i=1}^{N}\left[\int_{\mathbf{q}_{iI}}^{\mathbf{q}_{iF}}m_{i}\mathbf{v}_{i}'\cdot\mathbf{\mathit{d}q}_{i}(t)-\int_{t_{I}}^{t_{F}}\left(m_{i}c^{2}+\frac{1}{2m_{i}}\left[\mathbf{p}_{i}-\frac{e_{i}}{c}\mathbf{A}_{i}^{ext}\right]^{2}+m_{i}\Phi_{gi}^{ext}+e_{i}\Phi_{ei}^{ext}\right)dt-\hbar\phi_{i}\right]\right]\\
 & =\mathrm{E}\left[\int_{t_{I}}^{t_{F}}\sum_{i=1}^{N}\left\{ \frac{1}{2}m\mathbf{v}_{i}^{2}+\frac{e_{i}}{c}\mathbf{A}_{i}^{ext}\cdot\mathbf{v}_{i}-m_{i}c^{2}-m_{i}\Phi_{gi}^{ext}-e_{i}\Phi_{ei}^{ext}\right\} dt-\sum_{i=1}^{N}\hbar\phi_{i}\right],
\end{aligned}
\end{equation}
where the equated expressions are related by the usual Legendre transformation.
Imposing the variational constraint
\begin{equation}
J=extremal,
\end{equation}
a straightforward computation exactly along the lines of the Appendix
yields (3.69). And, upon replacing $\mathbf{q}_{i}(t)$ by $\mathbf{q}_{i}$,
we obtain the equation of motion for the acceleration field $\mathbf{a}(q,t)$:
\begin{equation}
\begin{aligned}m_{i}\mathbf{a}_{i} & =\left(\frac{\partial}{\partial t}+\mathbf{v}_{i}\cdot\nabla_{i}\right)\left[\nabla_{i}S-\frac{e_{i}}{c}\mathbf{A}_{i}^{ext}\right]\\
 & =-\nabla_{i}\left[m_{i}\Phi_{gi}^{ext}+e_{i}\Phi_{ei}^{ext}\right]-\frac{e_{i}}{c}\frac{\partial\mathbf{A}_{i}^{ext}}{\partial t}+\frac{e_{i}}{c}\mathbf{v}_{i}\times\mathbf{B}_{i}^{ext},
\end{aligned}
\end{equation}
where $\mathbf{v}_{i}=(1/m_{i})\nabla_{i}S-e_{i}\mathbf{A}_{i}^{ext}/m_{i}c$
corresponds to the kinetic velocity field associated with the \emph{i}-th
particle.

Integrating both sides of (3.73), summing over all $N$ terms, and
setting the integration constants equal to the rest masses, we then
obtain the classical $N$-particle Hamilton-Jacobi equation for (3.70)
\begin{equation}
-\partial_{t}S=\sum_{i=1}^{N}\frac{\left(\nabla_{i}S-\frac{e_{i}}{c}\mathbf{A}_{i}^{ext}\right)^{2}}{2m_{i}}+\sum_{i=1}^{N}\left[m_{i}c^{2}+m_{i}\Phi_{gi}^{ext}+e_{i}\Phi_{ei}^{ext}\right].
\end{equation}
Correspondingly, the probability density $\rho(q,t)$ now evolves
by the modified $N$-particle continuity equation
\begin{equation}
\frac{\partial\rho}{\partial t}=-\sum_{i=1}^{N}\nabla_{i}\cdot\left[\left(\frac{\nabla_{i}S}{m_{i}}-\frac{e_{i}}{m_{i}c}\mathbf{A}_{i}^{ext}\right)\rho\right],
\end{equation}
which preserves the normalization, $\int\rho_{0}d^{3N}q=1$. As in
the free particle case, since $S$ is a field over the possible positions
that the actual \emph{zbw} particles can occupy at a time \emph{t},
and since for each possible position the phase of each \emph{zbw}
particle satisfies the condition (3.63), it follows that $S$ is a
single-valued function of $q$ and $t$ (up to an additive integer
multiple of $2\pi$) and satisfies
\begin{equation}
\sum_{i=1}^{N}\oint_{L}\nabla_{i}S\cdot d\mathbf{q}_{i}=nh.
\end{equation}
Then we can combine (3.74-75) into the nonlinear Schr\"{o}dinger
equation
\begin{equation}
i\hbar\frac{\partial\psi}{\partial t}=\sum_{i=1}^{N}\left[\frac{\left[-i\hbar\nabla_{i}-\frac{e_{i}}{c}\mathbf{A}_{i}^{ext}\right]^{2}}{2m_{i}}+\frac{\hbar^{2}}{2m_{i}}\frac{\nabla_{i}^{2}|\psi|}{|\psi|}+m_{i}\Phi_{gi}^{ext}+e_{i}\Phi_{ei}^{ext}+m_{i}c^{2}\right]\psi,
\end{equation}
with $N$-particle wavefunction $\psi(q,t)=\sqrt{\rho(q,t)}e^{iS(q,t)/\hbar}$,
which is single-valued because of (3.76). We can also obtain the single-particle
versions of (3.74-77) in the case that $S$, $\rho$, and $\psi$
satisfy the factorization conditions (3.46), (3.48), and (3.53), respectively.

We are now ready to develop the more involved case of classically
interacting \emph{zbw} particles.

\subsection{Classically \textcolor{black}{interacting }\textcolor{black}{\emph{zbw}}\textcolor{black}{{}
particles }}

For simplicity we will consider just two \emph{zbw} particles classically
interacting through a scalar potential in the lab frame, under the
assumptions that $v_{i}\ll c$ and no external potentials are present.
(Restricting the particles to the non-relativistic regime also avoids
complications associated with potentials sourced by relativistic particles
\cite{Komar1978,Rohrlich1979}.) In particular, we suppose that the
particles interact through the Coulomb potential
\begin{equation}
V_{c}^{int}(\mathbf{q}_{1}(t),\mathbf{q}_{2}(t))=\sum_{i=1}^{2}e_{i}\Phi_{c}^{int}(\mathbf{q}_{1}(t),\mathbf{q}_{2}(t))=\frac{e_{1}e_{2}}{|\mathbf{q}_{1}(t)-\mathbf{q}_{2}(t)|},
\end{equation}
where we recall $\Phi_{c}^{int}(\mathbf{q}_{i}(t),\mathbf{q}_{j}(t))=\frac{1}{2}\sum_{j=1}^{2(j\neq i)}\frac{e_{j}}{|\mathbf{q}_{i}(t)-\mathbf{q}_{j}(t)|}$.
Note that we make the point-like interaction assumption $|\mathbf{q}_{1}(t)-\mathbf{q}_{2}(t)|\gg\lambda_{c}$.
So the motions of the particles are not physically independent in
the lab frame, and this implies that the \emph{zbw} oscillation of
particle 1 (particle 2) in the lab frame is physically dependent on
the position of particle 2 (particle 1), through the interaction potential
(3.78). We can represent this physical dependence of the \emph{zbw}
oscillations by a non-separable joint phase change, which involves
contributions from both particles in the form
\begin{equation}
\begin{aligned}\delta\theta_{joint}^{lab}(\mathbf{q}_{1}(t),\mathbf{q}_{2}(t),t) & =\left[\sum_{i=1}^{2}\omega_{ic}+\sum_{i=1}^{2}\omega_{ci}\frac{\mathbf{v}_{i}^{2}}{2c^{2}}+\sum_{i=1}^{2}\omega_{ci}\left(\frac{e_{i}\Phi_{c}^{int}}{m_{i}c^{2}}\right)\right]|_{\mathbf{q}_{j}=\mathbf{q}_{j}(t)}\\
 & \times\left(\delta t-\sum_{i=1}^{2}\frac{\mathbf{v}_{0i}}{c^{2}}\cdot\delta\mathbf{q}_{i}(t)\right)|_{\mathbf{q}_{j}=\mathbf{q}_{j}(t)}\\
 & =\sum_{i=1}^{2}\left[\omega_{ic}+\omega_{ci}\frac{\mathbf{v}_{i}^{2}}{2c^{2}}+\omega_{ci}\left(\frac{e_{i}\Phi_{c}^{int}}{m_{i}c^{2}}\right)\right]|_{\mathbf{q}_{j}=\mathbf{q}_{j}(t)}\delta t\\
 & -\sum_{i=1}^{2}\omega_{ci}\left(\frac{\mathbf{v}_{i}}{c^{2}}\right)|_{\mathbf{q}_{j}=\mathbf{q}_{j}(t)}\cdot\delta\mathbf{q}_{i}(t)\\
 & =\frac{1}{\hbar}\left[\left(\sum_{i=1}^{2}m_{i}c^{2}+\sum_{i=1}^{2}\frac{m_{i}\mathbf{v}_{i}^{2}}{2}+V_{c}^{int}\right)|_{\mathbf{q}_{j}=\mathbf{q}_{j}(t)}\delta t-\sum_{i=1}^{2}\mathbf{p}_{i}|_{\mathbf{q}_{j}=\mathbf{q}_{j}(t)}\cdot\delta\mathbf{q}_{i}(t)\right].
\end{aligned}
\end{equation}
Not surprisingly, when $|\mathbf{q}_{1}(t)-\mathbf{q}_{2}(t)|$ becomes
sufficiently great that $V_{c}^{int}$ is negligible, (3.79) reduces
to a sum of the physically independent phase changes associated with
particle 1 and particle 2, respectively.

Now, even though the particles don't have physically independent phases
because of $V_{c}^{int}$, it is clear that the \emph{zbw} oscillation
of particle 1 (particle 2) still has a well-defined individual phase
at all times. Moreover, we can deduce from (3.79) the individual (`conditional')
phase of a particle, given its physical interaction with the other
particle via (3.78), in much the same way that ``conditional wavefunctions''
for subsystems of particles can be deduced from the universal wavefunction
in the de Broglie-Bohm theory \cite{Duerr1992,Norsen2010}.

To motivate this, let us first ask: in the instantaneous rest frame
(IRF) of (say) particle 1, how will the phase associated with its
\emph{zbw} oscillation change in time for a co-moving observer that's
continously monitoring the oscillation? The phase change associated
with particle 1 in its IRF can be obtained from (3.79) simply by subtracting
$\omega_{c2}\delta t$ and setting $\mathbf{v}_{1}=0$, giving
\begin{equation}
\begin{aligned}\delta\theta_{1}^{rest}(\mathbf{q}_{01}(t),\mathbf{q}_{2}(t),t) & =\left[\omega_{c1}+\omega_{c2}\left(\frac{\mathbf{v}_{2}^{2}}{2c^{2}}\right)+\sum_{i=1}^{2}\omega_{ci}\left(\frac{e_{i}\Phi_{c}^{int}}{m_{i}c^{2}}\right)\right]|_{\mathbf{q}_{j}=\mathbf{q}_{j}(t)}\delta t-\omega_{c2}\left(\frac{\mathbf{v}_{2}}{c^{2}}\right)|_{\mathbf{q}_{j}=\mathbf{q}_{j}(t)}\cdot\delta\mathbf{q}_{2}(t)\\
 & =\frac{1}{\hbar}\left[\left(m_{1}c^{2}+\frac{m_{2}\mathbf{v}_{2}^{2}}{2}+V_{c}^{int}\right)|_{\mathbf{q}_{j}=\mathbf{q}_{j}(t)}\delta t-\mathbf{p}_{2}|_{\mathbf{q}_{j}=\mathbf{q}_{j}(t)}\cdot\delta\mathbf{q}_{2}(t)\right],
\end{aligned}
\end{equation}
where $\mathbf{q}_{01}(t)$ denotes the translational coordinate of
particle 1 in its IRF (which, of course, changes as a function of
time due to the Coulomb interaction). In other words, (3.80) tells
us how the Compton frequency of particle 1, $\omega_{c1}$, gets modulated
by the physical coupling of particle 1 to particle 2, in the IRF of
particle 1. Thus (3.80) represents the conditional phase change of
particle 1 in its IRF. We can also confirm that when $\Phi_{c}^{int}\approx0$
the velocity of particle 2 no longer depends on the position of particle
1 at time $t$, leaving $\delta\theta_{1}^{rest}=\omega_{c1}\delta t_{0}$.
Likewise we can obtain the conditional \emph{zbw} phase of particle
2 in its IRF.

The conditional \emph{zbw} phase of particle 1 in the lab frame where
$\mathbf{v}_{1}\neq0$ is just
\begin{equation}
\begin{aligned}\delta\theta_{1}^{lab}(\mathbf{q}_{1}(t),\mathbf{q}_{2}(t),t) & =\left[\omega_{c1}+\sum_{i=1}^{2}\omega_{ci}\left(\frac{\mathbf{v}_{i}^{2}}{2c^{2}}\right)+\sum_{i=1}^{2}\omega_{ci}\left(\frac{e_{i}\Phi_{c}^{int}}{m_{i}c^{2}}\right)\right]|_{\mathbf{q}_{j}=\mathbf{q}_{j}(t)}\delta t\\
 & -\sum_{i=1}^{2}\omega_{ci}\left(\frac{\mathbf{v}_{i}}{c^{2}}\right)|_{\mathbf{q}_{j}=\mathbf{q}_{j}(t)}\cdot\delta\mathbf{q}_{i}(t)\\
 & =\frac{1}{\hbar}\left[\left(m_{1}c^{2}+\sum_{i=1}^{2}\frac{m_{i}\mathbf{v}_{i}^{2}}{2}+V_{c}^{int}\right)|_{\mathbf{q}_{j}=\mathbf{q}_{j}(t)}\delta t-\sum_{i=1}^{2}\mathbf{p}_{i}|_{\mathbf{q}_{j}=\mathbf{q}_{j}(t)}\cdot\delta\mathbf{q}_{i}(t)\right].
\end{aligned}
\end{equation}
Equivalently, we can obtain (3.81) by just subtracting $\omega_{c2}\delta t$
from (3.79). And likewise for the conditional \emph{zbw} phase of
particle 2 in the lab frame.

Recall that, by hypothesis, each \emph{zbw} particle is essentially
a harmonic oscillator. This means that when $V_{c}^{int}\approx0$
each particle has its own well-defined phase at each point along its
space-time trajectory. Consistency with this hypothesis also means
that when $V_{c}^{int}>0$ the joint phase must be a well-defined
function of the space-time trajectories of \emph{both} particles (since
we posit that both particles remain harmonic oscillators despite having
their oscillations physically coupled by $V_{c}^{int}$). Then for
a closed loop \emph{L,} along which each particle can be physically
or virtually displaced, the joint phase in the lab frame will satisfy
\begin{equation}
\sum_{i=1}^{2}\oint_{L}\delta_{i}\theta_{joint}^{lab}=2\pi n,
\end{equation}
and for a loop in which time is held fixed,
\begin{equation}
\sum_{i=1}^{2}\oint_{L}\mathbf{p}_{i}\cdot\delta\mathbf{q}_{i}(t)=nh.
\end{equation}
It also follows from (3.82) and (3.83) that
\begin{equation}
\oint_{L}\delta_{1}\theta_{joint}^{lab}=2\pi n,
\end{equation}
and
\begin{equation}
\oint_{L}\mathbf{p}_{1}\cdot\delta\mathbf{q}_{1}(t)=nh,
\end{equation}
where this time the closed-loop integration involves keeping the coordinate
of particle 2 fixed while particle 1 is displaced along \emph{L}.
From (3.82-85), it will also be the case that
\begin{equation}
\sum_{i=1}^{2}\oint_{L}\delta_{i}\theta_{1}^{lab}=2\pi n,
\end{equation}
and
\begin{equation}
\oint_{L}\delta_{1}\theta_{1}^{lab}=2\pi n.
\end{equation}

Integrating (3.79) and multiplying through by $\hbar$ yields (using
$S_{joint}^{lab}\eqqcolon S)$ 
\begin{equation}
S=\sum_{i=1}^{2}\int_{\mathbf{q}_{i}(t_{i})}^{\mathbf{q}_{i}(t)}\mathbf{p}_{i}\cdot d\mathbf{q}_{i}(s)-\sum_{i=1}^{2}\int_{t_{i}}^{t}\left(m_{i}c^{2}+\frac{m_{i}\mathbf{v}_{i}^{2}}{2}+e_{i}\Phi_{c}^{int}\right)ds-\sum_{i=1}^{2}\hbar\phi_{i},
\end{equation}
and evolves by 
\begin{equation}
-\partial_{t}S|_{\mathbf{q}_{j}=\mathbf{q}_{j}(t)}=\sum_{i=1}^{2}m_{i}c^{2}+\sum_{i=1}^{2}\frac{\left(\nabla_{i}S\right)^{2}}{2m_{i}}|_{\mathbf{q}_{j}=\mathbf{q}_{j}(t)}+V_{c}^{int}.
\end{equation}
The conditional phase $S_{1}^{lab}=S_{1}$ and its equation of motion
only differ from (3.88-89) by subtracting $m_{2}c^{2}t-\hbar\phi_{2}$.
Analogous considerations apply to particle 2. Finally, the acceleration
of the \emph{i}-th particle is obtained from the equation of motion
\begin{equation}
m_{i}\ddot{\mathbf{q}}_{i}(t)=\left[\partial_{t}\mathbf{p}_{i}+\mathbf{v}_{i}\cdot\nabla_{i}\mathbf{p}_{i}\right]|_{\mathbf{q}_{j}=\mathbf{q}_{j}(t)}=-\nabla_{i}V_{c}^{int}|_{\mathbf{q}_{j}=\mathbf{q}_{j}(t)}.
\end{equation}

Another, more convenient way of modeling the case of two classically
interacting \emph{zbw} particles is by exploiting the well-known fact
that a two-particle system with an interaction potential of the form
(3.78) has an equivalent Hamiltonian of the form (ignoring the trivial
CM motion)
\begin{equation}
E_{rel}=\frac{p_{rel}^{2}}{2\mu}+V_{rel}(|\mathbf{q}_{rel}(t)|)+\mu c^{2},
\end{equation}
where the reduced mass $\mu=m_{1}m_{2}/(m_{1}+m_{2})$ and $V_{rel}(|\mathbf{q}_{rel}(t)|)=V_{c}^{int}(|\mathbf{q}_{1}(t)-\mathbf{q}_{2}(t)|)$.
In other words, (3.91) describes a fictitious \emph{zbw} particle
of mass $\mu$ and relative coordinate $\mathbf{q}_{rel}(t)$, moving
in an ``external'' potential $V_{rel}(|\mathbf{q}_{rel}(t)|)$.
This fictitious particle then has a Compton frequency, $\omega_{c}^{red}=\mu c^{2}/\hbar$,
and an associated phase change in the lab frame of the form
\begin{equation}
\begin{aligned}\delta\theta_{rel}(\mathbf{q}_{rel}(t)) & =\left(\omega{}_{c}^{red}+\omega{}_{c}^{red}\frac{\mathbf{v}_{rel}^{2}(\mathbf{q}_{rel}(t))}{2c^{2}}+\omega_{c}^{red}\frac{V_{rel}(|\mathbf{q}_{rel}(t)|)}{\mu c^{2}}\right)\left(\delta t-\frac{\mathbf{v}_{0rel}(\mathbf{q}_{rel}(t))\cdot\delta\mathbf{q}_{rel}(t)}{c^{2}}\right)\\
 & =\frac{1}{\hbar}\left[\left(\mu c^{2}+\frac{\mu\mathbf{v}_{rel}^{2}}{2}+V_{rel}\right)\delta t-\mathbf{p}_{rel}\cdot\delta\mathbf{q}_{rel}(t)\right].
\end{aligned}
\end{equation}
Upon integration, this of course gives
\begin{equation}
S_{rel}\coloneqq-\hbar\theta_{rel}=\int\left[\mathbf{p}_{rel}\cdot d\mathbf{q}_{rel}(t)-E_{rel}dt\right]-\hbar\phi_{rel},
\end{equation}
which evolves in time by the HJ equation 
\begin{equation}
-\partial_{t}S_{rel}|_{\mathbf{q}_{rel}=\mathbf{q}_{rel}(t)}=\mu c^{2}+\frac{\left(\nabla_{rel}S_{rel}\right)^{2}}{2\mu}|_{\mathbf{q}_{rel}=\mathbf{q}_{rel}(t)}+V_{rel},
\end{equation}
and gives the equation of motion
\begin{equation}
\mu\ddot{\mathbf{q}}_{rel}(t)=\left[\partial_{t}\mathbf{p}_{rel}+\mathbf{v}_{rel}\cdot\nabla_{rel}\mathbf{p}_{rel}\right]|_{\mathbf{q}_{rel}=\mathbf{q}_{rel}(t)}=-\nabla_{rel}V_{rel}|_{\mathbf{q}_{rel}=\mathbf{q}_{rel}(t)}.
\end{equation}

Since this situation is formally equivalent to the case of a single
\emph{zbw} particle moving in an external field, we can immediately
see that it follows
\begin{equation}
\oint_{L}\delta S_{rel}=nh,
\end{equation}
and
\begin{equation}
\oint_{L}\mathbf{p}_{rel}\cdot\delta\mathbf{q}_{rel}(t)=nh.
\end{equation}
Furthermore, the physical equivalence between this coordinatization
and the original two-particle coordinatization establishes that if
phase quantization holds in one coordinatization it must hold in the
other.

While we considered here only two \emph{zbw} particles classically
interacting through an electric scalar potential, all our considerations
straightforwardly generalize to the case of many \emph{zbw} particles
classically interacting through electric scalar potentials as well
as magnetic vector potentials (and likewise for the gravitational
analogues).

\subsection{Classical \textcolor{black}{Hamilton-Jacobi statistical mechanics
for two interacting }\textcolor{black}{\emph{zbw}}\textcolor{black}{{}
particles }}

For a statistical mechanical description of two classically interacting
\emph{zbw} particles, the trajectories $\{\mathbf{q}_{1}(t),\mathbf{q}_{2}(t)\}$
get replaced with the coordinates $\{\mathbf{q}_{1},\mathbf{q}_{2}\}$,
and the non-relativistic joint phase field in the lab frame is obtained
from (3.88) as
\begin{equation}
\begin{aligned}S(\mathbf{q}_{1},\mathbf{q}_{2},t) & =\sum_{i=1}^{2}\int_{\mathbf{q}_{i}(t_{i})}^{\mathbf{q}_{i}(t)}\mathbf{p}_{i}\cdot d\mathbf{q}_{i}(s)|_{\mathbf{q}_{j}(t)=\mathbf{q}_{j}}\\
 & -\sum_{i=1}^{2}\int_{t_{i}}^{t}\left[m_{i}c^{2}+\frac{m_{i}\mathbf{v}_{i}^{2}(\mathbf{q}_{1}(s),\mathbf{q}_{2}(s),s)}{2}+e_{i}\Phi_{c}^{int}(\mathbf{q}_{1}(s),\mathbf{q}_{2}(s))\right]ds|_{\mathbf{q}_{j}(t)=\mathbf{q}_{j}}-\sum_{i=1}^{2}\hbar\phi_{i},
\end{aligned}
\end{equation}
and evolves by 
\begin{equation}
-\partial_{t}S=\sum_{i=1}^{2}m_{i}c^{2}+\sum_{i=1}^{2}\frac{\left(\nabla_{i}S\right)^{2}}{2m_{i}}+V_{c}^{int},
\end{equation}
where $\mathbf{v}_{i}(\mathbf{q}_{1},\mathbf{q}_{2})=\nabla_{i}S(\mathbf{q}_{1},\mathbf{q}_{2},t)/m_{i}$
. Since (3.98) is a field over the possible positions of the actual
\emph{zbw} particles, and since for each possible initial position
the phase of each \emph{zbw} particle will satisfy relation (3.83),
it follows that
\begin{equation}
\sum_{i=1}^{2}\oint_{L}\nabla_{i}S\cdot d\mathbf{q}_{i}=nh,
\end{equation}
where \emph{L} is now a mathematical loop in the 2-particle configuration
space.

The two-particle probability density $\rho(\mathbf{q}_{1},\mathbf{q}_{2},t)\geq0$
evolves by the two-particle continuity equation
\begin{equation}
\frac{\partial\rho}{\partial t}=-\sum_{i=1}^{2}\nabla_{i}\cdot\left[\left(\mathbf{\frac{\nabla_{\mathit{i}}\mathrm{\mathit{S}}}{\mathit{m_{i}}}}\right)\rho\right],
\end{equation}
and the ensemble-averaged two-particle action is defined by
\begin{equation}
\begin{aligned}J & =\mathrm{E}\left[\sum_{i=1}^{2}\left[\int_{\mathbf{q}_{iI}}^{\mathbf{q}_{iF}}m_{i}\mathbf{v}_{i}\cdot\mathbf{\mathit{d}q}_{i}(t)-\int_{t_{I}}^{t_{F}}\left(m_{i}c^{2}+\frac{\mathbf{p}_{i}^{2}}{2m_{i}}+e_{i}\Phi_{ci}^{int}\right)dt-\hbar\phi_{i}\right]\right]\\
 & =\mathrm{E}\left[\int_{t_{I}}^{t_{F}}\sum_{i=1}^{N}\left(\frac{1}{2}m\mathbf{v}_{i}^{2}-m_{i}c^{2}-e_{i}\Phi_{ci}^{int}\right)dt-\sum_{i=1}^{N}\hbar\phi_{i}\right],
\end{aligned}
\end{equation}
where the equated expressions are related by the usual Legendre transformation.
Imposing
\begin{equation}
J=extremal,
\end{equation}
straightforward manipulations along the lines of those in the Appendix
yield (3.90). And, upon replacing $\mathbf{q}_{i}(t)$ with $\mathbf{q}_{i}$,
we obtain the classical Newtonian equation for the acceleration field
$\mathbf{a}_{i}(\mathbf{q}_{1},\mathbf{q}_{2},t)$: 
\begin{equation}
m_{i}\mathbf{a}_{i}=\partial_{t}\mathbf{p}_{i}+\mathbf{v}_{i}\cdot\nabla_{i}\mathbf{p}_{i}=-\nabla_{i}V_{c}^{int}.
\end{equation}

Now, we can obtain the conditional \emph{zbw} phase field for particle
1 by evaluating the joint phase field at the actual position of particle
2 at time $t$, i.e., $S(\mathbf{q}_{1},\mathbf{q}_{2}(t),t)\eqqcolon S_{1}(\mathbf{q}_{1},t)$.
Taking the total time derivative we have 
\begin{equation}
\partial_{t}S_{1}(\mathbf{q}_{1},t)=\partial_{t}S(\mathbf{q}_{1},\mathbf{q}_{2},t)|_{\mathbf{q}_{2}=\mathbf{q}_{2}(t)}+\frac{d\mathbf{q}_{2}(t)}{dt}\cdot\nabla_{2}S(\mathbf{q}_{1},\mathbf{q}_{2},t)|_{\mathbf{q}_{2}=\mathbf{q}_{2}(t)},
\end{equation}
where the conditional velocities 
\begin{equation}
\frac{d\mathbf{q}_{1}(t)}{dt}=\mathbf{v}_{1}(\mathbf{q}_{1},t)|_{\mathbf{q}_{1}=\mathbf{q}_{1}(t)}=\frac{\nabla_{1}S_{1}(\mathbf{q}_{1},t)}{m_{1}}|_{\mathbf{q}_{1}=\mathbf{q}_{1}(t)},
\end{equation}
and 
\begin{equation}
\frac{d\mathbf{q}_{2}(t)}{dt}=\mathbf{v}_{2}(\mathbf{q}_{2},t)|_{\mathbf{q}_{2}=\mathbf{q}_{2}(t)}=\frac{\nabla_{2}S_{2}(\mathbf{q}_{2},t)}{m_{2}}|_{\mathbf{q}_{2}=\mathbf{q}_{2}(t)},
\end{equation}
the latter defined from the conditional phase field $S_{2}(\mathbf{q}_{2},t)$
for particle 2. Inserting (3.105) into the left hand side of (3.99)
and adding the corresponding term on the right hand side, we then
find that the conditional phase field for particle 1 evolves by a
`conditional HJ equation', namely
\begin{equation}
-\partial_{t}S_{1}=m_{1}c^{2}+\frac{\left(\nabla_{1}S_{1}\right)^{2}}{2m_{1}}+\frac{\left(\nabla_{2}S\right)^{2}}{2m_{2}}|_{\mathbf{q}_{2}=\mathbf{q}_{2}(t)}-\frac{d\mathbf{q}_{2}(t)}{dt}\cdot\nabla_{2}S|_{\mathbf{q}_{2}=\mathbf{q}_{2}(t)}+V_{c}^{int}(\mathbf{q}_{1},\mathbf{q}_{2}(t)),
\end{equation}
where $V_{c}^{int}(\mathbf{q}_{1},\mathbf{q}_{2}(t))$ is the `conditional
potential' for particle 1; that is, the potential field that particle
1, at location $\mathbf{q}_{1}$, would `feel' given the actual location
of particle 2. The solution of (3.108) can be verified as 
\begin{equation}
S_{1}=\int\mathbf{p}_{1}\cdot d\mathbf{q}_{1}-\int\left[m_{1}c^{2}+\frac{m_{1}\mathbf{v}_{1}^{2}}{2}+\frac{m_{1}\mathbf{v}_{2}^{2}}{2}-\mathbf{p}_{2}\cdot\frac{d\mathbf{q}_{2}(t)}{dt}+V_{c}^{int}\right]dt-\hbar\phi_{1}.
\end{equation}
Notice here that the conditional phase field is a field on 3-D space.
This makes perfect sense since, after all, the conditional phase refers
to the phase associated to the \emph{zbw} oscillation of particle
1, a real physical oscillation in 3-D space. It can also be verified
that when (3.109) is evaluated at $\mathbf{q}_{1}=\mathbf{q}_{1}(t)$,
it is equivalent to $S_{joint}^{lab}(\mathbf{q}_{1}(t),\mathbf{q}_{2}(t),t)-m_{2}c^{2}t+\hbar\phi_{2}$. 

Once again, since the conditional \emph{zbw} phase field for particle
1 is a field over the possible positions that \emph{zbw} particle
1 could actually occupy at time $t$, it will be the case that
\begin{equation}
\oint_{L}\nabla_{1}S_{1}\cdot d\mathbf{q}_{1}=nh,
\end{equation}
where \emph{L} is a mathematical loop in 3-D space.

Likewise, we can obtain the conditional probability density for particle
1 by writing $\rho(\mathbf{q}_{1},\mathbf{q}_{2}(t),t)\eqqcolon\rho_{1}(\mathbf{q}_{1},t)$.
Taking the total time derivative gives 
\begin{equation}
\partial_{t}\rho_{1}(\mathbf{q}_{1},t)=\partial_{t}\rho(\mathbf{q}_{1},\mathbf{q}_{2},t)|_{\mathbf{q}_{2}=\mathbf{q}_{2}(t)}+\frac{d\mathbf{q}_{2}(t)}{dt}\cdot\nabla_{2}\rho(\mathbf{q}_{1},\mathbf{q}_{2},t)|_{\mathbf{q}_{2}=\mathbf{q}_{2}(t)}.
\end{equation}
Inserting this on the left hand side of (3.101) and adding the corresponding
term on the right hand side, we obtain the conditional continuity
equation for particle 1: 
\begin{equation}
\partial_{t}\rho_{1}=-\nabla_{1}\cdot\left[\left(\frac{\nabla_{1}S_{1}}{m_{1}}\right)\rho_{1}\right]-\nabla_{2}\cdot\left[\left(\frac{\nabla_{2}S}{m_{2}}\right)\rho\right]|_{\mathbf{q}_{2}=\mathbf{q}_{2}(t)}+\frac{d\mathbf{q}_{2}(t)}{dt}\cdot\nabla_{2}\rho|_{\mathbf{q}_{2}=\mathbf{q}_{2}(t)},
\end{equation}
which implies $\rho_{1}(\mathbf{q}_{1},t)\geq0$ and (upon suitable
redefinition of $\rho_{1}(\mathbf{q}_{1},t)$) preservation of the
normalization $\int_{\mathbb{R}^{3}}\rho_{1}(\mathbf{q}_{1},0)=1$. 

The ensemble-averaged conditional action for particle 1 is defined
as
\begin{equation}
\begin{aligned}J_{1} & =\mathrm{E}\left[\int_{\mathbf{q}_{1I}}^{\mathbf{q}_{1F}}m_{1}\mathbf{v}_{1}\cdot\mathbf{\mathit{d}q}_{1}(t)-\int_{t_{I}}^{t_{F}}\left(m_{1}c^{2}+\frac{m_{1}\mathbf{v}_{1}^{2}}{2}+\frac{m_{2}\mathbf{v}_{2}^{2}}{2}-\mathbf{p}_{2}\cdot\frac{d\mathbf{q}_{2}(t)}{dt}+V_{c}^{int}\right)dt-\hbar\phi_{1}\right]\\
 & =\mathrm{E}\left[\int_{t_{I}}^{t_{F}}\left[\frac{1}{2}m_{1}\mathbf{v}_{1}^{2}+\frac{1}{2}m_{2}\mathbf{v}_{2}^{2}-m_{1}c^{2}-V_{c}^{int}\right]dt-\hbar\phi_{1}\right],
\end{aligned}
\end{equation}
where it can be readily confirmed that the equated lines are related
by the Legendre transformation. Imposing
\begin{equation}
J_{1}=extremal,
\end{equation}
where the subscript 1 denotes that the variation is only with respect
to $\mathbf{q}_{1}(t)$, straightforward manipulations analogous to
those in the Appendix yield, upon replacing $\mathbf{q}_{1}(t)$ with
$\mathbf{q}_{1}$, the classical equation of motion for the conditional
acceleration field of particle 1: 
\begin{equation}
m_{1}\mathbf{a}_{1}(\mathbf{q}_{1},t)=\left[\partial_{t}\mathbf{p}_{1}+\mathbf{v}_{1}\cdot\nabla_{i}\mathbf{p}_{1}\right](\mathbf{q}_{1},t)=-\nabla_{1}V_{c}^{int}(\mathbf{q}_{1},\mathbf{q}_{2}(t)).
\end{equation}
The conditional phase field, probability density, etc., for particle
2, are developed analogously.

We now turn to the formulation of our classical statistical mechanics
in terms of the reduced mass \emph{zbw} particle. Replacing $\mathbf{q}_{rel}(t)$
with $\mathbf{q}_{rel}$, the reduced mass \emph{zbw} phase field
\begin{equation}
\begin{aligned}S_{rel}(\mathbf{q}_{rel},t) & =\int_{\mathbf{q}_{rel}(t_{i})}^{\mathbf{q}_{rel}(t)}\mathbf{p}_{rel}\cdot d\mathbf{q}{}_{rel}(s)|_{\mathbf{q}{}_{rel}(t)=\mathbf{q}_{rel}}\\
 & -\int_{t_{i}}^{t}\left(\mu c^{2}+\frac{\mathbf{p}_{rel}^{2}}{2\mu}+V_{rel}\right)ds|_{\mathbf{q}{}_{rel}(t)=\mathbf{q}_{rel}}-\hbar\phi_{rel},
\end{aligned}
\end{equation}
evolves by the reduced mass HJ equation 
\begin{equation}
-\partial_{t}S_{rel}=\mu c^{2}+\frac{\left(\nabla_{rel}S_{rel}\right)^{2}}{2\mu}+V_{rel},
\end{equation}
and satisfies
\begin{equation}
\oint_{L}\nabla_{rel}S_{rel}\cdot d\mathbf{q}_{rel}=nh,
\end{equation}
where \emph{L} is a mathematical loop in 3-D space. Introducing the
probability density for the reduced mass \emph{zbw} particle, $\rho_{rel}(\mathbf{q}_{rel},t)\geq0$,
it is straightforward to show it evolves by the continuity equation
\begin{equation}
\frac{\partial\rho_{rel}}{\partial t}=-\nabla_{rel}\cdot\left[\left(\mathbf{\frac{\nabla_{\mathit{rel}}\mathrm{\mathit{S_{rel}}}}{\mathit{m_{rel}}}}\right)\rho_{rel}\right],
\end{equation}
which preserves the normalization $\intop_{\mathbb{R}^{3}}d^{3}\mathbf{q}_{rel}\rho_{rel}(\mathbf{q}_{rel},0)=1$.
The corresponding ensemble-averaged action for the reduced mass particle
is defined by 
\begin{equation}
\begin{aligned}J_{rel} & =\mathrm{E}\left[\int_{\mathbf{q}_{relI}}^{\mathbf{q}_{relF}}\mu\mathbf{v}_{rel}\cdot\mathbf{\mathit{d}q}_{rel}(t)-\int_{t_{I}}^{t_{F}}\left(\mu c^{2}+\frac{\mathbf{p}_{rel}^{2}}{2\mu}+V_{rel}\right)dt-\hbar\phi_{rel}\right]\\
 & =\mathrm{E}\left[\int_{t_{I}}^{t_{F}}\left(\frac{1}{2}\mu\mathbf{v}_{rel}^{2}-\mu c^{2}-V_{rel}\right)dt-\hbar\phi_{rel}\right].
\end{aligned}
\end{equation}
Imposing the constraint 
\begin{equation}
J_{rel}=extremal,
\end{equation}
we obtain after manipulations (and replacing $\mathbf{q}_{rel}(t)$
by $\mathbf{q}_{rel}$) the equation of motion 
\begin{equation}
\mu\mathbf{a}_{rel}(\mathbf{q}_{rel},t)=\partial_{t}\mathbf{p}_{rel}+\mathbf{v}_{rel}\cdot\nabla_{rel}\mathbf{p}_{rel}=-\nabla_{rel}V_{rel}(|\mathbf{q}_{rel}|).
\end{equation}

Let us now recover the nonlinear Schr\"{o}dinger
equations for each of the three cases we've considered.

The combination of (3.99)-(3.101) gives
\begin{equation}
i\hbar\frac{\partial\psi}{\partial t}=\sum_{i=1}^{2}\left[-\frac{\hbar^{2}}{2m_{i}}\nabla_{i}^{2}+\frac{\hbar^{2}}{2m_{i}}\frac{\nabla_{i}^{2}|\psi|}{|\psi|}+m_{i}c^{2}\right]\psi+V_{c}^{int}\psi,
\end{equation}
where $\psi(\mathbf{q}_{1},\mathbf{q}_{2},t)=\sqrt{\rho(\mathbf{q}_{1},\mathbf{q}_{2},t)}e^{iS(\mathbf{q}_{1},\mathbf{q}_{2},t)/\hbar}$
is single-valued by (3.100).

Combining (3.108) and (3.112) gives the conditional nonlinear Schr\"{o}dinger
equation for particle 1:
\begin{equation}
\begin{aligned}i\hbar\frac{\partial\psi_{1}}{\partial t} & =-\frac{\hbar^{2}}{2m_{1}}\nabla_{1}^{2}\psi_{1}-\frac{\hbar^{2}}{2m_{1}}\nabla_{2}^{2}\psi|_{\mathbf{q}_{2}=\mathbf{q}_{2}(t)}+V_{c}^{int}(\mathbf{q}_{1},\mathbf{q}_{2}(t))\psi_{1}+m_{1}c^{2}\psi_{1}\\
 & +i\hbar\frac{d\mathbf{q}_{2}(t)}{dt}\cdot\nabla_{2}\psi|_{\mathbf{q}_{2}=\mathbf{q}_{2}(t)}+\left(\frac{\hbar^{2}}{2m_{1}}\frac{\nabla_{1}^{2}|\psi_{1}|}{|\psi_{1}|}\right)\psi_{1}+\left(\frac{\hbar^{2}}{2m_{2}}\frac{\nabla_{2}^{2}|\psi|}{|\psi|}\right)|_{\mathbf{q}_{2}=\mathbf{q}_{2}(t)}\psi_{1},
\end{aligned}
\end{equation}
where $\psi(\mathbf{q}_{1},\mathbf{q}_{2}(t),t)\eqqcolon\psi_{1}(\mathbf{q}_{1},t)=\sqrt{\rho_{1}(\mathbf{q}_{1},t)}e^{iS_{1}(\mathbf{q}_{1},t)/\hbar}$
is the conditional classical wavefunction for particle 1, and satisfies
single-valuedness as a consequence of (3.110). Here $d\mathbf{q}_{2}(t)/dt=(\hbar/m_{2})\mathrm{Im}\{\nabla_{2}ln(\psi_{2})\}|_{\mathbf{q}_{2}=\mathbf{q}_{2}(t)}$,
where $\psi_{2}=\psi_{2}(\mathbf{q}_{2},t)$ is the conditional wavefunction
for particle 2 and satisfies a conditional nonlinear Schr\"{o}dinger
equation analogous to (3.124). Note also that (3.124) can be obtained
by taking the total time derivative of the conditional wavefunction
for particle 1 
\begin{equation}
\partial_{t}\psi_{1}(\mathbf{q}_{1},t)=\partial_{t}\psi(\mathbf{q}_{1},\mathbf{q}_{2},t)|_{\mathbf{q}_{2}=\mathbf{q}_{2}(t)}+\frac{d\mathbf{q}_{2}(t)}{dt}\cdot\nabla_{2}\psi(\mathbf{q}_{1},\mathbf{q}_{2},t)|_{\mathbf{q}_{2}=\mathbf{q}_{2}(t)},
\end{equation}
inserting this on the left hand side of (3.123), adding the corresponding
term on the right hand side, and subtracting $m_{2}c^{2}\psi_{1}$.

Finally, combining (3.117-119) gives the nonlinear Schr\"{o}dinger
equation for the fictitious reduced mass particle:
\begin{equation}
i\hbar\frac{\partial\psi_{rel}}{\partial t}=\left[-\frac{\hbar^{2}}{2\mu}\nabla_{rel}^{2}+\frac{\hbar^{2}}{2\mu}\frac{\nabla_{rel}^{2}|\psi_{rel}|}{|\psi_{rel}|}+\mu c^{2}\right]\psi_{rel}+V(|\mathbf{q}_{rel}|)\psi_{rel},
\end{equation}
where $\psi_{rel}(\mathbf{q}_{rel},t)=\sqrt{\rho_{rel}(\mathbf{q}_{rel},t)}e^{iS_{rel}(\mathbf{q}_{rel},t)/\hbar}$
is a single-valued classical wavefunction. As with the linear Schr\"{o}dinger
equation of quantum mechanics, it is easily verified that (3.126)
can be obtained from (3.123) by transforming the two-particle Hamiltonian
operator to the center of mass and relative coordinates.

This completes the development of the classical HJ statistical mechanics
for two classically interacting \emph{zbw} particles. The generalization
to \emph{N} \emph{zbw} particles interacting through their electric
scalar and magnetic vector potentials (and the gravitational analogues
thereof) is straightforward, but will not be given here due to unnecessary
mathematical complexity.

\subsection{Remarks on close-range interactions}

Throughout we have assumed the point-like interaction case, $q_{rel}(t)=|\mathbf{q}_{1}(t)-\mathbf{q}_{2}(t)|\gg\lambda_{c}$.
But what changes when $q_{rel}(t)=|\mathbf{q}_{1}(t)-\mathbf{q}_{2}(t)|\sim\lambda_{c}$?
Not much. To show this, we adopt the approach of Zelevinsky \cite{Zelevinsky2011}
in modeling the deviation from point-like interactions with a Darwin
interaction term as follows. Consider the (hypothesized) 3-D \emph{zbw}
oscillation/fluctuation around the relative coordinate, $\mathbf{q}_{rel}(t)+\delta\mathbf{q}(t)$,
where $\delta q_{max}=|\mathbf{\delta q}_{max}(t)|=\lambda_{c}$.
Taylor expand the (Coulomb or Newtonian) interaction potential into
$V_{int}(|\mathbf{q}_{rel}(t)+\delta\mathbf{q}(t)|)\thickapprox V_{int}(|\mathbf{q}_{rel}(t)|)+\delta\mathbf{q}(t)\cdot\nabla V_{int}(|\mathbf{q}_{rel}(t)|)+\frac{1}{2}\sum_{i,j}\delta q^{i}(t)\delta q^{j}(t)\partial^{i}\partial^{j}V_{int}(|\mathbf{q}_{rel}(t)|)$.
Then, under the reasonable assumptions that the mean and variance
of the fluctuations are given by $<\delta\mathbf{q}(t)>=0$ and $<\delta q(t)^{i}\delta q(t)^{j}>=\frac{1}{3}<\delta q(t)^{2}>\delta_{ij}$,
respectively, the fluctuation-averaged potential $<V_{int}(|\mathbf{q}_{rel}(t)+\delta\mathbf{q}(t)|)>=V_{int}(|\mathbf{q}_{rel}(t)|)+\frac{1}{6}<\delta q(t)^{2}>\nabla^{2}V_{int}(|\mathbf{q}_{rel}(t)|).$
Finally, approximating $<\delta q(t)^{2}>=\frac{1}{2}\lambda_{c}^{2}$,
we find that the perturbation of the potential due to the fluctuations
is $\delta V\thickapprox\frac{1}{12}\lambda_{c}^{2}\nabla^{2}V_{int}=\frac{1}{12}\lambda_{c}^{2}4\pi K\delta(\mathbf{q})$,
if the interaction potential is of the general form, $V_{int}(q)=K\mathbf{\hat{q}}/q$,
where \emph{K} is a constant.

Note that because the \emph{zbw} oscillation is a (rheonomic) constraint
on each particle, the Coulomb interaction between them never causes
their oscillations to deviate from simple harmonic motion (even though
their oscillation frequencies can slightly shift by an amount of the
order $(\omega_{c}V_{int})/\hbar$); so phase/momentum quantization
for each particle is not altered, even when $q_{rel}(t)\sim\lambda_{c}$.
Alternatively, we could relax the \emph{zbw} constraint by assuming
that when $q_{rel}(t)\sim\lambda_{c}$, a slight deviation from simple
harmonic motion occurs because the Coulomb repulsion is sufficiently
strong to impart a nonlinear perturbation to the internal harmonic
potential of each \emph{zbw} particle; but this perturbation should
drop off rapidly as the particles move away from each other so that
simple harmonic motion is quickly restored and the momentum quantization
is stable again. Ideally, a physical model of the \emph{zbw} particle
would implement this latter possibility, but for the purposes of this
paper, it will simply be assumed throughout that the Coulomb interaction
does not alter the simple harmonic nature of the \emph{zbw} oscillations.

\section{Zitterbewegung Stochastic Mechanics}

\subsection{Free \emph{zbw} particles}

We take as our starting point the hypothesis that \emph{N} particles
of rest masses, $m_{i}$, and 3-D space positions, $\mathbf{q}_{i}(t)$,
are immersed in Nelson's hypothesized ether and undergo conservative
diffusion processes according to the stochastic differential equations
\begin{equation}
d\mathbf{q}_{i}(t)=\mathbf{b}_{i}(q(t),t)dt+d\mathbf{W}_{i}(t),
\end{equation}
and
\begin{equation}
d\mathbf{q}_{i}(t)=\mathbf{b}_{i*}(q(t),t)dt+d\mathbf{W}_{i*}(t),
\end{equation}
where the forward Wiener processes $d\mathbf{W}_{i}(t)$ satisfy $\mathrm{E}_{t}\left[d\mathbf{W}_{i}\right]=0$
and $\mathrm{E}_{t}\left[d\mathbf{W}_{i}^{2}\right]=\left(\hbar/m_{i}\right)dt$,
and analogously for the backward Wiener processes. Note that we take
the $\mathbf{b}_{i}$ $(\mathbf{b}_{i*})$ to be functions of all
the particle positions, $q(t)=\{\mathbf{q}_{1}(t),\mathbf{q}_{2}(t),...,\mathbf{q}_{N}(t)\}$
$\in$ $\mathbb{R}^{3N}$. The reasons for this are: (i) all the particles
are continuously exchanging energy-momentum with a common background
medium (Nelson's ether) and thus are in general physically connected
in their motions through the ether via $\mathbf{b}_{i}$ $(\mathbf{b}_{i*})$,
insofar as the latter are constrained by the physical properties of
the ether; and (ii) the dynamical equations and initial conditions
for the $\mathbf{b}_{i}$ $(\mathbf{b}_{i*})$ are what will determine
the specific situations under which the latter will be effectively
separable functions of the particle positions and when they cannot
be effectively separated. Hence, at this level, it is only sensible
to write $\mathbf{b}_{i}$ $(\mathbf{b}_{i*})$ as functions of all
the particle positions at a single time.

As in the single particle case, in order to incorporate the \emph{zbw}
oscillation as a property of each particle, we must amend Nelson's
original phenomenological hypotheses about his ether and particles
with the \emph{N}-particle generalizations of the new phenomenological
hypotheses we introduced in Part I: 
\begin{enumerate}
\item Nelson's ether is not only a stochastically fluctuating medium in
space-time, but an oscillating medium with a spectrum of angular frequencies
superposed at each point in 3-space. More precisely, we imagine the
ether as a continuous (or effectively continuous) medium composed
of a countably infinite number of fluctuating, stationary, spherical
waves superposed at each point in space, with each wave having a different
fixed angular frequency, $\omega_{0}^{k}$, where $k$ denotes the
\emph{k}-th ether mode. The relative phases between the modes are
taken to be random so that each mode is effectively uncorrelated with
every other mode. 
\item The particles of rest masses $m_{i}$, located at positions $\mathbf{q}_{0i}$
in their respective instantaneous mean forward translational rest
frames (IMFTRFs), i.e., the frames in which $D\mathbf{q}_{i}(t)=\mathbf{b}_{i}(\mathbf{q}(t),t)=0$,
are bounded to harmonic oscillator potentials with fixed natural frequencies
$\omega_{0i}=\omega_{ci}=\left(1/\hbar\right)m_{i}c^{2}$. In keeping
with the phenomenological approach of ZSM, and the approach taken
by de Broglie and Bohm with their \emph{zbw} models, we need not specify
the precise physical nature of these harmonic oscillator potentials;
this is task is left for a future physical model of the ZSM particle. 
\item Each particle's center of mass, as a result of being immersed in the
ether, undergoes approximately frictionless translational Brownian
motion (due to the homogeneous and isotropic ether fluctuations that
couple to the particles by possibly electromagnetic, gravitational,
or some other means), as modeled by Eqs. (3.127) and (3.128); and,
in their respective IMFTRFs, undergo driven oscillations about $\mathbf{q}_{0i}$
by coupling to a narrow band of ether modes that resonantly peak around
their natural frequencies. However, in order that the oscillation
of each particle doesn't become unbounded in kinetic energy, there
must be some mechanism by which the particles dissipate energy back
into the ether so that, on the average, a steady-state equilibrium
regime is reached for their oscillations. So we posit that on short
relaxation time-scales, $\tau$, which are identical for particles
of identical rest masses, the average energy absorbed from the driven
oscillation by the resonant ether modes equals the average energy
dissipated back to the ether by a given particle. The average, in
the present sense, would be over the random phases of the ether modes.
(Here we are taking inspiration from stochastic electrodynamics \cite{Boyer1975,Boyer1980},
where it has been shown that a classical charged harmonic oscillator
immersed in a classical electromagnetic zero-point field has a steady-state
condition where the phase-averaged power absorbed by the oscillator
balances the phase-averaged power radiated by the oscillator back
to the zero-point field; this yields a steady-state oscillation at
the natural frequency of the oscillator \cite{Boyer1975,Boyer1980,Puthoff1987,Puthoff2016}.
However, in keeping with our phenomenological approach, we do not
propose a specific mechanism for this energy exchange, only provisionally
assume that it occurs somehow.) Thus, in the steady-state regime,
each particle undergoes a steady-state\emph{ zbw} oscillation of angular
frequency $\omega_{ci}$ about its location $\mathbf{q}_{0i}$ in
its IMFTRF, as characterized by the `fluctuation-dissipation' relation,
$<H_{i}>_{steady-state}=\hbar\omega_{ci}=m_{i}c^{2}$, where $<H_{i}>_{steady-state}$
is the conserved random-phase-average energy due to the steady-state
oscillation of the \emph{i}-th particle. Accordingly, if, relative
to the ether, all the particles have zero mean translational motion,
then we will have $\sum_{i}^{N}<H_{i}>_{steady-state}=\sum_{i}^{N}\hbar\omega_{ci}=\sum_{i}^{N}m_{i}c^{2}=const$. 
\end{enumerate}
It follows then that, in the IMFTRF of the \emph{i}-th particle, the
mean forward steady-state \emph{zbw} phase change is given by
\begin{equation}
\delta\bar{\theta}_{i+}\coloneqq\omega_{ci}\delta t_{0}=\frac{m_{i}c^{2}}{\hbar}\delta t_{0},
\end{equation}
and the corresponding cumulative mean forward steady-state phase at
time $t_{0}$ is
\begin{equation}
\bar{\theta}_{i+}=\omega_{ci}t_{0}+\phi_{i}=\frac{m_{i}c^{2}}{\hbar}t_{0}+\phi_{i+}.
\end{equation}
Then the joint cumulative mean forward steady-state phase for all
the particles will just be
\begin{equation}
\bar{\theta}_{+}=\sum_{i=1}^{N}\bar{\theta}_{i+}=\sum_{i=1}^{N}\left(\omega_{ci}t_{0}+\phi_{i+}\right)=\sum_{i=1}^{N}\left(\frac{m_{i}c^{2}}{\hbar}t_{0}+\phi_{i+}\right).
\end{equation}

The reason for starting our analysis with the IMFTRFs goes back to
the fact that, before constraining the diffusion process to simultaneous
solutions of the forward and backward Fokker-Planck equations associated
to (3.127-128), neither the forward nor the backward stochastic differential
equations (3.127-128) have well-defined time reversals. So the forward
and backward stochastic differential equations describe independent,
time-asymmetric diffusion processes in opposite time directions, and
we must start by considering the steady-state \emph{zbw} phases in
each time direction separately. So we chose to start with the more
intuitive forward time direction. 

For the \emph{i}-th \emph{zbw} particle in its instantaneous mean
backward translational rest frame (IMBTRF), i.e., the frame defined
by $D_{*}\mathbf{q}_{i}(t)=\mathbf{b}_{i*}(q(t),t)=0$, its mean backward
steady-state \emph{zbw} phase change is given by
\begin{equation}
\delta\bar{\theta}_{i-}\coloneqq-\omega_{ci}\delta t_{0}=-\frac{m_{i}c^{2}}{\hbar}\delta t_{0},
\end{equation}
and
\begin{equation}
\bar{\theta}_{i-}=\left(-\omega_{ci}t_{0}\right)+\phi_{i-}=\left(-\frac{m_{i}c^{2}}{\hbar}t_{0}\right)+\phi_{i-}.
\end{equation}
Then the cumulative joint mean backward steady-state phase for all
the particles will just be
\begin{equation}
\bar{\theta}_{-}=\sum_{i=1}^{N}\bar{\theta}_{i-}=\sum_{i=1}^{N}\left(\omega_{ci}t_{0}+\phi_{i-}\right)=\sum_{i=1}^{N}\left(\frac{m_{i}c^{2}}{\hbar}t_{0}+\phi_{i-}\right).
\end{equation}

As in the single particle case, we note that both the diffusion coefficient
$\nu_{i}=\hbar/2m_{i}$ and the (reduced) \emph{zbw} period $T_{ci}=1/\omega_{ci}=\hbar/m_{i}c^{2}$
are scaled by $\hbar$. This is consistent with our hypothesis that
the ether is the common physical cause of both the frictionless diffusion
processes and the steady-state \emph{zbw} oscillations of the particles.
Had we not proposed Nelson's ether as the physical cause of the \emph{zbw}
oscillations as well as the frictionless diffusions, the occurrence
of $\hbar$ in both of these properties of the particles would be
inexplicable and compromising for the plausibility of our proposed
modification of NYSM.

As also in the single particle case, we cannot talk of the \emph{zbw}
phases in rest frames other than the IMFTRFs or IMBTRFs of the particles,
because we cannot transform to a frame in which $d\mathbf{q}_{i}(t)/dt=0$,
as this expression is undefined for the Wiener process.

Now suppose we Lorentz transform back to the lab frame. For the forward
time direction, this corresponds to a boost of (3.129) by $-\mathbf{b}_{i}(q(t),t)$.
Approximating the transformation for non-relativistic velocities so
that $\gamma=1/\sqrt{\left(1-\mathbf{b}_{i}^{2}/c^{2}\right)}\approx1+\mathbf{b}_{i}^{2}/2c^{2},$
the mean forward steady-state joint phase change becomes
\begin{equation}
\begin{aligned}\delta\bar{\theta}_{+}(q(t),t) & =\sum_{i=1}^{N}\frac{\omega_{ci}}{m_{i}c^{2}}\mathrm{E}_{t}\left[E_{i+}(D\mathbf{q}_{i}(t))\delta t-m_{i}D\mathbf{q}_{i}(t)\cdot\left(D\mathbf{q}_{i}(t)\right)\delta t\right]\\
 & =\frac{1}{\hbar}\mathrm{E}_{t}\left[\sum_{i=1}^{N}E_{i+}(D\mathbf{q}_{i}(t))\delta t-\sum_{i=1}^{N}m_{i}\mathbf{b}_{i}(q(t),t)\cdot\delta\mathbf{q}_{i+}(t)\right],
\end{aligned}
\end{equation}
where
\begin{equation}
E_{i+}(D\mathbf{q}_{i}(t))=m_{i}c^{2}+\frac{1}{2}m_{i}\left(D\mathbf{q}_{i}(t)\right)^{2}=m_{i}c^{2}+\frac{1}{2}m_{i}\mathbf{b}_{i}^{2},
\end{equation}
neglecting the momentum terms proportional to $\mathbf{b}_{i}^{3}/c^{2}$.
We emphasize that the $\delta\mathbf{q}_{i+}(t)$ in (3.135) corresponds
to the physical, translational, mean forward displacement of the \emph{i}-th
\emph{zbw} particle, defined by 
\begin{equation}
\delta\mathbf{q}_{i+}(t)=\left(D\mathbf{q}_{i}(t)\right)\delta t=\mathbf{b}_{i}(q(t),t)\delta t.
\end{equation}
This will be important later. 

For the backward time direction, the Lorentz transformation to the
lab frame corresponds to a boost of (3.132) by $-\mathbf{b}_{i*}(q(t),t)$.
Then the mean backward steady-state joint phase change becomes
\begin{equation}
\begin{aligned}\delta\bar{\theta}_{-}(q(t),t) & =\sum_{i=1}^{N}\frac{\omega_{ci}}{m_{i}c^{2}}\mathrm{E}_{t}\left[-E_{i-}(D_{*}\mathbf{q}_{i}(t))\delta t+m_{i}D_{*}\mathbf{q}_{i}(t)\cdot\left(D_{*}\mathbf{q}_{i}(t)\right)\delta t\right]\\
 & =\frac{1}{\hbar}\mathrm{E}_{t}\left[-\sum_{i=1}^{N}E_{i-}(D_{*}\mathbf{q}_{i}(t))\delta t+\sum_{i=1}^{N}m_{i}\mathbf{b}_{i*}(q(t),t)\cdot\delta\mathbf{q}_{i-}(t)\right],
\end{aligned}
\end{equation}
where
\begin{equation}
E_{i-}(D_{*}\mathbf{q}_{i}(t))=m_{i}c^{2}+\frac{1}{2}m_{i}\left(D_{*}\mathbf{q}_{i}(t)\right)^{2}=m_{i}c^{2}+\frac{1}{2}m_{i}\mathbf{b}_{i*}^{2}.
\end{equation}
The $\delta\mathbf{q}_{i-}(t)$ in (3.138) corresponds to the physical,
translational, mean backward displacement of the \emph{i}-th \emph{zbw}
particle, as defined by 
\begin{equation}
\delta\mathbf{q}_{i-}(t)=\left(D_{*}\mathbf{q}_{i}(t)\right)\delta t=\mathbf{b}_{i*}(q(t),t)\delta t.
\end{equation}
(Notice that $\delta\mathbf{q}_{i+}(t)$ and $\delta\mathbf{q}_{i-}(t)$
are not equal in general since $\delta\mathbf{q}_{i+}(t)-\delta\mathbf{q}_{i-}(t)=(\mathbf{b}_{i}-\mathbf{b}_{i*})\delta t\neq0$
in general.) Now since each \emph{zbw} particle is essentially a harmonic
oscillator, each particle has its own, effectively independent, well-defined
forward steady-state phase at each point along its forward space-time
trajectory, when $\mathbf{b}_{i}(q,t)\approx\sum_{i}^{N}\mathbf{b}_{i}(\mathbf{q}_{i},t)$.
Consistency with this hypothesis also means that when $\mathbf{b}_{i}(q,t)\neq\sum_{i}^{N}\mathbf{b}_{i}(\mathbf{q}_{i},t)$,
the forward steady-state joint phase must be a well-defined function
of the space-time trajectories of\emph{ all} \emph{the particles}
(since we posit that all particles remain harmonic oscillators despite
having their oscillations physically coupled through the common ether
medium they interact with). Furthermore, since, at this stage, the
forward and backward steady-state joint \emph{zbw} phase changes,
(3.135) and (3.138), are independent of one another, each must equal
$2\pi n$ when integrated along a closed loop $L$ in which both time
and position change. Otherwise, we will contradict our hypothesis
that the system of \emph{zbw} particles has a well-defined steady-state
joint phase in each time direction. 

In the lab frame, the forward and backward stochastic differential
equations for the translational motion are again given by (3.127)
and (3.128), and the corresponding forward and backward Fokker-Planck
equations take the form
\begin{equation}
\frac{\partial\rho(q,t)}{\partial t}=-\sum_{i=1}^{N}\nabla_{i}\cdot\left[\mathbf{b}_{i}(q,t)\rho(q,t)\right]+\sum_{i=1}^{N}\frac{\hbar}{2m_{i}}\nabla_{i}^{2}\rho(q,t),
\end{equation}
and
\begin{equation}
\frac{\partial\rho(q,t)}{\partial t}=-\sum_{i=1}^{N}\nabla_{i}\cdot\left[\mathbf{b}_{i*}(q,t)\rho(q,t)\right]-\sum_{i=1}^{N}\frac{\hbar}{2m_{i}}\nabla_{i}^{2}\rho(q,t).
\end{equation}
Restricting to simultaneous solutions of (3.137) and (3.138) entails
the current velocity field
\begin{equation}
\mathbf{v}_{i}(q,t)\coloneqq\frac{1}{2}\left[\mathbf{b}_{i}(q,t)+\mathbf{b}_{i*}(q,t)\right]=\frac{\nabla_{i}S(q,t)}{m_{i}},
\end{equation}
and the osmotic velocity field
\begin{equation}
\mathbf{u}_{i}(q,t)\coloneqq\frac{1}{2}\left[\mathbf{b}_{i}(q,t)-\mathbf{b}_{i*}(q,t)\right]=\frac{\hbar}{2m_{i}}\frac{\nabla_{i}\rho(q,t)}{\rho(q,t)}.
\end{equation}
Then (3.141) and (3.142) reduce to the continuity equation
\begin{equation}
\frac{\partial\rho({\normalcolor q},t)}{\partial t}=-\sum_{i=1}^{N}\nabla_{i}\cdot\left[\frac{\nabla_{i}S(q,t)}{m_{i}}\rho(q,t)\right],
\end{equation}
with $\mathbf{b}_{i}=\mathbf{v}_{i}+\mathbf{u}_{i}$ and $\mathbf{b}_{i*}=\mathbf{v}_{i}-\mathbf{u}_{i}$.

As we did for \emph{N}-particle NYSM, we now postulate here the presence
of an external (to the particle) osmotic potential, $U(q,t)$, which
couples to the $i$-th particle as $R(q(t),t)=\mu U(q(t),t)$ (assuming
that the coupling constant $\mu$ is identical for particles of the
same species), and imparts to the $i$-th particle a momentum, $\nabla_{i}R(q,t)|_{\mathbf{q}_{j}=\mathbf{q}_{j}(t)}$.
This momentum then gets counter-balanced by the ether fluid's osmotic
impulse pressure, $\left(\hbar/2m_{i}\right)\nabla_{i}\ln[n(q,t)]|_{\mathbf{q}_{j}=\mathbf{q}_{j}(t)}$,
leading to the equilibrium condition $\nabla_{i}R/m_{i}=\left(\hbar/2m_{i}\right)\nabla_{i}\rho/\rho$
(using $\rho=n/N$), which implies $\rho=e^{2R/\hbar}$ for all times.
As discussed in section 2, it is expected that $R$ generally depends
on the coordinates of all the other particles. The reasons, to remind
the reader, are that: (i) we argued, for reasons of consistency, that
$U$ should be sourced by the ether, and (ii) since the particles
continuously exchange energy-momentum with the ether, the functional
dependence of $U$ will be determined by the dynamical coupling of
the ether to the particles as well as the magnitude of the inter-particle
physical interactions (whether through a classical inter-particle
potential or, in the free particle case, just through the ether).
To make this last point more explicit, suppose two classically non-interacting
\emph{zbw} particles of identical mass, each initially driven in their
oscillations and translational motions by effectively independent
regions of oscillating ether, each region sourcing the osmotic potentials
$U_{1}(\mathbf{q}_{1},t)$ and $U_{2}(\mathbf{q}_{2},t)$, move along
trajectories that cause the spatial support of their dynamically relevant
regions of oscillating ether to significantly overlap; then the particles
will be exchanging energy-momentum with a common region of oscillating
ether modes, leading to an osmotic potential sourced by this common
region of oscillating ether that depends on the motions (hence positions)
of both particles, i.e., $U(\mathbf{q}_{1},\mathbf{q}_{2},t)$. Indeed,
this common region of oscillating ether will drive the subsequent
steady-state \emph{zbw} oscillations and translational Brownian motions
of both particles, leading (after the constraint of conservative diffusions
is imposed, as we will see) to a time-symmetrized steady-state joint
phase $S(\mathbf{q}_{1},\mathbf{q}_{2},t)$ whose gradient with respect
to the \emph{i}-th particle coordinate gives rise to the current velocity
of the \emph{i}-th particle, and to an osmotic counter-balancing of
$\nabla_{i}U(\mathbf{q}_{1},\mathbf{q}_{2},t)$, which gives rise
to the osmotic velocity of the \emph{i}-th particle (as we've already
seen). Mathematically, the non-linear coupling between the osmotic
potential and the evolution of the (conservative-diffusions-constrained)
time-symmetrized joint phase of the \emph{zbw} particles can be seen
by writing the solution to (3.145), which from section 2 is 
\begin{equation}
\rho(q,t)=\rho_{0}(q_{0})exp[-\int_{0}^{t}\left(\sum_{i}^{N}\nabla_{i}\cdot\mathbf{v}_{i}\right)dt'=\rho_{0}(q_{0})exp[-\int_{0}^{t}\left(\sum_{i}^{N}\frac{\nabla_{i}^{2}S}{m_{i}}\right)dt',
\end{equation}
giving
\begin{equation}
R(q,t)=R_{0}(q_{0})-(\hbar/2)\int_{0}^{t}\left(\sum_{i}^{N}\frac{\nabla_{i}^{2}S}{m_{i}}\right)dt',
\end{equation}
Then we can infer from (3.146) that if a narrow bandwidth of common
ether modes is driving the \emph{zbw} oscillations of both particles
(as described in hypothesis 3 above), the evolution of the osmotic
potential (sourced by the common ether modes) will develop functional
dependence on the positions of both particles. The precise form of
this functional dependence and how it evolves in time will depend
on the evolution equation for $S$, which we of course need to specify
(but already know will end up being the $N$-particle quantum HJ equation).

To obtain the 2nd-order time-symmetric dynamics for the mean translational
motions of the \emph{N} particles, we will define the ensemble-averaged
action Eq. (3.18) in terms of a symmetric combination of the forward
and backward steady-state joint \emph{zbw} phase changes (3.135) and
(3.138). This is natural to do since (3.135) and (3.138) correspond
to the same frame (the lab frame), and since (3.135) and (3.138) are
no longer independent of one another as a result of the constraints
(3.143-144). 

First, we take the difference between (3.135) and (3.138) to get (replacing
$\delta t\rightarrow dt$ and $\delta\mathbf{q}_{i+,-}(t)\rightarrow d\mathbf{q}_{i+,-}(t)$)
\begin{equation}
\begin{aligned}d\bar{\theta}(q(t),t) & \coloneqq\frac{1}{2}\left[d\bar{\theta}_{+}(q(t),t)-d\bar{\theta}_{-}(q(t),t)\right]\\
 & =\sum_{i=1}^{N}\frac{\omega_{ci}}{m_{i}c^{2}}\mathrm{E}_{t}\left[E_{i}(D\mathbf{q}_{i}(t),D_{*}\mathbf{q}_{i}(t))dt-\frac{m_{i}}{2}\left(\mathbf{b}_{i}(q(t),t)\cdot d\mathbf{q}_{i+}(t)+\mathbf{b}_{i*}(q(t),t)\cdot d\mathbf{q}_{i-}(t)\right)\right]+\phi\\
 & =\frac{1}{\hbar}\mathrm{E}_{t}\left[\sum_{i=1}^{N}E_{i}dt-\sum_{i=1}^{N}\frac{m_{i}}{2}\left(\mathbf{b}_{i}\cdot\frac{d\mathbf{q}_{i+}(t)}{dt}+\mathbf{b}_{i*}\cdot\frac{d\mathbf{q}_{i-}(t)}{dt}\right)dt\right]+\phi\\
 & =\frac{1}{\hbar}\mathrm{E}_{t}\left[\left(\sum_{i=1}^{N}E_{i}-\sum_{i=1}^{N}\frac{m_{i}}{2}\left(\mathbf{b}_{i}\cdot\frac{d\mathbf{q}_{i+}(t)}{dt}+\mathbf{b}_{i*}\cdot\frac{d\mathbf{q}_{i-}(t)}{dt}\right)\right)dt\right]+\phi\\
 & =\frac{1}{\hbar}\mathrm{E}_{t}\left[\left(\sum_{i=1}^{N}E_{i}-\sum_{i=1}^{N}\frac{m_{i}}{2}\left(\mathbf{b}_{i}^{2}+\mathbf{b}_{i*}^{2}\right)\right)dt\right]+\phi\\
 & =\frac{1}{\hbar}\mathrm{E}_{t}\left[\left(\sum_{i=1}^{N}E_{i}-\sum_{i=1}^{N}\left(m_{i}\mathbf{v}_{i}\cdot\mathbf{v}_{i}+m_{i}\mathbf{u}_{i}\cdot\mathbf{u}_{i}\right)\right)dt\right]+\phi\\
 & =\frac{1}{\hbar}\mathrm{E}_{t}\left[\sum_{i=1}^{N}\left(m_{i}c^{2}-\frac{1}{2}m_{i}\mathbf{v}_{i}^{2}-\frac{1}{2}m_{i}\mathbf{u}_{i}^{2}\right)dt\right]+\phi,
\end{aligned}
\end{equation}
where $\phi=\sum_{i=1}^{N}\left(\phi_{i+}-\phi_{i-}\right)$, and
from (3.136) and (3.139), we have
\begin{equation}
E_{i}(D\mathbf{q}_{i}(t),D_{*}\mathbf{q}_{i}(t))\coloneqq m_{i}c^{2}+\frac{1}{2}\left[\frac{1}{2}m_{i}\mathbf{b}_{i}^{2}+\frac{1}{2}m_{i}\mathbf{b}_{i*}^{2}\right]=m_{i}c^{2}+\frac{1}{2}m_{i}\mathbf{v}_{i}^{2}+\frac{1}{2}m_{i}\mathbf{u}_{i}^{2}.
\end{equation}
Equation (3.148) is the time-symmetrized steady-state joint phase
change of the \emph{zbw} particles in the lab frame, before the constraint
of conservative diffusions is imposed. Note that because $\bar{\theta}_{+}$
and $\bar{\theta}_{-}$ are no longer independent of one another,
it is no longer consistent to have that $\oint_{L}\delta\bar{\theta}_{+}$
and $\oint_{L}\delta\bar{\theta}_{-}$ both equal $2\pi n$. However,
the consistency of our theory does require that $\oint_{L}\delta\bar{\theta}=2\pi n$,
otherwise we will contradict our hypothesis that the system of \emph{N}
\emph{zbw} particles, after imposing (3.143-144) has a well-defined
and unique steady-state joint phase that functionally depends on the
3-space trajectories of the \emph{zbw} particles.

Now, defining the steady-state joint phase-principal function
\begin{equation}
I(q(t),t)=-\hbar\bar{\theta}(q(t),t)=\mathrm{E}\left[\int_{t_{I}}^{t}\sum_{i=1}^{N}\left(\frac{1}{2}m_{i}\mathbf{v}_{i}^{2}+\frac{1}{2}m_{i}\mathbf{u}_{i}^{2}-m_{i}c^{2}\right)dt'\left|\mathbf{q}_{j}(t)\right.\right]-\hbar\sum_{i=1}^{N}\left(\phi_{i+}-\phi_{i-}\right),
\end{equation}
we can use (3.150) to define the steady-state joint phase-action
\begin{equation}
\begin{aligned}J & =I_{IF}=\mathrm{E}\left[\int_{t_{I}}^{t_{F}}\sum_{i=1}^{N}\left(\frac{1}{2}m_{i}\mathbf{v}_{i}^{2}+\frac{1}{2}m_{i}\mathbf{u}_{i}^{2}-m_{i}c^{2}\right)dt'-\hbar\phi\right].\end{aligned}
\end{equation}
It is straightforward to see that (3.151) is just Eq. (3.18) in section
2, with the potentials set equal to zero, and modulo the rest-energy
terms and the time-symmetrized initial joint phase constant $\phi$.

Note, also, that from the second to last line of (3.148), we can write
the cumulative, time-symmetric, steady-state joint phase at time \emph{t}
as 
\begin{equation}
\begin{aligned}\bar{\theta}(q(t),t) & =\frac{1}{\hbar}\mathrm{E}\left[\int_{t_{I}}^{t}\left(\sum_{i=1}^{N}E_{i}-\sum_{i=1}^{N}\left(m_{i}\mathbf{v}_{i}\cdot\mathbf{v}_{i}+m_{i}\mathbf{u}_{i}\cdot\mathbf{u}_{i}\right)\right)dt'\left|\mathbf{q}_{j}(t)\right.\right]+\phi\\
 & =\frac{1}{\hbar}\mathrm{E}\left[\int_{t_{I}}^{t}\left(\sum_{i=1}^{N}\left(E_{i}-m_{i}\mathbf{u}_{i}\cdot\mathbf{u}_{i}\right)-\sum_{i=1}^{N}m_{i}\mathbf{v}_{i}\cdot\mathbf{v}_{i}\right)dt'\left|\mathbf{q}_{j}(t)\right.\right]+\phi\\
 & =\frac{1}{\hbar}\mathrm{E}\left[\int_{t_{I}}^{t}\left(H-\sum_{i=1}^{N}m_{i}\mathbf{v}_{i}\cdot\mathbf{v}_{i}\right)dt'\left|\mathbf{q}_{j}(t)\right.\right]+\phi\\
 & =\frac{1}{\hbar}\mathrm{E}\left[\int_{t_{I}}^{t}\left(H-\sum_{i=1}^{N}\frac{m_{i}}{4}\left(D\mathbf{q}_{i}(t')+D_{*}\mathbf{q}_{i}(t')\right)\cdot\left(D+D_{*}\right)\mathbf{q}_{i}(t')\right)dt'\left|\mathbf{q}_{j}(t)\right.\right]+\phi\\
 & =\frac{1}{\hbar}\mathrm{E}\left[\int_{t_{I}}^{t}Hdt'-\sum_{i=1}^{N}\frac{m_{i}}{2}\int_{\mathbf{q}_{i}(t_{I})}^{\mathbf{q}_{i}(t)}\left(D\mathbf{q}_{i}(t')+D_{*}\mathbf{q}_{i}(t')\right)\cdot\mathrm{D}\mathbf{q}_{i}(t')\left|\mathbf{q}_{j}(t)\right.\right]+\phi,
\end{aligned}
\end{equation}
where 
\begin{equation}
H\coloneqq\sum_{i=1}^{N}\left(E_{i}-m_{i}\mathbf{u}_{i}\cdot\mathbf{u}_{i}\right)=\sum_{i=1}^{N}\left(m_{i}c^{2}+\frac{1}{2}m_{i}\mathbf{v}_{i}^{2}-\frac{1}{2}m_{i}\mathbf{u}_{i}^{2}\right),
\end{equation}
and where we have used the fact that $0.5\left(D+D_{*}\right)\mathbf{q}_{i}(t)=\left(\partial_{t}+\sum_{j}\mathbf{v}_{j}(q(t),t)\cdot\nabla_{j}\right)\mathbf{q}_{i}(t)$,
and $\mathbf{v}_{i}(q(t),t)=\left(\partial_{t}+\sum_{j}\mathbf{v}_{j}\cdot\nabla_{j}\right)\mathbf{q}_{i}(t)\eqqcolon\mathrm{D}\mathbf{q}_{i}(t)/\mathrm{D}t$,
and $\mathrm{D}\mathbf{q}_{i}(t)=\left(\mathrm{D}\mathbf{q}_{i}(t)/\mathrm{D}t\right)dt$.
Now, consider an integral curve $\mathbf{Q}_{i}(t)$ of the \emph{i}-th
current velocity/momentum field, i.e., a solution of
\begin{equation}
m_{i}\frac{d\mathbf{Q}_{i}(t)}{dt}=m_{i}\mathbf{v}_{i}(Q(t),t)=\mathbf{p}_{i}(Q(t),t)=\nabla_{i}S(q,t)|_{\mathbf{q}_{j}=\mathbf{Q}_{j}(t)}.
\end{equation}
Then we can replace the functional dependence of (3.152) on $q(t)$
by $Q(t)$, obtaining 
\begin{equation}
\begin{aligned}\bar{\theta}(Q(t),t) & =\frac{1}{\hbar}\int_{t_{I}}^{t}\left[H-\sum_{i=1}^{N}m_{i}\mathbf{v}_{i}(Q(t'),t')\cdot\frac{d\mathbf{Q}_{i}(t')}{dt'}\right]dt'+\phi\\
 & =\frac{1}{\hbar}\left[\int_{t_{I}}^{t}Hdt'-\sum_{i=1}^{N}\int_{\mathbf{Q}_{i}(t_{I})}^{\mathbf{Q}_{i}(t)}\mathbf{p}_{i}\cdot d\mathbf{Q}_{i}(t')\right]+\phi,
\end{aligned}
\end{equation}
where it should be noticed that we've dropped the conditional expectation.
So (3.155) denotes the cumulative, time-symmetric, steady-state joint
phase of the \emph{zbw} particles, evaluated along the time-symmetric
mean trajectories of the \emph{zbw} particles, i.e., the integral
curves of (3.154). That the time-symmetric mean trajectories of the
\emph{zbw} particles should correspond to the integral curves of (3.154)
can be seen from the fact that the single-time joint probability density
$\rho(q,t)$, after imposing the time-symmetric constraints (3.143-144),
is a solution of the continuity equation (3.145), from which it follows
that the possible mean trajectories of the \emph{zbw} particles are
the flow lines of the probability current $\rho\mathbf{v}_{i}$, i.e.,
the solutions of (3.154) for all possible initial conditions $\mathbf{Q}_{i}(0)$.) 

Now, taking the total differential of the left hand side of (3.155)
gives
\begin{equation}
d\bar{\theta}=\sum_{i=1}^{N}\nabla_{i}\bar{\theta}|_{\mathbf{q}_{j}=\mathbf{Q}_{j}(t)}d\mathbf{Q}_{i}(t)+\partial_{t}\bar{\theta}|_{\mathbf{q}_{j}=\mathbf{Q}_{j}(t)}dt.
\end{equation}
This allows us to identify 
\begin{equation}
\mathbf{p}_{i}(Q(t),t)=-\hbar\nabla_{i}\bar{\theta}|_{\mathbf{q}_{j}=\mathbf{Q}_{j}(t)}=\nabla_{i}S|_{\mathbf{q}_{j}=\mathbf{Q}_{j}(t)},
\end{equation}
using (3.156) along with (3.155) and (3.143). Thus the \emph{i}-th
current velocity in the lab frame corresponds the gradient of the
time-symmetrized steady-state joint phase of the \emph{zbw} particles
at the location of the \emph{i}-th \emph{zbw} particle, and $S$ can
be identified with the cumulative, time-symmetric, steady-state joint
phase function of the \emph{zbw} particles in the lab frame. In addition,
we have 
\begin{equation}
H(Q(t),t)=\hbar\partial_{t}\bar{\theta}|_{\mathbf{q}_{j}=\mathbf{Q}_{j}(t)}=-\partial_{t}S|_{\mathbf{q}_{j}=\mathbf{Q}_{j}(t)}.
\end{equation}
From (3.158), (3.157), and (3.155), it follows that 
\begin{equation}
\begin{aligned}S(Q(t),t) & =\sum_{i=1}^{N}\int_{\mathbf{Q}_{i}(t_{I})}^{\mathbf{Q}_{i}(t)}\mathbf{p}_{i}\cdot d\mathbf{Q}_{i}(t')-\int_{t_{I}}^{t}Hdt'-\hbar\phi\\
 & =\int_{t_{I}}^{t}\sum_{i=1}^{N}\left[\frac{1}{2}m_{i}\mathbf{v}_{i}^{2}+\frac{1}{2}m_{i}\mathbf{u}_{i}^{2}-m_{i}c^{2}\right]dt'-\hbar\phi=I(Q(t),t),
\end{aligned}
\end{equation}
and 
\begin{equation}
\oint_{L}\delta S(Q(t),t)=\sum_{i=1}^{N}\oint_{L}\left[\mathbf{p}_{i}(Q(t),t)\cdot\delta\mathbf{Q}_{i}(t)-E_{i}(Q(t),t)\delta t\right]=nh.
\end{equation}
We shall use these last two expressions for later comparisons.

Recall that after restricting the forward and backward diffusions
to simultaneous solutions of (3.141-142), we have $\mathbf{b}_{i}=\mathbf{v}_{i}+\mathbf{u}_{i}$
and $\mathbf{b}_{i*}=\mathbf{v}_{i}-\mathbf{u}_{i}$. So the IMFTRF
and the IMBTRF will not coincide since, for $\mathbf{b}_{i}=\mathbf{v}_{i}+\mathbf{u}_{i}=0$,
it will generally not be the case that $\mathbf{b}_{i*}=\mathbf{v}_{i}-\mathbf{u}_{i}=0$.
Nevertheless, we can define an instantaneous mean (time-)symmetric
rest frame (IMSTRF) as the frame in which $\mathbf{b}_{i}+\mathbf{b}_{i*}=2\mathbf{v}_{i}=0$.
And the lab frame remains the lab frame. 

Applying the conservative diffusion constraint through the extremality
of (3.151), we obtain the mean acceleration equation
\begin{equation}
\sum_{i=1}^{N}\frac{m_{i}}{2}\left[D_{*}D+DD_{*}\right]\mathbf{q}_{i}(t)=0.
\end{equation}
Moreover, since the $\delta\mathbf{q}_{i}(t)$ are independent (as
shown in Appendix 7.1), it follows from (3.161) that we have the individual
equations of motion
\begin{equation}
m_{i}\mathbf{a}_{i}(q(t),t)=\frac{m_{i}}{2}\left[D_{*}D+DD_{*}\right]\mathbf{q}_{i}(t)=0.
\end{equation}
By applying the mean derivatives in (3.161), and using that $\mathbf{b}_{i}=\mathbf{v}_{i}+\mathbf{u}_{i}$
and $\mathbf{b}_{i*}=\mathbf{v}_{i}-\mathbf{u}_{i}$, straightforward
manipulations give
\begin{equation}
\sum_{i=1}^{N}m_{i}\left[\partial_{t}\mathbf{v}_{i}+\mathbf{v}_{i}\cdot\nabla_{i}\mathbf{v}_{i}-\mathbf{u}_{i}\cdot\nabla_{i}\mathbf{u}_{i}-\frac{\hbar}{2m_{i}}\nabla_{i}^{2}\mathbf{u}_{i}\right]|_{\mathbf{q}_{j}=\mathbf{q}_{j}(t)}=0.
\end{equation}
Using (3.143-144), (3.163) yields
\begin{equation}
\begin{aligned}\sum_{i=1}^{N}m_{i}\mathbf{a}_{i}(q(t),t) & =\sum_{i=1}^{N}m_{i}\left[\frac{\partial\mathbf{v}_{i}(q,t)}{\partial t}+\mathbf{v}_{i}(q,t)\cdot\nabla_{i}\mathbf{v}_{i}(q,t)\right.\\
 & \left.-\mathbf{u}_{i}(q,t)\cdot\nabla_{i}\mathbf{u}_{i}(q,t)-\frac{\hbar}{2m_{i}}\nabla_{i}^{2}\mathbf{u}_{i}(q,t)\right]|_{\mathbf{q}_{j}=\mathbf{q}_{j}(t)}\\
 & =\sum_{i=1}^{N}\nabla_{i}\left[\frac{\partial S(q,t)}{\partial t}+\frac{\left(\nabla_{i}S(q,t)\right)^{2}}{2m_{i}}-\frac{\hbar^{2}}{2m_{i}}\frac{\nabla_{i}^{2}\sqrt{\rho(q,t)}}{\sqrt{\rho(q,t)}}\right]|_{\mathbf{q}_{j}=\mathbf{q}_{j}(t)}=0.
\end{aligned}
\end{equation}
Integrating both sides of (3.164) and setting the arbitrary integration
constants equal to the rest energies, we then have the \emph{N}-particle
quantum Hamilton-Jacobi equation
\begin{equation}
\begin{aligned}\tilde{E}(q(t),t) & \coloneqq\sum_{i=1}^{N}\tilde{E}_{i}(q(t),t)\\
 & \coloneqq-\partial_{t}S(q(t),t)\\
 & =\sum_{i=1}^{N}m_{i}c^{2}+\sum_{i=1}^{N}\frac{\left(\nabla_{i}S(q,t)\right)^{2}}{2m_{i}}|_{\mathbf{q}_{j}=\mathbf{q}_{j}(t)}-\sum_{i=1}^{N}\frac{\hbar^{2}}{2m_{i}}\frac{\nabla_{i}^{2}\sqrt{\rho(q,t)}}{\sqrt{\rho(q,t)}}|_{\mathbf{q}_{j}=\mathbf{q}_{j}(t)},
\end{aligned}
\end{equation}
describing the total energy of the actual particles along their stochastic
trajectories $q(t)$. Alternatively, given the integral curves $\mathbf{Q}_{i}(t)$
of the reformulated mean acceleration equation
\begin{equation}
m_{i}\frac{d^{2}\mathbf{Q}_{i}(t)}{dt^{2}}=m_{i}\left(\partial_{t}\mathbf{v}_{i}+\mathbf{v}_{i}\cdot\nabla_{i}\mathbf{v}_{i}\right)|_{\mathbf{q}_{j}=\mathbf{Q}_{j}(t)}=-\nabla_{i}\left(-\frac{\hbar^{2}}{2m_{i}}\frac{\nabla_{i}^{2}\sqrt{\rho(q,t)}}{\sqrt{\rho(q,t)}}\right)|_{\mathbf{q}_{j}=\mathbf{Q}_{j}(t)},
\end{equation}
for $i=1,...,N$, we can replace $q(t)$ by $Q(t)$ and thereby obtain
the total energy $\tilde{E}(Q(t),t)$ of the actual \emph{zbw} particles
along their time-symmetric mean trajectories, the latter now given
by solutions of (3.166). The corresponding general solution of (3.165) is then
given by 
\begin{equation}
\begin{aligned}S(Q(t),t) & =\sum_{i=1}^{N}\int_{\mathbf{Q}_{i}(t_{I})}^{\mathbf{Q}_{i}(t)}\mathbf{p}_{i}(Q(t'),t')\cdot d\mathbf{Q}_{i}(t')-\sum_{i=1}^{N}\int_{t_{I}}^{t}\tilde{E}_{i}(Q(t'),t')dt'-\sum_{i=1}^{N}\hbar\phi_{i}\\
 & =\int_{t_{I}}^{t}\sum_{i=1}^{N}\left[\frac{1}{2}m_{i}\mathbf{v}_{i}^{2}-\left(-\frac{\hbar^{2}}{2m_{i}}\frac{\nabla_{i}^{2}\sqrt{\rho}}{\sqrt{\rho}}\right)-m_{i}c^{2}\right]dt'-\sum_{i=1}^{N}\hbar\phi_{i}\\
 & =\int_{t_{I}}^{t}\sum_{i=1}^{N}\left[\frac{1}{2}m_{i}\mathbf{v}_{i}^{2}+\frac{1}{2}m_{i}\mathbf{u}_{i}^{2}+\frac{\hbar}{2}\nabla_{i}\cdot\mathbf{u}_{i}-m_{i}c^{2}\right]dt'-\sum_{i=1}^{N}\hbar\phi_{i}.
\end{aligned}
\end{equation}
We identify (3.167) as the conservative-diffusion-constrained, time-symmetric,
steady-state joint phase associated with the \emph{zbw} particles
in the lab frame. Notice that the last line of (3.167) differs from
the last line of (3.159) only by addition of the terms involving $\nabla_{i}\cdot\mathbf{u}_{i}$. 

Notice also that the dynamics for (3.167) clearly differs from the
dynamics of the joint phase of the free classical \emph{zbw} particles
by the presence of the quantum kinetic in (3.165-166). As in the single-particle
case, the two phases are formally connected by the `classical limit'
$(\hbar/2m_{i})\rightarrow0$, but this is only formal since such
a limit corresponds to deleting the presence of the ether, thereby
also deleting the physical mechanism that causes the \emph{zbw} particles
to oscillate at their Compton frequencies. The physically realistic
`classical limit' for the phase (3.167) corresponds to situations
where the quantum kinetic and its gradient are negligible, which will
occur (as in the dBB theory) whenever the center of mass of a system
of interacting particles is sufficiently large and environmental decoherence
is appreciable \cite{Allori2001,Bowm2005,Oriols2016,Derakhshani2017b}.

Since each \emph{zbw} particle is posited to essentially be a harmonic
oscillator of (unspecified) identical type, each particle has its
own, effectively independent, well-defined phase at each point along
its time-symmetric mean space-time trajectory, when $\mathbf{v}_{i}(q,t)\approx\sum_{i}^{N}\mathbf{v}_{i}(\mathbf{q}_{i},t)$.
Consistency with this means that when $\mathbf{v}_{i}(q,t)\neq\sum_{i}^{N}\mathbf{v}_{i}(\mathbf{q}_{i},t)$,
the time-symmetric steady-state joint phase must be a well-defined
function of the time-symmetric mean trajectories of\emph{ all} the
particles (since we posit that all the particles remain harmonic oscillators,
despite having their oscillations physically coupled through the common
ether medium they interact with). Then, for a closed loop \emph{L}
along which each particle can be physically or virtually displaced,
it follows that
\begin{equation}
\oint_{L}\delta S(Q(t),t)=\sum_{i=1}^{N}\oint_{L}\left[\mathbf{p}_{i}(Q(t),t)\cdot\delta\mathbf{Q}_{i}(t)-\tilde{E_{i}}(Q(t),t)\delta t\right]=nh.
\end{equation}
And for a closed loop $L$ with $\delta t=0$, we have
\begin{equation}
\sum_{i=1}^{N}\oint_{L}\mathbf{p}_{i}\cdot\delta\mathbf{Q}_{i}(t)=\sum_{i=1}^{N}\oint_{L}\mathbf{\nabla}_{i}S(q,t)|_{\mathbf{q}_{j}=\mathbf{Q}_{j}(t)}\cdot\delta\mathbf{Q}_{i}(t)=nh.
\end{equation}
If we also consider the joint phase field $S(q,t)$, a field over
the possible positions of the \emph{zbw} particles, then, as a result
of the same physical reasoning applied to the \emph{i}-th particle
at any possible initial position it can occupy, we will have
\begin{equation}
\oint_{L}dS\left(q,t\right)=\sum_{i=1}^{N}\oint_{L}\mathbf{p}_{i}\cdot d\mathbf{q}_{i}=\sum_{i=1}^{N}\oint_{L}\nabla_{i}S(q,t)\cdot d\mathbf{q}_{i}=nh.
\end{equation}
Notice that (3.170) constrains the osmotic potential as well, due
to the coupling of $S$ to $R$ (hence $U$) via (3.147). This makes
physical sense since, as we observed earlier, the oscillating ether
drives the \emph{zbw} oscillations of the particles while also sourcing
the osmotic potential that imparts the osmotic velocities to the particles.

Combining (3.170), (3.165), and (3.145), we can construct the \emph{N}-particle
Schr\"{o}dinger equation
\begin{equation}
i\hbar\frac{\partial\psi(q,t)}{\partial t}=\sum_{i=1}^{N}\left[-\frac{\hbar^{2}}{2m_{i}}\nabla_{i}^{2}+m_{i}c^{2}\right]\psi(q,t),
\end{equation}
where the \emph{N}-particle wavefunction $\psi(q,t)=\sqrt{\rho(q,t)}e^{iS(q,t)/\hbar}$
is single-valued by (3.170). 

How does the interpretation of the ZSM wavefunction differ from that
of the NYSM wavefunction? The only difference comes from $S(q,t)$
being the conservative-diffusion-constrained, time-symmetrized, steady-state,
joint phase of the \emph{zbw} particles in ZSM, as opposed to being
an \emph{N}-particle velocity potential satisfying a law-like quantization
constraint of the form (3.170) in NYSM. This difference means that,
in ZSM, $S(q,t)$ reflects not only ontic aspects such as the irrotationality
of the ZSM version of the ether, and the influence of classical fields
on the \emph{zbw} particles, it also reflects the steady-state oscillations
of \emph{zbw} particles immersed in the ether, as well as the (hypothesized)
oscillations of the ether at each point in 3-D space. And it is a
consequence of these last two ontic aspects of an \emph{N}-particle
ZSM system that the quantization condition (3.170) follows; in other
words, the quantization condition is no longer a law-like constraint
on $S(q,t)$, but a consequence of certain ontic properties of an
\emph{N}-particle ZSM system.

Note that since the solution space of the combination of (3.170),
(3.165), and (3.145) is equivalent to the solution space of (3.171),
any non-factorizable wavefunctions that can be constructed as solutions
of (3.171) will also be solutions (in $\rho$ and $S$ variables)
of the combination of (3.170), (3.165), and (3.145). As an example,
let us consider two identical, classically non-interacting bosons
or fermions with initial wavefunction \footnote{The Nelsonian derivation of the symmetry postulates given by Bacciagaluppi
in \cite{Bacciagaluppi2003}, which allows us to write down a wavefunction
like (3.172) (or its anti-symmetric counterpart), is consistent with
the assumptions of ZSM and carries over without any change.}
\begin{equation}
\psi_{nf}(\mathbf{q}_{1},\mathbf{q}_{2})\coloneqq Norm_{\pm}\left[\psi_{A}(\mathbf{q}_{1})\psi_{B}(\mathbf{q}_{2})\pm\psi_{A}(\mathbf{q}_{2})\psi_{B}(\mathbf{q}_{1})\right],
\end{equation}
where particle 1 is associated with wavepacket $\psi_{A}$ and particle
2 is associated with packet $\psi_{B}$, and the wavepackets satisfy
$\psi_{A}\cap\psi_{B}\approx\varnothing$. Then, if the packets of
these particles move towards each other and overlap such that $\left(<\mathbf{q}_{1}>-<\mathbf{q}_{2}>\right)^{2}\leq\sigma_{A}^{2}+\sigma_{B}^{2}$,
the subsequent wavefunction of the 2-particle system will be (3.172)
but with $\psi_{A}\cap\psi_{B}\neq\varnothing$. Moreover, in terms
of $\rho$ and $S$ variables, we have
\begin{equation}
\begin{aligned}\rho_{nf}\left(\mathbf{q}_{1},\mathbf{q}_{2}\right) & \coloneqq|\psi_{nf}(\mathbf{q}_{1},\mathbf{q}_{2})|^{2}=Norm_{\pm}^{2}\left\{ e^{2\left(R_{A1}+R_{B2}\right)/\hbar}+e^{2\left(R_{A2}+R_{B1}\right)/\hbar}\right.\\
 & \pm e^{\left[\left(R_{A1}+R_{B2}+R_{A2}+R_{B1}\right)+i\left(S_{A2}+S_{B1}-S_{A1}-S_{B2}\right)\right]/\hbar}\\
 & \left.\pm e^{\left[\left(R_{A1}+R_{B2}+R_{A2}+R_{B1}\right)+i\left(S_{A1}+S_{B2}-S_{A2}-S_{B1}\right)\right]/\hbar}\right\} ,
\end{aligned}
\end{equation}
and 
\begin{equation}
S_{nf}(\mathbf{q}_{1},\mathbf{q}_{2},)\coloneqq-\frac{i\hbar}{2}\ln\left(\frac{\psi_{nf}(\mathbf{q}_{1},\mathbf{q}_{2})}{\psi_{nf}^{\ast}(\mathbf{q}_{1},\mathbf{q}_{2})}\right),
\end{equation}
where (3.173) satisfies (3.145), and (3.174) is a solution of (3.165)
and satisfies (3.170). That is, the two particles will be entangled
in their joint phase (3.174) and their joint osmotic potential obtained
from (3.172) or (3.173):
\begin{equation}
R_{nf}(\mathbf{q}_{1},\mathbf{q}_{2})\coloneqq\hbar\ln\left(|\psi_{nf}(\mathbf{q}_{1},\mathbf{q}_{2})|\right).
\end{equation}
This scenario of entanglement formation between two identical bosons
or fermions is essentially equivalent to the scenario we considered
earlier for our justification of why the osmotic potential should
have functional dependence on the positions of both particles: Eq.
(3.174) is the conservative-diffusion-constrained, time-symmetric,
steady-state joint phase that develops between the two particles from
having their \emph{zbw} oscillations driven by a common region of
oscillating ether that forms when $\left(<\mathbf{q}_{1}>-<\mathbf{q}_{2}>\right)^{2}\leq\sigma_{A}^{2}+\sigma_{B}^{2}$.
Likewise, (3.175) is the joint osmotic potential that arises from
this common region of oscillating ether sourcing the osmotic potential.

Additionally, Eqs. (3.170), (3.165), and (3.145) tell us how the non-local
functional dependence of (3.175) on the positions of the two particles
changes in time: for classically non-interacting particles, the non-local
correlations become negligible when the 3-D spatial separation between
the particles becomes sufficiently large, i.e., when the overlap of
the wavepackets in the summands of (3.172) becomes negligible. Of
course, the correlations never completely vanish because the overlap
of the wavepackets in the summands of (3.172) never completely vanishes,
implying that the common region of oscillating ether that physically
connects the steady-state \emph{zbw} oscillations and translational
Brownian motions of the particles must, in some sense, extend over
macroscopic distances in 3-D space. \footnote{More precisely, we have in mind that the regions of oscillating ether
immediately surrounding each particle will directly drive their respective
\emph{zbw} oscillations, while the ether in between the two particles
will nonlocally encode physical correlations between the immediate
regions of ether surrounding each particle, in a way consistent with
the conservative diffusion constraint $J=extremal$, even if the two
particles are macroscopically separated in 3-D space. Of course, the
exact details of how Nelson's ether (under the amendments 1-3) would
accomplish this await the construction of a physical model for it.} That is, if we view the ether as a medium in 3-D space and not in
3N-dimensional configuration space, even though (3.174-175) are non-separable
fields on configuration space. This last (TELB) view is indeed the
one we take, since, as we stated earlier, we think it's the most conceptually
plausible one among the present options.

To be sure, the interpretive issues we discussed in section 2 for
NYSM apply just as well to ZSM. To review the options, one might view
the mathematical non-factorizability of (3.174-175) as indicating
that the oscillating ether medium lives in 3N-dimensional configuration
space instead of 3D-space. Or, one might view the configuration space
representation (3.174-175) as a mathematically convenient encoding
of a much more complicated 3-D space representation of the joint phase
field and joint osmotic potential of the particles, making it conceptually
unproblematic to imagine the oscillating ether as a medium in 3-D
space. In the former case, we then have the options of: (i) viewing
the \emph{zbw} particles as living in 3-D space, and positing a law-like
dynamical relationship between the particles in 3-D space and the
oscillating ether in 3N-dimensional configuration space; and (ii)
viewing the particles in 3-D space as a fictitious representation
of a single real \emph{zbw} particle living at a definite point in
3N-dimensional configuration space, and taking the physical interactions
between this particle and the ether as occuring in the configuration
space. In the latter case, since both the particles and the oscillating
ether would live in 3-D space (the TELB view), their physical interactions
would occur there as well.

As with NYSM, the drawback of option 1 in the former case is that
it seems mysterious and implausible that two sets of beables, living
in completely independent physical spaces, should have a law-like
dynamical relationship between them (i.e., why should oscillations
of an ether medium in a 3N-dimensional configuration space `drive'
the steady-state \emph{zbw} oscillations of particles at definite
positions in a 3-D space?). The drawback of option 2 is that while
it's conceptually more plausible how oscillations of the ether could
drive the steady-state oscillations of the \emph{zbw} particles (since
they both live in the same physical space), it would then be necessary
to employ a complicated philosophical functionalist analysis of the
\emph{N}-particle quantum Hamilton-Jacobi equation, in order to derive
the image of \emph{N} \emph{zbw} particles moving in 3-D space; and
we would be in the seemingly paradoxical situation of having \emph{derived}
the \emph{N}-particle QHJ equation from an ensemble-averaged action
defined from \emph{N} contributions, under the starting hypothesis
that there really are \emph{N} particles diffusing in a 3-D space.
Of course, the main shortcoming of the TELB view is that it remains
speculative at the moment, since no such formulation of NYSM or ZSM
exists at present; but it is not implausible that such a formulation
can be constructed, and we have already sketched in section 2 one
way it could be done. Thus we assume, provisionally, that a TELB formulation
of ZSM exists and awaits discovery (unless shown otherwise), and base
our interpretation of the beables of ZSM on this provisional assumption.

It is interesting to observe that the existence of entangled solutions
such as (3.174-175) is a consequence of four physical constraints
we've used in our construction of ZSM: (i) time-reversal invariance
of the probability density via (3.145); (ii) the conservative diffusion
constraint on the ensemble-averaged \emph{N}-particle action (3.151);
(iii) single-valuedness of the conservative-diffusion-constrained,
time-symmetrized, joint phase field (up to an integer multiple of
$2\pi$) via (3.170); and (iv) the requirement that the particles,
under the time-evolution constraints (3.143-170), satisfy a natural
notion of identicality under exchange of their coordinates, thereby
yielding the symmetrization postulates associated with bosons and
fermions \cite{Bacciagaluppi2003} (though let us be clear that for
classically interacting non-identical particles, entangled solutions
can also arise by virtue of the previous three physical constraints).
So ZSM offers a novel way to understand the emergence of continuous-variable
entanglement nonlocality in terms of deeper `subquantum' principles.
One could then study how relaxing these physical constraints might
lead to experimentally testable differences from the entangled solutions
of the \emph{N}-particle Schr\"{o}dinger
equation, in experimental tests of Bell inequalities for continuous-variable
correlations \cite{Cavalcanti2007}.

Now, since we wish to view the particles as living at definite points
in 3-D space, and their \emph{zbw} oscillations as occuring in 3-D
space, we should find a way of constructing the phase field associated
with the \emph{i}-th particle's \emph{zbw} oscillation in 3-D space.
To do this, we can construct the conditional phase field and conditional
osmotic potential field for the \emph{i}-th particle from the solutions
of (3.165) and (3.145) using (3.170). For generality and to avoid
redundancy, we will give these constructions for the case of classically-interacting
\emph{zbw} particles in the next section.

\subsection{Classical fields interacting with \emph{zbw} particles}

For completeness, we will describe \emph{zbw} particles interacting
with each other through a scalar (Coulomb) potential and with external
vector and scalar potentials. For simplicity, we will restrict our
attention to only two \emph{zbw} particles.

We begin by supposing again that each particle undergoes a steady-state
\emph{zbw} oscillation in its IMFTRF, and that each \emph{zbw} particle
carries charge, $e_{i}$, making them classical charged harmonic oscillators
of some identical type. \footnote{Which we subject again to the hypothetical constraint of no electromagnetic
radiation emitted when there is no translational motion; or the constraint
that the oscillation of the charge is radially symmetric so that there
is no net energy radiated; or, if the ether turns out to be electromagnetic
in nature as Nelson suggested \cite{Nelson1985}, then that the steady-state
\emph{zbw} oscillations of the particles are due to a balancing between
the random-phase-averaged electromagnetic energy absorbed via the
driven oscillations of the particle charges, and the random-phase-averaged
electromagnetic energy radiated back to the ether by the particles.} So the classical interaction between the particles is described by
the interaction potential $\Phi_{c}^{int}(\mathbf{q}_{i}(t),\mathbf{q}_{j}(t))=\frac{1}{2}\sum_{j=1}^{2(j\neq i)}\frac{e_{j}}{|\mathbf{q}_{i}(t)-\mathbf{q}_{j}(t)|}$,
under the point-like interaction assumption, $|\mathbf{q}_{1}(t)-\mathbf{q}_{2}(t)|\gg\lambda_{c}$.
In addition, we allow coupling to an external electric potential $\Phi_{i}^{ext}(\mathbf{q}_{i}(t),t)$
(again making the point-like approximation $|\mathbf{q}_{i}|\gg\lambda_{c}$).
Then the mean forward, steady-state, joint phase change of the particles
in the lab frame is given by
\begin{equation}
\begin{aligned}\delta\bar{\theta}_{+}(\mathbf{q}_{1}(t),\mathbf{q}_{2}(t),t) & =\mathrm{E_{t}}\left[\sum_{i=1}^{2}\left(\omega_{ic}+\omega_{ci}\frac{\mathbf{b}_{i}^{2}}{2c^{2}}+\omega_{ci}\left(\frac{e_{i}\Phi_{i}^{ext}}{m_{i}c^{2}}+\frac{e_{i}\Phi_{c}^{int}}{m_{i}c^{2}}\right)\right)\left(\delta t-\sum_{i=1}^{2}\frac{\mathbf{b}_{0i}}{c^{2}}\cdot\delta\mathbf{q}_{i+}(t)\right)\right]\\
 & =\mathrm{E}_{t}\left[\sum_{i=1}^{2}\left(\omega_{ic}+\omega_{ci}\frac{\mathbf{b}_{i}^{2}}{2c^{2}}+\omega_{ci}\left(\frac{e_{i}\Phi_{i}^{ext}}{m_{i}c^{2}}+\frac{e_{i}\Phi_{c}^{int}}{m_{i}c^{2}}\right)\right)\delta t-\sum_{i=1}^{2}\omega_{ci}\left(\frac{\mathbf{b}_{i}}{c^{2}}\right)\cdot\delta\mathbf{q}_{i+}(t)\right]\\
 & =\frac{1}{\hbar}\mathrm{E}_{t}\left[\left(\sum_{i=1}^{2}m_{i}c^{2}+\sum_{i=1}^{2}\frac{m_{i}\mathbf{b}_{i}^{2}}{2}+\sum_{i=1}^{2}V_{i}^{ext}+V_{c}^{int}\right)\delta t-\sum_{i=1}^{2}m_{i}\mathbf{b}_{i}\cdot\delta\mathbf{q}_{i+}(t)\right].
\end{aligned}
\end{equation}
The mean backward joint phase change $\delta\bar{\theta}_{-}$
differs by $\mathbf{b}_{i}\rightarrow-\mathbf{b}_{i*}$, $\delta t\rightarrow-\delta t$,
and $\delta\mathbf{q}_{i+}(t)\rightarrow\delta\mathbf{q}_{i-}(t)$.
Incorporating coupling to an external vector potential, we then have
$\mathbf{b}_{i}=\mathbf{b}'_{i}-(e_{i}/m_{i}c)\mathbf{A}_{i}^{ext}$
and $\mathbf{b}_{i*}=\mathbf{b}'_{i*}-(e_{i}/m_{i}c)\mathbf{A}_{i}^{ext}$.
When $|\mathbf{q}_{1}(t)-\mathbf{q}_{2}(t)|$ becomes sufficiently
great that $V_{c}^{int}$ is negligible, (3.176) reduces to an effectively
separable sum of the forward steady-state phase changes associated
with particle 1 and particle 2, respectively. (Effectively, because
the ether will of course still physically correlate the phase changes
of the particles, even if negligibly.) We can then write 
\begin{equation}
\delta\bar{\theta}_{+}(q(t),t)=\frac{1}{\hbar}\mathrm{E}_{t}\left[E_{joint+}(q(t),Dq(t),t)\delta t-\sum_{i=1}^{N}m_{i}\mathbf{b}'_{i}(q(t),t)\cdot\delta\mathbf{q}_{i+}(t)\right],
\end{equation}
where
\begin{equation}
E_{joint+}=\sum_{i=1}^{2}m_{i}c^{2}+\sum_{i=1}^{2}\frac{m_{i}\mathbf{b}_{i}^{2}}{2}+\sum_{i=1}^{2}V_{i}^{ext}+V_{c}^{int}.
\end{equation}
Correspondingly,
\begin{equation}
\delta\bar{\theta}_{-}(q(t),t)=\frac{1}{\hbar}\mathrm{E}_{t}\left[-E_{joint-}(q(t),D_{*}q(t),t)\delta t+\sum_{i=1}^{2}m_{i}\mathbf{b}'_{i*}(q(t),t)\cdot\delta\mathbf{q}_{i-}(t)\right],
\end{equation}
where
\begin{equation}
E_{joint-}=\sum_{i=1}^{2}m_{i}c^{2}+\sum_{i=1}^{2}\frac{m_{i}\mathbf{b}_{i*}^{2}}{2}+\sum_{i=1}^{2}V_{i}^{ext}+V_{c}^{int}.
\end{equation}

As in the classical case, we can readily construct from (3.177) or
(3.179) the corresponding mean forward or backward conditional phase
change for particle 1 (particle 2), in the lab frame or IMFTRF/IMBTRF
of particle 1 (particle 2). Likewise for the backward conditional
phase change for particle 1 (particle 2). 

Because each \emph{zbw} particle is essentially a harmonic oscillator,
when $V_{c}^{int}\approx0$, each particle has its own well-defined
forward/backward steady-state phase at each point along its mean forward/backward
space-time trajectory. Consistency with this fact entails that for
$V_{c}^{int}>0$, the forward/backward steady-state joint phase must
be a well-defined function of the mean forward/backward space-time
trajectories of\emph{ }both particles (since we again posit that both
particles remain harmonic oscillators even when physically coupled
by $V_{c}^{int}$). Furthermore, we note that at this stage (3.177)
and (3.179) are independent of one another. Accordingly, for a closed
loop \emph{L} along which each particle can be physically or virtually
displaced, the forward steady-state joint phase in the lab frame will
satisfy
\begin{equation}
\oint_{L}\delta\bar{\theta}_{+}=2\pi n,
\end{equation}
and likewise for the steady-state backward joint phase. It also follows
from (3.181) that
\begin{equation}
\oint_{L}\delta_{1}\bar{\theta}_{+}=2\pi n,
\end{equation}
where the closed-loop integral here keeps the coordinate of particle
2 fixed while particle 1 is displaced along \emph{L}.

In the lab frame, the forward and backward stochastic differential
equations for the translational motion are then given by
\begin{equation}
d\mathbf{q}_{i}(t)=\left(\mathbf{b}_{i}'(q(t),t)-\frac{e_{i}}{m_{i}c}\mathbf{A}_{i}^{ext}(q(t),t)\right)dt+d\mathbf{W}_{i}(t),
\end{equation}
and
\begin{equation}
d\mathbf{q}_{i}(t)=\left(\mathbf{b}'_{i*}(q(t),t)-\frac{e_{i}}{m_{i}c}\mathbf{A}_{i}^{ext}(q(t),t)\right)dt+d\mathbf{W}_{i*}(t),
\end{equation}
with corresponding Fokker-Planck equations
\begin{equation}
\frac{\partial\rho(q,t)}{\partial t}=-\sum_{i=1}^{2}\nabla_{i}\cdot\left[\left(\mathbf{b}_{i}'(q,t)-\frac{e_{i}}{m_{i}c}\mathbf{A}_{i}^{ext}(q,t)\right)\rho(q,t)\right]+\sum_{i=1}^{2}\frac{\hbar}{2m_{i}}\nabla_{i}^{2}\rho(q,t),
\end{equation}
and
\begin{equation}
\frac{\partial\rho(q,t)}{\partial t}=-\sum_{i=1}^{2}\nabla_{i}\cdot\left[\left(\mathbf{b}'_{i*}(q,t)-\frac{e_{i}}{m_{i}c}\mathbf{A}_{i}^{ext}(q,t)\right)\rho(q,t)\right]-\sum_{i=1}^{2}\frac{\hbar}{2m_{i}}\nabla_{i}^{2}\rho(q,t).
\end{equation}
Restricting to simultaneous solutions of (3.185-186) leads us to the
modified current velocity
\begin{equation}
\mathbf{v}_{i}\coloneqq\frac{1}{2}\left[\mathbf{b}_{i}+\mathbf{b}_{i*}\right]=\frac{\nabla_{i}S}{m_{i}}-\frac{e_{i}}{m_{i}c}\mathbf{A}_{i}^{ext},
\end{equation}
and the usual osmotic velocity
\begin{equation}
\mathbf{u}_{i}\coloneqq\frac{1}{2}\left[\mathbf{b}_{i}-\mathbf{b}_{i*}\right]=\frac{\hbar}{2m_{i}}\frac{\nabla_{i}\rho}{\rho}.
\end{equation}
Then (3.185) and (3.186) reduce to
\begin{equation}
\frac{\partial\rho}{\partial t}=-\sum_{i=1}^{2}\nabla_{i}\cdot\left[\left(\frac{\nabla_{i}S}{m_{i}}-\frac{e_{i}}{m_{i}c}\mathbf{A}_{i}^{ext}\right)\rho\right],
\end{equation}
where $\mathbf{b}_{i}'=\mathbf{v}_{i}'+\mathbf{u}_{i}$ and $\mathbf{b}'_{i*}=\mathbf{v}_{i}'-\mathbf{u}_{i}$
since $\mathbf{v}_{i}'=\mathbf{v}_{i}+(e_{i}/m_{i}c)\mathbf{A}_{i}^{ext}$,
and $\mathbf{b}_{i}=\mathbf{b}_{i}'-(e_{i}/m_{i}c)\mathbf{A}_{i}^{ext}$,
and $\mathbf{b}_{i*}=\mathbf{b}'_{i*}-(e_{i}/m_{i}c)\mathbf{A}_{i}^{ext}$.
The solution of (3.189) is just 
\begin{equation}
\rho(q,t)=\rho_{0}(q_{0})exp[-\int_{0}^{t}\left[\sum_{i}^{2}\nabla_{i}\cdot\mathbf{v}_{i}\right]dt'=\rho_{0}(q_{0})exp[-\int_{0}^{t}\left[\sum_{i}^{2}\left(\frac{\nabla_{i}^{2}S}{m_{i}}-\frac{e_{i}}{m_{i}c}\nabla_{i}\cdot\mathbf{A}_{i}^{ext}\right)\right]dt'.
\end{equation}
Here again we postulate an osmotic potential to which each particle
couples via $R(q(t),t)=\mu U(q(t),t)$, which imparts momentum $\nabla_{i}R(q,t)|_{\mathbf{q}_{j}=\mathbf{q}_{j}(t)}$
that is counter-balanced by the osmotic impulse $\left(\hbar/2m_{i}\right)\nabla_{i}\ln[n(q,t)]|_{\mathbf{q}_{j}=\mathbf{q}_{j}(t)}$,
giving the equilibrium condition $\nabla_{i}R/m_{i}=\left(\hbar/2m_{i}\right)\nabla_{i}\rho/\rho$.
Thus $\rho=e^{2R/\hbar}$ for all times and
\begin{equation}
R(q,t)=R_{0}(q_{0})-(\hbar/2)\int_{0}^{t}\left[\sum_{i}^{2}\left(\frac{\nabla_{i}^{2}S}{m_{i}}-\frac{e_{i}}{m_{i}c}\nabla_{i}\cdot\mathbf{A}_{i}^{ext}\right)\right]dt',
\end{equation}
where $S$ will end up playing the role of the conservative-diffusion-constrained,
time-symmetrized, steady-state joint phase of the \emph{zbw} particles.

As in the free particle case, we will obtain the 2nd-order time-symmetric
mean dynamics for the \emph{zbw} particles from Yasue's variational
principle. 

Since (3.177) and (3.179) correspond to the same (lab) frame and are
no longer independent because of (3.187-188), it is natural to define
the time-symmetrized steady-state joint \emph{zbw} particle phase
in the lab frame by taking the difference between (3.177) and (3.179)
(under the replacements $\mathbf{b}_{i}\rightarrow\mathbf{b}'_{i}$
and $\mathbf{b}_{i*}\rightarrow\mathbf{b}'_{i*}$ in the mean forward
and mean backward momentum contributions to the phases):
\begin{equation}
\begin{aligned}d\bar{\theta}(q(t),t) & \coloneqq\frac{1}{2}\left[d\bar{\theta}_{+}(q(t),t)-d\bar{\theta}_{-}(q(t),t)\right]\\
 & =\frac{1}{\hbar}\mathrm{E}_{t}\left[\sum_{i=1}^{2}\left(E_{i}(q(t),D\mathbf{q}_{i}(t),D_{*}\mathbf{q}_{i}(t),t)dt-\frac{m_{i}}{2}\left(\mathbf{b}'_{i}\cdot d\mathbf{q}_{i+}(t)+\mathbf{b}'_{i*}\cdot d\mathbf{q}_{i-}(t)\right)\right)+\phi\right]\\
 & =\frac{1}{\hbar}\mathrm{E}_{t}\left[\sum_{i=1}^{2}E_{i}dt-\sum_{i=1}^{2}\frac{m_{i}}{2}\left(\mathbf{b}'_{i}\cdot\frac{d\mathbf{q}_{i+}(t)}{dt}+\mathbf{b}'_{i*}\cdot\frac{d\mathbf{q}_{i-}(t)}{dt}\right)dt\right]+\phi\\
 & =\frac{1}{\hbar}\mathrm{E}_{t}\left[\left(E_{joint}-\sum_{i=1}^{2}\frac{m_{i}}{2}\left(\mathbf{b}'_{i}\cdot\frac{d\mathbf{q}_{i+}(t)}{dt}+\mathbf{b}'_{i*}\cdot\frac{d\mathbf{q}_{i-}(t)}{dt}\right)\right)dt\right]+\phi\\
 & =\frac{1}{\hbar}\mathrm{E}_{t}\left[\left(E_{joint}-\sum_{i=1}^{2}\frac{m_{i}}{2}\left(\mathbf{b}'_{i}\cdot\mathbf{b}_{i}+\mathbf{b}'_{i*}\cdot\mathbf{b}_{i*}\right)\right)dt\right]+\phi\\
 & =\frac{1}{\hbar}\mathrm{E}_{t}\left[\left(E_{joint}-\sum_{i=1}^{2}\frac{m_{i}}{2}\left(\mathbf{b}{}_{i}^{2}+\frac{e_{i}}{m_{i}c}\mathbf{b}_{i}\cdot\mathbf{A}_{i}^{ext}+\mathbf{b}_{i*}^{2}+\frac{e_{i}}{m_{i}c}\mathbf{b}_{i*}\cdot\mathbf{A}_{i}^{ext}\right)\right)dt\right]+\phi\\
 & =\frac{1}{\hbar}\mathrm{E}_{t}\left[\left(E_{joint}-\sum_{i=1}^{2}\frac{m_{i}}{2}\left(\mathbf{b}_{i}{}^{2}+\mathbf{b}_{i*}^{2}\right)-\sum_{i=1}^{2}\frac{e_{i}}{c}\left(\frac{\mathbf{b}_{i}+\mathbf{b}_{i*}}{2}\right)\cdot\mathbf{A}_{i}^{ext}\right)dt\right]+\phi\\
 & =\frac{1}{\hbar}\mathrm{E}_{t}\left[\left(E_{joint}-\sum_{i=1}^{2}\left(m_{i}\mathbf{v}_{i}\cdot\mathbf{v}_{i}+m_{i}\mathbf{u}_{i}\cdot\mathbf{u}_{i}\right)-\sum_{i=1}^{2}\frac{e_{i}}{c}\mathbf{v}_{i}\cdot\mathbf{A}_{i}^{ext}\right)dt\right]+\phi\\
 & =\frac{1}{\hbar}\mathrm{E}_{t}\left[\left(\sum_{i=1}^{2}\left(m_{i}c^{2}-\frac{1}{2}m_{i}\mathbf{v}_{i}^{2}-\frac{1}{2}m_{i}\mathbf{u}_{i}^{2}-\frac{e_{i}}{c}\mathbf{v}_{i}\cdot\mathbf{A}_{i}^{ext}\right)+\sum_{i=1}^{2}V_{i}^{ext}+V_{c}^{int}\right)dt\right]+\phi.
\end{aligned}
\end{equation}
where $\phi=\sum_{i=1}^{2}\left(\phi_{i+}-\phi_{i-}\right)$ and,
using (3.178) and (3.180), along with the constraints (3.187) and
(3.188), we have
\begin{equation}
\begin{aligned}E_{joint} & \coloneqq\sum_{i=1}^{2}E_{i}\\
 & \coloneqq\frac{1}{2}\left[E_{joint+}+E_{joint-}\right]\\
 & =\sum_{i=1}^{2}m_{i}c^{2}+\sum_{i=1}^{2}\frac{1}{2}\left[\frac{1}{2}m_{i}\mathbf{b}_{i}{}^{2}+\frac{1}{2}m\mathbf{b}_{i*}{}^{2}\right]+\sum_{i=1}^{2}V_{i}^{ext}+V_{c}^{int}\\
 & =\sum_{i=1}^{2}m_{i}c^{2}+\sum_{i=1}^{2}\left[\frac{1}{2}m_{i}\mathbf{v}{}_{i}{}^{2}+\frac{1}{2}m_{i}\mathbf{u}_{i}^{2}\right]+\sum_{i=1}^{2}V_{i}^{ext}+V_{c}^{int}.
\end{aligned}
\end{equation}
As in the free particle case, the consistency of our theory requires
that (3.192) satisfies 
\begin{equation}
\oint_{L}\delta\bar{\theta}=2\pi n.
\end{equation}
Otherwise we would contradict our hypothesis that, after imposing
(3.187-188), the \emph{zbw} particles have a well-defined, unique,
steady-state joint phase at the 3-space locations that they can occupy
at a time \emph{t}.

Defining the steady-state joint phase-principal function
\begin{equation}
I=-\hbar\bar{\theta}=\mathrm{E}\left[\int_{t_{I}}^{t}\left(\sum_{i=1}^{2}\left[\frac{1}{2}m_{i}\mathbf{v}_{i}^{2}+\frac{1}{2}m_{i}\mathbf{u}_{i}^{2}+\frac{e_{i}}{c}\mathbf{v}_{i}\cdot\mathbf{A}_{i}^{ext}-m_{i}c^{2}-V_{i}^{ext}\right]-V_{c}^{int}\right)dt'\left|\mathbf{q}_{j}(t)\right.\right]-\hbar\phi,
\end{equation}
allows us to define the joint phase-action
\begin{equation}
\begin{aligned}J & =I_{IF}=\mathrm{E}\left[\int_{t_{I}}^{t_{F}}\left(\sum_{i=1}^{2}\left[\frac{1}{2}m_{i}\mathbf{v}_{i}^{2}+\frac{1}{2}m_{i}\mathbf{u}_{i}^{2}+\frac{e_{i}}{c}\mathbf{v}_{i}\cdot\mathbf{A}_{i}^{ext}-m_{i}c^{2}-V_{i}^{ext}\right]-V_{c}^{int}\right)dt'\right]-\hbar\phi.\end{aligned}
\end{equation}
Equation (3.196) is just Eq. (3.18) in section 2, with the addition
of the rest-energy terms and the time-symmetrized initial joint phase
constant $\phi$.

From the second to last line of (3.192), we can use Fubini's theorem
in stochastic calculus to write the cumulative, time-symmetric, steady-state
joint phase at time \emph{t} as
\begin{equation}
\begin{aligned}\bar{\theta}(q(t),t) & =\frac{1}{\hbar}\mathrm{E}\left[\int_{t_{I}}^{t}\left(E_{joint}-\sum_{i=1}^{2}\left(m_{i}\mathbf{v}_{i}\cdot\mathbf{v}_{i}+m_{i}\mathbf{u}_{i}\cdot\mathbf{u}_{i}\right)-\sum_{i=1}^{2}\frac{e_{i}}{c}\mathbf{v}_{i}\cdot\mathbf{A}_{i}^{ext}\right)dt'\left|\mathbf{q}_{j}(t)\right.\right]+\phi\\
 & =\frac{1}{\hbar}\mathrm{E}\left[\int_{t_{I}}^{t}\left(\left(E_{joint}-\sum_{i=1}^{2}m_{i}\mathbf{u}_{i}\cdot\mathbf{u}_{i}\right)-\sum_{i=1}^{2}m_{i}\mathbf{v}_{i}\cdot\mathbf{v}_{i}-\sum_{i=1}^{2}\frac{e_{i}}{c}\mathbf{v}_{i}\cdot\mathbf{A}_{i}^{ext}\right)dt'\left|\mathbf{q}_{j}(t)\right.\right]+\phi\\
 & =\frac{1}{\hbar}\mathrm{E}\left[\int_{t_{I}}^{t}\left(H-\sum_{i=1}^{2}m_{i}\mathbf{v}_{i}\cdot\mathbf{v}_{i}-\sum_{i=1}^{2}\frac{e_{i}}{c}\mathbf{v}_{i}\cdot\mathbf{A}_{i}^{ext}\right)dt'\left|\mathbf{q}_{j}(t)\right.\right]+\phi\\
 & =\frac{1}{\hbar}\mathrm{E}\left[\int_{t_{I}}^{t}\left(H-\sum_{i=1}^{2}\left(\mathbf{p}_{i}+\frac{e_{i}}{c}\mathbf{A}_{i}^{ext}\right)\cdot\mathbf{v}_{i}\right)dt'\left|\mathbf{q}_{j}(t)\right.\right]+\phi\\
 & =\frac{1}{\hbar}\mathrm{E}\left[\int_{t_{I}}^{t}Hdt'-\sum_{i=1}^{2}\int_{\mathbf{q}_{i}(t_{I})}^{\mathbf{q}_{i}(t)}\left(\mathbf{p}_{i}+\frac{e_{i}}{c}\mathbf{A}_{i}^{ext}\right)\cdot\mathrm{D}\mathbf{q}_{i}(t')\left|\mathbf{q}_{j}(t)\right.\right]+\phi,
\end{aligned}
\end{equation}
where 
\begin{equation}
H\coloneqq E_{joint}-\sum_{i=1}^{2}m_{i}\mathbf{u}_{i}\cdot\mathbf{u}_{i}=\sum_{i=1}^{2}m_{i}c^{2}+\sum_{i=1}^{2}\left[\frac{1}{2}m_{i}\mathbf{v}{}_{i}{}^{2}-\frac{1}{2}m_{i}\mathbf{u}_{i}^{2}\right]+\sum_{i=1}^{2}V_{i}^{ext}+V_{c}^{int}.
\end{equation}
Now, consider an integral curve $\mathbf{Q}_{i}(t)$ obtained from
\begin{equation}
m_{i}\frac{d\mathbf{Q}_{i}(t)}{dt}=m_{i}\mathbf{v}_{i}(Q(t),t)=\mathbf{p}_{i}(Q(t),t)=\nabla_{i}S(q,t)|_{\mathbf{q}_{j}=\mathbf{Q}_{j}(t)}.
\end{equation}
Then we can replace the functional dependence of (3.197) on $q(t)$
by $Q(t)$, obtaining 
\begin{equation}
\begin{aligned}\bar{\theta}(Q(t),t) & =\frac{1}{\hbar}\int_{t_{I}}^{t}\left[H-\sum_{i=1}^{2}\left(m_{i}\mathbf{v}_{i}+\frac{e_{i}}{c}\mathbf{A}_{i}^{ext}\right)\cdot\frac{d\mathbf{Q}_{i}(t')}{dt'}\right]dt'+\phi\\
 & =\frac{1}{\hbar}\left[\int_{t_{I}}^{t}Hdt'-\sum_{i=1}^{2}\int_{\mathbf{Q}_{i}(t_{I})}^{\mathbf{Q}_{i}(t)}\left(\mathbf{p}_{i}+\frac{e_{i}}{c}\mathbf{A}_{i}^{ext}\right)\cdot d\mathbf{Q}_{i}(t')\right]+\phi,
\end{aligned}
\end{equation}
where we've dropped the conditional expectation. 

The total differential of the left hand side of (3.200) gives
\begin{equation}
d\bar{\theta}=\sum_{i=1}^{2}\nabla_{i}\bar{\theta}|_{\mathbf{q}_{j}=\mathbf{Q}_{j}(t)}d\mathbf{Q}_{i}(t)+\partial_{t}\bar{\theta}|_{\mathbf{q}_{j}=\mathbf{Q}_{j}(t)}dt,
\end{equation}
hence, 
\begin{equation}
\mathbf{p}_{i}(Q(t),t)+\frac{e_{i}}{c}\mathbf{A}_{i}^{ext}(\mathbf{Q}_{i}(t),t)=-\hbar\nabla_{i}\bar{\theta}|_{\mathbf{q}_{j}=\mathbf{Q}_{j}(t)}=\nabla_{i}S|_{\mathbf{q}_{j}=\mathbf{Q}_{j}(t)}.
\end{equation}
Thus the \emph{i}-th current velocity in the lab frame, plus the correction
due to the \emph{i}-th external vector potential, corresponds the
gradient of the time-symmetrized steady-state joint phase at the location
of the \emph{i}-th \emph{zbw} particle, and $S$ can again be identified
with the cumulative, time-symmetric, steady-state joint phase function
of the \emph{zbw} particles in the lab frame. Along with 
\begin{equation}
H(Q(t),t)=\hbar\partial_{t}\bar{\theta}|_{\mathbf{q}_{j}=\mathbf{Q}_{j}(t)}=-\partial_{t}S|_{\mathbf{q}_{j}=\mathbf{Q}_{j}(t)},
\end{equation}
it follows that 
\begin{equation}
\begin{aligned}S(Q(t),t) & =\sum_{i=1}^{2}\int_{\mathbf{Q}_{i}(t_{I})}^{\mathbf{Q}_{i}(t)}\left(\mathbf{p}_{i}+\frac{e_{i}}{c}\mathbf{A}_{i}^{ext}\right)\cdot d\mathbf{Q}_{i}(t')-\int_{t_{I}}^{t}Hdt'-\hbar\phi\\
 & =\int_{t_{I}}^{t}\left\{ \sum_{i=1}^{2}\left[\frac{1}{2}m_{i}\mathbf{v}_{i}^{2}+\frac{1}{2}m_{i}\mathbf{u}_{i}^{2}+\frac{e_{i}}{c}\mathbf{v}_{i}\cdot\mathbf{A}_{i}^{ext}-m_{i}c^{2}-V_{i}^{ext}\right]-V_{c}^{int}\right\} dt'-\hbar\phi=I(Q(t),t).
\end{aligned}
\end{equation}

The restriction to simultaneous solutions of (3.185-186) means that
the IMFTRF and the IMBTRF of the \emph{i}-th zbw particle will not
coincide since $\mathbf{b}_{i}=\mathbf{v}_{i}+\mathbf{u}_{i}=0$ will
generally not entail $\mathbf{b}_{i*}=\mathbf{v}_{i}-\mathbf{u}_{i}=0$.
So we define an instantaneous mean (time-)symmetric rest frame (IMSTRF)
as the frame in which $\mathbf{b}_{i}+\mathbf{b}_{i*}=2\mathbf{v}_{i}=0$,
and the lab frame remains unchanged. 

Applying $J=extremal$, we have
\begin{equation}
\sum_{i=1}^{2}\frac{m_{i}}{2}\left[D_{*}D+DD_{*}\right]\mathbf{q}_{i}(t)=\sum_{i=1}^{2}e_{i}\left[-\frac{1}{c}\partial_{t}\mathbf{A}_{i}^{ext}-\nabla_{i}\left(\Phi_{i}^{ext}+\Phi_{c}^{int}\right)+\frac{\mathbf{v}_{i}}{c}\times\left(\nabla_{i}\times\mathbf{A}_{i}^{ext}\right)\right]|_{\mathbf{q}_{j}=\mathbf{q}_{j}(t)},
\end{equation}
and from the independent $\delta\mathbf{q}_{i}(t)$, the individual
equations of motion
\begin{equation}
\begin{aligned}m_{i}\mathbf{a}_{i}(q(t),t) & =\frac{m_{i}}{2}\left[D_{*}D+DD_{*}\right]\mathbf{q}_{i}(t)\\
 & =\left[-\frac{e_{i}}{c}\partial_{t}\mathbf{A}_{i}^{ext}-e_{i}\nabla_{i}\left(\Phi_{i}^{ext}+\Phi_{c}^{int}\right)+\frac{e_{i}}{c}\mathbf{v}_{i}\times\left(\nabla_{i}\times\mathbf{A}_{i}^{ext}\right)\right]|_{\mathbf{q}_{j}=\mathbf{q}_{j}(t)}.
\end{aligned}
\end{equation}
Applying the mean derivatives and using that $\mathbf{b}_{i}=\mathbf{v}_{i}+\mathbf{u}_{i}$
and $\mathbf{b}_{i*}=\mathbf{v}_{i}-\mathbf{u}_{i}$, (3.206) becomes
\begin{equation}
\begin{aligned} & \sum_{i=1}^{2}m_{i}\left[\partial_{t}\mathbf{v}_{i}+\mathbf{v}_{i}\cdot\nabla_{i}\mathbf{v}_{i}-\mathbf{u}_{i}\cdot\nabla_{i}\mathbf{u}_{i}-\frac{\hbar}{2m_{i}}\nabla_{i}^{2}\mathbf{u}_{i}\right]|_{\mathbf{q}_{j}=\mathbf{q}_{j}(t)}\\
 & =\sum_{i=1}^{2}\left[-\frac{e_{i}}{c}\partial_{t}\mathbf{A}_{i}^{ext}-e_{i}\nabla_{i}\left(\Phi_{i}^{ext}+\Phi_{c}^{int}\right)+\frac{e_{i}}{c}\mathbf{v}_{i}\times\left(\nabla_{i}\times\mathbf{A}_{i}^{ext}\right)\right]|_{\mathbf{q}_{j}=\mathbf{q}_{j}(t)}.
\end{aligned}
\end{equation}
Integrating and setting the integration constants equal to the particle
rest energies, we then get
\begin{equation}
\begin{aligned}\tilde{E}(q(t),t) & =\sum_{i=1}^{2}\tilde{E}_{i}(q(t),t)\\
 & =-\partial_{t}S(q(t),t)\\
 & =\sum_{i=1}^{2}m_{i}c^{2}+\sum_{i=1}^{2}\frac{\left[\nabla_{i}S(q,t)-\frac{e_{i}}{c}\mathbf{A}_{i}^{ext}(\mathbf{q}_{i},t)\right]^{2}}{2m_{i}}|_{\mathbf{q}_{j}=\mathbf{q}_{j}(t)}\\
 & +\sum_{i=1}^{2}e_{i}\left[\Phi_{i}^{ext}(\mathbf{q}_{i}(t),t)+\Phi_{c}^{int}(\mathbf{q}_{i}(t),\mathbf{q}_{j}(t))\right]-\sum_{i=1}^{2}\frac{\hbar^{2}}{2m_{i}}\frac{\nabla_{i}^{2}\sqrt{\rho(q,t)}}{\sqrt{\rho(q,t)}}|_{\mathbf{q}_{j}=\mathbf{q}_{j}(t)},
\end{aligned}
\end{equation}
where the $q(t)$ in (3.208) corresponds to the solution set of the
stochastic differential equations (3.183-184). Alternatively, given
the integral curves $\mathbf{Q}_{i}(t)$ of the reformulated mean
acceleration equation (3.206),
\begin{equation}
\begin{aligned}m_{i}\frac{d^{2}\mathbf{Q}_{i}(t)}{dt^{2}} & =m_{i}\left(\partial_{t}\mathbf{v}_{i}+\mathbf{v}_{i}\cdot\nabla_{i}\mathbf{v}_{i}\right)|_{\mathbf{q}_{j}=\mathbf{Q}_{j}(t)}=-\nabla_{i}\left(-\frac{\hbar^{2}}{2m_{i}}\frac{\nabla_{i}^{2}\sqrt{\rho(q,t)}}{\sqrt{\rho(q,t)}}\right)|_{\mathbf{q}_{j}=\mathbf{Q}_{j}(t)}\\
 & +e_{i}\left[-\frac{1}{c}\partial_{t}\mathbf{A}_{i}^{ext}-\nabla_{i}\left(\Phi_{i}^{ext}+\Phi_{c}^{int}\right)+\frac{\mathbf{v}_{i}}{c}\times\mathbf{B}_{ext}^{i}\right]|_{\mathbf{q}_{j}=\mathbf{Q}_{j}(t)},
\end{aligned}
\end{equation}
we can also obtain $\tilde{E}(Q(t),t)$. The general solution of (3.208),
written in terms of $Q(t)$, is given by 
\begin{equation}
\begin{aligned}S(Q(t),t) & =\sum_{i=1}^{2}\int_{\mathbf{Q}_{i}(t_{I})}^{\mathbf{Q}_{i}(t)}\mathbf{p}_{i}'\cdot d\mathbf{Q}_{i}(s)-\sum_{i=1}^{2}\int_{t_{I}}^{t}\tilde{E}_{i}ds-\hbar\phi\\
 & =\int_{t_{I}}^{t}\left\{ \sum_{i=1}^{2}\left[\frac{1}{2}m_{i}\mathbf{v}_{i}^{2}-\left(-\frac{\hbar^{2}}{2m_{i}}\frac{\nabla_{i}^{2}\sqrt{\rho}}{\sqrt{\rho}}\right)+\frac{e_{i}}{c}\mathbf{v}_{i}\cdot\mathbf{A}_{i}^{ext}-m_{i}c^{2}-V_{i}^{ext}\right]-V_{c}^{int}\right\} ds-\hbar\phi\\
 & =\int_{t_{I}}^{t}\left\{ \sum_{i=1}^{2}\left[\frac{1}{2}m_{i}\mathbf{v}_{i}^{2}+\frac{1}{2}m_{i}\mathbf{u}_{i}^{2}+\frac{\hbar}{2}\nabla_{i}\cdot\mathbf{u}_{i}+\frac{e_{i}}{c}\mathbf{v}_{i}\cdot\mathbf{A}_{i}^{ext}-m_{i}c^{2}-V_{i}^{ext}\right]-V_{c}^{int}\right\} ds-\hbar\phi,
\end{aligned}
\end{equation}
and corresponds to the conservative-diffusion-constrained, time-symmetric,
steady-state joint phase for the \emph{zbw} particles in the lab frame
(hereafter, just the steady-state joint phase), evaluated along the
time-symmetric mean trajectory of the \emph{zbw} particles, i.e.,
solutions of (3.209) for initial conditions $\mathbf{Q}_{i}(0)$,
and for $i=1,..,N$. Replacing $Q(t)$ with $q$ on both sides of
(3.210) yields the steady-state joint phase field over the possible
positions of the \emph{zbw} particles. Note the difference between
the last lines of (3.210) and (3.204) via the terms involving $\nabla_{i}\cdot\mathbf{u}_{i}$.

As in the classical model, we make the natural assumption that the
presence of classical external potentials doesn't alter the harmonic
nature of the steady-state \emph{zbw} oscillations. Moreover, since
each \emph{zbw} particle is a harmonic oscillator, each particle has
its own well-defined steady-state phase at each point along its time-symmetric
mean trajectory. Accordingly, when $V_{c}^{int}$ is not negligible,
the steady-state joint phase must be a well-defined function of the
mean trajectories of\emph{ }both particles (since we posit that all
particles remain harmonic oscillators despite having their oscillations
physically coupled through $\Phi_{c}^{int}$ and through the common
ether medium they interact with). So for a closed loop \emph{L} along
which each particle can be physically or virtually displaced, it follows
that
\begin{equation}
\oint_{L}\delta S=\sum_{i=1}^{2}\oint_{L}\left[\mathbf{p}_{i}'\cdot\delta\mathbf{Q}_{i}(t)-\tilde{E}_{i}\delta t\right]=nh,
\end{equation}
and
\begin{equation}
\sum_{i=1}^{2}\oint_{L}\mathbf{p}_{i}'\cdot\delta\mathbf{Q}_{i}(t)=\sum_{i=1}^{2}\oint_{L}\mathbf{\nabla}_{i}S|_{\mathbf{q}_{j}=\mathbf{Q}_{j}(t)}\cdot\delta\mathbf{Q}_{i}(t)=nh,
\end{equation}
for a closed loop $L$ with $\delta t=0$. For the steady-state joint
phase field $S(q,t)$, we can apply the same physical reasoning above
to each \emph{zbw} particle for each possible 3-space position that
can be occupied at time \emph{t}, thereby implying
\begin{equation}
\oint_{L}dS\left(q,t\right)=\sum_{i=1}^{2}\oint_{L}\mathbf{p}_{i}'\cdot d\mathbf{q}_{i}=\sum_{i=1}^{2}\oint_{L}\mathbf{\nabla}_{i}S\cdot d\mathbf{q}_{i}=nh.
\end{equation}
Clearly (3.212-213) implies `phase quantization' for each individual
\emph{zbw} particle, upon keeping all but the \emph{i}-th coordinate
fixed and performing the closed-loop integration. Combining (3.213),
(3.208), and (3.189), we can construct the 2-particle Schr\"{o}dinger
equation for classically interacting \emph{zbw} particles in the presence
of external fields
\begin{equation}
i\hbar\frac{\partial\psi(\mathbf{q}_{1},\mathbf{q}_{2},t)}{\partial t}=\sum_{i=1}^{2}\left[\frac{\left[-i\hbar\nabla_{i}-\frac{e_{i}}{c}\mathbf{A}_{i}^{ext}(\mathbf{q}_{i},t)\right]^{2}}{2m_{i}}+m_{i}c^{2}+e_{i}\left(\Phi_{i}^{ext}(\mathbf{q}_{i},t)+\Phi_{c}^{int}(\mathbf{q}_{i},\mathbf{q}_{j})\right)\right]\psi(\mathbf{q}_{1},\mathbf{q}_{2},t),
\end{equation}
where $\psi(\mathbf{q}_{1},\mathbf{q}_{2},t)=\sqrt{\rho(\mathbf{q}_{1},\mathbf{q}_{2},t)}e^{iS(\mathbf{q}_{1},\mathbf{q}_{2},t)/\hbar}$
is single-valued via (3.213).

We would now like to specify the evolution of the conditional steady-state
phase field and conditional probability density associated to each\emph{
zbw} particle. For simplicity, we first set $\mathbf{A}_{i}^{ext}=\Phi_{i}^{ext}=0$.
We then obtain the conditional steady-state phase field for particle
1 by writing $S(\mathbf{q}_{1},\mathbf{Q}_{2}(t),t)\eqqcolon S_{1}(\mathbf{q}_{1},t)$.
Taking the total time derivative gives 
\begin{equation}
\partial_{t}S_{1}(\mathbf{q}_{1},t)=\partial_{t}S(\mathbf{q}_{1},\mathbf{q}_{2},t)|_{\mathbf{q}_{2}=\mathbf{Q}_{2}(t)}+\frac{d\mathbf{Q}_{2}(t)}{dt}\cdot\nabla_{2}S(\mathbf{q}_{1},\mathbf{q}_{2},t)|_{\mathbf{q}_{2}=\mathbf{Q}_{2}(t)},
\end{equation}
where the conditional velocities 
\begin{equation}
\mathbf{v}_{1}(\mathbf{q}_{1},t)|_{\mathbf{q}_{1}=\mathbf{Q}_{1}(t)}\coloneqq\frac{\nabla_{1}S_{1}(\mathbf{q}_{1},t)}{m_{1}}|_{\mathbf{q}_{1}=\mathbf{Q}_{1}(t)}=\frac{d\mathbf{Q}_{1}(t)}{dt},
\end{equation}
and 
\begin{equation}
\mathbf{v}_{2}(\mathbf{q}_{2},t)|_{\mathbf{q}_{2}=\mathbf{Q}_{2}(t)}\coloneqq\frac{\nabla_{2}S_{2}(\mathbf{q}_{2},t)}{m_{2}}|_{\mathbf{q}_{2}=\mathbf{Q}_{2}(t)}=\frac{d\mathbf{Q}_{2}(t)}{dt},
\end{equation}
the latter defined from the conditional steady-state phase field,
$S_{2}(\mathbf{q}_{2},t)$, for particle 2. Likewise, for the conditional
density for particle 1, $\rho(\mathbf{q}_{1},\mathbf{Q}_{2}(t),t)\eqqcolon\rho_{1}(\mathbf{q}_{1},t)$
and 
\begin{equation}
\partial_{t}\rho_{1}(\mathbf{q}_{1},t)=\partial_{t}\rho(\mathbf{q}_{1},\mathbf{q}_{2},t)|_{\mathbf{q}_{2}=\mathbf{Q}_{2}(t)}+\frac{d\mathbf{Q}_{2}(t)}{dt}\cdot\nabla_{2}\rho(\mathbf{q}_{1},\mathbf{q}_{2},t)|_{\mathbf{q}_{2}=\mathbf{Q}_{2}(t)}.
\end{equation}
Inserting (3.218) on the left hand side of (3.189) and adding the
corresponding term on the right hand side, we obtain the conditional
continuity equation for particle 1: 
\begin{equation}
\partial_{t}\rho_{1}=-\nabla_{1}\cdot\left[\left(\frac{\nabla_{1}S_{1}}{m_{1}}\right)\rho_{1}\right]-\nabla_{2}\cdot\left[\left(\frac{\nabla_{2}S}{m_{2}}\right)\rho\right]|_{\mathbf{q}_{2}=\mathbf{Q}_{2}(t)}+\frac{d\mathbf{Q}_{2}(t)}{dt}\cdot\nabla_{2}\rho|_{\mathbf{q}_{2}=\mathbf{Q}_{2}(t)},
\end{equation}
which implies $\rho_{1}(\mathbf{q}_{1},t)\geq0$ and (upon suitable
redefinition of $\rho_{1}(\mathbf{q}_{1},t)$) preservation of the
normalization $\int_{\mathbb{R}^{3}}\rho_{1}(\mathbf{q}_{1},0)=1$.
Similarly, inserting (3.215) into the left hand side of (3.208) and
adding the corresponding term on the right hand side, we find that
the conditional steady-state phase field for particle 1 evolves by
the conditional quantum Hamilton-Jacobi equation
\begin{equation}
\begin{aligned}-\partial_{t}S_{1} & =m_{1}c^{2}+\frac{\left(\nabla_{1}S_{1}\right)^{2}}{2m_{1}}+\frac{\left(\nabla_{2}S\right)^{2}}{2m_{2}}|_{\mathbf{q}_{2}=\mathbf{Q}_{2}(t)}-\frac{d\mathbf{Q}_{2}(t)}{dt}\cdot\nabla_{2}S|_{\mathbf{q}_{2}=\mathbf{Q}_{2}(t)}\\
 & +V_{c}^{int}(\mathbf{q}_{1},t)-\frac{\hbar^{2}}{2m_{1}}\frac{\nabla_{1}^{2}\sqrt{\rho_{1}}}{\sqrt{\rho_{1}}}-\frac{\hbar^{2}}{2m_{2}}\frac{\nabla_{2}^{2}\sqrt{\rho}}{\sqrt{\rho}}|_{\mathbf{q}_{2}=\mathbf{Q}_{2}(t)}
\end{aligned}
\end{equation}
where $V_{c}^{int}(\mathbf{q}_{1},t)$ is the `conditional interaction
potential' for particle 1. The solution of (3.219) can be verified
as 
\begin{equation}
\rho_{1}=\rho_{01}exp[-\int_{0}^{t}\left[\nabla_{1}\cdot\mathbf{v}_{1}(\mathbf{q}_{1},t)+\nabla_{2}\cdot\mathbf{v}_{2}(\mathbf{q}_{1},\mathbf{q}_{2},t)|_{\mathbf{q}_{2}=\mathbf{Q}_{2}(t)}\right]dt',
\end{equation}
from which we extract the conditional osmotic potential
\begin{equation}
R_{1}=R_{01}-(\hbar/2)\int_{0}^{t}\left[\nabla_{1}\cdot\mathbf{v}_{1}(\mathbf{q}_{1},t)+\nabla_{2}\cdot\mathbf{v}_{2}(\mathbf{q}_{1},\mathbf{q}_{2},t)|_{\mathbf{q}_{2}=\mathbf{Q}_{2}(t)}\right]dt',
\end{equation}
while the solution of (3.220) is
\begin{equation}
\begin{aligned}S_{1} & =\int_{\mathbf{Q}_{1}(t_{I})}^{\mathbf{Q}_{1}(t)}\mathbf{p}_{1}\cdot d\mathbf{Q}_{1}(t')|_{\mathbf{Q}_{1}(t)=\mathbf{q}_{1}}\\
 & -\int_{0}^{t}\left[m_{1}c^{2}+\frac{m_{1}\mathbf{v}_{1}^{2}}{2}+\frac{m_{1}\mathbf{v}_{2}^{2}}{2}-\mathbf{p}_{2}\cdot\frac{d\mathbf{Q}_{2}(t)}{dt}+V_{c}^{int}\right.\\
 & \left.+\frac{\hbar^{2}}{2m_{1}}\frac{\nabla_{1}^{2}\sqrt{\rho_{1}}}{\sqrt{\rho_{1}}}+\frac{\hbar^{2}}{2m_{2}}\frac{\nabla_{2}^{2}\sqrt{\rho}}{\sqrt{\rho}}|_{\mathbf{q}_{2}=\mathbf{Q}_{2}(t)}\right]dt'|_{\mathbf{Q}_{1}(t)=\mathbf{q}_{1}}-\hbar\phi_{1}.
\end{aligned}
\end{equation}
Hence (3.222) allows us to consistently ascribe a region of oscillating
ether in 3-D space that sources a local (i.e., in 3-D space) osmotic
potential that imparts the osmotic momentum to particle 1. Likewise,
(3.223) lets us ascribe a region of oscillating ether in 3-D space
that directly drives the steady-state \emph{zbw} oscillation of particle
1 in 3-D space. Note that when (3.223) is evaluated at $\mathbf{q}_{1}=\mathbf{Q}_{1}(t)$,
it is equivalent to $S(\mathbf{q}_{1}(t),\mathbf{Q}_{2}(t),t)-m_{2}c^{2}t+\hbar\phi_{2}$.
As in the classical model, since the conditional steady-state phase
field for particle 1 is a field over the possible positions of the
\emph{zbw} particles, it follows that
\begin{equation}
\oint_{L}\nabla_{1}S_{1}\cdot d\mathbf{q}_{1}=nh,
\end{equation}
where \emph{L} is a mathematical loop in 3-D space.

With these results in hand, the conditional forward and backward stochastic
differential equations for particle 1 can be straightforwardly obtained
by writing $\mathbf{b}_{1}=\mathbf{v}_{1}+\mathbf{u}_{1}$, $\mathbf{b}_{1*}=\mathbf{v}_{1}-\mathbf{u}_{1}$,
and inserting these expressions into (3.183) and (3.184), respectively.

Also like in the classical model, we can define the steady-state conditional
phase-action
\begin{equation}
\begin{aligned}J_{1} & =I_{1}^{IF}=\mathrm{E}\left[\int_{t_{I}}^{t_{F}}\left[m_{1}c^{2}+\frac{1}{2}m_{1}\mathbf{v}_{1}^{2}+\frac{1}{2}m_{2}\mathbf{v}_{2}^{2}+\frac{1}{2}m_{1}\mathbf{u}_{1}^{2}+\frac{1}{2}m_{2}\mathbf{u}_{2}^{2}-V_{c}^{int}\right]dt-\hbar\phi_{1}\right],\end{aligned}
\end{equation}
and then impose
\begin{equation}
J_{1}=extremal,
\end{equation}
we get the conditional mean acceleration for particle 1:
\begin{equation}
m_{1}\mathbf{a}_{1}(\mathbf{q}_{1}(t),t)=\frac{m_{1}}{2}\left[D_{*}D+DD_{*}\right]\mathbf{q}_{1}(t)=-\nabla_{1}V_{c}^{int}(\mathbf{q}_{1},\mathbf{q}_{2}(t))|_{\mathbf{q}_{1}=\mathbf{q}_{1}(t)},
\end{equation}
thus
\begin{equation}
\begin{aligned}m_{1}\frac{\mathrm{D}\mathbf{v}_{1}(\mathbf{Q}_{1}(t),t)}{\mathrm{D}t} & =\left[\partial_{t}\mathbf{p}_{1}+\mathbf{v}_{1}\cdot\nabla_{i}\mathbf{p}_{1}\right](\mathbf{q}_{1},t)|_{\mathbf{q}_{1}=\mathbf{Q}_{1}(t)}\\
 & =-\nabla_{1}\left[V_{c}^{int}(\mathbf{q}_{1},\mathbf{Q}_{2}(t))-\frac{\hbar^{2}}{2m_{1}}\frac{\nabla_{1}^{2}\sqrt{\rho_{1}(\mathbf{q}_{1},t)}}{\sqrt{\rho_{1}(\mathbf{q}_{1},t)}}\right]|_{\mathbf{q}_{1}=\mathbf{Q}_{1}(t)},
\end{aligned}
\end{equation}
and likewise for particle 2. Equation (3.228) is what we would obtain
from computing the derivatives in (3.227) for $i=1$ (modulo the external
potentials) and subtracting the $\mathbf{u}_{i}$ dependent terms
on both sides. Of course, it should be said that we cannot obtain
(3.210) simply by integrating (3.220) and the analogous expression
for particle 2, and then summing up the terms. This is because we
obtained (3.220) directly from the full configuration space fields
$S$ and $\rho$, themselves obtained from extremizing (3.196).

For particle 2, the conditional steady-state phase field, probability
density, etc., are defined analogously.

Finally, combining (3.224), (3.220), and (3.219) gives us the conditional
Schr\"{o}dinger equation for particle 1:
\begin{equation}
\begin{aligned}i\hbar\frac{\partial\psi_{1}}{\partial t} & =-\frac{\hbar^{2}}{2m_{1}}\nabla_{1}^{2}\psi_{1}-\frac{\hbar^{2}}{2m_{1}}\nabla_{2}^{2}\psi|_{\mathbf{q}_{2}=\mathbf{Q}_{2}(t)}+V_{c}^{int}(\mathbf{q}_{1},\mathbf{Q}_{2}(t))\psi_{1}\\
 & +m_{1}c^{2}\psi_{1}+i\hbar\frac{d\mathbf{Q}_{2}(t)}{dt}\cdot\nabla_{2}\psi|_{\mathbf{q}_{2}=\mathbf{Q}_{2}(t)},
\end{aligned}
\end{equation}
where $\psi_{1}(\mathbf{q}_{1},t)=\sqrt{\rho_{1}(\mathbf{q}_{1},t)}e^{iS_{1}(\mathbf{q}_{1},t)/\hbar}$
is the single-valued conditional wavefunction for particle 1, and
$d\mathbf{Q}_{2}(t)/dt=(\hbar/m_{2})\mathrm{Im}\{\nabla_{2}\ln(\psi_{2})\}|_{\mathbf{q}_{2}=\mathbf{Q}_{2}(t)}$,
where $\psi_{2}=\psi_{2}(\mathbf{q}_{2},t)$ is the conditional wavefunction
for particle 2, satisfying the analogous conditional Schr\"{o}dinger
equation. Like in the classical case, (3.229) can also be obtained
from writing 
\begin{equation}
\partial_{t}\psi_{1}(\mathbf{q}_{1},t)=\partial_{t}\psi(\mathbf{q}_{1},\mathbf{q}_{2},t)|_{\mathbf{q}_{2}=\mathbf{Q}_{2}(t)}+\frac{d\mathbf{Q}_{2}(t)}{dt}\cdot\nabla_{2}\psi(\mathbf{q}_{1},\mathbf{q}_{2},t)|_{\mathbf{q}_{2}=\mathbf{Q}_{2}(t)},
\end{equation}
inserting this on the left hand side of (3.214), adding the corresponding
term on the right hand side, and subtracting $m_{2}c^{2}\psi_{1}$
(again, modulo the external potentials).

The development of ZSM in relative coordinates is formally identical
to the case of a single \emph{zbw} particle in an external potential,
and need not be explicitly given here.

This completes the formulation of ZSM for \emph{N}-particles interacting
with classical fields.

\subsection{Remark on on close-range interactions}

Since the quantum kinetic doesn't depend on the inter-particle separation,
its presence in the equation of motion (3.209) doesn't introduce any
fundamentally new complications for the description of two-particle
scattering in ZSM. So the account we gave of two-particle scattering
in section 4.7 carries over to classically interacting particles in
ZSM.

\section{Plausibility of the Zitterbewegung Hypothesis}

Ultimately, the plausibility of our suggested answer to Wallstrom
hinges (in no particular order) on the plausibility of the \emph{zbw}
hypothesis, its incorporation into NYSM, and the generalizability
of ZSM. So we should ask if: 1) ZSM can be consistently generalized
to relativistic flat and curved spacetimes; 2) the \emph{zbw} hypothesis
can be generalized to incorporate electron spin; 3) ZSM has a conceivable
field-theoretic extension; 4) a self-consistent physical model of
the \emph{zbw} particle, Nelson's ether (suitably amended for ZSM),
and dynamical interaction between the two, can be constructed; and
5) ZSM suggests testable new predictions and/or offers novel solutions
to open problems in the foundations of quantum mechanics that justify
its mathematical and conceptual complexity (relative to other hidden
variable approaches to solving the measurement problem, such as the
dBB theory).

Can ZSM be consistently generalized to relativistic flat and curved
spacetimes? We have implicitly assumed throughout our paper that this
is possible, based on our repeated use of the next-to-leading order
approximation of the Lorentz transformation. But there is also good
reason to expect that relativistic generalizations of ZSM to flat
and curved spacetimes do exist. Stochastic mechanics based on the
Guerra-Morato variational principle has already been given a consistent
generalization to the case of relativistic spacetimes (flat and curved)
by Dohrn and Guerra \cite{Dohrn1978,Dohrn1979,Dohrn1985} as well
as Serva \cite{Serva1988}. An attempt was made by Zastawniak to give
a relativistic flat-spacetime generalization of Yasue's variational
principle \cite{Zastawniak1990}, but it seems problematic since it
doesn't address the problem of not having a normalizable spacetime
probability density when the metric is not positive-definite. Fortunately,
this problem can be resolved in the approaches of Dohrn-Guerra and
Serva, and there seems to be no obstacle in adapting Dohrn and Guerra's
methods or Serva's method to extend Yasue's variational principle
to flat and curved spacetimes (currently in progress by us). Once
done, we see no fundamental reason why a corresponding generalization
of ZSM cannot be given.

Can the \emph{zbw} hypothesis be generalized to incorporate electron
spin? It seems plausible to us that it can. As is well-known, in standard
relativistic quantum mechanics for spin-1/2 particles, the Dirac spinor
satisfying the Dirac equation implies \emph{zbw} of the corresponding
velocity operator \cite{Greiner2000}. What's more, realist versions
of relativistic quantum mechanics for spin-1/2 particles - the Bohm-Dirac
theory \cite{Holland1992,Holland1993}, the ``zig-zag'' model of
de Broglie-Bohm theory by Colin \& Wiseman \cite{Colin2011} and Struyve
\cite{Struyve2012}, and the stochastic mechanical models of the Dirac
electron by de Angelis et al. \cite{Angelis1986} and Garbaczewski
\cite{Garbaczewski1992} - all predict \emph{zbw} as a real, continuous
oscillation of the particle beable. In the de Broglie-Bohm theories,
the \emph{zbw} arises from imposing Lorentz invariance and the Dirac
spinor algebra on the dynamics of the wavefunction (described by Dirac
spinors in the Bohm-Dirac theory, or Weyl spinors in the zig-zag model),
and then using this wavefunction in the definition of the guiding
equation for the de Broglie-Bohm particle. Likewise, in the stochastic
mechanical theories, the \emph{zbw} beable arises from constructing
Nelsonian diffusion processes from the Dirac wavefunction. The description
of a physically real spin-based \emph{zbw} can also be implemented
in classical physics, namely in the Barut-Zangh\`{i} model of a classical
Dirac electron \cite{Barut1984,Barut1987,Barut1989,Barut1990}, which
turns into the usual flat-space and curved-space versions of the Dirac
equation (in the proper-time formulation) upon first-quantization
by the standard methods \cite{BarutnPavsic1987,BarutnDuru1989}. Here
it is the imposition of relativistic covariance and the Dirac spinor
algebra that leads to classical equations of motion for a massless
(non-radiating) point charge circularly orbiting a center of mass,
the former moving with speed $c$ and the latter moving translationally
with sub-luminal relativistic speeds. So it is plausible to imagine
a relativistic generalization of ZSM in which the Barut-Zangh\`{i} model
of a \emph{zbw} particle is implemented into a relativistic version
of the Nelson-Yasue diffusion process (e.g., along the lines of Dohrn
and Guerra), under the hypothesis that Nelson's ether has vorticity
that imparts to the massless point charge a mean rotational motion
of speed $c$ and angular momentum $\hbar/2$, and derive from this
spin-based \emph{zbw} a relativistic generalization of the quantization
condition, along with the Dirac equation for a double-valued Dirac
spinor wavefunction. (The approaches of de Angelis et al. and Garbaczweski
don't seem adequate for this task because they don't actually derive
the zitterbewegung and Dirac equation from Nelson-Yasue diffusions;
rather, they start from the Dirac equation and Dirac spinor wavefunction,
and show that Nelsonian diffusions can be associated to them.) The
non-relativistic limit of this ZSM theory should presumably then recover
non-relativistic ZSM for a spinning \emph{zbw} particle with angular
momentum magnitude $\hbar/2$, along with a vorticity term added to
the current velocity (as is known to arise from the non-relativistic
limit of the relativistic guiding equation under Gordon decomposition
in the Bohm-Dirac theory \cite{HollandPhilipp2003,Bacciagaluppi1999}).
Alternatively, we might try deducing a non-relativistic ZSM theory
directly from Takabayasi's non-relativistic generalization of the
Madelung fluid to spin-1/2 motion \cite{Takabayasi1983}. These tasks
remain for a future paper.

Does ZSM have a field-theoretic generalization that recovers the predictions
of relativistic quantum field theory for fermions and bosons? A generalization
of ZSM to massive scalar or spinor fields seems in-principle unproblematic,
but a generalization to massless fields (such as to describe the photon
or gluon, which have no measured rest mass) would seem, at first sight,
difficult (though not necessarily impossible \footnote{For example, we might consider introducing small rest masses for the
photon and gluon consistent with experimental bounds, which for the
photon is $<10^{-14}eV/c^{2}$ \cite{Adelberger2007} and for the
gluon $<0.0002eV/c^{2}$ \cite{Yndurain1995}, if both masses are
to be produced by the Higgs mechanism. This would, of course, change
the gauge symmetries of QED and QCD, but not in a way that can be
experimentally discerned at energy scales above these lower-bounds
\cite{GoldhaberNieto2010}.}). Another possibility is to note that one can reproduce nearly all
\footnote{The single different prediction appears to be that this Dirac sea
pilot-wave model predicts fermion number conservation, whereas the
Standard Model predicts a violation of fermion number for sufficiently
high energies (so-called anomalies of the Standard Model). To the
best of our knowledge, no evidence has been found for fermion number
violation thus far \cite{Durieux2013}. But as Colin and Struyve point
out \cite{Colin2007}, even if fermion number violation is eventually
observed, it may still be possible to model it in a Dirac sea picture.} the predictions of the Standard Model (SM) with a pilot-wave model
for point-like fermions in which the Dirac sea is taken seriously
(i.e., taken as ontological) \cite{Colin2007}. In this model, no
beables are introduced for the massless bosons, yet it recovers nearly
all the predictions of the SM. So we might try constructing a version
of relativistic ZSM for spin-1/2 particles in which the Dirac sea
for fermions is taken seriously, and check if it can recover nearly
all the predictions of the SM as well. If one insists on adding beables
for the bosons, perhaps one could adapt the approach of Nielsen et
al. \cite{Nielsen98,Habara2008}, who show how to introduce a Dirac
sea for bosons in second-quantized field theory based on massive hypermultiplets.
Finally, it seems plausible that one could make a ZSM generalization
of bosonic string theory by constructing a Nelson-Yasue version of
the model of Santos and Escobar \cite{Santos99}, who use the Guerra-Morato
variational principle to construct a stochastic mechanics of the open
bosonic string (the idea being that the open bosonic string's instantaneous-rest-frame
oscillations would play the role of the \emph{zbw}, and would be hypothesized
to be dynamically driven by resonant coupling to the ZSM version of
Nelson's ether). All this remains for future work.

Can a self-consistent dynamical model of the \emph{zbw} particle,
Nelson's ether, and the physical interaction between the two, be constructed?
We see no principled obstacle to this possibility. Furthermore, physical
models of a real classical \emph{zbw} particle have been constructed
in the context of stochastic electrodynamics (SED), by Rueda \& Cavelleri
\cite{Rueda1983}, Rueda \cite{Rueda1993,Rueda1993a}, de la Pe\~{n}a
\& Cetto \cite{Pena1996}, and Haisch \& Rueda \cite{Haisch2000}.
These models involve treating the electron as a structured object
composed of a point charge with negligible (or zero) mass, harmonically
bound to some non-charged center of mass, and driven to oscillate
at near or equal to the speed of light (i.e., Compton frequency) by
resonant modes of a classically fluctuating electromagnetic zero-point
field. Additionally, in Rueda's model \cite{Rueda1993,Rueda1993a},
not only does the classical zero-point field drive the \emph{zbw}
oscillations, but the frequency cut-off generated by the \emph{zbw}
results in a non-dissipative, (effectively) Markovian diffusion process
with diffusion coefficient $\hbar/2m$. Of course, these SED-based
approaches should be cautioned; SED is know to have difficulties as
a viable theory of quantum electrodynamical phenomena \cite{Pope2000,Genovese2007},
and it is not clear that these difficulties can be resolved (but see
\cite{Valdes-Hernandez2011,Pena2012a,Pena2012,Cetto2012,Cetto2014}
for recent counter-arguments). Furthermore, we expect that any realistic
physical model of the \emph{zbw} particle should consistently incorporate
the Higgs mechanism (or some subquantum generalization thereof) \cite{Penrose2005}
as the process by which the self-stable \emph{zbw} harmonic potential
of rest-mass $m$ is formed in the first place. Nevertheless, these
SED-based models can at least be viewed as proofs of principle that
the \emph{zbw} hypothesis can be implemented in a concrete model;
and, in a future paper, we will show how one of these SED-based models
can in fact recover the quantization condition as an effective condition.
But the task of constructing a physical model of the \emph{zbw} particle,
the ZSM version of Nelson's ether, and the physical/dynamical interaction
between the two, which also incorporates spin and can be used to recover
the Dirac/Pauli/Schr\"{o}dinger  equation, remains for future work.

Lastly, does ZSM suggest testable new predictions and/or novel solutions
to open foundational problems in quantum mechanics? We claim it does.
Since the equilibrium density $\rho=|\psi|^{2}$, ZSM's statistical
predictions in equilibrium will agree with all the statistical predictions
of non-relativistic quantum mechanics. But if $\rho\neq|\psi|^{2}$,
we should expect differences, such as position and momentum measurements
with more precision than allowed by Heisenberg's uncertainty principle
\cite{Pearle2006}. \footnote{Everything we have said here is of course also true of the dBB theory
\cite{Pearle2006}. However, in our view, a proper understanding of
the origin of randomness in the dBB theory (the `typicality' approach
of D\"{u}rr-Goldstein-Zangh\`{i} \cite{Duerr1992}) entails that the existence
of quantum nonequilibrium subsystems in the observable universe is
extremely improbable, even in the context of early universe cosmology
(for a different view, see \cite{Valentini2010}). By contrast, we
will suggest here that this limitation of the dBB theory does not
necessarily apply to ZSM.} Accordingly, it would be possible, in principle, to experimentally
detect the stochasticity of the particle trajectories, hence deviations
from the mean trajectories satisfying the quantization condition.
Under what physical conditions might we see nonequilibrium fluctuations?
The most obvious possibility is by measuring the position or momentum
of a Nelsonian particle on time-scales comparable to or shorter than
the correlation time of the ether fluctuations. For ZSM, insofar as
it's based on Nelson's white-noise diffusion process, the correlation
timescale of the fluctuations is infinitesimal because of the assumption
that the noise is white. Nelson stressed, however, that his white-noise
(Markovian) assumption was only a simplifying one \cite{Nelson1985};
so one could instead consider a colored-noise (non-Markovian) description
of conservative diffusions, to which Nelson's white-noise description
is a long-time approximation (as is the case with all other known
statistical fluctuation phenomena in nature \cite{Hanggi1995a}).
\footnote{Of course, this idea could also be explored in NYSM with the quantization
condition imposed ad-hoc. The advantage of ZSM, though, is that it
makes the idea worth taking seriously as a possibility since ZSM gives
an independent justification for the (more basic) quantization condition,
without which the stochastic mechanics approach would be neither empirically
viable nor plausible.} Then the true fluctuation timescale would be finite and one could
work out the expected experimental signatures of the nonequilibrium
dynamics on timescales comparable to some hypothetical finite correlation
time $\tau_{noise}$ (work on this is currently underway by us). In
this connection, Montina's theorem \cite{Montin2008} says that any
ontic theory compatible with the predictions of a quantum system with
Hilbert space dimensionality $k$ must contain at least $2k-2$ continuous
ontic variables, assuming that the theory has deterministic or stochastic
Markovian dynamics (i.e., a dynamics that is local in time). Correspondingly,
$2k-2$ turns out to be the minimum number of real-valued parameters
required to describe a pure quantum state. On the other hand, Montina's
theorem implies that an ontic theory with non-Markovian dynamics (i.e.,
dynamics which is nonlocal in time) could have fewer continuous ontic
variables than $2k-2$. Montina has demonstrated this in a toy model
of a single ontic variable with stochastic evolution driven by time-correlated
(colored) noise that exactly reproduces any unitary evolution of a
qubit ($\psi$ for a qubit has two degrees of freedom) \cite{Montin2011,Montin2012}.
Extrapolating the implications of Montina's theorem to stochastic
mechanics, we should expect that a non-Markovian extension of stochastic
mechanics would recover an $N$-particle wavefunction that can be
described by fewer than $2k-2$ real-valued parameters, where $k$
would be the dimensionality of the Hilbert space associated to the
$N$-particle wavefunction of Markovian stochastic mechanics (Markovian
stochastic mechanics would be the $\tau_{noise}\rightarrow0$ limit
of non-Markovian stochastic mechanics). And insofar as the $N$-particle
wavefunction can be polar decomposed into $N$-particle $R$ and $S$
fields, the $N$-particle $R$ and $S$ fields of non-Markovian ZSM
would presumably also require fewer real-valued parameters to describe
than the $N$-particle $R$ and $S$ fields of Markovian ZSM. Moreover,
since $R$ and $S$ directly reflect ontological elements of ZSM (see
sections 3 and 5.1), the reduced complexity of the $R$ and $S$ fields
of non-Markovian ZSM would imply that the ontological complexity of
non-Markovian ZSM will be less than that of Markovian ZSM. It seems
conceivable, then, that if we make a TELB \cite{Norsen2010,Norsen2014}
version of non-Markovian ZSM by decomposing the $N$-particle $R$
and $S$ fields into $N$ single-particle $R$ and $S$ fields (a
pair for each particle), we may only require a finite number of (or
perhaps zero) supplementary continuous ontic variables on 3-space
in order to encode non-local correlations arising between \emph{zbw}
particles that are classically interacting and coupling to the common
oscillating ether. If so, we would (arguably) then have a reasonably
ontologically parsimonious TELB version of ZSM. This TELB version
of ZSM would considerably strengthen the justification for our viewing
the joint \emph{zbw} phase $S$ for an $N$-particle system as the
joint phase of real physical oscillations about the actual 3-space
locations of the \emph{zbw} particles, while supporting our hypothesis
that the ether is a medium that fundamentally lives in 3-space instead
of configuration space.

\section{Comparison to Other Answers }

Several other answers to Wallstrom's criticism have been offered in
the context of stochastic mechanics \cite{Carlen1989,Wallstrom1994,Smolin06,Fritsche2009,Schmelzer(2011),Groessing2011}.
Here we briefly review and assess each approach, and compare them
to ZSM.

Smolin proposed \cite{Smolin06} that Wallstrom's criticism could
be answered by allowing discontinuities in the wavefunction - that
is, for a given multi-valued wavefunction, one could introduce discontinuities
at the multi-valued points to make it single-valued. The example he
used is stochastic mechanics on $\textrm{S}^{1}$, where he argued
that although the resultant wavefunction is not single-valued and
smooth, it is well-known that almost every wavefunction in the Hilbert
space $\mathcal{L}^{2}(\mathrm{S}^{1})$ is discontinuous at one or
many points, and yet each wavefunction is normalizable and gives well-defined
(i.e., single-valued) current velocities. Smolin's proposal seems
incomplete, however. Even if his proposal works for the multiply connected
configuration space of the unit circle, how will it work in the more
general cases of simply connected configuration spaces of dimensionality
3N? Wallstrom emphasizes, after all, that the inequivalence between
the HJM equations and Schr\"{o}dinger 's equation applies to simply connected
configuration spaces of two dimensions or greater \cite{Wallstrom1994}.
(See also \cite{Valentini2010} for a critique of Smolin's approach.)
To compare with ZSM, these concerns don't arise - the derived wavefunctions
are single-valued and smooth, and ZSM works for the general case of
simply connected 3N-dimensional configuration space.

Carlen \& Loffredo \cite{Carlen1989} considered stochastic mechanics
on $\textrm{S}^{1}$ and suggested to introduce a stochastic analogue
of the quantization condition, which they argue is related in a natural
way to the topological properties of $\textrm{S}^{1}$. They then
showed that this stochastic analogue of the quantization condition
establishes mathematical equivalence between stochastic mechanics
and quantum mechanics on $\textrm{S}^{1}$. However, the difficulty
with taking their proposal as a general answer is that it seems to
only work in the special case of $\textrm{S}^{1}$, whereas Wallstrom's
criticism applies to simply connected configuration spaces of two
dimensions or greater, as mentioned earlier.

Fritsche \& Haugk \cite{Fritsche2009} attempted to answer Wallstrom
by motivating the quantization condition from the physical requirement
that the probability density, $|\psi|^{2}$, should always be normalizable.
To accomplish this, they first required that the velocity potential,
$S$, be single-valued on a closed loop (in analogy with the definition
of a single-valued magnetic scalar potential) via jump discontinuities.
Constructing the wavefunction from this $S$ function through an approach
equivalent to Nelson's Newtonian formulation of stochastic mechanics,
they then argued that the only way $|\psi|^{2}$ can remain normalizable
for a superposition of two eigenstates is if the phase difference
between the eigenstates satisfies the quantization condition. The
main problem with their approach lies in the their non-trivial assumption
that $S$ can have jump discontinuities. As pointed out by Wallstrom
\cite{Wallstrom1989,Wallstrom1994}, allowing jump discontinuities
in $S$ implies that $\nabla\psi=\left(\frac{1}{\hbar}\right)\left(\nabla R+i\nabla S\right)\psi$
develops a singularity, which is physically inadmissible. Accordingly,
the same technical concerns we raised towards Smolin's proposal apply
here as well. We note, by contrast, that in ZSM, $\nabla S$ is always
continuous even though $S$ is in general discontinuous (e.g., at
nodal points of the probability density).

Wallstrom made the observation \cite{Wallstrom1994} that if one takes
the quantization condition as an initial condition on the current
velocity, then the time-evolution of the HJM equations will ensure
that it is valid for all future times, in analogy with Kelvin's circulation
theorem from classical fluid mechanics. So one might think to use
this as a justification for the quantization condition in the context
of the HJM equations. As he pointed out, however, this seems to require
an extreme form of fine-tuning (why should the initial condition on
the current velocity correspond exactly to the quantization condition?),
and it is not clear that this initial condition would be stable for
interacting particles. By contrast, we saw in ZSM that the \emph{zbw}
hypothesis combined with the Lorentz transformation implies the quantization
condition so that it is not the result of fine-tuning (other than
the assumption that the steady-state oscillation frequency in the
IMFTRF/IMBTRF/IMSTRF is of fixed Compton magnitude). Moreover, we
showed that in the case of classically interacting \emph{zbw} particles,
it can be plausibly argued that the quantization condition remains
stable.

Bacciagaluppi \cite{Bacciagaluppi2005} suggested that when the external
potential $V$ has time-dependence, the complement of the nodal set
of $\rho$ may become simply connected in a neighborhood of a given
time $t$. In other words, the time-dependence of $V$ may make it
possible to eliminate the nodes of $\rho$ around which a multi-valued
$S$ accumulates values other than $nh$ (because $S$ would have
to be single-valued in that neighborhood of $t$). While Bacciagaluppi's
suggestion was intended as an abstract, mathematical argument, it
is interesting to note that his proposal seems relevant to measurement
situations when the interaction of a system with a pointer apparatus
entails a time-dependent $V$; in other words, Bacciagaluppi's suggestion
might be used to argue that energy-momentum quantization arises as
a dynamical effect of measurement interactions, as opposed to a measurement-independent
property of particles in bound states (as in ZSM). We find this an
intriguing possibility, but the technical details need to be developed
for it to become a serious proposal.

Gr\"{o}ssing et al. \cite{Groessing2011} constructed a model of a classical
``walking bouncer'' particle (essentially a harmonic oscillator
of natural frequency $\omega_{0}$) coupled to a dissipative thermal
environment which imparts a stochastic, periodic, driving force. They
then showed that in the large friction limit the mean stochastic dynamics
of the bouncer satisfies what amounts to the quantization condition.
They claim ``this condition resolves the problem discussed by Wallstrom
{[}20{]} about the single-valuedness of the quantum mechanical wavefunctions
and eliminates possible contradictions arising from Nelson-type approaches
to model quantum mechanics.'' It is unclear to us that their model
involves physically consistent assumptions; \footnote{They assume that their dissipative thermal environment corresponds
to a classical ``zero-point field'' of Ornstein-Uhlenbeck statistical
type, unknown positive temperature, and that imparts to the bouncer
a total energy of $\hbar\omega_{0}/2$. But the zero-point fields
of QED and SED are, by construction, frequency-cubed-dependent in
their spectral density, non-dissipative in that they produce no Einstein-Hopf
drag force, and \emph{non-thermal} in that the zero-point motion they
induce on charged particles persists at zero temperature \cite{Boyer1980,Milonni1994}. } but setting aside this concern, the main difficulty we see with their
claim is that they don't show how to derive the HJM equations from
their model (although they do show that their model yields the energy
spectrum of the quantum harmonic oscillator), which is the context
in which Wallstrom's critique applies. In addition, it is unclear
how their model is consistent with NYSM since Nelson's diffusion process
is a conservative one while their model assumes a dissipative diffusion
process in a thermal environment. No such (apparent) inconsistency
exists for ZSM, since we implemented the \emph{zbw} hypothesis into
NYSM in a manner consistent with Nelson's (suitably generalized) ether
hypothesis. Nevertheless, in our view, Gr\"{o}ssinget al.'s model (if
it can be shown physically consistent) has value as a proof-of-principle
that one can construct a physical model of a classical, harmonically
oscillating particle coupled to some fluctuating, oscillating, ether-like
background medium, and dynamically obtain the quantization condition.

Schmelzer \cite{Schmelzer(2011)} argued that in order to obtain empirical
equivalence with quantum mechanics, it is sufficient for stochastic
mechanics to only recover wavefunctions with simple zeros. He then
showed that if one invokes the postulate, $0<\Delta\rho(\mathbf{x})<\infty$
almost everywhere when $\rho(\mathbf{x})=0$, one obtains the quantization
condition for simple zeros, i.e., where $n=\pm1$. He also showed
that this postulate corresponds to an ``energy balance'' constraint,
namely, that the total energy density of the Nelsonian particle remains
finite. Schmelzer suggested that it remains for subquantum theories
to somehow dynamically justify the energy balance constraint. In our
view, Schmelzer does not adequately justify his claim that simple
zeros are sufficient to recover empirical equivalence with quantum
mechanics (e.g., how can this account for energy level shifts in the
hydrogen atom described by the Rydberg formula?); but if this can
be shown, then we would concur that his proposal seems to be a non-circular,
non-ad\emph{-}hoc, empirically adequate justification for a limited
version of the quantization condition. In ZSM, by contrast, the full
quantization condition is obtained from the phase of the hypothesized
\emph{zbw} particle(s), with the proviso that it should be understood
as a phenomenological stepping-stone to a physical theory of Nelson's
(suitable modified) ether, the \emph{zbw} particle, and the dynamical
interaction between the two.

Caticha and his collaborators \cite{Catich2011,BartolomeoCaticha2015}
have offered two routes to answering Wallstrom within the context
of his ``entropic dynamics'' (ED) framework (essentially, a Bayesian
inference version of stochastic mechanics). In the first route, Caticha
appeals to Pauli \cite{Pauli1980}, who suggested that the criterion
for admissibility for wavefunctions is that they must form a basis
for a representation of the transformation group for a given eigenvalue
problem. He then suggests that this criterion is ``extremely natural''
from the perspective of a theory of inference since ``in any physical
situation symmetries constitute the most common and most obviously
relevant pieces of information'' \cite{Catich2011}. However, it
should be noted that Pauli's criterion, more precisely, is that ``repeated
actions of the operators corresponding to physical quantities should
not lead outside the domain of square-integrable eigenfunctions''
\cite{Pauli1980}. In other words, Pauli's criterion just requires
that wavefunctions continue to satisfy the linearity of Schr\"{o}dinger 's
equation (i.e., the superposition principle), even after being acted
upon by operators for physical quantities. But insofar as ED attempts
to recover the Schr\"{o}dinger  equation from the HJM equations, such a
criterion cannot be invoked in entropic dynamics without begging the
question. In the second route, Bartolomeo and Caticha \cite{BartolomeoCaticha2015}
take inspiration from Takabayasi's generalization of the HJM equations
to a spinning fluid \cite{Takabayasi1983}; they propose to interpret
their postulated ``drift potential'', $\phi(\mathbf{x},t)$, as
an angle describing particle spin, and thereby argue that the change
of $\phi$ along a closed loop in space must equal $2\pi n$. In fact,
this argument is conceptually equivalent to the ones given by de Broglie
\cite{Broglie1925,Darrigol1994} and Bohm \cite{Bohm1957,BohmHiley1982,Bohm2002},
and which we've used in ZSM. On the other hand, it should be noted
that Bartolomeo and Caticha don't actually model spin in ED, nor do
they suggest to connect spin to the dynamical influence of an ether
or background field (in contrast to ZSM). Indeed, Bartolomeo and Caticha
admit that ``ED is a purely epistemic theory. It does not attempt
to describe the world.... In fact ED is silent on the issue of what
causative power is responsible for the peculiar motion of the particles''
\cite{BartolomeoCaticha2015}. From our point of view, this makes
their argument for the quantization condition less compelling than
the one offered by ZSM, and ED less compelling as a satisfactory theory
of quantum phenomena compared to the (programmatic) ontological approach
offered by ZSM. Nevertheless, to whatever extent one views the Bayesian
inference approach to physics as valuable and interesting, it appears
that one can give a somewhat non-ad-hoc justification for the quantization
condition via ED.

\section{Conclusion}

We have extended our classical \emph{zbw} model and ZSM to the cases
of free particles, particles in external fields, and classically interacting
particles. Along the way, we have made explicit the beables of ZSM
and suggested three possible approaches for parsing the beables into
local vs. nonlocal types. In addition, we have given arguments for
the plausibility of the \emph{zbw} hypothesis and suggested new lines
of research that could be pursued from the foundation provided here.
We have also reviewed and compared several other proposals for answering
the Wallstrom criticism, arguing that ZSM is the most general and
viable approach of all of them presently.

We wish to emphasize once more that ZSM should not be viewed as a
proposal for a fundamental physical theory of non-relativistic quantum
phenomena; rather, it should be viewed as a provisional, phenomenological
theory that provides the conceptual and mathematical scaffolding for
an eventual physical theory of Nelson's ether (amended for ZSM), the
\emph{zbw} particle, and the dynamical coupling between the two.

In his 1994 paper \cite{Wallstrom1994}, Wallstrom wrote: ``There
seems to be nothing within the particle-oriented world of stochastic
mechanics which can lead to what is effectively a condition on the
`wave function'''. We would suggest that, given the example of ZSM,
Wallstrom's claim can no longer be sustained for all formulations
of stochastic mechanics.

\section{Acknowledgments}

It is a pleasure to thank Guido Bacciagaluppi, Dieter Hartmann, and
Herman Batelaan for helpful discussions and encouragement throughout
this work. Special thanks to Guido for a careful reading of this paper
and several helpful suggestions for improvements.

\chapter{Semiclassical Newtonian Field Theories Based On Stochastic Mechanics
I}

This is the first in a two-part series in which we extend non-relativistic
stochastic mechanics, in the ZSM formulation \cite{Derakhshani2016a,Derakhshani2016b},
to semiclassical Newtonian gravity (ZSM-Newton) and semiclassical
Newtonian electrodynamics (ZSM-Coulomb), under the assumption that
the gravitational and electromagnetic fields are fundamentally classical
(i.e., not independently quantized fields). Our key findings are:
(1) a derivation of the usual $N$-particle Schr\"{o}dinger  equation
for many particles interacting through q-number gravitational or Coulomb
potentials, and (2) recovery of the `single-body' Schr\"{o}dinger -Newton
and Schr\"{o}dinger -Coulomb equations as mean-field equations valid for
systems of gravitationally and electrostatically interacting identical
particles, respectively, in the weak-coupling large \emph{N}  limit. We also
compare ZSM-Newton/Coulomb to semiclassical Newtonian gravity and electrodynamics
approaches based on standard quantum theory, dynamical collapse theories,
and the de Broglie-Bohm theory.

\section{Introduction}

Semiclassical theories \footnote{As in field theories where the matter sector is described within the
framework of quantum mechanics or quantum field theory, and the gravity
sector is described by classical (c-number) fields.} of gravity and electrodynamics, based on the formalism of standard
quantum theory, have been thoroughly studied over the past 55 years
\cite{Moeller1962,Rosenfeld1963,KibbleRandjbarDaemi1980,PageGeilker1981,HartleHorowitz1981,BirrellDavies1982,Ford1982,Wald1994,Visser2002,Ford2005,HuRouraVerdaguer2004,Hu2008,Kiefer2012,Hu2014,AnastopoulosHu2014,Barut1988a,Barut1988b,BarutDowling1990a,BarutDowling1990b,Milonni1994}.
In the past 20 years or so, semiclassical Newtonian gravity based
on the Schr\"{o}dinger -Newton (SN) equation \cite{Diosi1984,Penrose1996,Penrose1998,Marshall2003,Guzman2003,Salzman2005,Diosi2007,Adler2007,Carlip2008,vanWezel2008,Boughn2009,Meter2011,Giulini2011,Kiefer2012,Salcedo2012,Giulini2012,Bassi2013,Giulini2013,Yang2013,AnastopoulosHu2014,Hu2014,Colin2014,Giulini2014,BahramiBassi2014a,Anastopoulos2015,Grossardt2015,Bahrami2015,Bera2015,DerakProbingGravCat2016,DerAnaHu2016,Grossardt2016,Pino2016,Gan2016,Diosi2016,DerakNewtLimStoGra2017}
has become a popular focus of discussions in the foundations of quantum
mechanics \cite{Penrose1996,Penrose1998,Adler2007,vanWezel2008,Bassi2013,Hu2014,Derakhshani2014,AnastopoulosHu2014,BahramiBassi2014a,Bahrami2015,Diosi2016,DerakNewtLimStoGra2017},
quantum gravity phenomenology \cite{Penrose1996,Penrose1998,Marshall2003,Salzman2005,Carlip2008,Boughn2009,Bassi2013,Giulini2013,Yang2013,BahramiBassi2014a,Giulini2014,Colin2014,Bahrami2015,Anastopoulos2015,Bera2015,Grossardt2015,Grossardt2016,Gan2016,DerakProbingGravCat2016,DerakNewtLimStoGra2017},
and state-of-the-art AMO experimental physics \cite{Marshall2003,RomeroIsart2012,Yang2013,Grossardt2015,Bahrami2015,Grossardt2016,Gan2016,Pino2016,DerAnaHu2016,DerakProbingGravCat2016}.
Variants of the SN equation, based on alternative formulations of
quantum theory, have also been developed \cite{Prezhdo2001,Diosi2007,Landau2012,Diosi2012A,Diosi2012B,Adler2013,Bassi2013,Derakhshani2014,Kafri2014,Nimmrichter2015,Struyve2015,Tilloy2016,DerakProbingGravCat2016},
mostly in the context of dynamical collapse theories \cite{Adler2013,Derakhshani2014,DerakProbingGravCat2016,Diosi2007,Diosi2012A,Diosi2012B,Nimmrichter2015,Landau2012,Tilloy2016}.
Less discussion has been given to the possibility of semiclassical
theories of gravity/electrodynamics based on `hidden-variables' \footnote{This phrase is somewhat misleading for the theories in question, but
we will use it abusingly due to its already widespread use in the
literature.} theories; the only instances we know of are Struyve \cite{Struyve2015},
Kiessling \cite{Kiessling2006}, and Prezhdo-Brooksby \cite{Prezhdo2001}
in the context of the de Broglie-Bohm (dBB) pilot-wave theory \cite{HollandBook1993,Goldstein2013,Duerr2009,DGZbook2012}.
Until now, no such discussion has been given in the context of stochastic
mechanical hidden-variables theories \cite{Nelson1967,Nelson1966,Nelson1985,Bohm1989,Kyprianidis1992,Derakhshani2016a,Derakhshani2016b}. 

A central reason for considering formulations of semiclassical gravity
based on alternative quantum theories is that the SN equation, whether
understood as a mean-field approximation to the standard exact quantum
description of matter-gravity coupling \cite{HartleHorowitz1981,HuRouraVerdaguer2004,Hu2008,Kiefer2012,Hu2014,AnastopoulosHu2014,DerakNewtLimStoGra2017}
or as a fundamental theory describing matter-gravity coupling consistent
with standard quantum theory \cite{KibbleRandjbarDaemi1980,PageGeilker1981,Adler2007,Kiefer2012,Derakhshani2014,AnastopoulosHu2014,BahramiBassi2014a,Giulini2014,Bahrami2015,Bera2015,DerakProbingGravCat2016},
is either very limited in applicability or fatally problematic \cite{PageGeilker1981,HuRouraVerdaguer2004,Ford2005,Adler2007,Hu2008,vanWezel2008,AnastopoulosHu2014,Hu2014,Derakhshani2014,Anastopoulos2015,Diosi2016,DerakNewtLimStoGra2017}. 

Understood as a mean-field theory, the nonlinearity of the SN equation
(or the stochastic SN equation, as will be discussed by us in \cite{DerakNewtLimStoGra2017})
means that its solutions lack consistent Born-rule interpretations
\cite{Adler2007,vanWezel2008,Derakhshani2014,Hu2014,Diosi2016} (see
section 4 and subsection 5.1); instead, the SN solutions must be understood
as describing self-gravitating \emph{classical matter fields} that
approximate quantum systems involving large numbers of identical particles
that weakly interact \footnote{In the sense of the coupling scaling as $1/N$, where N is the number
of particles.} quantum-gravitationally \cite{BardosGolseMauser2000,BardosErdosGolseMauserYau2002,Golse2003,Hu2014,AnastopoulosHu2014,DerakNewtLimStoGra2017}.
Moreover, only SN solutions with `small quantum fluctuations' (i.e.,
solutions which don't correspond to superpositions of effectively
orthogonal classical field states, but rather Gaussian quantum states
such as a coherent state) can have this physical interpretation \cite{Ford1982,Diosi1984,Ford2005,Derakhshani2014,Anastopoulos2015,DerakProbingGravCat2016,DerakNewtLimStoGra2017},
implying that the vast majority of SN solutions are (physically) superfluous. 

Understood as a fundamental theory, the nonlinearity of the SN equation
is fatal because the consequent lack of consistent Born-rule interpretations
for the SN solutions destroys the standard quantum interpretation
of the matter sector of fundamentally-semiclassical gravity based
the SN equation (see subsection 5.1). A fundamentally-semiclassical
description of matter-gravity coupling, based on the SN equation,
would actually be a nonlinear classical field theory that makes empirical
predictions (such as macroscopic semiclassical gravitational cat states;
see section 4 for an example) grossly inconsistent with standard quantum
mechanics and the world of lived experience \cite{Ford1982,Diosi1984,Derakhshani2014,Ford2005,BahramiBassi2014a,DerakProbingGravCat2016}.
(Analogous comments apply to semiclassical electrodynamics based on
the Schr\"{o}dinger -Coulomb (SC) equation \cite{Barut1988a,Barut1988b};
see subsection 5.1 for a discussion.) 

Another key motivation for considering formulations of semiclassical
gravity based on alternative quantum theories is that while the standard
exact quantum description of matter-gravity coupling yields semiclassical
gravity as a consistent mean-field approximation, the matter sector
of the standard exact quantum description is afflicted by the quantum
measurement problem \cite{BellAgainstMeasure,HollandBook1993,Bassi2000,AdlerMP2003,Schlosshauer2004a,Duerr2009,Bassi2013}.
This puts a fundamental limitation on the domain of applicability
of the standard exact quantum description (whether at the Newtonian
level or the fully relativistic level), hence a fundamental limitation
on the domain of applicability of semiclassical gravity (whether at
the Newtonian level or the fully relativistic level). Namely, the
standard exact quantum description and the mean-field semiclassical-gravitational
description are only applicable to laboratory experiments involving
the coupling of gravity to quantum matter, since laboratory experiments
are the only places where the standard quantum formalism can be sensibly
applied. 

Thus it stands to reason that a formulation of quantum theory convincingly
free of the measurement problem might, when extended to a semiclassical
description of gravity (whether as a fundamental theory or a mean-field
theory), yield a superior formulation of semiclassical gravity than
the options based on standard quantum theory. Arguably, this suggestion
has already been confirmed (at least at the Newtonian level) by dynamical
collapse versions of fundamentally-semiclassical Newtonian gravity,
insofar as the models of Derakhshani \cite{Derakhshani2014,DerakProbingGravCat2016}
and Tilloy-Di\'{o}si \cite{Tilloy2016,TilloyDiosi2017} seem to have consistent
statistical interpretations while adequately suppressing gravitational
cat state solutions at the macroscopic scale. In addition, the works
of Struyve \cite{Struyve2015} and Prezhdo-Brooksby \cite{Prezhdo2001}
suggest that the dBB theory offers a more empirically accurate semiclassical
approximation scheme than does standard quantum theory, at least for
simple examples considered at the relativistic level \cite{Struyve2015}
and the Newtonian level \cite{Prezhdo2001,Struyve2015} (see subsection
5.2 for more detail); however, Struyve has pointed out \cite{Struyve2015}
that the naive extension of dBB theory to fundamentally-semiclassical
Einstein gravity does not yield a mathematically and physically consistent
model \footnote{The reason, basically, is that a classical stress-energy tensor built
out of a dBB field beable $\phi_{B}(\mathbf{x},t)$ entails covariant
non-conservation of stress-energy, i.e., $\nabla^{\mu}T_{\mu\nu}\left(\phi_{B}\right)\neq0$,
since $\phi_{B}$ does not covariantly conserve total stress-energy
along its space-time trajectory (because it satisfies the non-covariant
wave equation $g^{nm}\nabla_{n}\nabla_{m}\phi_{B}=-(-detg_{nm})^{-1/2}\delta Q/\delta\phi_{B}$,
where $Q$ is the quantum potential for $\phi_{B}$). Thus $T_{\mu\nu}\left(\phi_{B}\right)$
cannot be used on the right-hand-side of the classical Einstein equations,
as this would violate the Bianchi identities $\nabla^{\mu}G_{\mu\nu}=0$.
Of course, alternative models of fundamentally-semiclassical Einstein
gravity based on dBB theory may be possible to construct that \emph{are}
consistent, whether by considering modified theories of Einstein gravity
that don't require covariant conservation of either the left hand
side or right hand side of the modified classical Einstein equations
(such as Scalar-Tensor theories in the `Einstein frame', $f(R)$ theories,
and Unimodular Einstein gravity) \cite{Clifton2012}, or by postulating
new matter degrees of freedom in addition to the dBB field beable,
where the new matter degrees of freedom exchange stress-energy-momentum
with the dBB field beable in such a way that ensures covariant conservation
of total stress-energy-momentum at all times. Such models have yet
to be developed and studied, however, so this remains conjecture at
the moment. Another possibility might be to derive a consistent semiclassical
Einstein gravity description from a consistent semiclassical approximation
to the dBB version of canonical quantum gravity, and then to simply
reinterpret that semiclassical Einstein gravity description as a fundamentally-semiclassical
description, as is sometimes done with the standard semiclassical
Einstein equation \cite{Kiefer2012,BahramiBassi2014a,Giulini2014}.
This latter possibility remains a work in progress, as it is still
an open question how to develop a consistent semiclassical approximation
scheme for the dBB version of canonical quantum gravity. We refer
the reader to the work of Struyve \cite{Struyve2015} for suggestions
along these lines.}. In our assessment (see section 5.2), the various options for
extending dBB theory to fundamentally-semiclassical Newtonian gravity/electrodynamics,
do yield consistent models, but the empirical predictions of these
models are either difficult to extract or demonstrably empirically inadequate.
It would seem, then, that there does not yet exist a compelling and
widely applicable model of semiclassical gravity based on a theory
of hidden-variables, whether in the form of a fundamental theory of
matter-gravity coupling or a mean-field approximation to an exact
`quantum' description of matter-gravity coupling.

The primary objectives of this two-part series are: (i) to construct
a fundamentally-semiclassical theory of Newtonian gravity/electrodynamics
within the framework of stochastic mechanics, in particular a new
formulation of stochastic mechanics we have recently proposed \cite{Derakhshani2016a,Derakhshani2016b}
to answer the long-standing ``Wallstrom criticism''; (ii) to show
that fundamentally-semiclassical Newtonian gravity/electrodynamics
based on our new formulation of stochastic mechanics - which we call
``zitterbewegung stochastic mechanics'' (ZSM), hence `ZSM-Newton'
and `ZSM-Coulomb' - has a consistent statistical interpretation and
recovers the standard exact quantum description of matter-gravity
coupling as an approximation (valid for all practical purposes), while
also being free of the measurement problem; (iii) to show that the
SN/SC equation and the stochastic SN/SC equation can be recovered
as mean-field approximations for large numbers of identical ZSM particles
that weakly interact \footnote{Also in the sense of coupling scaling as $1/N$.}
classical-gravitationally/electrostatically; and (iv) to show that
ZSM-Newton/Coulomb yields a new `large-\emph{N}' prescription that makes
it possible to: (a) accurately approximate the time-evolution of a
large number of identical ZSM particles that strongly interact classical-gravitationally/electrostatically,
within a consistent statistical interpretation; (b) avoid macroscopic
semiclassical gravitational cat states and recover classical Newtonian
gravity/electrodynamics for the center-of-mass descriptions of macroscopic
particles; and (c) recover classical Vlasov-Poisson mean-field theory
for macroscopic particles that weakly interact gravitationally/electrostatically. 

In the present paper, we will carry out objectives (i-iii), leaving
(iv) for Part II. We'll also compare ZSM-Newton/Coulomb to semiclassical Newtonian gravity/electrodynamics theories
based on standard quantum theory, dynamical collapse theories, and
the de Broglie-Bohm theory.

The paper organization is as follows. Section 2 reviews ZSM for the
case of many free particles. Section 3 formulates the basic equations
of ZSM-Newton/Coulomb, explicates the physical interpretation of those
equations, and shows how the standard exact quantum description of
matter-gravity coupling is recovered as a special case valid for all
practical purposes. Section 4 shows how to recover the SN/SC equation
and the stochastic SN/SC equation as mean-field approximations for
large numbers of identical ZSM particles that weakly interact gravitationally/electrostatically.
Finally, section 5 compares ZSM-Newton/Coulomb to extant theories
of semiclassical Newtonian gravity/electrodynamics, pointing out conceptual
and technical advantages entailed by ZSM-Newton/Coulomb, as well as
possibilities for experimental discrimination between ZSM-Newton/Coulomb
and these other semiclassical theories.

\section{Overview of ZSM for many free particles}

ZSM was developed in order to answer Wallstrom's criticism of stochastic
mechanical reconstructions of Schr\"{o}dinger's equation; namely, that
they don't give a plausible justification for the quantum mechanical
requirement that wavefunctions (for spinless particles) must always
be single-valued while allowing generally multi-valued phases \cite{Wallstrom1989,Wallstrom1994,Derakhshani2016a}.
In other words, why it should be that the wavefunction phase $S$
(in polar form) must change along a closed loop in configuration space
by integer multiples of Planck's constant. A formulation of stochastic
mechanics that plausibly answers this criticism is, in our view, a
necessary condition for seriously considering extensions of stochastic
mechanics to more general physical situations, hence why we will base
our approach on the ZSM formulation.

To prepare for the formulation of ZSM-Newton/Coulomb, it is useful
to first review ZSM for $N$ particles that are classically non-interacting
\cite{Derakhshani2016b}.

Our starting point is the following four phenomenological hypotheses.

First, we posit a Minkowski space-time that contains, on a $t=const$
hypersurface, N harmonic oscillators centered around 3-space positions
$\mathbf{q}_{0i}$ for $i=1,..,N$. As ZSM is a phenomenological framework,
we need not specify here the precise physical nature of these harmonic
oscillators (this is task is left for future work). However, we assume
that these oscillators have, in their respective translational rest
frames, natural frequencies $\omega_{ci}\coloneqq\left(1/\hbar\right)m_{i}c^{2}$,
where $c$ is the speed of light and the $m_{i}$ are mass parameters
that set the scales of the natural frequencies. It is reasonable to
call these natural frequencies ``Compton'' frequencies, hence the
label $\omega_{ci}$. We will refer to these oscillators hereafter
as\emph{ }``zitterbewegung (\emph{zbw}) particles'' \cite{Derakhshani2016a,Derakhshani2016b}. 

Second, we adapt Nelson's ether hypothesis \cite{Nelson1966,Nelson1967nopagelist,Nelson1985nopagelist,Nelson1986,Nelson2005}
by supposing now that the Minkowski space-time is pervaded by a frictionless
medium (which we will also call an ``ether''), with the qualitative
properties that (i) it is fluctuating everywhere with the same intensity,
and (ii) it is an oscillating medium with a spectrum of modes superposed
at each point in 3-space. More precisely, we imagine the ether as
a continuous (or effectively continuous) medium composed of a countably
infinite number of fluctuating, stationary, spherical waves superposed
at each point in space, with each wave having a different fixed angular
frequency, $\omega_{0}^{k}$, where $k$ denotes the \emph{k}-th ether
mode. The relative phases between the modes are taken to be random
so that each mode is effectively uncorrelated with every other mode.
Again, since ZSM is a phenomenological framework, specifying the precise
physical nature of this ether is left to future work.

Third, we follow Nelson \cite{Nelson1966,Nelson1967nopagelist,Nelson1985}
in hypothesizing that each particle's center of mass, as a result
of being immersed in the ether, undergoes an approximately \emph{frictionless}
translational Brownian motion (due to the homogeneous and isotropic
ether fluctuations that couple to the particles by possibly electromagnetic,
gravitational, or some other means), as modeled by the first-order
stochastic differential equations
\begin{equation}
d\mathbf{q}_{i}(t)=\mathbf{b}_{i}(q(t),t)dt+d\mathbf{W}_{i}(t).
\end{equation}
Here the index $i=1,...,N$, the particle trajectories $q(t)=\{\mathbf{q}_{1}(t),\mathbf{q}_{2}(t),...,\mathbf{q}_{N}(t)\}$
$\in$ $\mathbb{R}^{3N}$, $\mathbf{b}_{i}(q(t),t)$ are the mean
forward drift velocities, and $\mathbf{W}_{i}(t)$ are Wiener processes
modeling each particle's interaction with the ether fluctuations.
The Wiener increments $d\mathbf{W}_{i}(t)$ are assumed to be Gaussian
with zero mean, independent of $d\mathbf{q}_{i}(s)$ for $s\leq t$,
and with variance
\begin{equation}
\mathrm{E}_{t}\left[d\mathbf{W}_{in}(t)d\mathbf{W}_{im}(t)\right]=2\nu_{i}\delta_{nm}dt,
\end{equation}
where $\mathrm{E}_{t}$ denotes the conditional expectation at time
\emph{t}. We then follow Nelson in hypothesizing that the magnitude
of the diffusion coefficients $\nu_{i}$ are defined by 
\begin{equation}
\nu_{i}\coloneqq\frac{\hbar}{2m_{i}}.
\end{equation}
Along with (4.1), we also have the backward equations
\begin{equation}
d\mathbf{q}_{i}(t)=\mathbf{b}_{i*}(q(t),t)+d\mathbf{W}_{i*}(t),
\end{equation}
where $\mathbf{b}_{i*}(q(t),t)$ are the mean backward drift velocities,
and $d\mathbf{W}_{i*}(t))$ are the backward Wiener processes. As
in the single-particle case, the $d\mathbf{W}_{i*}(t)$ have all the
properties of $d\mathbf{W}_{i}(t)$ except that they are independent
of the $d\mathbf{q}_{i}(s)$ for $s\geq t$. With these conditions
on $d\mathbf{W}_{i}(t)$ and $d\mathbf{W}_{i*}(t)$, equations (4.1)
and (4.4) respectively define forward and backward Markov processes
for $N$ particles on $\mathbb{R}^{3}$ (or, equivalently, for a single
particle on $\mathbb{R}^{3N}$). Having introduced $\mathbf{b}_{i}(q(t),t)$
and $\mathbf{b}_{i*}(q(t),t)$, let us also define the mean forward
and mean backward derivatives:
\begin{equation}
D\mathbf{q}_{i}(t)\coloneqq\underset{_{\Delta t\rightarrow0^{+}}}{lim}\mathrm{E}_{t}\left[\frac{q_{i}(t+\Delta t)-q_{i}(t)}{\Delta t}\right]=\mathbf{b}_{i}(q(t),t),
\end{equation}
and
\begin{equation}
D_{*}\mathbf{q}_{i}(t)\coloneqq\underset{_{\Delta t\rightarrow0^{+}}}{lim}\mathrm{E}_{t}\left[\frac{q_{i}(t)-q_{i}(t-\Delta t)}{\Delta t}\right]=\mathbf{b}_{i*}(q(t),t),
\end{equation}
where we used the Gaussianity of $d\mathbf{W}_{i}(t)$ and $d\mathbf{W}_{i*}(t)$
in equations (4.1) and (4.4). Finding $D\mathbf{b}_{i}(q(t),t)$ (or
$D_{*}\mathbf{b}_{i}(q(t),t)$) is straightforward: expand $\mathbf{b}_{i}$
in a Taylor series up to terms of order two in $d\mathbf{q}_{i}(t)$,
replace $dq_{i}(t)$ by $dW_{i}(t)$ in the last term, and replace
$d\mathbf{q}_{i}(t)\cdot\nabla_{i}\mathbf{b}_{i}|_{\mathbf{q}_{j}=\mathbf{q}_{j}(t)}$
by $\mathbf{b}_{i}(\mathbf{q}(t),t)\cdot\nabla_{i}\mathbf{b}_{i}|_{\mathbf{q}_{j}=\mathbf{q}_{j}(t)}$
when taking the conditional expectation at time \emph{t} (since $d\mathbf{W}_{i}(t)$
is independent of $\mathbf{q}_{i}(t)$ and has mean 0). We then have
\begin{equation}
D\mathbf{b}_{i}(q(t),t)=\left[\frac{\partial}{\partial t}+\sum_{i=1}^{N}\mathbf{b}_{i}(q(t),t)\cdot\nabla_{i}+\sum_{i=1}^{N}\frac{\hbar}{2m_{i}}\nabla_{i}^{2}\right]\mathbf{b}_{i}(q(t),t),
\end{equation}
and
\begin{equation}
D_{*}\mathbf{b}_{i*}(q(t),t)=\left[\frac{\partial}{\partial t}+\sum_{i=1}^{N}\mathbf{b}_{i*}(q(t),t)\cdot\nabla_{i}-\sum_{i=1}^{N}\frac{\hbar}{2m_{i}}\nabla_{i}^{2}\right]\mathbf{b}_{i*}(q(t),t).
\end{equation}
Note that we take the $\mathbf{b}_{i}$ $(\mathbf{b}_{i*})$ to be
functions of $q(t)=\{\mathbf{q}_{1}(t),\mathbf{q}_{2}(t),...,\mathbf{q}_{N}(t)\}$
$\in$ $\mathbb{R}^{3N}$. The reasons are: (i) all the particles
are continuously exchanging energy-momentum with a common background
medium (the ether) and thus are in general physically connected in
their translational motions via $\mathbf{b}_{i}$ $(\mathbf{b}_{i*})$,
insofar as the latter are constrained by the physical properties of
the ether; and (ii) the dynamical equations and initial conditions
for the $\mathbf{b}_{i}$ $(\mathbf{b}_{i*})$ are what will determine
the specific situations under which the latter will be effectively
factorizable functions of the particle positions and when they cannot
be effectively factorized. Hence, at this level, it is only sensible
to write $\mathbf{b}_{i}$ $(\mathbf{b}_{i*})$ as functions of all
the particle positions at a single time. 

Fourth, we suppose that, in their respective IMFTRFs, i.e., the frames
in which $D\mathbf{q}_{i}(t)=\mathbf{b}_{i}(\mathbf{q}(t),t)=0$,
the \emph{zbw} particles undergo driven oscillations about $\mathbf{q}_{0i}$
by coupling to a narrow band of ether modes that resonantly peak around
their natural frequencies. However, in order that the oscillation
of each particle doesn't become unbounded in kinetic energy, there
must be some mechanism by which the particles dissipate energy back
into the ether so that, on the average, a steady-state equilibrium
regime is reached for their oscillations. So we posit that on short
relaxation time-scales, $\tau$, which are identical for particles
of identical rest masses, the average energy absorbed from the driven
oscillation by the resonant ether modes equals the average energy
dissipated back to the ether by a given particle. (The average, in
the present sense, would be over the random phases of the ether modes.)
Thus, in the steady-state regime, each particle undergoes a steady-state\emph{
zbw} oscillation of angular frequency $\omega_{ci}$ about its location
$\mathbf{q}_{0i}$ in its IMFTRF, as characterized by the `fluctuation-dissipation'
relation, $<H_{i}>_{steady-state}=\hbar\omega_{ci}=m_{i}c^{2}$, where
$<H_{i}>_{steady-state}$ is the conserved (random-phase-)average
energy due to the steady-state oscillation of the \emph{i}-th particle.
Accordingly, if, relative to the ether, all the particles have zero
mean translational motion, then we will have $\sum_{i}^{N}<H_{i}>_{steady-state}=\sum_{i}^{N}\hbar\omega_{ci}=\sum_{i}^{N}m_{i}c^{2}=const$. 

Now, as a consequence of this last hypothesis, it follows that in
the IMFTRF of the \emph{i}-th particle, the forward steady-state \emph{zbw}
phase change is given by
\begin{equation}
\delta\bar{\theta}_{i+}\coloneqq\omega_{ci}\delta t_{0}=\frac{m_{i}c^{2}}{\hbar}\delta t_{0},
\end{equation}
and the corresponding absolute forward steady-state phase is
\begin{equation}
\bar{\theta}_{i+}=\omega_{ci}t_{0}+\phi_{i}=\frac{m_{i}c^{2}}{\hbar}t_{0}+\phi_{i+}.
\end{equation}
Then the joint forward steady-state phase for all the particles will
just be
\begin{equation}
\bar{\theta}_{+}=\sum_{i=1}^{N}\bar{\theta}_{i+}=\sum_{i=1}^{N}\left(\omega_{ci}t_{0}+\phi_{i+}\right)=\sum_{i=1}^{N}\left(\frac{m_{i}c^{2}}{\hbar}t_{0}+\phi_{i+}\right).
\end{equation}
The reason for starting our analysis with the IMFTRFs is that, at
this stage, neither (4.1) nor (4.4) have well-defined time reversals
\cite{Bacciagaluppi2005,Derakhshani2016a,Derakhshani2016b}, so the
forward and backward stochastic differential equations (4.1) and (4.4)
describe independent, time-asymmetric diffusion processes in opposite
time directions. Hence we must start by considering the steady-state
\emph{zbw} phases in each time direction separately, and it is natural
to start with the more intuitive forward time direction. 

For the \emph{i}-th \emph{zbw} particle in its instantaneous mean
backward translational rest frame (IMBTRF), i.e., the frame defined
by $D_{*}\mathbf{q}_{i}(t)=\mathbf{b}_{i*}(q(t),t)=0$, its backward
steady-state \emph{zbw} phase change is given by
\begin{equation}
\delta\bar{\theta}_{i-}\coloneqq-\omega_{ci}\delta t_{0}=-\frac{m_{i}c^{2}}{\hbar}\delta t_{0},
\end{equation}
and
\begin{equation}
\bar{\theta}_{i-}=\left(-\omega_{ci}t_{0}\right)+\phi_{i-}=\left(-\frac{m_{i}c^{2}}{\hbar}t_{0}\right)+\phi_{i-}.
\end{equation}
Then the joint backward steady-state phase for all the particles will
just be
\begin{equation}
\bar{\theta}_{-}=\sum_{i=1}^{N}\bar{\theta}_{i-}=\sum_{i=1}^{N}\left(\omega_{ci}t_{0}+\phi_{i-}\right)=\sum_{i=1}^{N}\left(\frac{m_{i}c^{2}}{\hbar}t_{0}+\phi_{i-}\right).
\end{equation}

We note that both the diffusion coefficient $\nu_{i}=\hbar/2m_{i}$
and the (reduced) \emph{zbw} period $T_{ci}=1/\omega_{ci}=\hbar/m_{i}c^{2}$
are scaled by $\hbar$. This is consistent with our hypothesis that
the ether is the common physical cause of both the frictionless diffusion
processes and the steady-state \emph{zbw} oscillations of the particles.
Had we not proposed Nelson's ether as the physical cause of the \emph{zbw}
oscillations as well as the frictionless diffusions, the occurrence
of $\hbar$ in both of these properties of the particles would be
inexplicable and compromising for the plausibility of our proposed
modification of NYSM.

Note also that we cannot transform to a frame in which $d\mathbf{q}_{i}(t)/dt=0$,
as this expression is undefined for the Wiener process. So we cannot
talk of the \emph{zbw} phases in rest frames other than in the IMFTRFs
or IMBTRFs of the particles

If we Lorentz transform back to the lab frame in the forward time
direction, this corresponds to a boost of (4.9) by $-\mathbf{b}_{i}(q(t),t)$.
Approximating the transformation for non-relativistic velocities so
that $\gamma=1/\sqrt{\left(1-\mathbf{b}_{i}^{2}/c^{2}\right)}\approx1+\mathbf{b}_{i}^{2}/2c^{2},$
the mean forward steady-state joint phase change becomes
\begin{equation}
\begin{aligned}\delta\bar{\theta}_{+}(q(t),t) & =\sum_{i=1}^{N}\frac{\omega_{ci}}{m_{i}c^{2}}\mathrm{E}_{t}\left[E_{i+}(D\mathbf{q}_{i}(t))\delta t-m_{i}D\mathbf{q}_{i}(t)\cdot\left(D\mathbf{q}_{i}(t)\right)\delta t\right]\\
 & =\frac{1}{\hbar}\mathrm{E}_{t}\left[\sum_{i=1}^{N}E_{i+}(D\mathbf{q}_{i}(t))\delta t-\sum_{i=1}^{N}m_{i}\mathbf{b}_{i}(q(t),t)\cdot\delta\mathbf{q}_{i+}(t)\right],
\end{aligned}
\end{equation}
where
\begin{equation}
E_{i+}(D\mathbf{q}_{i}(t))=m_{i}c^{2}+\frac{1}{2}m_{i}\left(D\mathbf{q}_{i}(t)\right)^{2}=m_{i}c^{2}+\frac{1}{2}m_{i}\mathbf{b}_{i}^{2},
\end{equation}
neglecting the momentum terms proportional to $\mathbf{b}_{i}^{3}/c^{2}$.
The $\delta\mathbf{q}_{i+}(t)$ in (4.15) corresponds to the physical,
translational, mean forward displacement of the \emph{i}-th \emph{zbw}
particle, defined by 
\begin{equation}
\delta\mathbf{q}_{i+}(t)=\left[D\mathbf{q}_{i}(t)\right]\delta t=\mathbf{b}_{i}(q(t),t)\delta t.
\end{equation}
For the backward time direction, the Lorentz transformation to the
lab frame corresponds to a boost of (4.12) by $-\mathbf{b}_{i*}(q(t),t)$.
Then the mean backward steady-state joint phase change becomes
\begin{equation}
\begin{aligned}\delta\bar{\theta}_{-}(q(t),t) & =\sum_{i=1}^{N}\frac{\omega_{ci}}{m_{i}c^{2}}\mathrm{E}_{t}\left[-E_{i-}(D_{*}\mathbf{q}_{i}(t))\delta t+m_{i}D_{*}\mathbf{q}_{i}(t)\cdot\left(D_{*}\mathbf{q}_{i}(t)\right)\delta t\right]\\
 & =\frac{1}{\hbar}\mathrm{E}_{t}\left[-\sum_{i=1}^{N}E_{i-}(D_{*}\mathbf{q}_{i}(t))\delta t+\sum_{i=1}^{N}m_{i}\mathbf{b}_{i*}(q(t),t)\cdot\delta\mathbf{q}_{i-}(t)\right],
\end{aligned}
\end{equation}
where
\begin{equation}
E_{i-}(D_{*}\mathbf{q}_{i}(t))=m_{i}c^{2}+\frac{1}{2}m_{i}\left(D_{*}\mathbf{q}_{i}(t)\right)^{2}=m_{i}c^{2}+\frac{1}{2}m_{i}\mathbf{b}_{i*}^{2}.
\end{equation}
The $\delta\mathbf{q}_{i-}(t)$ in (4.18) corresponds to the physical,
translational, mean backward displacement of the \emph{i}-th \emph{zbw}
particle, as defined by 
\begin{equation}
\delta\mathbf{q}_{i-}(t)=\left(D_{*}\mathbf{q}_{i}(t)\right)\delta t=\mathbf{b}_{i*}(q(t),t)\delta t.
\end{equation}
(Notice that $\delta\mathbf{q}_{i+}(t)-\delta\mathbf{q}_{i-}(t)=(\mathbf{b}_{i}-\mathbf{b}_{i*})\delta t\neq0$
in general.) Now since each \emph{zbw} particle is essentially a harmonic
oscillator, each particle has its own, effectively independent, well-defined
forward steady-state phase at each point along its forward space-time
trajectory, when $\mathbf{b}_{i}(q,t)\approx\sum_{i}^{N}\mathbf{b}_{i}(\mathbf{q}_{i},t)$.
Consistency with this hypothesis also means that when $\mathbf{b}_{i}(q,t)\neq\sum_{i}^{N}\mathbf{b}_{i}(\mathbf{q}_{i},t)$,
the forward steady-state joint phase must be a well-defined function
of the space-time trajectories of\emph{ all} \emph{the particles}
(since we posit that all particles remain harmonic oscillators despite
having their oscillations physically coupled through the common ether
medium they interact with). Furthermore, since, at this stage, the
forward and backward steady-state joint \emph{zbw} phase changes,
(4.15) and (4.18), are independent of one another, each must equal
$2\pi n$ when integrated along a closed loop $L$ in which both time
and position change. Otherwise, we will contradict our hypothesis
that the system of \emph{zbw} particles has a well-defined steady-state
joint phase in each time direction. 

Associated to (4.1) and (4.4) in the lab frame are the forward and
backward Fokker-Planck equations
\begin{equation}
\frac{\partial\rho(q,t)}{\partial t}=-\sum_{i=1}^{N}\nabla_{i}\cdot\left[\mathbf{b}_{i}(q,t)\rho(q,t)\right]+\sum_{i=1}^{N}\frac{\hbar}{2m_{i}}\nabla_{i}^{2}\rho(q,t),
\end{equation}
and
\begin{equation}
\frac{\partial\rho(q,t)}{\partial t}=-\sum_{i=1}^{N}\nabla_{i}\cdot\left[\mathbf{b}_{i*}(q,t)\rho(q,t)\right]-\sum_{i=1}^{N}\frac{\hbar}{2m_{i}}\nabla_{i}^{2}\rho(q,t),
\end{equation}
where $\rho(q,t)$ is the probability density for the particle trajectories
and satisfies the normalization condition 
\begin{equation}
\int_{\mathbb{R}^{3N}}\rho_{0}(q)d^{3N}q=1.
\end{equation}
Restricting to simultaneous solutions of (4.21-22) entails the current
velocity field
\begin{equation}
\mathbf{v}_{i}(q,t)\coloneqq\frac{1}{2}\left[\mathbf{b}_{i}(q,t)+\mathbf{b}_{i*}(q,t)\right]=\frac{\nabla_{i}S(q,t)}{m_{i}},
\end{equation}
and the osmotic velocity field
\begin{equation}
\mathbf{u}_{i}(q,t)\coloneqq\frac{1}{2}\left[\mathbf{b}_{i}(q,t)-\mathbf{b}_{i*}(q,t)\right]=\frac{\hbar}{2m_{i}}\frac{\nabla_{i}\rho(q,t)}{\rho(q,t)}.
\end{equation}
Hence (4.21-22) reduce to the continuity equation
\begin{equation}
\frac{\partial\rho({\normalcolor q},t)}{\partial t}=-\sum_{i=1}^{N}\nabla_{i}\cdot\left[\frac{\nabla_{i}S(q,t)}{m_{i}}\rho(q,t)\right],
\end{equation}
with $\mathbf{b}_{i}=\mathbf{v}_{i}+\mathbf{u}_{i}$ and $\mathbf{b}_{i*}=\mathbf{v}_{i}-\mathbf{u}_{i}$.

To give (4.25) a coherent physical interpretation, we introduce the
presence of an external (to the particle) osmotic potential $U(q,t)$
which couples to the $i$-th particle as $R(q(t),t)\coloneqq\mu U(q(t),t)$
(assuming that the coupling constant $\mu$ is identical for particles
of the same species), and imparts a momentum, $\nabla_{i}R(q,t)|_{\mathbf{q}_{j}=\mathbf{q}_{j}(t)}$.
This momentum then gets counter-balanced by the ether fluid's osmotic
impulse pressure, $\left(\hbar/2m_{i}\right)\nabla_{i}\ln[n(q,t)]|_{\mathbf{q}_{j}=\mathbf{q}_{j}(t)}$,
leading to the equilibrium condition $\nabla_{i}R/m_{i}=\left(\hbar/2m_{i}\right)\nabla_{i}\rho/\rho$
(using $\rho=n/N$), which implies $\rho=e^{2R/\hbar}$ for all times.
It is assumed that $R$ generally depends on the coordinates of all
the other particles because: (i) if $U$ was an independently existing
field on configuration space, rather than sourced by the ether, then
the diffusions of the particles through the ether would not be conservative
(i.e., energy conserving), in contradiction with Nelson's hypothesis
that the diffusions \emph{are} conservative, and (ii) since the particles
continuously exchange energy-momentum with the ether, the functional
dependence of $U$ should be determined by the dynamical coupling
of the ether to the particles as well as the magnitude of the inter-particle
physical interactions (whether through a classical inter-particle
potential or, in the free particle case, just through the ether). 

To obtain the 2nd-order time-symmetric dynamics for the mean translational
motions of the \emph{N} particles, we must first define the time-symmetric,
steady-state joint phase of the \emph{zbw} particles, in terms of
a symmetric combination of (4.15) and (4.18). This is natural to do
since (4.15) and (4.18) correspond to the same frame (the lab frame),
and since (4.15) and (4.18) are no longer independent of one another
as a result of constraints (4.24-25). From there, we can define Yasue's
ensemble-averaged action \cite{Derakhshani2016,Derakhshani2016b}
and apply the stochastic variational principle.

Taking the difference between (4.15) and (4.18) yields (replacing
$\delta t\rightarrow dt$ and $\delta\mathbf{q}_{i+,-}(t)\rightarrow d\mathbf{q}_{i+,-}(t)$)
\begin{equation}
\begin{aligned}d\bar{\theta}(q(t),t) & \coloneqq\frac{1}{2}\left[d\bar{\theta}_{+}(q(t),t)-d\bar{\theta}_{-}(q(t),t)\right]\\
 & =\frac{1}{\hbar}\sum_{i=1}^{N}\mathrm{E}_{t}\left[E_{i}(D\mathbf{q}_{i}(t),D_{*}\mathbf{q}_{i}(t))dt-\frac{m_{i}}{2}\left(\mathbf{b}_{i}(q(t),t)\cdot d\mathbf{q}_{i+}(t)+\mathbf{b}_{i*}(q(t),t)\cdot d\mathbf{q}_{i-}(t)\right)\right]+\phi\\
 & =\frac{1}{\hbar}\mathrm{E}_{t}\left[\sum_{i=1}^{N}E_{i}dt-\sum_{i=1}^{N}\frac{m_{i}}{2}\left(\mathbf{b}_{i}\cdot\frac{d\mathbf{q}_{i+}(t)}{dt}+\mathbf{b}_{i*}\cdot\frac{d\mathbf{q}_{i-}(t)}{dt}\right)dt\right]+\phi\\
 & =\frac{1}{\hbar}\mathrm{E}_{t}\left[\left(\sum_{i=1}^{N}E_{i}-\sum_{i=1}^{N}\frac{m_{i}}{2}\left(\mathbf{b}_{i}\cdot\frac{d\mathbf{q}_{i+}(t)}{dt}+\mathbf{b}_{i*}\cdot\frac{d\mathbf{q}_{i-}(t)}{dt}\right)\right)dt\right]+\phi\\
 & =\frac{1}{\hbar}\mathrm{E}_{t}\left[\left(\sum_{i=1}^{N}E_{i}-\sum_{i=1}^{N}\frac{m_{i}}{2}\left(\mathbf{b}_{i}^{2}+\mathbf{b}_{i*}^{2}\right)\right)dt\right]+\phi\\
 & =\frac{1}{\hbar}\mathrm{E}_{t}\left[\left(\sum_{i=1}^{N}E_{i}-\sum_{i=1}^{N}\left(m_{i}\mathbf{v}_{i}\cdot\mathbf{v}_{i}+m_{i}\mathbf{u}_{i}\cdot\mathbf{u}_{i}\right)\right)dt\right]+\phi\\
 & =\frac{1}{\hbar}\mathrm{E}_{t}\left[\sum_{i=1}^{N}\left(m_{i}c^{2}-\frac{1}{2}m_{i}\mathbf{v}_{i}^{2}-\frac{1}{2}m_{i}\mathbf{u}_{i}^{2}\right)dt\right]+\phi,
\end{aligned}
\end{equation}
where $\phi=\sum_{i=1}^{N}\left(\phi_{i+}-\phi_{i-}\right)$, and
from (4.16) and (4.19) we have
\begin{equation}
E_{i}(D\mathbf{q}_{i}(t),D_{*}\mathbf{q}_{i}(t))\coloneqq m_{i}c^{2}+\frac{1}{2}\left[\frac{1}{2}m_{i}\mathbf{b}_{i}^{2}+\frac{1}{2}m_{i}\mathbf{b}_{i*}^{2}\right]=m_{i}c^{2}+\frac{1}{2}m_{i}\mathbf{v}_{i}^{2}+\frac{1}{2}m_{i}\mathbf{u}_{i}^{2}.
\end{equation}
Equation (4.27) is the time-symmetrized steady-state joint phase change
of the \emph{zbw} particles in the lab frame, before the constraint
of conservative diffusions is imposed. Note that because $\bar{\theta}_{+}$
and $\bar{\theta}_{-}$ are no longer independent of one another,
it is no longer consistent to have that $\oint_{L}\delta\bar{\theta}_{+}$
and $\oint_{L}\delta\bar{\theta}_{-}$ both equal $2\pi n$. However,
the consistency of our theory does require that $\oint_{L}\delta\bar{\theta}=2\pi n$,
otherwise we will contradict our hypothesis that the system of \emph{N}
\emph{zbw} particles, after imposing (4.24-25) has a well-defined
and unique steady-state joint phase that functionally depends on the
3-space trajectories of the \emph{zbw} particles.

From the second to last line of (4.27), we can apply the stochastic
calculus version of Fubini's theorem to obtain the cumulative, time-symmetric,
steady-state joint phase at time \emph{t}: 
\begin{equation}
\begin{aligned}\bar{\theta}(q(t),t) & =\frac{1}{\hbar}\mathrm{E}\left[\int_{t_{I}}^{t}\left(\sum_{i=1}^{N}E_{i}-\sum_{i=1}^{N}\left(m_{i}\mathbf{v}_{i}\cdot\mathbf{v}_{i}+m_{i}\mathbf{u}_{i}\cdot\mathbf{u}_{i}\right)\right)dt'\left|\mathbf{q}_{j}(t)\right.\right]+\phi\\
 & =\frac{1}{\hbar}\mathrm{E}\left[\int_{t_{I}}^{t}\left(\sum_{i=1}^{N}\left(E_{i}-m_{i}\mathbf{u}_{i}\cdot\mathbf{u}_{i}\right)-\sum_{i=1}^{N}m_{i}\mathbf{v}_{i}\cdot\mathbf{v}_{i}\right)dt'\left|\mathbf{q}_{j}(t)\right.\right]+\phi\\
 & =\frac{1}{\hbar}\mathrm{E}\left[\int_{t_{I}}^{t}\left(H-\sum_{i=1}^{N}m_{i}\mathbf{v}_{i}\cdot\mathbf{v}_{i}\right)dt'\left|\mathbf{q}_{j}(t)\right.\right]+\phi\\
 & =\frac{1}{\hbar}\mathrm{E}\left[\int_{t_{I}}^{t}\left(H-\sum_{i=1}^{N}\frac{m_{i}}{4}\left(D\mathbf{q}_{i}(t')+D_{*}\mathbf{q}_{i}(t')\right)\cdot\left(D+D_{*}\right)\mathbf{q}_{i}(t')\right)dt'\left|\mathbf{q}_{j}(t)\right.\right]+\phi\\
 & =\frac{1}{\hbar}\mathrm{E}\left[\int_{t_{I}}^{t}Hdt'-\sum_{i=1}^{N}\frac{m_{i}}{2}\int_{\mathbf{q}_{i}(t_{I})}^{\mathbf{q}_{i}(t)}\left(D\mathbf{q}_{i}(t')+D_{*}\mathbf{q}_{i}(t')\right)\cdot\mathrm{D}\mathbf{q}_{i}(t')\left|\mathbf{q}_{j}(t)\right.\right]+\phi,
\end{aligned}
\end{equation}
where 
\begin{equation}
H\coloneqq\sum_{i=1}^{N}\left(E_{i}-m_{i}\mathbf{u}_{i}\cdot\mathbf{u}_{i}\right)=\sum_{i=1}^{N}\left(m_{i}c^{2}+\frac{1}{2}m_{i}\mathbf{v}_{i}^{2}-\frac{1}{2}m_{i}\mathbf{u}_{i}^{2}\right),
\end{equation}
and where $0.5\left(D+D_{*}\right)\mathbf{q}_{i}(t)=\left(\partial_{t}+\sum_{j}\mathbf{v}_{j}(q(t),t)\cdot\nabla_{j}\right)\mathbf{q}_{i}(t)$,
and $\left(\partial_{t}+\sum_{j}\mathbf{v}_{j}\cdot\nabla_{j}\right)\mathbf{q}_{i}(t)\eqqcolon\mathrm{D}\mathbf{q}_{i}(t)/\mathrm{D}t$,
and $\mathrm{D}\mathbf{q}_{i}(t)=\left(\mathrm{D}\mathbf{q}_{i}(t)/\mathrm{D}t\right)dt$.
From (4.29), we can define the steady-state joint phase-principal
function
\begin{equation}
I(q(t),t)=-\hbar\bar{\theta}(q(t),t)=\mathrm{E}\left[\int_{t_{I}}^{t}\sum_{i=1}^{N}\left(\frac{1}{2}m_{i}\mathbf{v}_{i}^{2}+\frac{1}{2}m_{i}\mathbf{u}_{i}^{2}-m_{i}c^{2}\right)dt'\left|\mathbf{q}_{j}(t)\right.\right]-\phi,
\end{equation}
and we can use (4.31) to define the steady-state joint phase-action
\begin{equation}
\begin{aligned}J & =I_{IF}=\mathrm{E}\left[\int_{t_{I}}^{t_{F}}\sum_{i=1}^{N}\left(\frac{1}{2}m_{i}\mathbf{v}_{i}^{2}+\frac{1}{2}m_{i}\mathbf{u}_{i}^{2}-m_{i}c^{2}\right)dt'-\hbar\phi\right].\end{aligned}
\end{equation}
Now, consider an integral curve $\mathbf{Q}_{i}(t)$ of the \emph{i}-th
current velocity/momentum field, i.e., a solution of
\begin{equation}
m_{i}\frac{d\mathbf{Q}_{i}(t)}{dt}=m_{i}\mathbf{v}_{i}(Q(t),t)=\mathbf{p}_{i}(Q(t),t)=\nabla_{i}S(q,t)|_{\mathbf{q}_{j}=\mathbf{Q}_{j}(t)}.
\end{equation}
Then we can replace (4.29)'s functional depedence on $q(t)$ by $Q(t)$,
obtaining 
\begin{equation}
\begin{aligned}\bar{\theta}(Q(t),t) & =\frac{1}{\hbar}\int_{t_{I}}^{t}\left[H-\sum_{i=1}^{N}m_{i}\mathbf{v}_{i}(Q(t'),t')\cdot\frac{d\mathbf{Q}_{i}(t')}{dt'}\right]dt'+\phi\\
 & =\frac{1}{\hbar}\left[\int_{t_{I}}^{t}Hdt'-\sum_{i=1}^{N}\int_{\mathbf{Q}_{i}(t_{I})}^{\mathbf{Q}_{i}(t)}\mathbf{p}_{i}\cdot d\mathbf{Q}_{i}(t')\right]+\phi,
\end{aligned}
\end{equation}
where it should be noticed that we've dropped the conditional expectation.
So (4.34) denotes the cumulative, time-symmetric, steady-state joint
phase of the \emph{zbw} particles, evaluated along the time-symmetric
mean trajectories of the \emph{zbw} particles, i.e., the integral
curves of (4.33). 

Now, taking the total differential of the left hand side of (4.34)
gives
\begin{equation}
d\bar{\theta}=\sum_{i=1}^{N}\nabla_{i}\bar{\theta}|_{\mathbf{q}_{j}=\mathbf{Q}_{j}(t)}d\mathbf{Q}_{i}(t)+\partial_{t}\bar{\theta}|_{\mathbf{q}_{j}=\mathbf{Q}_{j}(t)}dt.
\end{equation}
Then, comparing (4.35) with (4.34-33), we can identify 
\begin{equation}
\mathbf{p}_{i}(Q(t),t)=-\hbar\nabla_{i}\bar{\theta}|_{\mathbf{q}_{j}=\mathbf{Q}_{j}(t)}=\nabla_{i}S|_{\mathbf{q}_{j}=\mathbf{Q}_{j}(t)}.
\end{equation}
Thus the \emph{i}-th current velocity in the lab frame corresponds
the gradient of the time-symmetrized steady-state joint phase of the
\emph{zbw} particles at the location of the \emph{i}-th \emph{zbw}
particle, and $S$ can be identified with the cumulative, time-symmetric,
steady-state joint phase function of the \emph{zbw} particles in the
lab frame. 

Applying the variational principle to (4.32), i.e., the conservative
diffusion constraint,
\begin{equation}
J=extremal,
\end{equation}
straightforward computation shows that this yields
\begin{equation}
\sum_{i=1}^{N}\frac{m_{i}}{2}\left[D_{*}D+DD_{*}\right]\mathbf{q}_{i}(t)=0.
\end{equation}
Moreover, since the $\delta\mathbf{q}_{i}(t)$ are independent (as
shown in Appendix 7.1), it follows from (4.38) that we have the individual
equations of motion
\begin{equation}
m_{i}\mathbf{a}_{i}(q(t),t)=\frac{m_{i}}{2}\left[D_{*}D+DD_{*}\right]\mathbf{q}_{i}(t)=0.
\end{equation}
By applying the mean derivatives in (4.38), and using that $\mathbf{b}_{i}=\mathbf{v}_{i}+\mathbf{u}_{i}$
and $\mathbf{b}_{i*}=\mathbf{v}_{i}-\mathbf{u}_{i}$, straightforward
manipulations give
\begin{equation}
\sum_{i=1}^{N}m_{i}\left[\partial_{t}\mathbf{v}_{i}+\mathbf{v}_{i}\cdot\nabla_{i}\mathbf{v}_{i}-\mathbf{u}_{i}\cdot\nabla_{i}\mathbf{u}_{i}-\frac{\hbar}{2m_{i}}\nabla_{i}^{2}\mathbf{u}_{i}\right]|_{\mathbf{q}_{j}=\mathbf{q}_{j}(t)}=0,
\end{equation}
hence
\begin{equation}
\begin{aligned}\sum_{i=1}^{N}m_{i}\mathbf{a}_{i}(q(t),t) & =\sum_{i=1}^{N}m_{i}\left[\frac{\partial\mathbf{v}_{i}}{\partial t}+\mathbf{v}_{i}\cdot\nabla_{i}\mathbf{v}_{i}-\mathbf{u}_{i}\cdot\nabla_{i}\mathbf{u}_{i}-\frac{\hbar}{2m_{i}}\nabla_{i}^{2}\mathbf{u}_{i}\right](q,t)|_{\mathbf{q}_{j}=\mathbf{q}_{j}(t)}\\
 & =\sum_{i=1}^{N}\nabla_{i}\left[\frac{\partial S(q,t)}{\partial t}+\frac{\left(\nabla_{i}S(q,t)\right)^{2}}{2m_{i}}-\frac{\hbar^{2}}{2m_{i}}\frac{\nabla_{i}^{2}\sqrt{\rho(q,t)}}{\sqrt{\rho(q,t)}}\right]|_{\mathbf{q}_{j}=\mathbf{q}_{j}(t)}=0.
\end{aligned}
\end{equation}
Integrating both sides of (4.41) and setting the arbitrary integration
constants equal to the rest energies, we then have the \emph{N}-particle
quantum Hamilton-Jacobi equation
\begin{equation}
\begin{aligned}\tilde{E}(q(t),t) & \coloneqq\sum_{i=1}^{N}\tilde{E}_{i}(q(t),t)\\
 & \coloneqq-\partial_{t}S(q(t),t)\\
 & =\sum_{i=1}^{N}m_{i}c^{2}+\sum_{i=1}^{N}\frac{\left(\nabla_{i}S(q,t)\right)^{2}}{2m_{i}}|_{\mathbf{q}_{j}=\mathbf{q}_{j}(t)}-\sum_{i=1}^{N}\frac{\hbar^{2}}{2m_{i}}\frac{\nabla_{i}^{2}\sqrt{\rho(q,t)}}{\sqrt{\rho(q,t)}}|_{\mathbf{q}_{j}=\mathbf{q}_{j}(t)},
\end{aligned}
\end{equation}
describing the total energy of the actual particles along their stochastic
trajectories $q(t)$. Alternatively, given the integral curves $\mathbf{Q}_{i}(t)$
of the reformulated mean acceleration equation
\begin{equation}
m_{i}\frac{d^{2}\mathbf{Q}_{i}(t)}{dt^{2}}=m_{i}\left(\partial_{t}\mathbf{v}_{i}+\mathbf{v}_{i}\cdot\nabla_{i}\mathbf{v}_{i}\right)|_{\mathbf{q}_{j}=\mathbf{Q}_{j}(t)}=-\nabla_{i}\left(-\frac{\hbar^{2}}{2m_{i}}\frac{\nabla_{i}^{2}\sqrt{\rho(q,t)}}{\sqrt{\rho(q,t)}}\right)|_{\mathbf{q}_{j}=\mathbf{Q}_{j}(t)},
\end{equation}
for $i=1,...,N$, we can replace $q(t)$ by $Q(t)$ and thereby obtain
the total energy $\tilde{E}(Q(t),t)$ of the actual \emph{zbw} particles
along their time-symmetric mean trajectories, the latter now specified
by (4.43). The corresponding general solution of (4.42) is then
\begin{equation}
\begin{aligned}S(Q(t),t) & =\sum_{i=1}^{N}\int_{\mathbf{Q}_{i}(t_{I})}^{\mathbf{Q}_{i}(t)}\mathbf{p}_{i}(Q(t'),t')\cdot d\mathbf{Q}_{i}(t')-\sum_{i=1}^{N}\int_{t_{I}}^{t}\tilde{E}_{i}(Q(t'),t')dt'-\sum_{i=1}^{N}\hbar\phi_{i}\\
 & =\int_{t_{I}}^{t}\sum_{i=1}^{N}\left[\frac{1}{2}m_{i}\mathbf{v}_{i}^{2}-\left(-\frac{\hbar^{2}}{2m_{i}}\frac{\nabla_{i}^{2}\sqrt{\rho}}{\sqrt{\rho}}\right)-m_{i}c^{2}\right]dt'-\sum_{i=1}^{N}\hbar\phi_{i}\\
 & =\int_{t_{I}}^{t}\sum_{i=1}^{N}\left[\frac{1}{2}m_{i}\mathbf{v}_{i}^{2}+\frac{1}{2}m_{i}\mathbf{u}_{i}^{2}+\frac{\hbar}{2}\nabla_{i}\cdot\mathbf{u}_{i}-m_{i}c^{2}\right]dt'-\sum_{i=1}^{N}\hbar\phi_{i},
\end{aligned}
\end{equation}
which is the conservative-diffusion-constrained, time-symmetric, steady-state
joint phase associated with the \emph{zbw} particles in the lab frame. 

Since each \emph{zbw} particle is posited to essentially be a harmonic
oscillator of (unspecified) identical type, each particle has its
own, effectively independent, well-defined phase at each point along
its time-symmetric mean space-time trajectory, when $\mathbf{v}_{i}(q,t)\approx\sum_{i}^{N}\mathbf{v}_{i}(\mathbf{q}_{i},t)$.
Consistency with this means that when $\mathbf{v}_{i}(q,t)\neq\sum_{i}^{N}\mathbf{v}_{i}(\mathbf{q}_{i},t)$,
the time-symmetric steady-state joint phase must be a well-defined
function of the time-symmetric mean trajectories of\emph{ all} the
particles (since we posit that all the particles remain harmonic oscillators,
despite having their oscillations physically coupled through the common
ether medium they interact with). Then, for a closed loop \emph{L}
along which each particle can be physically or virtually displaced,
it follows that
\begin{equation}
\oint_{L}\delta S(Q(t),t)=\sum_{i=1}^{N}\oint_{L}\left[\mathbf{p}_{i}(Q(t),t)\cdot\delta\mathbf{Q}_{i}(t)-\tilde{E_{i}}(Q(t),t)\delta t\right]=nh.
\end{equation}
And for a closed loop $L$ with $\delta t=0$, we have
\begin{equation}
\sum_{i=1}^{N}\oint_{L}\mathbf{p}_{i}\cdot\delta\mathbf{Q}_{i}(t)=\sum_{i=1}^{N}\oint_{L}\mathbf{\nabla}_{i}S(q,t)|_{\mathbf{q}_{j}=\mathbf{Q}_{j}(t)}\cdot\delta\mathbf{Q}_{i}(t)=nh.
\end{equation}
If we also consider the joint phase field $S(q,t)$, a field over
the possible positions of the \emph{zbw} particles, then, as a result
of the same physical reasoning applied to the \emph{i}-th particle
at any possible initial position it can occupy, we will have
\begin{equation}
\sum_{i=1}^{N}\oint_{L}\mathbf{p}_{i}\cdot d\mathbf{q}_{i}=\sum_{i=1}^{N}\oint_{L}\nabla_{i}S(q,t)\cdot d\mathbf{q}_{i}=nh.
\end{equation}
Combining (4.47), (4.42), and (4.26), we can construct the \emph{N}-particle
Schr\"{o}dinger equation
\begin{equation}
i\hbar\frac{\partial\psi(q,t)}{\partial t}=\sum_{i=1}^{N}\left[-\frac{\hbar^{2}}{2m_{i}}\nabla_{i}^{2}+m_{i}c^{2}\right]\psi(q,t),
\end{equation}
where the \emph{N}-particle wavefunction $\psi(q,t)=\sqrt{\rho(q,t)}e^{iS(q,t)/\hbar}$
is single-valued by (4.47). 

We should observe that the solutions of (4.48) are generally non-factorizable
fields on 3N-dimensional configuration space, which implies non-separability
of $S$ and $R$ (hence non-factorizability of $\rho$) in general.
Insofar as ZSM starts with the heuristic hypothesis of an ontic ether
that lives in 3-space and couples to ontic \emph{zbw} particles in
3-space, this would seem \emph{prima facie} paradoxical, assuming
one takes the mathematical representation of $S$ and $R$ as a literal
indication of the ontic nature of the hypothesized ether (i.e., that
if $S$ and $R$ live in configuration space, then so must the ether).
As discussed at length in \cite{Derakhshani2016b}, there are three
possible ways to resolve this apparent inconsistency: (i) postulate
that the ether lives in configuration space, but, as a matter of physical
law, determines the motion of \emph{N} \emph{zbw} particles in 3-space;
(ii) postulate that the ether lives in configuration space along with
a \emph{zbw} `world particle' (in analogy with Albert's formulation
of the de Broglie-Bohm theory \cite{Albert2015}), and employ a philosophical
functionalist analysis to deduce the emergence of \emph{N} \emph{zbw}
particles floating in a common 3-space; and (iii) view the $S$ and
$R$ fields on configuration space as convenient mathematical representations
of some corresponding ontic fields on 3-space (in analogy with Norsen's
``TELB'' approach to the de Broglie-Bohm theory \cite{Norsen2010,Norsen2014})
which couple to \emph{N} \emph{zbw} particles in 3-space. As also
discussed in \cite{Derakhshani2016b}, we view option (iii) to be
the most natural and fruitful one for ZSM, and we will implicitly
assume this viewpoint throughout this paper.

With the overview completed, we can now develop ZSM-Newton/Coulomb.

\section{ZSM-Newton/Coulomb: Basic equations}

ZSM-Newton/Coulomb is just the generalization of $N$-particle ZSM
to include classical Newtonian gravitational and Coulomb interactions
between the $zbw$ particles. 

We suppose again that each particle undergoes a steady-state \emph{zbw}
oscillation in its IMFTRF or IMBTRF, and now also that each \emph{zbw}
particle carries charge $e_{i}$, making them classical charged harmonic
oscillators of identical type. (We subject these particles to the
hypothetical constraint of no electromagnetic radiation emitted when
there is no translational motion; or the constraint that the oscillation
of the charge is radially symmetric so that there is no net energy
radiated; or, if the ether turns out to be electromagnetic in nature
as Nelson suggested \cite{Nelson1985nopagelist,Nelson1986}, then
that the steady-state \emph{zbw} oscillations of the particles are
due to a balancing between the random-phase-average electromagnetic
energy absorbed via the driven oscillations of the particle charges,
and the random-phase-average electromagnetic energy radiated back
to the ether by the particles.) So the classical Newtonian gravitational
and Coulomb interactions between the \emph{zbw} particles are defined
by the gravitational potential (in CGS units) 
\begin{equation}
V_{g}^{int}(\mathbf{q}_{i}(t),\mathbf{q}_{j}(t))=\sum_{i=1}^{N}\frac{m_{i}\Phi_{g}}{2}=-\sum_{i=1}^{N}\frac{m_{i}}{2}\sum_{j=1}^{N(j\neq i)}\frac{m_{j}}{|\mathbf{q}_{i}(t)-\mathbf{q}_{j}(t)|},
\end{equation}
and the Coulomb potential 
\begin{equation}
V_{c}^{int}(\mathbf{q}_{i}(t),\mathbf{q}_{j}(t))=\sum_{i=1}^{N}\frac{e_{i}\Phi_{c}}{2}=\sum_{i=1}^{N}\frac{e_{i}}{2}\sum_{j=1}^{N(j\neq i)}\frac{e_{j}}{|\mathbf{q}_{i}(t)-\mathbf{q}_{j}(t)|}
\end{equation}
respectively, under the point-like interaction assumption $|\mathbf{q}_{i}(t)-\mathbf{q}_{j}(t)|\gg\lambda_{c}$. 

Then the forward joint \emph{zbw} phase change of the particles in
the lab frame (in the $\left|\mathbf{b}_{i}\right|\ll c$ approximation)
is given by
\begin{equation}
\begin{aligned}\delta\bar{\theta}_{+}(\mathbf{q}_{i}(t),\mathbf{q}_{j}(t),t) & =\mathrm{E}_{t}\left[\sum_{i=1}^{N}\left(\omega_{ic}+\omega_{ci}\frac{\mathbf{b}_{i}^{2}}{2c^{2}}+\omega_{ci}\left(\frac{\Phi_{g}}{2c^{2}}+\frac{e_{i}\Phi_{c}}{m_{i}c^{2}}\right)\right)\left(\delta t-\sum_{i=1}^{N}\frac{\mathbf{b}_{0i}}{c^{2}}\cdot\delta\mathbf{q}_{i+}(t)\right)\right]\\
 & =\mathrm{E}_{t}\left[\sum_{i=1}^{N}\left(\omega_{ic}+\omega_{ci}\frac{\mathbf{b}_{i}^{2}}{2c^{2}}+\omega_{ci}\left(\frac{\Phi_{g}}{2c^{2}}+\frac{e_{i}\Phi_{c}}{m_{i}c^{2}}\right)\right)\delta t-\sum_{i=1}^{N}\omega_{ci}\left(\frac{\mathbf{b}_{i}}{c^{2}}\right)\cdot\delta\mathbf{q}_{i+}(t)\right]\\
 & =\frac{1}{\hbar}\mathrm{E}_{t}\left[\left(\sum_{i=1}^{N}m_{i}c^{2}+\sum_{i=1}^{N}\frac{m_{i}\mathbf{b}_{i}^{2}}{2}+V_{g}^{int}+V_{c}^{int}\right)\delta t-\sum_{i=1}^{N}m_{i}\mathbf{b}_{i}\cdot\delta\mathbf{q}_{i+}(t)\right].
\end{aligned}
\end{equation}
The backward joint \emph{zbw} phase change $\delta\bar{\theta}_{-}$
differs by $\mathbf{b}_{i}\rightarrow-\mathbf{b}_{i*}$, $\delta t\rightarrow-\delta t$,
and $\delta\mathbf{q}_{i+}(t)\rightarrow\delta\mathbf{q}_{i-}(t)$. 

Note that when $|\mathbf{q}_{i}(t)-\mathbf{q}_{j}(t)|$ becomes sufficiently
great that $V_{g,c}^{int}$ is negligible, (4.51) reduces to an effectively
separable sum of the forward steady-state phase changes for all the
\emph{zbw} particles. (Effectively, because the ether will of course
still physically correlate the phase changes of the particles, even
if negligibly.) 

We can rewrite (4.51) in the form
\begin{equation}
\delta\bar{\theta}_{+}(q(t),t)=\frac{1}{\hbar}\mathrm{E}_{t}\left[E_{joint+}(q(t),Dq(t),t)\delta t-\sum_{i=1}^{N}m_{i}\mathbf{b}_{i}(q(t),t)\cdot\delta\mathbf{q}_{i+}(t)\right],
\end{equation}
where
\begin{equation}
E_{joint+}=\sum_{i=1}^{N}m_{i}c^{2}+\sum_{i=1}^{N}\frac{m_{i}\mathbf{b}_{i}^{2}}{2}+V_{g}^{int}+V_{c}^{int}.
\end{equation}
Correspondingly,
\begin{equation}
\delta\bar{\theta}_{-}(q(t),t)=\frac{1}{\hbar}\left[-E_{joint-}(q(t),D_{*}q(t),t)\delta t+\sum_{i=1}^{N}m_{i}\mathbf{b}_{i*}(q(t),t)\cdot\delta\mathbf{q}_{i-}(t)\right],
\end{equation}
where
\begin{equation}
E_{joint-}=\sum_{i=1}^{N}m_{i}c^{2}+\sum_{i=1}^{N}\frac{m_{i}\mathbf{b}_{i*}^{2}}{2}+V_{g}^{int}+V_{c}^{int}.
\end{equation}
Because each \emph{zbw} particle is essentially a harmonic oscillator,
when $V_{g,c}^{int}\approx0$, each particle has its own well-defined
(forward/backward) steady-state phase at each point along its mean
(forward/backward) space-time trajectory. Consistency with this fact
entails that for $V_{g,c}^{int}>0$, the (forward/backward) joint
phase must be a well-defined function of the mean (forward/backward)
space-time trajectories of\emph{ }both particles (since we again posit
that both particles remain harmonic oscillators even when physically
coupled by $V_{g,c}^{int}$). Furthermore, we note that at this stage
(4.52) and (4.54) are independent of one another. Accordingly, for
a closed loop \emph{L} along which each particle can be physically
or virtually displaced, the forward steady-state joint phase in the
lab frame will satisfy
\begin{equation}
\oint_{L}\delta\bar{\theta}_{+}=2\pi n,
\end{equation}
and likewise for the steady-state backward joint phase. It also follows
from (4.56) that
\begin{equation}
\oint_{L}\delta_{1}\bar{\theta}_{+}=2\pi n,
\end{equation}
where the closed-loop integral here keeps the coordinates of all the
particles fixed except for particle 1, the latter of which is displaced
along \emph{L}. 

In the lab frame, the forward and backward stochastic differential
equations for the translational motion are 
\begin{equation}
d\mathbf{q}_{i}(t)=\mathbf{b}_{i}(q(t),t)+d\mathbf{W}_{i}(t),
\end{equation}
and
\begin{equation}
d\mathbf{q}_{i}(t)=\mathbf{b}_{i*}(q(t),t)+d\mathbf{W}_{i*}(t),
\end{equation}
with corresponding Fokker-Planck equations
\begin{equation}
\frac{\partial\rho(q,t)}{\partial t}=-\sum_{i=1}^{N}\nabla_{i}\cdot\left[\left(\mathbf{b}_{i}(q,t)\right)\rho(q,t)\right]+\sum_{i=1}^{N}\frac{\hbar}{2m_{i}}\nabla_{i}^{2}\rho(q,t),
\end{equation}
and
\begin{equation}
\frac{\partial\rho(q,t)}{\partial t}=-\sum_{i=1}^{N}\nabla_{i}\cdot\left[\left(\mathbf{b}_{i*}(q,t)\right)\rho(q,t)\right]-\sum_{i=1}^{N}\frac{\hbar}{2m_{i}}\nabla_{i}^{2}\rho(q,t).
\end{equation}
Restricting to simultaneous solutions of (4.60-61) leads us to
\begin{equation}
\mathbf{v}_{i}\coloneqq\frac{1}{2}\left[\mathbf{b}_{i}+\mathbf{b}_{i*}\right]=\frac{\nabla_{i}S}{m_{i}},
\end{equation}
and
\begin{equation}
\mathbf{u}_{i}\coloneqq\frac{1}{2}\left[\mathbf{b}_{i}-\mathbf{b}_{i*}\right]=\frac{\hbar}{2m_{i}}\frac{\nabla_{i}\rho}{\rho}.
\end{equation}
Then (4.60-61) reduce to
\begin{equation}
\frac{\partial\rho}{\partial t}=-\sum_{i=1}^{N}\nabla_{i}\cdot\left[\frac{\nabla_{i}S}{m_{i}}\rho\right],
\end{equation}
where we have $\mathbf{b}_{i}=\mathbf{v}_{i}+\mathbf{u}_{i}$ and
$\mathbf{b}{}_{i*}=\mathbf{v}_{i}-\mathbf{u}_{i}$. 

Here again we postulate an osmotic potential to which each particle
couples via $R(q(t),t)=\mu U(q(t),t)$, which imparts momentum $\nabla_{i}R(q,t)|_{\mathbf{q}_{j}=\mathbf{q}_{j}(t)}$
that is counter-balanced by the osmotic impulse from the ether, $\left(\hbar/2m_{i}\right)\nabla_{i}\ln[n(q,t)]|_{\mathbf{q}_{j}=\mathbf{q}_{j}(t)}$,
giving the equilibrium condition $\nabla_{i}R/m_{i}=\left(\hbar/2m_{i}\right)\nabla_{i}\rho/\rho$.
Thus $\rho=e^{2R/\hbar}$ for all times

Now, taking the difference between $\delta\bar{\theta}_{+}$ and $\delta\bar{\theta}_{-}$
, we obtain
\begin{equation}
\begin{aligned}d\bar{\theta}(q(t),t) & \coloneqq\frac{1}{2}\left[d\bar{\theta}_{+}(q(t),t)-d\bar{\theta}_{-}(q(t),t)\right]\\
 & =\frac{1}{\hbar}\mathrm{E}_{t}\left[\sum_{i=1}^{N}\left(E_{i}(q(t),D\mathbf{q}_{i}(t),D_{*}\mathbf{q}_{i}(t),t)dt-\frac{m_{i}}{2}\left(\mathbf{b}_{i}\cdot d\mathbf{q}_{i+}(t)+\mathbf{b}_{i*}\cdot d\mathbf{q}_{i-}(t)\right)\right)\right]+\phi\\
 & =\frac{1}{\hbar}\mathrm{E}_{t}\left[\left(\sum_{i=1}^{N}E_{i}dt-\sum_{i=1}^{N}\frac{m_{i}}{2}\left(\mathbf{b}_{i}\cdot\frac{d\mathbf{q}_{i+}(t)}{dt}+\mathbf{b}_{i*}\cdot\frac{d\mathbf{q}_{i-}(t)}{dt}\right)dt\right)\right]+\phi\\
 & =\frac{1}{\hbar}\mathrm{E}_{t}\left[E_{joint}-\sum_{i=1}^{N}\frac{m_{i}}{2}\left(\mathbf{b}{}_{i}\cdot\frac{d\mathbf{q}_{i+}(t)}{dt}+\mathbf{b}_{i*}\cdot\frac{d\mathbf{q}_{i-}(t)}{dt}\right)\right]dt+\phi\\
 & =\frac{1}{\hbar}\mathrm{E}_{t}\left[E_{joint}-\sum_{i=1}^{N}\frac{m_{i}}{2}\left(\mathbf{b}{}_{i}\cdot\mathbf{b}_{i}+\mathbf{b}{}_{i*}\cdot\mathbf{b}_{i*}\right)\right]dt+\phi\\
 & =\frac{1}{\hbar}\mathrm{E}_{t}\left[E_{joint}-\sum_{i=1}^{N}\frac{m_{i}}{2}\left(\mathbf{b}{}_{i}^{2}+\mathbf{b}_{i*}^{2}\right)\right]dt+\phi\\
 & =\frac{1}{\hbar}\mathrm{E}_{t}\left[E_{joint}-\sum_{i=1}^{N}\frac{m_{i}}{2}\left(\mathbf{b}_{i}{}^{2}+\mathbf{b}_{i*}^{2}\right)\right]dt+\phi\\
 & =\frac{1}{\hbar}\mathrm{E}_{t}\left[E_{joint}-\sum_{i=1}^{N}\left(m_{i}\mathbf{v}_{i}\cdot\mathbf{v}_{i}+m_{i}\mathbf{u}_{i}\cdot\mathbf{u}_{i}\right)\right]dt+\phi\\
 & =\frac{1}{\hbar}\mathrm{E}_{t}\left[\sum_{i=1}^{N}\left(m_{i}c^{2}-\frac{1}{2}m_{i}\mathbf{v}_{i}^{2}-\frac{1}{2}m_{i}\mathbf{u}_{i}^{2}\right)+V_{g}^{int}+V_{c}^{int}\right]dt+\phi,
\end{aligned}
\end{equation}
where
\begin{equation}
\begin{aligned}E_{joint} & \coloneqq\sum_{i=1}^{N}E_{i}\\
 & \coloneqq\frac{1}{2}\left[E_{joint+}+E_{joint-}\right]\\
 & =\sum_{i=1}^{N}m_{i}c^{2}+\sum_{i=1}^{N}\frac{1}{2}\left[\frac{1}{2}m_{i}\mathbf{b}_{i}{}^{2}+\frac{1}{2}m\mathbf{b}_{i*}{}^{2}\right]+V_{g}^{int}+V_{c}^{int}\\
 & =\sum_{i=1}^{N}m_{i}c^{2}+\sum_{i=1}^{N}\left[\frac{1}{2}m_{i}\mathbf{v}{}_{i}{}^{2}+\frac{1}{2}m_{i}\mathbf{u}_{i}^{2}\right]+V_{g}^{int}+V_{c}^{int}.
\end{aligned}
\end{equation}
As in the free particle case, the consistency of our theory requires
that (4.65) satisfies 
\begin{equation}
\oint_{L}\delta\bar{\theta}=2\pi n.
\end{equation}
Otherwise we would contradict our hypothesis that the \emph{zbw} particle
has a well-defined, unique, time-symmetric steady-state joint phase
at each 3-space location it can occupy.

From the second to last line of (4.65), we can then obtain the cumulative,
time-symmetric, steady-state joint phase at time \emph{t} as 
\begin{equation}
\begin{aligned}\bar{\theta}(q(t),t) & =\frac{1}{\hbar}\mathrm{E}\left[\int_{t_{I}}^{t}\left(\sum_{i=1}^{N}E_{i}-\sum_{i=1}^{N}\left(m_{i}\mathbf{v}_{i}\cdot\mathbf{v}_{i}+m_{i}\mathbf{u}_{i}\cdot\mathbf{u}_{i}\right)\right)dt'\left|\mathbf{q}_{j}(t)\right.\right]+\phi\\
 & =\frac{1}{\hbar}\mathrm{E}\left[\int_{t_{I}}^{t}\left(\sum_{i=1}^{N}\left(E_{i}-m_{i}\mathbf{u}_{i}\cdot\mathbf{u}_{i}\right)-\sum_{i=1}^{N}m_{i}\mathbf{v}_{i}\cdot\mathbf{v}_{i}\right)dt'\left|\mathbf{q}_{j}(t)\right.\right]+\phi\\
 & =\frac{1}{\hbar}\mathrm{E}\left[\int_{t_{I}}^{t}\left(H-\sum_{i=1}^{N}m_{i}\mathbf{v}_{i}\cdot\mathbf{v}_{i}\right)dt'\left|\mathbf{q}_{j}(t)\right.\right]+\phi\\
 & =\frac{1}{\hbar}\mathrm{E}\left[\int_{t_{I}}^{t}\left(H-\sum_{i=1}^{N}\frac{m_{i}}{4}\left(D\mathbf{q}_{i}(t')+D_{*}\mathbf{q}_{i}(t')\right)\cdot\left(D+D_{*}\right)\mathbf{q}_{i}(t')\right)dt'\left|\mathbf{q}_{j}(t)\right.\right]+\phi\\
 & =\frac{1}{\hbar}\mathrm{E}\left[\int_{t_{I}}^{t}Hdt'-\sum_{i=1}^{N}\frac{m_{i}}{2}\int_{\mathbf{q}_{i}(t_{I})}^{\mathbf{q}_{i}(t)}\left(D\mathbf{q}_{i}(t')+D_{*}\mathbf{q}_{i}(t')\right)\cdot\mathrm{D}\mathbf{q}_{i}(t')\left|\mathbf{q}_{j}(t)\right.\right]+\phi
\end{aligned}
\end{equation}
where 
\begin{equation}
H\coloneqq\sum_{i=1}^{N}\left(E_{i}-m_{i}\mathbf{u}_{i}\cdot\mathbf{u}_{i}\right)=\sum_{i=1}^{N}\left(m_{i}c^{2}+\frac{1}{2}m_{i}\mathbf{v}_{i}^{2}-\frac{1}{2}m_{i}\mathbf{u}_{i}^{2}\right)+V_{g}^{int}+V_{c}^{int}.
\end{equation}
From (4.68), we can define the steady-state joint phase-principal
function
\begin{equation}
I(q(t),t)=-\hbar\bar{\theta}(q(t),t)=\mathrm{E}\left[\int_{t_{I}}^{t}\left(\sum_{i=1}^{N}\left(\frac{1}{2}m_{i}\mathbf{v}_{i}^{2}+\frac{1}{2}m_{i}\mathbf{u}_{i}^{2}-m_{i}c^{2}\right)-V_{g}^{int}-V_{c}^{int}\right)dt'\left|\mathbf{q}_{j}(t)\right.\right]-\phi,
\end{equation}
and we can use (4.70) to define the steady-state joint phase-action
\begin{equation}
\begin{aligned}J & =I_{IF}=\mathrm{E}\left[\int_{t_{I}}^{t_{F}}\left(\sum_{i=1}^{N}\left(\frac{1}{2}m_{i}\mathbf{v}_{i}^{2}+\frac{1}{2}m_{i}\mathbf{u}_{i}^{2}-m_{i}c^{2}\right)-V_{g}^{int}-V_{c}^{int}\right)dt'-\hbar\phi\right].\end{aligned}
\end{equation}
Now, given an integral curve $\mathbf{Q}_{i}(t)$ of the \emph{i}-th
current velocity/momentum field, i.e., a solution of
\begin{equation}
m_{i}\frac{d\mathbf{Q}_{i}(t)}{dt}=m_{i}\mathbf{v}_{i}(Q(t),t)=\mathbf{p}_{i}(Q(t),t)=\nabla_{i}S(q,t)|_{\mathbf{q}_{j}=\mathbf{Q}_{j}(t)},
\end{equation}
we can replace (4.68)'s functional depedence on $q(t)$ by $Q(t)$,
obtaining 
\begin{equation}
\begin{aligned}\bar{\theta}(Q(t),t) & =\frac{1}{\hbar}\int_{t_{I}}^{t}\left[H-\sum_{i=1}^{N}m_{i}\mathbf{v}_{i}(Q(t'),t')\cdot\frac{d\mathbf{Q}_{i}(t')}{dt'}\right]dt'+\phi\\
 & =\frac{1}{\hbar}\left[\int_{t_{I}}^{t}Hdt'-\sum_{i=1}^{N}\int_{\mathbf{Q}_{i}(t_{I})}^{\mathbf{Q}_{i}(t)}\mathbf{p}_{i}\cdot d\mathbf{Q}_{i}(t')\right]+\phi,
\end{aligned}
\end{equation}
Taking the total differential of the left hand side of (4.73) gives
\begin{equation}
d\bar{\theta}=\sum_{i=1}^{N}\nabla_{i}\bar{\theta}|_{\mathbf{q}_{j}=\mathbf{Q}_{j}(t)}d\mathbf{Q}_{i}(t)+\partial_{t}\bar{\theta}|_{\mathbf{q}_{j}=\mathbf{Q}_{j}(t)}dt.
\end{equation}
Then, comparing (4.74) with (4.73-72), we can once again identify
\begin{equation}
\mathbf{p}_{i}(Q(t),t)=-\hbar\nabla_{i}\bar{\theta}|_{\mathbf{q}_{j}=\mathbf{Q}_{j}(t)}=\nabla_{i}S|_{\mathbf{q}_{j}=\mathbf{Q}_{j}(t)}.
\end{equation}
Now, returning to (4.71), imposing extremality yields
\begin{equation}
\sum_{i=1}^{N}\frac{m_{i}}{2}\left[D_{*}D+DD_{*}\right]\mathbf{q}_{i}(t)=-\sum_{i=1}^{N}\nabla_{i}\frac{\left[m_{i}\Phi_{g}(\mathbf{q}_{i},\mathbf{q}_{j})+e_{i}\Phi_{c}(\mathbf{q}_{i},\mathbf{q}_{j})\right]}{2}|_{\mathbf{q}_{j}=\mathbf{q}_{j}(t)},
\end{equation}
for $i,j=1,...,N$ and $i\neq j$. And from the independent $\delta\mathbf{q}_{i}(t)$,
we have the individual equations of motion
\begin{equation}
m_{i}\mathbf{a}_{i}(q(t),t)=\frac{m_{i}}{2}\left[D_{*}D+DD_{*}\right]\mathbf{q}_{i}(t)=-\nabla_{i}\frac{\left[m_{i}\Phi_{g}(\mathbf{q}_{i},\mathbf{q}_{j})+e_{i}\Phi_{c}(\mathbf{q}_{i},\mathbf{q}_{j})\right]}{2}|_{\mathbf{q}_{j}=\mathbf{q}_{j}(t)}.
\end{equation}
Following the same steps as in the previous section, we then get from
(4.77) the quantum Hamilton-Jacobi equation
\begin{equation}
\begin{aligned}\tilde{E}(q(t),t) & =\sum_{i=1}^{N}\tilde{E}_{i}(q(t),t)\\
 & =-\partial_{t}S(q,t)|_{\mathbf{q}_{j}=\mathbf{q}_{j}(t)}\\
 & =\sum_{i=1}^{N}m_{i}c^{2}+\sum_{i=1}^{N}\frac{\left[\nabla_{i}S(q,t)\right]^{2}}{2m_{i}}|_{\mathbf{q}_{j}=\mathbf{q}_{j}(t)}\\
 & +\sum_{i=1}^{N}\frac{\left[m_{i}\Phi_{g}(\mathbf{q}_{i}(t),\mathbf{q}_{j}(t))+e_{i}\Phi_{c}(\mathbf{q}_{i}(t),\mathbf{q}_{j}(t))\right]}{2}-\sum_{i=1}^{N}\frac{\hbar^{2}}{2m_{i}}\frac{\nabla_{i}^{2}\sqrt{\rho(q,t)}}{\sqrt{\rho(q,t)}}|_{\mathbf{q}_{j}=\mathbf{q}_{j}(t)}.
\end{aligned}
\end{equation}
Or, in terms of the integral curves of the reformulated mean acceleration
equation,
\begin{equation}
\begin{aligned}m_{i}\frac{d^{2}\mathbf{Q}_{i}(t)}{dt^{2}} & =m_{i}\left(\partial_{t}\mathbf{v}_{i}+\mathbf{v}_{j}\cdot\nabla_{j}\mathbf{v}_{i}\right)|_{\mathbf{q}_{j}=\mathbf{Q}_{j}(t)}\\
 & =-\nabla_{i}\left(-\frac{\hbar^{2}}{2m_{i}}\frac{\nabla_{i}^{2}\sqrt{\rho(q,t)}}{\sqrt{\rho(q,t)}}\right)|_{\mathbf{q}_{j}=\mathbf{Q}_{j}(t)}-\nabla_{i}\frac{\left[m_{i}\Phi_{g}(\mathbf{q}_{i},\mathbf{q}_{j})+e_{i}\Phi_{c}(\mathbf{q}_{i},\mathbf{q}_{j})\right]}{2}|_{\mathbf{q}_{j}=\mathbf{Q}_{j}(t)},
\end{aligned}
\end{equation}
we can replace $\tilde{E}(q(t),t)$ by $\tilde{E}(Q(t),t)$, and write
the general solution of (4.78) as 
\begin{equation}
\begin{aligned}S(Q(t),t) & =\sum_{i=1}^{N}\int_{\mathbf{Q}_{i}(t_{I})}^{\mathbf{Q}_{i}(t)}\mathbf{p}_{i}(Q(t'),t')\cdot d\mathbf{Q}_{i}(t')-\sum_{i=1}^{N}\int_{t_{I}}^{t}\tilde{E}_{i}(Q(t'),t')dt'-\sum_{i=1}^{N}\hbar\phi_{i}\\
 & =\int_{t_{I}}^{t}\sum_{i=1}^{N}\left[\frac{1}{2}m_{i}\mathbf{v}_{i}^{2}-\left(-\frac{\hbar^{2}}{2m_{i}}\frac{\nabla_{i}^{2}\sqrt{\rho}}{\sqrt{\rho}}\right)-m_{i}c^{2}-\frac{\left(m_{i}\Phi_{g}+e_{i}\Phi_{c}\right)}{2}\right]dt'-\sum_{i=1}^{N}\hbar\phi_{i}\\
 & =\int_{t_{I}}^{t}\sum_{i=1}^{N}\left[\frac{1}{2}m_{i}\mathbf{v}_{i}^{2}+\frac{1}{2}m_{i}\mathbf{u}_{i}^{2}+\frac{\hbar}{2}\nabla_{i}\cdot\mathbf{u}_{i}-m_{i}c^{2}-\frac{\left(m_{i}\Phi_{g}+e_{i}\Phi_{c}\right)}{2}\right]dt'-\sum_{i=1}^{N}\hbar\phi_{i},
\end{aligned}
\end{equation}
which is the conservative-diffusion-constrained, time-symmetric, steady-state
joint phase associated with the classical interacting \emph{zbw} particles
in the lab frame.

We make the natural assumption that the presence of classical external
potentials doesn't alter the harmonic nature of the steady-state \emph{zbw}
oscillations. Moreover, since each \emph{zbw} particle is a harmonic
oscillator, each particle has its own well-defined steady-state phase
at each point along its time-symmetric mean trajectory. Accordingly,
when the classical interactions are not negligible, the steady-state
joint phase must be a well-defined function of the mean trajectories
of\emph{ }both particles (since we posit that all particles remain
harmonic oscillators despite having their oscillations physically
coupled through $\Phi_{g,c}$ and through the common ether medium
with which they interact). So for a closed loop \emph{L} along which
each particle can be physically or virtually displaced, it follows
that
\begin{equation}
\oint_{L}\delta S=\sum_{i=1}^{N}\oint_{L}\left[\mathbf{p}_{i}\cdot\delta\mathbf{Q}_{i}(t)-\tilde{E}_{i}\delta t\right]=nh,
\end{equation}
and
\begin{equation}
\sum_{i=1}^{2}\oint_{L}\mathbf{p}_{i}'\cdot\delta\mathbf{Q}_{i}(t)=\sum_{i=1}^{2}\oint_{L}\mathbf{\nabla}_{i}S|_{\mathbf{q}_{j}=\mathbf{Q}_{j}(t)}\cdot\delta\mathbf{Q}_{i}(t)=nh,
\end{equation}
for a closed loop $L$ with $\delta t=0$. For the steady-state joint
phase field $S(q,t)$, we can apply the same physical reasoning above
to each \emph{zbw} particle for each possible 3-space position that
can be occupied at time \emph{t}, thereby implying
\begin{equation}
\oint_{L}dS\left(q,t\right)=\sum_{i=1}^{N}\oint_{L}\mathbf{p}_{i}'\cdot d\mathbf{q}_{i}=\sum_{i=1}^{2}\oint_{L}\mathbf{\nabla}_{i}S\cdot d\mathbf{q}_{i}=nh.
\end{equation}
Applying the Madelung transformation to the combination of (4.83),
(4.78), and (4.64), we can construct the \emph{N}-particle Schr\"{o}dinger
equation for classically interacting \emph{zbw} particles in the presence
of external fields:
\begin{equation}
i\hbar\frac{\partial\psi(q,t)}{\partial t}=\sum_{i=1}^{N}\left[-\frac{\hbar^{2}}{2m_{i}}\nabla_{i}^{2}+m_{i}c^{2}+\frac{m_{i}\hat{\Phi}_{g}(\mathbf{\hat{q}}_{i},\mathbf{\hat{q}}_{j})}{2}+\frac{e_{i}\hat{\Phi}_{c}(\mathbf{\hat{q}}_{i},\mathbf{\hat{q}}_{j})}{2}\right]\psi(q,t),
\end{equation}
where $\psi(q,t)=\sqrt{\rho(q,t)}e^{iS(q,t)/\hbar}$ is single-valued
via (4.83).
Note the inclusion of hats on the interaction potentials and their
coordinates, in contrast to the quantum Hamilton-Jacobi (QHJ) equation
(4.78). As shown by Holland \cite{HollandBook1993} and Oriols \&
Mompart \cite{OriolsMompart2012}, there exists a correspondence between
quantum operators in the Schr\"{o}dinger
equation, and c-number variables in the QHJ equation. For example,
the quantum expectation value of the position operator corresponds
to the ensemble averaged position coordinate via $\left\langle \psi\right|\hat{\mathbf{q}}\left|\psi\right\rangle =\int_{\mathbb{R}^{3N}}d^{3N}\mathbf{q}\,\psi^{*}\,\mathbf{\hat{q}\,}\psi=\int_{\mathbb{R}^{3N}}d^{3N}\mathbf{q}\,\rho\,\mathbf{q}=<\mathbf{q}>$.
For another example, the quantum expectation value of the Hamiltonian
operator is equivalent to the ensemble average of the total energy
in the QHJ equation:
\begin{equation}
\begin{aligned}\left\langle \psi\right|\hat{H}\left|\psi\right\rangle  & =\int_{\mathbb{R}^{3N}}d^{3N}\mathbf{q}\,\psi^{*}(q,t)\left(\sum_{i=1}^{N}\left[-\frac{\hbar^{2}}{2m_{i}}\nabla_{i}^{2}+m_{i}c^{2}+\frac{m_{i}\hat{\Phi}_{g}(\mathbf{\hat{q}}_{i},\mathbf{\hat{q}}_{j})}{2}+\frac{e_{i}\hat{\Phi}_{c}(\mathbf{\hat{q}}_{i},\mathbf{\hat{q}}_{j})}{2}\right]\right)\psi(q,t)\\
 & =\int_{\mathbb{R}^{3N}}d^{3N}\mathbf{q}\,\rho(q,t)\left(\sum_{i=1}^{N}\left[m_{i}c^{2}+\frac{\left[\nabla_{i}S(q,t)\right]^{2}}{2m_{i}}+\frac{m_{i}\Phi_{g}(\mathbf{q}_{i},\mathbf{q}_{j})}{2}\right.\right.\\
 & \left.\left.+\frac{e_{i}\Phi_{c}(\mathbf{q}_{i},\mathbf{q}_{j})}{2}-\frac{\hbar^{2}}{2m_{i}}\frac{\nabla_{i}^{2}\sqrt{\rho(q,t)}}{\sqrt{\rho(q,t)}}\right]\right)=\,<H>.
\end{aligned}
\end{equation}
So the classical potentials are, in effect, `quantized' at the level
of the Schr\"{o}dinger equation, insofar
as they depend on q-number position coordinates and satisfy the Poisson
equations \footnote{The use of delta functions in the definitions of the mass and charge
densities is justified because we are using the point-like approximation
for interactions between the particles. In actuality, the mass and
charge densities should presumably depend on some form-factor $f(|\mathbf{x}-\mathbf{q}_{i}|)$
which distributes the mass or charge of the particle on its Compton
length-scale $\lambda_{c}$. Additionally, in scattering events where
$|\mathbf{q}_{i}(t)-\mathbf{q}_{j}(t)|\sim\lambda_{c}$ the point-like
approximation will no longer hold and it will become necessary to
include this form-factor in calculating the interactions. The precise
expression for this form-factor will depend on the specific physical
model used for the \emph{zbw} particle, which at present we do not
have (although see section 5 of \cite{Derakhshani2016b} for a discussion
of possibilities). Nevertheless, as we are only concerned here with
the non-relativistic regime, the point-like approximation will suffice.}
\begin{equation}
\nabla^{2}\hat{\Phi}_{g}=4\pi\sum_{i=1}^{N}m_{i}\delta^{3}\left(\mathbf{q}-\hat{\mathbf{q}}_{i}\right),
\end{equation}
 
\begin{equation}
\nabla^{2}\hat{\Phi}_{c}=-4\pi\sum_{i=1}^{N}e_{i}\delta^{3}\left(\mathbf{q}-\hat{\mathbf{q}}_{i}\right).
\end{equation}
Accordingly, the equation set (4.84-87) gives a statistical mechanical
description of $N$ \emph{zbw} particles undergoing Nelsonian diffusions,
while interacting both gravitationally and electrostatically through
the classical potentials (4.49-50). 

Equations (4.84-87) correspond to the standard quantum mechanical
equations for $N$ particles interacting gravitationally or electrostatically
in the Newtonian regime \cite{BardosGolseMauser2000,BardosErdosGolseMauserYau2002,Golse2003,AnastopoulosHu2014},
and that the standard quantum mechanical equations are the Newtonian
limits of the standard theories of perturbatively quantized gravity
and perturbative quantum electrodynamics \cite{AnastopoulosHu2014,BahramiBassi2014a}.
But because we derived (4.84-87) within the ZSM framework, we can
go further than the standard quantum description. That is, we can
use solutions of (4.84), or the equivalent solutions of the Madelung
equations, (4.64) and (4.78) with (4.83), to deduce an ensemble of
possible trajectories for the actual (\emph{zbw}) particles.

In particular, it is readily shown that the \emph{i}-th mean acceleration
\begin{equation}
\begin{aligned}m_{i}\mathbf{a}_{i}(q(t),t) & =\frac{m_{i}}{2}\left[D_{*}D+DD_{*}\right]\mathbf{q}_{i}(t)=-\nabla_{i}\frac{\left[m_{i}\Phi_{g}(\mathbf{q}_{i},\mathbf{q}_{j})+e_{i}\Phi_{c}(\mathbf{q}_{i},\mathbf{q}_{j})\right]}{2}|_{\mathbf{q}_{j}=\mathbf{q}_{j}(t)},\end{aligned}
\end{equation}
hence
\begin{equation}
\begin{aligned}m_{i}\frac{\mathrm{D}\mathbf{v}_{i}(q,t)}{\mathrm{D}t}|_{\mathbf{q}_{j}=\mathbf{Q}_{j}(t)} & =\left[\partial_{t}\mathbf{p}_{i}+\mathbf{v}_{i}\cdot\nabla_{i}\mathbf{p}_{i}\right](q,t)|_{\mathbf{q}_{j}=\mathbf{Q}_{j}(t)}\\
 & =-\nabla_{i}\left[\frac{m_{i}\Phi_{g}(\mathbf{q}_{i},\mathbf{q}_{j})}{2}+\frac{e_{i}\Phi_{c}(\mathbf{q}_{i},\mathbf{q}_{j})}{2}-\frac{\hbar^{2}}{2m_{i}}\frac{\nabla_{i}^{2}\sqrt{\rho(q,t)}}{\sqrt{\rho(q,t)}}\right]|_{\mathbf{q}_{j}=\mathbf{Q}_{j}(t)}.
\end{aligned}
\end{equation}
Integrating this last equation for different possible initial conditions
$\mathbf{Q}_{j}(0)$ allows us to construct an ensemble of mean trajectories,
only one of which is realized by the actual $i$-th\emph{ zbw} particle.
We can then find the mean gravitational and Coulomb potentials, i.e.,
the gravitational and Coulomb potentials sourced by the actual \emph{zbw}
particles along their mean trajectories, as follows: 
\begin{equation}
\nabla^{2}\Phi_{g}^{m.t.}=4\pi\sum_{i=1}^{N}m_{i}\delta^{3}\left(\mathbf{q}-\mathbf{Q}_{i}(t)\right),
\end{equation}
 
\begin{equation}
\nabla^{2}\Phi_{c}^{m.t.}=-4\pi\sum_{i=1}^{N}e_{i}\delta^{3}\left(\mathbf{q}-\mathbf{Q}_{i}(t)\right),
\end{equation}
where the superscript ``$m.t.$'' refers to the mean interaction
potentials sourced by the mean trajectories of the actual \emph{zbw}
particles. 

Actually, (4.90) doesn't contain all the terms that contribute to
the total mass-densities of the particles. The complete expression
is
\begin{equation}
\nabla^{2}\Phi_{g}^{m.t.}=4\pi\sum_{i=1}^{N}\left[m_{i}+\frac{\left[\nabla_{i}S\left(Q(t),t\right)\right]^{2}}{2m_{i}c^{2}}-\frac{\hbar^{2}}{2m_{i}c^{2}}\frac{\nabla_{i}^{2}\sqrt{\rho\left(Q(t),t\right)}}{\sqrt{\rho\left(Q(t),t\right)}}\right]\delta^{3}\left(\mathbf{q}-\mathbf{Q}_{i}(t)\right).
\end{equation}
But in the $v_{i}\ll c$ limit, the classical kinetic and quantum
kinetic \footnote{While the latter terms are referred to in the literature as ``quantum
potentials'' \cite{Broglie1930,Bohm1952I,Bohm1952II,BohmHiley1993,HollandBook1993,Duerr2009,OriolsMompart2012},
we prefer the term ``quantum kinetics'' \cite{Derakhshani2016a,Derakhshani2016b}
since, in stochastic mechanics, they arise from the kinetic energy
contributions of the osmotic velocities of the particles, as seen
from the left hand side of (4.41). } energy terms are negligible relative to the rest-energy terms, allowing
us to effectively neglect the contributions of the kinetic energies
to the total mass-energy density of the particle.

From the solutions of (4.84), we can also construct an ensemble of
possible stochastic trajectories for the $i$-th particle: 
\begin{equation}
d\mathbf{q}_{i}(t)=\left[\frac{\hbar}{m_{i}}\mathrm{Im}\frac{\nabla_{i}\psi(q,t)}{\psi(q,t)}+\frac{\hbar}{m_{i}}\mathrm{Re}\frac{\nabla_{i}\psi(q,t)}{\psi(q,t)}\right]|_{\mathbf{q}_{j}=\mathbf{q}_{j}(t)}dt+d\mathbf{W}_{i}(t),
\end{equation}

\begin{equation}
d\mathbf{q}_{i}(t)=\left[\frac{\hbar}{m_{i}}\mathrm{Im}\frac{\nabla_{i}\psi(q,t)}{\psi(q,t)}-\frac{\hbar}{m_{i}}\mathrm{Re}\frac{\nabla_{i}\psi(q,t)}{\psi(q,t)}\right]|_{\mathbf{q}_{j}=\mathbf{q}_{j}(t)}dt+d\mathbf{W}_{i*}(t).
\end{equation}
These stochastic trajectories can then be used in the definition of
the mass and charge densities, implying \emph{classically fluctuating}
mass and charge densities, hence \emph{classically fluctuating} gravitational
and Coulomb potentials satisfying the Poisson equations 
\begin{equation}
\nabla^{2}\Phi_{g}^{s.t.}=4\pi\sum_{i=1}^{N}m_{i}\delta^{3}\left(\mathbf{q}-\mathbf{q}_{i}(t)\right),
\end{equation}
 
\begin{equation}
\nabla^{2}\Phi_{c}^{s.t.}=-4\pi\sum_{i=1}^{N}e_{i}\delta^{3}\left(\mathbf{q}-\mathbf{q}_{i}(t)\right),
\end{equation}
where ``$s.t.$'' refers to the interaction potentials sourced by
the stochastic trajectories of the actual \emph{zbw} particles. 

Thus we see here that there are three `levels' of interaction potentials,
with $\Phi_{g,c}^{s.t.}$ being the most fundamental (in the sense
of being the potentials sourced by the actual, stochastic trajectories
of the actual \emph{zbw} particles), followed by $\Phi_{g,c}^{m.t.}$
(in the sense of being the potentials sourced by the mean trajectories
of the actual \emph{zbw} particles), and then $\Phi_{g,c}$ or $\hat{\Phi}_{g,c}$
(in the sense of being q-number potentials that reflect a statistical
ensemble of possible potentials sourced by the possible mean trajectories
of the actual \emph{zbw} particles). Indeed the q-number interaction
potentials $\hat{\Phi}_{g,c}$ have physical meaning inasmuch as 
\begin{equation}
\begin{aligned}\left\langle \psi\right|\hat{V}_{g}^{int}\left|\psi\right\rangle  & =\int_{\mathbb{R}^{3N}}d^{3N}\mathbf{q}\,\psi^{*}\hat{V}_{g}^{int}\,\psi=\int_{\mathbb{R}^{3N}}d^{3N}\mathbf{q}\,\rho\,V_{g}^{int}=<V_{g}^{int}>,\end{aligned}
\end{equation}
and
\begin{equation}
\begin{aligned}\left\langle \psi\right|\nabla^{2}\hat{\Phi}_{g}\left|\psi\right\rangle  & =4\pi\left\langle \psi\right|\sum_{i=1}^{N}m_{i}\delta^{3}\left(\mathbf{q}-\hat{\mathbf{q}}_{i}\right)\left|\psi\right\rangle \\
 & =4\pi\underset{i=1}{\overset{N}{\sum}}\int d^{3}\mathbf{q}'_{1}...d^{3}\mathbf{q}'_{N}|\psi(\mathbf{q}_{1}'...\mathbf{q}_{N}',t)|^{2}m_{i}\delta^{3}(\mathbf{q}-\mathbf{q}'_{i})\\
 & =4\pi\sum_{i=1}^{N}\int_{\mathbb{R}^{3N}}d^{3N}\mathbf{q}'\rho(q',t)\,m_{i}\delta^{3}\left(\mathbf{q}-\mathbf{Q}_{i}(t)\right)|_{\mathbf{Q}_{i}(t)=\mathbf{q}'_{i}}=<\nabla^{2}\Phi_{g}>,
\end{aligned}
\end{equation}
 and likewise for the Coulomb potentials.

Note the conceptual difference between the expected values of the
interaction potentials, equation (4.97-98), and the potentials obtained
from (4.90-91) (i.e., the potentials sourced by the mean trajectories
of the actual \emph{zbw} particles). The former are obtained from
averaging the interaction potentials over N statistical ensembles
of mean trajectories; the latter are obtained from using the integral
curves of (4.79) in (4.90-91).

It is interesting to compare (4.98) to the Poisson equation associated
with the \emph{N}-body Schr\"{o}dinger-Newton
(SN) gravitational potential \cite{Diosi1984,Adler2007,AnastopoulosHu2014,Derakhshani2014,BahramiBassi2014a,Giulini2014,DerakProbingGravCat2016}:
\begin{equation}
\nabla^{2}\Phi_{g}^{SN}=4\pi\underset{i=1}{\overset{N}{\sum}}\int d^{3}\mathbf{q}'_{1}...d^{3}\mathbf{q}'_{N}|\psi(\mathbf{q}_{1}'...\mathbf{q}_{N}',t)|^{2}m_{i}\delta^{3}(\mathbf{q}-\mathbf{q}'_{i}).
\end{equation}
In the SN equations, the solution of (4.99) describes the net interaction
potential sourced by \emph{N} matter density fields on space-time
(each field corresponding to an elementary `particle'), and this potential
feeds back into the Hamiltonian of the Schr\"{o}dinger
equation to generate a nonlinear Schr\"{o}dinger
evolution. In ZSM-Newton, by contrast, the solution of (4.99) describes
the ensemble-average of the net interaction potential sourced by the
\emph{N} point-like \emph{zbw} particles, and this potential \emph{does
not} feed back into the (derived) Schr\"{o}dinger
Hamiltonian. Everything said here also holds for the Coulombic analogue
of (4.99) and its comparison to the \emph{N}-body Schr\"{o}dinger-Coulomb
equations \cite{BardosGolseMauser2000,BardosErdosGolseMauserYau2002,Golse2003,Derakhshani2014,AnastopoulosHu2014}.
(See subsection 5.1 for a more detailed comparison of ZSM-Newton/Coulomb
to N-body Schr\"{o}dinger-Newton/Coulomb.) 

Earlier we observed that the complete expression for the mass densities
of the \emph{zbw} particles is given by the right hand side of (4.92).
While we also noted that the classical and quantum kinetic energy
terms can be neglected in the Newtonian regime, let us see what happens
if we do use the solution of (4.92) in the QHJ equation (4.78) and
the Schr\"{o}dinger equation (4.84).
For maximum clarity, we restrict to the two-particle case $q=\left\{ \mathbf{q}_{1},\mathbf{q}_{2}\right\} $
and drop the Coulomb potentials and rest-energy terms: 
\begin{equation}
\begin{aligned}-\partial_{t}S(q,t) & =\sum_{i=1}^{2}\left(\frac{\left[\nabla_{i}S(q,t)\right]^{2}}{2m_{i}}-\frac{\hbar^{2}}{2m_{i}}\frac{\nabla_{i}^{2}\sqrt{\rho(q,t)}}{\sqrt{\rho(q,t)}}\right)\\
 & -\frac{\left[m_{1}+\frac{T_{1}(q,t)}{c^{2}}+\frac{Q_{1}(q,t)}{c^{2}}\right]\left[m_{2}+\frac{T_{2}(q,t)}{c^{2}}+\frac{Q_{2}(q,t)}{c^{2}}\right]}{|\mathbf{q}_{1}-\mathbf{q}_{2}|}\\
 & =\sum_{i=1}^{2}\left(\frac{\left[\nabla_{i}S\right]^{2}}{2m_{i}}-\frac{\hbar^{2}}{2m_{i}}\frac{\nabla_{i}^{2}\sqrt{\rho}}{\sqrt{\rho}}\right)\\
 & -\frac{\left[m_{1}m_{2}+m_{1}\frac{T_{2}}{c^{2}}+m_{1}\frac{Q_{2}}{c^{2}}+m_{2}\frac{T_{1}}{c^{2}}+m_{2}\frac{Q_{1}}{c^{2}}+\frac{T_{1}T_{2}}{c^{4}}+\frac{T_{1}Q_{2}}{c^{4}}+\frac{Q_{1}T_{2}}{c^{4}}+\frac{Q_{1}Q_{2}}{c^{4}}\right]}{|\mathbf{q}_{1}-\mathbf{q}_{2}|},
\end{aligned}
\end{equation}
where $T_{i}(q,t)\coloneqq\frac{\left[\nabla_{i}S(q,t)\right]^{2}}{2m_{i}}$
and $Q_{i}(q,t)\coloneqq-\frac{\hbar^{2}}{2m_{i}}\frac{\nabla_{i}^{2}\sqrt{\rho(q,t)}}{\sqrt{\rho(q,t)}}$.
We can see that the gravitational interaction energy between the two
particles depends on their classical kinetic and quantum kinetic energy
terms, along with their rest masses. Furthermore, using the Madelung
transformation to combine the QHJ equation (4.100) with the continuity
equation (4.64), we obtain the \emph{nonlinear} two-particle Schr\"{o}dinger
equation 
\begin{equation}
\begin{aligned}i\hbar\frac{\partial\psi(q,t)}{\partial t} & =-\sum_{i=1}^{2}\frac{\hbar^{2}}{2m_{i}}\nabla_{i}^{2}\psi(q,t)\\
 & -\left(\frac{\left[m_{1}m_{2}+\frac{m_{1}T_{2}}{c^{2}}+\frac{m_{1}Q_{2}}{c^{2}}+\frac{m_{2}T_{1}}{c^{2}}+\frac{m_{2}Q_{1}}{c^{2}}+\frac{T_{1}T_{2}}{c^{4}}+\frac{T_{1}Q_{2}}{c^{4}}+\frac{T_{2}Q_{1}}{c^{4}}+\frac{Q_{1}Q_{2}}{c^{4}}\right]}{|\mathbf{q}_{1}-\mathbf{q}_{2}|}\right)\psi(q,t),
\end{aligned}
\end{equation}
where 
\begin{equation}
\frac{m_{i}T_{j}}{c^{2}}=\frac{m_{i}}{2m_{j}c^{2}}\left[\nabla_{j}S\right]^{2}=\frac{\hbar^{2}m_{i}}{2c^{2}m_{j}^{2}}\left(\nabla_{j}\ln\psi\right)^{2},
\end{equation}
 
\begin{equation}
\frac{m_{i}Q_{j}}{c^{2}}=-\frac{\hbar^{2}m_{i}}{2c^{2}m_{j}^{2}}\frac{\nabla_{j}^{2}\sqrt{\rho}}{\sqrt{\rho}}=-\frac{\hbar^{2}m_{i}}{2c^{2}m_{j}^{2}}\frac{\nabla_{j}^{2}|\psi|}{|\psi|},
\end{equation}
\begin{equation}
\frac{T_{1}T_{2}}{c^{4}}=\frac{\left[\nabla_{1}S\right]^{2}\left[\nabla_{2}S\right]^{2}}{4m_{1}m_{2}c^{4}}=\frac{\hbar^{2}}{4c^{4}m_{1}m_{2}}\left(\nabla_{1}\ln\psi\right)^{2}\left(\nabla_{2}\ln\psi\right)^{2},
\end{equation}
\begin{equation}
\frac{T_{i}Q_{j}}{c^{4}}=-\frac{\hbar^{2}}{4m_{i}m_{j}c^{4}}\left[\nabla_{i}S\right]^{2}\frac{\nabla_{j}^{2}\sqrt{\rho}}{\sqrt{\rho}}=-\frac{\hbar^{2}}{4m_{i}m_{j}c^{4}}\left(\nabla_{i}\ln\psi\right)^{2}\frac{\nabla_{j}^{2}|\psi|}{|\psi|},
\end{equation}
\begin{equation}
\frac{Q_{1}Q_{2}}{c^{4}}=\frac{\hbar^{4}}{4m_{1}m_{2}c^{4}}\frac{\left(\nabla_{1}^{2}\sqrt{\rho}\right)\left(\nabla_{2}^{2}\sqrt{\rho}\right)}{\rho}=\frac{\hbar^{4}}{4m_{1}m_{2}c^{4}}\frac{\left(\nabla_{1}^{2}|\psi|\right)\left(\nabla_{2}^{2}|\psi|\right)}{|\psi|^{2}},
\end{equation}
for $i\neq j$ and $\psi=\sqrt{\rho}e^{iS/\hbar}$. 

Because of the nonlinearity of (4.100), the 3-space coordinates $\mathbf{q}_{1}$
and $\mathbf{q}_{2}$ in the Green's function of the gravitational
potential in (4.100) can no longer be interpreted as linear operators
(hence why we don't put hats on them) and $\psi$ no longer has a
consistent Born-rule interpretation \cite{Diosi2016}. (That $\psi$
of (4.100) has no consistent Born-rule interpretation means that Salcedo's
``statistical consistency problem'' for quantum-classical hybrid
theories \cite{Salcedo2012,Salcedo2016} is not applicable in the
present context, since Salcedo's problem assumes the validity of the
Born rule and standard quantum measurement postulates for hybrid theories.)
Nevertheless, $\rho=|\psi^{2}|$ is still (by definition!) the stochastic
mechanical position probability density for the two-particle system
and still evolves by the continuity equation (4.64). The important
conceptual distinction here is that the Born-rule interpretation of
$|\psi|^{2}$ refers to the probability per unit volume of possible
outcomes of projective position measurements on the two-particle system,
while the stochastic mechanical definition of $|\psi|^{2}$ refers
to the probability per unit volume for the particles to \emph{be}
at 3-space positions $\{\mathbf{q}_{1},\mathbf{q}_{2}\}$ at time
$t$ as a result of their stochastic evolutions via (4.93-94). Thus,
a break-down of the Born-rule interpretation does not entail a break-down
of the stochastic mechanical meaning of $|\psi|^{2}$. 

Nonlinear Schr\"{o}dinger equations,
together with entangled states, are often said to imply superluminal
signaling \cite{Gisin1989,Polchinski1991,Bacciagaluppi2012}, due
to the well-known theorem of Gisin \cite{Gisin1989}. However, as
Bacciagaluppi has emphasized \cite{Bacciagaluppi2012}, superluminal
signaling only follows \emph{if} a theory with a nonlinear Schr\"{o}dinger
equation can also reproduce the usual phenomenology of wavefunction
collapse with Born-rule probabilities. Since said phenomenology does
not apply to solutions of (4.100), Gisin's theorem does not seem applicable
here. Of course, it may still be the case that the nonlinear Schr\"{o}dinger
equation (4.100) implies superluminal signaling, but determining this
depends on formulating a stochastic mechanical theory of measurement
consistent with (4.100). Since the nonlinearity of (4.100) makes naive
application of the standard stochastic mechanical theory of measurement
\cite{Blanchard1986,Goldstein1987,Jibu1990,Blanchard1992,Peruzzi1996}
unreliable, it remains an open question what variant of the stochastic
mechanical theory of measurement is consistent with (4.100). However,
it is expected that such a variant will yield empirical predictions
in close agreement with the empirical predictions of the standard
stochastic mechanical theory of measurement applied to the linear
counterpart of (4.100). The reason is that the nonlinear terms (4.102-106)
are ridiculously tiny in magnitude compared to the leading term proportional
to $m_{1}m_{2}$ in (4.100). So for all practical purposes, we can
ignore the nonlinear terms in modeling the Newtonian gravitational
interaction between the two \emph{zbw} particles, leaving us back
to the linear Schr\"{o}dinger equation
(4.84).

What about the ether's gravitational contribution? The answer to this
will depend on the details of an explicit physical model of the ether,
which we do not provide in this paper. Nevertheless, our phenomenological
hypotheses about the ether say that it is a medium in space-time with
superposed oscillations involving a countably infinite number of modes,
and that it continuously exchanges energy-momentum with the \emph{zbw}
particles. So it is reasonable to assume that there must be some stress-energy-momentum
associated with the ether. How this stress-energy-momentum gravitates
is an open question, but a couple possibilities can be noted: (i)
it doesn't gravitate at all, but rather the coupling of the ether
to massive \emph{zb}w particles somehow induces gravity on a Lorentzian
manifold, in analogy with Sakharov's `induced gravity' proposal \footnote{In Sakharov's approach, quantum vacuum fluctuations from matter fields
don't gravitate through their stress-energy-momentum tensor; rather,
one-loop vacuum fluctuations on a Lorentzian manifold (the latter
left to 'flap in the breeze') generate an effective action that contains
terms proportional to the Einstein-Hilbert action, the cosmological
constant, plus ``curvature-squared'' terms \cite{Visser2002}.} \cite{Visser2002}; (ii) it gravitates, but its overall contribution
to the total system energy density in the non-relativistic limit is
negligible compared to the rest-energy of a \emph{zbw} particle. In
our view, if the ether hypothesis of ZSM is correct, one of these
two possibilities must be correct, because all mass-energy quantities
experimentally measured in high energy scattering experiments and
nuclear binding/decay processes seem to come from three sources: (a)
the sum of the rest-masses of the particles, (b) the relativistic
kinetic energies of the particles, and (c) the mass-energy associated
with interactions between particles via the known fundamental forces.
We have provisional results that seem to support this view, in the
way of a semiclassical general relativistic extension of ZSM involving
a macroscopic model of the ether as a relativistic non-viscous fluid
that gravitates via the Einstein equations, but gives a negligible
contribution to the total rest-energy of a system of \emph{zbw} particles
in the non-relativistic limit. These results will be reported in a
future paper. Hence, for this paper, we shall continue with neglecting
the ether in gravitational effects (aside, of course, from the ether's
physical influence on the particles through their \emph{zbw} oscillations
and translational motions).

Finally, note that we have ignored the contribution of gravitational
and electrodynamical radiation reaction forces. In a separate paper,
we will show how these radiation reaction forces can be consistently
incorporated into ZSM-Newton/Coulomb through a stochastic generalization
of Galley's variational principle for nonconservative systems \cite{Galley2013}.

\section{Schr\"{o}dinger-Newton/Coulomb equations as mean-field theories}

We show in this section that ZSM-Newton/Coulomb recovers the `single-body'
Schr\"{o}dinger-Newton/Coulomb equations
\cite{Diosi1984,Guzman2003,Adler2007,Carlip2008,Giulini2011,Bassi2013,Hu2014,Derakhshani2014,AnastopoulosHu2014,BahramiBassi2014a,Bera2015,DerakNewtLimStoGra2017,BardosGolseMauser2000,BardosErdosGolseMauserYau2002,Golse2003}
as mean-field approximations when the number of \emph{zbw} particles
is sufficiently large. For clarity, we separate out the gravitational
and Coulomb interactions. 

The main idea of a `mean-field' (or `large \emph{N} ') theory is to approximate
the evolution of many particles interacting (gravitationally and/or
electromagnetically), when N is large (i.e., $N\rightarrow\infty$)
and the interactions are weak (in the sense that the gravitational
coupling between particles scales as $1/N$) \cite{Golse2003}. So
for example, if a system of identical particles has the mean-field
phase-space density $f(\mathbf{q},\mathbf{p},t)$, the mean-field
approximation says that the force exerted on a particle in the system
by the N other particles is approximated by averaging - with respect
to the phase-space density - the force exerted on the particle at
its 3-space location, from each point in the phase space. Mean-field
theory can also be used to approximate the net (gravitational and/or
electrostatic) force from a cloud of many weakly interacting identical
particles, on an external (macroscopic or mesoscopic or microscopic)
body such as a force-measurement probe. 

It is instructive to first discuss the mean-field approximation scheme
for a classical system of weakly interacting particles. Let us consider
a slight variation on the example discussed by Golse in \cite{Golse2003},
namely, a system of N identical classical point particles, weakly
interacting gravitationally, with 6N-dimensional Hamiltonian
\begin{equation}
H\left(\mathbf{q}_{1}(t),...,\mathbf{q}_{N}(t);\mathbf{p}_{1}(t),...,\mathbf{p}_{N}(t)\right)=\sum_{i=1}^{N}\frac{p_{i}^{2}}{2m}+\frac{1}{N}V_{g}^{int},
\end{equation}
where $V_{g}^{int}(\mathbf{q}_{i}(t),\mathbf{q}_{j}(t))=\frac{1}{2}\sum_{i,j=1}^{N(j\neq i)}\frac{m^{2}}{|\mathbf{q}_{i}(t)-\mathbf{q}_{j}(t)|}$
and the $1/N$ factor is the `weak-coupling scaling' \footnote{Without the scaling, $V_{g}^{int}$ diverge much faster than the total
kinetic energy (a sum of N terms) as $N\rightarrow\infty$, since
the sum in $V_{g}^{int}$ is composed of $0.5\,N\left(N-1\right)$
terms. With the scaling, however, $N^{-1}V_{g}^{int}$ scales as N
in the $N\rightarrow\infty$ limit. Thus the weak coupling scaling
ensures that $V_{g}^{int}$ and the total kinetic energy scale in
the same way in the $N\rightarrow\infty$ limit.} . Physically, the Hamiltonian (4.107) describes a collisionless dilute
gas of gravitationally interacting non-relativistic particles, and
is a special case of the Hamiltonian considered by Golse \cite{Golse2003}
and Bardos et al. \cite{BardosGolseMauser2000,BardosErdosGolseMauserYau2002}
(they considered (4.107) for an arbitrary, symmetric, smooth interaction
potential). The dynamics for the point particles is generated by (4.107)
via Hamilton's equations $\dot{\mathbf{q}}_{i}(t)=m^{-1}\nabla_{\mathbf{p}_{i}}H$
and $\dot{\mathbf{p}}_{i}(t)=-\nabla_{\mathbf{q}_{i}}H$. Consider
now the empirical distribution for the \emph{N} particles: $f_{N}(\mathbf{q},\mathbf{p},t)\coloneqq N^{-1}\sum_{i=1}^{N}\delta^{3}(\mathbf{q}-\mathbf{q}_{i}(t))\delta^{3}(\mathbf{p}-\mathbf{p}_{i}(t))$,
which satisfies (in the sense of distributions) the Vlasov equation
\begin{equation}
\partial_{t}f_{N}+\mathbf{p}\cdot\nabla_{\mathbf{q}}f_{N}+\nabla_{\mathbf{p}}\cdot\left[F_{N}\left(\mathbf{q},t\right)f_{N}\right]=\frac{1}{N^{2}}\sum_{i=1}^{N}\nabla_{\mathbf{p}}\cdot\left[\nabla_{\mathbf{q}}V\left(\mathbf{q}_{i},\mathbf{q}_{j}\right)\delta_{\mathbf{q}}^{3}\delta_{\mathbf{p}}^{3}\right],
\end{equation}
where
\begin{equation}
F_{N}\left(\mathbf{q},t\right)\coloneqq-\nabla_{\mathbf{q}}\int_{\mathbb{R}^{6}}\int_{\mathbb{R}^{6}}V\left(\mathbf{q},\mathbf{q}'\right)\,f_{N}\,d\mathbf{q}'d\mathbf{p}.
\end{equation}
Then, in the limit $N\rightarrow\infty$, the system described by
(4.107-109) is equivalent to a six-dimensional phase-space density
$f(\mathbf{q},\mathbf{p},t)$ (representing the density particles
of mass $m$ located at position $\mathbf{q}$ with momentum $\mathbf{p}$
at time $t$) evolving by the `large \emph{N} ' Vlasov equation
\begin{equation}
\partial_{t}f(\mathbf{q},\mathbf{p},t)+\left\{ H^{m.f.}(\mathbf{q},\mathbf{p},t),f(\mathbf{q},\mathbf{p},t)\right\} =0,
\end{equation}
where the time-dependent ``mean-field'' Hamiltonian $H^{m.f.}(\mathbf{q},\mathbf{p},t)$
is given by
\begin{equation}
H^{m.f.}(\mathbf{q},\mathbf{p},t)=\frac{p^{2}}{2m}+\int_{\mathbb{R}^{3}}\int_{\mathbb{R}^{3}}m\Phi(\mathbf{q},\mathbf{q}')f(\mathbf{q}',\mathbf{p},t)d^{3}\mathbf{p}d^{3}\mathbf{q}'.
\end{equation}
The last term on the right hand side of (4.111) is the ``mean-field''
potential energy, i.e., the phase-space averaged potential energy
of a particle of mass $m$ at position $\mathbf{q}$ at time \emph{t}.
It can be rewritten as 
\begin{equation}
\int_{\mathbb{R}^{3}}m\Phi(\mathbf{q},\mathbf{q}')\left(\int_{\mathbb{R}^{3}}f(\mathbf{q}',\mathbf{p},t)d^{3}\mathbf{p}\right)d^{3}\mathbf{q}',
\end{equation}
which tells us it should be interpreted, more precisely, as the sum
of the elementary potentials created at position \emph{$\mathbf{q}$}
by one particle located at position \emph{$\mathbf{q}'$} and distributed
according to the 3-space particle number density
\begin{equation}
\rho(\mathbf{q}',t)\coloneqq\int_{\mathbb{R}^{3}}f(\mathbf{q}',\mathbf{p},t)d^{3}\mathbf{p},
\end{equation}
with normalization
\begin{equation}
\int_{\mathbb{R}^{3}}\rho(\mathbf{q},t)d^{3}\mathbf{q}=\int_{\mathbb{R}^{3}}\int_{\mathbb{R}^{3}}f(\mathbf{q},\mathbf{p},t)d^{3}\mathbf{q}d^{3}\mathbf{p}=N.
\end{equation}
Since we are considering a system of identical particles interacting
gravitationally in the Newtonian approximation, the elementary potentials
are of the form 
\begin{equation}
\Phi(\mathbf{q},\mathbf{q}')=-\frac{m}{|\mathbf{q}-\mathbf{q}'|}.
\end{equation}
So with (4.113) and (4.115), we can rewrite (4.111) as 
\begin{equation}
H^{m.f.}(\mathbf{q},\mathbf{p},t)=\frac{p^{2}}{2m}-\int_{\mathbb{R}^{3}}\frac{m^{2}\rho(\mathbf{q}',t)}{|\mathbf{q}-\mathbf{q}'|}d^{3}\mathbf{q}',
\end{equation}
where $\rho(\mathbf{q},t)$ is the source in the Poisson equation
\begin{equation}
\nabla^{2}\Phi_{g}^{m.f.}=4\pi m\rho(\mathbf{q},t).
\end{equation}
Hence the mean-field Hamiltonian (4.116) describes, at time $t$,
the total energy of a particle with momentum $\mathbf{p}$ at position
$\mathbf{q}$ and with mean-field gravitational potential energy $\int_{\mathbb{R}^{3}}m\Phi(\mathbf{q},\mathbf{q}')\rho(\mathbf{q}',t)d^{3}\mathbf{q}'$.
Correspondingly, the position-space number density (4.113) can be
shown to evolve by the continuity equation
\begin{equation}
\partial_{t}\rho(\mathbf{q},t)=-\nabla\cdot\left(\frac{\mathbf{p}(\mathbf{q},t)}{m}\rho(\mathbf{q},t)\right),
\end{equation}
upon projecting the Liouville equation for $f(\mathbf{q},\mathbf{p},t)$
into position space, where $\mathbf{p}(\mathbf{q},t)$ is the mean
momentum
\begin{equation}
\mathbf{p}(\mathbf{q},t)\coloneqq\int_{\mathbb{R}^{3}}\frac{\mathbf{p}f(\mathbf{q},\mathbf{p},t)d^{3}\mathbf{p}}{\rho(\mathbf{q},t)}.
\end{equation}
It is interesting to consider the special case when
\begin{equation}
f(\mathbf{q},\mathbf{p},t)=\rho(\mathbf{q},t)\delta^{3}\left[\mathbf{p}-\nabla S_{cl}(\mathbf{q},t)\right],
\end{equation}
where $S_{cl}(\mathbf{q},t)$ is a single-valued classical velocity
potential associated to a particle at position $\mathbf{q}$ at time
$t$. (This special case will be of interest for comparison to the
mean-field description of ZSM-Newton/Coulomb.) We then have 
\begin{equation}
\mathbf{p}(\mathbf{q},t)=\int_{\mathbb{R}^{3}}\frac{\mathbf{p}f(\mathbf{q},\mathbf{p},t)d^{3}\mathbf{p}}{\rho(\mathbf{q},t)}=\nabla S_{cl}(\mathbf{q},t),
\end{equation}
and
\begin{equation}
H^{m.f.}(\mathbf{q},\nabla S_{cl},t)=\int_{\mathbb{R}^{3}}\frac{H^{m.f.}(\mathbf{q},\mathbf{p},t)f(\mathbf{q},\mathbf{p},t)d^{3}\mathbf{p}}{\rho(\mathbf{q},t)}=\frac{\left[\nabla S_{cl}(\mathbf{q},t)\right]^{2}}{2m}-\int_{\mathbb{R}^{3}}\frac{m^{2}\rho(\mathbf{q}',t)}{|\mathbf{q}-\mathbf{q}'|}d^{3}\mathbf{q}'.
\end{equation}
In this special (`Hamilton-Jacobi') case, (4.118) becomes
\begin{equation}
\partial_{t}\rho(\mathbf{q},t)=-\nabla\cdot\left(\frac{\nabla S_{cl}(\mathbf{q},t)}{m}\rho(\mathbf{q},t)\right),
\end{equation}
and (4.122) implies the Hamilton-Jacobi equation
\begin{equation}
H^{m.f.}(\mathbf{q},\nabla S_{cl},t)=-\partial_{t}S_{cl}(\mathbf{q},t)=\frac{\left[\nabla S_{cl}(\mathbf{q},t)\right]^{2}}{2m}-\int_{\mathbb{R}^{3}}\frac{m^{2}\rho(\mathbf{q}',t)}{|\mathbf{q}-\mathbf{q}'|}d^{3}\mathbf{q}'.
\end{equation}
Accordingly, the Madelung transformation on (4.123-124) yields the
nonlinear Schr\"{o}dinger equation
\begin{equation}
i\hbar\partial_{t}\chi_{cl}(\mathbf{q},t)=\left(-\frac{\hbar^{2}}{2m}\nabla^{2}-\int d^{3}\mathbf{q}'\frac{m^{2}|\chi_{cl}(\mathbf{q}',t)|^{2}}{|\mathbf{q}-\mathbf{q}'|}+\frac{\hbar^{2}}{2m}\frac{\nabla^{2}\sqrt{|\chi_{cl}|}}{\sqrt{|\chi_{cl}|}}\right)\chi_{cl}(\mathbf{q},t),
\end{equation}
with corresponding Poisson equation 
\begin{equation}
\nabla^{2}\Phi_{g}^{m.f.}=4\pi m|\chi_{cl}(\mathbf{q},t)|^{2}.
\end{equation}
Here, $\chi_{cl}(\mathbf{q},t)=\sqrt{\rho(\mathbf{q},t)}e^{iS_{cl}(\mathbf{q},t)/\hbar}$
is the classical mean-field `wavefunction', a collective variable
describing the evolution of a large number of identical particles
that weakly interact gravitationally. Note that the set (4.125-126)
looks formally just like the single-body SN equations, but with the
addition of an opposite-signed quantum kinetic defined in terms of
the classical mean-field wavefunction. (In this sense, (4.125) is
the mean-field generalization of the nonlinear Schr\"{o}dinger
equation of classical Hamilton-Jacobi mechanics \cite{Schiller1962,Rosen1964,Holland1993,Ghose2002,Nikolic2006,Nikolic2007,Bacciagaluppi2005,Oriols2016,Derakhshani2016a,Derakhshani2016b}.)
Likewise, if we had started with the description of N identical charged
particles weakly interacting electrostatically, with Hamiltonian (4.124)
under the replacement $V_{g}^{int}\rightarrow V_{c}^{int}$, then
by taking the large \emph{N}  limit and considering the Hamilton-Jacobi case,
we would obtain a nonlinear Schr\"{o}dinger-Coulomb-like
system identical to (4.125-126), with the charge $-e$ replacing the
mass $m$.

We shall now develop a similar mean-field approximation scheme for
ZSM-Newton.

To model a dilute `gas' of N identical ZSM particles interacting weakly
through Newtonian gravitational forces, we introduce the \emph{N}-particle
quantum Hamilton-Jacobi equation (for simplicity we drop the rest-energy
terms) with weak-coupling scaling:
\begin{equation}
\begin{aligned}-\partial_{t}S(q,t)|_{\mathbf{q}_{j}=\mathbf{Q}_{j}(t)} & =\sum_{i=1}^{N}\frac{\left[\nabla_{i}S(q,t)\right]^{2}}{2m}|_{\mathbf{q}_{j}=\mathbf{Q}_{j}(t)}+\frac{1}{N}V_{g}^{int}(\mathbf{Q}_{i}(t),\mathbf{Q}_{j}(t))-\sum_{i=1}^{N}\frac{\hbar^{2}}{2m}\frac{\nabla_{i}^{2}\sqrt{\rho(q,t)}}{\sqrt{\rho(q,t)}}|_{\mathbf{q}_{j}=\mathbf{Q}_{j}(t)},\end{aligned}
\end{equation}
where $S$ satisfies 
\begin{equation}
\sum_{i=1}^{N}\oint_{L}\mathbf{\nabla}_{i}S|_{\mathbf{q}_{j}=\mathbf{Q}_{j}(t)}\cdot\delta\mathbf{Q}_{i}(t)=nh,
\end{equation}
for a closed loop $L$ with $\delta t=0$. 

Now, it is well-known in classical mechanics \cite{Jehle1953,Bohm2002,JoseSaletan}
that when harmonic oscillators of the same natural frequency are nonlinearly
coupled, they eventually synchronize and oscillate in phase with each
other. (The relative phase does oscillate, but in the long run those
oscillations average out to zero.) Since the \emph{zbw} particles
are essentially harmonic oscillators of identical natural frequencies
and are nonlinearly coupled via $V_{g}^{int}$, it is reasonable to
expect that, after some time, their oscillations eventually come into
phase with each other. When this `phase-locking' occurs between the
\emph{zbw} particles, we can plausibly make the ansatz that
\begin{equation}
S(q,0)=\sum_{i=1}^{N}S(\mathbf{q}_{i},0),
\end{equation}
where all the $S(\mathbf{q}_{i},0)$ are identical. 

Furthermore, since the \emph{N}-particle continuity equation
\begin{equation}
\frac{\partial\rho({\normalcolor q},t)}{\partial t}=-\sum_{i=1}^{N}\nabla_{i}\cdot\left[\frac{\nabla_{i}S(q,t)}{m}\rho(q,t)\right],
\end{equation}
has the general solution
\begin{equation}
\rho(q,t)=e^{2R/\hbar}=\rho_{0}(q_{0})exp[-\int_{0}^{t}\left(\sum_{i}^{N}\nabla_{i}\cdot\frac{\nabla_{i}S}{m}\right)dt',
\end{equation}
the initial \emph{N}-particle osmotic potential takes the form
\begin{equation}
R(q,t)=R_{0}(q_{0})-(\hbar/2)\int_{0}^{t}\left(\sum_{i=1}^{N}\nabla_{i}\cdot\frac{\nabla_{i}S}{m}\right)dt'.
\end{equation}
So it is also plausible to make the ansatz
\begin{equation}
R(q,0)=\sum_{i=1}^{N}R(\mathbf{q}_{i},0),
\end{equation}
where all the $R(\mathbf{q}_{i},0)$ are identical, which implies
that the initial \emph{N}-particle probability density factorizes into a
product of identical single-particle densities:
\begin{equation}
\rho(q,0)=\prod_{i=1}^{N}\rho(\mathbf{q}_{i},0).
\end{equation}
From (4.134) it follows that (4.130) factorizes into N single-particle
continuity equations at $t=0$. Physically speaking, we can interpret
(4.133-134) as corresponding to the assumptions that, at $t=0$, the
way that the particle-ether coupling happens, in the local neighborhood
of each \emph{zbw} particle, is identical for all \emph{zbw} particles
(hence identical osmotic potentials sourced by the ether regions in
the local neighborhood of each \emph{zbw} particle), and that the
particles are interacting so weakly through $V_{g}^{int}$ and the
ether that they can be considered (effectively) physically independent
of one another.

Now, it is physically plausible to conjecture that, in the limit $N\rightarrow\infty$,
\footnote{Although we will not give a rigorous mathematical proof of this conjecture,
we will see later in this section that the conjecture is corroborated
by another large \emph{N}  argument that does have a rigorous mathematical
justification. Specifically, the large \emph{N}  limit prescription that leads
from the quantum N-body problem to the mean-field Schr\"{o}dinger-Poisson
equation that approximates a system of N quantum particles weakly
interacting by $1/r$ (e.g., Newtonian or Coulomb) potentials \cite{BardosGolseMauser2000,BardosErdosGolseMauserYau2002,Golse2003,AnastopoulosHu2014,BahramiBassi2014a,DerakNewtLimStoGra2017}. } the generation of correlations between the motions of the particles
gets suppressed (because of the weak-coupling scaling) so that time-evolution
by (4.130) yields
\begin{equation}
\rho(q,t)=\prod_{i=1}^{N}\rho(\mathbf{q}_{i},t),
\end{equation}
and time-evolution by (4.127) yields
\begin{equation}
S(q,t)=\sum_{i=1}^{N}S(\mathbf{q}_{i},t),
\end{equation}
where $\rho(\mathbf{q},t)$ satisfies
\begin{equation}
\partial_{t}\rho(\mathbf{q},t)=-\nabla\cdot\left(\frac{\nabla S(\mathbf{q},t)}{m}\rho(\mathbf{q},t)\right),
\end{equation}
and $S(\mathbf{q},t)$ satisfies
\begin{equation}
-\partial_{t}S(\mathbf{q},t)=\frac{\left[\nabla S(\mathbf{q},t)\right]^{2}}{2m}+\int_{\mathbb{R}^{3}}m\Phi(\mathbf{q},\mathbf{q}')\rho(\mathbf{q}',t)d^{3}\mathbf{q}'-\frac{\hbar^{2}}{2m}\frac{\nabla^{2}\sqrt{\rho(\mathbf{q},t)}}{\sqrt{\rho(\mathbf{q},t)}},
\end{equation}
along with
\begin{equation}
\oint_{L}\mathbf{\nabla}S\cdot d\mathbf{q}=nh.
\end{equation}

Although $S(\mathbf{q},t)$ and $\rho(\mathbf{q},t)$ look formally
like single-particle variables, they are, in fact, collective variables
in a mean-field description of the exact many-body description given
by (4.127-129) with (4.130) and (4.133). In particular, $\rho(\mathbf{q},t)$
has the physical meaning of the density of \emph{zbw} particles of
mass $m$ occupying position $\mathbf{q}$ at time $t$. Similarly,
$S(\mathbf{q},t)$ is the \emph{zbw} phase of a \emph{zbw} particle
at $\mathbf{q}$ at time \emph{t}. Accordingly, the last term on the
right side of (4.138) is the quantum kinetic energy of a \emph{zbw}
particle at $\mathbf{q}$ at $t$, and
\begin{equation}
V_{g}^{m.f.}(\mathbf{q},t)=m\Phi_{g}^{m.f.}(\mathbf{q},t)=\int_{\mathbb{R}^{3}}m\Phi(\mathbf{q},\mathbf{q}')\rho(\mathbf{q}',t)d^{3}\mathbf{q}'
\end{equation}
 is the mean-field gravitational potential energy of the \emph{zbw}
particle at $\mathbf{q}$ at $t$, where $\Phi$ is the elementary
potential given by (4.115) and $\Phi_{g}^{m.f.}$ satisfies the Poisson
equation 
\begin{equation}
\nabla^{2}\Phi_{g}^{m.f.}=4\pi m\rho(\mathbf{q},t).
\end{equation}

It is worth observing that (4.137) can also be viewed as the position-space
projection of the modified Vlasov equation 
\begin{equation}
\partial_{t}f(\mathbf{q},\mathbf{p},t)+\frac{\mathbf{p}}{m}\cdot\nabla_{\mathbf{q}}f(\mathbf{q},\mathbf{p},t)+\mathbf{F}(\mathbf{q},t)\cdot\nabla_{\mathbf{p}}f(\mathbf{q},\mathbf{p},t)=0,
\end{equation}
where the initial phase-space density is defined by $f_{0}(\mathbf{q},\mathbf{p})\coloneqq\rho_{0}(\mathbf{q})\delta^{3}\left[\mathbf{p}-\nabla S_{0}(\mathbf{q})\right]$
and 
\begin{equation}
\begin{array}{c}
f_{0}(\mathbf{q},\mathbf{p})\coloneqq\rho_{0}(\mathbf{q})\delta^{3}\left[\mathbf{p}-\nabla S_{0}(\mathbf{q})\right]\\
\Downarrow\\
f(\mathbf{q},\mathbf{p},t)=\rho(\mathbf{q},t)\delta^{3}\left[\mathbf{p}-\nabla S(\mathbf{q},t)\right],
\end{array}
\end{equation}
due time-evolution by (4.137), along with the normalization
\begin{equation}
\int_{\mathbb{R}^{3}}\rho(\mathbf{q},t)d^{3}\mathbf{q}=\int_{\mathbb{R}^{3}}\int_{\mathbb{R}^{3}}f(\mathbf{q},\mathbf{p},t)d^{3}\mathbf{q}d^{3}\mathbf{p}=N.
\end{equation}
From (4.143) it follows that the position-space projection of a \emph{zbw}
particle's 3-momentum $\mathbf{p}$ at position $\mathbf{q}$ yields
\begin{equation}
\mathbf{p}(\mathbf{q},t)=\int_{\mathbb{R}^{3}}\frac{\mathbf{p}f(\mathbf{q},\mathbf{p},t)d^{3}\mathbf{p}}{\rho(\mathbf{q},t)}=\nabla S(\mathbf{q},t)
\end{equation}
for all times, where $\rho=\int_{\mathbb{R}^{3}}f\,d^{3}\mathbf{p}$.
The force term in (4.142) is 
\begin{equation}
\begin{split}\mathbf{F}(\mathbf{q},t) & \coloneqq-\nabla_{\mathbf{q}}\left[\int_{\mathbb{R}^{3}}\int_{\mathbb{R}^{3}}m\Phi(\mathbf{q},\mathbf{q}')f(\mathbf{q}',\mathbf{p},t)d^{3}\mathbf{p}d^{3}\mathbf{q}-\frac{\hbar^{2}}{2m}\frac{\nabla^{2}\sqrt{\rho(\mathbf{q},t)}}{\sqrt{\rho(\mathbf{q},t)}}\right]\\
 & =-\nabla_{\mathbf{q}}\left[\int_{\mathbb{R}^{3}}m\Phi(\mathbf{q},\mathbf{q}')\rho(\mathbf{q}',t)d^{3}\mathbf{q}-\frac{\hbar^{2}}{2m}\frac{\nabla^{2}\sqrt{\rho(\mathbf{q},t)}}{\sqrt{\rho(\mathbf{q},t)}}\right],
\end{split}
\end{equation}
and has the physical interpretation of the net force on a \emph{zbw}
particle at $\mathbf{q}$ at $t$, due to spatial gradients of the
mean-field gravitational potential energy \emph{and} quantum kinetic
energy of the \emph{zbw} particle at $\mathbf{q}$ at $t$. Correspondingly,
it can be readily confirmed that the momentum-space projection of
(4.142), in conjunction with $f(\mathbf{q},\mathbf{p},t)=\rho(\mathbf{q},t)\delta^{3}\left[\mathbf{p}-\nabla S(\mathbf{q},t)\right]$,
yields \footnote{It is readily confirmed that the pressure tensor arising from the
momentum-space projection of (4.142) vanishes, because of the delta
function distribution in momentum in the definition of $f$. } 
\begin{equation}
\begin{split}\partial_{t}\mathbf{p}(\mathbf{q},t)+\mathbf{v}(\mathbf{q},t)\cdot\nabla\mathbf{p}(\mathbf{q},t) & =-\nabla_{\mathbf{q}}\left[\int_{\mathbb{R}^{3}}m\Phi(\mathbf{q},\mathbf{q}')\rho(\mathbf{q}',t)d^{3}\mathbf{q}-\frac{\hbar^{2}}{2m}\frac{\nabla^{2}\sqrt{\rho(\mathbf{q},t)}}{\sqrt{\rho(\mathbf{q},t)}}\right].\end{split}
\end{equation}

Now, applying the Madelung transformation to (4.137-139) yields the
mean-field nonlinear Schr\"{o}dinger
equation 
\begin{equation}
i\hbar\partial_{t}\chi(\mathbf{q},t)=\left(-\frac{\hbar^{2}}{2m}\nabla^{2}-\int d^{3}\mathbf{q}'\frac{m^{2}|\chi(\mathbf{q}',t)|^{2}}{|\mathbf{q}-\mathbf{q}'|}\right)\chi(\mathbf{q},t),
\end{equation}
with corresponding Poisson equation 
\begin{equation}
\nabla^{2}\Phi_{g}^{m.f.}=4\pi m|\chi(\mathbf{q},t)|^{2},
\end{equation}
where $\chi(\mathbf{q},t)=\sqrt{\rho(\mathbf{q},t)}e^{iS(\mathbf{q},t)/\hbar}$.
Here, the mean-field wavefunction is, like the classical mean-field
wavefunction, a collective variable describing the evolution of a
large number of identical \emph{zbw} particles that weakly interact
gravitationally. We note that, this time, the set (4.148-149) formally
looks \emph{exactly} like the single-body SN equations, but with the
very different physical meaning as a mean-field approximation in the
sense just explained. Similarly, if we had started with the description
of N identical charged \emph{zbw} particles interacting electrostatically,
with QHJ equation (4.127) under the replacement $V_{g}^{int}\rightarrow V_{c}^{int}$,
then by taking the large \emph{N}  limit as prescribed above, we would get
a nonlinear Schr\"{o}dinger-Coulomb
system identical to (4.148-149) with $-e$ replacing $m$. 

Note that when the quantum kinetic and its first $\nabla_{\mathbf{q}}$
are negligible relative to the mean-field gravitational potential
energy and mean gravitational force, (4.148) effectively becomes the
classical nonlinear Schr\"{o}dinger
equation (4.125), since (4.138) effectively becomes (4.124). This
observation seems to suggest a `quantum-classical' correspondence
between the Hamilton-Jacobi case of the classical Vlasov-Poisson mean-field
theory for a collisionless gas or plasma of non-relativistic interacting
particles, and the mean-field approximation for \emph{N}-particle ZSM-Newton/Coulomb.
However, such a correspondence is only formal; we will later see that
the reliability of (4.148-149), as a mean-field approximation, breaks
down for macroscopic superposition states.

To confirm the validity of our mean-field approximation proposal for
ZSM-Newton/Coulomb, let us reconsider the dilute gas of N identical
ZSM particles interacting through Newtonian gravitational forces,
but starting our description from the Schr\"{o}dinger
equation (4.84) (minus the rest-energy terms and the Coulomb potential)
with weak-coupling scaling:
\begin{equation}
i\hbar\frac{\partial\psi(q,t)}{\partial t}=\sum_{i=1}^{N}\left[-\frac{\hbar^{2}}{2m}+\frac{1}{N}\frac{m\hat{\Phi}_{g}(\mathbf{\hat{q}}_{i},\mathbf{\hat{q}}_{j})}{2}\right]\psi(q,t),
\end{equation}
where 
\begin{equation}
\nabla^{2}\hat{\Phi}_{g}=4\pi\sum_{i=1}^{N}m\delta^{3}\left(\mathbf{q}-\hat{\mathbf{q}}_{i}\right)
\end{equation}
and
\begin{equation}
\int_{\mathbb{R}^{3N}}|\psi(q,0)|^{2}d^{3N}\mathbf{q}=1.
\end{equation}
Supposing all the particles are in the same single-particle pure state
$\chi(\mathbf{q})$ at $t=0$, we can make the ``Hartree ansatz''
\begin{equation}
\psi(q,0)=\prod_{i=1}^{N}\chi(\mathbf{q}_{i},0),
\end{equation}
where the $\chi(\mathbf{q}_{i},0)$ are identical. Then, as shown
by Golse \cite{Golse2003} and Bardos et al. \cite{BardosGolseMauser2000,BardosErdosGolseMauserYau2002},
in the limit $N\rightarrow\infty$, the generation of correlations
between particles in time indeed gets suppressed (in the quantum BBGKY
hierarchy corresponding to (4.150-153)), and the time-dependent function
$\chi(\mathbf{q},t)$ satisfies (4.148-149). Likewise for the electrostatic
analogues of (4.150-151). Furthermore, we note that (4.150-151) is
equivalent to (4.127-134) by virtue of the Madelung transformation.
Presumably, then, there exists a Madelung BBGKY hierarchy corresponding
to (4.127-134), for which one can rigorously prove that in the limit
$N\rightarrow\infty$ the mean-field Madelung equations (4.137-141)
are recovered. We are unaware of such a proof in the mathematical
physics literature, however. 

Now, as a separate point, we can use the solution of (4.148-149) to
calculate the mean trajectory of a \emph{zbw} particle at position
$\mathbf{q}$ through the equations of motion
\begin{equation}
\frac{d\mathbf{Q}(t)}{dt}=\frac{\nabla S(\mathbf{q},t)}{m}|_{\mathbf{q}=\mathbf{Q}(t)}=\frac{\hbar}{m}\mathrm{Im}\frac{\nabla\chi(\mathbf{q},t)}{\chi(\mathbf{q},t)}|_{\mathbf{q}=\mathbf{Q}(t)},
\end{equation}
\begin{equation}
\begin{split}m\frac{d^{2}\mathbf{Q}(t)}{dt^{2}} & =\left[\partial_{t}\nabla S(\mathbf{q},t)+\frac{\nabla S(\mathbf{q},t)}{m}\cdot\nabla\left(\nabla S(\mathbf{q},t)\right)\right]|_{\mathbf{q}=\mathbf{Q}(t)}\\
 & =-\nabla\left[m\Phi_{g}^{m.f.}(\mathbf{q},t)-\frac{\hbar^{2}}{2m}\frac{\nabla^{2}\sqrt{|\chi(\mathbf{q},t)|)}}{\sqrt{|\chi(\mathbf{q},t)|}}\right]|_{\mathbf{q}=\mathbf{Q}(t)},
\end{split}
\end{equation}
 as well as the forward/backward stochastic trajectory through the
stochastic equations of motion
\begin{equation}
d\mathbf{q}(t)=\left[\frac{\hbar}{m}\mathrm{Im}\frac{\nabla\chi(\mathbf{q},t)}{\chi(\mathbf{q},t)}+\frac{\hbar}{m}\mathrm{Re}\frac{\nabla\chi(\mathbf{q},t)}{\chi(\mathbf{q},t)}\right]|_{\mathbf{q}=\mathbf{q}(t)}dt+d\mathbf{W}(t),
\end{equation}

\begin{equation}
d\mathbf{q}(t)=\left[\frac{\hbar}{m}\mathrm{Im}\frac{\nabla\chi(\mathbf{q},t)}{\chi(\mathbf{q},t)}-\frac{\hbar}{m}\mathrm{Re}\frac{\nabla\chi(\mathbf{q},t)}{\chi(\mathbf{q},t)}\right]|_{\mathbf{q}=\mathbf{q}(t)}dt+d\mathbf{W}_{*}(t).
\end{equation}
Considering that (4.148-149) is the leading-order large \emph{N}  approximation
to (4.150-153), trajectories calculated from (4.154-157) are expected
to only very roughly agree with the exact trajectories calculated
using the solutions of (4.150-151), whether for a dilute gas or plasma
of identical \emph{zbw} particles. Of course, in practice, it is impossible
to show this explicitly as it is a non-trivial problem to numerically
solve the system (4.150-151), even for just two particles.

Nonetheless, we can improve the mean-field approximation to (4.150-153)
by including the next-order terms in the large \emph{N}  limit. This will
be shown by us in full detail in \cite{DerakNewtLimStoGra2017}, but
the general reasoning can be sketched as follows: (i) take the Newtonian
limit of the Einstein-Langevin equation of semiclassical stochastic
Einstein gravity \cite{Hu2008}, (ii) describe the time evolution
of a bosonic quantum matter field coupled to the Newtonian-limited
Einstein-Langevin equation via the Heisenberg operator equation of
motion, and (iii) take the large \emph{N}  limit so as to justify replacing
the quantum operators in the Heisenberg equation of motion by their
quantum expectation values with respect to a coherent state. The result
is the `mean-field stochastic SN equations'
\begin{equation}
i\hbar\frac{\partial\chi(\mathbf{q},t)}{\partial t}=\left[-\frac{\hbar^{2}}{2m}\nabla^{2}+m\tilde{\Phi}_{g}^{m.f.}\right]\chi(\mathbf{q},t),
\end{equation}
\begin{equation}
\nabla^{2}\tilde{\Phi}_{g}^{m.f.}=4\pi\left[m|\chi(\mathbf{q},t)|^{2}+\frac{\xi(\mathbf{q},t)}{2c^{2}}\right],
\end{equation}
\begin{equation}
\begin{split}<\xi(\mathbf{q},t)>_{s}=0, & \qquad<\xi(\mathbf{q}_{A},t_{A})\xi(\mathbf{q}_{B},t_{A})>_{s}=N(\mathbf{q}_{A},\mathbf{q}_{B};t_{A},t_{B}),\end{split}
\end{equation}
\begin{equation}
\begin{split}N(\mathbf{q}_{A},\mathbf{q}_{B};t_{A},t_{B}) & \coloneqq\mathrm{Re}\left\{ m^{2}c^{4}\chi^{*}(\mathbf{q}_{A},t_{A})\chi(\mathbf{q}_{B},t_{B})\delta^{3}\left(\mathbf{q}_{A}-\mathbf{q}_{B}\right)\delta\left(t_{A}-t_{B}\right)\right.\\
 & \left.-m^{2}c^{4}|\chi(\mathbf{q}_{A},t_{A})|^{2}|\chi(\mathbf{q}_{B},t_{B})|^{2}\right\} .
\end{split}
\end{equation}

The bilocal field $N(\mathbf{q}_{A},\mathbf{q}_{B};t_{A},t_{B})$
is known as the ``noise kernel'', and essentially serves as a measure
of small (i.e., Gaussian) quantum fluctuations of the mass-energy
density of the \emph{N}-particle system, as described by (4.160-161), between
two nearby space-time points $\left\{ \mathbf{q}_{A},t_{A}\right\} $
and $\left\{ \mathbf{q}_{B},t_{B}\right\} $. (Technically, the noise
kernel defined by (4.160-161) is divergent due to the spatial delta
function. This can be remedied by replacing the delta function with
a smearing function \cite{AnastopoulosHu2014,Anastopoulos2015}, but
for our purposes this detail is inessential.) Furthermore, the noise
kernel plays the role of the diffusion coefficient for the classical
stochastic (colored) noise field $\xi(\mathbf{q},t)$ (where $<...>_{s}$
refers to the statistical average), the latter of which phenomenologically
models the back-reaction of the quantum fluctuations on the gravitational
field via $\Phi_{g}^{m.f.+}$. \footnote{The fact that the noise field is colored instead of white implies
that $\xi(\mathbf{q},t)$ is a smooth function, which further implies
that solutions of (4.158-159) are smooth functions. } In other words, the noise field in (4.159) reincorporates the quantum
coherence of the gravitational potential to first-order in the large
N approximation. To see this last point more explicitly, we can observe
that the stochastic correction to $\Phi_{g}^{m.f.}$ 
\begin{equation}
\Phi_{g}^{s}(\mathbf{q},t)\coloneqq-\frac{1}{c^{2}}\int d^{3}\mathbf{q}'\frac{\xi(\mathbf{q}',t)}{2|\mathbf{q}-\mathbf{q}'|},
\end{equation}
is known \cite{Hu2008} to formally reproduce the symmetrized two-point
correlation function for the quantized (q-number) gravitational potential:
\footnote{Equation (4.163) is deduced as follows. Start from the equality $2<h_{ab}(x_{A})h_{cd}(x_{B})>_{s}\begin{gathered}=\left\langle \Psi\right|\left\{ \hat{h}_{ab}(x_{A}),\hat{h}_{cd}(x_{B})\right\} \left|\Psi\right\rangle \end{gathered}
$ , where $h_{ab}(x_{A})$ is the classical stochastic metric perturbation
at spacetime point $x_{A}$ satisfying the regularized Einstein-Langevin
equation (see equation (3.14) of \cite{Hu2008}), $\hat{h}_{ab}(x_{A})$
is the quantum metric perturbation operator in the theory of perturbatively
quantized gravity (which is equivalent to the weak-field limit of
covariant path integral quantum gravity), and $\left|\Psi\right\rangle $
is the quantum state for a quantum field $\hat{\phi}(x)$ in the large
N expansion of covariant path integral quantum gravity \cite{HartleHorowitz1981,HuRouraVerdaguer2004,Hu2008}.
Implement the Newtonian limit by assuming $v\ll c$, $g_{ab}=\eta_{ab}+\delta\eta_{ab}$,
and $1\gg|T_{00}|/|T_{ij}|$; thus $\Phi_{g}^{s}\coloneqq\frac{1}{2}h_{00}$
and $\hat{\Phi}_{g}\coloneqq\frac{1}{2}\hat{h}_{00}$. Finally, take
$\left|\Psi\right\rangle $ to correspond to a coherent state with
the complex field eigenvalue $\chi$ \cite{DerakNewtLimStoGra2017}.
(We will show in \cite{DerakNewtLimStoGra2017} that taking $\left|\Psi\right\rangle $
to be a coherent state is equivalent to applying the Hartree ansatz
to the many-body wavefunction $\psi$ of the exact Newtonian quantized-gravitational
level of description, and then taking the large \emph{N}  limit.) The result
is (4.163).}
\begin{equation}
<\Phi_{g}^{s}(\mathbf{q}_{A},t_{A})\Phi_{g}^{s}(\mathbf{q}_{B},t_{B})>_{s}\begin{gathered}=\frac{1}{2}\left\langle \chi\right|\left\{ \hat{\Phi}_{g}(\mathbf{q}_{A},t_{A}),\hat{\Phi}_{g}(\mathbf{q}_{B},t_{B})\right\} \left|\chi\right\rangle \end{gathered}
.
\end{equation}

We say ``formally'' because the non-linear evolution (4.158-159)
implies failure of the Born-rule interpretation for $\chi$. Thus
the `expectation value' of the right hand side of (4.163) cannot be
understood as the standard quantum expectation value. However, since
$\chi$ does have a consistent stochastic mechanical statistical interpretation
(namely, $|\chi|^{2}$ corresponds to the number density of \emph{zbw}
particles at 3-space point $\mathbf{q}$ at time \emph{t}), we can
ascribe a stochastic mechanical statistical interpretation to the
right hand side of (4.163), in the sense that it is equivalent (by
the Madelung transformation) to the stochastic mechanical correlation
function: 
\begin{equation}
\left\langle \chi\right|\left\{ \hat{\Phi}_{g}(\mathbf{q}_{A},t_{A}),\hat{\Phi}_{g}(\mathbf{q}_{B},t_{B})\right\} \left|\chi\right\rangle =2\int_{-\infty}^{t}dt_{B}\int_{\mathbb{R}^{3}}d^{3}\mathbf{q}_{B}\rho(\mathbf{q}_{B},t_{B})\Phi_{g}(\mathbf{q}_{A},t_{A})\Phi_{g}(\mathbf{q}_{B},t_{B}),
\end{equation}
where $\Phi_{g}(\mathbf{q}_{A},t_{A})$ and $\Phi_{g}(\mathbf{q}_{B},t_{B})$
are solutions of the mean-trajectory Poisson equation (4.90).

Accordingly, if we use a solution of (4.158) in (4.154-157), the resulting
trajectories should slightly better approximate the exact trajectories
obtained from using the solutions of (4.150) for very large but finite
N. Note that with a solution of (4.158), the trajectories constructed
from integrating (4.154-155) contain classical (non-Markovian) stochastic
fluctuations through the stochasticity of the solution of (4.158).
On the other hand, the trajectories constructed from integrating (4.156-157)
contain classical stochastic fluctuations through the solution of
(4.158) \emph{and} the (Markovian) stochasticity encoded in the Wiener
process $d\mathbf{W}$ ($d\mathbf{W}_{*}$). Note, also, that even
though (4.158-161) are formulated for the case of a dilute system
of gravitationally interacting particles, they can also be applied
to dilute systems of electrostatically interacting particles, simply
by replacing $m\tilde{\Phi}_{g}^{m.f.}\rightarrow e\tilde{\Phi}_{c}^{m.f.}$
in (4.158-159), which implies the replacements $\xi/c^{2}\rightarrow-\xi/c$
in (4.159) and $m^{2}c^{4}\rightarrow e^{2}c^{2}$ in (4.159) and
(4.161). Then the `stochastic mean-field Schr\"{o}dinger-Coulomb
equations' provide a next-order correction to the large \emph{N}  limit of
the electrostatic analogue of (4.158-159), and thereby partially reincorporate
the quantum coherence of the \emph{N}-particle electrostatic potential operator.
(Presumably (4.163) holds in the electrostatic case as well, when
we replace $\Phi_{g}^{s}\rightarrow\Phi_{c}^{s}$ and $\hat{\Phi}_{g}\rightarrow\hat{\Phi}_{c}$,
but this has yet to be explicitly shown.) 

Finally, let us comment on the limitations of the mean-field approximations
considered here. 

First, the large \emph{N}  limit leading to (4.137-139) or (4.148-149) is
only applicable when the inter-particle interactions are sufficiently
weak that the independent-particle approximation is plausible. Some
example applications of (4.148-149) to self-gravitating \emph{N}-particle
systems that conform reasonably well to the independent-particle approximation,
are boson stars \cite{Ruffini1969,Guzman2003,Liebling2012} and (when
one includes short-range interactions between particles) Bose-Einstein
condensates \cite{Chavanis2012,Chavanis2016}; for electrostatically
self-interacting \emph{N}-particle systems, the electrostatic analogue of
(4.148-149) is widely used in condensed matter physics to model `jellium'
(i.e., homogeneous electron gas) systems \cite{Broglia2004,Giuliani2005}.
On the other hand, for strongly interacting \emph{N}-particle systems such
as (say) superconducting microspheres \cite{RomeroIsart2012,Pino2016,DerAnaHu2016,DerakProbingGravCat2016},
the independent-particle approximation is a poor one and the deterministic
or stochastic SN/SC equations cannot be used.

Second, even for dilute \emph{N}-particle systems, such as considered above,
the mean-field approximations provided by (4.148-149) and (4.158-161)
become empirically inadequate for calculating the gravitational force
on an external (macroscopic or mesoscopic or microscopic) probe mass,
when quantum fluctuations of the mass-energy density of the \emph{N}-particle
system become too large. As an example, for the dilute system of N
gravitationally interacting ZSM particles, with total mass $M=Nm$,
suppose that the solution of (4.148) or (4.158) takes the form of
a Schr\"{o}dinger cat state. In particular,
an equal-weighted superposition of two identical Gaussian packets
of width $\sigma$, where one is peaked at spatial location $\frac{1}{2}\mathbf{L}$,
the other at $-\frac{1}{2}\mathbf{L}$, with $\left|\mathbf{L}\right|\gg\sigma$
and both packets having zero mean momentum:
\begin{equation}
\chi_{cat}(\mathbf{x})=\frac{1}{\sqrt{2}}\left[\chi_{left}(\mathbf{x})+\chi_{right}(\mathbf{x})\right]=\frac{1}{\sqrt{2}}\frac{1}{\left(2\pi\sigma^{2}\right)^{3/4}}\left[e^{-\frac{(\mathbf{x}+\mathbf{L}/2)^{2}}{4\sigma^{2}}}+e^{-\frac{(\mathbf{x}-\mathbf{L}/2)^{2}}{4\sigma^{2}}}\right].
\end{equation}
Then the Poisson equation for the mass density corresponding to (4.149)
or (4.159) takes the form
\begin{equation}
\nabla^{2}\Phi_{g}^{m.f.}=4\pi M|\chi(\mathbf{x})|^{2}=4\pi\left[\frac{M}{2}|\chi_{left}|^{2}+\frac{M}{2}|\chi_{right}|^{2}\right],
\end{equation}
or
\begin{equation}
\nabla^{2}\widetilde{\Phi}_{g}^{m.f.}=4\pi M|\chi(\mathbf{x})|^{2}=4\pi\left[\frac{M}{2}|\chi_{left}|^{2}+\frac{M}{2}|\chi_{right}|^{2}+\frac{\xi(\mathbf{x},0)}{2c^{2}}\right],
\end{equation}
with
\begin{equation}
\begin{split}<\xi(\mathbf{x},t)>_{s}=0, & \qquad<\xi(\mathbf{x}_{A},t_{A})\xi(\mathbf{x}_{B},t_{A})>_{s}=N(\mathbf{x}_{A},\mathbf{x}_{B};t_{A},t_{B}),\end{split}
\end{equation}
\begin{equation}
\begin{split}N(\mathbf{x}_{A},\mathbf{x}_{B};t_{A},t_{B}) & =\mathrm{Re}\left\{ M^{2}c^{4}\chi_{cat}^{*}(\mathbf{x}_{A},t_{A})\chi_{cat}(\mathbf{x}_{B},t_{B})\delta^{3}\left(\mathbf{x}_{A}-\mathbf{x}_{B}\right)\delta\left(t_{A}-t_{B}\right)\right.\\
 & \left.-M^{2}c^{4}|\chi_{cat}(\mathbf{x}_{B},t_{B})|^{2}|\chi_{cat}(\mathbf{x}_{A},t_{A})|^{2}\right\} .
\end{split}
\end{equation}
If the spatial separation between the two Gaussians is macroscopic,
e.g., $\mathbf{L}=1m$, and if $M=1,000kg$, then the classical gravitational
field produced by (4.166) or (4.167-169) is totally unrealistic. For
example, a probe corresponding to a macroscopic test mass passing
through the mid-point of the two mass distributions will, according
to (4.166), go undeflected, or, according to (4.167-169), will oscillate
in between the two mass distributions before passing through with
no mean deflection (because of the Gaussian property of the noise
field). Both predictions are in stark contrast to what the exact \emph{N}-particle
description (4.150-151) would predict if $\psi(q)$ takes the form
of (4.165) and one applies the textbook quantum measurement postulates
\cite{Bohm1951,BellAgainstMeasure} or the stochastic mechanical theory
of measurement \cite{Blanchard1986,Goldstein1987,Jibu1990,Blanchard1992,Peruzzi1996};
namely, that the test mass will either deflect towards the left mass
distribution or the right mass distribution, with probability $\frac{1}{2}$
each. \footnote{Of course, the stochastic SN equations (and the Einstein-Langevin
equation more generally) are formulated to handle only dilute \emph{N}-particle
systems with small quantum fluctuations in the matter sector. Cat
state solutions clearly fall out of this regime, so it is not surprising
that the stochastic SN equations make an empirically inadequate prediction
in this case. In order to extend the stochastic SN equations to the
case of non-Gaussian fluctuations, we would (presumably) need to incorporate
into (4.158-159) the quantum coherence of the full \emph{n}-point
correlation function involving $\hat{\Phi}_{g}$, in terms of some
suitable generalization of the noise kernel. This remains an open
problem \cite{Hu2008,HuDICE2016,DerakNewtLimStoGra2017}.} Furthermore, apart from the fact that the solutions of (4.166) or
(4.167-169) don't have consistent Born-rule interpretations \cite{Adler2007,vanWezel2008,Hu2014,Derakhshani2014,Diosi2016},
the stochastic mechanical statistical interpretation of the solutions
of (4.166) or (4.167-169) doesn't predict a probed gravitational field
that's any more consistent with the prediction obtained from (4.150-151).
And, of course, all these issues with cat states apply as well in
the electrostatic case. 

As we will see in Part II, the limitations of the mean-field approximations
considered above can be circumvented by employing a center-of-mass
description of a large \emph{N}  system of ZSM-Newton/Coulomb particles. But
next let us compare ZSM-Newton/Coulomb, developed thus far, to other
semiclassical theories.

\section{Comparison to other semiclassical Newtonian field theories}

Here we compare ZSM-Newton/Coulomb, developed thus far, to other semiclassical
Newtonian field theories proposed in the literature. In particular,
we highlight conceptual advantages of the ZSM-Newton/Coulomb approach
and possibilities for experimental discrimination.

\subsection{Comparison to non-hidden-variable approaches}

Anastopoulos and Hu (AH) \cite{AnastopoulosHu2014} have shown that
the mean-field SN equations (4.148-149) can be derived from the standard
quantum field theoretic description of a scalar matter field interacting
with perturbatively quantized gravity (hereafter PQG): simply take
the Newtonian limit of PQG to obtain the \emph{N}-particle Schr\"{o}dinger
equation (4.84), consider the case of weakly-coupled systems of identical
particles, then apply the large \emph{N}  limit (as we did in (4.150-153)).
Complementing their analysis, we will show in \cite{DerakNewtLimStoGra2017}
that the mean-field SN equations follow from standard semiclassical
Einstein gravity (SCEG) \cite{Hu2008,Hu2014,AnastopoulosHu2014},
under the following prescription: (i) take the Newtonian limit of
the semiclassical Einstein equation (see (4.173) below) to obtain
the Poisson equation with the quantum expectation value of the mass
density operator as a source; (ii) describe the time evolution of
a bosonic quantum matter field coupled to the Poisson equation via
the Heisenberg equation of motion; and (iii) assume the large \emph{N}  limit
so as to justify replacing the quantum operators in the Heisenberg
equation of motion by their quantum expectation values with respect
to a coherent state.

Likewise AH have shown \cite{AnastopoulosHu2014} that the mean-field
SC equations follow from standard relativistic QED: take the non-relativistic
limit, consider a weakly-coupled system of identical particles, then
take the large \emph{N}  limit. As with the gravitational case, we will also
show in \cite{DerakNewtLimStoGra2017} that the mean-field SC equations
follow from analogously applying steps (i-iii) to standard semiclassical
relativistic electrodynamics (SCRED). \footnote{The semiclassical Maxwell equation of SCRED is given by $\nabla_{\mu}F^{\mu\nu}=\left\langle \psi\right|\hat{J}^{\nu}\left|\psi\right\rangle $,
where $\hat{J}^{\nu}$ is the charge four-current operator, $\left|\psi\right\rangle $
is some state-vector, and $\nabla_{\mu}$ is the covariant derivative
in case the background spacetime is curved. Taking the non-relativistic
limit to obtain the semiclassical Poisson equation, describing the
time evolution of a bosonic quantum matter field coupled to the semiclassical
Poisson equation via the Heisenberg equation of motion, and assuming
the large \emph{N}  limit so that the quantum operators in the Heisenberg
equation of motion can be replaced by their quantum expectation values
with respect to a coherent state, one obtains the mean-field SC system
\cite{DerakNewtLimStoGra2017}. }. 

Thus for weakly-coupled systems of identical particles, the large
N limit scheme used in ZSM-Newton/Coulomb can also be employed in
Newtonian PQG/QED; and in both cases one recovers the mean-field SN/SC
equations. These results also agree with the Newtonian limits of SCEG
and SCRED, when the latter are interpreted as mean-field theories
for weakly-coupled systems of identical particles.

It is notable that these correspondences follow despite ZSM-Newton/Coulomb
treating the gravitational/Coulomb potentials as fundamentally classical
fields sourced by point-like classical particles undergoing non-classical
motions in 3-space. In this respect, the ZSM approach is unique among
existing formulations of quantum theory that have been extended to
fundamentally-semiclassical gravity or electrodynamics.

For example, it is well known \cite{Diosi1984,Salzman2005,Carlip2008,Derakhshani2014,AnastopoulosHu2014,BahramiBassi2014a,Giulini2014,Bahrami2015,DerakNewtLimStoGra2017}
that if one formulates fundamentally-semiclassical gravity based on
the equations of either standard non-relativistic quantum mechanics
\cite{Diosi1984,Salzman2005,Carlip2008,Derakhshani2014,AnastopoulosHu2014,BahramiBassi2014a,Giulini2014,Bahrami2015,DerakNewtLimStoGra2017}
or non-relativistic many-worlds interpretations \cite{Yang2013,Derakhshani2014},
one obtains the \emph{N}-body SN equations
\begin{equation}
i\hbar\frac{\partial\psi(q,t)}{\partial t}=\sum_{i=1}^{N}\left[-\frac{\hbar^{2}}{2m_{i}}+\frac{m_{i}\Phi_{g}^{SN}}{2}\right]\psi(q,t),
\end{equation}
and 
\begin{equation}
\nabla^{2}\Phi_{g}^{SN}=4\pi m(\mathbf{q},t)=4\pi\underset{i=1}{\overset{N}{\sum}}\int d^{3}\mathbf{r}{}_{1}...d^{3}\mathbf{r}{}_{N}|\psi(\mathbf{r}_{1}...\mathbf{r}_{N},t)|^{2}m_{i}\delta^{(3)}(\mathbf{q}-\mathbf{r}{}_{i}),
\end{equation}
where 
\begin{equation}
\Phi_{g}^{SN}=-\sum_{j=1}^{N(j\neq i)}\int\frac{m_{j}(\mathbf{q}'_{j},t)}{|\mathbf{q}_{i}-\mathbf{q}'_{j}|}d^{3}\mathbf{q}'{}_{1}...d^{3}\mathbf{q}'_{N}.
\end{equation}
It is also well-known \cite{Derakhshani2014,AnastopoulosHu2014,BahramiBassi2014a,Giulini2014,DerakNewtLimStoGra2017}
that (4.170-171) can be obtained from the Newtonian limit of the semiclassical
Einstein equation
\begin{equation}
G_{nm}=\kappa\left\langle \psi\right|\hat{T}_{nm}\left|\psi\right\rangle ,
\end{equation}
\emph{if} one naively assumes that (4.173) is valid even when $\psi$
is a single-particle wavefunction, whether in a standard quantum theory
reading or a many-worlds interpretation (re: the latter context, see
\cite{PageGeilker1981,Yang2013,Derakhshani2014}). However, like the
mean-field SN equations, the solutions of (4.170-171) lack consistent
Born-rule interpretations \cite{Adler2007,vanWezel2008,Hu2014,Derakhshani2014,Diosi2016}
and include the macroscopic gravitational cat states discussed in
section 4. In other words, attempting to formulate fundamentally-semiclassical
gravity, based on either standard quantum theory or many-worlds interpretations,
results in a nonlinear classical-gravitational field theory that makes
absurd empirical predictions. As another example, it was shown in
\cite{Derakhshani2014,DerakProbingGravCat2016} that the N-body SN
equations (with stochastic corrections to dynamically induce intermittent
wavefunction collapse) arise naturally when one extends the GRW, CSL,
and DP theories to fundamentally-semiclassical gravity with a matter
density ontology (called GRWmN, CSLmN, and DPmN, respectively). In
contrast to SQM-Newton (where SQM = standard quantum mechanics) and
MW-Newton (where MW = many worlds), GRWmN/CSLmN/DPmN have been shown
to adequately suppress the empirically problematic macroscopic gravitational
cat states while also having consistent statistical interpretations
for pure states \cite{Derakhshani2014,DerakProbingGravCat2016}. Thus,
these dynamical collapse theories of fundamentally-semiclassical Newtonian
gravity are empirically viable. At the same time, these dynamical
collapse theories also make slightly different empirical predictions
from the Newtonian large \emph{N}  limit of PQG and SCEG; and given the empirical
equivalence between Newtonian-large-\emph{N} PQG and SCEG, and \emph{N}-particle
ZSM-Newton (when the nonlinear terms of the latter are neglected),
it will also be the case that these dynamical collapse theories make
slightly different empirical predictions from \emph{N}-particle ZSM-Newton
(see \cite{DerakProbingGravCat2016} for further elaboration on this
point). These slight differences in empirical predictions are entailed
by the collapse-inducing stochastic correction terms, and the fact
that these dynamical collapse theories still allow for stable gravitational
cat states in a mesoscopic regime of masses \cite{Derakhshani2014,DerakProbingGravCat2016}.
The slightly different empirical predictions of these collapse theories
in the semiclassical-gravitational context may be testable by the
next (or next-next) generation of state-of-the-art AMO experiments,
as argued by us in \cite{DerAnaHu2016,DerakProbingGravCat2016}. 

As yet another example, the Tilloy-Di\'{o}si (TD)
model of fundamentally-semiclassical gravity makes use of the flash
ontology within CSL or DP dynamics, to describe fundamentally-semiclassical
Newtonian gravitational interactions between \emph{N} particles, with
no nonlinear feedback from the wavefunction \cite{Tilloy2016,TilloyDiosi2017}.
(One can also make a GRW analogue of the TD model, as pointed out
by us in \cite{Derakhshani2014}.) TD's (stochastic) analogue of the
SN equations reads
\begin{equation}
\begin{aligned}\frac{d\left|\psi\right\rangle }{dt} & =-\frac{i}{\hbar}\left(\hat{H}+\hat{V}_{G}\right)\left|\psi\right\rangle \\
 & -\frac{1}{8\pi\hbar G}\int d\mathbf{r}\left(\nabla\hat{\Phi}(\mathbf{r})-\left\langle \nabla\hat{\Phi}(\mathbf{r})\right\rangle \right)^{2}\left|\psi\right\rangle \\
 & -\hbar\left(1+i\right)\int d\mathbf{r}\left(\hat{\Phi}(\mathbf{r})-\left\langle \nabla\hat{\Phi}(\mathbf{r})\right\rangle \right)\delta\rho(\mathbf{r})\left|\psi\right\rangle ,
\end{aligned}
\end{equation}
up to a fixed spatial cut-off $\sigma$. Here the potential $\hat{V}_{G}$
represents the usual Newtonian gravitational potential operator, while
the non-Hermitian terms on the right give rise to decoherence and
collapse of spatial superpositions of a massive particle. As shown
by TD \cite{Tilloy2016,TilloyDiosi2017}, their model adequately suppresses
macroscopic gravitational cat states and has a consistent statistical
interpretation. By virtue of the non-Hermitian terms in (4.174), the
TD model also makes slightly different predictions from both Newtonian-limited
PQG and ZSM-Newton. These differences might also be testable by the
next (or next-next) generation of state-of-the-art AMO experiments
\cite{DerakProbingGravCat2016}. A notable difference between the
TD model and ZSM-Newton is that the former predicts point-like mass
distributions (which source the classical gravitational field) that
discontinuously appear and disappear in space-time, because the flash
ontology is used as the means of defining the mass density sources
(we have previously made this point in regards to a GRW analogue of
the TD model \cite{Derakhshani2014}); by contrast, the mass density
sources in ZSM-Newton (the \emph{zbw} particles) involve no such discontinuities.
Another notable difference is that whereas the density operator corresponding
to the solution (4.174) evolves by a linear master equation (thus
retaining the usual statistical interpretation of mixtures of density
operators and not allowing for superluminal signaling) \cite{Tilloy2016,TilloyDiosi2017},
the von-Neumann-Liouville equation corresponding to the nonlinear
ZSM-Newton Schr\"{o}dinger equation
(4.101) will clearly be nonlinear, as will the master equation for
a reduced density matrix obtained from performing a partial trace
of the pure-state density matrix evolving by the aforementioned nonlinear
von-Neumann-Liouville equation. This means that the nonlinear ZSM-Newton
master equation will violate the interchangeability of mixing and
evolution required to sustain the usual statistical interpretation
of mixtures of density operators \cite{Diosi2016}. One would then
have to rely on `conditional wave functions' \cite{Duerr2009,Oriols2016,Derakhshani2016b,Derakhshani2017b}
(see section 5.2.2 of the present paper for examples) and `conditional
density matrices' \cite{DGTZ-2005,Duerr2009,Toros2016} in order to
describe the dynamics of sub-systems of a ZSM-Newton system described
by (4.101) or its \emph{N}-particle generalization, and to extract statistical
predictions for those sub-systems (assuming that a stochastic mechanical
theory of measurement can be developed consistent with solutions of
(4.101) and solutions of the corresponding nonlinear von-Neumann-Liouville
equation). Since both the conditional wave function, defined from
the solution of (4.101), and the conditional density matrix, defined
from the solution of the nonlinear von-Neumann-Liouville equation
corresponding to (4.101), have nonlinear evolutions, it is expected
that statistical predictions extracted for entangled sub-systems will
include superluminal signaling, assuming that the stochastic mechanical
theory of measurement consistent with solutions of (4.101) reproduces
the usual collapse phenomenology (as is required by Gisin's theorem)
\cite{Gisin1989,Bacciagaluppi2012}. Showing this explicitly remains
a task for future work. 

Concerning theories of fundamentally-semiclassical electrodynamics,
perhaps the best-known is Asim Barut's ``self-field QED'' \cite{Barut1988a,Barut1988b,BarutDowling1990a,BarutDowling1990b}.
This theory takes the Schr\"{o}dinger-Coulomb
(SC) analogue of (4.170-171) (and its relativistic generalization,
the Dirac-Maxwell system) as its starting point and purports to reproduce
the self-energy effects of non-relativistic and relativistic QED to
all orders of perturbation linear in alpha. However, there are more
basic predictions of the theory that were left (apparently) unaddressed
by Barut and his co-workers, and which seem to make the theory empirically
inadequate. First, just like the SN equations, the SC analogue of
(4.170-171) does not have a consistent Born-rule interpretation, thereby
preventing a naive application of the standard quantum measurement
postulates. Second, also just like the SN equations, the SC equations
admit macroscopic electrostatic cat states as solutions (i.e., the
electrostatic analogue of the cat state example discussed in section
4), and these solutions are clearly not seen in the real world (incidentally,
this rules out the possibility of many-worlds interpretations based
on the SC equations, since the worlds would electrostatically interact
with one another even for macroscopic superpositions). Third, even
if one attempts to add stochastic corrections to the SC equations
in the form of GRW/CSL/DP, numerical simulations of the SC equations
indicate that a free particle wavepacket would undergo Coulomb self-repulsion
(from the nonlinear electrostatic self-interaction), and this self-repulsion
effect would lead to maxima in the two-slit experiment much too broad
to be in agreement with existing experimental data \cite{GrossardtPersonalComm2013}.
As an alternative formulation of fundamentally-semiclassical electrodynamics
based on dynamical collapse theories, we might consider a straightforward
electrostatic analogue of TD's equation (4.174). Presumably such a
theory would be free of the problems entailed by the nonlinearity
of the SC equations (just as (4.174) is free of the problems entailed
by the nonlinearity of the SN equations, since the density operator
corresponding to (4.174) has a linear master equation evolution \cite{Tilloy2016,TilloyDiosi2017}),
but this remains to be shown. In any case, it would appear that, in
comparison to theories of fundamentally-semiclassical electrodynamics
based on standard quantum mechanics, many-worlds interpretations,
and dynamical collapse theories with matter density ontology, ZSM-Coulomb
is the only one among these that's empirically viable (within its
non-relativistic domain of validity) insofar as it's empirically equivalent
to the Newtonian limits of standard QED and SCRED (modulo the tiny
empirical differences entailed by the nonlinear correction terms (4.102-106)
discussed in section 3).

\subsection{Comparison to alternative hidden-variable approaches}

Other formulations of stochastic mechanics exist besides ZSM \cite{Fenyes1952,Davidson1979,Yasue1981a,Wallstrom1989,Wallstrom1994,Derakhshani2016a,Derakhshani2016b}.
Moreover, dBB pilot-wave theory is the most well-developed hidden-variables
formulation of quantum theory to date. Do these other hidden-variables
theories have consistent and empirically adequate extensions to semiclassical
Newtonian field theories, whether in the form of fundamentally-semiclassical
theories or semiclassical approximations? How do they compare and
contrast to ZSM-Newton/Coulomb?

As mentioned in section 2, all non-ZSM formulations of stochastic
mechanics are subject to Wallstrom's criticism \cite{Wallstrom1989,Wallstrom1994,Bacciagaluppi2005,Bacciagaluppi2012,Derakhshani2016a,Derakhshani2016b}
- they are all empirically inadequate because they either allow for
too many solutions or too few solutions, compared to the Schr\"{o}dinger
equation of standard quantum mechanics. For those formulations that
allow too many solutions, one can always impose by hand the quantization
condition needed in order to make the solution spaces of those formulations
isomorphic to the solution space of standard quantum mechanics \cite{Wallstrom1989,Wallstrom1994,Bacciagaluppi2005,Bacciagaluppi2012,Derakhshani2016a,Derakhshani2016b}.
This is, of course, an ad hoc move, but one might view it as provisional
until such a condition can be justified by some non-ZSM modification
of said formulations of stochastic mechanics. In this case, the amended
formulations of stochastic mechanics would result in exactly the same
mathematical descriptions of Newtonian gravity and electrodynamics
as we've found for ZSM, both at the exact (i.e., \emph{N}-particle Schr\"{o}dinger
equation) level and the level of the mean-field approximation schemes.
(Differences would arise, however, in physically motivating the mean-field
approximation, e.g., ansatz (4.120) in section 4; since the $S$ function
would not be interpretable as the phase of a periodic phenomenon localized
to the stochastic mechanical particle, such an ansatz would have to
be imposed ad hoc.)

Concerning semiclassical de Broglie-Bohm theories, let us consider
the possibilities separately.

\subsubsection{Comparison to fundamentally-semiclassical de Broglie-Bohm theories}

There is some ambiguity in how to construct a dBB-based theory of
fundamentally-semiclassical Newtonian gravity (or electrodynamics).
First, one has to make a choice about which version of dBB dynamics
to consider, i.e., the `first-order' version or `second-order' version
\cite{Bohm1952I,Bohm1952II,BohmHiley1993,HollandBook1993,Duerr2009,DGZbook2012,OriolsMompart2012,Goldstein2013}
(the choice one makes could potentially make a difference in how one
formulates a dBB-based theory of fundamentally-semiclassical Newtonian
gravity or electrodynamics). Second, one has to make a choice about
which part of the dBB ontology (depending on how it's interpreted)
- the wavefunction or the particles or both - plays the role of the
mass (or charge) density that sources the classical gravitational
(or electromagnetic) field; as it turns out, for versions of dBB in
which the wavefunction is part of the ontic variables, there is no
compelling reason why the particles (as opposed to the wavefunction,
or at the exclusion of the wavefunction) should be used to define
the mass (charge) density source for the classical gravitational (electromagnetic)
field, even though that might seem like a prima facie natural choice. 

Let us consider this last point in more detail for the gravitational
case first, under the first-order `dual space' version of non-relativistic
dBB \cite{Bohm1952I,Bohm1952II,BellQMCosmo2004,HollandBook1993,BohmHiley1993,Goldstein2013}.
In other words, the version of dBB theory that posits an ontic 3N-dimensional
configuration space, occupied by an ontic `universal wavefunction'
$\psi(\mathbf{q}_{1},,,\mathbf{q}_{N},t)$, and an ontic 3-dimensional
space (existing completely independently of the configuration space)
occupied by N (spinless) particles with configuration $Q(t)=\left\{ \mathbf{Q}_{1}(t),...,\mathbf{Q}_{N}(t)\right\} $.
The universal wavefunction \footnote{The universal wavefunction is required to satisfy the usual boundary
conditions of single-valuedness, smoothness, and finiteness.} evolves by the Schr\"{o}dinger equation 
\begin{equation}
i\hbar\frac{\partial\psi}{\partial t}=\left[-\sum_{i=1}^{N}\frac{\hbar^{2}}{2m_{i}}\nabla_{i}^{2}+V^{int}\right]\psi,
\end{equation}
where $V^{int}$ is some scalar interaction potential to be specified
and we assume the normalization $\int_{\mathbb{R}^{3N}}|\psi|^{2}d^{3N}q=1$.
The particles evolve by the guiding equation
\begin{equation}
\frac{d\mathbf{Q}_{i}(t)}{dt}=\frac{\hbar}{m_{i}}\mathrm{Im}\frac{\nabla_{i}\psi}{\psi}|_{\mathbf{q}_{j}=\mathbf{Q}_{j}(t)}=\frac{\nabla_{i}S}{m_{i}}|_{\mathbf{q}_{j}=\mathbf{Q}_{j}(t)},
\end{equation}
for all $i=1,..,N$, where the $\nabla S$ form follows if we write
$\psi=|\psi|e^{iS/\hbar}$. In addition, we have ``equivariance''
\cite{Duerr2009,OriolsMompart2012,Goldstein2013}, i.e., the statement
that if the initial particle configuration of the dBB system is distributed
as $\rho_{0}=|\psi_{0}|^{2}$, then this ``quantum equilibrium distribution''
\cite{Duerr2009,OriolsMompart2012,Goldstein2013} is preserved under
time-evolution by the quantum continuity equation implicit in (4.175).
In other words, the quantum continuity equation implicit in (4.175)
entails the map $|\psi_{0}|^{2}\rightarrow|\psi_{t}|^{2}$.

Notice that both the wavefunction and the particles `feel' the mass
parameters $\left\{ m_{1},...,m_{N}\right\} $. More specifically,
the time-evolution of $\psi$ (at every point in configuration space)
through (4.175) explicitly depends on all the mass parameters via
the kinetic energy operators, while the evolution of $\mathbf{Q}_{i}(t)$
depends explicitly on only $m_{i}$ but implicitly on all the other
mass parameters through the positions of all the other particles.
The dependence of the evolution of $\psi$ on the mass parameters
is made even more manifest by starting from the \emph{N}-particle Bohm-Dirac
theory \cite{Holland1992,HollandBook1993}, i.e., the most straightforward
relativistic \emph{N}-particle extension of (4.175-176), and then taking
the non-relativistic limit; we would find that the positive-energy
components of the Dirac spinor in the Bohm-Dirac theory evolve by
a corrected version of (4.175), where the correction terms are rest-energy
terms $\sum_{i=1}^{N}m_{i}c^{2}$ in the Hamiltonian operator. 

One might think that since a classical gravitational field lives (by
definition!) in 3-space, and since only the particles live in 3-space,
this is why the particles should be the (point) sources for the gravitational
field. However, recall that the right hand side of (4.171) gives a
natural definition of a 3-space mass density in terms of $\psi$ in
configuration space. 

Consequently, it would seem that inertial mass is a property of both
the wavefunction and the particles, and there seems to be no justification
for assuming that the particles \emph{must} \emph{be used} \emph{solely}
as the mass density sources for a classical gravitational field, if
one wants to make a fundamentally-semiclassical Newtonian gravitational
theory out of the present version of dBB. Not only that, if one allows
$\psi$ to have properties such as energy density, momentum density,
etc., one can define the Hamiltonian density 
\begin{equation}
\mathcal{H}=\psi^{\ast}\left[-\sum_{i=1}^{N}\frac{\hbar^{2}}{2m_{i}}\nabla_{i}^{2}+V^{int}+m_{i}c^{2}\right]\psi.
\end{equation}
This Hamiltonian density has the physical interpretation of the energy
density stored in the ontic wavefunction, and indicates that the rest-energy
terms, hence the $m_{i}$, compose the total mass-energy density of
the wavefunction in configuration space. To be sure, nothing in the
first-order version of dBB or the dual space version thereof \emph{requires}
that $\psi$ have additional properties like energy density; but nothing
excludes these additional properties either. In any case, \emph{if}
one allows $\psi$ to have properties like energy density, then the
present version of dBB theory seems to make a compelling case for
(at least) taking $\psi$ to be the mass density source for the classical
gravitational field. 

Given that the dBB theory under consideration is ambiguous about which
part of its ontology should be used (or is most natural to use) as
the mass density source for a classical gravitational field, let us
consider the empirical consequences of using either the wavefunction
or the particles or both.

If $\psi$ is used as a source, then the Poisson equation for the
classical gravitational field takes the SN form (4.171), and the Schr\"{o}dinger
equation (4.175) takes the SN form (4.170). Because the SN system
(4.170-171) predicts that the components of a macroscopic superposition
of position wavefunctions gravitationally interact with one another,
this means that the evolution of the dBB particle configuration occupying
one component of the macroscopic superposition will be influenced
by the classical gravitational field sourced by the matter density
associated with the other components. This will lead to evolutions
of the dBB configuration that grossly disagree with experience. (For
example, the evolution of the dBB configuration of the sun would be
influenced by the classical gravitational field produced by a sun-mass
density corresponding to an empty wave packet corresponding to a macroscopically
distinct alternative spatial location for the sun). So we must conclude
that this version of fundamentally-semiclassical dBB Newtonian gravity
(hereafter, dBBfsc-Newton1 where ``fsc'' = ``fundamentally-semiclassical'')
is not empirically viable.

If the particles are used as point sources, then the Poisson equation
takes the form 
\begin{equation}
\nabla^{2}\Phi_{g}=4\pi\sum_{i=1}^{N}m_{i}\delta^{3}\left(\mathbf{q}-\mathbf{Q}_{i}(t)\right),
\end{equation}
where the $\mathbf{Q}_{i}(t)$ are solutions of the guiding equation
(4.176) for all $i=1,..,N$. The solution of (4.169) then yields the
inter-particle gravitational potential energy, which depends on the
actual positions of all the dBB particles at a single time, and feeds
back into the Schr\"{o}dinger equation
(4.175), giving
\begin{equation}
i\hbar\frac{\partial\psi}{\partial t}=\left[-\sum_{i=1}^{N}\frac{\hbar^{2}}{2m_{i}}\nabla_{i}^{2}+V_{g}^{int}(\mathbf{Q}_{i}(t),\mathbf{Q}_{j}(t))\right]\psi,
\end{equation}
where 
\begin{equation}
V_{g}^{int}(\mathbf{Q}_{i}(t),\mathbf{Q}_{j}(t))=\sum_{i=1}^{N}\frac{m_{i}\Phi_{g}}{2}=-\sum_{i=1}^{N}\frac{m_{i}}{2}\sum_{j=1}^{N(j\neq i)}\frac{m_{j}}{|\mathbf{Q}_{i}(t)-\mathbf{Q}_{j}(t)|}.
\end{equation}
This version of dBBfsc-Newton (dBBfsc-Newton2) was also considered
by Struyve \cite{Struyve2015}, who suggested that it might constitute
a viable alternative to the SN equations. (Kiessling considered the
electrostatic analogue in \cite{Kiessling2006}.) However, it is not
yet clear what the empirical predictions are of the system (4.179-180).
Since in dBBfsc-Newton2 the evolution of $|\psi_{t}|^{2}$ depends
on the actual positions of all the particles at each time, the equivariance
property breaks down and the standard means of showing that dBB theory
is empirically adequate cannot be applied. The reason is that the
derivation of the standard quantum formalism from standard dBB relies
on the equivariance of $|\psi_{t}|^{2}$; but since equivariance can't
even be formulated in dBBfsc-Newton2, it is no longer consistent to
assume that the dBB particle configuration is $|\psi_{t}|^{2}$ distributed. 

More precisely, in standard dBB, typicality needs to be time-independent
(for many reasons) and one needs to prove the law of large numbers
for subsystems of a dBB universe (i.e., that an ensemble of N subsystems
of a dBB universe, each with the same effective wavefunction, will
have relative frequencies of configuration coordinates that converges
to the $|\psi_{t}|^{2}$ distribution as N becomes large). \cite{Duerr1992,Duerr2009}
Equivariance is necessary to ensure the time-independence of typicality,
and equivariance plays a crucial technical role in proving the law
of large numbers. Without equivariance, typicality will not be time-independent
and one cannot prove the law of large numbers for said subsystems. 

Nevertheless, it has been suggested by Goldstein {[}Struyve, personal
communication{]} that it should be possible to extract statistical
predictions from (4.179-180) by another argument. The argument suggested
is as follows: given a stationary measure on Hilbert space $d\mu$,
then the measure $d\mu|\psi_{0}|^{2}$ will be preserved by the dynamics
on the product space of Hilbert space and configuration space. Therefore,
starting with an ensemble of systems all with (approximately) the
same effective wavefunction, it is natural to take $|\psi_{0}|^{2}$
as the initial position distribution (although, as time evolves, it
is expected that the position distribution will deviate from the $|\psi_{0}|^{2}$
distribution, due to the nonlinear dynamics). Assuming this works,
it is worth noting that the nonlinearity of (4.179) entails a break-down
of the general validity of the superposition principle. That is, the
nonlinear evolution for $\psi$ means that a linear superposition
of two solutions to (4.179) at the same space-time point does not
in general form a new solution, i.e., the solution space of (4.179)
doesn't form a linear space. (As exceptions, $\psi_{1}+\psi_{2}$
will be a solution of a nonlinear wave equation if either $\psi_{2}=a\psi_{1}$,
where $a$ is a constant, or if $\psi_{1}$ and $\psi_{2}$ have no
common support \cite{Holland1993}.) So, at the moment, it seems unclear
how dBBfsc-Newton2 will be able to account for quantum phenomena that
rely on the general validity of the superposition principle, e.g.,
electron two-slit (or N-slit) interference and decoherence of two-state
systems. Perhaps it can be argued that, because of the small magnitude
of the gravitational interaction energy in (4.179), relative to the
kinetic energy term and any external potentials that might arise in
a given physical situation, the nonlinear time-evolution of solutions
of (4.179) will still closely approximate the time-evolution of solutions
of the usual linear Schr\"{o}dinger
equation in the aforementioned situations. In any case, the empirical
predictions dBBfsc-Newton2 remain to be worked out.

Using both the wavefunction and the particles as mass density sources
for classical gravitational fields (dBBfsc-Newton3) would entail the
same ambiguities as dBBfsc-Newton2, and (even worse) the same empirically-inadequate
predictions of dBBfsc-Newton1 (involving gravitational interactions
between components of a macroscopic superposition of position wavefunctions).

We must therefore conclude that there does not appear to be, at present,
an empirically viable formulation of dBBfsc-Newton that's based on
the first-order dual-space version of dBB. (By ``empirically viable''',
we mean `shown to be consistent with' or `likely to be consistent
with' existing non-relativistic quantum mechanical experiments.) Moreover,
we do not see how to obtain an empirically viable formulation of dBBfsc-Newton
using other versions of first-order dBB theory, whether Albert's `world
particle' formulation \footnote{Albert's formulation takes as fundamental ontological postulates (i)
configuration space $\mathbb{R}^{3N}$, (ii) $\psi$ in configuration
space evolving by the \emph{N}-particle Schr\"{o}dinger
equation, the latter defined in terms of a Hamiltonian that includes
an \emph{N}-particle interaction potential $\hat{V}^{int}(\hat{\mathbf{q}}_{i},\mathbf{\hat{q}}_{j})$
that's written in a preferred coordinate system, and (iii) a single
configuration point (the world particle) in $\mathbb{R}^{3N}$, evolving
by the guidance equation. 3-space, and a configuration of particles
in 3-space, are claimed to be emergent ontologies in the sense that
they are claimed to arise from a philosophical-functionalist analysis
of $\hat{V}^{int}$ and the latter's influence on the motion of the
world particle through $\psi$. Thus a fundamentally-semiclassical
gravity version of dBB theory in Albert's formulation would correspond
to just equations (4.170-171) of dBBfsc-Newton2, with the interpretation
that the `world particle' backreacts on $\psi$ via the former's gravitational
`self-energy'. } \cite{Albert2015}, Norsen's TELB formulation \footnote{The TELB (Theory with Exclusively Local Beables) formulation differs
from the dual space formulation in that 3-space is the only ontic
space. This approach is (mathematically) motivated by Taylor-expanding
$\psi$ in configuration space into an infinite hierarchy of nonlocally
coupled fields in 3-space; more precisely, each particle has a single-particle
wavefunction pushing it around via the guidance equation, but the
single-particle wavefunction is coupled to an infinite hierarchy of
3-space ``entanglement fields'', which are themselves nonlocally
coupled to the entanglement fields of every other particle (hence
why they are called ``entanglement'' fields). The postulate $\rho_{0}=|\psi_{0}|^{2}$
is still imposed on the single-particle wavefunctions, and equivariance
still holds. One could then define classical gravitational fields
directly in terms of mass density sources built out of the single-particle
wavefunctions, but this would just lead to a TELB version of the SN
equation, which would entail all the empirically problematic predictions
of the SN equation discussed in relation to dBBfsc-Newton1. And if
one were to use the particles as point sources, instead (or in tandem
with the single-particle wavefunctions), the ambiguities associated
with dBBfsc-Newton2 would arise.} \cite{Norsen2010,Norsen2014}, or D\"{u}rr-Goldstein-Zangh\`{i}'s
(and Vassallo et al.'s `Quantum Humeanist' \cite{VassalloDeckertEsfeld2016})
nomological formulation \footnote{The nomological formulation of D\"{u}rr-Goldstein-Zangh\`{i}
(DGZ) is still conjectural, but the basic idea is that the `fundamental'
wavefunction is the time-independent Wheeler-DeWitt wavefunctional
$\Psi\left(h,\phi\right)$, interpreted as part of \emph{physical
law} rather than physical ontology. Time-dependent wavefunctions are
suggested to be derived, effective descriptions for `subsystems' of
the universe, and not part of physical ontology either. Only 3-space
and particles living in 3-space constitute physical ontology. Accordingly,
one cannot not use time-dependent wavefunctions in the definition
of an SN-type classical mass-density source in 3-space, as this would
be inconsistent with the expected Newtonian limit of the Wheeler-DeWitt
equation (i.e., the usual linear Schr\"{o}dinger
equation involving a q-number gravitational interaction potential)
\cite{DGZ1995,GoldsteinZanghi2011}. Nor could one use the dBB particles
as point sources for a classical gravitational field coupling back
to the time-dependent wavefunction, as this is inconsistent with the
expected Newtonian limits of the Wheeler-DeWitt equation and the guiding
equations for $h$ and $\phi$ (i.e., the equations of dBB-Newton).
So the DGZ nomological formulation of dBB doesn't seem compatible
with fundamentally-semiclassical Newtonian gravity.

Vassallo et al.'s `Quantum Humeanist' formulation of dBB gives $\psi$,
evolving by the time-dependent Schr\"{o}dinger
equation, a nomological interpretation in the Humean sense (i.e.,
``being a variable in the law that achieves the simplest and most
informative description of the change in the primitive ontology (e.g.
relative particle positions) throughout the history of the universe''
\cite{VassalloDeckertEsfeld2016}), and takes as the fundamental (and
only) ontology the distance relations among dBB point particles along
with the `Humean mosaic' traced out by those distance relations in
time. Thus the Quantum Humeanist formulation is compatible with dBBfsc-Newton2. }, \cite{DGZ1995,Duerr2009,GoldsteinZanghi2011,DGZbook2012,Goldstein2013}.
As it turns out, second-order formulations of dBB, namely the ``ontological
interpretation'' advocated by Bohm-Hiley \cite{BohmHiley1993} and
Holland \cite{HollandBook1993}, don't seem to change the situation
either: their only difference from first-order formulations of dBB
is that the Schr\"{o}dinger equation
and wavefunction are replaced by the Madelung equations for $|\psi|$
and $S$, with the quantization condition imposed on the latter. 

By comparison, while the ontology of ZSM-Newton involves more than
just particles, it is clear from the very formulation of ZSM-Newton
that the particles must be understood as possessors of inertial mass.
This is manifest from (i) the definition of the rest-mass of a \emph{zbw}
particle as corresponding to the energy associated with the Compton
frequency oscillation of the \emph{zbw} particle in its rest frame,
and (ii) the definition of the \emph{i}-th Wiener process, which describes
the stochastic evolution of the \emph{i}-th particle position and
depends on the \emph{i}-th mass parameter through the diffusion coefficient
$\hbar/m_{i}$. Furthermore, as we argued in section 3, while the
ether of ZSM is expected to carry stress-energy, it is expected to
be negligible in the Newtonian regime as well as conceptually different
from the mass parameters that appear in the diffusion coefficient
and the equations of motion for $S$ and $\rho$. So, in contrast
to dBB, ZSM seems to make the choice of the particles as mass density
sources for a classical gravitational field, inevitable. Another difference
from dBB is the following: recall from section 3 that, because the
Schr\"{o}dinger equation and wavefunction
are derived in ZSM, the use of the particles as sources for a classical
gravitational field doesn't entail the nonlinear coupling in (4.170-171);
rather, as we saw in section 3, the gravitational field that does
couple to the Schr\"{o}dinger equation/wavefunction
corresponds (to leading order) to $\hat{V}_{g}^{int}(\mathbf{\hat{q}}_{i},\mathbf{\hat{q}}_{j})$.
This is why ZSM-Newton avoids a break-down of the equivariance property.
So despite ZSM and dBB sharing many equations in common - the Schr\"{o}dinger
equation (4.174), the guiding equation (4.175), and equivariance of
$\rho_{0}=|\psi_{0}|^{2}$ - and despite both theories sharing in
common a ``primitive ontology'' \footnote{Primitive ontology is defined by Allori et al. \cite{Allori2012}
as ``variables describing the distribution of matter in 4-dimensional
space-time''.} involving particles with definite 3-space trajectories, the different
axioms on which ZSM and dBB are based lead to significantly different
conclusions about how to formulate a theory of fundamentally-semiclassical
Newtonian gravity, and the empirical viability thereof. Of course
the empirical predictions entailed by the ZSM-Newton nonlinear Schr\"{o}dinger
equation (4.101) remain to be studied, and our comments about the
inapplicability of the superposition principle to solutions of (4.179)
apply just as well to ZSM-Newton with (4.101). So the empirical viability
of ZSM-Newton with (4.101) is also an open question. This being said,
one notable advantage of ZSM-Newton over dBBfsc-Newton2 is that one
can still justifiably use the usual linear Schr\"{o}dinger
equation with $\hat{V}_{g}^{int}(\mathbf{\hat{q}}_{i},\mathbf{\hat{q}}_{j})$
in the context of ZSM-Newton, because the contributions to the gravitational
interactions from the nonlinear terms in (4.101) are miniscule compared
to the leading-order term (i.e., $\hat{V}_{g}^{int}(\mathbf{\hat{q}}_{i},\mathbf{\hat{q}}_{j})$).
So, at least at this level of approximation, ZSM-Newton already makes
clearly testable predictions for Newtonian-gravitational experiments
involving coherent (or decohered) quantum systems, and these predictions
are extracted using the usual theory of measurement in conventional
stochastic mechanics and standard dBB theory \cite{DerakProbingGravCat2016}.
Moreover, those predictions will exactly coincide with the predictions
of SQM-Newton, Newtonian PQG, and dBB-Newton (in quantum equilibrium). 

It is a straightforward exercise to demonstrate that analogous conclusions
follow from consideration of the electrodynamical case, i.e., dBBfsc-Coulomb
theories vs. ZSM-Coulomb, except that ZSM-Coulomb doesn't have an
analogue of the nonlinear Schr\"{o}dinger
equation (4.101), so the issues related to (4.101) don't arise in
the ZSM-Coulomb case. All this said, we now wish to evaluate a well-known
peculiarity of standard dBB theory involving charge-field coupling
(i.e., dBB-Coulomb), from the viewpoint of ZSM-Coulomb. 

For a single-particle dBB system, in the presence of an external magnetic
vector potential $\mathbf{A}_{ext}(\mathbf{q},t)$, the momentum operator
in the Schr\"{o}dinger equation gets
a correction $\mathbf{\hat{p}}\rightarrow\mathbf{\hat{p}}-e\mathbf{A}_{ext}$.
Now, consider the magnetic Aharonov-Bohm (AB) effect in dBB \cite{PhilippidisBohmKaye1982,BohmHiley1993,HollandBook1993},
where $\mathbf{A}_{sol}=\left(\Phi/2\pi r\right)\hat{\theta}$ is
the magnetic vector potential sourced by an infinitely long cylindrical
solenoid with flux $\Phi$. For an electron wavepacket split into
two partial packets passing on either side of the solenoid, where
the paths $P_{1}$ and $P_{2}$ traversed by the packets form a loop
$C$ encircling the solenoid, the correction to the momentum operator
entails a phase shift $\psi\rightarrow\psi'=N'\left[\psi_{1}+\psi_{2}e^{ie\Phi/\hbar}\right]e^{\left(ie/\hbar\right)\int_{P_{1}}\mathbf{A}_{sol}\cdot d\mathbf{q}}$,
when the packets are recombined to form an interference pattern ($N'$
is a normalization constant). Correspondingly, the position probability
density associated to the interference pattern gets shifted as $|\psi'|^{2}=\rho'=N'^{2}\left\{ \rho_{1}+\rho_{2}+2\sqrt{\rho_{1}}\sqrt{\rho_{2}}cos\left[(S_{1}-S_{2})/\hbar-\delta\right]\right\} $,
where $\delta=e\Phi/\hbar$. Note that while the dBB particle moves
along with only one of the packets around the solenoid, say the packet
traversing path $P_{1}$, with modified momentum $\mathbf{p}=\nabla S_{1}-e\mathbf{A}_{sol}$,
both packets `feel' $\mathbf{A}_{sol}$ since each picks up a phase
factor $\psi_{a}\rightarrow\psi_{a}e^{(ie/\hbar)\int_{P_{a}}\mathbf{A}_{sol}\cdot d\mathbf{q}}$
such that $\Phi=\oint_{C}\mathbf{A}_{sol}\cdot d\mathbf{q}=\int_{P_{1}}\mathbf{A}_{sol}\cdot d\mathbf{q}-\int_{P_{2}}\mathbf{A}_{sol}\cdot d\mathbf{q}$
and $\oint_{C}\mathbf{p}\cdot d\mathbf{q}=nh-e\Phi$. In other words,
even though the motion of the dBB particle is altered by the presence
of the vector potential, suggesting (seemingly) that the charge $e$
is a property localized to the dBB particle (like in classical electrodynamics),
the fact that the `empty' packet (i.e., the packet moving along $P_{2}$)
also picks up a phase factor, and that this phase factor contributes
to the shift in the interference pattern of the recombined packets,
suggests that charge is \emph{also} a property carried by the (spatially
delocalized) wavefunction \cite{PhilippidisBohmKaye1982,BohmHiley1993,HollandBook1993,BrownDewdneyHorton1995}.
A completely analogous situation arises for the gravitational analogue
of the magnetic AB effect, where $\mathbf{A}_{sol}$ is the gravitomagnetic
vector potential sourced by a solenoid carrying a mass (instead of
charge) current, and all other expressions are identical except for
the replacement $e\rightarrow m$ \cite{Hohensee2012}. Analogous
considerations apply to the case of the electric/gravitoelectric AB
effect.

Since the dBB treatment of the AB effect is formally the same as the
ZSM-Coulomb/Newton treatment of the AB effect, this might seem to
conflict with the ZSM-Coulomb/Newton hypothesis that the charge (rest
mass) of a system is a property localized to \emph{zbw} particles.
However, there is no inconsistency. In ZSM-Coulomb/Newton, the finding
that the empty packet in the AB effect picks up a phase factor that
contributes to the shift in the interference pattern intensity is
a \emph{consequence} of the following set of postulates: (i) rest-mass
and charge are intrinsic properties of \emph{zbw} particles; (ii)
the \emph{zbw} particles, whose oscillations are dynamically driven
by the ether medium, always have well-defined mean phases along their
3-space trajectories; and (iii) the diffusion process for the \emph{zbw}
particles in the ether satisfies the global constraint of being conservative.
It might then be asked if ZSM-Coulomb/Newton gives physical insight
into what it means, in terms of its proposed underlying ontological
picture of the world, for empty packets to electromagnetically (or
gravitationally) couple to external fields, even though it is the
\emph{zbw} particles that carry the rest-mass and charge of a system.
We can sketch an answer as follows.

As discussed in \cite{Schroedinger1938,Bohm1951,Derakhshani2016a},
the superposition principle for wavefunctions is a consequence of
the single-valuedness condition, and the single-valuedness condition
on wavefunctions in ZSM follows from the union of postulates (ii)
and (iii). And as we've discussed in \cite{Derakhshani2016b}, an
empty packet describes possible alternative histories of a Nelsonian/\emph{zbw}
particle through a different region of the ether (the different region
corresponding to the spatial support of the empty wavepacket in 3-space),
while also indirectly reflecting spatio-temporal variations in that
different region of the ether (because the ether-sourced osmotic potential
$U(q,t)$ changes as a function of space and time via the continuity
equation and is constrained by boundary conditions in the environment).
Thus the empty packet traversing path $P_{2}$ reflects (indirectly)
a region of the ether that's (spatio-temporally) varying along $P_{2}$,
and the interference of the recombined packets reflects (indirectly)
two regions of ether recombining and interfering while satisfying
postulates (ii) and (iii). Since the ether medium is presumed to pervade
all of 3-space, and since all components of the ether are presumed
to be nonlocally connected to each other, the ether region corresponding
to the empty packet is actually not physically independent of the
ether region corresponding to the occupied packet. In other words,
for the ether to maintain the quantization condition $\oint_{L}\nabla S\cdot d\mathbf{q}=nh$
on the \emph{zbw} particle, while maintaining that the diffusion of
the \emph{zbw} particle through the ether is conservative, it must
know to compensate for the phase shift experienced by the \emph{zbw}
particle passing around the solenoid along $P_{1}$, by correspondingly
shifting phase in the region that's spatio-temporally varying along
$P_{2}$. How exactly this works (assuming the ZSM framework is correct)
will presumably require developing an explicit physical model of the
ether, the \emph{zbw} particle, and the dynamical coupling of the
two, in accord with postulates (i-iii). This is left for future work.

\subsubsection{Comparison to semiclassical approximations in de Broglie-Bohm theory}

As we've seen, there does not appear to exist an empirically viable
formulation of dBBfsc-Newton/Coulomb. Nevertheless, it is possible
to formulate semiclassical approximation schemes for the `fully quantum'
formulation of dBB Newtonian gravity/electrodynamics (hereafter, dBB-Newton/Coulomb). 

The dBB-Newton/Coulomb theory corresponds to (4.175-176) with $V^{int}=\hat{V}_{g,e}^{int}(\mathbf{\hat{q}}_{i},\mathbf{\hat{q}}_{j})$
(for simplicity, we neglect vector potentials). In other words the
\emph{N}-particle Schr\"{o}dinger equation
of dBB-Newton/Coulomb is identical to the \emph{N}-particle Schr\"{o}dinger
equation of ZSM-Newton/Coulomb, when the nonlinear correction terms
predicted by the latter are neglected. The physical interpretation,
however, is different. 

In ZSM-Newton/Coulomb, the \emph{zbw} particles carry rest-mass/charge
and interact with one another through the classical gravitational/electrostatic
fields they source. In dBB-Newton/Coulomb the particles are just points
at definite locations, and $\hat{V}_{g,e}^{int}(\mathbf{\hat{q}}_{i},\mathbf{\hat{q}}_{j})$
is a potential energy function on configuration space that influences
the evolution of $\psi$ in configuration space; so, to the extent
that the particles `interact' gravitationally or electrostatically,
they only do so indirectly via the influence of $\hat{V}_{g,e}^{int}(\mathbf{\hat{q}}_{i},\mathbf{\hat{q}}_{j})$
on $\psi$ through the Schr\"{o}dinger
equation (4.175), and the influence of $\psi$ on the evolution of
the particles through the guiding equation (4.176). Thus the mean-field
approximation scheme discussed in section 4 applies just as well to
dBB-Newton/Coulomb. 

Another dBB-based semiclassical approximation scheme has been suggested
by Prezhdo-Brooksby \cite{Prezhdo2001} and elaborated on by Struyve
\cite{Struyve2015}. Consider, for simplicity, the dBB theory with
two-particle Schr\"{o}dinger equation
\begin{equation}
i\hbar\frac{\partial\psi\left(\mathbf{q}_{1},\mathbf{q}_{2},t\right)}{\partial t}=\left[-\left(\frac{\hbar^{2}}{2m_{1}}\nabla_{1}^{2}+\frac{\hbar^{2}}{2m_{2}}\nabla_{2}^{2}\right)+\hat{V}_{g,e}^{int}(\mathbf{\hat{q}}_{1},\mathbf{\hat{q}}_{2})\right]\psi\left(\mathbf{q}_{1},\mathbf{q}_{2},t\right).
\end{equation}
The guiding equations for each particle are again given by (4.176),
and the 2nd-order equations of motion are
\begin{equation}
\begin{split}m_{1}\ddot{\mathbf{Q}}_{1}(t)=-\nabla_{1}\left[V_{g}^{int}\left(\mathbf{q}_{1},\mathbf{Q}_{2}(t)\right)+Q\left(\mathbf{q}_{1},\mathbf{Q}_{2}(t)\right)\right]|_{\mathbf{q}_{1}=\mathbf{Q}_{1}(t)},\end{split}
\end{equation}
\begin{equation}
m_{2}\ddot{\mathbf{Q}}_{2}(t)=-\nabla_{2}\left[V_{g}^{int}\left(\mathbf{Q}_{1}(t),\mathbf{q}_{2}\right)+Q\left(\mathbf{Q}_{1}(t),\mathbf{q}_{2}\right)\right]|_{\mathbf{q}_{2}=\mathbf{Q}_{2}(t)},
\end{equation}
where $Q\left(\mathbf{q}_{1},\mathbf{q}_{2}\right)$ is the total
quantum potential of the two-particle system. 

Now, the conditional wavefunction for particle 1, defined as $\psi_{1}(\mathbf{q}_{1},t)=\psi(\mathbf{q}_{1},\mathbf{Q}_{2}(t),t)$,
satisfies the conditional Schr\"{o}dinger
equation
\begin{equation}
i\hbar\frac{\partial\psi_{1}\left(\mathbf{q}_{1},t\right)}{\partial t}=\left[-\frac{\hbar^{2}}{2m_{1}}\nabla_{1}^{2}+V_{g,e}^{int}\left(\mathbf{q}_{1},\mathbf{Q}_{2}(t)\right)\right]\psi_{1}\left(\mathbf{q}_{1},t\right)+K(\mathbf{q}_{1},t),
\end{equation}
where 
\begin{equation}
K(\mathbf{q}_{1},t)=-\frac{\hbar^{2}}{2m_{2}}\nabla_{2}^{2}\psi(\mathbf{q}_{1},\mathbf{q}_{2},t)|_{\mathbf{q}_{2}=\mathbf{Q}_{2}(t)}+i\hbar\frac{d\mathbf{Q}_{2}(t)}{dt}\cdot\nabla_{2}\psi(\mathbf{q}_{1},\mathbf{q}_{2},t)|_{\mathbf{q}_{2}=\mathbf{Q}_{2}(t)}.
\end{equation}
Correspondingly, the conditional guiding equation for particle 1 is
\begin{equation}
\frac{d\mathbf{Q}_{1}(t)}{dt}=\frac{\hbar}{m_{1}}\mathrm{Im}\frac{\nabla_{1}\psi_{1}}{\psi_{1}}|_{\mathbf{q}_{1}=\mathbf{Q}_{1}(t)}=\frac{\nabla_{1}S_{1}}{m_{1}}|_{\mathbf{q}_{1}=\mathbf{Q}_{1}(t)},
\end{equation}
where $S_{1}=S_{1}(\mathbf{q}_{1},t)$. The Newtonian equation of
motion for particle 2 is then 
\begin{equation}
m_{2}\ddot{\mathbf{Q}}_{2}(t)=-\nabla_{2}\left[V_{g,e}^{int}\left(\mathbf{Q}_{1}(t),\mathbf{q}_{2}\right)+Q\left(\mathbf{Q}_{1}(t),\mathbf{q}_{2}\right)\right]|_{\mathbf{q}_{2}=\mathbf{Q}_{2}(t)}.
\end{equation}
The semiclassical approximation is when $m_{2}\gg m_{1}$ and $\psi$
varies slowly in $\mathbf{q}_{2}$ (compared to $\mathbf{q}_{1}$).
Then $K\approx0$ and $-\nabla_{2}Q\approx0$. In other words the
time-evolution of particle 2 depends (approximately) only on the classical
interaction potential $V_{g,e}^{int}$, evaluated at the actual position
of particle 1. And the time-evolution of particle 1 depends on $\psi_{1}$
satisfying (approximately) (4.183) with $K\approx0$, i.e., particle
1's effective Schr\"{o}dinger equation
that takes into account the back-reaction of particle 2 through $V_{g,e}^{int}\left(\mathbf{q}_{1},\mathbf{Q}_{2}(t)\right)$.
Note that, unlike models of dBBfsc-Newton/Coulomb, this semiclassical
approximation scheme defines a consistent back-reaction between the
two particles in the following sense: the conditional wavefunction
of particle 1, in the semiclassical approximation, just corresponds
to the effective wavefunction of particle 1, for which $|\psi_{1}|^{2}$
satisfies an equivariance-like property (through the conditional quantum
continuity equation implicit in (4.183)), even though the (semiclassically
approximated) evolution of $\psi_{1}$ still depends on the actual
position of particle 2 through $V_{g,e}^{int}$.

By contrast, the standard QM semiclassical approximation scheme for
two interacting particles \cite{Prezhdo2001,Struyve2015} is defined
by 
\begin{equation}
i\hbar\frac{\partial\psi\left(\mathbf{q}_{1},t\right)}{\partial t}=\left[-\frac{\hbar^{2}}{2m_{1}}\nabla_{1}^{2}+V_{g,e}^{int}\left(\mathbf{q}_{1},\mathbf{\overline{Q}}_{2}(t)\right)\right]\psi\left(\mathbf{q}_{1},t\right),
\end{equation}
\begin{equation}
m_{2}\ddot{\overline{\mathbf{Q}}}_{2}(t)=\int_{\mathbb{R}^{3}}d^{3}\mathbf{q}_{1}|\psi(\mathbf{q}_{1},t)|^{2}\left[-\nabla_{2}V_{g,e}^{int}\left(\mathbf{q}_{1},\mathbf{q}_{2}\right)\right]|_{\mathbf{q}_{2}=\mathbf{\overline{Q}}_{2}(t)},
\end{equation}
where $\psi(\mathbf{q}_{1},t)$ is a single-particle wavefunction
(as opposed to a conditional or effective wavefunction), and the back-reaction
from particle 2 on particle 1 is via the average trajectory $\mathbf{\overline{Q}}_{2}(t)$
inserted into $V_{g,e}^{int}$ in (4.187). 

Prezhdo and Brooksby \cite{Prezhdo2001} have compared the dBB-based
semiclassical approximation scheme to this standard QM scheme, for
the case of a light particle scattering off a heavy particle, where
the heavy particle is bound to a fixed surface. They found that the
dBB scheme is superior at tracking the scattering probability as a
function of time (when compared to the exact quantum dynamics description),
in addition to being computationally simpler to implement than the
standard QM scheme. 

Struyve \cite{Struyve2015} has applied the dBB-based scheme to a
dBB version of scalar electrodynamics, as well as to a dBB version
of canonical quantum gravity under the minisuperspace approximation
\footnote{Canonical quantum gravity under the minisuperspace approximation refers
to the Wheeler-DeWitt equation $\mathcal{H}\Psi\left(h,\phi\right)=0$
(and momentum constraint $\mathcal{H}_{i}\Psi\left(h,\phi\right)=0$),
under the restriction that the 3-metric $h$ and matter field $\phi$
are homogeneous and isotropic \cite{HollandBook1993,Kiefer2012,Struyve2015}.
This corresponds to a time-dependent homogeneous matter field $\phi(t)$
in an FLRW metric with homogeneous scale factor $a(t)$. The Wheeler-DeWitt
equation then takes the form $\left(H_{metric}+H_{matter}\right)\psi\left(a,\phi\right)=0$.
In the dBB version \cite{Struyve2015}, this latter form of the Wheeler-DeWitt
equation is accompanied by guidance equations for the field beables
$a(t)$ and $\phi(t)$, which turn out to be coupled to each other
via the phase $S$ of $\psi$. In this way, the metric and matter
field beables back-react on each other. It is worth mentioning that
the minisuperspace approximation is also referred to in the literature
as a `semiclassical' approximation; it should not be confused with
the dBB-based semiclassical approximation scheme, the latter of which
is applied by Struyve \emph{on top of} the minisuperspace approximation. }. In the latter case, he has compared the dBB-based scheme to the
standard scheme (applied to standard canonical quantum gravity under
the minisuperspace approximation) for cases involving macroscopic
superpositions of two Gaussians wavepackets. As it turns out, the
dBB-based scheme yields better agreement with the exact dBB version
of canonical quantum gravity under the minisuperspace approximation,
than does the standard scheme \footnote{Struyve did not compare the standard scheme to the standard quantum
interpretation of the Wheeler-DeWitt equation, the reasoning being
that the ``problem of time'' makes the standard quantum interpretation
of the Wheeler-DeWitt equation incoherent. Nevertheless, Struyve pointed
out that for approaches to quantum theory that associate approximately
classical dynamics to macroscopic superpositions of Gaussian states
(such as many-worlds interpretations \cite{Kiefer2012}), the standard
scheme is expected to do worse than the dBB scheme in approximating
exact solutions of the Wheeler-DeWitt equation (assuming those non-dBB
approaches to quantum theory yield consistent quantum interpretations
of the Wheeler-DeWitt solutions in the first place).}. 

The dBB semiclassical approximation scheme for two interacting particles
can, of course, be imported into ZSM-Newton/Coulomb. In this sense,
the results obtained by Prezhdo-Brooksby are also results that follow
from ZSM-Newton/Coulomb. However, in ZSM-Newton/Coulomb, we also have
the option of implementing the back-reaction from particle 1 onto
particle 2 via solutions of $d\mathbf{q}_{1}(t)=\left(\mathrm{Im}+\mathrm{Re}\right)m_{1}^{-1}\hbar\nabla_{1}\ln\psi_{1}|_{\mathbf{q}_{1}=\mathbf{q}_{1}(t)}dt+d\mathbf{W}(t)$, the conditional stochastic differential equation for
particle 1.
Since the trajectories predicted by this stochastic differential equation
differ from the trajectories predicted by the conditional guidance
equation (4.185), we would expect differences in the predictions of
the ZSM-Newton/Coulomb version as compared to the dBB version. Although,
considering that the semiclassical approximation requires the mass
of particle 2 to be much greater than particle 1, we would expect
any differences to be very slight. Nevertheless, it would be interesting
to revisit the cases studied by Prezhdo-Brooksby and Struyve, to see
if the differences might be amenable to experimental/observational
discrimination. (Revisiting Struyve's analyses from the viewpoint
of ZSM will of course require extending ZSM to relativistic field
theories in flat and curved spacetimes, and to the spacetime metric
itself. Future work will show how this can be done.)

\section{Conclusion}

We have shown how to formulate fundamentally-semiclassical Newtonian
gravity/electrodynamics based on stochastic mechanics in the ZSM formulation.
In addition, we have shown that ZSM-Newton/Coulomb has a consistent
statistical interpretation, recovers the standard exact quantum description
of matter-gravity coupling as a special case valid for all practical
purposes (even though gravity remains fundamentally classical in the
ZSM approach), and recovers the SN/SC and stochastic SN/SC equations
as mean-field approximations. We have also compared ZSM-Newton/Coulomb
to theories of semiclassical Newtonian gravity based on standard quantum
theory, dynamical collapse theories, other possible formulations of
stochastic mechanics, and the dBB pilot-wave theory. In doing so,
we have highlighted conceptual and technical advantages entailed by
ZSM-Newton/Coulomb, and indicated possibilities for experimentally
testable differences. 

In Part II, we will use ZSM-Newton/Coulomb to formulate a new `large-\emph{N}'
prescription that makes it possible to consistently describe large
numbers of identical (ZSM) particles \emph{strongly interacting} classical-gravitationally/electrostatically.
This new large-\emph{N} prescription will also make it possible to recover
classical Newtonian gravity/electrodynamics for macroscopic particles,
as well as classical Vlasov-Poisson mean-field theory for macroscopic
particles weakly interacting gravitationally/electrostatically.

We wish to emphasize once more the two key results of the present
paper: (i) while ZSM-Newton and ZSM-Coulomb treat the gravitational
and Coulomb potentials, respectively, as fundamentally classical fields
sourced by point-like classical particles undergoing non-classical
(stochastic mechanical) motions in 3-space, these semiclassical theories
nevertheless recover the standard quantum descriptions of Newtonian/non-relativistic
gravitational/Coulombic interactions between particles; and (ii) the
large \emph{N}  limit scheme of Golse \cite{Golse2003} and Bardos et al.
\cite{BardosGolseMauser2000,BardosErdosGolseMauserYau2002}, applied
to ZSM-Newton/Coulomb, makes it possible to recover the same mean-field
approximations as obtained from standard Newtonian PQG/SCEG and standard
non-relativistic QED/SCRED (the SN/SC and stochastic SN/SC equations).

In a forthcoming standalone paper, we will show how to consistently
incorporate gravitational and electrodynamical radiation reaction
effects within ZSM-Newton and ZSM-Coulomb, respectively, through a
stochastic mechanical generalization of Galley's variational principle
for nonconservative systems \cite{Galley2013}. Further down the road,
we will show how to extend ZSM to particles and fields in relativistic
spacetimes, and then use that framework to formulate consistent hidden-variables
theories of semiclassical Einstein gravity and semiclassical relativistic
electrodynamics; we will then show that the Newtonian limits of these
two theories yield ZSM-Newton and ZSM-Coulomb, respectively.

\section{Acknowledgments}

It is a pleasure to thank Guido Bacciagaluppi, Ward Struyve, Roderich
Tumulka, Bei-Lok Hu, and Dieter Hartmann for many helpful discussions.
I am especially thankful to Guido and Ward for reading earlier drafts
of this paper and making several useful suggestions to improve the
paper.

\chapter{Semiclassical Newtonian Field Theories Based On Stochastic Mechanics
II}

Continuing the development of the ZSM-Newton/Coulomb approach to semiclassical
Newtonian gravity/electrodynamics \cite{Derakhshani2017}, we formulate
a ZSM-Newton/Coulomb version of the large \emph{N}  approximation scheme proposed
by Oriols et al. \cite{Oriols2016}. We show that this new large \emph{N} 
scheme makes it possible to self-consistently describe the center-of-mass
evolution of a large number of gravitationally/electrostatically interacting,
identical, \emph{zbw} particles, without assuming that the particles
are weakly coupled, and without entailing the problematic macroscopic
semiclassical gravitational/electrostatic cat states characteristic
of the mean-field Schr\"{o}dinger-Newton/Coulomb
equations. We also show how to recover \emph{N}-particle classical Newtonian
gravity/electrodynamics for many gravitationally/electrostatically
interacting macroscopic particles (composed of many interacting \emph{zbw}
particles), as well as classical Vlasov-Poisson mean-field theory
for macroscopic particles weakly interacting gravitationally/electrostatically.
Finally, we outline an explicit model of environmental decoherence
that can be incorporated into Oriols et al. scheme as applied to ZSM-Newton/Coulomb.

\section{Introduction}

This paper is a direct continuation of Part I \cite{Derakhshani2017}.
There, we formulated fundamentally-semiclassical Newtonian gravity/electrodynamics
based on stochastic mechanics in the ZSM formulation (ZSM-Newton/Coulomb).
Our key results were: (i) ZSM-Newton/Coulomb has a consistent statistical
interpretation; (ii) ZSM-Newton/Coulomb recovers the standard quantum
description of non-relativistic matter-gravity/charge-field coupling
as a special case valid for all practical purposes, even though the
gravitational/electrostatic interaction between \emph{zbw} particles
is fundamentally classical; and (iii) ZSM-Newton/Coulomb recovers
the `single-body' Schr\"{o}dinger-Newton/Coulomb
(SN/SC) and stochastic SN/SC equations as mean-field approximations
for systems of gravitationally/electrostatically interacting, identical,
\emph{zbw} particles, in the weak-coupling large \emph{N}  limit. 

We also discussed some limitations of the mean-field SN/SC and stochastic
SN/SC equations: (i) they are based on the assumption that interactions
between \emph{zbw} particles are sufficiently weak that the independent
particle approximation is plausible; and (ii) the single-body SN/SC
and stochastic SN/SC equations admit solutions corresponding to macroscopic
semiclassical gravitational/electrostatic cat states, and these cat
states predict unphysical gravitational/electrostatic forces on external
probe masses. (We also pointed out that the latter difficulty afflicts
any formulation of fundamentally-semiclassical Newtonian gravity/electrodynamics
based on the many-body SN/SC and stochastic SN/SC equations, as these
equations also allow for such cat states.)

The primary objective of the present paper is to develop a new large
N scheme for ZSM-Newton/Coulomb, that bypasses the limitations of
the mean-field SN/SC and stochastic SN/SC equations.

Our scheme will be based on the one developed recently by Oriols et
al. \cite{Oriols2016}, who consider the center-of-mass (CM) motion
of a system of N identical, non-relativistic, de Broglie-Bohm (dBB)
particles coupled through interaction potentials of the form $\hat{U}_{int}(\hat{x}_{j}-\hat{x}_{k})$
and to external potentials of the form $\hat{U}_{ext}(\hat{x}_{j})$.
They show that, in the limit $N\rightarrow\infty$, the CM motion
becomes effectively indistinguishable from classical Hamilton-Jacobi
mechanics for a single massive particle in an external field. 

Essentially, we will import the Oriols et al. scheme into ZSM-Newton/Coulomb.
In doing so, we will find that it is possible to: (i) self-consistently
describe the CM motion of large numbers of classically-gravitationally
and/or classically-electrostatically interacting, identical, \emph{zbw}
particles, without an independent particle approximation; (ii) avoid
macroscopic semiclassical gravitational and electrostatic cat states
and recover many-particle classical Newtonian gravity and/or electrodynamics
for the CM descriptions of gravitationally and/or electrostatically
interacting macroscopic particles (where the macroscopic particles
are composed of many interacting \emph{zbw} particles); and (iii)
recover classical Vlasov-Poisson mean-field theory for macroscopic
particles that interact gravitationally and/or electrostatically,
in the weak-coupling large particle number limit. We will also be
led to suggest an explicit model of environmental decoherence that's
consistent with the Oriols et al. scheme, and which could justify
a crucial assumption of the scheme.

The outline of the paper is as follows. Section 2 implements the Oriols
et al. scheme into ZSM-Newton/Coulomb, and shows how classical Newtonian
dynamics for the center-of-mass of a many-particle system is recovered
in the large \emph{N}  limit. Section 3 shows how to derive the classical
nonlinear Schr\"{o}dinger equation
for the large \emph{N}  center-of-mass motion. Section 4 shows how to recover
classical Newtonian gravity/electrodynamics for many gravitationally/electrostatically
interacting macroscopic particles. Section 5 shows how to recover
classical Vlasov-Poisson mean-field theory. Section 6 sketches an
explicit model of environmental decoherence that's consistent with
the Oriols et al. scheme applied to ZSM-Newton/Coulomb.

\section{Large \emph{N}  center-of-mass approximation in ZSM-Newton/Coulomb}

\subsection{General approach}

We begin by considering ZSM for N identical \emph{zbw} particles in
(for simplicity) 1-dimensional space, with configuration $X(t)=\left\{ x_{1}(t),...,x_{N}(t)\right\} $
and the ensemble-averaged, time-symmetric, joint \emph{zbw} phase
\begin{equation}
\begin{aligned}J(X) & \coloneqq\mathrm{E}\left[\int_{t_{I}}^{t_{F}}\left\{ \sum_{i=1}^{N}\frac{1}{2}\left[2mc^{2}+\frac{1}{2}m\left(Dx_{i}(t)\right)^{2}+\frac{1}{2}m\left(D_{*}x_{i}(t)\right)^{2}\right]-U(X(t),t)\right\} dt+\sum_{i=1}^{N}\phi_{i}\right]\\
 & =\mathrm{E}\left[\int_{t_{I}}^{t_{F}}\left\{ \sum_{i=1}^{N}\left[mc^{2}+\frac{1}{2}mv_{i}^{2}(X(t),t)+\frac{1}{2}mu_{i}^{2}(X(t),t)\right]-U(X(t),t)\right\} dt+\sum_{i=1}^{N}\phi_{i}\right],
\end{aligned}
\end{equation}
where
\begin{equation}
Dx_{i}(t)=\left[\frac{\partial}{\partial t}+\sum_{i=1}^{N}\mathbf{b}_{i}(X(t),t)\frac{\partial}{\partial x_{i}}+\sum_{i=1}^{N}\frac{\hbar}{2m}\frac{\partial^{2}}{\partial x_{i}^{2}}\right]x_{i}(t),
\end{equation}

\begin{equation}
D_{*}x_{i}(t)=\left[\frac{\partial}{\partial t}+\sum_{i=1}^{N}\mathbf{b}_{i*}(X(t),t)\frac{\partial}{\partial x_{i}}-\sum_{i=1}^{N}\frac{\hbar}{2m}\frac{\partial^{2}}{\partial x_{i}^{2}}\right]x_{i}(t).
\end{equation}
The potential $U(X(t),t)$ is assumed to take the general form 
\begin{equation}
U(X(t),t)\coloneqq\sum_{j=1}^{N}U_{ext}(x_{j}(t))+\frac{1}{2}\sum_{j,k=1}^{N(j\neq k)}U_{int}(x_{j}(t)-x_{k}(t)),
\end{equation}
and we assume the usual constraints 
\begin{equation}
\mathbf{v}_{i}\coloneqq\frac{1}{2}\left[\mathbf{b}_{i}+\mathbf{b}_{i*}\right]=\frac{1}{m}\frac{\partial S}{\partial x_{i}},
\end{equation}

\begin{equation}
\mathbf{u}_{i}\coloneqq\frac{1}{2}\left[\mathbf{b}_{i}-\mathbf{b}_{i*}\right]=\frac{\hbar}{2m}\frac{1}{\rho}\frac{\partial\rho}{\partial x_{i}}.
\end{equation}
As a consequence of (5.5-6), the time-reversal invariant joint probability
density $\rho(X,t)$ evolves by
\begin{equation}
\frac{\partial\rho}{\partial t}=-\sum_{i=1}^{N}\frac{\partial}{\partial x_{i}}\left(\frac{\rho}{m}\frac{\partial S}{\partial x_{i}}\right),
\end{equation}
and satisfies the normalization
\begin{equation}
\int_{\mathbb{R}^{N}}\rho_{0}(X)d^{N}X=1.
\end{equation}
The stochastic differential equations of motion for $x_{i}(t)$ take
the form
\begin{equation}
dx_{i}(t)=\left[\frac{1}{m}\frac{\partial S}{\partial x_{i}}+\frac{\hbar}{2m}\frac{1}{\rho}\frac{\partial\rho}{\partial x_{i}}\right]|_{x_{j}=x_{j}(t)}dt+dW_{i}(t),
\end{equation}

\begin{equation}
dx_{i}(t)=\left[\frac{1}{m}\frac{\partial S}{\partial x_{i}}-\frac{\hbar}{2m}\frac{1}{\rho}\frac{\partial\rho}{\partial x_{i}}\right]|_{x_{j}=x_{j}(t)}dt+dW_{i*}(t),
\end{equation}
where the $dW_{i}$ are 1-dimensional Wiener processes satisfying
Gaussianity, independence of $dx_{i}(s)$ for $s\leq t$, and variance
\begin{equation}
\mathrm{E}_{t}\left[dW_{i}^{2}\right]=\frac{\hbar}{m}dt.
\end{equation}
Analogous conditions apply to the backward Wiener processes $dW_{i*}$. 

Note that, since we are considering the case of particle motion in
a 1-dimensional space, we can disregard the quantization condition
for (5.5) (we will come back to it later, though, when we consider
the case of particle motion in a 3-dimensional Euclidean space).

Now, following Oriols et al. \cite{Oriols2016}, we would like to
redefine (5.1) in terms of the CM position $x_{cm}(t)$ and relative
positions $\mathbf{y}(t)=\left\{ y_{2}(t),...,y_{N}(t)\right\} $
such that no cross terms arise from the Laplacians in $D$ and $D_{*}$.
As shown by Oriols et al. \cite{Oriols2016}, the coordinate transformation
\begin{equation}
x_{cm}\coloneqq\frac{1}{N}\sum_{i=1}^{N}x_{i},
\end{equation}
 
\begin{equation}
y_{j}\coloneqq x_{j}-\frac{\left(\sqrt{N}x_{cm}+x_{1}\right)}{\sqrt{N}+1},
\end{equation}
makes it possible to rewrite the \emph{N}-particle Schr\"{o}dinger
equation, with potential (5.4), in terms of $x_{cm}$ and $\mathbf{y}=\left\{ y_{2},...,y_{N}\right\} $
without cross terms arising from the Laplacian in the Schr\"{o}dinger
Hamiltonian. Thus, applying (5.12-13) to (5.1), we obtain \footnote{The proof of this goes along the same lines as Appendix A.1 of Oriols
et al. \cite{Oriols2016}.}
\begin{equation}
\begin{aligned}J(x_{cm},\mathbf{y}) & \coloneqq\mathrm{E}\left[\int_{t_{I}}^{t_{F}}\Bigl\{ Mc^{2}+\frac{m}{4}\left[\left(\tilde{D}x_{cm}(t)\right)^{2}+\left(\tilde{D}_{cm*}x_{cm}(t)\right)^{2}\right]\right.\\
 & \left.+\frac{m}{4}\sum_{j=2}^{N}\left[\left(\tilde{D}y_{j}(t)\right)^{2}+\left(\tilde{D}_{*}y_{j}(t)\right)^{2}\right]-U\Bigr\} dt+\phi_{cm}+\phi_{rel}\right]\\
 & =\mathrm{E}\left[\int_{t_{I}}^{t_{F}}\left(Mc^{2}+\frac{1}{2}M\left(v_{cm}^{2}+u_{cm}^{2}\right)+\frac{1}{2}m\sum_{j=2}^{N}\left(v_{j}^{2}+u_{j}^{2}\right)-U\right)dt+\phi_{cm}+\phi_{rel}\right],
\end{aligned}
\end{equation}
where $M=Nm$, the CM velocities are given by
\begin{equation}
\mathbf{v}_{cm}\coloneqq\frac{1}{2}\left[\mathbf{b}_{cm}+\mathbf{b}_{cm*}\right]=\frac{1}{m}\frac{\partial S(x_{cm},\mathbf{y}(t),t)}{\partial x_{cm}}|_{x_{cm}=x_{cm}(t)},
\end{equation}

\begin{equation}
\mathbf{u}_{cm}\coloneqq\frac{1}{2}\left[\mathbf{b}_{cm}-\mathbf{b}_{cm*}\right]=\frac{\hbar}{2m}\frac{1}{\rho(x_{cm},\mathbf{y}(t),t)}\frac{\partial\rho(x_{cm},\mathbf{y}(t),t)}{\partial x_{cm}}|_{x_{cm}=x_{cm}(t)},
\end{equation}
the relative velocities are given by
\begin{equation}
\mathbf{v}_{j}\coloneqq\frac{1}{2}\left[\mathbf{b}_{j}+\mathbf{b}_{j*}\right]=\frac{1}{m}\frac{\partial S(x_{cm}(t),\mathbf{y},t)}{\partial y_{j}}|_{\mathbf{y}=\mathbf{y}(t)},
\end{equation}

\begin{equation}
\mathbf{u}_{j}\coloneqq\frac{1}{2}\left[\mathbf{b}_{j}-\mathbf{b}_{j*}\right]=\frac{\hbar}{2m}\frac{1}{\rho(x_{cm}(t),\mathbf{y},t)}\frac{\partial\rho(x_{cm}(t),\mathbf{y},t)}{\partial y_{j}}|_{\mathbf{y}=\mathbf{y}(t)},
\end{equation}
and the transformed mean forward/backward derivatives take the form
\begin{equation}
\tilde{D}x_{cm}(t)=\left[\frac{\partial}{\partial t}+\mathbf{b}_{cm}\frac{\partial}{\partial x_{cm}}+\sum_{j=2}^{N}\mathbf{b}_{j}\frac{\partial}{\partial y_{j}}+\frac{\hbar}{2M}\frac{\partial^{2}}{\partial x_{cm}^{2}}+\sum_{j=2}^{N}\frac{\hbar}{2m}\frac{\partial^{2}}{\partial y_{j}^{2}}\right]x_{cm}(t)=\mathbf{b}_{cm},
\end{equation}

\begin{equation}
\tilde{D}_{*}x_{cm}(t)=\left[\frac{\partial}{\partial t}+\mathbf{b}_{cm*}\frac{\partial}{\partial x_{cm}}+\sum_{j=2}^{N}\mathbf{b}_{j*}\frac{\partial}{\partial y_{j}}-\frac{\hbar}{2M}\frac{\partial^{2}}{\partial x_{cm}^{2}}-\sum_{j=2}^{N}\frac{\hbar}{2m}\frac{\partial^{2}}{\partial y_{j}^{2}}\right]x_{cm}(t)=\mathbf{b}_{cm*},
\end{equation}
and
\begin{equation}
\tilde{D}y_{j}(t)=\mathbf{b}_{j},
\end{equation}

\begin{equation}
\tilde{D}_{*}y_{j}(t)=\mathbf{b}_{j*}.
\end{equation}
Accordingly, the continuity equation (5.7) becomes 
\begin{equation}
\frac{\partial\rho}{\partial t}=-\frac{\partial}{\partial x_{cm}}\left(\rho v_{cm}\right)-\sum_{j=2}^{N}\frac{\partial}{\partial y_{j}}\left[\rho v_{j}\right],
\end{equation}
and the forward stochastic differential equations of motion for $x_{cm}(t)$
and $y_{j}(t)$, respectively, take the form
\begin{equation}
dx_{cm}(t)=\left[\frac{1}{M}\frac{\partial S(x_{cm},\mathbf{y}(t),t)}{\partial x_{cm}}+\frac{\hbar}{2M}\frac{1}{\rho(x_{cm},\mathbf{y}(t),t)}\frac{\partial\rho(x_{cm},\mathbf{y}(t),t)}{\partial x_{cm}}\right]|_{x_{cm}=x_{cm}(t)}dt+dW_{cm}(t),
\end{equation}
and
\begin{equation}
dy_{j}(t)=\left[\frac{1}{m}\frac{\partial S(x_{cm}(t),\mathbf{y},t)}{\partial y_{j}}+\frac{\hbar}{2m}\frac{1}{\rho(x_{cm}(t),\mathbf{y},t)}\frac{\partial\rho(x_{cm}(t),\mathbf{y},t)}{\partial y_{j}}\right]|_{\mathbf{y}=\mathbf{y}(t)}dt+dW_{j}(t).
\end{equation}
The $dW_{cm}$ and $dW_{j}$ are 1-dimensional Wiener processes satisfying
Gaussianity, independence of $dx_{cm}(s)$ and $dy_{j}(s)$ for $s\leq t$,
and variances 
\begin{equation}
\mathrm{E}_{t}\left[dW_{cm}^{2}\right]=\frac{\hbar}{M}dt,
\end{equation}
\begin{equation}
\mathrm{E}_{t}\left[dW_{j}^{2}\right]=\frac{\hbar}{m}dt,
\end{equation}
respectively. Analogous relations for the backward stochastic differential
equations can be written down as well.

We emphasize that (5.14) is equivalent to (5.1), the two being related
by the coordinate transformations (5.12-13). Thus, applying 
\begin{equation}
J(x_{cm},\mathbf{y})=extremal,
\end{equation}
we obtain 
\begin{equation}
\begin{aligned}\frac{M}{2}\left[\tilde{D}_{*}\tilde{D}+\tilde{D}\tilde{D}_{*}\right]x_{cm}(t)+\frac{m}{2}\sum_{j=2}^{N}\left[\tilde{D}_{*}\tilde{D}+\tilde{D}\tilde{D}_{*}\right]y_{j}(t)\\
=-\left[\frac{\partial}{\partial x_{cm}}U(X,t)|_{X=X(t)}+\sum_{j=2}^{N}\frac{\partial}{\partial y_{j}}U(X,t)|_{X=X(t)}\right].
\end{aligned}
\end{equation}
By D'Alembert's principle, the variations $\delta x_{cm}(t)$ and
$\delta\mathbf{y}(t)$ are independent of each other, and the $\delta y_{j}(t)$
are independent for all $j$. So (5.29) separates into the pair
\begin{equation}
\frac{M}{2}\left[\tilde{D}_{*}\tilde{D}+\tilde{D}\tilde{D}_{*}\right]x_{cm}(t)=-\frac{\partial}{\partial x_{cm}}U(X,t)|_{X=X(t)}=-\sum_{i=1}^{N}\frac{\partial}{\partial x_{i}}U_{ext}(x_{i})|_{x_{i}=x_{i}(t)},
\end{equation}
\begin{equation}
\frac{m}{2}\sum_{j=2}^{N}\left[\tilde{D}_{*}\tilde{D}+\tilde{D}\tilde{D}_{*}\right]y_{j}(t)=-\sum_{j=2}^{N}\frac{\partial}{\partial y_{j}}U(X,t)|_{X=X(t)}=-\frac{1}{2}\sum_{j=2}^{N}\sum_{j,k=1}^{N(j\neq k)}\frac{\partial}{\partial x_{j}}U_{int}(x_{j}-x_{k})|_{X=X(t)},
\end{equation}
and (5.31) separates into 
\begin{equation}
\frac{m}{2}\left[\tilde{D}_{*}\tilde{D}+\tilde{D}\tilde{D}_{*}\right]y_{j}(t)=-\frac{\partial}{\partial y_{j}}U(X,t)|_{X=X(t)},
\end{equation}
for all $j$ from $2,...,N$. The last equality on the right hand
side (rhs) of (5.30) follows from the fact that the symmetry of $U_{int}$
implies no net force on the CM, and the observation that $\partial x_{i}/\partial x_{cm}=1$
which follows from inverting (5.12); the last equality on the rhs
of (5.31-32) follows from the fact that the forces on the relative
degrees of freedom come only from $U_{int}$. 

Computing the derivatives on the left sides of (5.30-31), and removing
the evaluation at $X=X(t)$ on both sides, we obtain
\begin{equation}
\begin{aligned}M\left[\partial_{t}\mathbf{v}_{cm}+\mathbf{v}_{cm}\frac{\partial}{\partial x_{cm}}\mathbf{v}_{cm}-\mathbf{u}_{cm}\frac{\partial}{\partial x_{cm}}\mathbf{u}_{cm}-\frac{\hbar}{2M}\frac{\partial^{2}}{\partial x_{cm}^{2}}\mathbf{u}_{cm}\right.\\
\left.+\sum_{j=2}^{N}\left(\mathbf{v}_{j}\frac{\partial}{\partial y_{j}}\mathbf{v}_{cm}-\mathbf{u}_{j}\frac{\partial}{\partial y_{j}}\mathbf{u}_{cm}-\frac{\hbar}{2m}\frac{\partial^{2}}{\partial y_{j}^{2}}\mathbf{u}_{cm}\right)\right]=-\frac{\partial}{\partial x_{cm}}U,
\end{aligned}
\end{equation}

\begin{equation}
\begin{aligned}m\sum_{j=2}^{N}\left[\partial_{t}\mathbf{v}_{j}+\sum_{j=2}^{N}\left(\mathbf{v}_{j}\frac{\partial}{\partial y_{j}}\mathbf{v}_{j}-\mathbf{u}_{j}\frac{\partial}{\partial y_{j}}\mathbf{u}_{j}\right)-\frac{\hbar}{2M}\frac{\partial^{2}}{\partial x_{cm}^{2}}\mathbf{u}_{j}-\frac{\hbar}{2m}\sum_{j=2}^{N}\frac{\partial^{2}}{\partial y_{j}^{2}}\mathbf{u}_{j}\right.\\
\left.-\frac{\hbar}{2m}\sum_{j=2}^{N}\frac{\partial^{2}}{\partial y_{j}^{2}}\mathbf{u}_{j}+\mathbf{v}_{cm}\frac{\partial}{\partial x_{cm}}\mathbf{v}_{j}-\mathbf{u}_{cm}\frac{\partial}{\partial x_{cm}}\mathbf{u}_{j}\right]=-\sum_{j=2}^{N}\frac{\partial}{\partial y_{j}}U,
\end{aligned}
\end{equation}
where $\mathbf{v}_{cm}$ ($\mathbf{v}_{j}$) and $\mathbf{u}_{cm}$
($\mathbf{u}_{j}$) are now velocity fields over the possible positions
of the (CM and relative) particles. Thus, by integrating the positional
derivatives on both sides of (5.33) and (5.34), respectively, and
setting the arbitrary integration constants equal to zero (for simplicity),
each equation yields the quantum Hamilton-Jacobi equation in CM and
relative coordinates: 
\begin{equation}
-\partial_{t}S=U+\frac{1}{2M}\left(\frac{\partial S}{\partial x_{cm}}\right)^{2}-\frac{\hbar^{2}}{2M}\frac{1}{\sqrt{\rho}}\frac{\partial^{2}}{\partial x_{cm}^{2}}\sqrt{\rho}+\sum_{j=2}^{N}\left[\frac{1}{2m}\left(\frac{\partial S}{\partial y_{j}}\right)^{2}-\frac{\hbar^{2}}{2m}\frac{1}{\sqrt{\rho}}\frac{\partial^{2}}{\partial y_{j}^{2}}\sqrt{\rho}\right].
\end{equation}
Combining with (5.35) with (5.23), the Madelung transformation yields
the coordinate-transformed Schr\"{o}dinger
equation 
\begin{equation}
i\hbar\frac{\partial\psi}{\partial t}=\left[-\frac{\hbar^{2}}{2M}\frac{\partial^{2}}{\partial x_{cm}^{2}}-\frac{\hbar^{2}}{2m}\sum_{j=2}^{N}\frac{\partial^{2}}{\partial y_{j}^{2}}+U\right]\psi,
\end{equation}
where $\psi(x_{cm},\mathbf{y},t)=\sqrt{\rho(x_{cm},\mathbf{y},t)}e^{iS(x_{cm},\mathbf{y},t)/\hbar}$
is single-valued and smooth (because we're restricted to the configuration
space of dimension $\mathbb{R}^{N}$). As shown by Oriols et al. \cite{Oriols2016},
(5.36) corresponds to the \emph{N}-particle Schr\"{o}dinger
equation 
\begin{equation}
i\hbar\frac{\partial\psi}{\partial t}=\left[-\frac{\hbar^{2}}{2m}\sum_{i=1}^{N}\frac{\partial^{2}}{\partial x_{i}^{2}}+U\right]\psi,
\end{equation}
where $\psi(X,t)=\sqrt{\rho(X,t)}e^{iS(X,t)/\hbar}$, under the coordinate
transformations (5.12-13). 

From the solution of (5.36), we can rewrite (5.24-25) as
\begin{equation}
dx_{cm}(t)=\left[\frac{\hbar}{M}\mathrm{Im}\frac{\partial}{\partial x_{cm}}\ln\psi(x_{cm},\mathbf{y}(t),t)+\frac{\hbar}{M}\mathrm{Re}\frac{\partial}{\partial x_{cm}}\ln\psi(x_{cm},\mathbf{y}(t),t)\right]|_{x_{cm}=x_{cm}(t)}dt+dW_{cm}(t),
\end{equation}

\begin{equation}
dy_{j}(t)=\left[\frac{\hbar}{m}\mathrm{Im}\frac{\partial}{\partial y_{j}}\ln\psi(x_{cm}(t),\mathbf{y},t)+\frac{\hbar}{m}\mathrm{Re}\frac{\partial}{\partial y_{j}}\ln\psi(x_{cm}(t),\mathbf{y},t)\right]|_{\mathbf{y}=\mathbf{y}(t)}dt+dW_{j}(t).
\end{equation}
Given an initial wavefunction $\psi(x_{cm},\mathbf{y},0)$ and initial
trajectories $\left\{ x_{cm}^{h}(0),\mathbf{y}^{h}(0)\right\} $,
where the $h$ index labels a particular set of possible initial trajectories,
the stochastic evolution of the CM and relative coordinates can be
determined completely. 

Let's now consider the 2nd-order time-evolution of the mean trajectories
of the CM. Defining 
\begin{equation}
Q_{cm}\coloneqq-\frac{\hbar^{2}}{2M}\frac{1}{\sqrt{\rho}}\frac{\partial^{2}}{\partial x_{cm}^{2}}\sqrt{\rho},
\end{equation}
\begin{equation}
\sum_{j=2}^{N}Q_{j}\coloneqq-\frac{\hbar^{2}}{2m}\sum_{j=2}^{N}\frac{1}{\sqrt{\rho}}\frac{\partial^{2}}{\partial y_{j}^{2}}\sqrt{\rho},
\end{equation}
we can rewrite (5.33) as 
\begin{equation}
\begin{aligned}M\frac{d^{2}\bar{x}_{cm}(t)}{dt^{2}} & =M\left[\partial_{t}\mathbf{v}_{cm}+\mathbf{v}_{cm}\frac{\partial}{\partial x_{cm}}\mathbf{v}_{cm}+\sum_{j=2}^{N}\mathbf{v}_{j}\frac{\partial}{\partial y_{j}}\mathbf{v}_{cm}\right]|_{x_{cm}=\bar{x}_{cm}(t)}^{\mathbf{y}=\mathbf{\bar{y}}(t)}\\
 & =-\frac{\partial}{\partial x_{cm}}\left[U+Q_{cm}+\sum_{j=2}^{N}Q_{j}\right]|_{x_{cm}=\bar{x}_{cm}(t)}^{\mathbf{y}=\mathbf{\bar{y}}(t)},
\end{aligned}
\end{equation}
and the $j$-th component of (5.34) as
\begin{equation}
\begin{aligned}m\frac{d^{2}\bar{y}_{j}(t)}{dt^{2}} & =m\left[\partial_{t}\mathbf{v}_{j}+\sum_{j=2}^{N}\mathbf{v}_{j}\frac{\partial}{\partial y_{j}}\mathbf{v}_{j}+\mathbf{v}_{cm}\frac{\partial}{\partial x_{cm}}\mathbf{v}_{j}\right]|_{x_{cm}=\bar{x}_{cm}(t)}^{\mathbf{y}=\mathbf{\bar{y}}(t)}\\
 & =-\frac{\partial}{\partial y_{j}}\left[U+Q_{cm}+\sum_{j=2}^{N}Q_{j}\right]|_{x_{cm}=\bar{x}_{cm}(t)}^{\mathbf{y}=\mathbf{\bar{y}}(t)},
\end{aligned}
\end{equation}
where the bars denote that the solutions of (5.42-43) are mean trajectories,
as opposed to the stochastic trajectories obtained from solutions
of (5.38-39). We note that equation (5.42) corresponds to Equation
(14) of Oriols et al. \cite{Oriols2016}. 

Now, consider M experimental preparations \footnote{`Experimental' could refer to an actual laboratory experiment or a
natural physical process outside laboratories.} of a system of N identical particles, described by (5.36), each with
the same initial wavefunction $\psi(x_{cm},\mathbf{y},0)$. For each
preparation, there will be a different set of ``$h$-trajectories''
\cite{Oriols2016}, and because of the identicality of the particles,
they will all have the same marginal probability distribution:
\begin{equation}
\overline{\rho}(y_{k},0)\coloneqq\frac{1}{M}\sum_{h=1}^{M}\delta(y-\bar{y}_{k}^{h}(0)),
\end{equation}
where M is a very large number of preparations. When $N\rightarrow\infty,$
the distribution of initial particle positions in a single $h$-preparation,
\begin{equation}
P(y_{k},0)\coloneqq\frac{1}{N}\sum_{h=1}^{N}\delta(y-\bar{y}_{k}^{h}(0)),
\end{equation}
fills the entire support of (5.44), thereby giving 
\begin{equation}
\overline{\rho}(y_{k},0)\approx P(y_{k},0)
\end{equation}
for the vast majority of the M preparations, where `vast majority'
refers to the possible sets of N initial mean trajectories $\bar{X}^{h}(t)=\left\{ \bar{x}_{1}^{h}(t),...,\bar{x}_{N}^{h}(t)\right\} $
selected according to the initial probability density $\rho(X,0)=|\psi(X,0)|^{2}$.
The fact that the possible set of initial trajectories is selected
randomly according to $|\psi(X,0)|^{2}$ ensures that possible sets
of initial trajectories which don't satisfy (5.46) will be extremely
rare; and because the $|\psi(X,0)|^{2}$ distribution is preserved
in time by the equivariant evolution given by (5.23), such possible
sets of initial trajectories not satisfying (5.46) will be extremely
rare for all times. Thus, Oriols et al. \cite{Oriols2016} refer to
wavefunctions with probability densities satisfying (5.46) as ``wavefunctions
full of particles'' (WFPs). 

As noted by Oriols et al. \cite{Oriols2016}, however, there are \emph{N}-particle
wavefunctions which don't satisfy (5.46). For example, a factorizable
wavefunction $\psi(X,0)=\prod_{i=1}^{N}\phi(x_{i},0)$ in general
won't be a WFP because it won't have the necessary bosonic or fermionic
symmetry requirements to justify the independence of the marginal
probability distributions for each $x_{i}$ (the only exception being
a bosonic wavefunction where all the $\phi_{i}$ are equal, such as
in the mean-field approximation). For another example, wavefunctions
with strong quantum correlations between the particles (see equation
D3 of Oriols et al. \cite{Oriols2016} for an example involving an
unphysical macroscopic superposition state) won't have a single $h$-preparation
which, in the limit $N\rightarrow\infty$, fills the entire support
of (5.44); however, Oriols et al. \cite{Oriols2016} argue that ``most
of the wave functions associated to macroscopic objects fulfill the
requirements of a wave function full of particles, i.e. they do not
include strong quantum correlations between particles'' (page 12).
While they don't explain why they argue that most wavefunctions associated
to macroscopic objects don't include strong quantum correlations between
particles, their expectation can be justified from the following observation:
in dBB and stochastic mechanics, macroscopic superposition states
(in the real world) arise as a result of decoherence from system-environment
interactions \cite{Goldstein1987,Jibu1990,Blanchard1992,Peruzzi1996,Duerr2009,Goldstein2013},
and such decoherence is always accompanied by ``effective collapse''
\cite{Goldstein1987,Jibu1990,Blanchard1992,Peruzzi1996,Duerr2009,Goldstein2013}.
Effective collapse being the process whereby a dBB/Nelsonian/ZSM particle
(or collection of such particles) composing the system dynamically
evolves into one of the effective system wavefunction components of
a system-environment entangled state, the latter formed during an
environmental decoherence process. Thus effective collapse ensures
that, for all practical purposes, only \emph{one} of the components
of a system-environment entangled state (i.e., a macroscopic quantum
superposition state) will be dynamically relevant to the future motion
of a single-particle or multi-particle system coupled to a macroscopic
environment. In other words, an environmentally decohered macroscopic
object (composed of dBB/Nelsonian/ZSM particles), which are virtually
all of the macroscopic objects in the real world (according to dBB
and stochastic mechanics), can always be expected to have a many-particle
effective wavefunction associated to it corresponding to a WFP. In
section 6, we will say more about how environmental decoherence and
effective collapse of large \emph{N}  systems might be modeled within the
Oriols et al. scheme. In the mean time, we will continue with assuming
pure states that satisfy (5.46) and thus correspond to WFPs.

Focusing now on the CM motion given by (5.42) and (5.38), we shall
specify the conditions under which its classical limit is obtained.
For convenience, we rewrite (5.42) as 
\begin{equation}
M\frac{d^{2}\bar{x}_{cm}(t)}{dt^{2}}=F_{U}+F_{cm}+F_{rel}.
\end{equation}
Classicality conditions will be obtained from comparing the N-dependences
of the three forces on the rhs of (5.47). 

First we recall that, because of the symmetry of $U_{int}$, its net
force on the CM is zero, leaving the only non-zero net force coming
from $U_{ext}$: 
\begin{equation}
F_{U}\coloneqq-\frac{\partial}{\partial x_{cm}}\sum_{i=1}^{N}U_{ext}(x_{i})=-\sum_{i=1}^{N}\frac{\partial U_{ext}(x_{i})}{\partial x_{i}}.
\end{equation}
Furthermore, if spatial variations of $U_{ext}$ are much larger than
the size of the \emph{N}-particle system under consideration \footnote{More precisely, if, for an \emph{N}-particle system with number density of
width $d$, in the presence of an external potential $U$ with scale
of spatial variation given by $L(U)=\sqrt{|\frac{U'}{U'''}|}$, we
have that $d\ll L(U)$. This statement is closely related to the classicality
condition used in the Ehrenfest theorem and in the quantum-classical
limit scheme of Allori et al. \cite{Allori2001,Allori2002,Duerr2009},
i.e., that the de Broglie wavelength $\lambda$ of a single-particle
wavepacket of width $\sigma$ (where $\sigma\geq\lambda$) satisfies
$\lambda\ll L(U)$.} (which will typically be the case for classical external potentials
on macroscopic lengthscales), then (5.48) can be approximated as (using
$\partial x_{i}/\partial x_{cm}=1$) 
\begin{equation}
F_{U}=F_{ext}\approx-N\frac{\partial U_{ext}(x_{cm})}{\partial x_{cm}},
\end{equation}
which is exact for linear and quadratic potentials as pointed out
by Oriols et al. \cite{Oriols2016}. Thus we have that $F_{U}\propto N$. 

Second, Oriols et al. \cite{Oriols2016} note that the conditional
probability distribution for the CM position can be found by considering
the probability distribution of $\bar{x}_{cm}^{h}(t)$ for a large
number of different $h$-trajectories given by $\bar{X}^{h}(t)=\left\{ \bar{x}_{1}^{h}(t),...,\bar{x}_{N}^{h}(t)\right\} $.
For a WFP in the limit $N\rightarrow\infty$, the second and third
moments of the distribution are zero (see Theorems 9-10 of Appendix
D of Oriols et al. \cite{Oriols2016}). Hence, for very large but
finite N, one can expect a normal distribution for the CM position:
\begin{equation}
\rho(\bar{x}_{cm}^{h}(t))\approx\frac{1}{\sqrt{2\pi}\sigma_{cm}}exp\left(-\frac{\left[\overline{x}-\bar{x}_{cm}^{h}(t)\right]^{2}}{2\sigma_{cm}^{2}}\right),
\end{equation}
where $\sigma_{cm}$ is estimated (see Theorem 10 in Appendix D of
Oriols et al. \cite{Oriols2016}) to be given by 
\begin{equation}
\sigma_{cm}^{2}\leq\frac{\sigma^{2}}{N},
\end{equation}
where $\sigma^{2}$ is the variance of the marginal distributions,
and where $\bar{x}$ is the mean position of the CM. The relation
between these last two variables can be seen as follows. First, we
have (from Corollary 1 and Theorem 4 in Appendix B of Oriols et al.
\cite{Oriols2016}) that 
\begin{equation}
\overline{x}\equiv\overline{x}_{i}=\int_{-\infty}^{\infty}dx\,x\,\overline{\rho}(x),
\end{equation}
where $\bar{\rho}(x)$ is the marginal probability density of the
\emph{i}-th particle \footnote{This is defined as $\overline{\rho}(x)\equiv\overline{\rho}_{i}(x_{i})\coloneqq\int_{-\infty}^{\infty}dx{}_{1}...\int_{-\infty}^{\infty}dx{}_{i-1}\int_{-\infty}^{\infty}dx_{i+1}...\int_{-\infty}^{\infty}dx_{N}\rho(X)$.
Furthermore, for identical particles, the marginal probability density
satisfies $\overline{\rho}_{i}(x_{i})=\overline{\rho}_{j}(x_{j})$
for $i\neq j$ (see the proof of Theorem 3 in Appendix B of Oriols
et al. \cite{Oriols2016}).} from which it follows (from Corollary 2 in Appendix B of Oriols et
al. \cite{Oriols2016}) that
\begin{equation}
\sigma^{2}\equiv\sigma_{i}^{2}=\int_{-\infty}^{\infty}dx\,\left(x-\overline{x}\right)\,\overline{\rho}(x).
\end{equation}
Now, calculating $Q_{cm}$ and $F_{cm}$ in terms of $\rho(\bar{x}_{cm}^{h}(t))$,
one obtains
\begin{equation}
Q_{cm}\approx\frac{\hbar}{2M\sigma_{cm}^{2}}\left(1-\frac{\left[\overline{x}-\bar{x}_{cm}(t)\right]^{2}}{\sigma_{cm}^{2}}\right),
\end{equation}
\begin{equation}
F_{cm}\approx-\frac{\partial Q_{cm}}{\partial x_{cm}}|_{x_{cm}=\bar{x}_{cm}(t)}^{\mathbf{y}=\mathbf{\bar{y}}(t)}\propto\frac{\hbar}{m\sigma^{3}}\sqrt{N},
\end{equation}
where it is used that $\overline{x}-\bar{x}_{cm}(t)\approx\sigma/\sqrt{N}$.
Thus $F_{cm}\propto\sqrt{N}$. 

Third, since we are dealing with identical particles, we have that
$\rho(x_{cm},y_{2},...,y_{j},...)=\rho(x_{cm},y_{j},...,y_{2},...)$,
and so the force $F_{rel}$ can be rewritten as
\begin{equation}
F_{rel}\coloneqq\frac{\hbar^{2}}{2m}\sum_{j=2}^{N}\left[\frac{\partial}{\partial x_{cm}}\left(\frac{1}{\sqrt{\rho}}\frac{\partial^{2}\sqrt{\rho}}{\partial y_{2}^{2}}\right)\right]|_{x_{cm}=\bar{x}_{cm}(t)}^{\mathbf{y}=\mathbf{\bar{y}}(t)}.
\end{equation}
As emphasized by Oriols et al., the exchange symmetry in $\rho$ means
that a single preparation with $N\rightarrow\infty$ is equivalent
to $h=\left\{ 1,...,N\right\} $ different preparations with $\bar{y}_{2}^{h}(t)$
approximately filling the entire support of $\rho$ in the $y_{2}$
3-space. Accordingly, the quantum equilibrium distribution for the
particles implies that the sum in (5.56) can be approximated by an
integral that's weighted by $\rho$: 
\begin{equation}
F_{rel}\approx N\frac{\hbar^{2}}{2m}\int_{y_{2}}\left[\rho\frac{\partial}{\partial x_{cm}}\left(\frac{1}{\sqrt{\rho}}\frac{\partial^{2}\sqrt{\rho}}{\partial y_{2}^{2}}\right)\right]|_{\bar{x}_{cm}(t)}^{\bar{y}_{3}(t),...,\bar{y}_{N}(t)}dy_{2}\rightarrow0.
\end{equation}
That (5.57) vanishes is due to a symmetric distribution of positive
and negative summands (for an explicit proof, see Appendix E of Oriols
et al. \cite{Oriols2016}). Thus $F_{rel}\approx0$ as $N\rightarrow\infty$.

To summarize, then, in the limit that $N\rightarrow\infty$ for identical
particles, we have 
\begin{equation}
\begin{array}{c}
F_{U}\propto N,\\
\\
F_{cm}\propto\sqrt{N},\\
\\
F_{rel}\rightarrow0.
\end{array}
\end{equation}
So it is clear that the classical external force $F_{U}$ grows much
faster (under the stated conditions) than the two quantum forces,
as the number of identical particles interacting through $U_{int}$
becomes very large. This conclusion does not hold, of course, for
the relative degrees of freedom, nor would we expect otherwise. In
fact, we should expect that the positions of the individual elementary
particles (or atoms) composing a macroscopic object will continue
to have non-classical (quantum) dynamics, even when the dynamics of
the CM position is approximatey classical. A useful and interesting
consequence of (5.58) is that quantum uncertainty becomes negligible:
between any two preparations of an \emph{N}-particle system, $\bar{X}^{h}(t)$
and $\bar{X}^{l}(t)$, the CM trajectories and velocities will be
very similar, i.e. $\bar{x}_{cm}^{h}(t)\approx\bar{x}_{cm}^{l}(t)$
and $\mathbf{v}_{cm}^{h}(t)\approx\mathbf{v}_{cm}^{l}(t)$. Thus it
is coherent to speak of fixing the initial position and velocity of
the CM position in the present context, as is done in classical mechanics. 

How large does N have to be for $F_{cm}$ and $F_{rel}$ to become
negligible relative to $F_{U}$? This was addressed by Oriols et al.
in numerical simulations \cite{Oriols2016}. 

In one simulation (Appendix F of Oriols et al. \cite{Oriols2016}),
an initial \emph{N}-particle wavefunction for identical particles was constructed
from pairs of Gaussian wave packets, with random dispersion and opposite
random momenta and central positions (in other words, the packets
move towards each other and eventually interfere), under the action
of an external linear potential. The linear potential spans a lengthscale
of $\sim10^{-7}m$, while the packet widths are only $\sim10^{-10}m$,
thereby satisfying the condition that the classical external potential
varies over lengthscales much greater than the size of the \emph{N}-particle
system. Half of the initial positions of the \emph{N} particles were selected
randomly according to the probability density of the left packet,
the other half according to the probability density of the right packet,
and then the evolution of the CM was computed under the influence
of the three forces in (5.47). As a comparison, the classical CM was
computed from Newton's law with the linear potential (i.e., $F_{U}$
alone), with the same initial CM position and velocity. The resulting
trajectories were compared for N = 1 through N = 20 (see Figure 1
of Oriols et al. \cite{Oriols2016}). For N = 1, the relative error
between the classical and quantum CM motions increases from zero to
45\% in 2 picoseconds; for N = 20, the relative error increases from
zero to less than 2\% in the same duration. In other words, for N
= 1, the classical and quantum CM motions significantly differ from
each other in a very short time, as expected, while for N = 20, the
two CM motions become effectively indistinguishable in a very short
time. Moreover, even for N distinguishable particles, under the same
conditions, Oriols et al. find that the relative error for N = 20
increases from zero to around 5\% in 2 picoseconds. It is remarkable
that, under the stated conditions, relatively few particles are needed
to reach the ``large \emph{N} '' regime. 

As a corollary to the above results, we note that, for the case of
a WFP, the CM osmotic velocity is given by
\begin{equation}
\mathbf{u}_{cm}=\frac{\hbar}{2M}\frac{1}{\rho}\frac{\partial\rho}{\partial x_{cm}}|_{x_{cm}=\bar{x}_{cm}(t)}\approx\frac{\hbar}{2m\sigma\sqrt{N}},
\end{equation}
while from (5.47) and (5.49) the CM current velocity is found to be
\begin{equation}
\mathbf{v}_{cm}=\frac{1}{M}\frac{\partial S}{\partial x_{cm}}|_{x_{cm}=\bar{x}_{cm}(t)}\approx-\frac{1}{Nm}\int_{t_{0}}^{t}\left[N\frac{\partial U_{ext}}{\partial x_{cm}}-F_{cm}\right]|_{x_{cm}=\bar{x}_{cm}(t)}dt'+\mathbf{v}_{cm0},
\end{equation}
where the contribution from $F_{rel}$ is neglected because, as we
saw from (5.57), it rapidly approaches zero in the large \emph{N}  limit.
Since $F_{cm}$ is the only N-dependent term in (5.60) and scales
like $\sqrt{N}$, we can see that in the large \emph{N}  limit, the dominant
contribution to the CM current velocity will come from $\partial U_{ext}/\partial x_{cm}$.
Accordingly, in the large \emph{N} limit, the CM current velocity will dominate
over the CM osmotic velocity (5.59). Thus, recalling the forward stochastic
differential equation
\begin{equation}
dx_{cm}(t)=\left[\frac{1}{M}\frac{\partial S}{\partial x_{cm}}+\frac{\hbar}{2M}\frac{1}{\rho}\frac{\partial\rho}{\partial x_{cm}}\right]|_{x_{cm}=x_{cm}(t)}dt+dW_{cm}(t),
\end{equation}
where $E_{t}\left[dW_{cm}^{2}\right]=\frac{\hbar}{M}dt$, we can see
that as $N\rightarrow\infty$, $E_{t}\left[dW_{cm}^{2}\right]\rightarrow0$
and (5.61) reduces to 
\begin{equation}
\frac{dx_{cm}(t)}{dt}\approx\frac{1}{M}\frac{\partial S_{cl}}{\partial x_{cm}}|_{x_{cm}=x_{cm}(t)}.
\end{equation}
The same follows, of course, for the backward stochastic differential
equation.

Extending the above approach to the case of 3-space is formally straightforward,
and entails the replacements $\partial/\partial x_{cm}\rightarrow\nabla_{cm}$,
$\partial/\partial y_{j}\rightarrow\nabla_{j}$, $x_{cm}\rightarrow\mathbf{R}_{cm}$,
$\mathbf{y}\rightarrow\mathbf{r}$, and inclusion of the quantization
relation for the phase field $S$:
\begin{equation}
\oint_{L}\nabla_{cm}S(\mathbf{R}_{cm},\mathbf{r},t)\cdot d\mathbf{R}_{cm}+\sum_{j=1}^{N-1}\oint_{L}\nabla_{j}S(\mathbf{R}_{cm},\mathbf{r},t)\cdot d\mathbf{r}_{j}=nh.
\end{equation}
This last ensures that the 3N-dimensional generalizations of (5.23)
and (5.35) are indeed equivalent to the 3N-dimensional generalization
of (5.36), and that $\psi(\mathbf{R}_{cm},\mathbf{r},t)$ is single-valued
with (generally) multi-valued phase.

\section{Classical nonlinear Schr\"{o}dinger
equation for large \emph{N} center-of-mass motion}

\subsection{Oriols et al.'s derivation}

What form does the time-dependent Schr\"{o}dinger
equation for $\psi(x_{cm},\mathbf{y},t)$ take in the large \emph{N} limit?
Before presenting our answer, let us first review and critique the
answer given by Oriols et al. \cite{Oriols2016}.

Introduce the conditional $S$ and $\rho$ functions for the CM by
the following definitions:
\begin{equation}
\begin{array}{ccc}
S_{cm}(x_{cm},t)\coloneqq S(x_{cm},\mathbf{\bar{y}}(t),t), &  & \rho_{cm}(x_{cm},t)\coloneqq\rho(x_{cm},\mathbf{\bar{y}}(t),t),\end{array}
\end{equation}
where $S_{cm}$ satisfies 
\begin{equation}
-\partial_{t}S_{cm}=\frac{1}{2M}\left(\frac{\partial S_{cm}}{\partial x_{cm}}\right)^{2}+U(x_{cm},\mathbf{\bar{y}}(t),t)+A,
\end{equation}
with 
\begin{equation}
A\coloneqq Q_{cm}+\sum_{j=2}^{N}\left[\frac{1}{2m}\left(\frac{\partial S_{cm}}{\partial y_{j}}\right)^{2}+Q_{j}-v_{j}^{h}(t)\frac{\partial S_{cm}}{\partial y_{j}}\right],
\end{equation}
and where $\rho_{cm}$ satisfies 
\begin{equation}
-\partial_{t}\rho_{cm}=\frac{\partial}{\partial x_{cm}}\left(\frac{1}{M}\frac{\partial S_{cm}}{\partial x_{cm}}\rho_{cm}\right)+B,
\end{equation}
with 
\begin{equation}
B\coloneqq-\sum_{j=2}^{N}\left[\frac{\partial\rho_{cm}}{\partial y_{j}}v_{j}^{h}(t)-\frac{\partial}{\partial y_{j}}\left(\frac{1}{m}\frac{\partial S_{cm}}{\partial y_{j}}\rho_{cm}\right)\right].
\end{equation}
Using the Madelung transformation, (5.65) and (5.67) can then be combined
into the `conditional Schr\"{o}dinger
equation' \cite{Derakhshani2016b}
\begin{equation}
\begin{aligned}i\hbar\partial_{t}\psi_{cm} & =-\frac{\hbar^{2}}{2M}\frac{\partial^{2}\psi_{cm}}{\partial x_{cm}^{2}}-\frac{\hbar^{2}}{2m}\sum_{j=2}^{N}\frac{\partial^{2}\Psi(x_{cm},\mathbf{y},t)}{\partial y_{j}^{2}}|_{\mathbf{y}=\mathbf{\bar{y}}^{h}(t)}\\
 & +i\hbar\sum_{j=2}^{N}v_{j}^{h}(t)\frac{\partial\Psi(x_{cm},\mathbf{y},t)}{\partial y_{j}}|_{\mathbf{y}=\mathbf{\bar{y}}^{h}(t)}+U(x_{cm},\mathbf{\bar{y}}(t),t)\psi,
\end{aligned}
\end{equation}
where $\psi_{cm}(x_{cm},t)=\sqrt{\rho_{cm}(x_{cm},t)}e^{iS_{cm}(x_{cm},t)/\hbar}$
is the `conditional wavefunction' in polar form.

Now, from the earlier observation that the large \emph{N} limit implies 
\begin{equation}
\frac{\partial V}{\partial x_{cm}}|_{x_{cm}^{h}=\bar{x}_{cm}^{h}(t)}\gg\frac{\partial\left(Q_{cm}+\sum_{j=2}^{N}Q_{j}\right)}{\partial x_{cm}}|_{x_{cm}=\bar{x}_{cm}(t)}^{\mathbf{y}=\mathbf{\bar{y}}(t)},
\end{equation}
and noting that 
\begin{equation}
0=\left[\frac{\partial}{\partial x_{cm}}\left(\frac{1}{2m}\left(\frac{\partial S_{cm}}{\partial y_{j}}\right)^{2}-v_{j}^{h}(t)\frac{\partial S_{cm}}{\partial y_{j}}\right)\right]|_{x_{cm}^{h}=\bar{x}_{cm}^{h}(t)}^{\mathbf{y}=\mathbf{\bar{y}}(t)},
\end{equation}
it follows that $A\approx0$ along the CM trajectory. So (5.65) effectively
corresponds to the classical Hamilton-Jacobi equation for the CM,
in the large \emph{N} limit. 

Oriols et al. assert that it is reasonable to assume $B=0$, since
this turns (5.67) into the standard continuity equation for the large
N CM Gaussian density (5.50). Thus (5.65) and (5.67) can be combined
via the Madelung transformation to get the classical nonlinear Schr\"{o}dinger
equation 
\begin{equation}
i\hbar\partial_{t}\psi_{cl}=\left(-\frac{\hbar^{2}}{2M}\frac{\partial^{2}}{\partial x_{cm}^{2}}+U(x_{cm},t)-Q_{cm}\right)\psi_{cl},
\end{equation}
where $\psi_{cl}(x_{cm},t)=\sqrt{\rho_{cl}(x_{cm},t)}e^{iS_{cl}(x_{cm},t)/\hbar}$
is the `classical wavefunction' for the CM. 

The problem with this derivation, in our view, is that no physical
justification is given for why it is reasonable to take $B=0$. The
fact that such an assumption turns (5.67) into the standard continuity
equation is of course true, but this doesn't constitute an explanation
for \emph{why} it should be true. So let us turn now to our explanation
for how (5.72) can be derived within the Oriols et al. framework,
without having to assume that $B=0$ in (5.67).

\subsection{Conditional Madelung equations}

To demonstrate effective decoupling of the CM and relative coordinates
in the large \emph{N} limit, it is convenient to focus first on the relative
coordinates.

Consider the conditional Madelung variables $\rho_{rel}(\mathbf{y},t)\coloneqq\rho(\bar{x}_{cm}(t),\mathbf{y},t)$
and $S_{rel}(\mathbf{y},t)\coloneqq S(\bar{x}_{cm}(t),\mathbf{y},t)$,
with evolution equations 
\begin{equation}
\begin{split}\partial_{t}\rho_{rel} & =-\sum_{j=2}^{N}\frac{\partial}{\partial y_{j}}\end{split}
\left[\rho_{rel}\frac{\partial S_{rel}}{\partial y_{j}}\frac{1}{m}\right]-\frac{\partial}{\partial x_{cm}}\left[\rho\frac{\partial S}{\partial x_{cm}}\frac{1}{M}\right]|_{\bar{x}_{cm}(t)}+\left(v_{cm}(t)\frac{\partial\rho}{\partial x_{cm}}\right)|_{\bar{x}_{cm}(t)},
\end{equation}
\begin{equation}
\begin{split}-\partial_{t}S_{rel} & =\sum_{j=2}^{N}\frac{1}{2m}\left(\frac{\partial S_{rel}}{\partial y_{j}}\right)^{2}+\sum_{j=2}^{N}\left(-\frac{\hbar^{2}}{2m}\frac{1}{\sqrt{\rho_{rel}}}\frac{\partial^{2}\sqrt{\rho_{rel}}}{\partial y_{j}^{2}}\right)\\
 & +\frac{1}{2M}\left(\frac{\partial S}{\partial x_{cm}}\right)|_{\bar{x}_{cm}(t)}-\left(\frac{\hbar^{2}}{2M}\frac{1}{\sqrt{\rho}}\frac{\partial^{2}\sqrt{\rho}}{\partial x_{cm}^{2}}\right)|_{\bar{x}_{cm}(t)}-v_{cm}(t)\frac{\partial S}{\partial x_{cm}}|_{\bar{x}_{cm}(t)}+U|_{\bar{x}_{cm}(t)},
\end{split}
\end{equation}
where again 
\begin{equation}
v_{cm}(t)=\frac{d\bar{x}_{cm}(t)}{dt}=\frac{1}{M}\frac{\partial S}{\partial x_{cm}}|_{\bar{x}_{cm}(t),\mathbf{\bar{y}}(t)},
\end{equation}
and
\begin{equation}
\begin{aligned}U|_{\bar{x}_{cm}(t)} & =\left[\sum_{j=1}^{N}U_{ext}(x_{j})+\frac{1}{2}\sum_{j,k=1;j\neq k}^{N}U_{int}(x_{j}-x_{k})\right]|_{\bar{x}_{cm}(t)}\\
 & =NU_{ext}(\bar{x}_{cm}(t))+\frac{1}{2}\sum_{j,k=1;j\neq k}^{N}U_{int}(x_{j}-x_{k}).
\end{aligned}
\end{equation}

We will argue that, in the limit $N\rightarrow\infty$, the `global'
$S$ and $\rho$ variables effectively decouple in $\mathbf{y}$ and
$x_{cm}$, thereby reducing (5.73-74) to the corresponding effective
Madelung equations for the relative coordinates, and likewise for
the conditional CM Madelung equations.

\subsubsection{Conditional-to-effective continuity equation}

Equation (5.73) can be rewritten as follows:
\begin{equation}
\begin{split}\partial_{t}\rho_{rel} & =-\sum_{j=2}^{N}\frac{\partial}{\partial y_{j}}\left[\rho_{rel}\frac{\partial S_{rel}}{\partial y_{j}}\frac{1}{m}\right]-\left[\frac{\partial}{\partial x_{cm}}\left(\rho\frac{\partial S}{\partial x_{cm}}\frac{1}{M}\right)-v_{cm}(t)\frac{\partial\rho}{\partial x_{cm}}\right]|_{\bar{x}_{cm}(t)}\\
 & =-\sum_{j=2}^{N}\frac{\partial}{\partial y_{j}}\left[\rho_{rel}\frac{\partial S_{rel}}{\partial y_{j}}\frac{1}{m}\right]-\left[\frac{1}{M}\frac{\partial S}{\partial x_{cm}}\frac{\partial\rho}{\partial x_{cm}}+\frac{1}{M}\rho\frac{\partial^{2}S}{\partial x_{cm}^{2}}-\frac{1}{M}\frac{\partial S}{\partial x_{cm}}|_{\mathbf{\bar{y}}(t)}\frac{\partial\rho}{\partial x_{cm}}\right]|_{\bar{x}_{cm}(t)}.
\end{split}
\end{equation}
We claim that, in the limit $N\rightarrow\infty$, all the terms in
the last bracket on the rhs of (5.77) contribute only as time-dependent
correction factors, and therefore can be dropped. 

To see this, recall that when $N\rightarrow\infty$ we have 
\begin{equation}
M\frac{d^{2}\bar{x}_{cm}(t)}{dt^{2}}\approx-\frac{\partial\left(NU_{ext}(x_{cm})\right)}{\partial x_{cm}}|_{\bar{x}_{cm}(t)}=-N\frac{\partial U_{ext}(x_{cm})}{\partial x_{cm}}|_{\bar{x}_{cm}(t)},
\end{equation}
where $U_{ext}$ spatially varies on macroscopic scales. Integrating
(5.78) gives 
\begin{equation}
\frac{d\bar{x}_{cm}(t)}{dt}=\frac{1}{M}\frac{\partial S(x_{cm},\mathbf{y},t)}{\partial x_{cm}}|_{\bar{x}_{cm}(t),\mathbf{\bar{y}}(t)}\approx-\frac{1}{m}\int_{t_{0}}^{t}\left(\frac{\partial U_{ext}}{\partial x_{cm}}\right)|_{\bar{x}_{cm}(t')}dt'+v_{cm}(0)\eqqcolon\frac{1}{M}\frac{\partial S_{cl}(x_{cm},t)}{\partial x_{cm}}|_{\bar{x}_{cm}(t)},
\end{equation}
and thus 
\begin{equation}
\frac{1}{M}\frac{\partial^{2}S(x_{cm},\mathbf{y},t)}{\partial x_{cm}^{2}}|_{\bar{x}_{cm}(t),\mathbf{\bar{y}}(t)}\approx-\frac{1}{m}\int_{t_{0}}^{t}\left(\frac{\partial^{2}U_{ext}}{\partial x_{cm}^{2}}\right)|_{\bar{x}_{cm}(t')}dt'\eqqcolon\frac{1}{M}\frac{\partial^{2}S_{cl}(x_{cm},t)}{\partial x_{cm}^{2}}|_{\bar{x}_{cm}(t)},
\end{equation}
which we see are effectively independent of $\mathbf{y}$ and only
depend on time.

Note, also, that the equation of motion for the relative positions
is given by 
\begin{equation}
m\frac{d^{2}\bar{y}_{j}(t)}{dt^{2}}=-\frac{\partial}{\partial y_{j}}\left[\frac{1}{2}\sum_{j,k=1;j\neq k}^{N}U_{int}+Q_{cm}+\sum_{j=2}^{N}Q_{j}\right].
\end{equation}
where 
\begin{equation}
\frac{d\bar{y}_{j}(t)}{dt}=-\frac{1}{m}\int_{t_{0}}^{t}\frac{\partial}{\partial y_{j}}\left[\frac{1}{2}\sum_{j,k=1;j\neq k}^{N}U_{int}+Q_{cm}+\sum_{j=2}^{N}Q_{j}\right]|_{\mathbf{\bar{y}}(t'),\bar{x}_{cm}(t')}dt'+v_{j}(0)=\frac{1}{m}\frac{\partial S}{\partial y_{j}}|_{\bar{x}_{cm}(t),\mathbf{\bar{y}}(t)}.
\end{equation}
In the limit $N\rightarrow\infty$, we have
\begin{equation}
m\frac{d^{2}\bar{y}_{j}(t)}{dt^{2}}\approx-\frac{\partial}{\partial y_{j}}\left[\frac{1}{2}\sum_{j,k=1;j\neq k}^{N}U_{int}+\sum_{j=2}^{N}Q_{j}\right]
\end{equation}
and
\begin{equation}
\begin{split}\frac{d\bar{y}_{j}(t)}{dt} & =\frac{1}{m}\frac{\partial S(x_{cm},\mathbf{y},t)}{\partial y_{j}}|_{\bar{x}_{cm}(t),\mathbf{\bar{y}}(t)}\\
 & \approx-\frac{1}{m}\int_{t_{0}}^{t}\frac{\partial}{\partial y_{j}}\left[\frac{1}{2}\sum_{j,k=1;j\neq k}^{N}U_{int}+\sum_{j=2}^{N}Q_{j}\right]|_{\mathbf{\bar{y}}(t')}dt'+v_{j}(0)=\frac{1}{m}\frac{\partial S_{rel}(\mathbf{y},t)}{\partial y_{j}}|_{\mathbf{\bar{y}}(t)},
\end{split}
\end{equation}
since the large \emph{N} CM density (corresponding to a WFP) takes the form
(5.50), implying the effective factorization 
\begin{equation}
\underset{N\rightarrow\infty}{lim}\:\rho(x_{cm},\mathbf{y},t)\approx\rho_{cl}(x_{cm},t)\rho_{rel}(\mathbf{y},t),
\end{equation}
which leads to $Q_{cm}$ taking the $\mathbf{y}$-independent form
(5.54). Correspondingly, for all $j=2,..,N$, equation (5.85) implies
\begin{equation}
Q_{j}(x_{cm},\mathbf{y},t)\approx-\left(\frac{\hbar^{2}}{2m}\frac{1}{\sqrt{\rho_{rel}(\mathbf{y},t)}}\frac{\partial^{2}\sqrt{\rho_{rel}(\mathbf{y},t)}}{\partial y_{j}^{2}}\right)=Q_{j}(\mathbf{y},t),
\end{equation}
which is effectively independent of $x_{cm}$. 

In other words, in the large \emph{N} limit, the relative coordinates evolve
in time (effectively) independently of the CM coordinate. 

Accordingly, it follows that (5.80) only contributes to (5.77) an
uninteresting time-dependent factor of the form 
\begin{equation}
\frac{1}{M}\left[\rho(x_{cm},\mathbf{y},t)\frac{\partial^{2}S(x_{cm},\mathbf{y},t)}{\partial x_{cm}^{2}}\right]|_{\bar{x}_{cm}(t)}\approx\frac{1}{M}\left[\rho_{cl}(x_{cm},t)\frac{\partial^{2}S_{cl}(x_{cm},t)}{\partial x_{cm}^{2}}\right]|_{\bar{x}_{cm}(t)}\rho_{rel}(\mathbf{y},t),
\end{equation}
while (5.79) along with (5.85) imply the time-dependent factors 
\begin{equation}
\frac{1}{M}\left(\frac{\partial S}{\partial x_{cm}}\frac{\partial\rho}{\partial x_{cm}}\right)|_{\bar{x}_{cm}(t)}\approx\frac{1}{M}\left(\frac{\partial S_{cl}(x_{cm},t)}{\partial x_{cm}}\frac{\partial\rho_{cl}(x_{cm},t)}{\partial x_{cm}}\right)|_{\bar{x}_{cm}(t)}\rho_{rel}(\mathbf{y},t),
\end{equation}
and 
\begin{equation}
\frac{1}{M}\left(\frac{\partial S}{\partial x_{cm}}|_{\mathbf{y}=\mathbf{\bar{y}}(t)}\frac{\partial\rho}{\partial x_{cm}}\right)|_{\bar{x}_{cm}(t)}\approx\frac{1}{M}\left(\frac{\partial S_{cl}(x_{cm},t)}{\partial x_{cm}}\frac{\partial\rho_{cl}(x_{cm},t)}{\partial x_{cm}}\right)|_{\bar{x}_{cm}(t)}\rho_{rel}(\mathbf{y},t).
\end{equation}
Hence, terms (5.87-89) might as well be dropped from (5.77), leaving
\begin{equation}
\begin{split}\partial_{t}\rho_{rel} & \approx-\sum_{j=2}^{N}\frac{\partial}{\partial y_{j}}\end{split}
\left[\rho_{rel}\frac{\partial S_{rel}}{\partial y_{j}}\frac{1}{m}\right],
\end{equation}
which is just the effective continuity equation for $\rho_{rel}$.

\subsubsection{Conditional-to-effective quantum Hamilton-Jacobi equation}

Equation (5.74) can be rewritten as
\begin{equation}
\begin{split}-\partial_{t}S_{rel} & =\sum_{j=2}^{N}\frac{1}{2m}\left(\frac{\partial S_{rel}}{\partial y_{j}}\right)^{2}+\frac{1}{2M}\left(\frac{\partial S}{\partial x_{cm}}\right)^{2}|_{\bar{x}_{cm}(t)}-\frac{1}{M}\frac{\partial S}{\partial x_{cm}}|_{\bar{x}_{cm}(t),\mathbf{\bar{y}}(t)}\cdot\frac{\partial S}{\partial x_{cm}}|_{\bar{x}_{cm}(t)}\\
 & +\sum_{j=2}^{N}\left(-\frac{\hbar^{2}}{2m}\frac{1}{\sqrt{\rho_{rel}}}\frac{\partial^{2}\sqrt{\rho_{rel}}}{\partial y_{j}^{2}}\right)-\left(\frac{\hbar^{2}}{2M}\frac{1}{\sqrt{\rho}}\frac{\partial^{2}\sqrt{\rho}}{\partial x_{cm}^{2}}\right)|_{\bar{x}_{cm}(t)}+U|_{\bar{x}_{cm}(t)}.
\end{split}
\end{equation}
From the arguments in section 3.2.1, the large \emph{N} limit entails that
we can neglect the terms involving $(\partial S/\partial x_{cm})|_{\bar{x}_{cm}(t)}$,
since they contribute only as time-dependent factors in (5.91) and drop
out of the equations of motion for the relative coordinates.

Similarly, as noted in subsection 2.1, for large \emph{N} the center-of-mass
quantum kinetic takes the form (5.54), which means it contributes only
an uninteresting time-dependent phase shift to $S_{rel}$ in (5.91).
And, as we showed in subsection 2.1, that the CM quantum kinetic takes
the $\mathbf{y}$-independent form (5.54) means that the CM quantum
kinetic drops out of the equations of motion for the relative positions,
i.e., equations (5.81-82), and might as well also be dropped from
(5.91).

Likewise, in $U|_{x_{cm}(t)}$, the external potential component $\sum_{j=1}^{N}U_{ext}(x_{j})=NU_{ext}(x_{cm})$
will also contribute to $S_{rel}$ only a time-dependent phase shift,
and thus can be dropped as well. 

We are thereby left with the effective quantum Hamilton-Jacobi equation
\begin{equation}
\begin{split}-\partial_{t}S_{rel} & \approx\sum_{j=2}^{N}\frac{1}{2m}\left(\frac{\partial S_{rel}}{\partial y_{j}}\right)^{2}+\sum_{j=2}^{N}\left(-\frac{\hbar^{2}}{2m}\frac{1}{\sqrt{\rho_{rel}}}\frac{\partial^{2}\sqrt{\rho_{rel}}}{\partial y_{j}^{2}}\right)+\frac{1}{2}\sum_{j,k=1;j\neq k}^{N}U_{int}.\end{split}
\end{equation}
Accordingly, we conclude that the `global' $S$ function effectively
decomposes as 
\begin{equation}
\underset{N\rightarrow\infty}{lim}\:S(x_{cm},\mathbf{y},t)\approx S_{cl}(x_{cm},t)+S_{rel}(\mathbf{y},t),
\end{equation}
where $S_{cl}(x_{cm},t)$ evolves autonomously by its effective classical
Hamilton-Jacobi equation (equation (5.65) with $A\approx0$), and
likewise for $\rho_{cl}(x_{cm},t)$ (equation (5.67) with $B\approx0$).
The Madelung transformation involving $S_{cl}(x_{cm},t)$ and $\rho_{cl}(x_{cm},t)$
then yields the classical nonlinear Schr\"{o}dinger
equation (5.72).

\subsection{Comments on the classical nonlinear Schr\"{o}dinger
equation}

As is well known \cite{Schiller1962,Rosen1964,HollandBook1993,Ghose2002,Nikolic2006,Nikolic2007,Oriols2016,Derakhshani2016a,Derakhshani2016b},
(5.72) can also be formally derived (with $\hbar$ as a free parameter)
from classical statistical mechanics of a single particle in an external
scalar potential, in the Hamilton-Jacobi representation. What's different
here is that (5.72) is an approximate description of the Schr\"{o}dinger
evolution for the CM of an \emph{N}-particle system, with potential $U$
(where the external component spatially varies on scales larger than
the size of the \emph{N}-particle system), in the limit that $N\rightarrow\infty$. 

In order to verify the robustness of (5.72) as an approximation to
classical dynamics, Oriols et al. \cite{Oriols2016} numerically simulated
a Gaussian wavepacket, defined by taking the square root of (5.50)
and multiplying by $exp\left(ik_{0}x_{cm}\right)$, evolving by (5.72)
for two cases: a packet in free fall in external potential $U=2x_{cm}$,
and a packet oscillating in a harmonic oscillator potential $U=x_{cm}^{2}/2$.
In both cases (Figures 2 and 3 of \cite{Oriols2016}), their simulations
confirm that the packets do not disperse over time, and the CM trajectories
(for different initial positions) closely mimic the CM trajectories
one expects from classical mechanics. 

Extending our derivation of (5.72) to the 3-dimensional case is formally
straightforward, and requires inclusion of the quantization condition
on the 3-dimensional generalization of the CM conditional phase field
as follows:
\begin{equation}
\oint_{L}\mathbf{\nabla}_{cm}S(\mathbf{R}_{cm},\mathbf{r}(t),t)\cdot d\mathbf{R}_{cm}=\oint_{L}\mathbf{\nabla}_{cm}S_{cm}\cdot d\mathbf{R}_{cm}=nh.
\end{equation}
This assures that the 3-dimensional version of $\psi_{cm}$ is single-valued
with (generally) multi-valued phase.

A notable advantage of (5.72) as a `large \emph{N}' approximation is that,
in contrast to the mean-field SN and stochastic SN equations, (5.72)
does not admit macroscopic superpositions of CM position states, and
so does not predict macroscopic semiclassical gravitational/electrostatic
cat states in the case that $U_{int}$ corresponds to an N-body Newtonian
gravitational/Coulomb potential (e.g., such as in a neutron star or
the sun). Basically, this is because the nonlinearity of (5.72) means
that any pair of solutions, $\psi_{1}^{cl}$ and $\psi_{2}^{cl}$,
cannot, in general, be superposed to form a more complex solution.
Thus, only one positional wavepacket is associated to the evolution
of the CM at any time.

\section{Recovering classical Newtonian gravity for many macro particles}

Suppose now that we have K many-particle systems, with CM masses $\left\{ M^{i},...,M^{K}\right\} $,
where the \emph{i}-th CM `particle' is described by a pair of CM Madelung
variables $\left\{ \rho_{cl}^{i},S_{cl}^{i}\right\} $ evolving by
their own effective Madelung equations. Suppose, further, that these
CM particles classically interact via macroscopically long-range classical
gravitational (or electrostatic) potentials (i.e., potentials spatially
varying on scales much larger than the sizes of the \emph{N}-particle systems
composing the CM particles). Then the K-body effective Madelung equations
for these gravitationally interacting CM particles, are given by
\begin{equation}
-\partial_{t}\rho_{cl}^{K}\approx\sum_{i=1}^{K}\frac{\partial}{\partial x_{cm}^{i}}\left(\frac{1}{M^{i}}\frac{\partial S_{cl}^{K}}{\partial x_{cm}^{i}}\rho_{cl}^{K}\right),
\end{equation}
\begin{equation}
-\partial_{t}S_{cl}^{K}\approx\sum_{i=1}^{K}\frac{1}{2M^{i}}\left(\frac{\partial S_{cl}^{K}}{\partial x_{cm}^{i}}\right)^{2}+\sum_{i=1}^{K}U^{i},
\end{equation}
where the solution of (5.95) is a product state of narrow Gaussians
\begin{equation}
\rho_{cl}^{K}\approx\prod_{i=1}^{K}\frac{1}{\sqrt{2\pi}\sigma_{cm}^{i}}exp\left(-\frac{\left[\overline{x}^{i}-x_{cm}^{i}\right]^{2}}{2\left[\sigma_{cm}^{i}\right]^{2}}\right)\eqqcolon\prod_{i=1}^{K}\rho_{cl}^{i},
\end{equation}
the solution of (5.96) takes the form 
\begin{equation}
S_{cl}^{K}\approx\left[\sum_{i=1}^{K}\int p_{cm}^{i}dx_{cm}^{i}-\int\left(\sum_{i=1}^{K}\frac{1}{2M^{i}}\left(p_{cm}^{i}\right)^{2}+\sum_{i=1}^{K}U^{i}\right)dt\right]-\sum_{i=1}^{K}\hbar\phi_{cm}^{i}\eqqcolon\sum_{i=1}^{K}S_{cl}^{i},
\end{equation}
and the potential 
\begin{equation}
U^{i}\coloneqq\sum_{n=1}^{N^{i}}U_{ext}^{i}(x_{n}^{i})+\frac{1}{2}\sum_{j,k=1}^{N^{i}(j\neq k)}U_{int}^{i}(x_{j}-x_{k}),
\end{equation}
where 
\begin{equation}
\sum_{n=1}^{N^{i}}U_{ext}^{i}(x_{n}^{i})\approx N^{i}U_{ext}^{i}(x_{cm}^{i})\coloneqq-\frac{M^{i}}{2}\sum_{l=1}^{K(l\neq i)}\frac{M^{l}}{|x_{cm}^{i}-x_{cm}^{l}|},
\end{equation}
using $(\partial x_{n}^{i}/\partial x_{cm}^{i})=1$.

Notice that, despite the CM `particles' gravitationally interacting
via (5.100), the large-\emph{N} CM densities form a product state (5.97).
This follows from our assumption that the gravitational potentials
sourced by the CM `particles' are (macroscopically) long-range, and
therefore vary on distance scales much larger than the sizes of the
\emph{N}-particle systems composing the CM particles (i.e., $\sigma_{cm}^{i}\ll\sqrt{|U_{ext}^{i}\prime/U_{ext}^{i}\prime\prime\prime|}$
\cite{Allori2002,Allori2001,Duerr2009}). Recall from subsection 2.1
that this was a necessary condition for showing that the large-\emph{N} density,
corresponding to a WFP, takes the (approximately) Gaussian form of
the factors in (5.97). Moreover, although the \emph{zbw} phases of
the CM `particles' are not physically independent, due to the non-separable
potential (5.100) which physically influences each CM particle via
the (approximately) classical equations of motion
\begin{equation}
M^{i}\frac{dx_{cm}^{i}(t)}{dt}\approx\frac{\partial}{\partial x_{cm}^{i}}S_{cl}^{K}|_{x_{cm}^{i}=x_{cm}^{i}(t)},
\end{equation}

\begin{equation}
M^{i}\frac{d^{2}x_{cm}^{i}(t)}{dt^{2}}\approx-N^{i}\frac{\partial}{\partial x_{cm}^{i}}U_{ext}^{i}(x_{cm}^{i},t)|_{x_{cm}^{i}=x_{cm}^{i}(t)}=M^{i}\frac{\partial}{\partial x_{cm}^{i}}\sum_{l=1}^{K(l\neq i)}\frac{M^{l}}{2|x_{cm}^{i}-x_{cm}^{l}|}|_{x_{cm}^{i}=x_{cm}^{i}(t)}^{x_{cm}^{l}=x_{cm}^{l}(t)},
\end{equation}
it is still meaningful to speak of the \emph{zbw} phase of an individual
CM `particle' in the lab frame; the \emph{i}-th CM `particle', in
the lab frame, has an associated \emph{zbw} phase $S_{cl}^{i}$ that
depends on the sum of all the potentials sourced by the $K-1$ other
CM `particles', at the space-time location of the \emph{i}-th CM `particle'.
Indeed, the net potential `seen' by an individual CM particle, from
the $K-1$ CM particles, looks like a slowly varying external potential
as a consequence of $\sigma_{cm}^{i}\ll\sqrt{|U_{ext}^{i}\prime/U_{ext}^{i}\prime\prime\prime|}$.
Thus $S_{cl}^{i}$ varies slowly as a function of $U_{ext}^{i}$ for
all $i=1,...,K$, much like the phase of a light wave moving through
a medium of slowly (spatially) varying refractive index.

If we employ the Madelung transformation, (5.95-96) can be combined
into the K-body version of (5.72): 
\begin{equation}
i\hbar\partial_{t}\psi_{cl}^{K}=\sum_{i=1}^{K}\left(-\frac{\hbar^{2}}{2M^{i}}\frac{\partial^{2}}{\partial x_{cm}^{i2}}+U^{i}-Q_{cm}^{i}\right)\psi_{cl}^{K},
\end{equation}
where
\begin{equation}
\psi_{cl}^{K}\approx\prod_{i=1}^{K}\sqrt{\rho_{cl}^{i}}e^{iS_{cl}^{i}/\hbar}\eqqcolon\prod_{i=1}^{K}\psi_{cl}^{i},,
\end{equation}
\begin{equation}
Q_{cm}^{i}\coloneqq-\frac{\hbar^{2}}{2M^{i}}\frac{1}{\sqrt{\rho_{cl}^{K}}}\frac{\partial^{2}}{\partial x_{cm}^{i2}}\sqrt{\rho_{cl}^{K}}\approx\frac{\hbar}{2M^{i}\left(\sigma_{cm}^{i}\right)^{2}}\left(1-\frac{\left[\overline{x}^{i}-x_{cm}^{hi}\right]^{2}}{\left[\sigma_{cm}^{i}\right]^{2}}\right).
\end{equation}
So the Madelung variables for each large-\emph{N} CM particle define narrow
(in position space) classical wavepackets $\psi_{cl}^{i}$ satisfying
the nonlinear Schr\"{o}dinger equation
(5.103). 

An important property of the K-body system of large-\emph{N} CM `particles'
is that the CM particle trajectories can cross in configuration space.
To see this, let us recall what the exact dBB/Nelsonian dynamics predict
for a CM `particle' associated to a pure state $\Psi$, when $\Psi$
is a superposition of two (not necessarily narrow) Gaussian wavepackets
in position space moving with fixed speeds in opposite directions
towards each other. When the packets overlap in configuration space,
the exact description says that an ensemble of identical CM particle
trajectories, corresponding to each packet, will not cross but rather
will abruptly (but not discontinuously) change directions and exit
the overlapping region with the packets they did not initially occupy
\cite{HollandBook1993,Bohm1995,Duerr2009}. The physical reason for
this non-classical behavior is that the pure state defines a single-valued
momentum field in configuration space through $p=\hbar\mathrm{Im}\nabla\ln\Psi$,
which means that there will be a unique momentum for each point in
the overlap region. Equivalently, the quantum forces from the quantum
kinetic associated to $\Psi$ in the overlap region push the trajectories
away from each other and causes them to abruptly change directions.
In the case of the K-body system, a superposition of two wavepackets
can't be applied since the packets associated to the large-\emph{N} CM `particles'
evolve by coupled nonlinear Schr\"{o}dinger
equations (5.103), and any superposition of two packets doesn't form
a new solution of (5.103). Nevertheless, we can consider two, identical,
large-\emph{N} CM particles, associated to two narrow Gaussian wavepackets
moving in opposite directions towards each other and ask if their
trajectories will cross (assume the two particles don't classically
interact or only negligibly so). Yes, because (i) the narrowness of
the two packets (recall that $\sigma_{cm}\equiv\frac{\sigma^{2}}{N}$,
and we have $N\rightarrow\infty$, implying that the amplitudes of
the packets are effectively Dirac delta functions) ensures that they
are effectively disjoint (hence don't interfere) in position space,
and (ii) the quantum force is absent from the large \emph{N} equations of
motion (5.101-102). So the large-\emph{N} CM `particles' indeed move like
classical mechanical particles, since classical mechanics predicts
that particle trajectories can cross in configuration space (but not
in phase space). Similar observations have been made by Benseny et
al. in \cite{Benseny2016} and D\"{u}rr
et al. in \cite{Duerr2009}.

\section{Recovering classical Vlasov-Poisson mean-field theory}

We can now connect the K-body system of gravitationally interacting,
large-\emph{N} CM `particles' to the classical Vlasov-Poisson mean-field
theory. 

Assuming the special case of identical CM `particles', multiplying
the first term on the right hand side in (5.99) by $1/K$ (the weak-coupling
scaling \cite{BardosGolseMauser2000,BardosErdosGolseMauserYau2002,Golse2003}),
and subtracting out the second term on the right hand side in (5.99)
(since it will only yield a global phase factor), the K-body effective
CM Madelung equations become 
\begin{equation}
-\partial_{t}\rho_{cl}^{K}\approx\sum_{i=1}^{K}\frac{\partial}{\partial x_{cm}^{i}}\left(\frac{1}{M}\frac{\partial S_{cl}^{K}}{\partial x_{cm}^{i}}\rho_{cl}^{K}\right),
\end{equation}
\begin{equation}
H_{cl}^{K}\coloneqq-\partial_{t}S_{cl}^{K}\approx\sum_{i=1}^{K}\frac{1}{2M}\left(\frac{\partial S_{cl}^{K}}{\partial x_{cm}^{i}}\right)^{2}-\frac{M^{2}}{K}\sum_{i=1}^{K}\sum_{l=1}^{K(l\neq i)}\frac{1}{2|x_{cm}^{i}-x_{cm}^{l}|},
\end{equation}
with solutions given by (5.97-98) for $M^{i}=M$. The classical equations
of motion are just
\begin{equation}
p_{cm}^{i}(t)\coloneqq M\frac{dx_{cm}^{i}(t)}{dt}\approx\frac{\partial}{\partial x_{cm}^{i}}S_{cl}^{K}|_{x_{cm}^{i}=x_{cm}^{i}(t)},
\end{equation}

\begin{equation}
\frac{dp_{cm}^{i}(t)}{dt}=M\frac{d^{2}x_{cm}^{i}(t)}{dt^{2}}\approx\frac{M^{2}}{K}\frac{\partial}{\partial x_{cm}^{i}}\sum_{l=1}^{K(l\neq i)}\frac{1}{2|x_{cm}^{i}-x_{cm}^{l}|}|_{x_{cm}^{i}=x_{cm}^{i}(t)}^{x_{cm}^{l}=x_{cm}^{l}(t)}.
\end{equation}
Now, consider the empirical distribution for the \emph{K} particles $f_{K}(x_{cm},p_{cm},t)\coloneqq K^{-1}\sum_{i=1}^{K}\delta(x_{cm}-x_{cm}^{i}(t))\delta(p_{cm}-p_{cm}^{i}(t))$
satisfying (in the sense of distributions) the Vlasov equation
\begin{equation}
\begin{split}\partial_{t}f_{K}+p_{cm}\frac{\partial}{\partial x_{cm}}f_{K} & +\frac{\partial}{\partial p_{cm}}\left[F_{K}\left(x_{cm},t\right)f_{K}\right]\\
 & =\frac{1}{K^{2}}\sum_{i=1}^{K}\frac{\partial}{\partial p_{cm}}\left[\left(\frac{\partial}{\partial x_{cm}^{i}}\sum_{l=1}^{K(l\neq i)}\frac{M^{2}}{2|x_{cm}^{i}-x_{cm}^{l}|}|_{x_{cm}^{i}=x_{cm}^{i}(t)}^{x_{cm}^{l}=x_{cm}^{l}(t)}\right)\right.\\
 & \left.\times\delta(x_{cm}-x_{cm}^{i}(t))\delta(p_{cm}-p_{cm}^{i}(t))\right],
\end{split}
\end{equation}
 where
\begin{equation}
F_{K}\left(x_{cm},t\right)\coloneqq-\frac{\partial}{\partial x_{cm}}\int_{\mathbb{R}}\int_{\mathbb{R}}\frac{M^{2}}{|x_{cm}-x_{cm}'|}\,f_{K}\,dx_{cm}'dp_{cm}.
\end{equation}

We recall that Golse \cite{Golse2003} and Bardos et al. \cite{BardosGolseMauser2000,BardosErdosGolseMauserYau2002}
considered a D-dimensional generalization of (5.106-111) \footnote{Regarding (5.106), Golse \cite{Golse2003} and Bardos et al. \cite{BardosGolseMauser2000,BardosErdosGolseMauserYau2002}
take the empirical position distributions for the particles to be
Dirac delta functions. The 1-dimensional Dirac delta function in position
space is indeed a solution of (5.106), and note that the factors of
(5.97) approach 1-dimensional Dirac delta functions as $K\rightarrow\infty$.}, for an arbitrary, symmetric, smooth interaction potential $V$,
and showed that for $K\rightarrow\infty$ one obtains the D-dimensional
Vlasov-Poisson mean-field equations (see also section 4 of Part I).
Thus the system (5.106-111), in the limit $K\rightarrow\infty$, is
equivalent to the 2-dimensional Vlasov-Poisson mean-field equations
(hereafter, writing $x_{cm}=x$ and $p_{cm}=p$):
\begin{equation}
\partial_{t}f(x,p,t)+\left\{ H^{m.f.}(x,p,t),f(x,p,t)\right\} =0,
\end{equation}

\begin{equation}
H_{cl}^{m.f.}(x,p,t)\coloneqq\frac{p^{2}}{2M}+\int_{\mathbb{R}}M\Phi_{g}^{m.f.}(x,x',t)\,dx'.
\end{equation}

\begin{equation}
\frac{\partial^{2}\Phi_{g}^{m.f.}}{\partial x^{2}}=4\pi M\int_{\mathbb{R}}f(x,p,t)dp=4\pi M\rho(x,t),
\end{equation}
\begin{equation}
F(x,t)\coloneqq-\frac{\partial}{\partial x}\int_{\mathbb{R}}M\Phi_{g}^{m.f.}(x,x',t)\,dx'.
\end{equation}
Extending the above results to the D-dimensional case is formally
straightforward.

\section{Incorporating environmental decoherence}

As discussed in subsection 2.1, environmental decoherence accompanied
by effective collapse ensures that wavefunctions associated to macroscopic
objects (composed of dBB/Nelsonian/ZSM particles) in the real world
will not correspond to macroscopic quantum superpositions (i.e., involve
strong quantum correlations). Thus it is reasonable to expect that
wavefunctions associated to macroscopic objects in the real world
will in general be WFPs. Though this expectation seems reasonable
on general grounds, it would be even more convincing if we could demonstrate
it in an explicit model of environmental decoherence in ZSM (or dBB).
Here we sketch a suggestion for an explicit model. 

There exists a well-known model of generalized Brownian motion in
classical nonequilibrium statistical mechanics called the Kac-Zwanzig
(KZ) model \cite{FordKacMazur1965,Zwanzig1973,EberlingSokolov2005}
(the quantum mechanical analogue is the well-known Caldeira-Leggett
model \cite{CaldeiraLeggett1983a,CaldeiraLeggett1983b}). The KZ model
describes a heavy particle coupled to an external field and a heat
bath, the bath modeled as an \emph{N}-particle system of light harmonic oscillators,
where the system particle couples bilinearly to each bath particle,
with possibly frequency-dependent coupling strength. The classical
Newtonian equations of motion for the system particle and bath particles
are thereby coupled, and if one integrates out the bath variables,
one finds, under the assumptions that the bath is at thermal equilibrium
at temperature T and has arbitrary spectral density, a non-Markovian
Langevin equation describing the time-evolution of the system particle. 

Relatedly, Chou et al. \cite{Chou2008} have shown that if one replaces
the heavy probe particle of the KZ model with a system of N interacting
identical harmonic oscillators, and if one assumes bilinear coupling
of identical strength between the system and bath position coordinates,
then there exists a canonical transformation that makes it possible
to separate out the CM of the system from its relative degrees of
freedom in the system-bath Hamiltonian. In other words, the transformed
Hamiltonian, in the system degrees of freedom, is of the same form
as the Schr\"{o}dinger Hamiltonian
in (5.36), the latter obtained from Oriols et al.'s coordinate transformation
(5.12-13). Moreover, the transformed Hamiltonian entails that only
the CM couples to the bath particles. Under the assumptions that (i)
the system and bath are initially uncorrelated, (ii) the heat bath
is initially at thermal equilibrium at temperature T, and (iii) the
spectral density of the bath is arbitrary, Chou et al. then use the
transformed Hamiltonian to define the unitary evolution of a system-bath
density matrix. Tracing over the bath degrees of freedom, they find
that the reduced density matrix for the system evolves by a non-Markovian
master equation of Hu-Paz-Zhang type \cite{HuPazZhang1992}. Such
a master equation is, of course, well-known in the theory of quantum
Brownian motion for open systems \cite{Schlosshauer2008}.

Our proposal, then, is to construct a KZ-type model from ZSM-Newton/Coulomb
(or dBB-Newton/Coulomb), using the same starting assumptions as Chou
et al., and applying the Oriols et al. scheme to the system and bath,
respectively. This should make it possible to show that decoherence
of the system wavefunction via interaction with the bath leads, under
unitary evolution, to a macroscopic superposition of effectively orthogonal
system-bath product states, and that such an evolution is accompanied
by effective collapse of the system-bath configuration into one of
the system-bath product states. In addition, the evolution of the
system's CM particle position, with the bath variables integrated
out, should be described by a non-Markovian modified Langevin equation,
where the modifying terms are the quantum forces from the CM's quantum
kinetic and the quantum kinetics of the relative degrees of freedom,
and where both types of quantum kinetics are constructed from the
effective system wavefunction to which the system configuration has
collapsed. Then, taking the large particle number limits simultaneously
for system and bath, it should be possible to show, by applying the
arguments in section 3 of the present paper, that the equations of
motion for the system and bath CM positions become effectively classical.
In other words, we should recover the classical non-Markovian Langevin
equation for the heavy particle in the classical KZ model.

The details of this proposal will be worked out in a stand-alone paper.

\section{Conclusion}

We have applied Oriols et al.'s large-\emph{N}-CM approximation scheme to
a system of N identical, non-relativistic, \emph{zbw} particles interacting
via potentials $\hat{U}_{int}(\hat{x}_{j}-\hat{x}_{k})$ and with
external potentials $\hat{U}_{ext}(\hat{x}_{j})$. This made it possible
to: (i) self-consistently describe large numbers of identical \emph{zbw}
particles interacting classical-gravitationally/electrostatically,
without an independent particle approximation; (ii) avoid macroscopic
semiclassical gravitational/electrostatic cat states and recover \emph{K}-particle
classical Newtonian gravity/electrodynamics for the CM descriptions
of gravitationally/electrostatically interacting macroscopic particles
(where the macroscopic particles are built out of interacting \emph{zbw}
particles); and (iii) recover classical Vlasov-Poisson mean-field
theory for macroscopic particles that interact gravitationally/electrostatically,
in the weak-coupling large \emph{K} limit. In addition, we have sketched
a proposal for an explicit model of environmental decoherence consistent
with the Oriols et al. large-\emph{N}-CM approximation scheme, the purpose
of which is to explicitly demonstrate our claim that environmental
decoherence plus effective collapse entails WFPs associated to real-world
macroscopic objects.

We leave for future work the task of extending the ZSM-based large-\emph{N}-CM
approximation scheme to relativistic massive particles and fields,
in flat and curved spacetimes.

\chapter{Appendix to Chapter 2}

\section{Proof of the 1-particle Stochastic Variational Principle }

Following Yasue's presentation \cite{Yasue1981a}, let $\mathbf{q}'(t)=\mathbf{q}(t)+\delta\mathbf{q}(t)$
be a variation of the sample path $\mathbf{q}(t)$, with end-point
constraints $\delta\mathbf{q}(t_{i})=\delta\mathbf{q}(t_{f})=0$.
Let us also assume, for the sake of generality, that the particle
has charge \emph{e} and couples to the external magnetic vector potential,
$\mathbf{A}_{ext}(\mathbf{q}(t),t)$, as well as the external electric
scalar potential, $\Phi_{e}(\mathbf{q}(t),t)$. Then the condition 
\begin{equation}
\begin{aligned}J & =\mathrm{E}\left[\int_{t_{i}}^{t_{f}}\left\{ \frac{1}{2}\left[\frac{1}{2}m\mathbf{b}(\mathbf{q}(t),t)^{2}+\frac{1}{2}m\mathbf{b}_{*}(\mathbf{q}(t),t)^{2}\right]+\frac{e}{c}\mathbf{A}_{ext}(\mathbf{q}(t),t)\cdot\mathbf{v}(\mathbf{q}(t),t)-e\Phi_{e}(\mathbf{q}(t),t)\right\} dt\right]\\
 & =\mathrm{E}\left[\int_{t_{i}}^{t_{f}}\left\{ \frac{1}{2}\left[\frac{1}{2}m\left(D\mathbf{q}(t)\right)^{2}+\frac{1}{2}m\left(D_{*}\mathbf{q}(t)\right)^{2}\right]+\frac{e}{c}\mathbf{A}_{ext}\cdot\mathbf{v}-e\Phi_{e}\right\} dt\right]=extremal,
\end{aligned}
\end{equation}
where $\mathrm{E}\left[...\right]$ is the absolute expectation, is
equivalent to the variation,
\begin{equation}
\delta J(\mathbf{q})=J(\mathbf{q}')-J(\mathbf{q}),
\end{equation}
up to first order in $||\delta\mathbf{q}(t)||$. So (6.2) gives 
\begin{equation}
\begin{aligned}\delta J & =\mathrm{E}\left[\int_{t_{i}}^{t_{f}}\left\{ \left[\frac{1}{2}m\left(D\mathbf{q}(t)\cdot D\delta\mathbf{q}(t)+D_{*}\mathbf{q}(t)\cdot D_{*}\delta\mathbf{q}(t)\right)\right]\right.\right.\\
 & \left.\left.+\frac{e}{c}\mathbf{A}_{ext}\cdot\frac{1}{2}\left(D\delta\mathbf{q}(t)+D_{*}\delta\mathbf{q}(t)\right)+\frac{e}{c}\left(\delta\mathbf{q}(t)\cdot\nabla\mathbf{A}_{ext}\right)\frac{1}{2}\left(D\mathbf{q}(t)+D_{*}\mathbf{q}(t)\right)-e\nabla\Phi_{e}\cdot\delta\mathbf{q}(t)\right\} |_{\mathbf{q}=\mathbf{q}(t)}dt\right],
\end{aligned}
\end{equation}
where we note that $\mathbf{v}=\frac{1}{2}\left(D+D_{*}\right)\mathbf{q}(t)$
and is constrained by Eq. (2.10). Now, observing that for an arbitrary
function, $f(\mathbf{q}(t),t)$, we have
\begin{equation}
\mathrm{E}\left[\int_{t_{i}}^{t_{f}}\left[f(\mathbf{q}(t),t)D\delta\mathbf{q}(t)\right]dt\right]=-\mathrm{E}\left[\int_{t_{i}}^{t_{f}}\left[\delta\mathbf{q}(t)D_{*}f(\mathbf{q}(t),t)\right]dt\right],
\end{equation}
and
\begin{equation}
\mathrm{E}\left[\int_{t_{i}}^{t_{f}}\left[f(\mathbf{q}(t),t)D_{*}\delta\mathbf{q}(t)\right]dt\right]=-\mathrm{E}\left[\int d^{3}\mathbf{q}\rho\int_{t_{i}}^{t_{f}}\left[\delta\mathbf{q}(t)Df(\mathbf{q}(t),t)\right]dt\right],
\end{equation}
and
\begin{equation}
\frac{1}{2}\left(D+D_{*}\right)f(\mathbf{q}(t),t)=\left\{ \frac{\partial}{\partial t}+\frac{1}{2}\left[D\mathbf{q}(t)+D_{*}\mathbf{q}(t)\right]\cdot\nabla\right\} f(\mathbf{q},t)|_{\mathbf{q}=\mathbf{q}(t)},
\end{equation}
we then obtain 
\begin{equation}
\begin{aligned}\delta J & =\mathrm{E}\left[\int_{t_{i}}^{t_{f}}\left\{ \frac{m}{2}\left[D_{*}D+DD_{*}\right]\mathbf{q}(t)\right.\right.\\
 & \left.\left.-\frac{e}{c}\mathbf{v}\times\left(\nabla\times\mathbf{A}_{ext}\right)+\frac{e}{c}\frac{\partial\mathbf{A}_{ext}}{\partial t}+e\nabla\Phi_{e}\right\} |_{\mathbf{q}=\mathbf{q}(t)}\delta\mathbf{q}(t)dt\right]+\vartheta(||\delta\mathbf{q}||).
\end{aligned}
\end{equation}
From the variational constraint (6.1-2), it follows that the first-order
variation of $J$ must be zero for arbitrary sample-wise variation
$\delta\mathbf{q}(t)$. Moreover, since the expectation is a positive
linear functional, we will have the equation of motion
\begin{equation}
\frac{m}{2}\left[D_{*}D+DD_{*}\right]\mathbf{q}(t)=-e\left[\nabla\Phi_{e}+\frac{1}{c}\frac{\partial\mathbf{A}_{ext}}{\partial t}\right]|_{\mathbf{q}=\mathbf{q}(t)}+\frac{e}{c}\mathbf{v}\times\left(\nabla\times\mathbf{A}_{ext}\right)|_{\mathbf{q}=\mathbf{q}(t)}
\end{equation}
for each time $t$ $\in$ $\left[t_{i},t_{f}\right]$ with probability
one.

\section{\textcolor{black}{Classical Zitterbewegung in the Central Potential}}

Suppose that the non-relativistic \emph{zbw} particle in the lab frame
is moving in a circular orbit about some central potential, $V(\mathbf{r}),$
where $\mathbf{r}$ is the radius of the orbit. In this case, for
the spherical coordinates $(r,\alpha,\beta)$, $r$ is fixed, $\alpha$
is varies with time, and $\beta$ has the constant value $\pi/2$,
giving translational velocities $\mathbf{v}_{r}=\dot{r}=0$ (and we
require $\ddot{r}=0$), $\mathbf{v}_{\alpha}=r\dot{\alpha}$, and
$\mathbf{v}_{\beta}=r\dot{\beta}sin\alpha=0$. The $v\ll c$ approximated
\emph{zbw} phase change in the lab frame is then
\begin{equation}
\begin{aligned}\delta\theta(\alpha(t),t) & =\left(\omega_{c}+\omega_{\alpha}+\kappa(\mathbf{r})\right)\delta t-\frac{v_{\alpha}r\delta\alpha(t)}{c^{2}}\\
 & =\frac{\omega_{c}}{mc^{2}}\left[\left(mc^{2}+\frac{p_{\alpha}^{2}}{2mr^{2}}+V(\mathbf{r})\right)\delta t-v_{\alpha}r\delta\alpha(t)\right],
\end{aligned}
\end{equation}
where $p_{\alpha}=mr^{2}\dot{\alpha}$. Because the total energy of
the system is constant, integrating this gives
\begin{equation}
\theta=\frac{\omega_{c}}{mc^{2}}\left[\left(mc^{2}+\frac{p_{\alpha}^{2}}{2mr^{2}}+V(\mathbf{r})\right)t-p_{\alpha}\alpha(t)\right]+C,
\end{equation}
or
\begin{equation}
\begin{aligned}S & =p_{\alpha}\alpha(t)-\left(mc^{2}+\frac{p_{\alpha}^{2}}{2mr^{2}}+V(\mathbf{r})\right)t+C\\
 & =p_{\alpha}\alpha(t)-Et+C.
\end{aligned}
\end{equation}
Incidentally, we could have also obtained (6.11) by starting with
the non-relativistic Lagrangian
\begin{equation}
L(\mathbf{\alpha}(t),t)=\frac{1}{2}mr{}^{2}\dot{\alpha}(t)^{2}-V(\mathbf{r})-mc^{2},
\end{equation}
and using the Legendre transformation,
\begin{equation}
E=p_{\alpha}\dot{\mathbf{\alpha}}-L=\frac{p_{\alpha}^{2}}{2mr^{2}}+V(\mathbf{r})+mc^{2},
\end{equation}
to get
\begin{equation}
S=\int Ldt+C=\int\left(p_{\alpha}\dot{\mathbf{\alpha}}-E\right)dt+C=p_{\alpha}\alpha-Et+C.
\end{equation}
Clearly (6.11) satisfies the classical Hamilton-Jacobi equation
\begin{equation}
-\frac{\partial S}{\partial t}=\frac{1}{2mr^{2}}\left(\frac{\partial S}{\partial\alpha}\right)^{2}+V(\mathbf{r}),
\end{equation}
where $-\partial S/\partial t=E$ and $\partial S/\partial\alpha=\mathbf{p}_{\alpha}=\mathbf{L}_{\alpha}$,
the latter being the constant angular momentum of the particle in
the $\hat{z}$-direction.

Because the \emph{zbw} oscillation is simply harmonic and the phase
is a well-defined function of the particle position, the change in
$S$ will now be quantized upon fixed time integration around a closed
(circular) orbit \emph{L}. In other words, we will have
\begin{equation}
\oint_{L}p_{\alpha}\delta\alpha=2\pi mv_{\alpha}r\mathrm{=}nh,
\end{equation}
or
\begin{equation}
L_{\alpha}=mv_{\alpha}r=n\hbar,
\end{equation}
where $n$ is an integer. From (6.17) and the force balance equation
(assuming a Coulomb force), $mv_{\alpha}^{2}/r=(1/4\pi\epsilon_{0})e^{2}/r^{2}$,
it follows that the radius is quantized as
\begin{equation}
r_{n}=\frac{4\pi\epsilon_{0}\hbar^{2}}{m_{e}e^{2}}n^{2},
\end{equation}
where for $n=1$, (6.18) gives the Bohr radius. Inserting (6.18) into
the force balance equation and recognizing that $E=V/2$, we then
obtain the quantized energy states
\begin{equation}
E_{n}=\frac{E_{1}}{n^{2}},
\end{equation}
where $E_{1}=-e^{2}/8\pi\epsilon_{0}r_{1}=-13.6eV$ is precisely the
magnitude of the ground state energy of the Bohr hydrogen atom.

We wish to emphasize that, whereas Bohr simply assumed a condition
equivalent to (6.16) in order to stabilize the electron's circular
orbit in the classical hydrogen atom, we \textit{\textcolor{black}{obtained}}
(6.16) just from the zitterbewegung hypothesis in the particle's instantaneous
translational rest frame combined with the usual Lorentz transformation.
In other words, in Bohr's model, (6.16) is imposed ad hoc while in
our model it arises as a direct consequence of a relativistic (\emph{zbw})
constraint on the particle's motion.

\chapter{Appendix to Chapter 3}

\section{Proof of the \emph{N}-particle Stochastic Variational Principle }

Let $\mathbf{q}_{i}'(t)=\mathbf{q}_{i}(t)+\delta\mathbf{q}_{i}(t)$
be variations of the sample paths $\mathbf{q}_{i}(t)$, with end-point
constraints $\delta\mathbf{q}_{i}(t_{I})=\delta\mathbf{q}_{i}(t_{F})=0$.
Then, using $\mathbf{b}_{i}=D\mathbf{q}_{i}(t)$ and $\mathbf{b}_{i*}=D_{*}\mathbf{q}_{i}(t)$,
the condition
\begin{equation}
\begin{aligned}J & =\mathrm{E}\left[\int_{t_{I}}^{t_{F}}\sum_{i=1}^{N}\left\{ \frac{1}{2}\left[\frac{1}{2}m_{i}\left(D\mathbf{q}_{i}(t)\right)^{2}+\frac{1}{2}m_{i}\left(D_{*}\mathbf{q}_{i}(t)\right)^{2}\right]\right.\right.\\
 & \left.\left.+\frac{e_{i}}{c}\mathbf{A}_{i}^{ext}\cdot\frac{1}{2}\left(D+D_{*}\right)\mathbf{q}_{i}(t)-e_{i}\left(\Phi_{i}^{ext}+\Phi_{c}^{int}\right)\right\} dt\right]\\
 & =\mathrm{E}\left[\int_{t_{I}}^{t_{F}}\sum_{i=1}^{N}\left\{ \frac{1}{2}m_{i}v_{i}^{2}+\frac{1}{2}m_{i}u_{i}^{2}+\frac{e_{i}}{c}\mathbf{A}_{i}^{ext}\cdot\mathbf{v}_{i}-e_{i}\left[\Phi_{i}^{ext}+\Phi_{c}^{int}\right]\right\} dt\right]=extremal,
\end{aligned}
\end{equation}
is equivalent to the variation,
\begin{equation}
\delta J(q)=J(q')-J(q),
\end{equation}
up to first order in $||\delta\mathbf{q}_{i}(t)||$. So (7.2) gives 
\begin{equation}
\begin{aligned}\delta J & =\mathrm{E}\left[\int_{t_{I}}^{t_{F}}\sum_{i=1}^{N}\left\{ \left[\frac{1}{2}m_{i}\left(D\mathbf{q}_{i}(t)\cdot D\delta\mathbf{q}_{i}(t)+D_{*}\mathbf{q}_{i}(t)\cdot D_{*}\delta\mathbf{q}_{i}(t)\right)\right]\right.\right.\\
 & \left.\left.+\frac{e_{i}}{c}\mathbf{A}_{i}^{ext}\cdot\frac{1}{2}\left(D\delta\mathbf{q}_{i}(t)+D_{*}\delta\mathbf{q}_{i}(t)\right)+\frac{e_{i}}{c}\left(\delta\mathbf{q}_{i}(t)\cdot\nabla_{i}\mathbf{A}_{i}^{ext}\right)\mathbf{v}_{i}-e_{i}\nabla_{i}\left[\Phi_{i}^{ext}+\Phi_{c}^{int}\right]\cdot\delta\mathbf{q}_{i}(t)\right\} |_{\mathbf{q}_{j}=\mathbf{q}_{j}(t)}dt\right].
\end{aligned}
\end{equation}

Now, for an arbitrary function $f_{i}(q(t),t)$, we have the relations
\begin{equation}
\mathrm{E}\left[\int_{t_{I}}^{t_{F}}\sum_{i=1}^{N}\left[f_{i}(q(t),t)D\delta\mathbf{q}_{i}(t)\right]dt\right]=-\mathrm{E}\left[\int_{t_{I}}^{t_{F}}\sum_{i=1}^{N}\left[\delta\mathbf{q}_{i}(t)D_{*}f_{i}(q(t),t)\right]dt\right],
\end{equation}
and
\begin{equation}
\mathrm{E}\left[\int_{t_{I}}^{t_{F}}\sum_{i=1}^{N}\left[f_{i}(q(t),t)D_{*}\delta\mathbf{q}_{i}(t)\right]dt\right]=-\mathrm{E}\left[\int_{t_{I}}^{t_{F}}\sum_{i=1}^{N}\left[\delta\mathbf{q}_{i}(t)Df_{i}(q(t),t)\right]dt\right],
\end{equation}
and
\begin{equation}
\frac{1}{2}\left(D+D_{*}\right)f_{i}(q(t),t)=\left\{ \partial_{t}+\frac{1}{2}\left[D\mathbf{q}_{i}(t)+D_{*}\mathbf{q}_{i}(t)\right]\cdot\nabla_{i}\right\} f_{i}(q,t)|_{\mathbf{q}_{j}=\mathbf{q}_{j}(t)}.
\end{equation}
So, using Eq. (3.9) in section 3.2, the integrand of (7.3) becomes 
\begin{equation}
\begin{aligned}\delta J & =\mathrm{E}\left[\int_{t_{I}}^{t_{F}}\sum_{i=1}^{N}\left\{ \frac{m_{i}}{2}\left[D_{*}D+DD_{*}\right]\mathbf{q}_{i}(t)\right.\right.\\
 & \left.\left.-\frac{e_{i}}{c}\mathbf{v}_{i}\times\left(\nabla_{i}\times\mathbf{A}_{i}^{ext}\right)+\frac{e_{i}}{c}\partial_{t}\mathbf{A}_{i}^{ext}+e_{i}\nabla_{i}\left[\Phi_{i}^{ext}+\Phi_{c}^{int}\right]\right\} |_{\mathbf{q}_{j}=\mathbf{q}_{j}(t)}\delta\mathbf{q}_{i}(t)dt\right]+\vartheta(||\delta\mathbf{q}_{i}||).
\end{aligned}
\end{equation}

From the variational constraint (7.1-2), and using the fact that the
arbitrary variations (i.e., the virtual displacements in the generalized
coordinates) $\delta\mathbf{q}_{i}(t)$ are independent for all \emph{i}
by D'Alembert's principle \cite{Ray2006}, it follows that the first-order
variation of $J$ must be zero for each $\delta\mathbf{q}_{i}(t)$.
Moreover, since the expectation is a positive linear functional, we
will have the equations of motion
\begin{equation}
\sum_{i=1}^{N}\frac{m_{i}}{2}\left[D_{*}D+DD_{*}\right]\mathbf{q}_{i}(t)=\sum_{i=1}^{N}e_{i}\left[-\frac{1}{c}\partial_{t}\mathbf{A}_{i}^{ext}-\nabla_{i}\left(\Phi_{i}^{ext}+\Phi_{c}^{int}\right)+\left(\frac{\mathbf{v}_{i}}{c}\right)\times\left(\nabla_{i}\times\mathbf{A}_{i}^{ext}\right)\right]|_{\mathbf{q}_{j}=\mathbf{q}_{j}(t)},
\end{equation}
and
\begin{equation}
\frac{m_{i}}{2}\left[D_{*}D+DD_{*}\right]\mathbf{q}_{i}(t)=\left[-\frac{e_{i}}{c}\partial_{t}\mathbf{A}_{i}^{ext}-e_{i}\nabla_{i}\left(\Phi_{i}^{ext}+\Phi_{c}^{int}\right)+\frac{e_{i}}{c}\mathbf{v}_{i}\times\left(\nabla_{i}\times\mathbf{A}_{i}^{ext}\right)\right]|_{\mathbf{q}_{j}=\mathbf{q}_{j}(t)},
\end{equation}
for each time $t$ $\in$ $\left[t_{I},t_{F}\right]$ with probability
one.

\chapter{Summary and Outlook}

In this dissertation we carried out two objectives. For the first
objective, we proposed an answer to Wallstrom's criticism of stochastic
mechanics by reformulating Nelson's stochastic mechanics (or, more
precisely, Nelson-Yasue stochastic mechanics) so as to consistently
incorporate the ``zitterbewegung'' particle model(s) of de Broglie
and Bohm, and thereby explain how the quantization condition $\oint_{L}\nabla S\cdot d\mathbf{x}=nh$
could arise naturally instead of by ad hoc imposition or by making
logically-circular appeals to the single-valuedeness of wavefunctions
in standard quantum mechanics. This was done in Chapters 2 and 3,
where we developed ``zitterbewegung stochastic mechanics'' (ZSM)
for the single particle case and the \emph{N}-particle case. 

For the second objective, we (i) used ZSM to formulate fundamentally-semiclassical
theories of Newtonian gravity and electrodynamics, which we termed
``ZSM-Newton'' and ``ZSM-Coulomb'', respectively; (ii) compared
ZSM-Newton and ZSM-Coulomb to existing formulations of semiclassical
Newtonian gravity and electrodynamics; (iii) demonstrated that ZSM-Newton/Coulomb
are consistent, empirically viable theories of semiclassical Newtonian
gravity and electrodynamics, with conceptual and technical advantages
over extant semiclassical theories based on either standard quantum
theory or measurement-problem-free alternative quantum theories; and
(iv) demonstrated that ZSM-Newton can recover classical Newtonian
gravity (and classical mechanics more generally) in the large \emph{N}
limit of the center of mass description of a system of \emph{N} gravitationally
strongly interacting stochastic mechanical particles. This was done
in Chapters 4 and 5. 

Of course, as we have made clear throughout, there is much more work
to be done. In terms of developing ZSM, it remains to be shown that
it can be generalized to particles with spin, relativistic particles
and fields (accounting for all of the Standard Model) on flat and
curved spacetimes, open quantum systems, and
non-Markovian conservative diffusions. It also remains to be shown
that a physical (i.e., non-phenomenological) theory of the ZSM ether,
the \emph{zbw} particle, and the dynamical coupling between the two,
can be developed, and that such a theory can justify the phenomenological
assumptions of ZSM. 

In terms of developing ZSM-Newton and ZSM-Coulomb, the former needs
to be generalized to (at least) semiclassical Einstein gravity and
perturbative quantum gravity, both for relativistic \emph{zbw} particles
and relativistic \emph{zbw} fields; the latter needs to be generalized
to perturbative QED for relativistic \emph{zbw} particles and \emph{zbw}
fields. And all of the results shown for ZSM-Newton and ZSM-Coulomb
should be shown for the aforementioned generalizations of ZSM-Newton
and ZSM-Coulomb.

So although this thesis is the culmination of a nearly decade-long journey,
it is also the start of another journey, one that could well take
(at least) another decade. If I am fortunate enough, it won't be
a solo journey, but if so, the work will be done; that is, until I
reach the absolute limits of my intellectual abilities or experimental
tests confirm a different version of quantum theory or compelling
new theoretical arguments are posed against stochastic mechanical
theories (including ZSM).

\chapter{Samenvatting}

De voornaamste taak van deze these is het herformuleren van een benaderingswijze van de kwantummechanica gebaseerd op de stochastisch-mechanische theorie van Edward Nelson en Kunio Yasue (hierna, Nelson-Yasue stochastische mechanica), op zo een manier dat de stochastisch-mechanische $S$-functie voldoet aan de Bohr-Sommerfeld-achtige kwantisatieconditie $\oint_{L}\nabla S\cdot d\mathbf{x}=nh$, zonder deze ad hoc aan te nemen noch te baseren op het logisch-circulaire beroep op de \'{e}\'{e}nwaardigheid van de kwantummechanische golffunctie. Ik stel een antwoord voor op de zogenaamde ``Wallstrom kritiek" jegens stochastisch-mechanische theorie\"en, die luidt dat de kwantisatieconditie ofwel ad hoc wordt gepostuleerd, ofwel wordt ingebracht door een logisch-circulair beroep op de \'{e}\'{e}nwaardige golffuncties, en dat beide methoden de claim ondermijnen  dat bestaande stochastisch-mechanische theorie\"{e}n  de kwantummechanica afleiden of onderliggen. Deze Wallstrom kritiek is lange tijd beschouwd als een definitieve weerlegging van de mogelijkheid om stochastisch-mechanische theorie\"{e}n te gebruiken als grondslag voor kwantummechanica.

De voorgestelde herformulering van Nelson-Yasue stochastische mechanica, die ik ``zitterbewegung stochastische mechanica" (ZSM) noem, is ontwikkeld door Nelson-Yasue stochastische mechanica op zorgvuldige en consistente wijze te combineren met het klassieke ``zitterbewegung" deeltjesmodel van Louis de Broglie en David Bohm, waarin de kwantisatieconditie voortkomt uit het combineren van de aannames dat (i) een elementair deeltje correspondeert met een gelocaliseerd periodiek proces met  constante Compton frequentie in het ruststelsel, en dat (ii) de Lorentztransformaties uit de speciale relativiteitstheorie kunnen worden toegepast. Met andere woorden, in ZSM wordt het puntdeeltje van Nelson-Yasue stochastische mechanica vervangen door het zitterbewegungsdeeltje van de Broglie en Bohm (met enkele kleine aanpassingen). Het zitterbewegungsdeeltje ondergaat het stochastische-mechanische diffusie proces uit de Nelson-Yasue theorie en vinden we de $S$-functie als een ``evenwichtsfase" van het gelokaliseerde periodieke proces van het deeltje in het labstelsel, waarbij wordt voldaan aan de kwantisatieconditie. Dit is beschreven in de hoofdstukkken (Chapters) 2 en 3, waarbij het geval voor \'{e}\'{e}n enkel deeltje met een uitgewerkt voorbeeld wordt ge\"{i}llusteerd in hoofdstuk 2 en waarbij hoofdstuk 3 het meerdere-deeltjes geval en mogelijke uitbreidingen van ZSM bekijkt, en deze vergelijkt met andere suggesties om de kwantisatieconditie in stochastische mechanica te rechtvaardigen. In ieder hoofdstuk is een samenvatting opgenomen.

Een tweede doel van deze these is om (i) ZSM te gebruiken om fundamenteel-semi-klassieke theorie\"{e}n van Newtoniaanse zwaartekracht en electrodynamica te formuleren; (ii) deze te vergelijken met   bestaande formuleringen van semi-klassieke Newtoniaanse zwaartekracht en electrodynamica; (iii) te laten zien dat de theorie\"{e}n gebaseerd op ZSM consistente, empirisch levensvatbare theorie\"{e}n zijn van (fundamenteel-)semi-klassieke Newtoniaanse zwaartekracht en electrodynamica, met zekere conceptuele en technische voordelen ten opzichte van reeds bestaande semi-klassieke theorie\"{e}n die ofwel gebaseerd zijn op orthodoxe kwantummechanica ofwel op alternatieve formuleringen van kwantummechanica die het meetprobleem omzeilen; en (iv) te laten zien dat theorie\"{e}n die gebaseerd zijn op ZSM klassieke Newtoniaanse zwaartekracht kunnen reproduceren als er wordt voldaan aan bepaalde (fysisch acceptabele) voorwaarden. De motivatie voor het nastreven van deze doelen (naast de intrinsieke waarde die het voor mij heeft) is dat semi-klassieke Newtoniaanse zwaartekracht vandaag de dag een populair onderwerp is binnen de natuurkundige gemeenschap, waarbij er veel aandacht is voor de Schr\"{o}dinger-Newton vergelijkingen en modellen van fundamenteel-semi-klassieke zwaartekracht gebaseerd op dynamisch verval. Er is ook de brandende vraag of het mogelijk is om een consistente en empirisch vruchtbare versie van fundamenteel-semi-klassieke zwaartekracht te ontwikkellen (en, daarmee, of het wel nodig is om zwaartekracht \"{u}berhaupt te kwantiseren). Door te laten zien dat stochastische mechanica een orginele en nuttige bijdrage kan leveren aan deze debatten, bovenop het aangeven dat Wallstroms kritiek niet langer een (op het oog) definitieve tegenwerping is, zou (hopelijk) een algemene interesse in stochastische mechanica moeten opwekken bij zowel natuurkundigen als filosofen. Deze onderwerpen worden aangesneden in hoofdstukken 4 en 5, wederom elk voorafgegaan door wetenschappelijke samenvattingen. 

De these sluit af met een volledige samenvatting en een sectie gewijd aan verder onderzoek, en (naast deze Nederlandse samenvatting) een kort curriculum vitae.

\chapter{About the author}

Maaneli (Max) Derakhshani was born in Rancho Palos Verdes, California, on October 3rd, 1985. He was raised in Westchester County, New York, and lived there until the age of 18, graduating from Eastchester High School. Between ages 5 and 6, Maaneli underwent 5 neurosurgeries to correct two neurological conditions, Arnold Chiari Malformation and Hydrocephalus. This experience, and the after-effects of it, forever changed the course of his life. At the age of 9, Maaneli discovered physics, fell in love with the subject, and committed himself to someday becoming a physicist. Along the way, in his high school years, Maaneli discovered philosophy as his second love and became committed to synthesizing physics and philosophy in his university studies. However, these were no easy tasks, as he first had to pull himself out of the special education programs he was placed in after his surgeries. Although Maaneli faced resistance from some of his teachers in reaching his goal, eventually he succeeded; and, despite a setback at the age of 17 (when he was forced to undergo two more corrective neurosurgeries), he went on to complete an Intel Science Talent Search project on stochastic electrodynamics (with Prof. Daniel Cole at Boston University) before studying physics and philosophy at Stony Brook University. At Stony Brook, Maaneli worked on a sonoluminescence experiment in the Laser Teaching Center of John No\'{e}, and on experimental quantum optics in the AMO group of Harold Metcalf. Along the way, he studied the foundations of quantum mechanics with ever-greater depth and interest, often traveling to Rutgers University to be tutored by Shelly Goldstein and Rodi Tumulka, and attending conferences such as the New Directions series in Maryland. Maaneli also initiated and helped design a course on the philosophy of quantum mechanics that was jointly taught by a physicist (Alfred Goldhaber) and a philosopher of science (Robert Crease); since then, the course has become officially part of the physics and philosophy programs, and a book has since been published based on it. Maaneli graduated from Stony Brook with the John S. Toll award for the ``most outstanding physics senior", along with other accolades. After taking a couple years off from academia to do physics research and have other life experiences, Maaneli enrolled in the physics graduate program of Clemson University. He earned his M.S. degree with a thesis in theoretical foundations of physics (under Prof. Dieter Hartmann), and then enrolled in the physics Ph.D program of the University of Nebraska at Lincoln (under the supervision of Prof. Herman Batelaan). After a two year stint, Maaneli switched to Utrecht University to complete his Ph.D in the theoretical foundations of physics, under the supervision of Dr. Guido Bacciagaluppi, Prof. Robb Mann, and Prof. Bert Theunissen.

\bibliographystyle{unsrt}
\bibliography{PhDthesisRefscopy}
 
\end{document}